\documentclass[british,traditabstract,longauth]{aa}

\pdfoutput=1    
\usepackage{etex}\reserveinserts{28}
\usepackage[nonamebreak,authoryear]{natbib} 
\bibpunct{(}{)}{;}{a}{}{,}
\usepackage{fixltx2e}
\usepackage[utf8]{inputenc}
\usepackage[T1]{fontenc}
\usepackage[usenames,dvipsnames]{color}
\usepackage{babel}
\usepackage{url}
\usepackage{amsmath}
\usepackage{amssymb}
\usepackage[mathscr]{eucal}
\usepackage{array}
\usepackage{graphicx} \graphicspath{{./figures/}{.figures/cspec/} } 
\usepackage{morefloats}
\usepackage{xspace}
\usepackage{txfonts}
\usepackage{multirow}
\usepackage{ifthen}
\usepackage[breaklinks,colorlinks,citecolor=blue,backref=section]{hyperref}
\usepackage{bbold}
\usepackage{soul} 
\makeatletter
\pdfoutput=1

\newcommand{\mksym}[1]{\ifmmode {\rm #1}\else #1\fi}

\newcommand{\plikTT}{\texttt{Plik}\rm TT}
\newcommand{\plikEE}{\texttt{Plik}\rm EE}
\newcommand{\plikTE}{\texttt{Plik}\rm TE}
\newcommand{\pliklite}{\texttt{Plik\_lite}}

\newcommand{\WP}{{\rm WP}}
\newcommand{\highL}{{\rm highL}}

\newcommand{\TTTEEE}{\mksym{{\rm TT{,}TE{,}EE}}}
\newcommand{\planckTTonly}{{\rm PlanckTT}}

\newcommand{\lowTEB}{\mksym{{\rm lowTEB}}}
\newcommand{\lowEB}{\mksym{{\rm lowP}}}

\newcommand{\planckTT}{{\rm PlanckTT{+}\lowEB}}
\newcommand{\planckall}{{\rm PlanckTT,TE,EE{+}\lowEB}}
\newcommand{\shortTT}{{\plikTT{+}\lowTEB}}
\newcommand{\shortTE}{{\plikTE{+}\lowTEB}}
\newcommand{\shortEE}{{\plikEE{+}\lowTEB}}
\newcommand{\shortall}{{\plikTTTEEE{+}\lowTEB}}

\newcommand{\planckTTTEEE}{{\rm PlanckTT,TE,EE}}

\newcommand{\As}{A_{\rm s}}
\newcommand{\lnAs}{\ln(10^{10} A_{\rm s})}

\newcommand{\ns}{n_{\rm s}}
\newcommand{\lcdm}{$\Lambda$CDM}

\newcommand{\Alens}{A_{\rm L}}

\newcommand{\nrun}{d \ns / d\ln k}
\newcommand{\zre}{z_{\text{re}}}
\newcommand{\yhe}{Y_{\text{P}}}

\newcommand{\nnu}{N_{\rm eff}}
\newcommand{\neff}{N_{\rm eff}}
\newcommand{\mnu}{\sum m_\nu}

\newcommand{\rstar}{r_{\ast}}

\newcommand{\fixme}[1]{{\color{Red}{ FIXME: #1}}}

\providecommand{\text}[1]{\rm{#1}}

\newcommand{\Mpc}{\text{Mpc}}

\providecommand{\muK}{\mu\rm{K}}

\newcommand{\muKsq}{\mu\rm{K}^2}   

\newcommand{\lmax}{l_{\text{max}}}

\providecommand{\Oml}{\Omega_{\Lambda}}

\providecommand{\Omb}{\Omega_{\mathrm{b}}}
\providecommand{\Omc}{\Omega_{\mathrm{c}}}
\providecommand{\Omm}{\Omega_{\mathrm{m}}}
\providecommand{\omb}{\omega_{\mathrm{b}}}
\providecommand{\omc}{\omega_{\mathrm{c}}}

\newcommand{\Omch}{\Omega_{\mathrm{c}}h^2}
\newcommand{\Ombh}{\Omega_{\mathrm{b}}h^2}
\newcommand{\clamp}{\As \exp{(-2\tau)}}

\providecommand{\healpix}{\texttt{HEALPix}}
 
\providecommand{\CAMB}{\texttt{CAMB}}

\providecommand{\PICO}{\texttt{PICO}}

\providecommand{\LCDM}{{$\rm{\Lambda CDM}$}}

\newcommand{\begm}{\begin{pmatrix}}
\newcommand{\enm}{\end{pmatrix}}

\newcommand\ba{\begin{eqnarray}}
\newcommand\ea{\end{eqnarray}}
\newcommand\bea{\begin{eqnarray}}
\newcommand\eea{\end{eqnarray}}

\newcommand\be{\begin{equation}}
\newcommand\ee{\end{equation}}

\def\setsymbol#1#2{\expandafter\def\csname #1\endcsname{#2}}
\def\getsymbol#1{\csname #1\endcsname}

\def\Planck{\textit{Planck}}

\def\HeJT{$^4$He-JT}

\def\alltwentyfifteenresultspapers{\nocite{planck2014-a01, planck2014-a03, planck2014-a04, planck2014-a05, planck2014-a06, planck2014-a07, planck2014-a08, planck2014-a09, planck2014-a11, planck2014-a12, planck2014-a13, planck2014-a14, planck2014-a15, planck2014-a16, planck2014-a17, planck2014-a18, planck2014-a19, planck2014-a20, planck2014-a22, planck2014-a24, planck2014-a26, planck2014-a28, planck2014-a29, planck2014-a30, planck2014-a31, planck2014-a35, planck2014-a36, planck2014-a37, planck2014-ES}}

\newbox\tablebox    \newdimen\tablewidth
\def\leaderfil{\leaders\hbox to 5pt{\hss.\hss}\hfil}
\def\endPlancktable{\tablewidth=\columnwidth 
    $$\hss\copy\tablebox\hss$$
    \vskip-\lastskip\vskip -2pt}
\def\endPlancktablewide{\tablewidth=\textwidth 
    $$\hss\copy\tablebox\hss$$
    \vskip-\lastskip\vskip -2pt}
\def\tablenote#1 #2\par{\begingroup \parindent=0.8em
    \abovedisplayshortskip=0pt\belowdisplayshortskip=0pt
    \noindent
    $$\hss\vbox{\hsize\tablewidth \hangindent=\parindent \hangafter=1 \noindent
    \hbox to \parindent{$^#1$\hss}\strut#2\strut\par}\hss$$
    \endgroup}
\def\doubleline{\vskip 3pt\hrule \vskip 1.5pt \hrule \vskip 5pt}

\def\L2{\ifmmode L_2\else $L_2$\fi}

\def\DeltaT{\ifmmode \Delta T\else $\Delta T$\fi}
\def\deltat{\ifmmode \Delta t\else $\Delta t$\fi}
\def\fknee{\ifmmode f_{\rm knee}\else $f_{\rm knee}$\fi}
\def\Fmax{\ifmmode F_{\rm max}\else $F_{\rm max}$\fi}
\def\solar{\ifmmode{\rm M}_{\mathord\odot}\else${\rm M}_{\mathord\odot}$\fi}
\def\Msolar{\ifmmode{\rm M}_{\mathord\odot}\else${\rm M}_{\mathord\odot}$\fi}
\def\Lsolar{\ifmmode{\rm L}_{\mathord\odot}\else${\rm L}_{\mathord\odot}$\fi}
\def\inv{\ifmmode^{-1}\else$^{-1}$\fi}
\def\mo{\ifmmode^{-1}\else$^{-1}$\fi}
\def\sup#1{\ifmmode ^{\rm #1}\else $^{\rm #1}$\fi}
\def\expo#1{\ifmmode \times 10^{#1}\else $\times 10^{#1}$\fi}
\def\,{\thinspace}
\def\lsim{\mathrel{\raise .4ex\hbox{\rlap{$<$}\lower 1.2ex\hbox{$\sim$}}}}
\def\gsim{\mathrel{\raise .4ex\hbox{\rlap{$>$}\lower 1.2ex\hbox{$\sim$}}}}

\def\simprop{\mathrel{\raise .4ex\hbox{\rlap{$\propto$}\lower 1.2ex\hbox{$\sim$}}}}
\def\deg{\ifmmode^\circ\else$^\circ$\fi}
\def\pdeg{\ifmmode $\setbox0=\hbox{$^{\circ}$}\rlap{\hskip.11\wd0 .}$^{\circ}
          \else \setbox0=\hbox{$^{\circ}$}\rlap{\hskip.11\wd0 .}$^{\circ}$\fi}
\def\arcs{\ifmmode {^{\scriptstyle\prime\prime}}
          \else $^{\scriptstyle\prime\prime}$\fi}
\def\arcm{\ifmmode {^{\scriptstyle\prime}}
          \else $^{\scriptstyle\prime}$\fi}
\newdimen\sa  \newdimen\sb
\def\parcs{\sa=.07em \sb=.03em
     \ifmmode \hbox{\rlap{.}}^{\scriptstyle\prime\kern -\sb\prime}\hbox{\kern -\sa}
     \else \rlap{.}$^{\scriptstyle\prime\kern -\sb\prime}$\kern -\sa\fi}
\def\parcm{\sa=.08em \sb=.03em
     \ifmmode \hbox{\rlap{.}\kern\sa}^{\scriptstyle\prime}\hbox{\kern-\sb}
     \else \rlap{.}\kern\sa$^{\scriptstyle\prime}$\kern-\sb\fi}
\def\ra[#1 #2 #3.#4]{#1\sup{h}#2\sup{m}#3\sup{s}\llap.#4}
\def\dec[#1 #2 #3.#4]{#1\deg#2\arcm#3\arcs\llap.#4}
\def\deco[#1 #2 #3]{#1\deg#2\arcm#3\arcs}
\def\rra[#1 #2]{#1\sup{h}#2\sup{m}}

\def\dots{\relax\ifmmode \ldots\else $\ldots$\fi}
\def\WHzsr{\ifmmode $W\,Hz\mo\,sr\mo$\else W\,Hz\mo\,sr\mo\fi}
\def\mHz{\ifmmode $\,mHz$\else \,mHz\fi}
\def\GHz{\ifmmode $\,GHz$\else \,GHz\fi}
\def\mKs{\ifmmode $\,mK\,s$^{1/2}\else \,mK\,s$^{1/2}$\fi}
\def\muKs{\ifmmode \,\mu$K\,s$^{1/2}\else \,$\mu$K\,s$^{1/2}$\fi}
\def\muKRJs{\ifmmode \,\mu$K$_{\rm RJ}$\,s$^{1/2}\else \,$\mu$K$_{\rm RJ}$\,s$^{1/2}$\fi}
\def\muKHz{\ifmmode \,\mu$K\,Hz$^{-1/2}\else \,$\mu$K\,Hz$^{-1/2}$\fi}
\def\MJysr{\ifmmode \,$MJy\,sr\mo$\else \,MJy\,sr\mo\fi}
\def\MJysrmK{\ifmmode \,$MJy\,sr\mo$\,mK$_{\rm CMB}\mo\else \,MJy\,sr\mo\,mK$_{\rm CMB}\mo$\fi}
\def\microns{\ifmmode \,\mu$m$\else \,$\mu$m\fi}

\def\muK{\ifmmode \,\mu$K$\else \,$\mu$\hbox{K}\fi}
\def\microK{\ifmmode \,\mu$K$\else \,$\mu$\hbox{K}\fi}
\def\muW{\ifmmode \,\mu$W$\else \,$\mu$\hbox{W}\fi}
\def\kms{\ifmmode $\,km\,s$^{-1}\else \,km\,s$^{-1}$\fi}
\def\kmsMpc{\ifmmode $\,\kms\,Mpc\mo$\else \,\kms\,Mpc\mo\fi}
\providecommand{\sorthelp}[1]{}

\newcommand{\rev}[1]{{#1}}  	

\newcommand{\quickbeam}{\texttt{QuickBeam}\xspace}
\newcommand{\febecop}{\texttt{FEBeCoP}\xspace}

\newcommand{\commander}{\texttt{Commander}\xspace}

\newcommand{\camspec}{\texttt{CamSpec}\xspace} 
\newcommand{\plik}{\texttt{Plik}\xspace} 
\newcommand{\mspec}{\texttt{Mspec}\xspace}
\newcommand{\hil}{\texttt{Hillipop}\xspace}

\newcommand{\plikTTtau}{\texttt{Plik}\rm TT+tauprior}
\newcommand{\plikEEtau}{\texttt{Plik}\rm EE+tauprior}
\newcommand{\plikTEtau}{\texttt{Plik}\rm TE+tauprior}

\newcommand{\plikTTEETEtau}{{\texttt{Plik}\rm TT,TE,EE+tauprior}}
\newcommand{\plikTTEETE}{{\texttt{Plik}\rm TT,TE,EE}}

\newcommand{\plikTTTEEE}{\texttt{Plik}\rm TT,EE,TE}

\newcommand{\plikALL}{{\texttt{Plik}\rm TT,TE,EE+\lowTEB}}

\newcommand{\polspice}{\texttt{PolSpice}\xspace} 
\newcommand{\xfaster}{\texttt{Xfaster}\xspace} 
\newcommand{\smica}{\texttt{SMICA}\xspace}

\newcommand{\ie}{{i.e., \xspace}}
\newcommand{\eg}{{e.g.,  \xspace}} 

\newcommand{\degree}{\ensuremath{^\circ}}

\newcommand{\planck}{\textit{Planck}\xspace} 
\newcommand{\GM}{B} 

\newcommand{\HFI}{{HFI}\xspace} 
\newcommand{\LFI}{{LFI}\xspace} 

\newcommand{\WMAP}{{WMAP}\xspace} 
\newcommand{\COBE}{{COBE}\xspace} 
\newcommand{\act}{{ACT}\xspace} 
\newcommand{\spt}{{SPT}\xspace} 
 
\newcommand{\fsky}{\ensuremath{f_{\rm sky}}}
\newcommand{\calibM}{\ensuremath{y}}
\newcommand{\calibC}{\ensuremath{c}}

\def\Var{\mathop{\rm Var}}

\def\GHz{\ifmmode $GHz$\else \,GHz\fi}
\def\ghz{\ifmmode $GHz$\else \,GHz\fi}
\def\MHz{\ifmmode $MHz$\else \,MHz\fi}
\def\Hz{\ifmmode $Hz$\else \,Hz\fi}

\def\muK{\ifmmode \,\mu$K$\else \,$\mu$\hbox{K}\fi}
\def\microK{\ifmmode \,\mu$K$\else \,$\mu$\hbox{K}\fi}
\def\muW{\ifmmode \,\mu$W$\else \,$\mu$\hbox{W}\fi}
\def\kms{\ifmmode $\,km\,s$^{-1}\else \,km\,s$^{-1}$\fi}
\def\kmsmpc{\ifmmode $\,km\,s$^{-1}\,$Mpc$^{-1}\else \,km\,s$^{-1}$Mpc$^{-1}$\fi}
 
\newcommand{\ellp}{\ell^{\prime}}
\newcommand{\none}{\emptyset \emptyset}

\newcommand{\hiTTij}{^{TT \, i, j}}
\newcommand{\hiTTpq}{^{TT \, p, q}}
\newcommand{\hiTTip}{^{TT \, i, p}}
\newcommand{\hiTTiq}{^{TT \, i, q}}
\newcommand{\hiTTjp}{^{TT \, j, p}}
\newcommand{\hiTTjq}{^{TT \, j, q}}
\newcommand{\hiEE}{^{EE}}
\newcommand{\hiEEij}{^{EE \, i, j}}
\newcommand{\hiEEpq}{^{EE \, p, q}}
\newcommand{\hiEEip}{^{EE \, i, p}}
\newcommand{\hiEEiq}{^{EE \, i, q}}
\newcommand{\hiEEjp}{^{EE \, j, p}}
\newcommand{\hiEEjq}{^{EE \, j, q}}
\newcommand{\hiTE}{^{TE}}
\newcommand{\hiTEij}{^{TE \, i, j}}
\newcommand{\hiTEpq}{^{TE \, p, q}}
\newcommand{\hiTEip}{^{TE \, i, p}}
\newcommand{\hiTEiq}{^{TE \, i, q}}
\newcommand{\hiTEjp}{^{TE \, j, p}}
\newcommand{\hiTEjq}{^{TE \, j, q}}
\newcommand{\hiET}{^{ET}}

\newcommand{\hiTP}{^{TP}} 
\newcommand{\hiPT}{^{PT}} 
\newcommand{\hiPP}{^{PP}}
\newcommand{\hiTT}{^{TT}}
\newcommand{\singleT}{T} 
\newcommand{\singleP}{P}
\newcommand{\TT}{{TT}} 
\newcommand{\TE}{{TE}} 
\newcommand{\ET}{{ET}} 
\newcommand{\EE}{{EE}} 
\newcommand{\BB}{{BB}}
\newcommand{\PP}{PP} 
\newcommand{\TP}{TP} 
\newcommand{\PT}{PT}
\newcommand{\II}{II} 
\newcommand{\QQ}{QQ} 
\newcommand{\UU}{UU}

\newcommand{\wigner}[6]{\ensuremath{\begin{pmatrix} #1 & #2 & #3 \\
      #4 & #5 & #6 \end{pmatrix}}}

\defcitealias{planck2013-p08}{Like13}

\newbox\tablebox    \newdimen\tablewidth
\def\leaderfil{\leaders\hbox to 5pt{\hss.\hss}\hfil}
\def\endPlancktable{\tablewidth=\columnwidth 
    $$\hss\copy\tablebox\hss$$
    \vskip-\lastskip\vskip -2pt}
\def\endPlancktablewide{\tablewidth=\textwidth 
    $$\hss\copy\tablebox\hss$$
    \vskip-\lastskip\vskip -2pt}
\def\tablenote#1 #2\par{\begingroup \parindent=0.8em
    \abovedisplayshortskip=0pt\belowdisplayshortskip=0pt
    \noindent
    $$\hss\vbox{\hsize\tablewidth \hangindent=\parindent \hangafter=1 \noindent
    \hbox to \parindent{$^#1$\hss}\strut#2\strut\par}\hss$$
    \endgroup}
\def\doubleline{\vskip 3pt\hrule \vskip 1.5pt \hrule \vskip 5pt}

\makeatother

\graphicspath{{figures/}{figures-arxiv/}}

\begin{document}
\global\long\def\pmb#1{\setbox0=\hbox{#1}
}
 \global\long\def\ltsima{$\frac{\;\buildrel<}{\sim\;}$}
 \global\long\def\gtsima{$\frac{\;\buildrel>}{\sim\;}$}
 \global\long\def\simlt{\lower.5ex\hbox{\ltsima}}
 \global\long\def\simgt{\lower.5ex\hbox{\gtsima}}
 \global\long\def\muk{$(\mu{\rm K})^{2}$ }
 \global\long\def\etal{{\it et al.}}
 \global\long\def\etals{{\it et al. }}
 \def\p 2Y{\;_2Y} \def\m 2Y{\;_{-2}Y} \global\long\def\beglet{ \addtocounter{equation}{1}
}
 \global\long\def\endlet{ \setcounter{equation}{\value{parentequation}}}

\global\long\def\adj{^{\dagger}}
\global\long\def\inv{^{-1} }

\global\long\def\nmc{\ensuremath{n_{{\rm MC}}}}

 \global\long\def\nmodes{\ensuremath{n_{{\rm modes}}}}
 \global\long\def\Bmean{\ensuremath{B_{{\rm mean}}}}
 \global\long\def\Wmean{\ensuremath{W_{{\rm mean}}}}
 \global\long\def\lmax{\ensuremath{\ell_{{\rm max}}}}
 \global\long\def\lmin{\ensuremath{\ell_{{\rm min}}}}
 \global\long\def\lhyb{\ensuremath{\ell_{{\rm hyb}}}}
 \global\long\def\fixme#1{{\bf {\it #1}}}
\global\long\def\FRB#1{{\bf \textcolor{red}{\textbf{\ensuremath{\triangleleft}#1 \ensuremath{\triangleright}}}}}

\global\long\def\matthreethree#1#2#3#4#5#6#7#8#9{\left( \begin{array}{ccc}
\!#1\!  &  \!#2\!  &  \!#3\!\!\\
\!#4\!  &  \!#5\!  &  \!#6\!\!\\
\!#7\!  &  \!#8\!  &  \!#9\!\!%
\end{array}%
\right)}

\selectlanguage{british}%
\titlerunning{CMB power spectra,  likelihoods, and parameters} \authorrunning{\Planck\ collaboration}

\author{\small
Planck Collaboration: N.~Aghanim\inst{63}
\and
M.~Arnaud\inst{78}
\and
M.~Ashdown\inst{74, 6}
\and
J.~Aumont\inst{63}
\and
C.~Baccigalupi\inst{91}
\and
A.~J.~Banday\inst{103, 10}
\and
R.~B.~Barreiro\inst{69}
\and
J.~G.~Bartlett\inst{1, 71}
\and
N.~Bartolo\inst{32, 70}
\and
E.~Battaner\inst{105, 106}
% ...
}
\institute{\small
APC, AstroParticule et Cosmologie, Universit\'{e} Paris Diderot, CNRS/IN2P3, CEA/lrfu, Observatoire de Paris, Sorbonne Paris Cit\'{e}, 10, rue Alice Domon et L\'{e}onie Duquet, 75205 Paris Cedex 13, France\goodbreak
\and
Aalto University Mets\"{a}hovi Radio Observatory and Dept of Radio Science and Engineering, P.O. Box 13000, FI-00076 AALTO, Finland\goodbreak
\and
African Institute for Mathematical Sciences, 6-8 Melrose Road, Muizenberg, Cape Town, South Africa\goodbreak
\and
Agenzia Spaziale Italiana Science Data Center, Via del Politecnico snc, 00133, Roma, Italy\goodbreak
\and
Aix Marseille Universit\'{e}, CNRS, LAM (Laboratoire d'Astrophysique de Marseille) UMR 7326, 13388, Marseille, France\goodbreak
\and
Astrophysics Group, Cavendish Laboratory, University of Cambridge, J J Thomson Avenue, Cambridge CB3 0HE, U.K.\goodbreak
\and
Astrophysics \& Cosmology Research Unit, School of Mathematics, Statistics \& Computer Science, University of KwaZulu-Natal, Westville Campus, Private Bag X54001, Durban 4000, South Africa\goodbreak
\and
CGEE, SCS Qd 9, Lote C, Torre C, 4$^{\circ}$ andar, Ed. Parque Cidade Corporate, CEP 70308-200, Bras\'{i}lia, DF, Brazil\goodbreak
\and
CITA, University of Toronto, 60 St. George St., Toronto, ON M5S 3H8, Canada\goodbreak
\and
CNRS, IRAP, 9 Av. colonel Roche, BP 44346, F-31028 Toulouse cedex 4, France\goodbreak
\and
CRANN, Trinity College, Dublin, Ireland\goodbreak
\and
California Institute of Technology, Pasadena, California, U.S.A.\goodbreak
\and
Centre for Theoretical Cosmology, DAMTP, University of Cambridge, Wilberforce Road, Cambridge CB3 0WA, U.K.\goodbreak
\and
Centro de Estudios de F\'{i}sica del Cosmos de Arag\'{o}n (CEFCA), Plaza San Juan, 1, planta 2, E-44001, Teruel, Spain\goodbreak
\and
Computational Cosmology Center, Lawrence Berkeley National Laboratory, Berkeley, California, U.S.A.\goodbreak
\and
DSM/Irfu/SPP, CEA-Saclay, F-91191 Gif-sur-Yvette Cedex, France\goodbreak
\and
DTU Space, National Space Institute, Technical University of Denmark, Elektrovej 327, DK-2800 Kgs. Lyngby, Denmark\goodbreak
\and
D\'{e}partement de Physique Th\'{e}orique, Universit\'{e} de Gen\`{e}ve, 24, Quai E. Ansermet,1211 Gen\`{e}ve 4, Switzerland\goodbreak
\and
Departamento de Astrof\'{i}sica, Universidad de La Laguna (ULL), E-38206 La Laguna, Tenerife, Spain\goodbreak
\and
Departamento de F\'{\i}sica, Universidad de Oviedo, Avda. Calvo Sotelo s/n, Oviedo, Spain\goodbreak
\and
Department of Astrophysics/IMAPP, Radboud University Nijmegen, P.O. Box 9010, 6500 GL Nijmegen, The Netherlands\goodbreak
\and
Department of Physics \& Astronomy, University of British Columbia, 6224 Agricultural Road, Vancouver, British Columbia, Canada\goodbreak
\and
Department of Physics and Astronomy, Dana and David Dornsife College of Letter, Arts and Sciences, University of Southern California, Los Angeles, CA 90089, U.S.A.\goodbreak
\and
Department of Physics and Astronomy, University College London, London WC1E 6BT, U.K.\goodbreak
\and
Department of Physics and Astronomy, University of Sussex, Brighton BN1 9QH, U.K.\goodbreak
\and
Department of Physics, Florida State University, Keen Physics Building, 77 Chieftan Way, Tallahassee, Florida, U.S.A.\goodbreak
\and
Department of Physics, Gustaf H\"{a}llstr\"{o}min katu 2a, University of Helsinki, Helsinki, Finland\goodbreak
\and
Department of Physics, Princeton University, Princeton, New Jersey, U.S.A.\goodbreak
\and
Department of Physics, University of California, One Shields Avenue, Davis, California, U.S.A.\goodbreak
\and
Department of Physics, University of California, Santa Barbara, California, U.S.A.\goodbreak
\and
Department of Physics, University of Illinois at Urbana-Champaign, 1110 West Green Street, Urbana, Illinois, U.S.A.\goodbreak
\and
Dipartimento di Fisica e Astronomia G. Galilei, Universit\`{a} degli Studi di Padova, via Marzolo 8, 35131 Padova, Italy\goodbreak
\and
Dipartimento di Fisica e Scienze della Terra, Universit\`{a} di Ferrara, Via Saragat 1, 44122 Ferrara, Italy\goodbreak
\and
Dipartimento di Fisica, Universit\`{a} La Sapienza, P. le A. Moro 2, Roma, Italy\goodbreak
\and
Dipartimento di Fisica, Universit\`{a} degli Studi di Milano, Via Celoria, 16, Milano, Italy\goodbreak
\and
Dipartimento di Fisica, Universit\`{a} degli Studi di Trieste, via A. Valerio 2, Trieste, Italy\goodbreak
\and
Dipartimento di Matematica, Universit\`{a} di Roma Tor Vergata, Via della Ricerca Scientifica, 1, Roma, Italy\goodbreak
\and
Discovery Center, Niels Bohr Institute, Blegdamsvej 17, Copenhagen, Denmark\goodbreak
\and
Discovery Center, Niels Bohr Institute, Copenhagen University, Blegdamsvej 17, Copenhagen, Denmark\goodbreak
\and
European Space Agency, ESAC, Planck Science Office, Camino bajo del Castillo, s/n, Urbanizaci\'{o}n Villafranca del Castillo, Villanueva de la Ca\~{n}ada, Madrid, Spain\goodbreak
\and
European Space Agency, ESTEC, Keplerlaan 1, 2201 AZ Noordwijk, The Netherlands\goodbreak
\and
Gran Sasso Science Institute, INFN, viale F. Crispi 7, 67100 L'Aquila, Italy\goodbreak
\and
HGSFP and University of Heidelberg, Theoretical Physics Department, Philosophenweg 16, 69120, Heidelberg, Germany\goodbreak
\and
Haverford College Astronomy Department, 370 Lancaster Avenue, Haverford, Pennsylvania, U.S.A.\goodbreak
\and
Helsinki Institute of Physics, Gustaf H\"{a}llstr\"{o}min katu 2, University of Helsinki, Helsinki, Finland\goodbreak
\and
INAF - Osservatorio Astrofisico di Catania, Via S. Sofia 78, Catania, Italy\goodbreak
\and
INAF - Osservatorio Astronomico di Padova, Vicolo dell'Osservatorio 5, Padova, Italy\goodbreak
\and
INAF - Osservatorio Astronomico di Roma, via di Frascati 33, Monte Porzio Catone, Italy\goodbreak
\and
INAF - Osservatorio Astronomico di Trieste, Via G.B. Tiepolo 11, Trieste, Italy\goodbreak
\and
INAF/IASF Bologna, Via Gobetti 101, Bologna, Italy\goodbreak
\and
INAF/IASF Milano, Via E. Bassini 15, Milano, Italy\goodbreak
\and
INFN, Sezione di Bologna, viale Berti Pichat 6/2, 40127 Bologna, Italy\goodbreak
\and
INFN, Sezione di Ferrara, Via Saragat 1, 44122 Ferrara, Italy\goodbreak
\and
INFN, Sezione di Roma 1, Universit\`{a} di Roma Sapienza, Piazzale Aldo Moro 2, 00185, Roma, Italy\goodbreak
\and
INFN, Sezione di Roma 2, Universit\`{a} di Roma Tor Vergata, Via della Ricerca Scientifica, 1, Roma, Italy\goodbreak
\and
INFN/National Institute for Nuclear Physics, Via Valerio 2, I-34127 Trieste, Italy\goodbreak
\and
IPAG: Institut de Plan\'{e}tologie et d'Astrophysique de Grenoble, Universit\'{e} Grenoble Alpes, IPAG, F-38000 Grenoble, France, CNRS, IPAG, F-38000 Grenoble, France\goodbreak
\and
IUCAA, Post Bag 4, Ganeshkhind, Pune University Campus, Pune 411 007, India\goodbreak
\and
Imperial College London, Astrophysics group, Blackett Laboratory, Prince Consort Road, London, SW7 2AZ, U.K.\goodbreak
\and
Infrared Processing and Analysis Center, California Institute of Technology, Pasadena, CA 91125, U.S.A.\goodbreak
\and
Institut N\'{e}el, CNRS, Universit\'{e} Joseph Fourier Grenoble I, 25 rue des Martyrs, Grenoble, France\goodbreak
\and
Institut Universitaire de France, 103, bd Saint-Michel, 75005, Paris, France\goodbreak
\and
Institut d'Astrophysique Spatiale, CNRS, Univ. Paris-Sud, Universit\'{e} Paris-Saclay, B\^{a}t. 121, 91405 Orsay cedex, France\goodbreak
\and
Institut d'Astrophysique de Paris, CNRS (UMR7095), 98 bis Boulevard Arago, F-75014, Paris, France\goodbreak
\and
Institut f\"ur Theoretische Teilchenphysik und Kosmologie, RWTH Aachen University, D-52056 Aachen, Germany\goodbreak
\and
Institute of Astronomy, University of Cambridge, Madingley Road, Cambridge CB3 0HA, U.K.\goodbreak
\and
Institute of Theoretical Astrophysics, University of Oslo, Blindern, Oslo, Norway\goodbreak
\and
Instituto de Astrof\'{\i}sica de Canarias, C/V\'{\i}a L\'{a}ctea s/n, La Laguna, Tenerife, Spain\goodbreak
\and
Instituto de F\'{\i}sica de Cantabria (CSIC-Universidad de Cantabria), Avda. de los Castros s/n, Santander, Spain\goodbreak
\and
Istituto Nazionale di Fisica Nucleare, Sezione di Padova, via Marzolo 8, I-35131 Padova, Italy\goodbreak
\and
Jet Propulsion Laboratory, California Institute of Technology, 4800 Oak Grove Drive, Pasadena, California, U.S.A.\goodbreak
\and
Jodrell Bank Centre for Astrophysics, Alan Turing Building, School of Physics and Astronomy, The University of Manchester, Oxford Road, Manchester, M13 9PL, U.K.\goodbreak
\and
Kavli Institute for Cosmological Physics, University of Chicago, Chicago, IL 60637, USA\goodbreak
\and
Kavli Institute for Cosmology Cambridge, Madingley Road, Cambridge, CB3 0HA, U.K.\goodbreak
\and
Kazan Federal University, 18 Kremlyovskaya St., Kazan, 420008, Russia\goodbreak
\and
LAL, Universit\'{e} Paris-Sud, CNRS/IN2P3, Orsay, France\goodbreak
\and
LERMA, CNRS, Observatoire de Paris, 61 Avenue de l'Observatoire, Paris, France\goodbreak
\and
Laboratoire AIM, IRFU/Service d'Astrophysique - CEA/DSM - CNRS - Universit\'{e} Paris Diderot, B\^{a}t. 709, CEA-Saclay, F-91191 Gif-sur-Yvette Cedex, France\goodbreak
\and
Laboratoire Traitement et Communication de l'Information, CNRS (UMR 5141) and T\'{e}l\'{e}com ParisTech, 46 rue Barrault F-75634 Paris Cedex 13, France\goodbreak
\and
Laboratoire de Physique Subatomique et Cosmologie, Universit\'{e} Grenoble-Alpes, CNRS/IN2P3, 53, rue des Martyrs, 38026 Grenoble Cedex, France\goodbreak
\and
Laboratoire de Physique Th\'{e}orique, Universit\'{e} Paris-Sud 11 \& CNRS, B\^{a}timent 210, 91405 Orsay, France\goodbreak
\and
Lebedev Physical Institute of the Russian Academy of Sciences, Astro Space Centre, 84/32 Profsoyuznaya st., Moscow, GSP-7, 117997, Russia\goodbreak
\and
Leung Center for Cosmology and Particle Astrophysics, National Taiwan University, Taipei 10617, Taiwan\goodbreak
\and
Max-Planck-Institut f\"{u}r Astrophysik, Karl-Schwarzschild-Str. 1, 85741 Garching, Germany\goodbreak
\and
National University of Ireland, Department of Experimental Physics, Maynooth, Co. Kildare, Ireland\goodbreak
\and
Nicolaus Copernicus Astronomical Center, Bartycka 18, 00-716 Warsaw, Poland\goodbreak
\and
Niels Bohr Institute, Blegdamsvej 17, Copenhagen, Denmark\goodbreak
\and
Niels Bohr Institute, Copenhagen University, Blegdamsvej 17, Copenhagen, Denmark\goodbreak
\and
Nordita (Nordic Institute for Theoretical Physics), Roslagstullsbacken 23, SE-106 91 Stockholm, Sweden\goodbreak
\and
Optical Science Laboratory, University College London, Gower Street, London, U.K.\goodbreak
\and
SISSA, Astrophysics Sector, via Bonomea 265, 34136, Trieste, Italy\goodbreak
\and
SMARTEST Research Centre, Universit\`{a} degli Studi e-Campus, Via Isimbardi 10, Novedrate (CO), 22060, Italy\goodbreak
\and
School of Physics and Astronomy, Cardiff University, Queens Buildings, The Parade, Cardiff, CF24 3AA, U.K.\goodbreak
\and
Sorbonne Universit\'{e}-UPMC, UMR7095, Institut d'Astrophysique de Paris, 98 bis Boulevard Arago, F-75014, Paris, France\goodbreak
\and
Space Research Institute (IKI), Russian Academy of Sciences, Profsoyuznaya Str, 84/32, Moscow, 117997, Russia\goodbreak
\and
Space Sciences Laboratory, University of California, Berkeley, California, U.S.A.\goodbreak
\and
Special Astrophysical Observatory, Russian Academy of Sciences, Nizhnij Arkhyz, Zelenchukskiy region, Karachai-Cherkessian Republic, 369167, Russia\goodbreak
\and
Sub-Department of Astrophysics, University of Oxford, Keble Road, Oxford OX1 3RH, U.K.\goodbreak
\and
Sydney Institute for Astronomy, School of Physics A28, University of Sydney, NSW 2006, Australia\goodbreak
\and
The Oskar Klein Centre for Cosmoparticle Physics, Department of Physics,Stockholm University, AlbaNova, SE-106 91 Stockholm, Sweden\goodbreak
\and
Theory Division, PH-TH, CERN, CH-1211, Geneva 23, Switzerland\goodbreak
\and
UPMC Univ Paris 06, UMR7095, 98 bis Boulevard Arago, F-75014, Paris, France\goodbreak
\and
Universit\'{e} de Toulouse, UPS-OMP, IRAP, F-31028 Toulouse cedex 4, France\goodbreak
\and
University Observatory, Ludwig Maximilian University of Munich, Scheinerstrasse 1, 81679 Munich, Germany\goodbreak
\and
University of Granada, Departamento de F\'{\i}sica Te\'{o}rica y del Cosmos, Facultad de Ciencias, Granada, Spain\goodbreak
\and
University of Granada, Instituto Carlos I de F\'{\i}sica Te\'{o}rica y Computacional, Granada, Spain\goodbreak
% ...
}


\title{\Planck~2015 results. XI. CMB power spectra, likelihoods, and robustness of parameters}


\abstract{
This paper presents the \Planck\ 2015 likelihoods, statistical descriptions of the 2-point correlation functions of the cosmic microwave background (CMB) temperature and polarization fluctuations that account for relevant uncertainties, both instrumental and astrophysical in nature. 
They are based on the same hybrid approach used for the previous release, \ie  a pixel-based likelihood at low multipoles 
\rev{($\ell < 30$)} 
and a Gaussian approximation to the distribution of cross-power spectra at higher multipoles. The main improvements are the use of more and better processed data and of \Planck\ polarization information, along with more detailed models of foregrounds and instrumental uncertainties. The increased redundancy brought by more than doubling the amount of data analysed enables further consistency checks and enhanced immunity to systematic effects. It also improves the constraining power of \Planck, in particular with regard to small-scale foreground properties. Progress in the modelling of foreground emission enables the retention of a larger fraction of the sky to determine the properties of the CMB, which also contributes to the enhanced precision of the spectra. Improvements in data processing and instrumental modelling further reduce uncertainties.  
Extensive tests establish the robustness and accuracy of the likelihood results, from temperature alone, from polarization alone, and from their combination. 

\rev{For temperature, we also perform a full likelihood analysis of realistic end-to-end simulations of the instrumental response to the sky, which were fed into the actual data processing pipeline; this does not reveal biases from residual low-level instrumental systematics.}
\rev{Even with the increase in precision and robustness, the \LCDM\ cosmological model continues to offer a very good fit to the \Planck\ data. The slope of the primordial scalar fluctuations, $\ns$, is confirmed smaller than unity at more than 5\,$\sigma$ from \Planck\ alone.}
We further validate \rev{the robustness of the likelihood results} against specific extensions to the baseline cosmology, which are particularly sensitive to data at high multipoles. \rev{For instance, the effective number of neutrino species remains compatible with the canonical value of 3.046.}

For this first detailed analysis of \Planck\ polarization spectra, we concentrate at high multipoles on the $E$ modes, leaving the analysis of the weaker $B$ modes to future work. At low multipoles we use temperature maps at all \Planck\ frequencies along with a subset of
polarization data. These data take advantage of \Planck's wide frequency coverage to improve the separation of CMB and foreground emission. Within the baseline \LCDM\ cosmology this requires $\tau=0.078\pm0.019$ for the reionization optical depth, which is significantly lower than estimates without the use of high-frequency data for explicit monitoring of dust emission.
At high multipoles we detect residual systematic errors in $E$
polarization, typically at the $\muKsq$ level; we therefore choose to retain temperature information alone for high multipoles as the recommended baseline, in particular for testing non-minimal models. Nevertheless, the high-multipole polarization spectra from \Planck\ are already good enough to enable a separate high-\rev{precision} determination of the parameters of the \LCDM\ model, showing consistency with those established independently from temperature information alone. 
}

\keywords{cosmic background radiation -- cosmology: observations -- cosmological parameters -- methods: data analysis
}


\maketitle


\newpage

\section{Introduction}

This paper presents the angular power spectra of the cosmic microwave background (CMB) and the related likelihood functions, calculated from \Planck\footnote{\Planck\ 
(\url{http://www.esa.int/Planck}) is a project of the European Space Agency (ESA) with instruments provided by two scientific consortia funded by ESA member states and led by Principal Investigators from France and Italy, telescope reflectors provided through a collaboration between ESA and a scientific consortium led and funded by Denmark, and 
additional contributions from NASA (USA).}
2015 data, which consists of intensity maps from the full mission, along with a subset of the polarization data. 

The CMB power spectra contain all of the information available if the CMB is statistically isotropic and distributed as a multivariate Gaussian. For realistic data, these must be augmented with models of instrumental noise, of other instrumental systematic effects, and of contamination from astrophysical foregrounds. 

The power spectra are, in turn, uniquely determined by the underlying cosmological model and its parameters. In temperature, the power spectrum has been measured over large fractions of the sky by the Cosmic Background Explorer \citep[\COBE;][]{Wright1996DMR} and the Wilkinson Microwave Anistropy Probe \citep[\WMAP;][]{bennett2012}, and in smaller regions by a host of balloon- and ground-based telescopes \citep[\eg][]{Netterfield97,hanany00,GraingeVSA2002,Pearson03,tristram/etal:2005,Jones06,reichardt/etal:2009,fowler/etal:2010,das/etal:2011,2011ApJ...743...28K,story/etal:prep,das/etal:prep}. The \Planck\ 2013 power spectrum and likelihood were discussed in \citet[][hereafter \citetalias{planck2013-p08}]{planck2013-p08}.

The distribution of temperature and polarization on the sky is further affected by gravitational lensing by the inhomogeneous mass distribution along the line of sight between the last scattering surface and the observer. This introduces correlations between large and small scales, which can be estimated by computing the expected contribution of lensing to the 4-point function (\ie the trispectrum). This can in turn be used to determine the power spectrum of the lensing potential, as is done in \citet{planck2014-a17} for this \planck\ release, and to further constrain the cosmological parameters via a separate likelihood function \citep{planck2014-a15}.

Over the last decade, CMB intensity (temperature) has been augmented by linear polarization data \citep[\eg][]{DASI_pol02,kogut2003,Sievers07,dunkley:2009,Pryke09,Quiet12,Polarbear14a}.   
Because linear polarization is given by both an amplitude and direction, it can, in turn, be decomposed into two coordinate-independent quantities, each with a different dependence on the cosmology \citep[\eg][]{Seljak1996,KamionkowskiKS1997,ZaldarriagaS1997}. 
One, the so-called $E$ mode, is determined by much the same physics as the intensity, and therefore enables an independent measurement of the background cosmology, as well as a determination of some new parameters (\eg the reionization optical depth). The other polarization observable, the $B$ mode, is only sourced at early times by gravitational radiation, as produced, for example, during an inflationary epoch. The $E$ and $B$ components are also conventionally taken to be isotropic Gaussian random fields, with only $E$ expected to be correlated with intensity. Thus we expect to be able to measure four independent power spectra, namely the three auto-spectra $C_\ell^{TT}$, $C_\ell^{EE}$, and $C_\ell^{BB}$, along with the cross-spectrum $C_\ell^{TE}$.

Estimating these spectra from the likelihood requires cleaned and calibrated maps for all \Planck\ detectors, along with a quantitative description of their noise properties.
The required data processing is discussed in \citet{planck2014-a03}, \citet{planck2014-a04}, \citet{planck2014-a05}, \citet{planck2014-a06}, and \citet{planck2014-a07} for the low-frequency instrument (LFI; 30, 44, and 70\,GHz) and \citet{planck2014-a08} and \citet{planck2014-a09} for the high-frequency instrument (HFI; 100, 143, 217, 353, 585, and 857\,GHz). Although the CMB is brightest over 70--217\,GHz, the full range of \Planck\ frequencies is crucial to distinguish between the cosmological component and sources of astrophysical foreground emission, present in even the cleanest regions of sky. We therefore use measurements from those \Planck\ bands dominated by such emission as a template to model the foreground in the bands where the CMB is most significant.

This paper presents the $C_\ell^{TT}$, $C_\ell^{EE}$, and $C_\ell^{TE}$ spectra, likelihood functions, and basic cosmological parameters from the \Planck\ 2015 release. A complete analysis in the context of an extended \LCDM\ cosmology of these and other results from \planck regarding the lensing power spectrum results, as well as constraints from other observations, is given in \citet{planck2014-a15}. Wider extensions to the set of models are discussed in other \Planck\ 2015 papers; for example, \citet{planck2014-a16} examines specific models for the dark energy component \rev{and extensions to general relativity,} and \citet{planck2014-a24} discusses inflationary models. 

This paper shows that the contribution of high-$\ell$ systematic errors to the polarization spectra are at quite a low level (of the order of a few $\muKsq$), therefore enabling an interesting comparison of the polarization-based cosmological results with those derived from $C_\ell^{TT}$ alone. We therefore discuss the results for $C_\ell^{TE}$ and $C_\ell^{EE}$ at high multipoles. However, the technical difficulties involved with polarization measurements and subsequent data analysis, along with the inherently lower signal-to-noise ratio (especially for $B$ modes), thus require a careful understanding of the random noise and instrumental and astrophysical systematic effects. For this reason, at large angular scales (\ie low multipoles $\ell$) the baseline results use only a subset of \Planck\ polarization data.

Because of these different sensitivities to systematic errors at different angular scales, as well as the increasingly Gaussian behaviour of the likelihood function at smaller angular scales, we adopt a hybrid approach to the likelihood calculation \citep{Efstathiou2004,Efstathiou2006}, splitting between a direct calculation of the likelihood on large scales and the use of pseudo-spectral estimates at smaller scales, as we did for the previous release. 

The plan of the paper reflects this hybrid approach along with the importance of internal tests and cross-validation. 
In Sect.~\ref{sec:low-ell}, we present the low-multipole ($\ell<30$) likelihood and its validation. At these large scales, we compute the likelihood function directly in pixel space; the temperature map is obtained by a Gibbs sampling approach in the context of a parameterized foreground model, while the polarized maps are cleaned of foregrounds by a template removal technique. 

In Sect.~\ref{sec:hil}, we introduce the high-multipole ($\ell\ge30$) likelihood and present its main results. At these smaller scales, we employ a pseudo-$C_\ell$ approach, beginning with a numerical spherical harmonic transform of the full-sky map, debiased and deconvolved to account for the mask and noise.

\begin{table*}[ht!]                 
\begingroup
\newdimen\tblskip \tblskip=5pt
\caption{Likelihood codes and datasets. We use these designations throughout the text to refer to specific likelihood codes and implementations that use different input data. \rev{A sum of spectra in the description column designates the joint likelihood of these spectra.}}                          
\label{tab:datasets}                            
\nointerlineskip
\vskip -3mm
\footnotesize
\setbox\tablebox=\vbox{
   \newdimen\digitwidth 
   \setbox0=\hbox{\rm 0} 
   \digitwidth=\wd0 
   \catcode`*=\active 
   \def*{\kern\digitwidth}
   \newdimen\signwidth 
   \setbox0=\hbox{+} 
   \signwidth=\wd0 
   \catcode`!=\active 
   \def!{\kern\signwidth}
\halign{\hbox to 1.2in{#\leaderfil}\hfil\tabskip 2em&                            
	#\hfil\tabskip 0pt\cr
\noalign{\doubleline}
\omit \hfil Name\hfil & \omit\hfil Description\hfil\cr
\noalign{\vskip 3pt\hrule\vskip 5pt}
\planckTTonly & Full \Planck\ temperature-only $C_\ell^{TT}$ likelihood\cr
\planckTTTEEE & \planckTTonly\ combined with high-$\ell$ $C_\ell^{TE}$+$C_\ell^{EE}$ likelihood\cr
\lowEB & Low-$\ell$  polarization $C_\ell^{TE}+C_\ell^{EE}+C_\ell^{BB}$ likelihood\cr
\lowTEB & Low-$\ell$ temperature-plus-polarization likelihood\cr
\plikTT & High-$\ell$ $C_\ell^{TT}$-only likelihood\cr
\plikEE & High-$\ell$ $C_\ell^{EE}$-only likelihood\cr
\plikTE & High-$\ell$ $C_\ell^{TE}$-only likelihood\cr
\plikTTEETE & High-$\ell$ $C_\ell^{TT}+C_\ell^{TE}+C_\ell^{EE}$ likelihood\cr
\pliklite & High-$\ell$ $C_b^{TT}+C_b^{TE}+C_b^{EE}$, foreground-marginalized bandpower likelihood\cr
{\rm tauprior} & Gaussian prior, $\tau = 0.07 \pm 0.02$ \cr
\highL & ACT+SPT high-$\ell$ likelihood \cr
\WP & WMAP low-$\ell$ polarization likelihood${}^a$\cr						
\noalign{\vskip 5pt\hrule\vskip 3pt}}}
\endPlancktablewide                 
\tablenote a ``Low-$\ell$'' refers to $\ell<23$ for WP, but $\ell<30$ for the \Planck\ likelihoods.\par
\endgroup
\end{table*}                        

Section~\ref{sec:hil-ass} is devoted to the detailed assessment of this high-$\ell$ likelihood. One technical difference between \citetalias{planck2013-p08} and the present work is the move from the \camspec\ code to \plik\ for high-$\ell$ results as well as the released software \citep{planck2014-ES}. The \rev{main reason for this change is that} the structure of \plik\ allows more fine-grained tests on the polarization spectra for individual detectors or subsets of detectors. We are able to compare the effect of different cuts on \Planck\ and external data, as well as using methods that take different approaches to estimate the maximum-likelihood spectra from the input maps; these illustrate the small impact of  differences in methodology and data preparation, which are difficult to assess otherwise.

We then combine the low- and high-$\ell$ algorithms to form the full \Planck\ likelihood in Sect.~\ref{sec:hal}, \rev{assessing there the choice of $\ell=30$ for the hybridization scale and} establishing the basic cosmological results from \Planck\ 2015 data alone. 

Finally, in Sect.~\ref{sec:conc} we conclude. A series of Appendices discusses sky masks and gives more detail on the individual likelihood codes, both the released version and a series of other codes used to validate the overall methodology. 

To help distinguish the many different likelihood codes, which are functions of different parameters and use different input data, Table~\ref{tab:datasets} summarizes the designations used throughout the text.

\section{Low-multipole likelihood}\label{sec:low-ell}

At low multipoles, the current \Planck\ release implements a standard joint pixel-based likelihood including both temperature and polarization for multipoles $\ell\le29$. Throughout this paper, we denote this likelihood ``\lowTEB'', while ``\lowEB'' denotes the polarization part of this likelihood.
For temperature, the formalism uses the \rev{CMB maps cleaned with} \commander\ \citep{Eriksen2004,Eriksen2008a} maps, while for polarization we use the 70\GHz\ LFI maps and explicitly marginalize over the 30\GHz\ and 353\GHz\ maps taken as tracers of synchrotron and dust emission, respectively (see Sect.~\ref{sec:70ghz_pol}), accounting in both cases for the induced noise covariance in the likelihood.

This approach is somewhat different from the \Planck\ 2013 low-$\ell$
likelihood. As described in \citetalias{planck2013-p08}, this comprised two nearly independent components, covering temperature and polarization information, respectively. The temperature likelihood employed a Blackwell-Rao estimator
\citep{chu2005} at $\ell\le49$, averaging over Monte Carlo samples
drawn from the exact power spectrum posterior using
\commander. For polarization, we
had adopted the pixel-based 9-year \WMAP\ polarization likelihood,
covering multipoles $\ell\le23$ \citep{bennett2012}.

The main advantage of the exact joint approach now employed is mathematical rigour and consistency to higher $\ell$, while the main disadvantage is a slightly higher computational expense due to the higher pixel resolution required to extend the calculation to $\ell = 29$ in polarization. However, after implementation of the Sherman--Morrison--Woodbury formula to reduce computational costs (see Appendix \ref{app:lol-speed}), the two approaches perform similarly, both with respect to speed and accuracy, and our choice is primarily a matter of implementational convenience and flexibility, rather than actual results or performance. 
\subsection{Statistical description and algorithm}
\label{sec:lowl_algorithm}

We start by reviewing the general CMB likelihood formalism for the analysis
of temperature and polarization at low $\ell$, as described for instance by
\citet{Tegmark2001}, \citet{page2007}, and in \citetalias{planck2013-p08}. 
We begin with maps of the three Stokes parameters $\{T,Q,U\}$ 
for the observed CMB intensity and linear polarization in some set of
\healpix\footnote{\url{http://healpix.sourceforge.org}} \citep{gorski2005} pixels on
the sky. In order to use multipoles $\ell\le\ell_{\textrm{cut}}=29$ in the likelihood, we adopt a \healpix\
resolution of $N_{\textrm{side}}=16$ which has 3072 pixels (of area $13.6\,\textrm{deg}^2$) per map; this accommodates multipoles up to $\ell_{\rm max}=3N_{\textrm{side}}-1=47$, and, considering separate maps of $T$, $Q$, and $U$, corresponds to a maximum of $N_{\textrm{pix}}=3\times3072=9216$ pixels in any given calculation, not accounting for any masking. 

After component separation, the data vector may be modelled as a sum
of cosmological CMB signal and instrumental noise,
$\vec{m}^{X}=\vec{s}^{X}+\vec{n}^{X}$, where $\vec{s}$ is assumed to be a set of
statistically isotropic and Gaussian-distributed random fields on the sky, indexed by pixel or spherical-harmonic indices $(\ell m)$, with $X=\{T,E,B\}$ selecting the appropriate intensity or polarization component. The signal fields $\vec{s}^{X}$ have auto- 
and cross-power spectra $C_{\ell}^{XY}$ and a pixel-space covariance matrix
\begin{equation}
  \tens{S}(C_{\ell})=\sum_{\ell=2}^{\ell_{\rm max}} \sum_{XY} C_{\ell}^{XY}
\tens{P}_{\ell}^{XY}.
\end{equation}
Here we restrict the spectra to ${XY}=\{TT, EE, BB, TE\}$, with
$N_{\textrm{side}}=16$ pixelization, and $\tens{P}^{XY}_{\ell}$ is a
beam-weighted sum over (associated) Legendre polynomials. For
temperature, the explicit expression is
\begin{equation}
(\tens{P}_{\ell}^{TT})_{i,j} = \frac{2\ell+1}{4\pi} \, B_\ell^2 \, P_\ell (\vec{\hat{n}}_i \cdot \, \vec{\hat{n}}_j),
\end{equation}
where $\hat{\vec{n}}_i$ is a unit vector pointing towards pixel $i$,
$B_{\ell}$ is the product of the instrumental beam Legendre transform
and the \healpix\ 
pixel window, and $P_{\ell}$ is the Legendre polynomial of order
$\ell$; for corresponding polarization components, see, \eg
\citet{Tegmark2001}. The instrumental noise is also assumed to be
Gaussian distributed, with a covariance matrix $\tens{N}$ that depends
on the \Planck\ detector sensitivity and scanning strategy, and the
full data covariance is therefore $\tens{M} = \tens{S} +
\tens{N}$. With these definitions, the full likelihood expression
reads
\begin{equation}
\mathcal{L}(C_{\ell}) =
\mathcal{P}(\vec{m}|C_{\ell})=\frac{1}{2\pi|\tens{M}|^{1/2}}
\exp\left(-\frac{1}{2}\vec{m}^{\tens{T}}\,\tens{M}^{-1}\vec{m}\right)\, ,
\label{eq:pbLike}
\end{equation}
where the conditional probability 
$\mathcal{P}(\vec{m}|C_{\ell})$ defines the likelihood $\mathcal{L}(C_{\ell})$.

The computational cost of this expression is driven by the presence of
the matrix inverse and determinant operations, both of which scale
computationally as $\mathcal{O}(N_{\textrm{pix}}^3)$. For this reason,
the direct approach is only computationally feasible at large angular
scales, where the number of pixels is low. In practice, we only
analyse multipoles below or equal to $\ell_{\textrm{cut}}=29$ with this formalism,
requiring maps with $N_{\textrm{side}}=16$. Multipoles between $\ell_{\textrm{cut}}+1$ and
$\ell_{\textrm{max}}$ are fixed to the best-fit $\Lambda$CDM spectrum
when calculating $\tens{S}$.
This division between varying and fixed multipoles speeds
up the evaluation of Eq.~(\ref{eq:pbLike}) through the Sherman--Morrison--Woodbury formula and the related matrix determinant lemma, as described in Appendix~\ref{app:lol-speed}. This results in an order-of-magnitude speed-up compared to the
brute-force computation.

\subsection{Low-$\ell$ temperature map and mask}
\label{sec:commander_lowl}

\begin{figure}[t]
    \centering
    \includegraphics[width=\columnwidth]{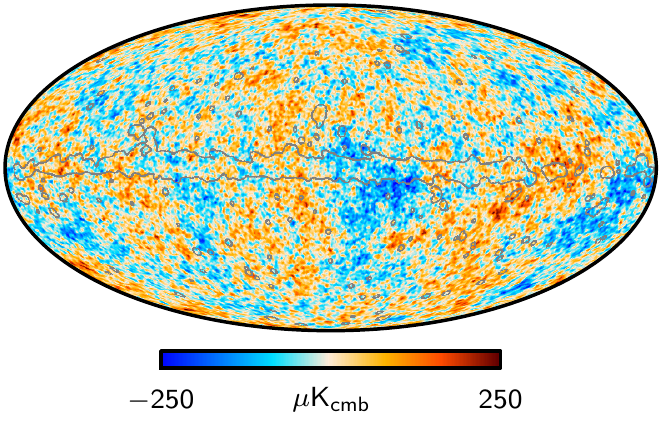}
    \includegraphics[width=\columnwidth]{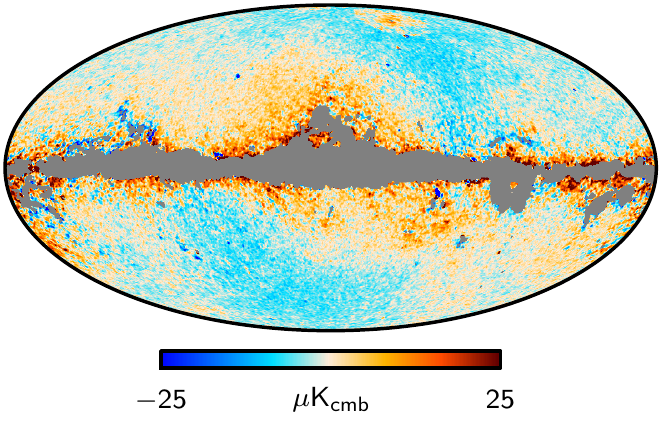}
    \includegraphics[width=\columnwidth]{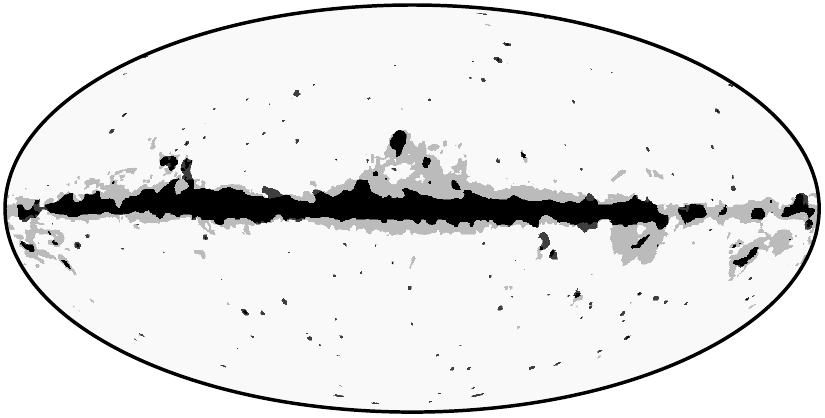}
    \caption{\emph{Top}: \commander\ CMB temperature map derived from
      the \Planck\ 2015, 9-year \WMAP, and 408$\,$MHz Haslam
      et al.\ observations, as described in
      \citet{planck2014-a12}. The gray boundary indicates the 2015
      likelihood temperature mask, covering a total of 7\,\% of the
      sky. The masked area has been filled with a constrained Gaussian
      realization. \emph{Middle}: difference between the 2015 and 2013
      \commander\ temperature maps. The masked region indicates the
      2013 likelihood mask, removing 13\,\% of the sky. \emph{Bottom}:
      comparison of the 2013 (gray) and 2015 (black) temperature
      likelihood masks. }
    \label{fig:lowlTmap}
\end{figure}

\begin{figure}[t]
    \centering
    \includegraphics[width=\columnwidth]{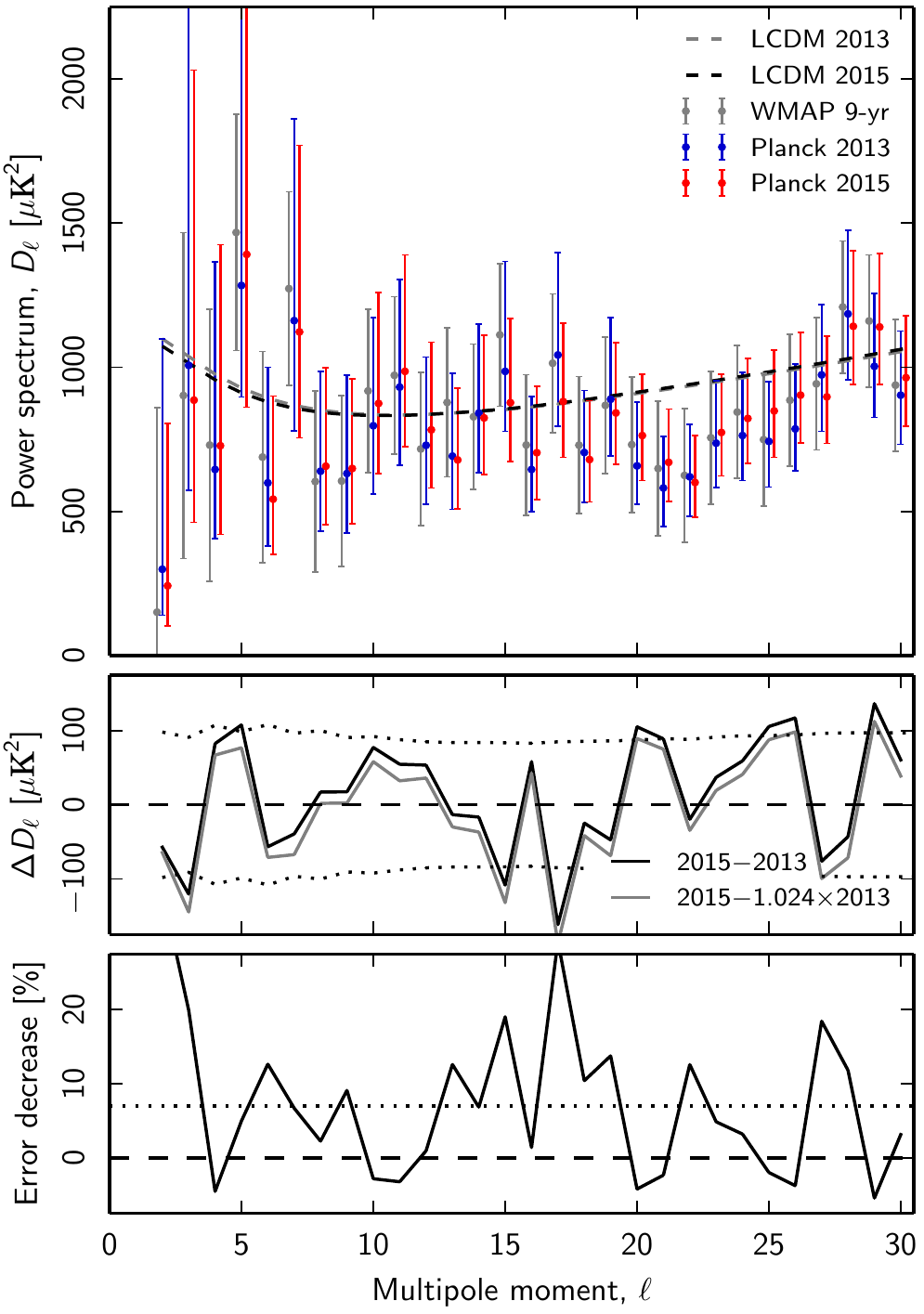}
    \caption{\emph{Top}: comparison of the \Planck\ 2013 (blue points) and 2015 (red points) posterior-maximum low-$\ell$
      temperature power spectra, as derived with \commander. Error
      bars indicate asymmetric marginal posterior 68\,\% confidence
      regions. For reference, we also show the final 9-year
      \WMAP\ temperature spectrum in light gray points, as presented by \citet{bennett2012}; note that the error bars indicate symmetric Fisher uncertainties in this case.  The dashed lines
      show the best-fit $\Lambda$CDM spectra derived from the
      respective data sets, including high-multipole and polarization
      information. \emph{Middle:} difference between the 2015 and 2013
      maximum-posterior power spectra (solid black line). The gray
      shows the same difference after scaling the 2013 spectrum up by
      2.4\,\%. Dotted lines indicate the expected $\pm1\,\sigma$ confidence
      region, accounting only for the sky fraction
      difference. \emph{Bottom}: reduction in marginal error bars
      between the 2013 and 2015 temperature spectra; see main text for
      explicit definition. The dotted line shows the reduction
      expected from increased sky fraction alone. }
    \label{fig:lowlTpowspec}
\end{figure}

Next, we consider the various data inputs that are required to
evaluate the likelihood in Eq.~(\ref{eq:pbLike}), and we start our
discussion with the temperature component. As in 2013, we employ the
\commander\ algorithm for component separation. This is a Bayesian Monte Carlo
method that either samples from or maximizes a global posterior
defined by some explicit parametric data model and a set of
priors. The data model adopted for the \Planck\ 2015 analysis is
described in detail in \citet{planck2014-a12}, and reads
\begin{equation}
  \vec{s}_{\nu}(\theta) = g_{\nu} \sum_{i=1}^{N_{\textrm{comp}}}
  \tens{F}_\nu^i(\beta_i, \Delta_\nu)\,\vec{a}_{i} + \sum_{j=1}^{N_\textrm{template}}{T}^j_{\nu}\vec{b}^\nu_{j},
\end{equation}
where $\theta$ denotes the full set of unknown parameters determining the signal at frequency $\nu$. The first sum runs over $N_\textrm{comp}$ independent astrophysical components \rev{including the CMB itself};
$\vec{a}_i$ is the corresponding amplitude map for each component
at some given reference frequency; $\beta_i$ is a general set of
spectral parameters for the same component; $g_\nu$ is a
multiplicative calibration factor for frequency $\nu$; $\Delta_\nu$ is
a linear correction of the bandpass central frequency; and the function $\tens{F}_\nu^i(\beta_i, \Delta_\nu)$  gives the frequency dependence for component $i$ (which can vary pixel-by-pixel and is hence most generally an $N_\textrm{pix}\times N_\textrm{pix}$ matrix). In the second sum, ${T}^j_{\nu}$ is one of a set of $N_\textrm{template}$ correction template amplitudes, accounting for known effects such as monopole, dipole, or zodiacal light, with template maps $\vec{b}^{\nu}_j$.

In 2013, only \Planck\ observations between 30 and 353$\,$GHz were
employed in the corresponding fit. In the updated analysis, we broaden
the frequency range considerably, by including the \Planck\ 545 and
857$\,$GHz channels, the 9-year \WMAP\ observations between 23 and
94$\,$GHz \citep{bennett2012}, and the \citet{haslam1982} 408$\,$MHz survey.
We can then separate the low-frequency
foregrounds into separate synchrotron, free-free, and spinning-dust
components, as well as to constrain the thermal dust temperature
pixel-by-pixel. In addition, in the updated analysis we employ
individual detector and detector-set maps rather than co-added
frequency maps, and this gives stronger constraints on both
line emission (primarily CO) processes and bandpass measurement
uncertainties. For a comprehensive discussion of all these results, we
refer the interested reader to \citet{planck2014-a12}.

For the purposes of the present paper, the critical output from this
process is the maximum-posterior CMB temperature sky map, shown in the
top panel of Fig.~\ref{fig:lowlTmap}. This map is natively produced at
an angular resolution of $1\deg$ FWHM, determined by the instrumental
beams of the \WMAP\ 23$\,$GHz and 408$\,$MHz frequency channels. In
addition, the \commander\ analysis provides a direct goodness-of-fit
measure per pixel in the form of the $\chi^2$ map shown in 
\citet[][figure~22]{planck2014-a12}. Thresholding this $\chi^2$ map results in a
confidence mask that may be used for likelihood analysis, and the
corresponding masked region is indicated in the top panel of
Fig.~\ref{fig:lowlTmap} by a gray boundary. Both the map and mask are
downgraded from their native \healpix\ $N_{\textrm{side}}=256$ pixel
resolution to $N_{\textrm{side}}=16$ before insertion into the
likelihood code, and the map is additionally smoothed to an effective
angular resolution of $440\arcm$ FWHM.

The middle panel of Fig.~\ref{fig:lowlTmap} shows the difference
between the \Planck\ 2015 and 2013 \commander\ maximum-posterior maps,
where the gray region now corresponds to the 2013 confidence
mask. Overall, there are large-scale differences at the
$10\,\mu\textrm{K}$ level at high Galactic latitudes, while at
low Galactic latitudes there are a non-negligible number of pixels
that saturate the colour scale of $\pm25\,\mu\textrm{K}$. 
These differences are well understood. First, the most striking red and blue large-scale features at high latitudes are dominated by destriping errors in our 2013 analysis, due to bandpass
mismatch in a few frequency channels effectively behaving as
correlated noise during map making. As discussed in section~3 of
\citet{planck2014-a12} and illustrated in figure 2 therein, the most significant outliers have been
removed from the updated 2015 analysis, and, consequently, the pattern
is clearly visible from the difference map in
Fig.~\ref{fig:lowlTmap}. Second, the differences near the Galactic plane and close to the mask
boundary are dominated by negative CO residuals near the
Fan region, at Galactic coordinates $(l,b)\approx (110\deg,20\deg)$;
by negative free-free residuals near the Gum nebula at
$(l,b)\approx(260\deg,15\deg)$; and by thermal dust residuals along
the plane. Such differences are expected because of the wider
frequency coverage and improved foreground model in the new fit. In
addition, the updated model also includes the thermal
Sunyaev-Zeldovich (SZ) effect near the Coma and Virgo clusters in the
northern hemisphere, and this may be seen as a roughly circular patch
near the Galactic north pole.

Overall, the additional frequency range provided by the \WMAP\ and
408$\,$MHz observations improves the component separation, and combining these data
sets makes more
sky effectively available for CMB analysis. The bottom panel of Fig.~\ref{fig:lowlTmap} compares the two
$\chi^2$-based confidence masks. In total, 7\,\% of the sky is removed
by the 2015 confidence mask, compared with 13\,\% in the 2013 version.

The top panel in Fig.~\ref{fig:lowlTpowspec} compares the marginal
posterior low-$\ell$ power spectrum,
$D_{\ell}\equiv C_{\ell}\,\ell(\ell+1)/(2\pi)$, derived from the updated map
and mask using the Blackwell--Rao estimator \citep{chu2005} with the corresponding 2013
spectrum \citepalias{planck2013-p08}. 
The middle panel shows their difference. The dotted lines
indicate the expected variation between the two spectra,
$\sigma_{\ell}$, accounting only for their different sky
fractions.\footnote{These rms estimates were computed with the \polspice 
  power-spectrum estimator \citep{chon2004} by averaging over 1000
  noiseless simulations.} From this, we can compute 
\begin{equation}
\chi^2 = \sum_{\ell=2}^{29} \left(\frac{D_{\ell}^{2015}-D_{\ell}^{2013}}{\sigma_{\ell}}\right)^2,
\end{equation}
and we find this to be 21.2 for the current data set. With 28 degrees
of freedom, and assuming both Gaussianity and statistical independence
between multipoles, this corresponds formally to a
probability-to-exceed (PTE) of 82\,\%. According to these tests, the observed differences are
consistent with random fluctuations due to increased sky fraction
alone.

As discussed in \citet{planck2014-a01}, the absolute calibration of
the \Planck\ sky maps has been critically reassessed in the new
release. The net outcome of this process was an effective
recalibration of +1.2$\,$\% in map domain, or +2.4$\,$\% in terms of
power spectra. The gray line in the middle panel of
Fig.~\ref{fig:lowlTpowspec} shows the same difference as discussed
above, but after rescaling the 2013 spectrum up by 2.4$\,$\%. At the
precision offered by these large-scale observations, the difference is
small, and either calibration factor is consistent with expectations.

Finally, the bottom panel compares the size of the statistical error
bars of the two spectra, in the form of 
\begin{equation}
r_{\ell} \equiv \frac{\left.\left(\sigma_{\ell}^{\textrm{l}} +
  \sigma_{\ell}^{\textrm{u}}\right)\right|_{2013}}{\left.\left(\sigma_{\ell}^{\textrm{l}} +
  \sigma_{\ell}^{\textrm{u}}\right)\right|_{2015}} - 1,
\end{equation}
where $\sigma_{\ell}^{\textrm{u}}$ and $\sigma_{\ell}^{\textrm{l}}$ denote upper and lower asymmetric
68\,\% error bars, respectively. Thus, this quantity measures the
decrease in error bars between the 2013 and 2015 spectra, averaged
over the upper and lower uncertainties. Averaging over 1000 ideal
simulations and multipoles between $\ell=2$ and 29, we find that the
expected change in the error bar due to sky fraction alone is 7$\,$\%, in
good agreement with the real data. Note that because the net
uncertainty of a given multipole is dominated by cosmic variance, its
magnitude depends on the actual power spectrum value. Thus, multipoles
with a positive power difference between 2015 and 2013 tend to have a
smaller uncertainty reduction than points with a negative power
difference. Indeed, some multipoles have a negative uncertainty
reduction because of this effect.

For detailed discussions and higher-order statistical analyses of the
new \commander\ CMB temperature map, we refer the interested reader
to \citet{planck2014-a12} and \citet{planck2014-a18}.

\subsection{70\GHz\ Polarization low-resolution solution}
\label{sec:70ghz_pol}

The likelihood in polarization uses only a subset of the full \Planck\ polarization data, chosen to have well-characterized noise properties and negligible contribution from foreground contamination and unaccounted-for systematic errors. Specifically, we use data from the 70\GHz\ channel of the LFI instrument, for the full mission except for Surveys~2 and 4, which are conservatively removed because they stand as $3\,\sigma$ outliers in survey-based null tests \citep{planck2014-a03}. 
While the reason for this behaviour is not completely understood, it is likely related to the fact that these two surveys exhibit the
deepest minimum in the dipole modulation amplitude \citep{planck2014-a03,planck2014-a05}, leading to an increased
vulnerability to gain uncertainties and to contamination from diffuse polarized foregrounds.

To account for foreground contamination, the \Planck\ $Q$ and $U$ 70\GHz\ maps are cleaned using 30\GHz\ maps to generate a template for low-frequency foreground contamination, and 353\GHz\ maps to generate a template for polarized dust emission \citep{planck2014-XIX,planck2014-XXX,planck2014-a11}. Linear polarization maps are downgraded from high resolution to $N_\mathrm{side}=16$ employing an inverse-noise-weighted averaging procedure, without applying any smoothing \citep{planck2014-a07}.

\begin{figure}[htbp]
    \centering
        \includegraphics[width=\columnwidth]{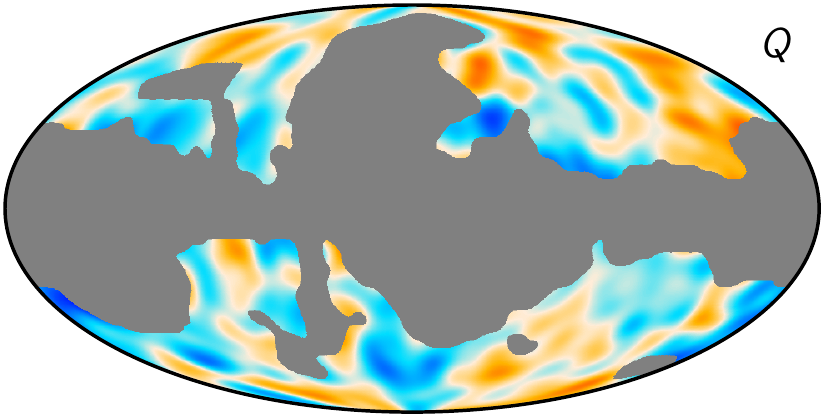}
        \includegraphics[width=\columnwidth]{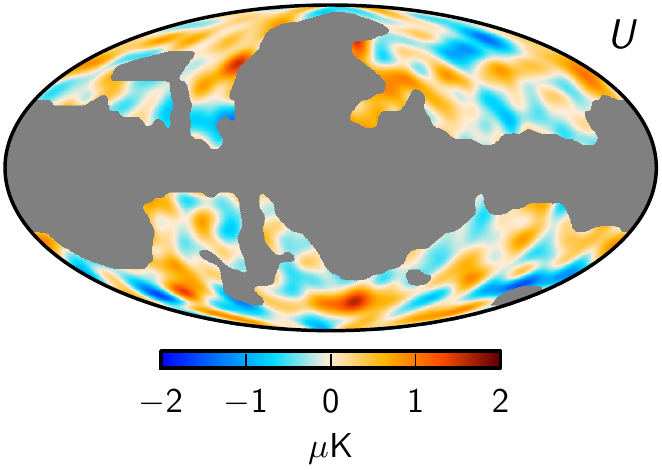}
    \caption{Foreground-cleaned, 70\,GHz $Q$ (top) and $U$ (bottom) maps used for the low-$\ell$ polarization part of the likelihood. Each of the maps covers $46\,\%$ of the sky.}
    \label{fig:lowlPmap}
\end{figure}

The final cleaned $Q$ and $U$ maps, shown in Fig.~\ref{fig:lowlPmap}, retain a fraction $\fsky=0.46$ of the sky, masking out the Galactic plane and the ``spur regions'' to the north and south of the Galactic centre. 

At multipoles $\ell < 30$, we model the likelihood assuming that the maps follow a Gaussian distribution with known covariance, as in Eq.~(\ref{eq:pbLike}). For polarization, however, we use foreground-cleaned maps, explicitly taking into account the induced increase in variance through an effective noise correlation matrix.

\begin{figure}[htbp]
    \centering
        \includegraphics[width=\columnwidth]{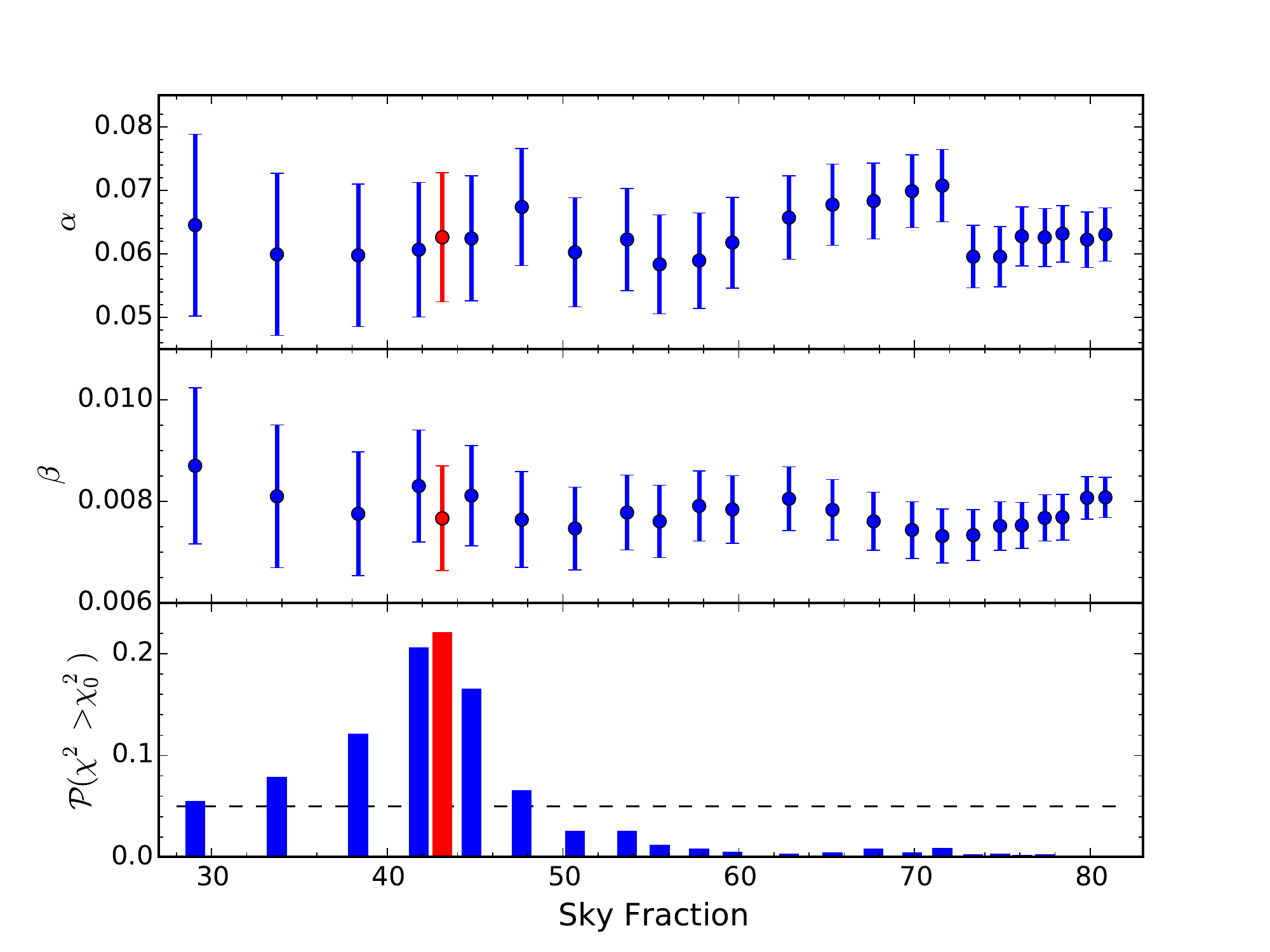}
       \caption{\textit{Upper panels}: estimated best-fit scaling coefficients for synchrotron ($\alpha$) and dust ($\beta$), for several masks, whose sky fractions are displayed along the bottom horizontal axis (see text). \textit{Lower panel}: the probability to exceed, $\mathcal{P}(\chi^2 > \chi^2_0)$. The red symbols identify the mask from which the final scalings are estimated, but note how the latter are roughly stable over the range of sky fractions. Choosing such a  large ``processing'' mask ensures that the associated errors are conservative.}
    \label{fig:lowl-alphabeta}
\end{figure}

To clean the 70\,GHz $Q$ and $U$ maps we use a template-fitting procedure. Restricting $\vec{m}$ to the $Q$ and $U$ maps (\ie\  $\vec{m} \equiv [Q, U]$) we write
\begin{equation}
\vec{m} = \frac{1}{1-\alpha-\beta}\left ( \vec{m}_{70} - \alpha \vec{m}_{30} - \beta \vec{m}_{353} \right)\,,
\end{equation}
where $\vec{m}_{70}$, $\vec{m}_{30}$, and $\vec{m}_{353}$ are bandpass-corrected versions of the 70, 30, and 353\,\GHz\ maps \citep{planck2014-a04,planck2014-a08}, and $\alpha$ and $\beta$ are the scaling coefficients for  synchrotron and dust emission, respectively. The latter can be estimated by minimizing the quantity 
\begin{equation}
\chi^2 =  (1-\alpha-\beta)^2 \vec{m}^\tens{T} \tens{C}_{\tens{S}+\tens{N}}^{-1} \vec{m}\,, \label{chi2_alphabeta}
\end{equation}
where
\begin{equation}
\tens{C}_{\tens{S}+\tens{N}} \equiv  (1-\alpha-\beta)^2 \langle \vec{m} \vec{m}^\tens{T} \rangle =    (1-\alpha-\beta)^2 \tens{S}(C_\ell)+ \tens{N}_{70} \,.\label{CSN_alphabeta}
\end{equation}
Here $ \tens{N}_{70}$ is the pure polarization part of the 70\,GHz noise covariance matrix\footnote{We assume here, and have checked in the data, that the noise-induced $TQ$ and $TU$ correlations are negligible.}  \citep{planck2014-a07}, and $C_\ell$ is taken as the \Planck\ 2015 fiducial model \citep{planck2014-a15}. We have verified that using the \Planck\ 2013 model has negligible impact on the results describe below. Minimization 
of the quantity in Eq. (\ref{chi2_alphabeta}) using the form of the covariance matrix given in Eq. (\ref{CSN_alphabeta}) is numerically demanding,
since it would require inversion of the covariance matrix at every step of the minimization procedure. However, the signal-to-noise ratio
 in the 70 GHz maps is relatively low, and we may neglect the dependence on the $\alpha$ and $\beta$ of the covariance matrix in Eq. (\ref{chi2_alphabeta}) using instead:
\begin{equation}
\tens{C}_{\tens{S}+\tens{N}} =  \tens{S}(C_\ell)+ \tens{N}_{70} \, ,\label{CSN_alphabeta2} 
\end{equation}
so that the matrix needs to be inverted only once. We have verified for a test case that accounting  for the dependence on the scaling parameters in the covariance matrix yields consistent results.
We find $\alpha = 0.063 $ and $\beta = 0.0077$, with $3\,\sigma$ uncertainties $\delta_\alpha\equiv3\,\sigma_\alpha = 0.025$ and $\delta_\beta\equiv3\,\sigma_\beta = 0.0022$.  The best-fit values quoted correspond to a polarization mask using 46\,\% of the sky and correspond to spectral indexes (with $2\,\sigma$ errors) $n_\mathrm{synch} = -\rev{3.16} \pm 0.40$ and $n_\mathrm{dust} =  1.50 \pm 0.16 $, for synchrotron and dust emission respectively \citep[see][for a definition of the foreground spectral indexes]{planck2014-a12}. To select the cosmological analysis mask, the following scheme is employed. We scale to 70\GHz\ both  $\vec{m}_{30}$ and  $\vec{m}_{353}$,  assuming fiducial spectral indexes $n_\mathrm{synch} = -3.2$ and $n_\mathrm{dust} = 1.6$, respectively. In this process, we do not include bandpass correction templates. From either rescaled template we compute the polarized intensities $P=\sqrt{Q^2+U^2}$ and sum them. We clip the resulting template at equally spaced thresholds to generate a set of 24 masks, with unmasked fractions in the range from 30\,\% to 80\,\% of the sky. Finally, for each mask, we estimate the best-fit scalings and evaluate the probability to exceed, $\mathcal{P}(\chi^2 > \chi^2_0)$, where $\chi^2_0$ is the value achieved by minimizing Eq.\ (\ref{chi2_alphabeta}). The $f_\mathrm{sky}=43\,\%$ processing mask is chosen as the tightest mask (i.e., the one with the greatest $f_\mathrm{sky}$) satisfying the requirement $\mathcal{P}>5\,\%$ (see Fig.~\ref{fig:lowl-alphabeta}). We use a slightly smaller mask ($f_\mathrm{sky}=46\,\%$) for the cosmological analysis, which is referred to as the R1.50 mask in what follows. 

We define the final polarization noise covariance matrix used in Eq.~(\ref{eq:pbLike}) as
\begin{equation}
\tens{N} = \frac{1}{(1-\alpha-\beta)^2}\left ( \tens{N}_{70} + \delta^2_\alpha \vec{m}_{30} \vec{m}^\tens{T}_{30} +  \delta^2_\beta \vec{m}_{353} \vec{m}^\tens{T}_{353}\right)\,. 
\label{lowl_covmat}
\end{equation}
We use $3\,\sigma$ uncertainties, $\delta_\alpha$ and $\delta_\beta$, to define the covariance matrix, conservatively increasing the errors due to foreground estimation.
We have verified that the external (column to row) products involving the foreground templates are sub-dominant corrections. We do not include further correction terms arising from the bandpass leakage error budget since they are completely negligible. \rev{Intrinsic noise from the templates also proved negligible.}

\subsection{Low-$\ell$ \Planck\ power spectra and parameters}\label{sec:lowl_pse_params}

We use the foreground-cleaned $Q$ and $U$ maps derived in the previous section along with the \texttt{Commander} temperature map to derive angular power spectra. For the polarization part, we use the noise covariance matrix given in Eq.~(\ref{lowl_covmat}), while assuming only $1\,\muK^2$ diagonal regularization noise for temperature. Consistently, a white noise realization of the corresponding variance is added to the \texttt{Commander} map. By adding regularization noise, we ensure that the noise covariance matrix is  numerically well conditioned.
 
\begin{figure}[htbp]
    \centering
        \includegraphics[width=\columnwidth]{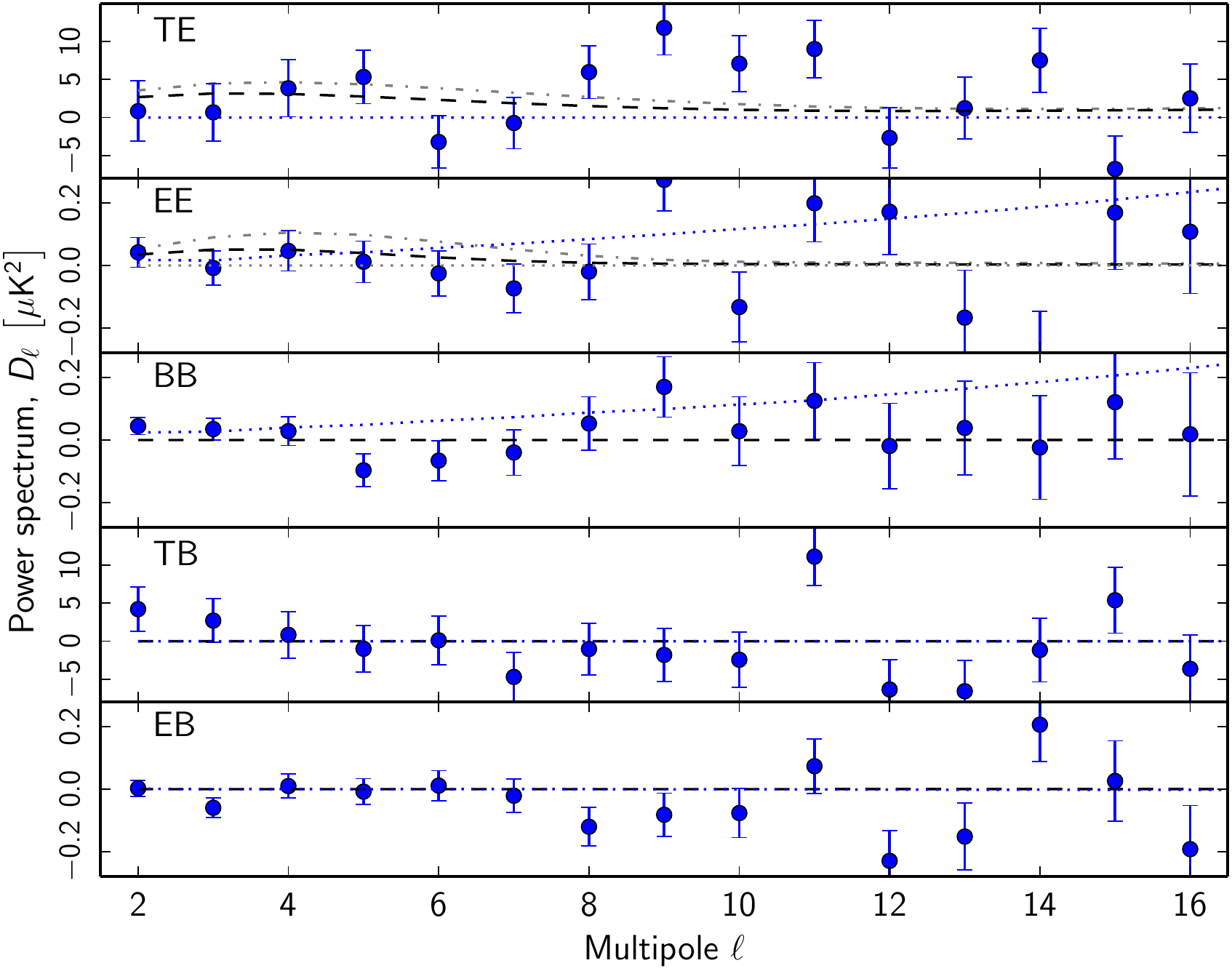}
     
    \caption{Polarized QML spectra from foreground-cleaned maps. Shown are the 2013 \Planck\ best-fit model ($\tau = 0.089$, dot-dashed) and the 2015 model ($\tau = 0.067$, dashed), as well as the 70\,GHz noise bias computed from Eq.\ (\ref{lowl_covmat}) (blue dotted).
    }
    \label{fig:lowl-polarized_aps}
\end{figure}

For power spectra, we employ the \texttt{BolPol} code \citep{Gruppuso2009}, an
implementation of the quadratic maximum likelihood (QML) power
spectrum estimator \citep{Tegmark1997,Tegmark2001}. Figure~\ref{fig:lowl-polarized_aps} presents all five polarized power spectra. The errors shown
in the plot are derived from the Fisher matrix. In the case of  $\EE$ and $\TE$ we plot the \Planck\ 2013 best-fit power spectrum model, which has an optical depth $\tau = 0.089$, as derived from low-$\ell$ \WMAP-9 polarization maps, along with the \Planck\ 2015 best model, which has $\tau =0.067$ as discussed below.\footnote{The models considered have been derived by fixing all parameters except $\tau$ and $A_\textrm{s}$ to their full multipole range 2015 best-fit values} Since the $\EE$ power spectral amplitude scales with $\tau$ as $\tau^2$ (and $\TE$ as $\tau$), the 2015 model exhibits a markedly lower reionization bump, which is a better description of \Planck\ data. There is a $2.7\,\sigma$ outlier in the $\EE$ spectrum at $\ell=9$, not unexpected given the number of low-$\ell$ multipole estimates involved.

To estimate cosmological parameters, we couple the machinery described in Sect.\ \ref{sec:lowl_algorithm} to \texttt{cosmomc}\footnote{\url{http://cosmologist.info/cosmomc/}} \citep{cosmomc}. We fix all parameters that are not sampled to their \Planck\ 2015 $\Lambda$CDM best-fit value \citep{planck2014-a15} and concentrate on those that have the greatest effect at low $\ell$: the reionization optical depth $\tau$, the scalar amplitude $A_\textrm{s}$, and the tensor-to-scalar ratio $r$. Results are shown in Table~\ref{table:lowl_params} for the combinations $(\tau, A_\textrm{s})$ and $(\tau, A_\textrm{s},r)$.

\begin{table}[ht!]                 
\begingroup
\newdimen\tblskip \tblskip=5pt
\caption{Parameters estimated from the low-$\ell$ likelihood.$^{\rm a}$}  
\label{table:lowl_params}                        
\nointerlineskip
\vskip -3mm
\footnotesize
\setbox\tablebox=\vbox{
   \newdimen\digitwidth 
   \setbox0=\hbox{\rm 0} 
   \digitwidth=\wd0 
   \catcode`*=\active 
   \def*{\kern\digitwidth}
   \newdimen\signwidth 
   \setbox0=\hbox{+} 
   \signwidth=\wd0 
   \catcode`!=\active 
   \def!{\kern\signwidth}
\openup 4pt
\halign{\hbox to 1.0in{#\leaderfil}\tabskip=2em&
   \hfil#\hfil& 
   \hfil#\hfil\tabskip=0pt\cr
\noalign{\doubleline}
\omit\hfil Parameter\hfil& $\Lambda$CDM& $\Lambda$CDM+$r$\cr
\noalign{\vskip 4pt\hrule\vskip 5pt}
$\tau$& $0.067\pm0.023$& $0.064\pm0.022$\cr
$\log[10^{10}A_\textrm{s}]$& $2.952\pm0.055$& $2.788^{+0.19}_{-0.09}$\cr
$r$& $0$& $[0,\,0.90]$\cr
\noalign{\vskip 8pt}
$z_\textrm{re}$& $8.9^{+2.5}_{-2.0}$& $8.5^{+2.5}_{-2.1}$\cr
$10^9 A_\textrm{s}$& $1.92^{+0.10}_{-0.12}$& $1.64^{+0.29}_{-0.17}$\cr
$A_\textrm{s} e^{-2\tau}$& $1.675^{+0.082}_{-0.093}$& $1.45^{+0.24}_{-0.14}$\cr
\noalign{\vskip 5pt\hrule\vskip 3pt}}}
\endPlancktable
\tablenote {{\rm a}} For the centre column the set of parameters $(\tau, A_\textrm{s})$ was sampled, while it was the set $(\tau, A_\textrm{s},r)$ for the right column. Unsampled parameters are fixed to their $\Lambda$CDM 2015 best-fit fiducial values. All errors are 68\,\% CL (confidence level), while the upper limit on $r$ is 95\,\%. The bottom portion of the table shows a few additional derived parameters for information.\par
\endgroup
\end{table}

It is interesting to disentangle the cosmological information provided by low-$\ell$ polarization from that derived from temperature. 
Low-$\ell$ temperature mainly contains information on the combination $A_\textrm{s} e^{-2 \tau}$, at least at multipoles corresponding to angular scales
smaller than the scale subtended by the horizon at reionization (which itself depends on $\tau$). The lowest temperature multipoles, however, are directly sensitive to $A_\textrm{s}$.
On the other hand, large-scale polarization is sensitive to the combination $A_\textrm{s} \tau^2$. Thus, neither low-$\ell$ temperature nor polarization can separately constrain
$\tau$ and $A_\textrm{s}$. Combining temperature and polarization  breaks the degeneracies and puts tighter constraints on these parameters.

In order to disentangle the temperature and polarization contributions to the constraints, we consider four versions of the low-resolution likelihood.
\begin{enumerate}
\item The standard version described above, which considers the full set of $T$, $Q$, and $U$ maps, along with their  covariance matrix, and is sensitive to the $\TT$, $\TE$, $\EE$, and $\BB$ spectra. 
\item A temperature-only version, which considers the temperature map and its regularization noise covariance matrix. It is only sensitive to $\TT$.
\item A polarization-only version, considering only the $Q$ and $U$ maps and the $QQ$, $QU$, and $UU$ blocks of the covariance matrix. This is sensitive to the $\EE$ and $\BB$ spectra.
\item A mixed temperature-polarization version, which uses the previous polarization-only likelihood but multiplies it by the temperature-only likelihood. This is different from the standard $T, Q, U$ version in that it assumes vanishing temperature-polarization correlations.	

\end{enumerate}

\begin{figure}[htbp]
\begin{center}
        \includegraphics[width=0.45\columnwidth]{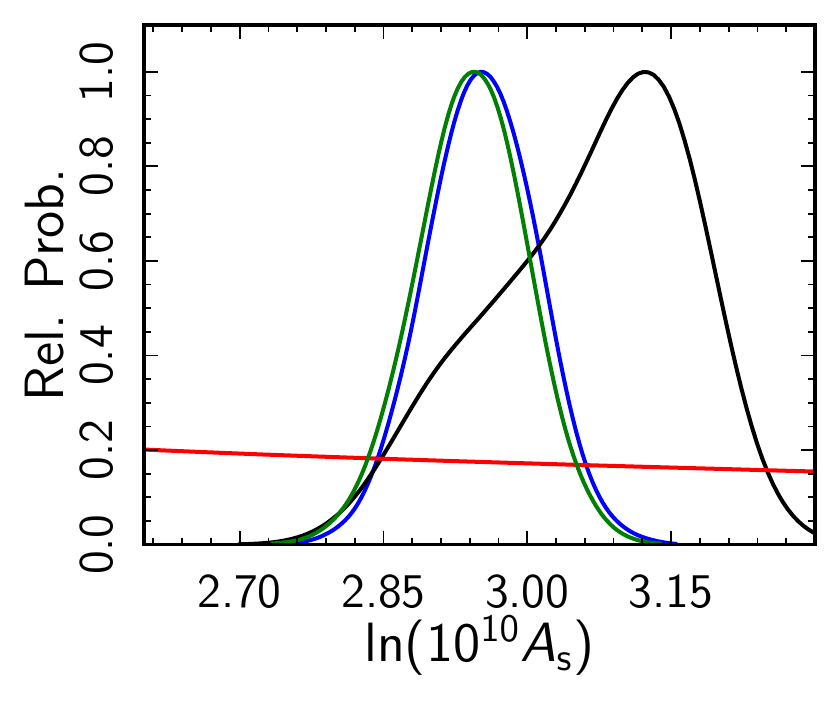}
        \includegraphics[width=0.45\columnwidth]{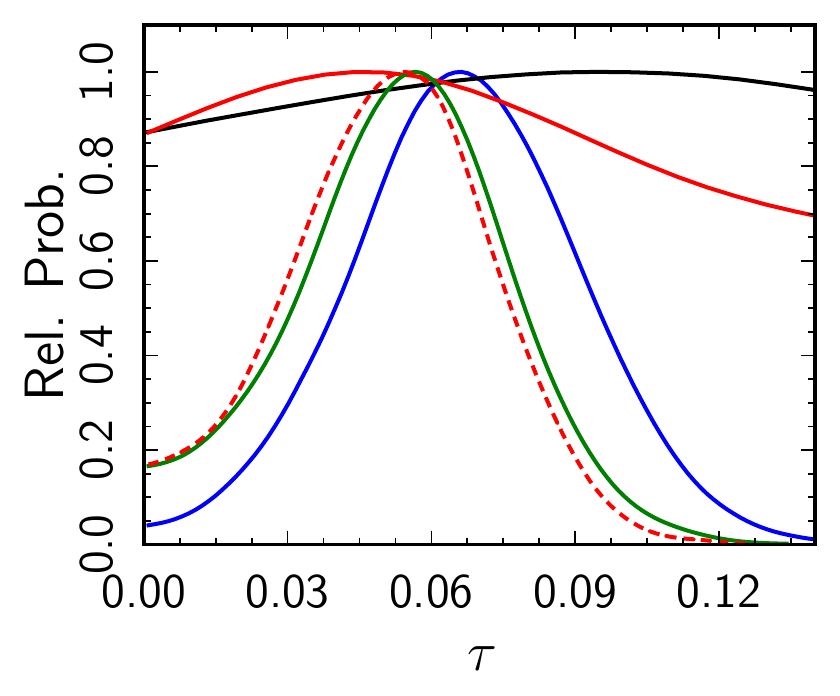}
        \includegraphics[width=0.45\columnwidth]{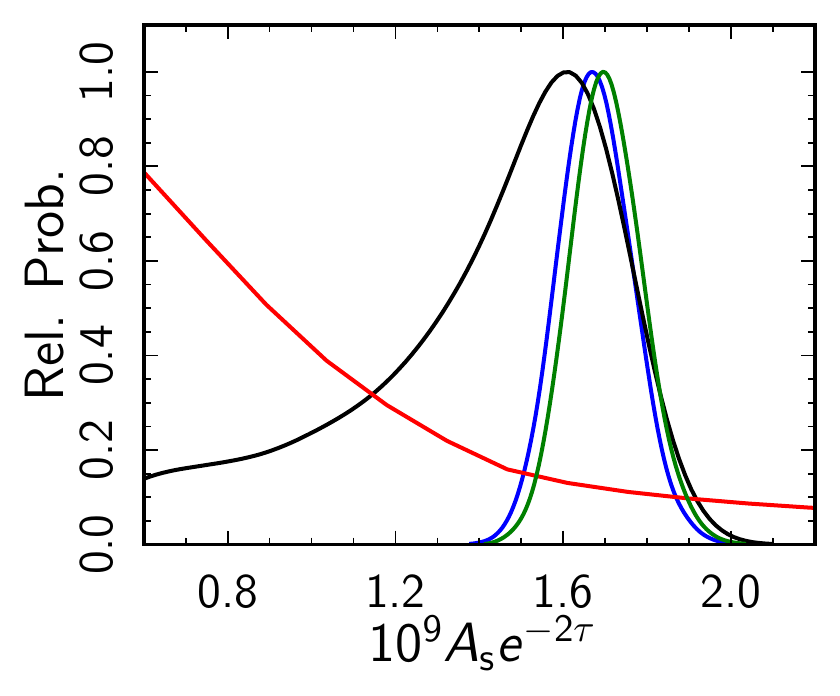}
        \includegraphics[width=0.465\columnwidth]{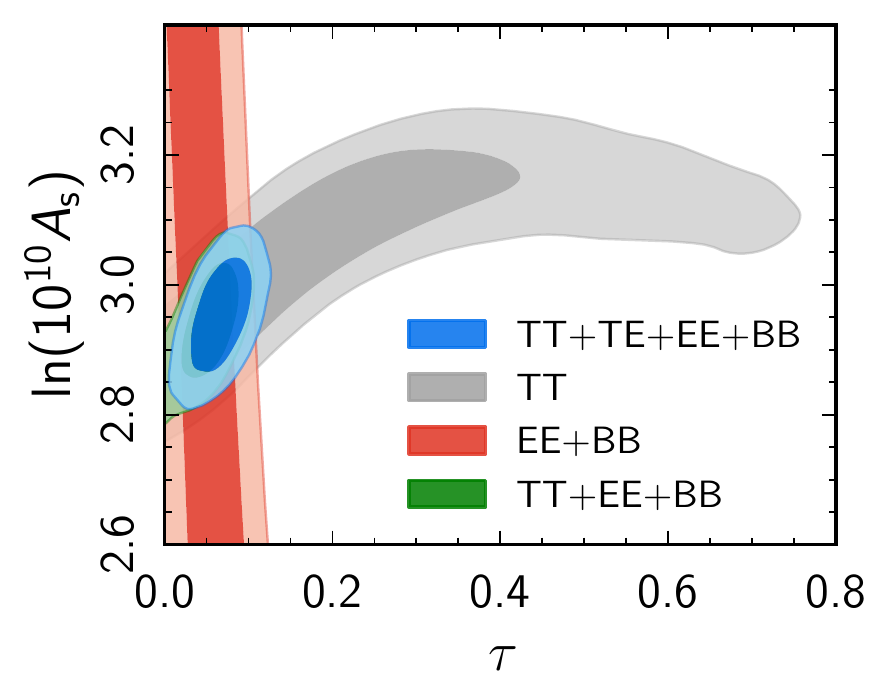}

    \caption{Likelihoods for parameters from low-$\ell$ data. \textit{Panels 1--3}: One-dimensional posteriors for $\log[10^{10} A_\textrm{s}]$, $\tau$, and $A_\textrm{s} e^{-2\tau}$ for the several sub-blocks of the likelihood, for cases 1 (blue), 2 (black), 3 (red), and 4 (green) --- see text for definitions;  dashed red is the same as case 3 but imposes a sharp prior $10^9 A_\textrm{s} e^{-2\tau} = 1.88$. \textit{Panel 4}: Two-dimensional posterior for $\log[10^{10} A_\textrm{s}]$ and $\tau$ for the same data combinations; shading indicates  the 68\,\% and 95\,\% confidence regions.}
    \label{fig:lowl-pol_cuts}
\end{center}
\end{figure}

The posteriors derived from these four likelihood versions are displayed in Fig.\ \ref{fig:lowl-pol_cuts}. These plots show how temperature 
and polarization 
nicely combine to break the degeneracies and provide joint constraints on the two parameters. The degeneracy directions for cases (2) and (3) are as expected from the discussion above; 
the degeneracy in case (2) flattens for increasing values of $\tau$ because for such values the scale corresponding to the horizon at reionization
is pulled forward to $\ell > 30$. By construction, the posterior for case 4 must be equal to the product of the temperature-only (2) and polarization-only (3) posteriors. This is indeed the case at the level of the two-dimensional posterior (see lower right panel of Fig.\ \ref{fig:lowl-pol_cuts}). It is not immediately evident in the one-dimensional distributions because this property does not survive the final marginalization over the non-Gaussian shape of the temperature-only posterior. It is also apparent from Fig.\ \ref{fig:lowl-pol_cuts} that $\EE$ and $\BB$ alone do not constrain $\tau$. This is to be expected, and is due to the inverse degeneracy of $\tau$ with $A_\textrm{s}$, which is almost completely unconstrained without temperature information, and not to the lack of $\EE$ signal. By assuming a sharp prior $10^9 A_\textrm{s} e^{-2\tau} = 1.88$, corresponding to the best estimate obtained when also folding in the high-$\ell$ temperature information \citep{planck2014-a15}, 
the polarization-only analysis yields $\tau = 0.051^{+0.022}_{-0.020}$ (red dashed curve in Fig.\ \ref{fig:lowl-pol_cuts}). 
The latter bound does not differ much from having $A_\textrm{s}$ constrained by including $\TT$ in the analysis, which yields $\tau = 0.054^{+0.023}_{-0.021}$ (green curves).  Finally, the inclusion of 
non-vanishing temperature-polarization correlations \rev{(blue curves)} increases the significance of the $\tau$ detection at $\tau = 0.067 \pm 0.023$.
We have also performed a three-parameter fit, considering $\tau$, $A_\textrm{s}$, and $r$ for all four likelihood versions described above, finding consistent results.

\subsection{Consistency analysis}\label{sec:lowl_consistency}

Several tests have been carried out to validate the 2015 low-$\ell$ likelihood. Map-based validation and simple spectral tests are discussed  extensively in \citet{planck2014-a11} for temperature, and in \citet{planck2014-a03} for \Planck\ 70\,GHz polarization. We focus here on tests based on QML and likelihood analyses, respectively employing spectral estimates and cosmological parameters as benchmarks. 

\begin{figure}[htbp]
    \centering
       \includegraphics[width=0.45\columnwidth]{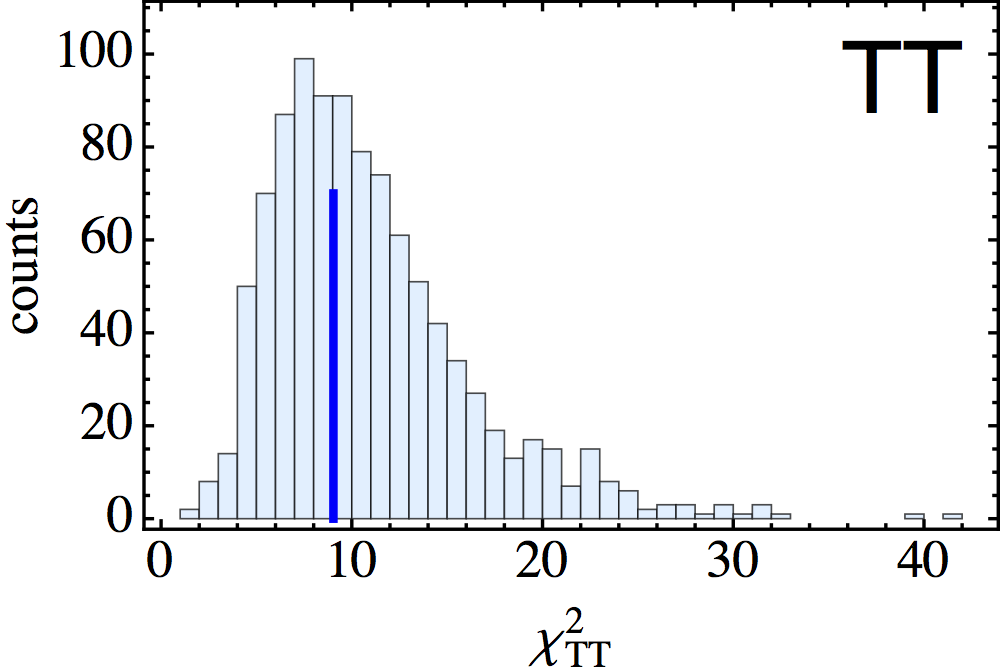}
       \includegraphics[width=0.45\columnwidth]{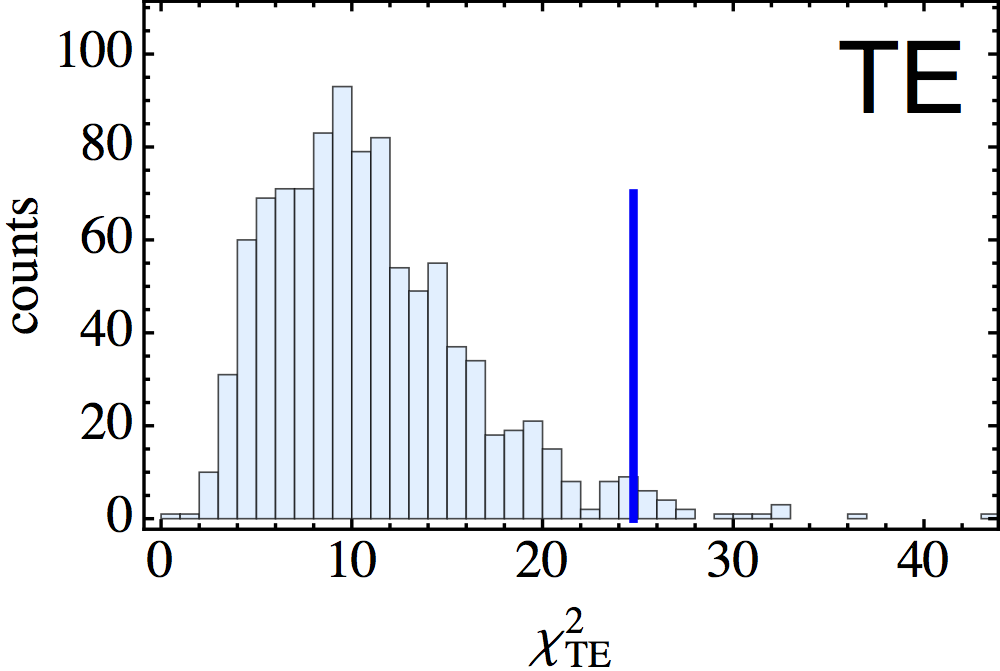}
       \includegraphics[width=0.45\columnwidth]{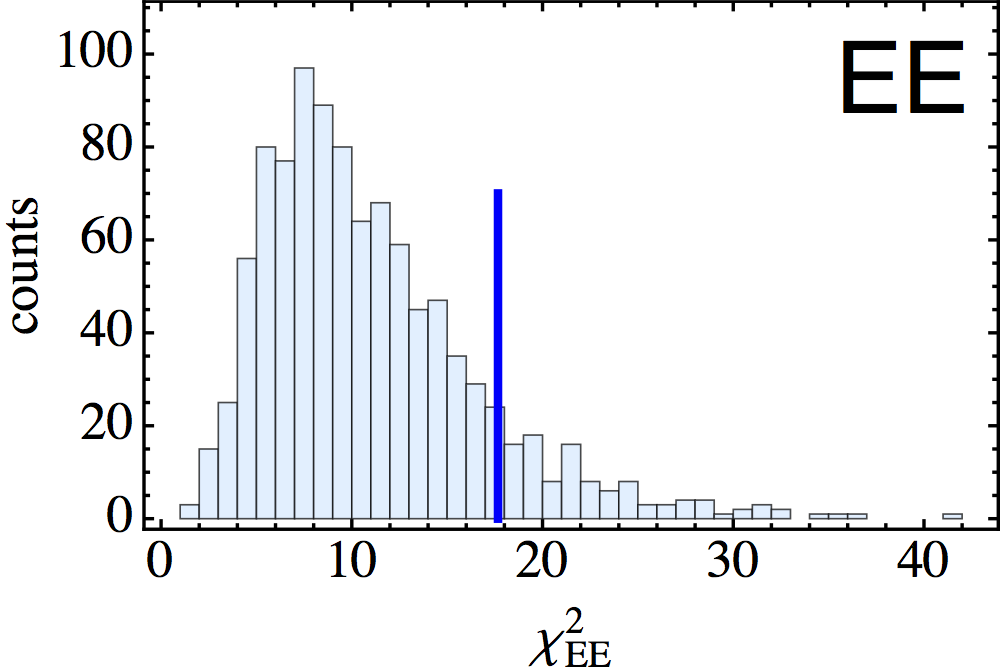}
       \includegraphics[width=0.45\columnwidth]{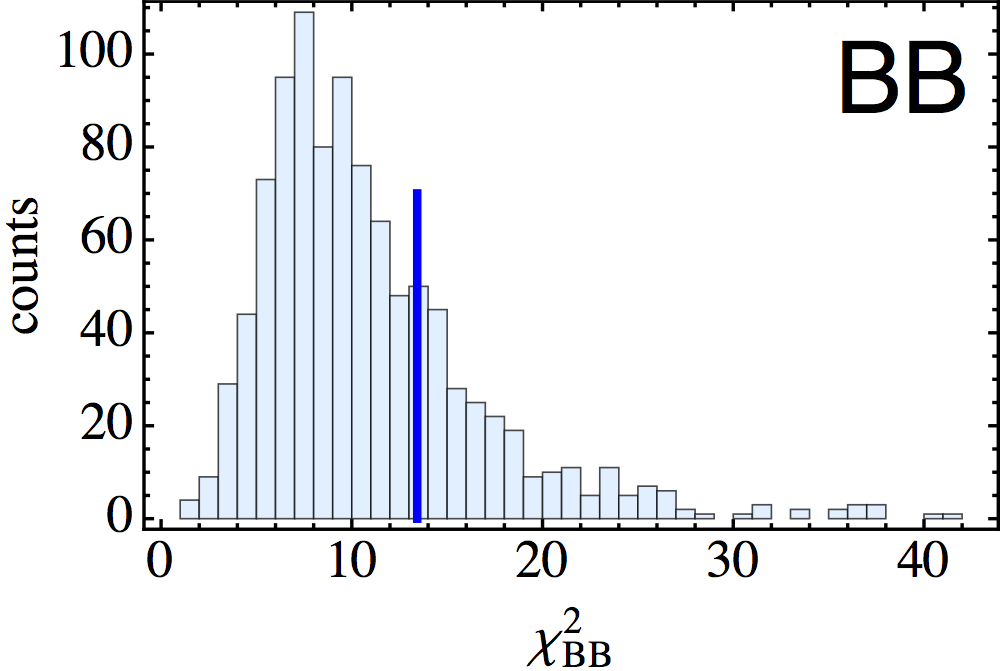}
       \includegraphics[width=0.45\columnwidth]{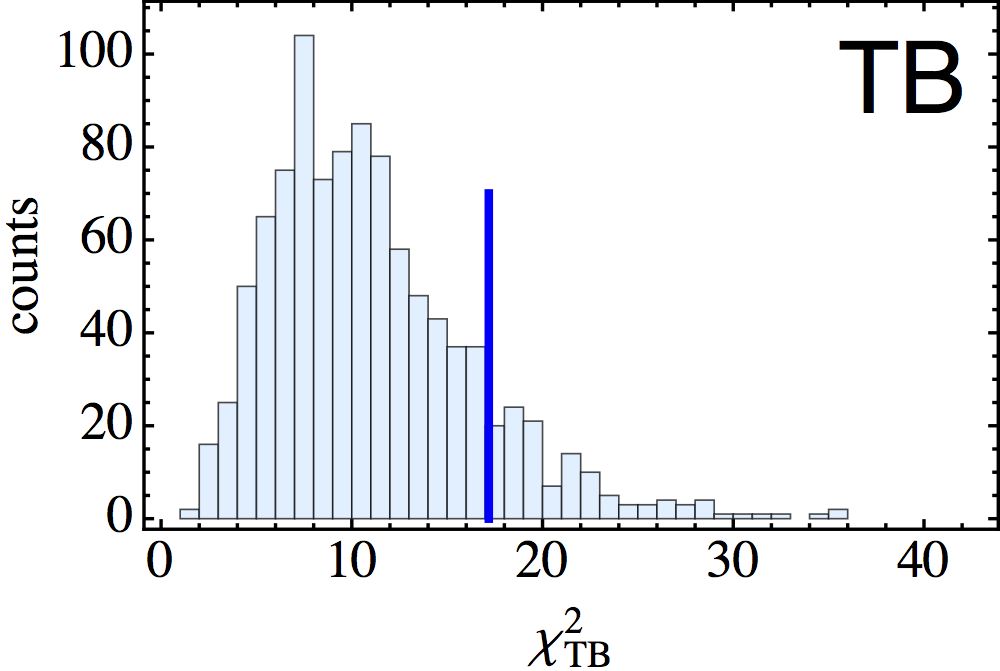}
       \includegraphics[width=0.45\columnwidth]{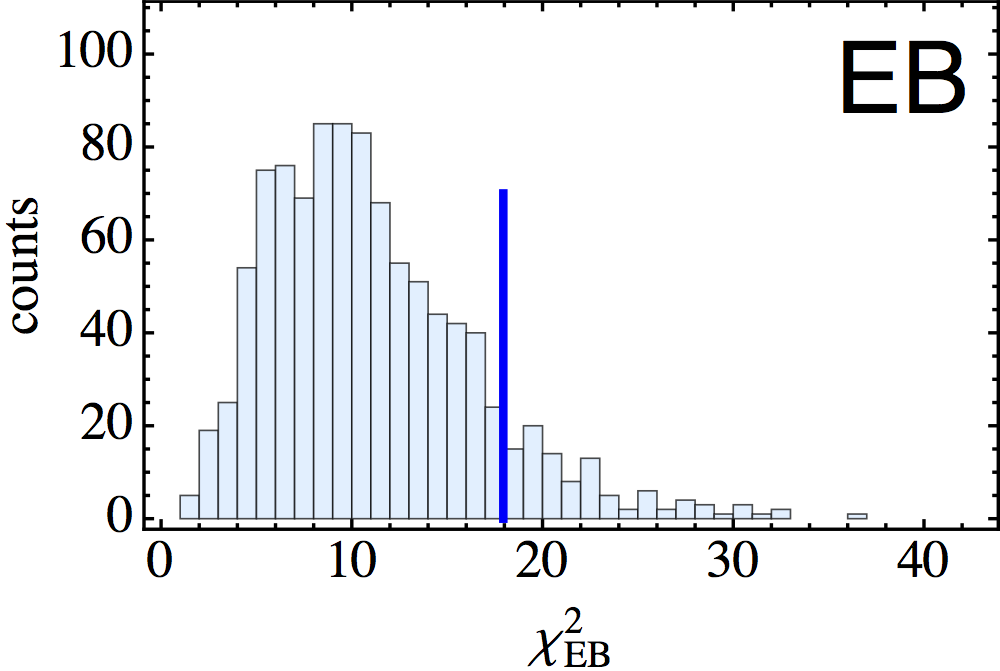}
    \caption{Empirical distribution of $\chi_\textrm{h}^2$ derived from 1000 simulations, for the case $\ell_\textrm{max}=12$ (see text). Vertical bars reindicate the observed values.}
    \label{fig:lowl_harmchi2}
\end{figure}

We first consider QML spectral estimates $C_\ell$ derived using \texttt{BolPol}. To test their consistency, we consider the following quantity:
\begin{equation}
\chi_\textrm{h}^2  = \sum_{\ell=2}^{\ell_\textrm{max}} (C_{\ell} - C_{\ell}^\textrm{th}) \, \tens{M}_{\ell \ell^{\prime}}^{-1} \, (C_{\ell} - C_{\ell}^\textrm{th}) 
\, ,
\label{chi2hred}
\end{equation}
where $\tens{M}_{\ell \ell^{\prime}} = \langle (C_{\ell} - C_{\ell}^\textrm{th}) (C_{\ell^{\prime}} - C_{\ell^{\prime}}^\textrm{th}) \rangle $,
 $C_{\ell}^\textrm{th}$ represents the fiducial \Planck\ 2015 $\Lambda$CDM model, and the average is taken over 1000 signal and noise simulations. The latter were generated using the noise covariance matrix given in Eq.\ (\ref{lowl_covmat}). We also use the simulations to sample the empirical distribution for $\chi_\textrm{h}^2$, considering both $\ell_\textrm{max}=12$ (shown in Fig.~\ref{fig:lowl_harmchi2}, along with the corresponding values obtained from the data) and $\ell_\textrm{max}=30$, for each of the six CMB polarized spectra. We report in Table~\ref{table:lowl_harmPTE} the empirical probability of observing a value of $\chi_\textrm{h}^2$ greater than for the data (hereafter, PTE). This test  supports the hypothesis that the observed polarized spectra are consistent with \Planck's best-fit cosmological model and the propagated instrumental uncertainties. We verified that the low PTE values obtained for $TE$ are related to the unusually high (but not intrinsically anomalous) estimates $9\le \ell \le11$, a range that does not contribute significantly to constraining $\tau$. For spectra involving $B$, the fiducial model is null, making this, in fact, a null test, probing instrumental characteristics and data processing independent of any cosmological assumptions.


\begin{table}[ht!]
\begingroup
\newdimen\tblskip \tblskip=5pt
\caption{Empirical probability of observing a value of $\chi_h^2$ greater than that calculated from the data.}
	\label{table:lowl_harmPTE}
\nointerlineskip
\vskip -3mm
\footnotesize
\setbox\tablebox=\vbox{
   \newdimen\digitwidth 
   \setbox0=\hbox{\rm 0} 
   \digitwidth=\wd0 
   \catcode`*=\active 
   \def*{\kern\digitwidth}
   \newdimen\signwidth 
   \setbox0=\hbox{+} 
   \signwidth=\wd0 
   \catcode`!=\active 
   \def!{\kern\signwidth}
\halign{\hbox to 0.8in{#\leaderfil}\tabskip=2em&
   \hfil#\hfil&
   \hfil#\hfil\tabskip=0pt\cr
\noalign{\doubleline}
\omit&\multispan2\hfil PTE [\%]\hfil\cr
\noalign{\vskip -3pt}
\omit&\multispan2\hrulefill\cr
\noalign{\vskip 3pt}
\omit\hfil Spectrum\hfil& $\ell_\textrm{max}=12$& $\ell_\textrm{max}=30$\cr
\noalign{\vskip 3pt\hrule\vskip 5pt}
$\TT$& 57.6& 94.2\cr
$\EE$& 12.0& 50.8\cr
$\TE$& *2.2& *2.3\cr
$\BB$& 24.7& 20.6\cr
$TB$& 12.3& 35.2\cr
$EB$& 10.2& *4.5\cr
\noalign{\vskip 5pt\hrule\vskip 3pt}}}
\endPlancktable 
\endgroup
\end{table}

\

\begin{figure}[htbp]
    \centering
        \includegraphics[width=0.9\columnwidth]{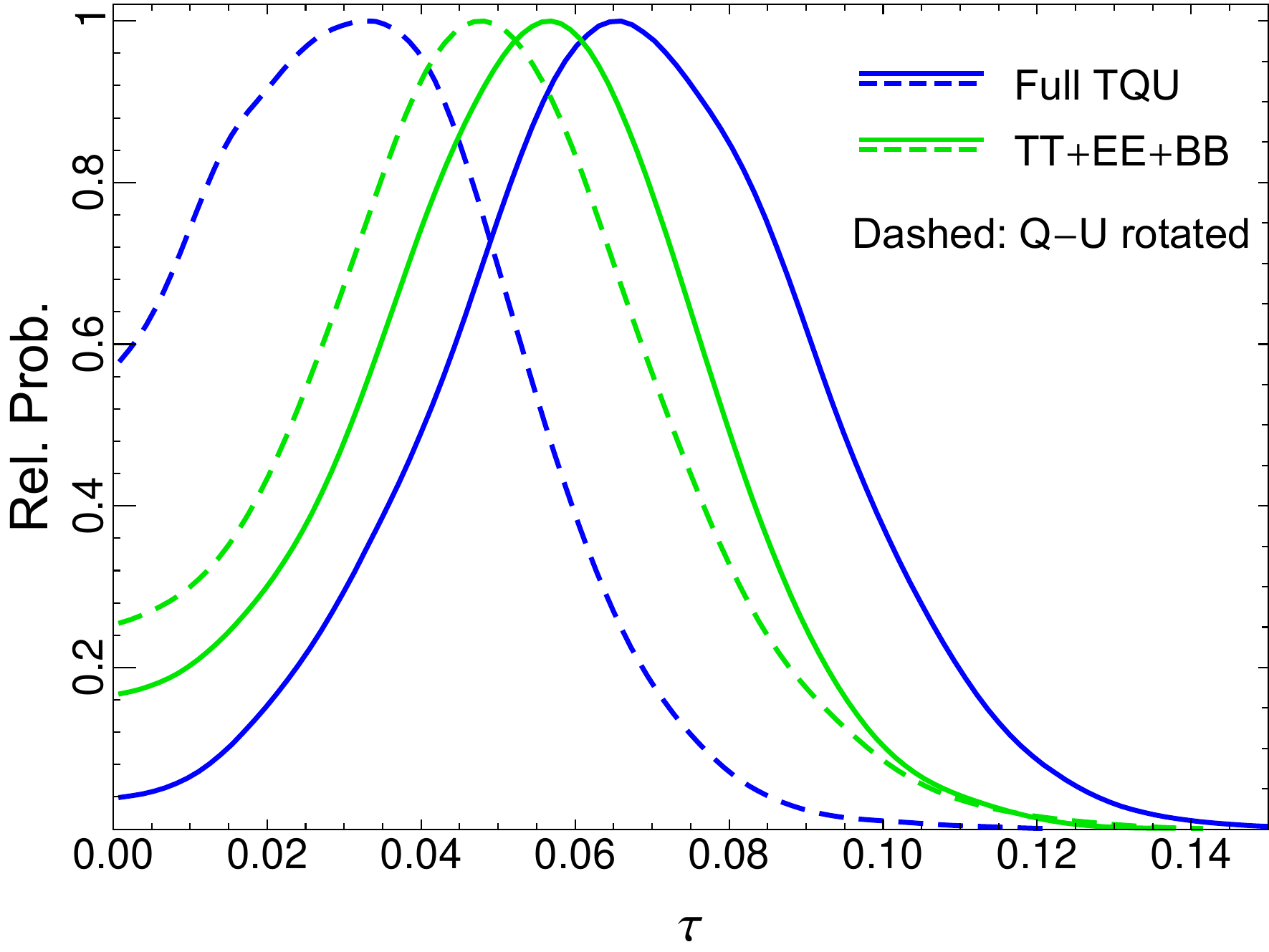}
     
    \caption{Posterior for $\tau$ for both rotated and unrotated likelihoods. The definition and colour convention of the datasets shown are the same as in the previous section (see Fig.\ \ref{fig:lowl-pol_cuts}), while solid and dashed lines distinguish the unrotated and rotated likelihood, respectively.}
    \label{fig:lowl-rotparams}
\end{figure}

In order to test the likelihood module, we first perform a $45\deg$ rotation of the reference frame. This leaves the $T$ map unaltered, while sending $Q \rightarrow -U$ and  $U \rightarrow Q$ (and, hence, $E \rightarrow -B$ and $B \rightarrow E$). The sub-blocks of the noise covariance matrix are rotated accordingly. We should not be able to detect a $\tau$ signal under these circumstances. Results are shown in Fig.\ \ref{fig:lowl-rotparams} for all the full $TQU$ and the $TT$+$EE$+$BB$ sub-block likelihoods presented in the previous section.
Indeed, rotating polarization reduces only slightly the constraining power in $\tau$ for the $TT$+$EE$+$BB$ case, suggesting the presence of comparable power in the latter two. On the other hand, $\tau$ is not detected at all when rotating the full $T,Q,U$ set, which includes $TE$ and $TB$. We interpret these results as further evidence that the $TE$ signal is relevant for constraining $\tau$, a result that cannot be reproduced by substituting $TB$ for $TE$. These 
findings appear consistent with the visual impression of the low-$\ell$ spectra of Fig.~\ref{fig:lowl-polarized_aps}. We have also verified that our results stand when $r$ is sampled. 
 
\begin{figure}[htbp]
    \centering
    \includegraphics[width=\columnwidth]{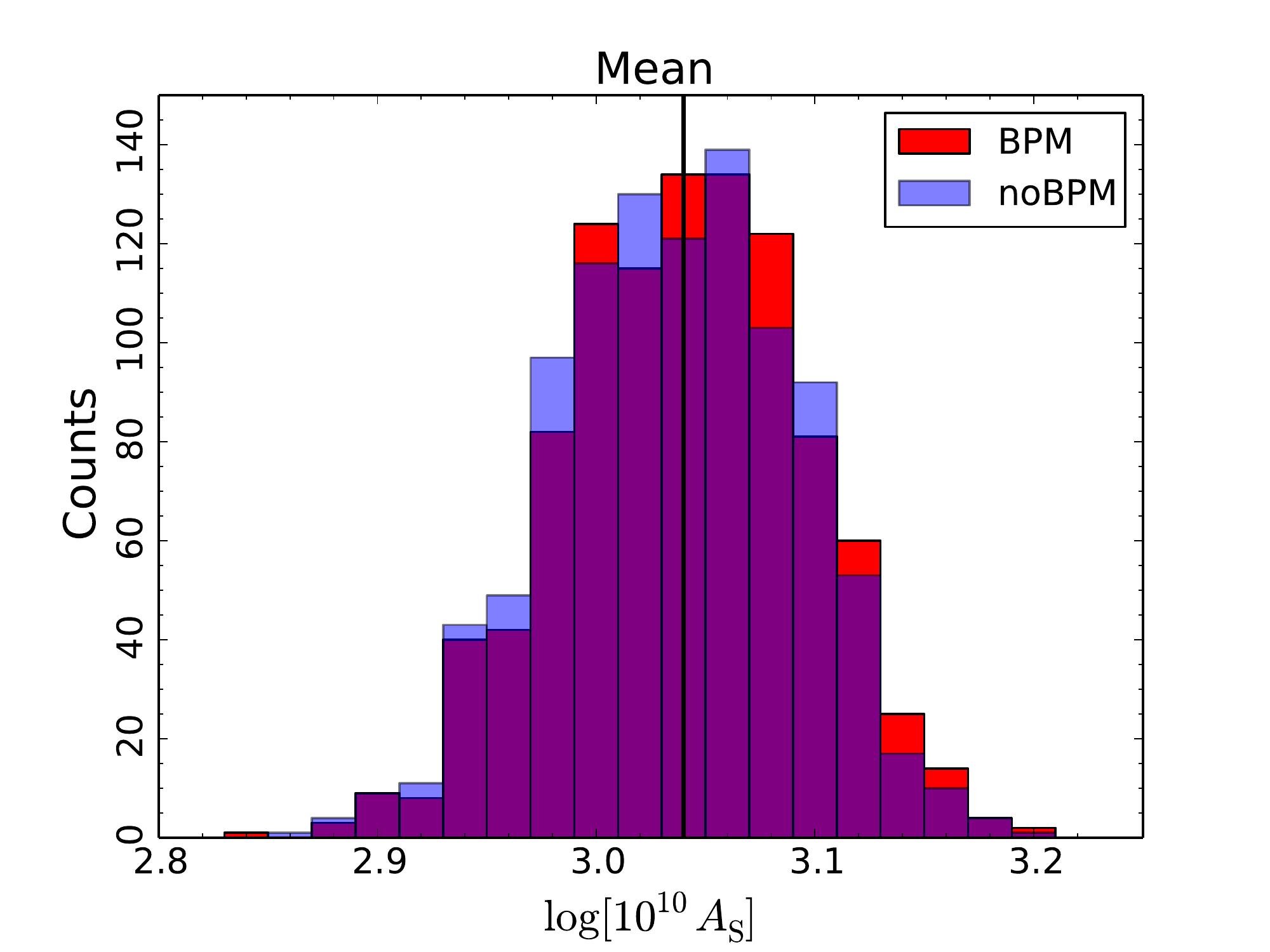}
        \includegraphics[width=\columnwidth]{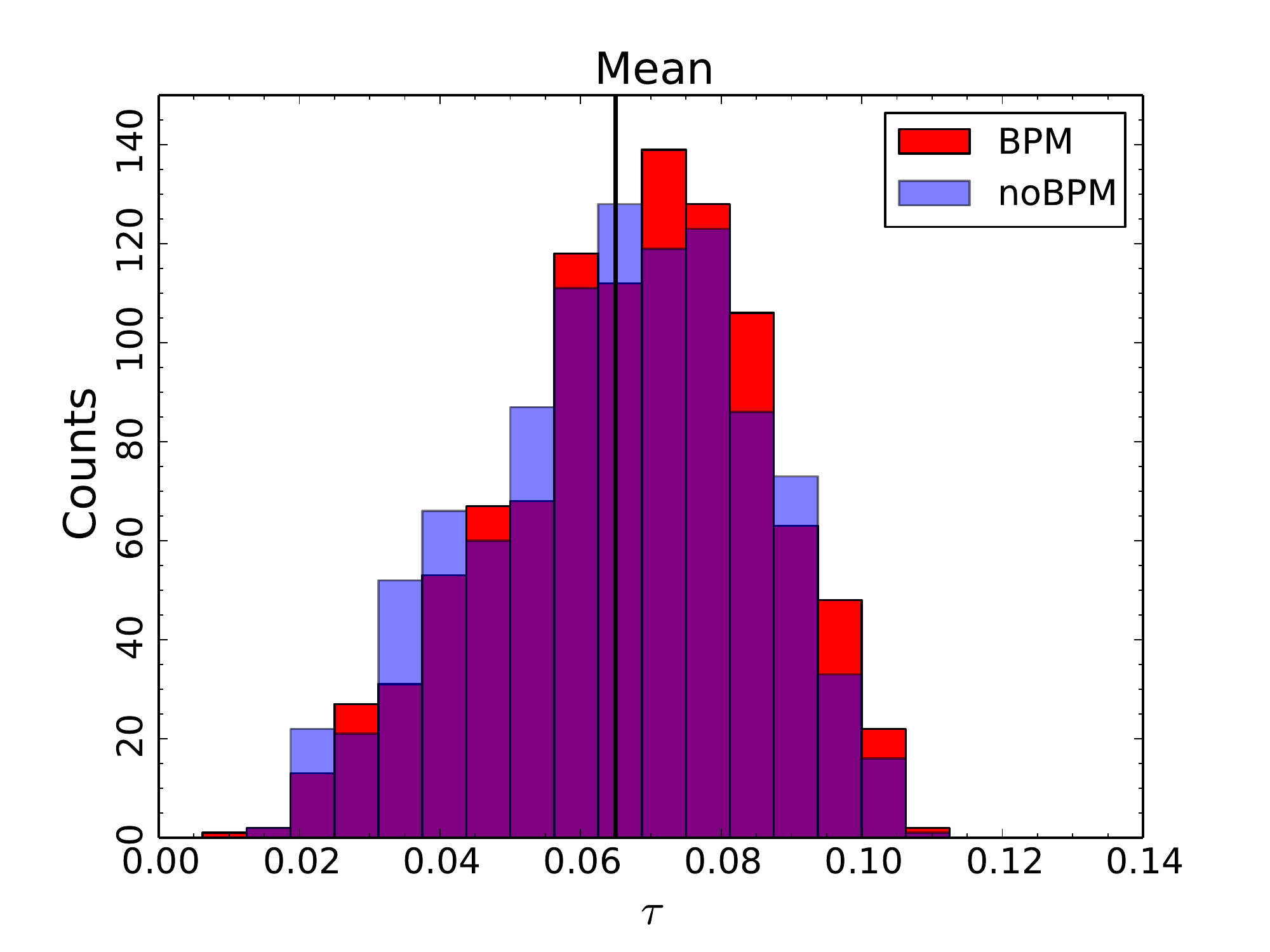}     
    \caption{Empirical distribution of the mean estimated values for $\log[10^{10}A_\textrm{s}]$ (top)  and $\tau$ (bottom), derived from 1000 FFP8 simulations (see text). For each simulation, we perform a full end-to-end run, including foreground cleaning and parameter estimation. Blue bars refer to simulations that do not include the instrumental bandpass mismatch, while red bars do. The violet bars flag the overlapping area, while the vertical black lines show the input parameters. We note that the (uncorrected) bandpass mismatch effect hardly changes the estimated parameters.}
    \label{fig:lowl-hist_tau}
\end{figure}

As a final test of the 2015 \Planck\ low-$\ell$ likelihood, we perform a full end-to-end Monte Carlo validation of its polarization part. For this, we use 1000 signal and noise \rev{full focal plane (FFP8) simulated maps \citep{planck2014-a14}}, whose resolution has been downgraded to $N_\mathrm{side} = 16$ using the same procedure as that applied to the data. We make use of a custom-made simulation set for the \Planck\ 70\,GHz channel, which does not include Surveys 2 and 4. For each simulation, we perform the foreground-subtraction procedure described in Sect.~\ref{sec:70ghz_pol} above, deriving foreground-cleaned maps and covariance matrices, which we use to feed the low-$\ell$ likelihood. As above, we sample only $\log[10^{10}A_\textrm{s}]$  and $\tau$, with all other parameters kept to their \Planck\ best-fit fiducial values. We consider two sets of polarized foreground simulations, with and without the instrumental bandpass mismatch at 30 and 70\,GHz. To emphasize the impact of bandpass mismatch, we do not attempt to correct the polarization maps for bandpass leakage. This choice marks a difference from what is done to real data, where the correction is performed \citep{planck2014-a03}; thus, the simulations that include the bandpass mismatch effect should be considered as a worst-case scenario. This notwithstanding, the impact of bandpass mismatch on estimated parameters is very small, as shown in Fig.\ \ref{fig:lowl-hist_tau} and detailed in Table~\ref{tab:tau_As_bpm_nobpm}. Even without accounting for bandpass mismatch, the bias is at most $1/10$ of the final $1\,\sigma$ error estimated from real data posteriors. The Monte Carlo analysis also enables us to validate the (Bayesian) confidence intervals estimated by \texttt{cosmomc} on data by comparing their empirical counterparts observed from the simulations. We find excellent agreement (see Table~\ref{tab:tau_As_bpm_nobpm}).

The validation described above only addresses the limited number of instrumental systematic effects that are modelled in the FFP8 simulations, i.e., the bandpass mismatch. Other systematics may in principle affect the measurement of polarization at large angular scales. To address this issue, we have carried out a detailed analysis to quantify the possible impact of LFI-specific instrumental effects in the 70\,GHz map
\citep[see][for details]{planck2014-a04}. Here we just report the main conclusion of that analysis, which estimates the final bias on  $\tau$ due to all known instrumental systematics to be at most  $0.005$, i.e., about \rev{0.25\,$\sigma$,  well below the final error budget}.

\begin{table*}[ht!]                 
\begingroup
\newdimen\tblskip \tblskip=5pt
\caption{Statistics for the empirical distribution of estimated cosmological parameters from the FFP8 simulations.$^{\rm a}$}                   
\label{tab:tau_As_bpm_nobpm}
\nointerlineskip
\vskip -3mm
\footnotesize
\setbox\tablebox=\vbox{
   \newdimen\digitwidth 
   \setbox0=\hbox{\rm 0} 
   \digitwidth=\wd0 
   \catcode`*=\active 
   \def*{\kern\digitwidth}
   \newdimen\signwidth 
   \setbox0=\hbox{+} 
   \signwidth=\wd0 
   \catcode`!=\active 
   \def!{\kern\signwidth}
%
\halign{\hbox to 1.0in{#\leaderfil}\tabskip=2em& 
   \hfil#\hfil\tabskip=1em&
   \hfil#\hfil\tabskip=1em& 
   \hfil#\hfil\tabskip=2em& 
   \hfil#\hfil\tabskip=1em& 
   \hfil#\hfil& 
   \hfil#\hfil\tabskip=2em& 
   \hfil#\hfil\tabskip=1em& 
   \hfil#\hfil\tabskip=0pt\cr
\noalign{\doubleline}
\omit&\multispan3\hfil{\tt Cosmomc} best-fit\hfil&\multispan3\hfil{\tt Cosmomc} mean \hfil&\multispan2\hfil Standard deviation\hfil\cr
\noalign{\vskip -3pt}
\omit&\multispan3\hrulefill&\multispan3\hrulefill&\multispan2\hrulefill\cr
\noalign{\vskip 1pt}
\noalign{\vskip 2pt}
\omit\hfil Parameter\hfil& mean& $\sigma$& $\Delta$& mean&$\sigma$& $\Delta $&mean& $\sigma$\cr
\noalign{\vskip 5pt\hrule\vskip 5pt}
$\tau$&                         $0.0641\pm0.0007$& $0.0227$& $-4.1\%$& $0.0650\pm0.0006$& $0.0190$& $*-0.1\%$& $0.0186\pm0.0001$& $0.0030$\cr
$\tau^\ast$&                    $0.0665\pm0.0007$& $0.0226$& $+6.4\%$& $0.0672\pm0.0006$& $0.0189$& $+11.0\%$& $0.0185\pm0.0001$& $0.0031$\cr
$\log[10^{10}A_\textrm{s}]$&     $3.035\pm0.002$& $0.059$&    $-9.4\%$& $3.036\pm0.002$&  $0.055$&  $*-8.0\%$& $0.0535\pm0.0001$& $0.0032$\cr
$\log[10^{10}A_\textrm{s}^\ast]$&$3.039\pm0.002$& $0.059$&    $-1.7\%$& $3.040\pm0.002$&  $0.056$&  $*-0.3\%$& $0.0533\pm0.0001$& $0.0033$\cr
\noalign{\vskip 3pt\hrule\vskip 3pt}
}}
\endPlancktablewide
\tablenote {{\rm a}} Mean and standard deviation for cosmological parameters, computed over the empirical distributions for the estimated best-fit (left columns) and mean  (centre columns) values, as obtained from the FFP8 simulation set. Asterisked parameters flag the presence of (untreated) bandpass mismatch in the simulated maps. The columns labeled $\Delta$ give the bias from the input values in units of the empirical standard deviation. This bias always remains small, being at most $0.1\,\sigma$. Also, note how the empirical standard deviations for the estimated parameters measured from the simulations are very close to the standard errors inferred from \texttt{cosmomc} posteriors on real data.  The rightmost columns show statistics of the standard errors for parameter posteriors, estimated from each \texttt{cosmomc} run. 
The input FFP8 values are $\tau_\mathrm{input} = 0.0650$ and $\log[10^{10}A_\textrm{s}]_\mathrm{input} = 3.040$.\par
\endgroup
\end{table*}

\subsection{Comparison with \WMAP-9 polarization cleaned with \Planck\ 353\,GHz}\label{Planck_wmap_lowP}

In \citetalias{planck2013-p08}, we attempted to clean the \WMAP-9 low resolution maps using a preliminary version of \Planck\ 353\,GHz polarization. This resulted in an approximately $1\,\sigma$ shift towards lower values of $\tau$, providing the first evidence based on CMB observations that the \WMAP\ best-fit value for the optical depth may have been biased high. We repeat the analysis here with the 2015 \Planck\ products. We employ the  
procedure described in \citet{bennett2012}, which is  similar to that described above for \Planck\ 2015. However, in contrast to the \Planck\ 70\,GHz foreground cleaning, we do not attempt to optimize the foreground mask based on a goodness-of-fit analysis, but stick to the processing and analysis masks made available by the \WMAP\ team. \WMAP's P06 mask is significantly smaller than the 70\,GHz mask used in the \Planck\ likelihood, leaving $73.4\,\%$ of the sky. Specifically we minimize the quadratic form of Eq.~(\ref{chi2_alphabeta}), separately for the Ka, Q, and V channels from the \WMAP-9 release, but using \WMAP-9's own K channel as a synchrotron tracer rather than \Planck\ 30\,GHz.\footnote{To exactly mimic the procedure followed by the \WMAP team, we exclude the signal correlation matrix from the noise component of the $\chi^2$ form. We have checked, however, that the impact of this choice is negligible for \WMAP.} The purpose of the latter choice is to minimize the differences with respect to \WMAP's own analysis. However, unlike the \WMAP-9 native likelihood products, which operate at $N_\mathrm{side} = 8$ in polarization, we use
$N_\mathrm{side} = 16$ in $Q$ and $U$, for consistency with the \Planck\ analysis. The scalings we find are consistent with those from \WMAP \citep{bennett2012} for $\alpha$ in both Ka and Q.  However, we find less good agreement for the higher-frequency V channel, where our scaling is roughly $25\,\%$ lower than that reported in \WMAP's own analysis.\footnote{There is little point in comparing the scalings obtained for dust, as \WMAP employs a model which is not calibrated to physical units.}  We combine the three cleaned channels in a noise-weighted average to obtain a three-band map and an associated covariance matrix. 
\begin{table}[ht!]
\begingroup
\caption{Scalings for synchrotron ($\alpha$) and dust ($\beta$) obtained for \WMAP, when \WMAP K band and \Planck\ 353\,GHz data are used as templates.}
\label{table:lowl_WMAPscalings}
\nointerlineskip
\vskip -3mm
\footnotesize
\setbox\tablebox=\vbox{
   \newdimen\digitwidth 
   \setbox0=\hbox{\rm 0} 
   \digitwidth=\wd0 
   \catcode`*=\active 
   \def*{\kern\digitwidth}
   \newdimen\signwidth 
   \setbox0=\hbox{+} 
   \signwidth=\wd0 
   \catcode`!=\active 
   \def!{\kern\signwidth}
\halign{\hbox to 0.8in{#\leaderfil}\tabskip=2em&
  \hfil#\hfil& 
  \hfil#\hfil\tabskip=0pt\cr
\noalign{\doubleline}
\omit\hfil Band\hfil& $\alpha$& $\beta$\cr
\noalign{\vskip 3pt\hrule\vskip 5pt}
Ka& $0.3170\pm0.0016$& $0.0030\pm0.0002$\cr
 Q& $0.1684\pm0.0014$& $0.0031\pm0.0003$\cr
V& $0.0436\pm0.0017$& $0.0079 \pm 0.0003$\cr
\noalign{\vskip 5pt\hrule\vskip 3pt}}}
\endPlancktable
\endgroup
\end{table}

We evaluate the consistency of the low-frequency \WMAP\ and \Planck\ 70\GHz\ low-$\ell$ maps. Restricting the analysis to the intersection of the \WMAP\ P06 and \Planck\ R1.50 masks ($f_
\mathrm{sky}=45.3\,\%$), we evaluate half-sum and half-difference $Q$ and $U$ maps. We then compute the quantity $\chi^2_\mathrm{sd} = \vec{m}^\mathrm{T} \tens{N}^{-1} \vec{m}$ where $\vec{m}$ is 
either the half-sum or the half-difference $[Q,U]$ combination  and $N$ is the corresponding noise covariance matrix. Assuming that $\chi^2_\mathrm{sd}$ is $\chi^2$ distributed with $2786$ 
degrees of freedom we find a $\textrm{PTE}(\chi^2>\chi^2_\mathrm{sd}) = 1.3 \times 10^{-5}$ 
(reduced $\chi^2=1.116$) for the half-sum, and $\textrm{PTE} = 0.84$ (reduced $\chi^2=0.973$)  for 
the half-difference. This strongly suggests that the latter is consistent with the assumed noise, and that the common signal present in the half-sum map is wiped out in the difference. 

\begin{figure}[htbp]
    \centering
        \includegraphics[width=\columnwidth]{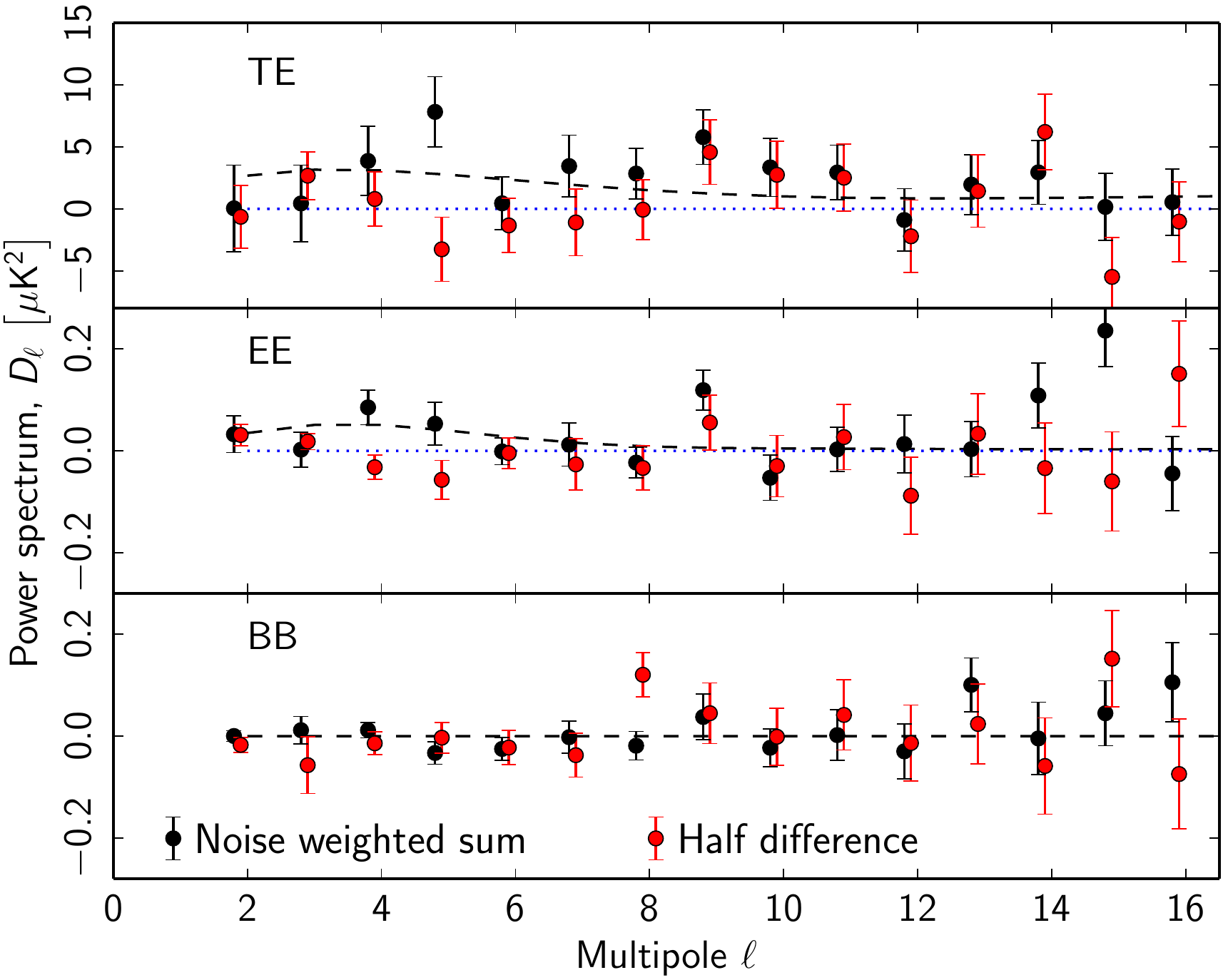}
     
    \caption{\texttt{BolPol} spectra for the noise-weighted sum (black) and half-difference (red) \WMAP\ and \Planck\ combinations. The temperature map employed is always the  \texttt{Commander} map described in Sect.\ \ref{sec:commander_lowl} above. The fiducial model shown has $\tau=0.065$. 
    } 
       \label{fig:qml_WMAP_hs_hd}
\end{figure}

We also produce noise-weighted sums of the low-frequency \WMAP\ and \Planck\ 70\GHz\ low-resolution $Q$ and $U$ maps,
evaluated in the union of the \WMAP\ P06 and \Planck\ R1.50 masks ($f_\mathrm{sky}=73.8\,\%$).
We compute \texttt{BolPol} spectra for the noise-weighted sum and half-difference combinations. 
These $EE$, $TE$, and $BB$ spectra are shown in Fig.\ \ref{fig:qml_WMAP_hs_hd}
and are evaluated in the intersection of the P06 and R1.50 masks. The spectra also support the hypothesis that there is a common signal between the two experiments in the typical 
multipole range of the reionization bump. In fact, considering multipoles up to $\ell_\mathrm{max} = 12$ we find an empirical PTE for the spectra of the half-difference map of $6.8\,\%$ for $EE$ 
and 9.5\,\% for $TE$, derived from the analysis of $10000$ simulated noise maps. Under the same hypothesis, but considering the noise-weighted sum, the PTE for $EE$ drops to $0.8\,\%$, while 
that for $TE$ is below the resolution allowed by the simulation set (PTE $< 0.1\,\%$). 
The $BB$ spectrum, on the other hand, is compatible with a null signal in both the noise-weighted sum map ($\mathrm{PTE}=47.5\,\%$) and the half-difference map ($\mathrm{PTE}=36.6\,\%$). 

\begin{figure}[htbp]
    \centering
        \includegraphics[width=\columnwidth]{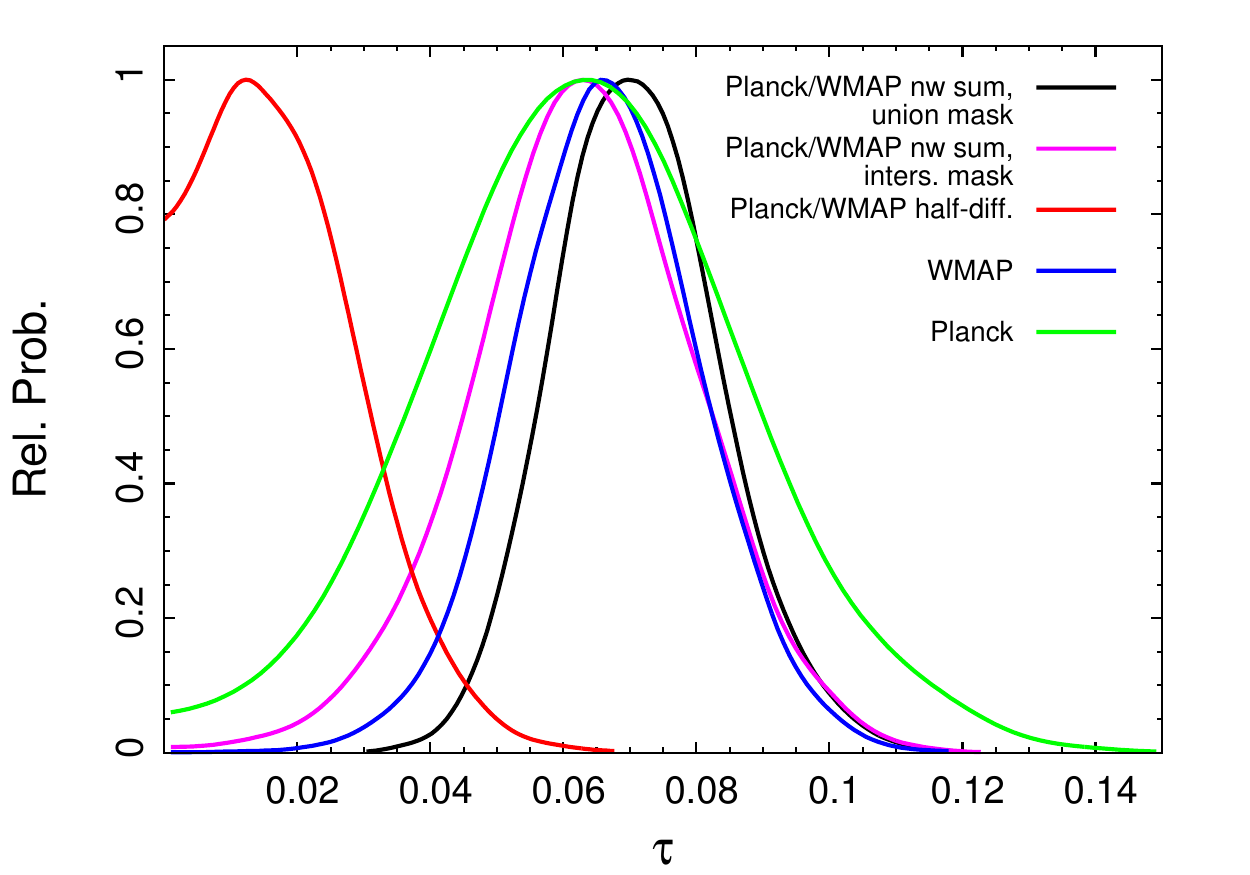}
    \caption{Posterior probabilities for $\tau$ from the \WMAP (cleaned with \Planck\ 353 GHz as a dust template) and \Planck\ combinations listed in the legend. Results are presented for the noise-weighted sum both in the union and the intersection of the two analysis masks. The half-difference map is consistent with a null detection, as expected.}
    \label{fig:tau_WMAP_LFI}
\end{figure}

We use the \Planck\ and \WMAP\ map combinations to perform parameter estimates from low-$\ell$ data only. We show here results from sampling $\log[10^{10}A_\textrm{s}]$, $\tau$, and the tensor-to-scalar ratio $r$, with all other parameters kept to the \Planck\ 2015 best fit (the case with $r=0$ produces similar results). Figure\ \ref{fig:tau_WMAP_LFI} shows the posterior probability for $\tau$ for several \Planck\ and \WMAP\ combinations. They are all consistent, except the \Planck\ and \WMAP\ half-difference case, which yields a null detection for $\tau$ --- as it should. As above, we always employ the \texttt{Commander} map in temperature. Table~\ref{table:params_WMAP_LFI} gives the mean values for the sampled parameters, and for the derived parameters $z_\textrm{re}$ (mean redshift of reionization) and $A_\textrm{s}e^{-2\tau}$. Results from a joint analysis of the \WMAP-based low-$\ell$ polarization likelihoods presented here and the \Planck\ high-$\ell$ likelihood are discussed in Sect.~\ref{sec:wmap9}. 

\begin{table}[ht!]
\begingroup
\caption{Selected parameters estimated from the low-$\ell$ likelihood, for  \Planck, \WMAP and their noise-weighted combination.$^{\rm a}$}
	\label{table:params_WMAP_LFI}
\nointerlineskip
\vskip -3mm
\footnotesize
\setbox\tablebox=\vbox{
   \newdimen\digitwidth 
   \setbox0=\hbox{\rm 0} 
   \digitwidth=\wd0 
   \catcode`*=\active 
   \def*{\kern\digitwidth}
   \newdimen\signwidth 
   \setbox0=\hbox{+} 
   \signwidth=\wd0 
   \catcode`!=\active 
   \def!{\kern\signwidth}
 \openup 4pt
\halign{\hbox to 0.9in{#\leaderfil}\tabskip=2em&
  \hfil#\hfil& 
  \hfil#\hfil\tabskip=1.2em&
  \hfil#\hfil \tabskip=0pt\cr
\noalign{\doubleline}
\omit\hfil Parameter\hfil& \Planck& \WMAP& \Planck/\WMAP\cr
\noalign{\vskip 3pt\hrule\vskip 5pt}
$\tau$& $0.064^{+0.022}_{-0.023}$& $0.067^{+0.013}_{-0.013}$& $0.071^{+0.011}_{-0.013}$\cr
$z_\textrm{re}$& $8.5^{+2.5}_{-2.1}$& $8.9^{+1.3}_{-1.3}$&  $9.3^{+1.1}_{-1.1}$\cr
 $\log[10^{10}A_\textrm{s}]$& $2.79^{+0.19}_{-0.09}$& $2.87^{+0.11}_{-0.06}$& $2.88^{+0.10}_{-0.06}$\cr
 $r$& $[0,\,0.90]$&  $[0,\,0.52]$& $[0,\,0.48]$\cr
 $A_\textrm{s} e^{-2\tau}$& $1.45^{+0.24}_{-0.14}$& $1.55^{+0.16}_{-0.10}$& $1.55^{+0.14}_{-0.11}$\cr 
\noalign{\vskip 5pt\hrule\vskip 3pt}}}
\endPlancktable
\tablenote {{\rm a}} The temperature map used is always \Planck\ \texttt{Commander}. Only $\log[10^{10}A_\textrm{s}]$, $\tau$, and $r$ are sampled. The other $\Lambda$CDM parameters are kept fixed to the \Planck\ 2015 fiducial. The likelihood for the noise-weighted combination is evaluated in the union of the WMAP P06 and Planck R1.50 masks.\par
\endgroup
\end{table}

\section{High-multipole likelihood}\label{sec:hil}

At high multipoles ($\ell>29$), as in \citetalias{planck2013-p08}, we use a likelihood function based on pseudo-$C_\ell$s calculated from \Planck\ \HFI data, as well as further parameters describing the contribution of foreground astrophysical emission and instrumental effects (e.g., calibration, beams). Aside from the data themselves, the main advances over 2013 include the use of high-$\ell$ polarization information along with more detailed models of foregrounds and instrumental effects. 

Section~\ref{sec:highl:intro} introduces the high-$\ell$ statistical description, Sect.~\ref{sec:highl:data} describes the data we use, Sects.~\ref{sec:foreground_modelling} and \ref{sec:instrument} describe foreground and instrumental modelling, and Sect.~\ref{sec:covariance} describes the covariance matrix between multipoles and spectra. Section~\ref{sec:valid-sims} validates the overall approach on realistic simulations, \rev{while Sect.~\ref{sec:e2e} addresses the question of the potential impact of low-level instrumental systematics imperfectly corrected by the DPC processing}. The reference results generated with the high multipole likelihood are described in Sect.~\ref{sec:highlbase}. A detailed assessment of these results is presented in Sect.~\ref{sec:hil-ass}.

\subsection{Statistical description}
\label{sec:highl:intro}

Assuming a Gaussian distribution for the CMB temperature anisotropies and polarization, all of the statistical information contained in the \Planck\ maps can be compressed into the likelihood of the temperature and polarization auto- and cross-power spectra. In the case of a perfect CMB observation of the full sky (with spatially uniform noise and isotropic beam-smearing), we know the joint distribution of the empirical temperature and polarization power spectra and can build an exact likelihood, which takes the simple form of an inverse Wishart distribution, uncorrelated between multipoles. For a single power spectrum (\ie ignoring polarization and temperature cross-spectra between detectors) the likelihood for each multipole $\ell$ simplifies to an inverse $\chi^2$ distribution with $2\ell+1$ degrees of freedom. At high enough $\ell$, the central limit theorem ensures that the shape of the likelihood is very close to that of a Gaussian distributed variable. This remains true for the inverse Wishart generalization to multiple spectra, where, for each $\ell$, the shape of the joint spectra and cross-spectra likelihood approaches that of a correlated Gaussian \citep{HL08, 2012A&A...542A..60E}. 
In the simple full-sky case, the correlations are easy to compute \citep{HL08}, and only depend on the theoretical CMB $\TT$, $\TE$, and $\EE$ spectra. For small excursions around a fiducial cosmology, as is the case here given the constraining power of the \Planck\ data, one can show that computing the covariance matrix at a fiducial model is sufficient \citep{HL08}. 

The data, however, differ from the idealized case. In particular, foreground astrophysical processes contribute to the temperature and polarization maps. As we see in Sect.~\ref{sec:foreground_modelling}, the main foregrounds in the frequency range we use are emission from dust in our Galaxy, the clustered and Poisson contributions from the cosmic infrared background (CIB), and radio point sources. Depending on the scale and frequency, foreground emission can be a significant contribution to the data, or even exceed the CMB. This is particularly true for dust near the Galactic plane, and for the strongest point sources. We excise the most contaminated regions of the sky (see Sect.~\ref{sec:masks}). The remaining foreground contamination is taken into account in our model, using the fact that CMB and foregrounds have different emission laws; this enables them to be separated while estimating parameters.

Foregrounds also violate the Gaussian approximation assumed above. The dust distribution, in particular, is clearly non-Gaussian. Following \citetalias{planck2013-p08}, however, we assume that outside the masked regions we can neglect non-Gaussian features and assume that, as for the CMB, all the  relevant statistical information about the foregrounds is encoded in the spatial power spectra. This assumption is verified to be sufficient for our purposes in Sect.~\ref{sec:valid-sims}, where we assess the accuracy of the cosmological parameter constraints in realistic Monte Carlo simulations that include data-based (non-Gaussian) foregrounds.

Cutting out the foreground-contaminated regions from our maps biases the empirical power spectrum estimates. We de-bias them using the \polspice\footnote{http://www2.iap.fr/users/hivon/software/PolSpice/} algorithm
\citep{chon2004} and, following \citetalias{planck2013-p08}, we take the correlation between multipoles induced by the mask and de-biasing into account when computing our covariance matrix. The masked-sky covariance matrix is computed using the equations in \citetalias{planck2013-p08}, which are extended to the case of polarization in Appendix~\ref{app:hil_eqs}. Those equations also take into account the inhomogeneous distribution of coloured noise on the sky using a heuristic approach.
The approximation of the covariance matrix that can be obtained from those equations is only valid for some specific mask properties, and for high enough multipoles. In particular, as discussed in Appendix~\ref{app:hil_pts_mask_correction}, correlations induced by  point sources cannot be faithfully described in our approximation. Similarly, Monte Carlo simulations have shown that our analytic approximation loses accuracy around $\ell=30$. We correct for both of those effects using empirical estimates from Monte Carlo simulations. 
The computation of the covariance matrix requires knowledge of both the CMB and foreground power spectra, as well as the map characteristics (beams, noise, sky coverage). The CMB and foreground power spectra are obtained iteratively from previous, less accurate versions of the likelihood. 

At this stage, we would thus construct our likelihood approximation by compressing all of the individual \Planck\ detector data into mask-corrected (pseudo-) cross-spectra, and build a grand likelihood using these spectra and the corresponding analytical covariance matrix:
\begin{equation}
-\ln{\cal L}(\vec{\hat C} | \vec{C}(\theta)) = \frac{1}{2} \left[\vec{\hat C} - \vec{C}(\theta)\right]^{\tens{T}} \tens{C}^{-1}
 \left[\vec{\hat C} - \vec{C}(\theta)\right] + {\rm const}\;,
 \label{eq:basic-likelihood}
 \end{equation}
 where  $\vec{\hat C}$ is the data vector, $\vec{C}(\theta)$ is the model with parameters $\theta$, and $\tens{C}$ is the covariance matrix.
This formalism enables us to separately marginalize over or condition upon different components of the model vector, separately treating cases such as individual frequency-dependent spectra, or temperature and polarization spectra. 
Obviously, \Planck\ maps at different frequencies have different constraining powers on the underlying CMB, and following \citetalias{planck2013-p08} we use this to impose and assess various cuts to keep only the most relevant data.  

We therefore consider only the three best CMB \Planck\ channels, \ie 100\,GHz, 143\,GHz, and 217\,GHz, in the multipole range where they have  significant CMB contributions and low enough foreground contamination after masking; \rev{we therefore did not directly include the adjacent channels at 70~GHz and 350~GHz in the analysis. In particular, including the 70~GHz data would not bring much at large scales where the results are already cosmic variance limited, and would entail additional complexity in foreground modelling (synchrotron at large scales, additional radio sources excisions at small scales)}. The cuts \rev{in multipole ranges} is be described in detail in Sect.~\ref{sec:cuts}. 
Further, in order to achieve a significant reduction in the covariance matrix size (and computation time), we compress the data vector (and accordingly the covariance matrix), both by co-adding the individual detectors for each frequency and by binning the combined power spectra.
We also co-add the two different $\TE$ and $\ET$ inter-frequency cross-spectra into a single $\TE$ spectrum for each pair of frequencies. This compression is lossless in the case without foregrounds. The exact content of the data vector is discussed in Sect.~\ref{sec:highl:data}.

The model vector $\vec{C}(\theta)$ must represent the content of the data vector. It can be written schematically as
\begin{eqnarray}
\left. C^{XY}_{\nu \times \nu'} \right|_\ell(\theta) &=& \left. M^{XY}_{ZW,\nu \times \nu'}\right|_\ell (\theta_{\mathrm{inst}})\  \left. C^{ZW,\mathrm{sky}}_{\nu \times \nu'} \right|_\ell(\theta) + \left. N^{XY}_{\nu \times \nu'} \right|_\ell(\theta_{\mathrm{inst}}),\nonumber\\
\left. C^{ZW,\mathrm{sky}}_{\nu \times \nu'} \right|_\ell(\theta) &=& \left. C^{ZW,\mathrm{cmb}} \right|_\ell(\theta) + \left. C^{ZW,\mathrm{fg}}_{\nu \times \nu'} \right|_\ell(\theta),
\label{eq:spectrum-model}
\end{eqnarray}
where $\left. C^{XY}_{\nu \times \nu'} \right|_\ell(\theta)$ is the element of the model vector corresponding to the multipole $\ell$ of the $XY$ cross-spectra ($X$ and $Y$ being either $T$ or $E$) between the pair of frequencies $\nu$ and $\nu'$. This element of the model originates from the sum of the microwave emission of the sky, \ie the CMB ($\left. C^{ZW,\mathrm{cmb}} \right|_\ell(\theta)$) which does not depend of the pair of frequencies (all maps are in units of  $K_\mathrm{cmb}$), and foreground ($\left. C^{ZW,\mathrm{fg}}_{\nu \times \nu'} \right|_\ell(\theta)$). Section~\ref{sec:foreground_modelling} describes the foreground modelling.
The mixing matrix $\left. M^{XY}_{ZW,\nu \times \nu'}\right|_\ell (\theta_{\mathrm{inst}})$ accounts for imperfect calibration, imperfect beam correction, and possible leakage between temperature and polarization. It does depend on the pair of frequencies and can depend on the multipole\footnote{\rev{We assume an $\ell$-diagonal mixing matrix here. This is not necessarily the case, as sub-pixel beam effects, for example, can induce mode couplings. As discussed in Sect.~\ref{sec:beam-uncert}, those were estimated in \citetalias{planck2013-p08} and found to be negligible for temperature. They are not investigated further in this paper.}} when accounting for imperfect beams and leakages.  Finally, the noise term $\left. N^{XY}_{\nu \times \nu'} \right|_\ell(\theta_{\mathrm{inst}})$
accounts for the possible correlated noise in the $XY$ cross-spectra for the pair of frequencies $\nu \times \nu'$.   Sections~\ref{sec:beams} and \ref{sec:instrument} describe our 
instrument model.

 \subsection{Data}  \label{sec:highl:data}
 
The data vector $\hat{C}$ in the likelihood equation
(Eq.~\ref{eq:basic-likelihood}) is constructed from concatenated
temperature and polarization components,
\begin{equation}
\vec{\hat{C}} = \left( \vec{\hat{C}}^{TT}, \vec{\hat{C}}^{EE}, \vec{\hat{C}}^{TE} \right) \, ,
\end{equation}
which, in turn, comprise the following frequency-averaged spectra:
\begin{align}
\vec{\hat{C}}^{TT} \! &= \left(\vec{\hat{C}}^{TT}_{100 \times 100},
\vec{\hat{C}}^{TT}_{143 \times 143}, \vec{\hat{C}}^{TT}_{143
  \times 217}, \vec{\hat{C}}^{TT}_{217 \times 217}\right) \\
  \vec{\hat{C}}^{EE} \! &= \left(\vec{\hat{C}}^{EE}_{100 \times 100},
  \vec{\hat{C}}^{EE}_{100 \times 143}, \vec{\hat{C}}^{EE}_{100
  \times 217}, \vec{\hat{C}}^{EE}_{143 \times 143}, \vec{\hat{C}}^{EE}_{143 \times 217}, \vec{\hat{C}}^{EE}_{217 \times 217}\right) \\
  \vec{\hat{C}}^{TE} \! &= \left(\vec{\hat{C}}^{TE}_{100 \times 100},
  \vec{\hat{C}}^{TE}_{100 \times 143}, \vec{\hat{C}}^{TE}_{100
  \times 217}, \vec{\hat{C}}^{TE}_{143 \times 143}, \vec{\hat{C}}^{TE
 }_{143 \times 217}, \vec{\hat{C}}^{TE}_{217 \times
  217}\right) .
\end{align}
The $\TT$ data selection is very similar to \citetalias{planck2013-p08}. We still discard the $100 \times 143$ and $100 \times 217$ cross-spectra in their entirety. They contain little extra information about the CMB, as they are strongly correlated with the high S/N maps at 143 and 217\,GHz. Including them, in fact, would only give information about the foreground contributions in these cross-spectra, at the expense of a larger covariance matrix with increased condition number. In $\TE$ and $\EE$, however, the situation is different since the overall S/N is significantly lower for all spectra, so a foreground model of comparatively low complexity can be used and it is beneficial to retain all the available cross-spectra.

We obtain cross power spectra at the frequencies $\nu \times \nu'$ using
weighted averages of the individual beam-deconvolved, mask-corrected
half-mission (HM) map power spectra,
\begin{equation}
\left. \hat{C}_{\nu \times \nu'}^{XY} \right|_{\ell} = \sum_{(i, j) \in
  (\nu,\nu')} \left. w^{XY}_{i,j}\right|_{\ell} \times \left.\hat{C}^{XY}_{i, j}\right|_{\ell} \, ,
\label{eq:hil_cl_freq_avg}
\end{equation}
where $XY \in \{ TT, TE, EE \}$, and $\left. w^{XY}_{i,j}\right|_{\ell}$ is the multipole-dependent inverse-variance weight for the detector-set map combination $(i, j)$, derived from its
covariance matrix (see Sect.~\ref{sec:covariance}). For $XY = TE$, we
further add the $\ET$ power spectra of the same frequency combination to
the sum of Eq.~(\ref{eq:hil_cl_freq_avg}); \ie the average includes
the correlation of temperature information from detector-set $i$ and
polarization information of detector-set $j$ and vice versa. 

We construct the \planck high-multipole likelihood solely from the
\HFI\ channels at 100, 143, and 217\,GHz. These perform best as they have high S/N combined with manageably low foreground contamination. As in
\citetalias{planck2013-p08}, we only employ 70\,GHz \LFI data for cross-checks (in the high-$\ell$ regime), while the \HFI\ 353\,GHz and 545\,GHz maps are used to determine the dust model.

\subsubsection{Detector combinations}
\label{sec:hil:detcomb}

Table~\ref{tab:detsets} summarizes the main characteristics of individual \HFI detector sets used in the construction of the likelihood function. As discussed in Sect.~\ref{sec:highl:intro}, the likelihood does not use the cross-spectra from individual detector-set maps; instead, we first combine all those contributing at each frequency to form weighted averages. As in 2013, we disregard all auto-power-spectra as the precision required to remove their noise bias is difficult to attain and even small residuals may hamper a robust inference of cosmological parameters \citepalias{planck2013-p08}.

In 2015, the additional data available from full-mission observations enables us to construct nearly independent full-sky maps from the first and the second halves of the mission duration. We constructed  cross-spectra by cross-correlating the two half-mission maps, ignoring the half-mission auto-spectra at the expense of a very small increase in the uncertainties. This differs from the procedure used in 2013, when we estimated cross-spectra between detectors or detector-sets, and has the advantage of minimizing possible contributions from systematic effects that are correlated in the time domain.

\begin{table}[ht!] 
\begingroup 
\newdimen\tblskip \tblskip=5pt
\caption{Detector sets used to make the maps for this analysis. \label{tab:detsets} }
\vskip -6mm
\footnotesize 
\setbox\tablebox=\vbox{
\newdimen\digitwidth
\setbox0=\hbox{\rm 0}
\digitwidth=\wd0
\catcode`*=\active
\def*{\kern\digitwidth}
\newdimen\signwidth
\setbox0=\hbox{+}
\signwidth=\wd0
\catcode`!=\active
\def!{\kern\signwidth}
\newdimen\decimalwidth
\setbox0=\hbox{.}
\decimalwidth=\wd0
\catcode`@=\active
\def@{\kern\signwidth}
\halign{ \hbox to 0.8in{#\leaderfil}\tabskip=1em& 
    \hfil#\hfil\tabskip=2em& 
    \hfil#\hfil& 
    #\hfil& 
    \hfil#\hfil \tabskip=0pt\cr
\noalign{\doubleline}
\omit& $\nu$&&&\cr
\omit\hfil Set\hfil& [GHz]& Type&\omit\hfil Detectors\hfil& FWHM\cr
\noalign{\vskip 3pt\hrule\vskip 5pt}
100-ds0& 100& PSB&  8 detectors& 9\parcm68\cr
\noalign{\vskip 4pt}
100-ds1& 100& PSB& 1a+1b + 4a+4b\cr
100-ds2& 100& PSB& 2a+2b + 3a+3b\cr
\noalign{\vskip 8pt}
143-ds0& 143& MIX& 11 detectors& 7\parcm30\cr
\noalign{\vskip 4pt}
143-ds1& 143& PSB& 1a+1b + 3a+3b\cr
143-ds2& 143& PSB& 2a+2b + 4a+4b\cr
143-ds3& 143& SWB& 143-5\cr
143-ds4& 143& SWB& 143-6\cr
143-ds5& 143& SWB& 143-7\cr
\noalign{\vskip 8pt}
217-ds0& 217& MIX& 12 detectors& 5\parcm02\cr
\noalign{\vskip 4pt}
217-ds1& 217& PSB& 5a+5b + 7a+7b\cr
217-ds2& 217& PSB& 6a+6b + 8a+8b\cr
217-ds3& 217& SWB& 217-1\cr
217-ds4& 217& SWB& 217-2\cr
217-ds5& 217& SWB& 217-3\cr
217-ds6& 217& SWB& 217-4\cr
\noalign{\vskip 8pt}
353-ds0& 353& MIX& 12 detectors& 4\parcm94\cr
545-ds0& 545& SWB& 3 detectors& 4\parcm83\cr
\noalign{\vskip 5pt\hrule\vskip 3pt}
}}
\endPlancktable 
\tablenote {{\rm a}} SWBs may be used individually; PSBs are used in pairs (denoted a and b), and we consider only the maps estimated from two pairs of PSBs. The FWHM quoted here correspond to a Gaussian whose solid angle is equivalent to that of the effective beam; see \citet{planck2014-a09} for details.\par
\endgroup
\end{table} 

The main motivation for this change from 2013 is that the correlated noise between detectors (at the same or different frequencies) is no longer small enough to be neglected (see Sect.~\ref{sec:noise_model}). 
And while the correction for the ``feature'' around $\ell = 1800$, which was (correctly) attributed to residual \HeJT\ cooler lines in 2013 \citep{planck2013-p03}, has been improved in the 2015 TOI processing pipeline \citep{planck2014-a08}, cross-spectra between the two half-mission periods can help to suppress time-dependent systematics, as argued by \citet{2013arXiv1312.3313S}. Still, in order to enable further consistency checks, we also build a likelihood based on cross-spectra between full-mission detector-set maps, applying a correction for the effect of correlated noise. The result illustrates that not much sensitivity is lost with half-mission cross-spectra (see the whisker labelled ``DS'' in Figs.~\ref{fig:wiskerTT}, \ref{fig:wiskerTText}, and \ref{fig:wiskerpol}).

\subsubsection{Masks}
\label{sec:masks}

Temperature and polarization masks are used to discard areas of the sky that are strongly contaminated by foreground emission. The choice of masks is a trade-off between maximizing the sky coverage to minimize sample variance, and the complexity and potentially insufficient accuracy of the foreground model needed in order to deal with regions of stronger foreground emission. The masks combine a Galactic mask, excluding mostly low Galactic-latitude regions, and a point-source mask. We aim to maximize the sky fraction with demonstrably robust results (see Sect.~\ref{sec:robust-fsky} for such a test).  

Temperature masks are obtained by merging the apodized Galactic, CO, and point-source masks described in Appendix~\ref{app:masks}. In polarization, as discussed in \citet{planck2014-XXX}, even at 100\,GHz foregrounds are dominated by the dust emission, so for polarization analysis we employ the same apodized Galactic masks as we  use for temperature, because they are also effective in reducing fluctuations in polarized dust emission at the relatively small scales covered by the high-$\ell$ likelihood (contrary to the large Galactic scales), but we do not include a compact-source mask because polarized emission from extragalactic foregrounds is negligible at the frequencies of interest \citep{actpol2014PS,sptpol2014PS}. 

\begin{table}[ht!] 
\begingroup 
\newdimen\tblskip \tblskip=5pt
\caption{Masks used for the high-$\ell$ analysis.$^{\rm a}$} 
\label{tab:mask_fsky}
\vskip -6mm
\footnotesize
\setbox\tablebox=\vbox{
\newdimen\digitwidth
\setbox0=\hbox{\rm 0}
\digitwidth=\wd0
\catcode`*=\active
\def*{\kern\digitwidth}
\newdimen\signwidth
\setbox0=\hbox{+}
\signwidth=\wd0
\catcode`!=\active
\def!{\kern\signwidth}
\newdimen\decimalwidth
\setbox0=\hbox{.}
\decimalwidth=\wd0
\catcode`@=\active
\def@{\kern\signwidth}
\halign{ 
\hbox to 1in{#\leaderfil}\tabskip=0em& 
    \hfil#\hfil\tabskip=1em& 
    \hfil#\hfil\tabskip=0pt\cr
\noalign{\doubleline}
\omit&\multispan2\hfil Mask\hfil\cr
\noalign{\vskip -3pt}
\omit\hfil Frequency\hfil&\multispan2\hrulefill\cr
\omit\hfil [GHz]\hfil&Temperature&Polarization\cr
\noalign{\vskip 3pt\hrule\vskip 5pt}
$100$&T66&P70\cr
$143$&T57&P50\cr
$217$&T47&P41\cr
\noalign{\vskip 5pt\hrule\vskip 3pt}
}}
\endPlancktable 
\tablenote {{\rm a}} Temperature and polarization masks used in the likelihood are identified by T and P, followed by two digits that  specify  the retained sky fraction (percent). As discussed in Appendix~\ref{app:masks}, T masks are derived by merging apodized Galactic, CO, and extragalactic sources masks. P masks, instead, are simply given by apodized Galactic masks.\par
\endgroup
\end{table}


Table~\ref{tab:mask_fsky} lists the masks used in the likelihood at each frequency channel. We refer throughout to the masks by explicitly indicating the percentage of the sky they {retain}: T66, T57, T47 for temperature and P70, P50, P41 for polarization. G70, G60, G50, and G41 denote the apodized Galactic masks. As noted above, the apodized P70, P50, and P41 polarization masks are identical to the G70, G50, and G41 Galactic masks.

\begin{figure}
\begin{centering}
\includegraphics[width=88.0mm,angle=0]{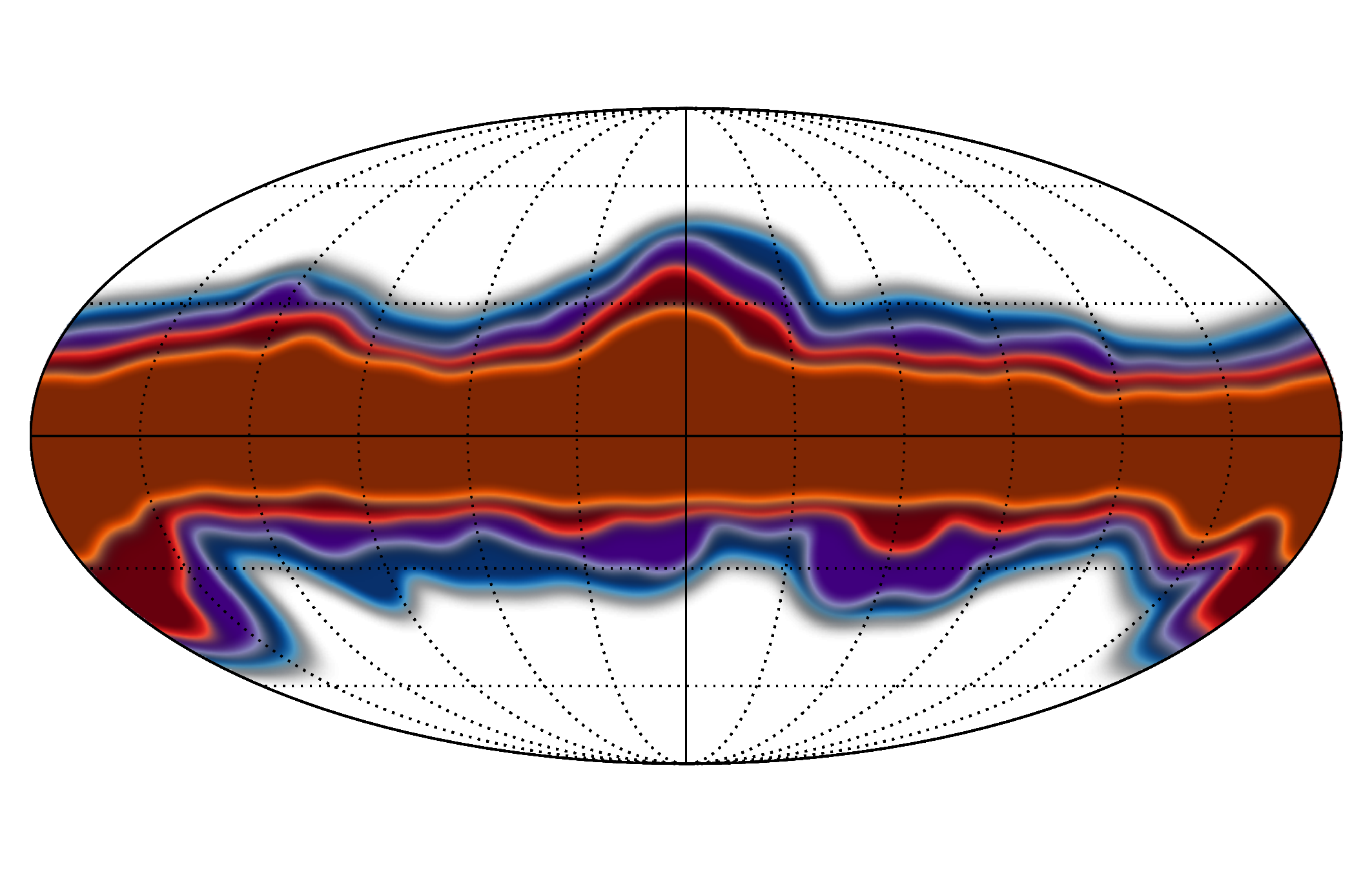}\\
\includegraphics[width=88.0mm,angle=0]{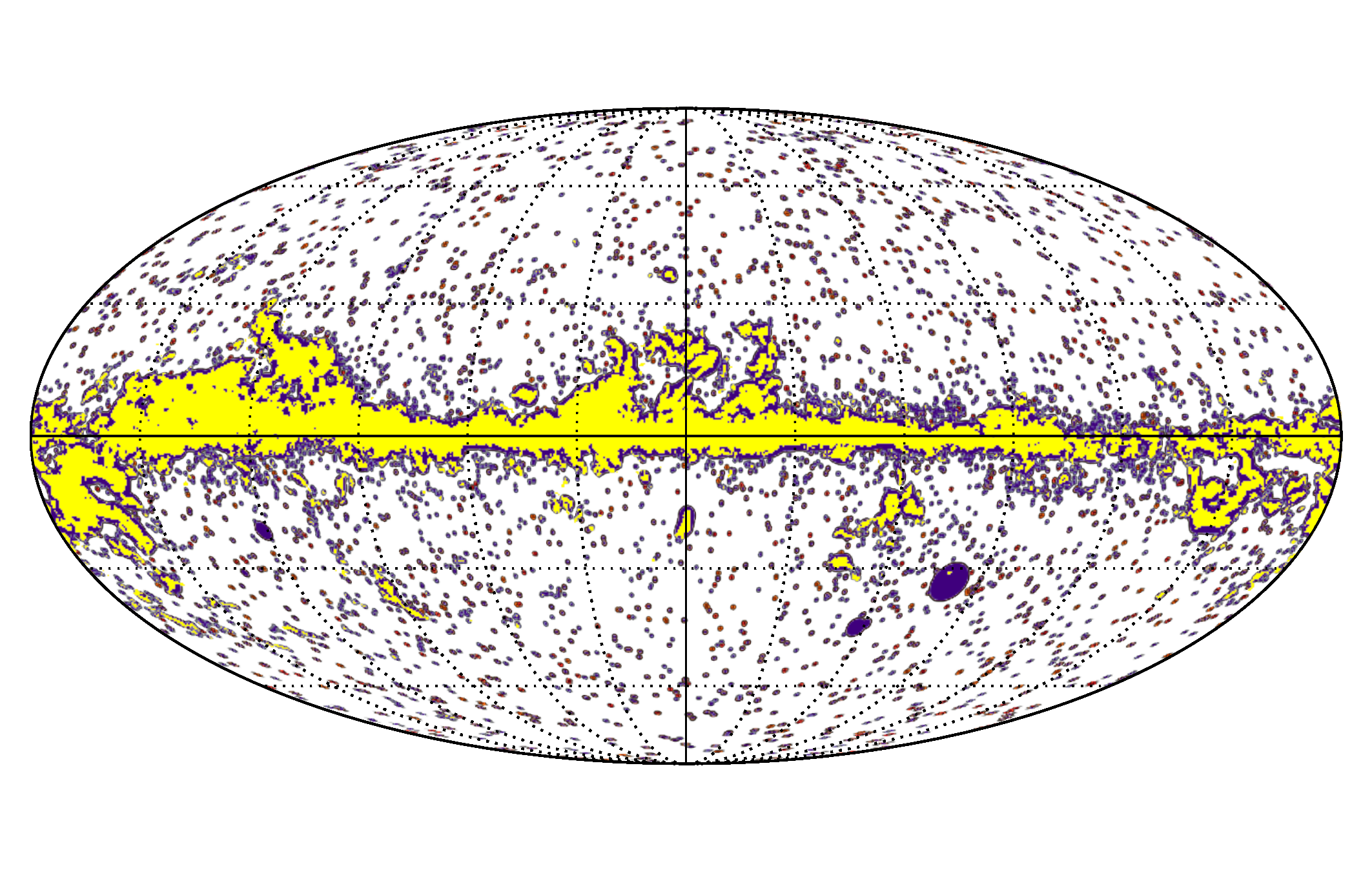}
\caption{\emph{Top}: Apodized Galactic masks: G41 (blue), G50 (purple), G60 (red), and G70 (orange); these are identical to the polarization masks P41 (used at 217\,GHz),  P50 (143\,GHz), P70 (100\,GHz). \emph{Bottom}: Extragalactic-object masks for 217\,GHz (purple), 143\,GHz (red), and 100\,GHz (orange);  the CO mask  is shown in yellow.}
\label{fig:hil:mask_stack}
\end{centering}
\end{figure}

The Galactic masks are obtained by thresholding the smoothed, CMB-cleaned 353\,GHz map at different levels to obtain different sky coverage. All of the Galactic masks are apodized with a 4\pdeg71 FWHM ($\sigma=2\degree$) Gaussian  window function to localize the mask power in multipole space. In order to adapt to the different relative strengths of signal, noise, and foregrounds, we use different sky coverage for temperature and polarization, ranging in effective sky fraction from 41\,\% to 70\,\% depending on the frequency. The Galactic masks are shown in Fig.~\ref{fig:hil:mask_stack}. 

\begin{figure}
\begin{centering}
\includegraphics[width=88.0mm,angle=0]{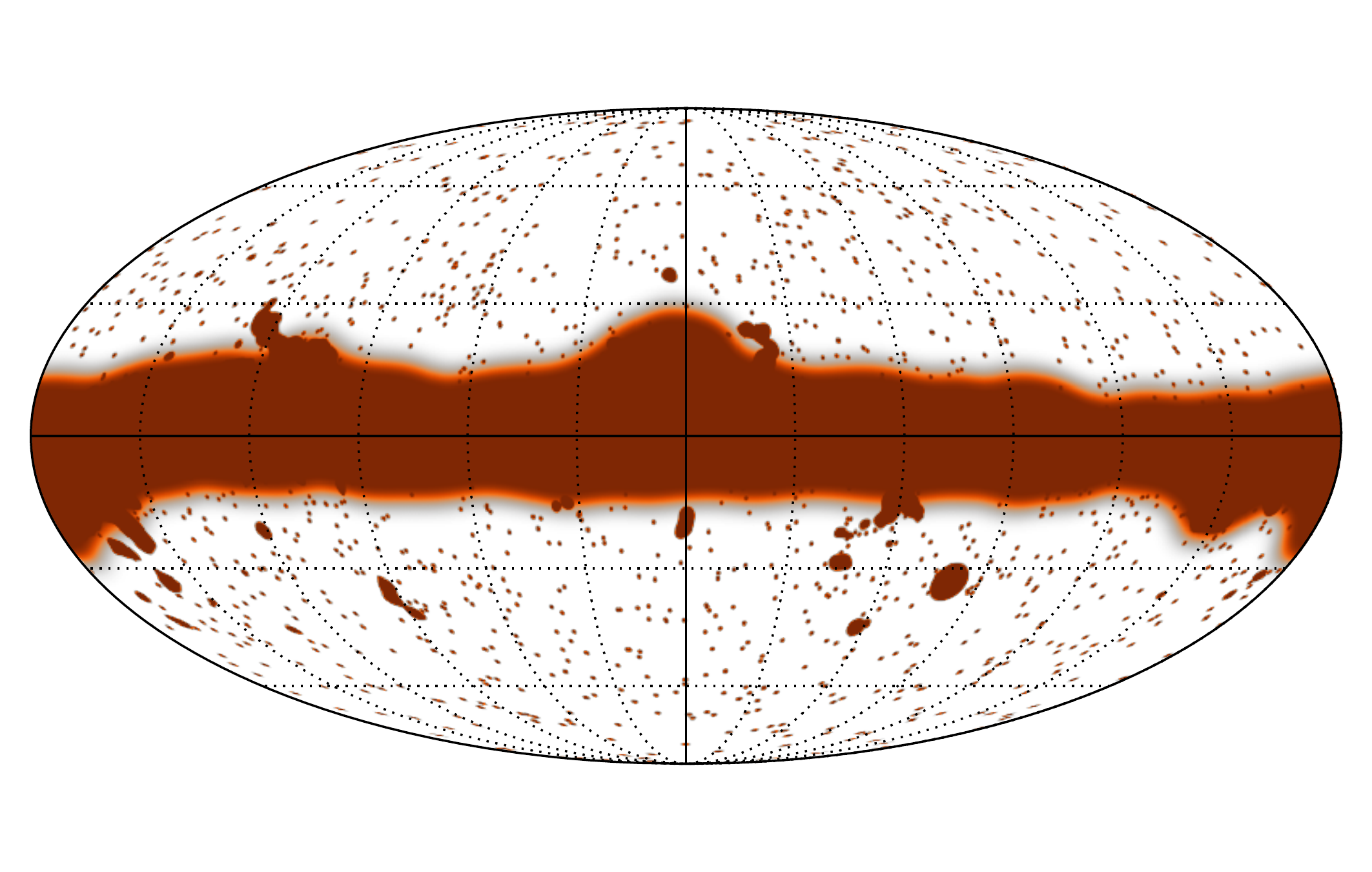}\\
\includegraphics[width=88.0mm,angle=0]{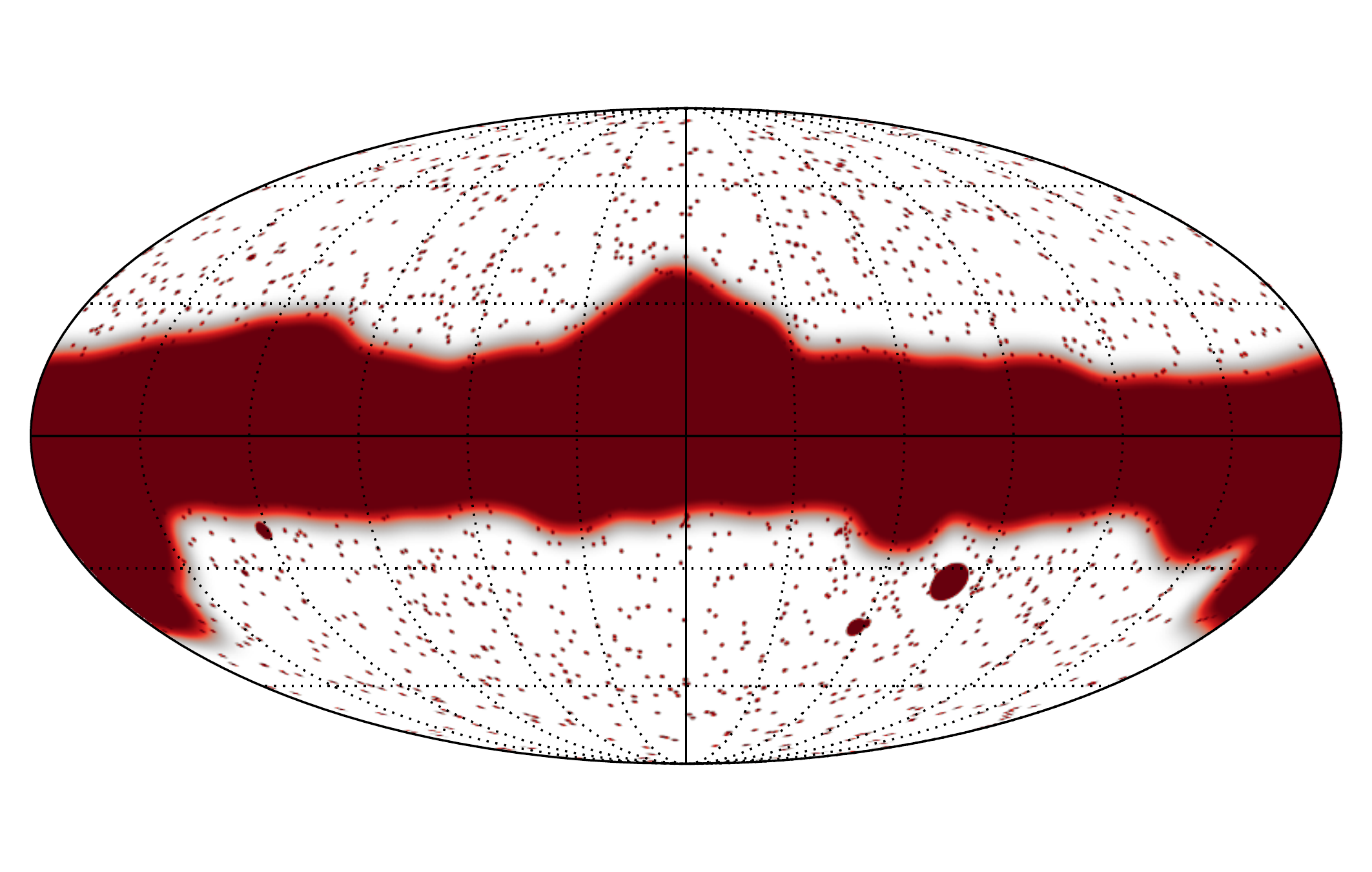}\\
\includegraphics[width=88.0mm,angle=0]{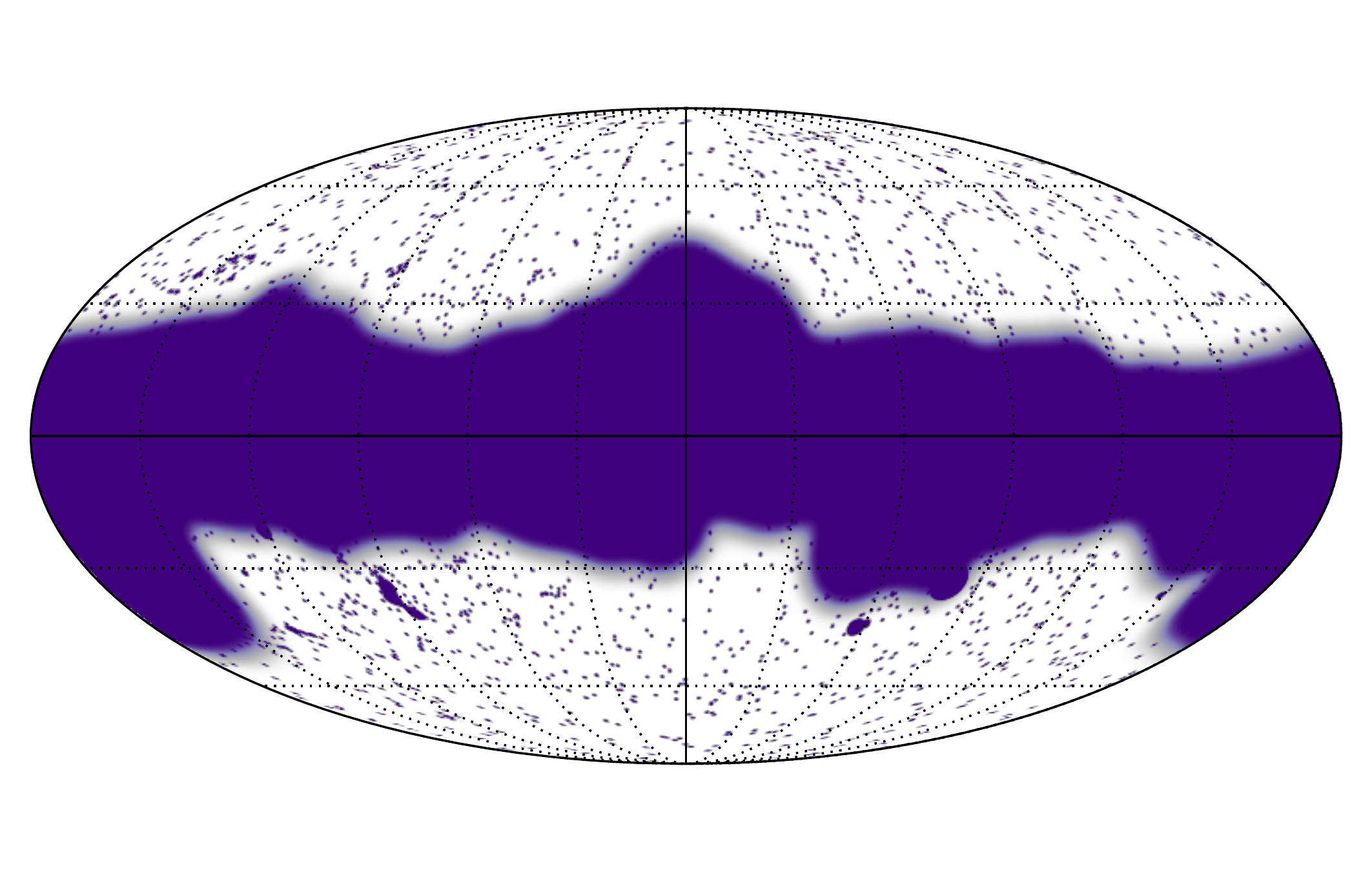}
\caption{Top to bottom: temperature masks for 100\,GHz (T66), 143\,GHz (T57) and 217\,GHz (T47). The colour scheme is the same as  in Fig.~\ref{fig:hil:mask_stack}.}
\label{fig:hil:mask_T}
\end{centering}
\end{figure}

For temperature we use the G70, G60, and G50 Galactic masks at (respectively) 100\,GHz, 143\,GHz, and 217\,GHz. For the first release of \Planck\ cosmological data \citep{planck2014-a13} we made more conservative choices of masks than in this paper (${f_{\rm sky}} = 49\,\%, 31\,\%$, and $31\,\%$ at, respectively, 100, 143, and 217\GHz, to be compared to ${f_{\rm sky}} = 66\,\%, 57\,\%$, and $47\,\%$). Admitting more sky into the analysis requires a thorough assessment of the robustness of the foreground modelling, and in particular of the Galactic dust model (see Sect.~\ref{sec:foreground_modelling}). When retaining more sky close to the Galactic plane at 100\,GHz, maps start to show contamination by CO emission that also needs to be masked. This was not the case in the \Planck\ 2013 analysis. We therefore build a CO mask as described in Appendix~\ref{app:masks}. Once we apply this mask, the residual foreground at 100\,GHz is consistent with dust and there is no evidence for other anisotropic foreground components, as shown by the double-difference spectra between the $100$\, GHz band  and the $143$\,GHz band where there is no CO line (Sect.~\ref{sec:dust}). We also use the CO mask at 217\,GHz, although we expect it to have a smaller impact  since at this frequency CO emission is fainter and the applied Galactic cut wider. 
The extragalactic ``point'' source masks in fact include both point sources and extended objects; they are used only with the temperature maps. Unlike in 2013, we use a different source mask for each frequency, taking into account different source selection and beam sizes (see Appendix~\ref{app:masks}).
Both the CO and the extragalactic object masks are apodized with a 30\arcm\ FWHM Gaussian window function.
The different extragalactic masks, as well as the CO mask, are shown in Fig.~\ref{fig:hil:mask_stack}. The resulting mask combinations for temperature are shown in Fig.~\ref{fig:hil:mask_T}.

\subsubsection{Beam and transfer functions}
\label{sec:beams}

The response to a point source is given by the combination of the optical response of the \Planck\ telescope and feed-horns (the optical beam) with the detector time response and electronic transfer function (whose effects are partially removed during the TOI processing). This response pattern is referred to as the ``scanning beam''. It is measured on planet transits \citep{planck2014-a09}.  However, the value in any pixel resulting from the map-making operation comes from a sum over many different elements of the timeline, each of which has hit the pixel in a different location and from a different direction. Furthermore, combined maps are weighted sums of individual detectors. All of these result in an ``effective beam'' window function encoding the multiplicative effect on the angular power spectrum. We note that beam non-circularity and the non-uniform scanning of the sky create differences between auto- and cross-detector beam window functions \citep{planck2013-p03c}.  

In the likelihood analysis, we correct for this by using the effective beam window function corresponding to each specific spectrum;  the window functions are calculated with the \quickbeam\ pipeline, except for one of the alternative analyses ({\xfaster}) which relied on the \febecop\ window functions (see \citealt{planck2014-a08}, \citealt{planck2013-p03c}, and references therein for details of these two codes). In Section~\ref{sec:beam-uncert} we discuss the model of their uncertainties.

\subsubsection{Multipole range}
\label{sec:cuts}

 Following the approach taken in \citetalias{planck2013-p08}, we use specifically tailored multipole ranges for each frequency-pair spectrum. In general, we exclude multipoles where either the S/N is too low for the data to contribute significant constraints on the CMB, or the level of foreground contamination is so high that the foreground contribution to the power spectra cannot be modelled sufficiently accurately; high foreground contamination would also require us to consider possible non-Gaussian terms in the estimation of the likelihood covariance matrix. We impose the same $\ell$ cuts for the detector-set and half-mission likelihoods for comparison, and we exclude the $\ell>1200$ range for the $100\times100$ spectra, where the correlated noise correction is rather uncertain. 

\begin{figure}[htb]
  \includegraphics[angle=0,width=0.5\textwidth]{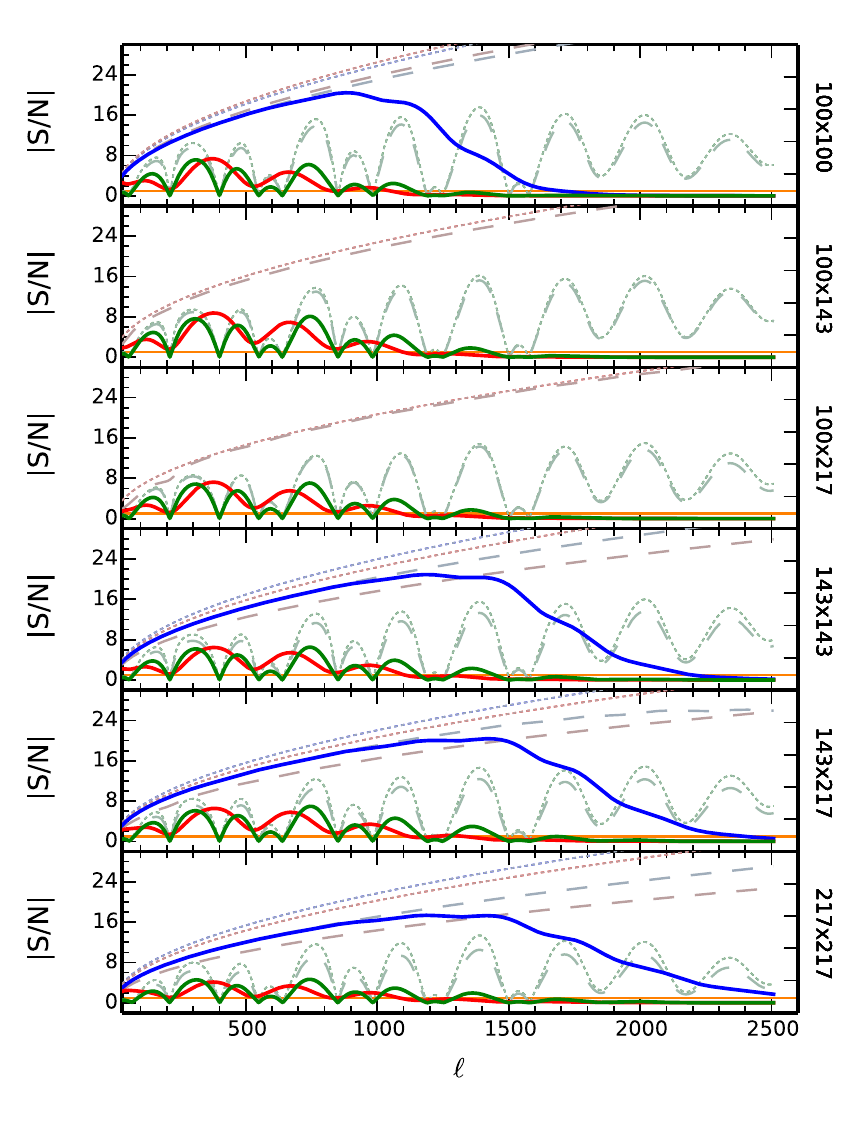}
  \caption{Unbinned S/N per frequency for $\TT$ (solid blue, for those detector combinations used in the estimate of the $\TT$ spectrum), $\EE$ (solid red), and $\TE$ (solid green). The horizontal orange line corresponds to $\mathrm{S/N}=1$. The dashed lines indicate the S/N in a cosmic-variance-limited case, obtained by forcing the instrumental noise terms to zero when calculating the power spectrum covariance matrix. The dotted lines indicate the cosmic-variance-limited case computed with the approximate formula of Eq.~(\ref{eq:cv}).}
  \label{fig:SN}
\end{figure}

Figure~\ref{fig:SN} shows the unbinned S/N per frequency for $\TT$, $\EE$, and $\TE$, where the  signal is given by the frequency-dependent CMB and foreground power spectra, while the noise term contains contributions from cosmic variance and instrumental noise and is given by the diagonal elements of the power-spectrum covariance matrix. The figure also shows the S/N assuming only cosmic variance (CV) in the noise term, obtained either by a full calculation of the covariance matrix with instrumental noise set to zero, or using the approximation
\begin{eqnarray}
\label{eq:cv}
\sigma_{CV}^{\{TT,EE\}}&=&\sqrt{\left(\frac{2}{(2\ell+1)\fsky}\right)\left(C^{\{TT,EE\}}_\ell\right)^2}\nonumber\\
\sigma_{CV}^{TE}&=&\sqrt{\left(\frac{2}{(2\ell+1)\fsky}\right)\frac{\left(C^{TE}_\ell\right)^2+C^{TT}_\ell C^{EE}_\ell}{2}} \, .
\end{eqnarray}
 \citep[see e.g.][]{PB06}.

This figure illustrates that the multipole cuts we apply ensure that the $|S/N| \gtrsim 1$. 
The $\TT$ multipole cuts are similar to those adopted in \citetalias{planck2013-p08}. While otherwise similar to the 2013 likelihood, the revised treatment of dust in the foreground model enables the retention of multipoles $\ell < 500$ of the $143 \times 217$ and $217 \times 217$\,GHz $\TT$ spectra. As discussed in detail in Sect.~\ref{sec:dust}, we are now marginalizing over a free amplitude parameter of the dust template, which was held constant for the 2013 release. Furthermore, the greater sky coverage at 100\,GHz maximizes its weight at low $\ell$, so that the best estimate of the  CMB signal on large scales is dominated by 100\,GHz data. We do not detect noticeable parameter shifts when removing or including multipoles at $\ell<500$.  
See Sect.~\ref{sec:hil_par_stability} for an in-depth analysis of the impact of different choices of multipole ranges on cosmological parameters.

\begin{table}[ht!] 
\begingroup 
\newdimen\tblskip \tblskip=5pt
 \caption{\rev{Multipole cuts for the \plik\ temperature and polarization spectra at high~$\ell$.}}
  \label{tab:highl:lrangeHERE}
\vskip -6mm
\footnotesize
\setbox\tablebox=\vbox{
\newdimen\digitwidth
\setbox0=\hbox{\rm 0}
\digitwidth=\wd0
\catcode`*=\active
\def*{\kern\digitwidth}
\newdimen\signwidth
\setbox0=\hbox{+}
\signwidth=\wd0
\catcode`!=\active
\def!{\kern\signwidth}
\newdimen\decimalwidth
\setbox0=\hbox{.}
\decimalwidth=\wd0
\catcode`@=\active
\def@{\kern\decimalwidth}
\openup 3pt
\halign{
\hbox to 0.85in{#\leaderfil}\tabskip=2em&
    \hfil#\hfil\tabskip=2em&
    \hfil#\hfil\tabskip=1.4em\cr
\noalign{\doubleline}
\omit&Multipole\cr
\omit\hfil Frequency [GHz]\hfil&range\cr
\noalign{\vskip 3pt\hrule\vskip 5pt}
\omit{\boldmath{$TT$}}\hfil\cr
\noalign{\vskip 3pt}
$\quad 100\times100$&*30--1197\cr
$\quad 143\times143$&*30--1996\cr
$\quad 143\times217$&*30--2508\cr
$\quad 217\times217$&*30--2508\cr
\noalign{\vskip 3pt}
\omit{\boldmath{$TE$}}\hfil\cr
\noalign{\vskip 3pt}
$\quad 100\times100$&*30--999*\cr
$\quad 100\times143$&*30--999*\cr
$\quad 100\times217$&505--999*\cr
$\quad 143\times143$&*30--1996\cr
$\quad 143\times217$&505--1996\cr
$\quad 217\times217$&505--1996\cr
\noalign{\vskip 3pt}
\omit{\boldmath{$EE$}}\hfil\cr
\noalign{\vskip 3pt}
$\quad 100\times100$&*30--999*\cr
$\quad 100\times143$&*30--999*\cr
$\quad 100\times217$&505--999*\cr
$\quad 143\times143$&*30--1996\cr
$\quad 143\times217$&505--1996\cr
$\quad 217\times217$&505--1996\cr
\noalign{\vskip 5pt\hrule\vskip 3pt}
}}
\endPlancktable
\endgroup
\end{table}

\begin{figure}[htb]
\centering
\includegraphics[width=0.495\textwidth]{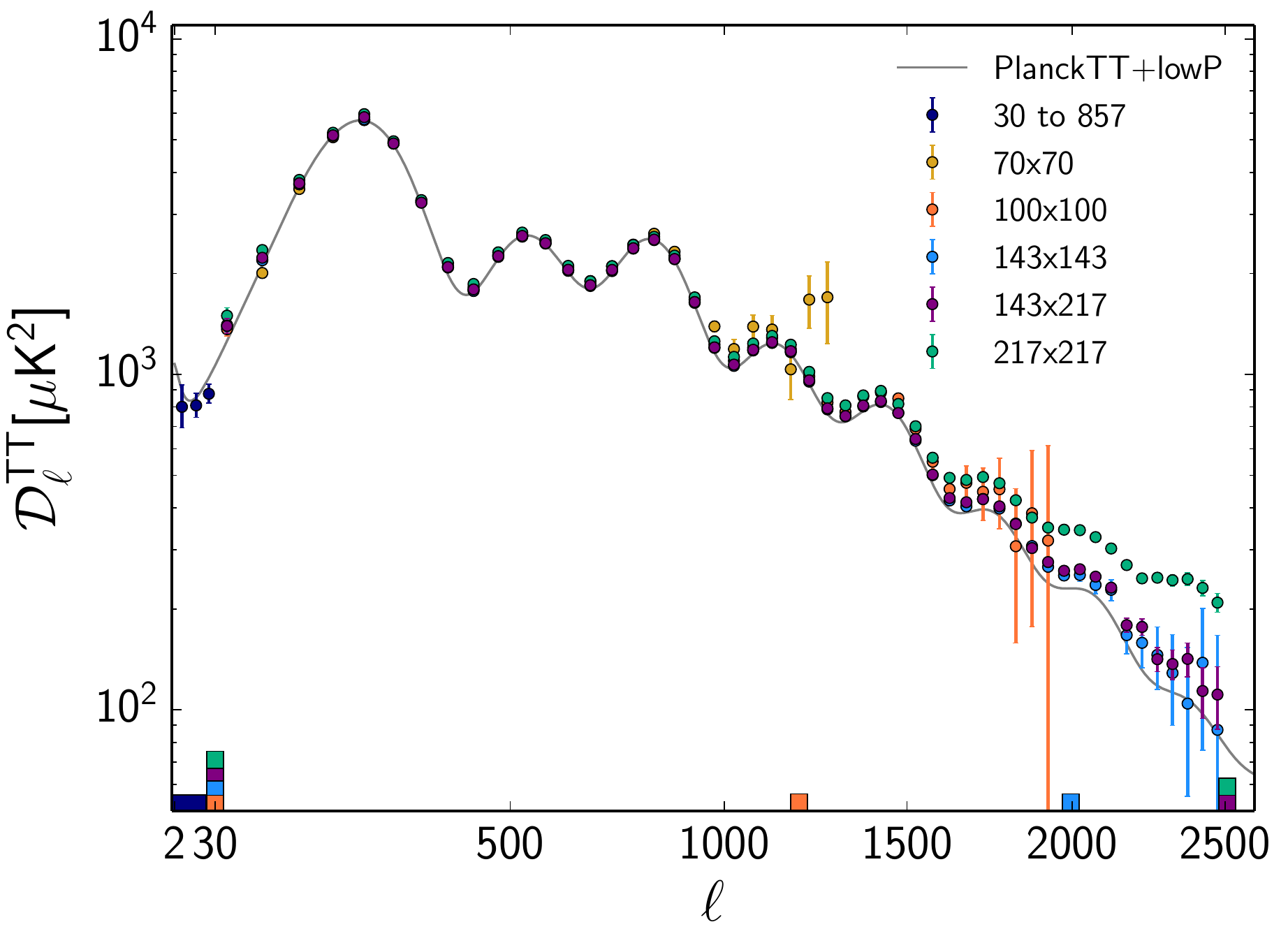}
\includegraphics[width=0.495\textwidth]{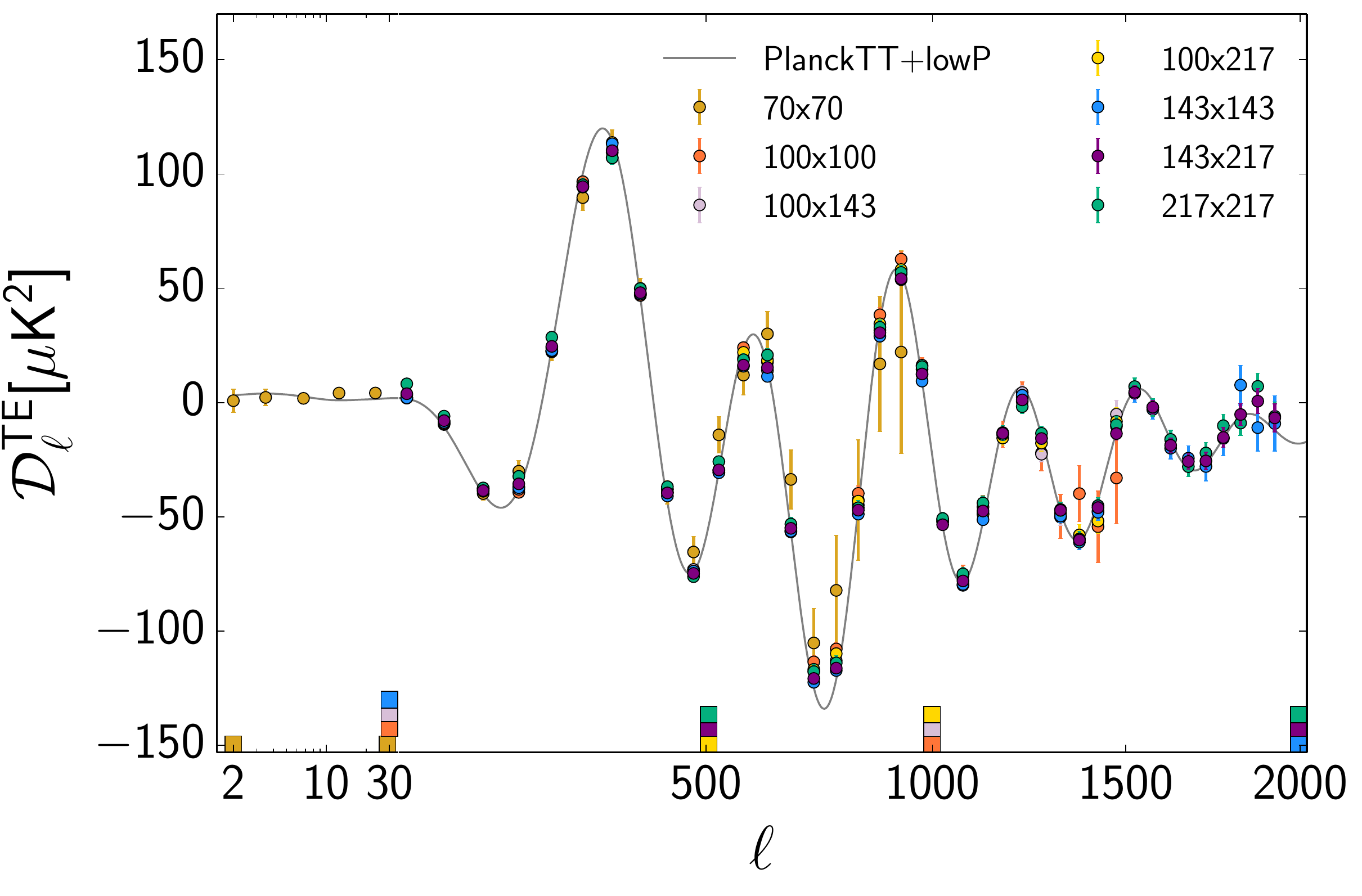}
\includegraphics[width=0.495\textwidth]{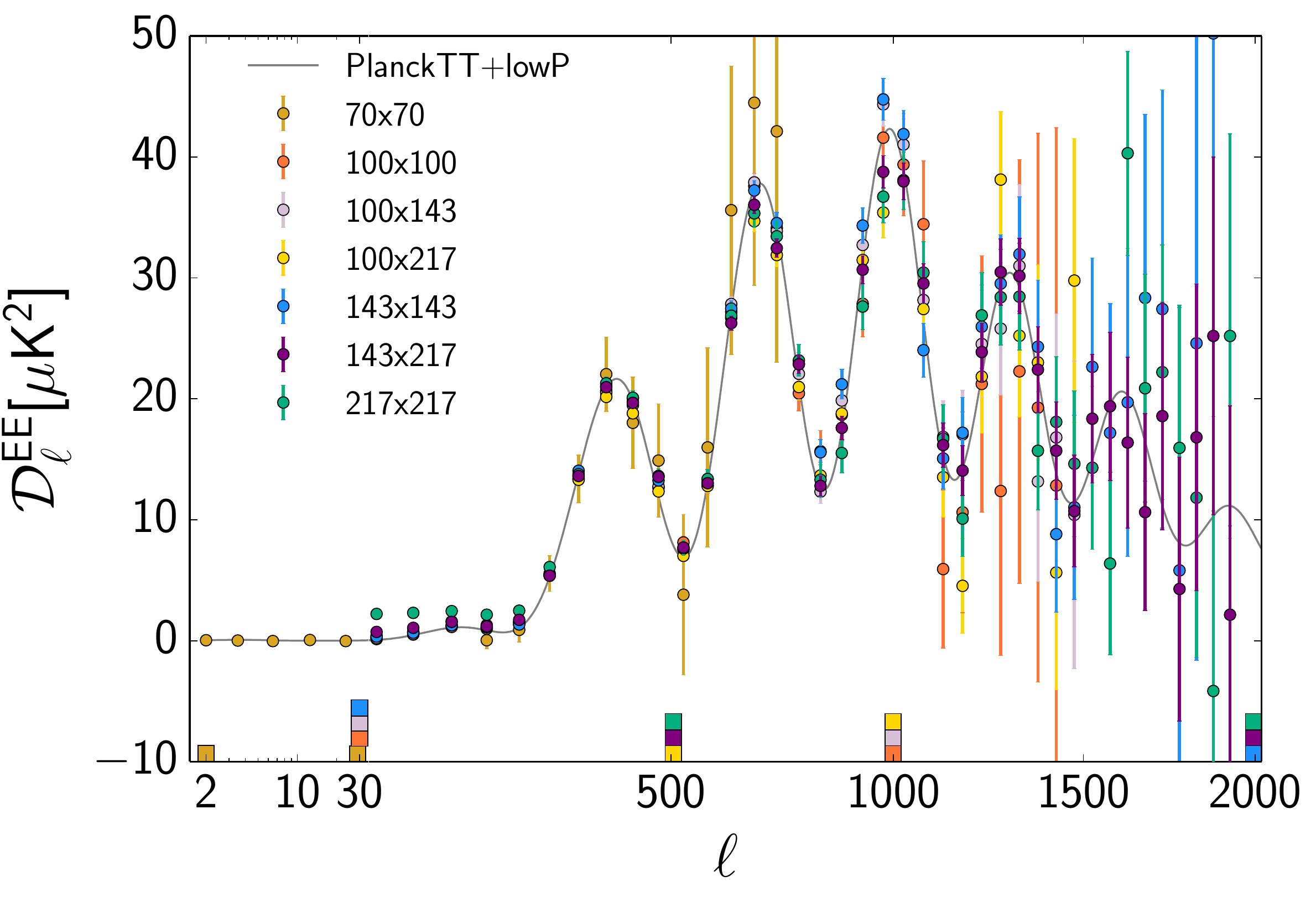}
\caption{\Planck\ power spectra (not yet corrected for foregrounds) and data selection. The coloured tick marks indicate the $\ell$-range of the cross-spectra included in the \Planck\ likelihood. Although not used in the high-$\ell$ likelihood, the 70\GHz\ spectra at $\ell > 29$ illustrate the consistency of the data. The grey line indicates the best-fit \Planck\ 2015 spectrum. The $\TE$ and $\EE$ plots have a logarithmic horizontal scale for $\ell < 30$.}
\label{fig:PS_ell-ranges}
\end{figure}

\begin{figure}[htb]
\centering
\includegraphics[width=\columnwidth]{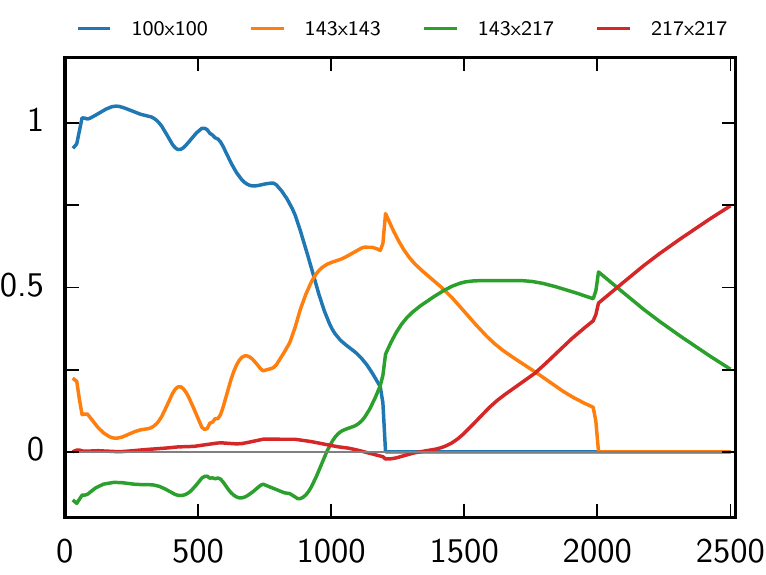}
\includegraphics[width=\columnwidth]{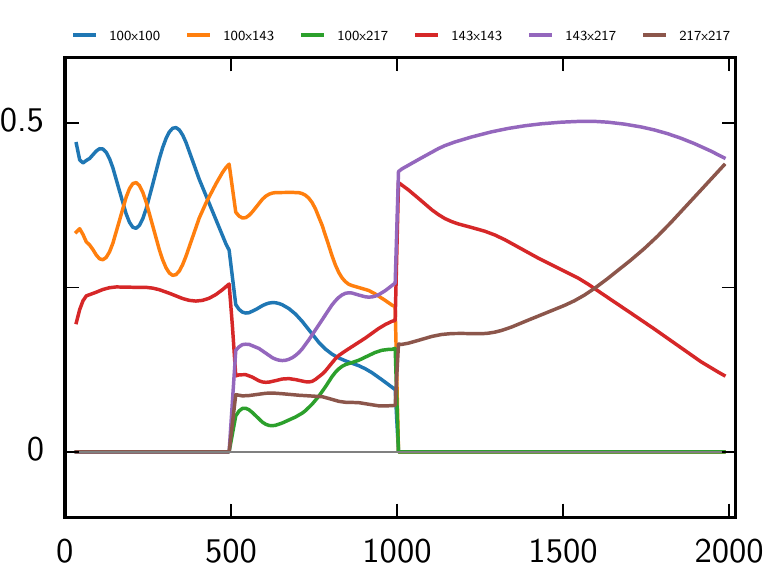}
\includegraphics[width=\columnwidth]{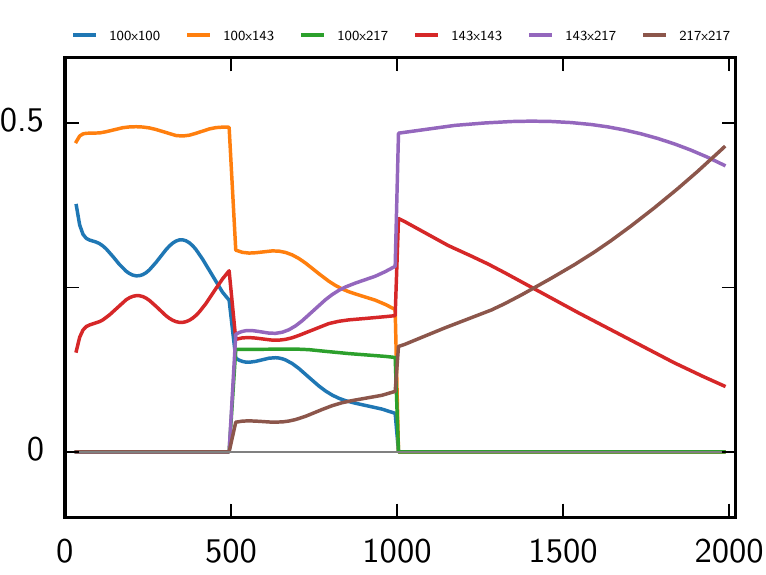}
\caption{\rev{The relative weights of each frequency cross-spectrum in the $\TT$ (top), $\TE$ (middle) and $\EE$ (bottom) best-fit solution. Sharp jumps are due to the multipole selection. Weights are normalized to sum to one.}}
\label{fig:mixing}
\end{figure}

For $\TE$ and $\EE$ we are more conservative, and cut the low S/N 100\,GHz data at small scales ($\ell>1000$), and the possibly dust-contaminated 217\,GHz at large scales ($\ell<500$). Only the $143 \times 143$ $\TE$ and $\EE$ spectra cover the full multipole range, restricted to $\ell<2000$. Retaining more multipoles would require more in-depth modelling of residual systematic effects, which is left to future work. All the cuts \rev{are summarized in Table~\ref{tab:highl:lrangeHERE} and} shown in Fig.~\ref{fig:PS_ell-ranges} .

Figure~\ref{fig:SN} also shows that each of the $\TT$ frequency power spectra is cosmic-variance dominated in a wide interval of multipoles. In particular, if we define as cosmic-variance dominated the ranges of multipoles where cosmic variance contributes more than half of the total variance, we find that the $100\times100$\,GHz spectrum is cosmic-variance dominated at $\ell\lesssim1156$, the $143\times143$\,GHz at $\ell\lesssim1528$, the $143\times 217$\,GHz at $\ell\lesssim1607$, and the $217\times 217$ GHz  at $\ell\lesssim1566$. To determine these ranges, we calculated the ratio of cosmic to total variance, where the cosmic variance is obtained  from the diagonal elements of the covariance matrix after setting the instrumental noise to zero.
Furthermore, we find that each of the $\TE$ frequency power spectra is cosmic-variance limited in some limited ranges of multipoles, below $\ell\lesssim 150$ ($\ell\lesssim 50$ for the $100\times100$),\footnote{Recall that these statements refer to the high-$\ell$ likelihood ($\ell\ge 30$).} in the range $\ell\approx 250$--$450$ and additionally in the range $\ell\approx 650$--$700$ only for the $100\times 143$\,GHz and the $143\times 217$\,GHz power spectra. 

 Finally, when we co-add the foreground-cleaned frequency spectra to provide the CMB spectra (see Appendix~\ref{app:co-added}), we find that the CMB $\TT$ power spectrum is cosmic-variance dominated at $\ell\lesssim 1586$, while $\TE$ is cosmic-variance dominated at $\ell\lesssim 158$ and $\ell\approx 257$--$464$. 

 \rev{Due to the different masks, multipole ranges, noise levels, and to a lesser extent differing foreground contamination, each cross-spectrum ends up contributing differently as a function of scale to the best CMB solution. The determination of the mixing weights is described in Appendix~\ref{app:co-added}. Figure~\ref{fig:mixing} presents the resulting (relative) weights of each cross-spectra. In temperature, the $100\times100$ spectrum dominates the solution until $\ell\approx800$, when the solution becomes driven by the $143\times143$ up to $\ell\approx1400$. The $143\times217$ and $217\times217$ provide the solution for the higher multipoles. In polarization, the $100\times143$ dominates the solution until $\ell\approx 800$ (with an equal contribution from $100\times100$ until $\ell\approx400$ in $TE$ only) while the  higher $\ell$ range is dominated by the $143\times217$ contribution. Not surprisingly, the weights of the higher frequencies tend to increase with $\ell$.}

\subsubsection{Binning} \label{sec:binning}

The 2013 baseline likelihood used unbinned temperature power spectra. For this release, we include polarization, which substantially increases  the size of the numerical task. The 2015 likelihood therefore uses  binned power spectra by default, downsizing the covariance matrix and speeding up likelihood computations. Indeed, even with the multipole-range cut just described, the unbinned data vector has around 23\,000 elements, two thirds of which correspond to $\TE$ and $\EE$. For some specific purposes (\eg searching for oscillatory features in the $\TT$ spectrum or testing $\chi^2$ statistics) we also produce an unbinned likelihood.

The spectra are binned into bins of width 
$\Delta\ell=5$ for $30 \le \ell \le 99$, 
$\Delta\ell=9$ for $100 \le \ell \le 1503$,
$\Delta\ell=17$ for $1504 \le \ell \le 2013$, and 
$\Delta\ell=33$ for $2014 \le \ell \le 2508$, 
with a weighting of the $C_\ell$ proportional to
$\ell(\ell+1)$ over the bin widths,
\begin{equation}
C_b = \sum_{\ell=\ell_b^{\mathrm{min}}}^{\ell_b^{\mathrm{max}}} w_b^\ell C_\ell, \ \mathrm{with}\ w_b^\ell = \frac{\ell(\ell+1)}{\sum_{\ell=\ell_b^{\mathrm{min}}}^{\ell_b^{\mathrm{max}}} \ell(\ell+1) } .
\end{equation}
The bin-widths are odd numbers, since for approximately azimuthal masks we expect a nearly symmetrical correlation function around the central multipole. It is shown explicitly in Sect.~\ref{sec:hil_par_stability} that the binning does not affect the determination of cosmological parameters in \LCDM-type models, which have smooth power spectra. 


\subsection{Foreground modelling}
\label{sec:foreground_modelling}

Most of the foreground elements in the model parameter vector are similar to those in \citetalias{planck2013-p08}. The main differences are in the dust templates, which have changed to accommodate the new masks. The $\TE$ and $\EE$ foreground model only takes into account the dust contribution and neglects any other Galactic polarized emission, in particular the synchrotron contamination. Nor do we mask out any extragalactic polarized foregrounds, as they have been found to be negligible by ground-based, small-scale experiments \citep{actpol2014PS,sptpol2014PS}.

\begin{figure*}
  \includegraphics[angle=0,width=0.5\textwidth]{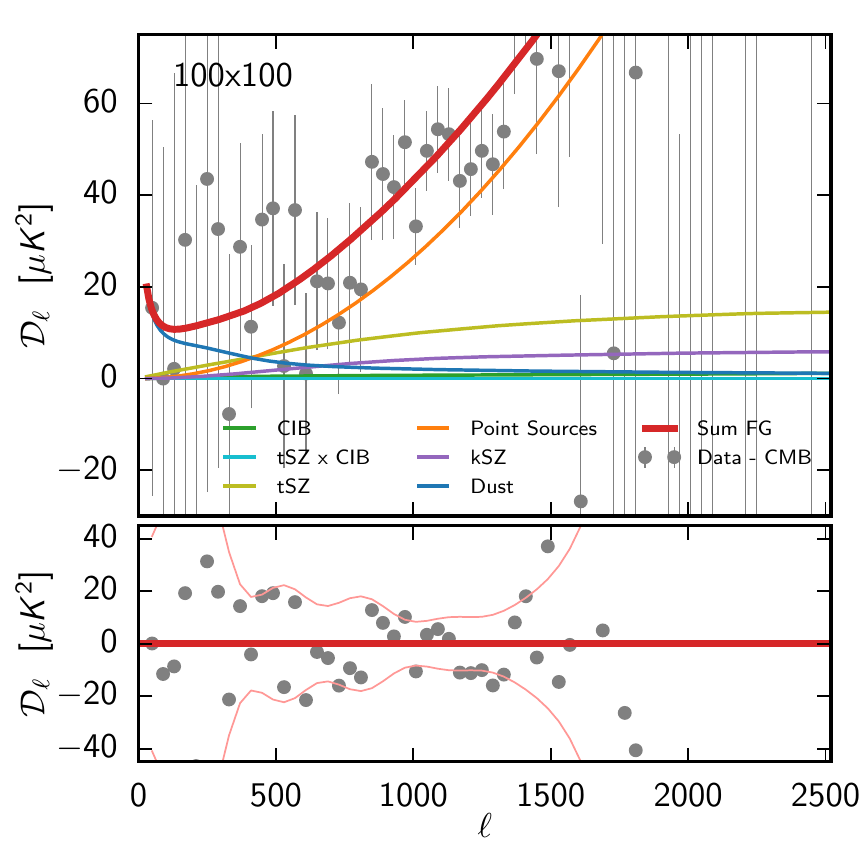}
  \includegraphics[angle=0,width=0.5\textwidth]{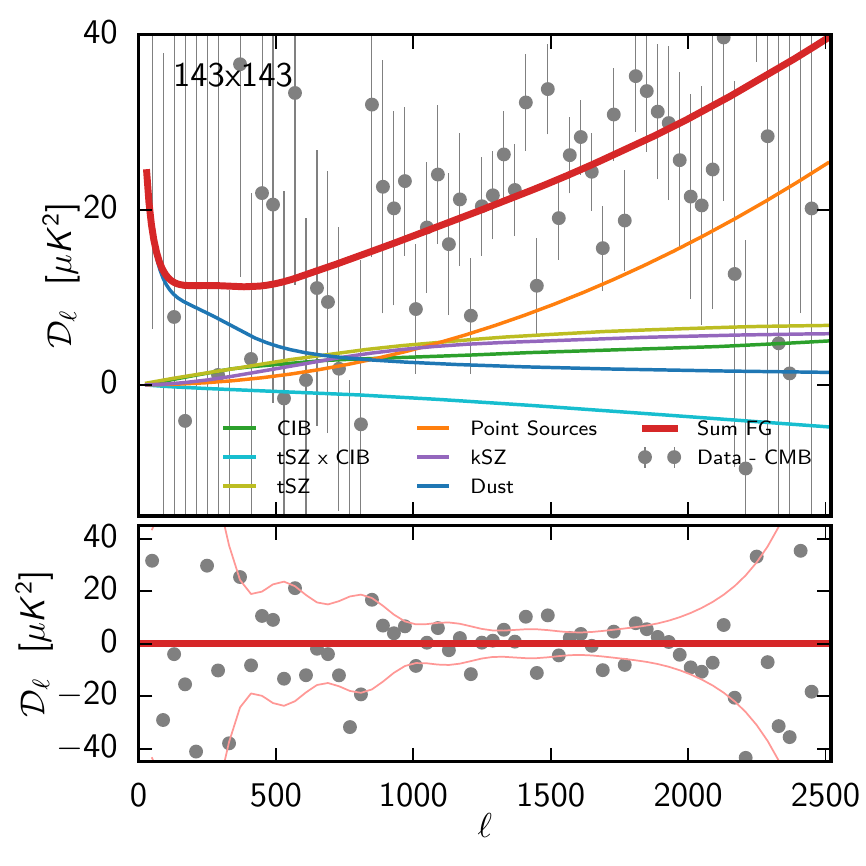}\\
  \includegraphics[angle=0,width=0.5\textwidth]{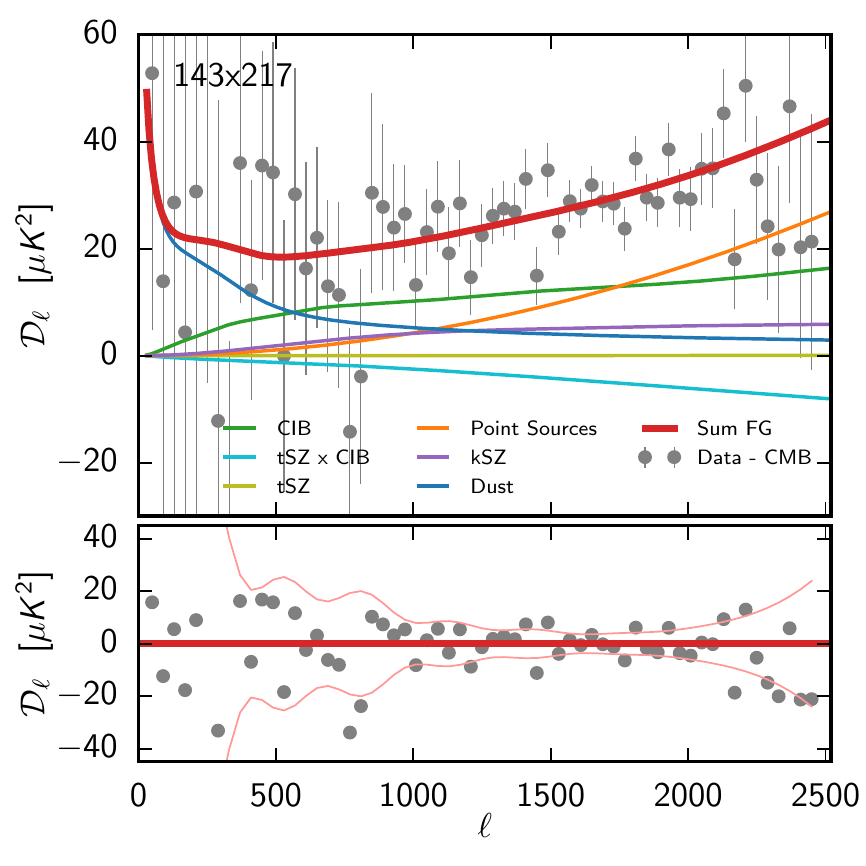}
  \includegraphics[angle=0,width=0.5\textwidth]{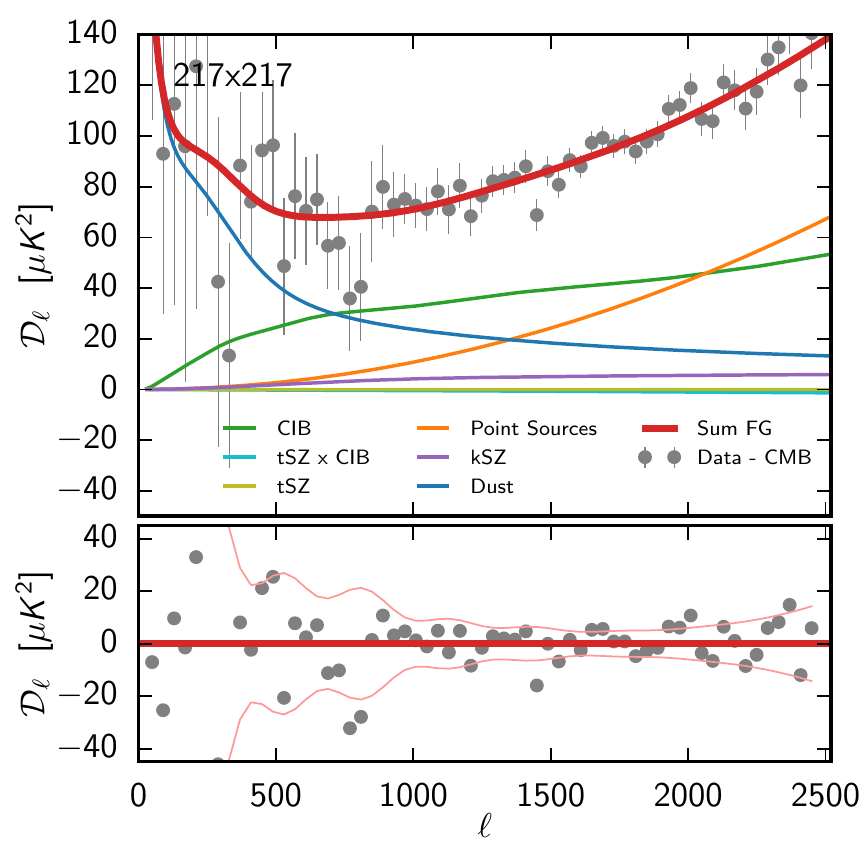}\\
  \caption{Best foreground model in each of the cross-spectra used for the temperature high-$\ell$ likelihood. The data corrected by the best theoretical CMB $C_\ell$ are shown in grey. The bottom panel of each plot shows the residual after foreground correction. The pink line shows the $1\,\sigma$ value from the diagonal of the covariance matrix (32\,\% of the unbinned points are out of this range). }
  \label{fig:hil:fg}
\end{figure*}

\begin{figure*}
  \includegraphics[angle=0,width=0.5\textwidth]{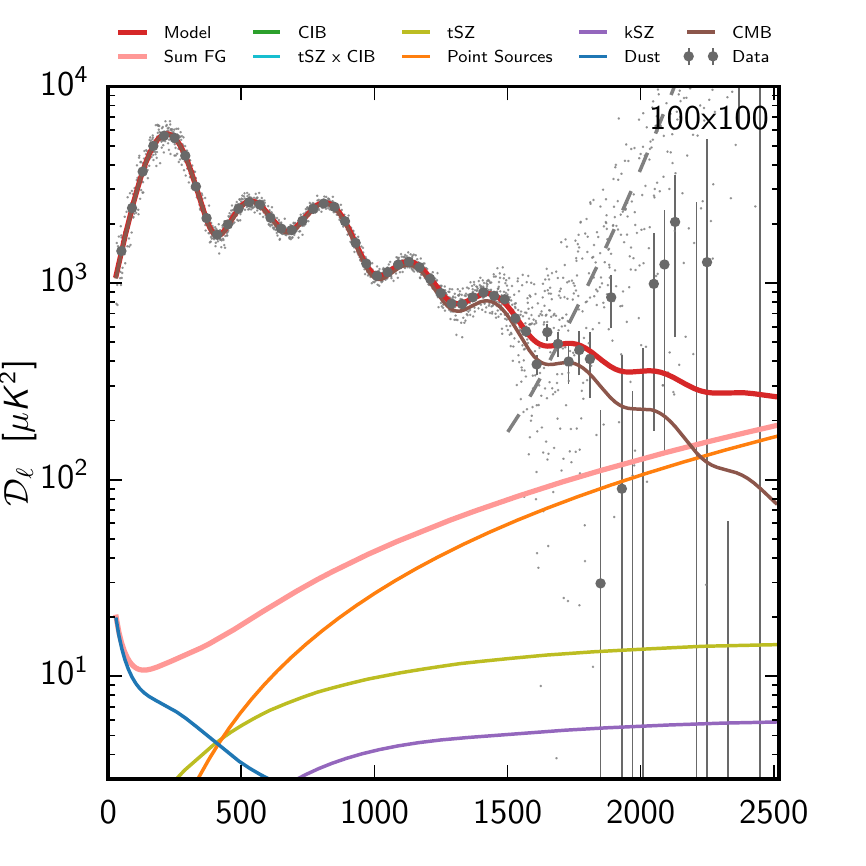}
  \includegraphics[angle=0,width=0.5\textwidth]{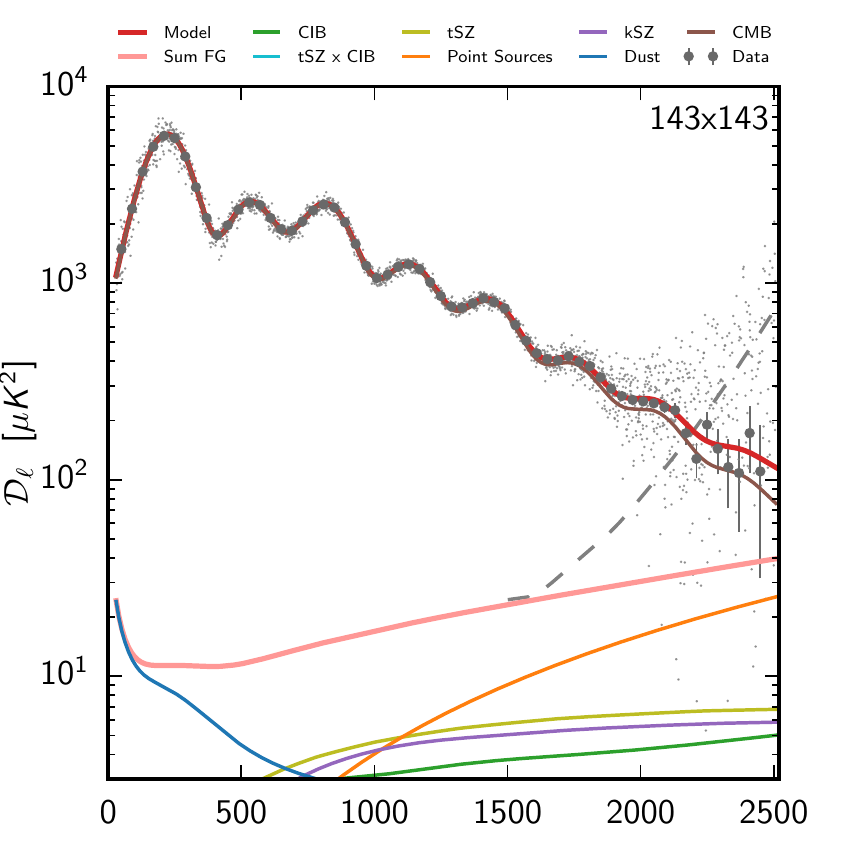}\\
  \includegraphics[angle=0,width=0.5\textwidth]{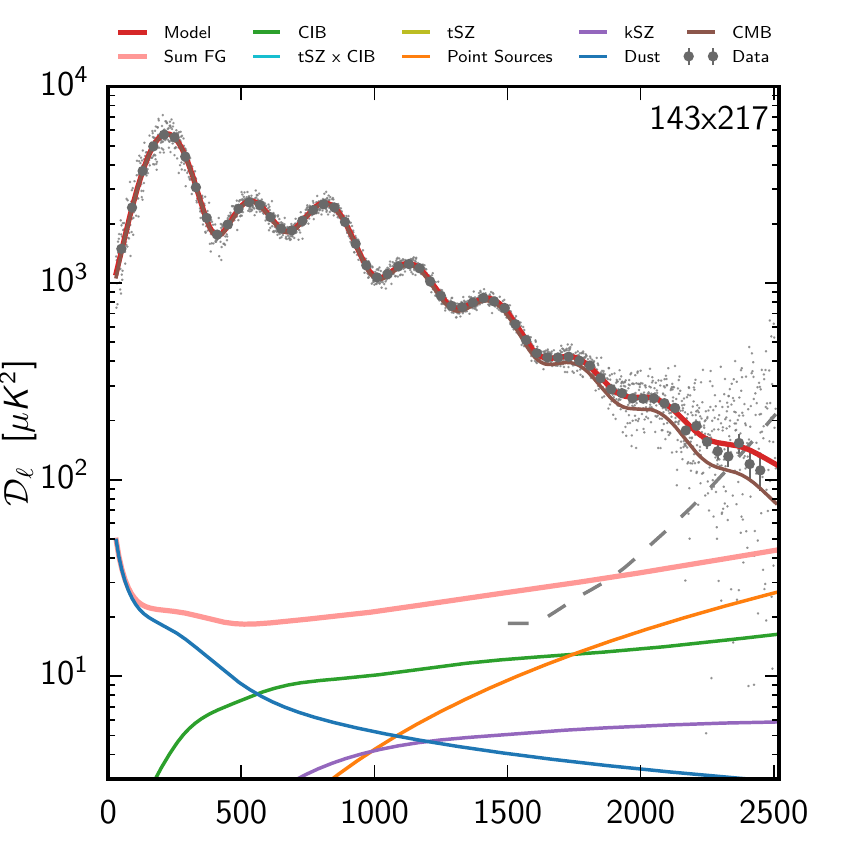}
  \includegraphics[angle=0,width=0.5\textwidth]{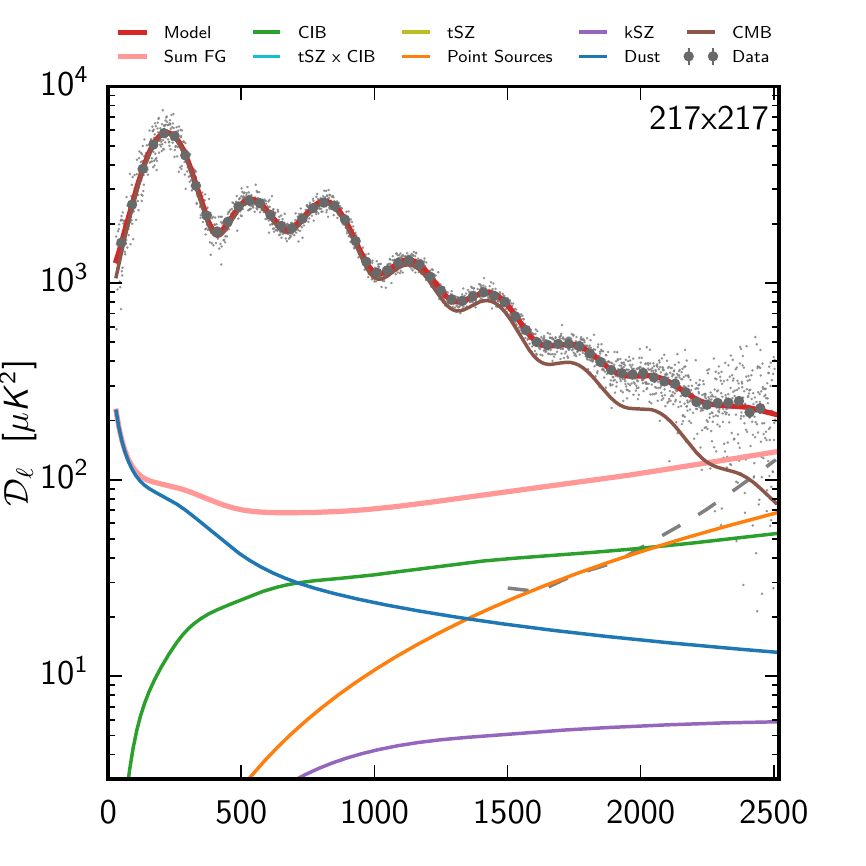}\\
  \caption{\rev{Best model (CMB and foreground) in each of the cross-spectra used for the temperature high-$\ell$ likelihood. The small light grey points show the unbinned data point, and the dashed grey line show the square root of the noise contribution to the diagonal of the unbinned covariance matrix.} }
  \label{fig:hil:fgcmb}
\end{figure*}

Figure~\ref{fig:hil:fg} shows the foreground decomposition in temperature for each of the cross-spectra combinations we use in the likelihood. The figure also shows the CMB-corrected data (\ie data minus the best-fit $\Lambda$CDM CMB model) as well as the residuals after foreground correction. In each spectrum, dust dominates the low-$\ell$ modes, while point sources dominate the smallest scales. For $217\times217$ and $143\times217$, the intermediate range has a significant CIB contribution. We note that for $100\times100$, even when including 66\,\% of the sky, the dust contribution is almost negligible and the point-source term is dominant well below $\ell=500$. The least foreground-contaminated spectrum is $143\times143$. \rev{For comparison, Fig.~\ref{fig:hil:fgcmb} shows the full model, including the CMB. The foreground contribution is a small fraction of the total power at large scales.}

Table~\ref{tab:fg-params} summarizes the parameters used for astrophysical foreground modelling and their associated priors.

\begin{table*}[ht!] 
\caption{Parameters used for astrophysical foregrounds and instrumental modelling.$^{\rm a}$} 
\label{tab:fg-params}
\begingroup 
\newdimen\tblskip \tblskip=5pt
\nointerlineskip
\vskip -3mm
\footnotesize
\setbox\tablebox=\vbox{
\newdimen\digitwidth
\setbox0=\hbox{\rm 0}
\digitwidth=\wd0
\catcode`*=\active
\def*{\kern\digitwidth}
\newdimen\signwidth
\setbox0=\hbox{+}
\signwidth=\wd0
\catcode`!=\active
\def!{\kern\signwidth}
\halign{\hbox to 2cm{$#$\leaderfil}\tabskip=2em& 
   $#$\hfil\tabskip=2em& 
   #\hfil\tabskip=0pt\cr
\noalign{\doubleline}
\omit\hfil Parameter\hfil&\omit\hfil Prior range\hfil&\omit\hfil Definition\hfil\cr
\noalign{\vskip 3pt\hrule\vskip 5pt}
A^{\rm PS}_{100}&                [0,400]&             Contribution of Poisson point-source power to $\mathcal{D}^{100\times 100}_{3000}$ for \planck\ (in $\mu\mathrm{K}^2$)\cr
A^{\rm PS}_{143}&                [0,400]&             As for $A^{\rm PS}_{100}$ but at $143\,$GHz\cr
A^{\rm PS}_{217}&                [0,400]&             As for $A^{\rm PS}_{100}$ but at $217\,$GHz\cr
A^{\rm PS}_{143\times217}&       [0,400]&             As for $A^{\rm PS}_{100}$ but at $143\times217\,$GHz\cr
A^{\mathrm{CIB}}_{217}&          [0,200]&             Contribution of CIB power to $\mathcal{D}^{217}_{3000}$ at the \planck\ CMB frequency for $217\,$GHz (in $\mu\mathrm{K}^2$)\cr 
A^{\mathrm{tSZ}}&                [0,10]&              Contribution of tSZ to $\mathcal{D}_{3000}^{143\times 143}$ at $143\,$GHz (in $\mu\mathrm{K}^2$)\cr
A^{\mathrm{kSZ}}&                [0,10]&              Contribution of kSZ to $\mathcal{D}_{3000}$ (in $\mu\mathrm{K}^2$)\cr
\xi^{{\rm tSZ}\times {\rm CIB}}& [0,1]&               Correlation coefficient between the CIB and tSZ\cr
A^{{\rm dust}TT}_{100}&          [0,50]&              Amplitude of Galactic dust power at $\ell=200$ at $100\,$GHz (in $\mu\mathrm{K}^2$)\cr
\omit&                           (7\pm 2)\cr
A^{{\rm dust}TT}_{143}&          [0,50]&              As for $A^{{\rm dust}TT}_{100}$ but at $143\,$GHz\cr
\omit&                           (9 \pm 2)\cr
A^{{\rm dust}TT}_{143\times 217}&[0,100]&             As for $A^{{\rm dust}TT}_{100}$ but at $143\times 217\,$GHz\cr
\omit&                           (21\pm 8.5)\cr
A^{{\rm dust}TT}_{217}&          [0,400]&             As for $A^{{\rm dust}TT}_{100}$ but at $217\,$GHz\cr
\omit&                           (80\pm 20)\cr
\noalign{\vskip 5pt\hrule\vskip 5pt}
c_{100}&                         [0,3]&               Power spectrum calibration for the $100\,$GHz\cr
\omit&                           (0.9990004\pm 0.001)\cr
c_{217}&                         [0,3]&               Power spectrum calibration for the $217\,$GHz\cr
\omit&                           (0.99501\pm 0.002)\cr
y_{\rm cal}&                     [0.9,1.1]&           Absolute map calibration for \planck\ \cr
\omit&                           (1\pm 0.0025)\cr
\noalign{\vskip 5pt\hrule\vskip 5pt}
A^{{\rm dust}EE}_{100}&          [0,10]&              Amplitude of Galactic dust power at $\ell=500$ at $100\,$GHz (in $\mu\mathrm{K}^2$)\cr
\omit&                           (0.06\pm 0.012)\cr
A^{{\rm dust}EE}_{100\times 143}&[0,10]&              As for $A^{{\rm dust}EE}_{100}$ but at $100\times 143\,$GHz\cr
\omit&                           (0.05\pm 0.015)\cr
A^{{\rm dust}EE}_{100\times 217}&[0,10]&              As for $A^{{\rm dust}EE}_{100}$ but at $100\times 217\,$GHz\cr
\omit&                           (0.11\pm 0.033)\cr
A^{{\rm dust}EE}_{143}&          [0,10]&               As for $A^{{\rm dust}EE}_{100}$ but at $143\,$GHz\cr
\omit&                           (0.1 \pm 0.02)\cr
A^{{\rm dust}EE}_{143\times 217}&[0,10]&              As for $A^{{\rm dust}EE}_{100}$ but at $143\times 217\,$GHz\cr
\omit&                           (0.24\pm 0.048)\cr
A^{{\rm dust}EE}_{217}&          [0,10]&               As for $A^{{\rm dust}EE}_{100}$ but at $217\,$GHz\cr
\omit&                           (0.72\pm 0.14)\cr
\noalign{\vskip 5pt\hrule\vskip 5pt}
A^{{\rm dust}TE}_{100}&          [0,10]&              Amplitude of Galactic dust power at $\ell=500$ at $100\,$GHz (in $\mu\mathrm{K}^2$)\cr
\omit&                           (0.14\pm 0.042)\cr
A^{{\rm dust}TE}_{100\times 143}&[0,10]&              As for $A^{{\rm dust}TE}_{100}$ but at $100\times 143\,$GHz\cr
\omit&                           (0.12\pm 0.036)\cr
A^{{\rm dust}TE}_{100\times 217}&[0,10]&              As for $A^{{\rm dust}TE}_{100}$ but at $100\times 217\,$GHz\cr
\omit&                           (0.3\pm 0.09)\cr
A^{{\rm dust}TE}_{143}&          [0,10]&              As for $A^{{\rm dust}TE}_{100}$ but at $143\,$GHz\cr
\omit&                           (0.24 \pm 0.072)\cr
A^{{\rm dust}TE}_{143\times 217}&[0,10]&              As for $A^{{\rm dust}TE}_{100}$ but at $143\times 217\,$GHz\cr
\omit&                           (0.6\pm 0.18)\cr
A^{{\rm dust}TE}_{217}&          [0,10]&               As for $A^{{\rm dust}TE}_{100}$ but at $217\,$GHz\cr
\omit&                           (1.8\pm 0.54)\cr
\noalign{\vskip 5pt\hrule\vskip 3pt}
}}
\endPlancktablewide
\tablenote {{\rm a}} The columns indicate the symbol for each parameter, the prior used for exploration (square brackets denote uniform priors, parentheses indicate Gaussian priors), and definitions. Beam eigenmode amplitudes require a correlation matrix to fully describe their joint prior and so do not appear in the table; they are internally marginalized over rather than explicitly sampled.
This table only lists the instrumental parameters that are explored in the released version, but we do consider more parameters to assess the effects of beam uncertainties and beam leakage; see Sect.~\ref{sec:beam-uncert}.\par
\endgroup
\end{table*}

\subsubsection{Galactic dust emission}
\label{sec:dust}

\begin{figure}
  \includegraphics[angle=0,width=0.5\textwidth]{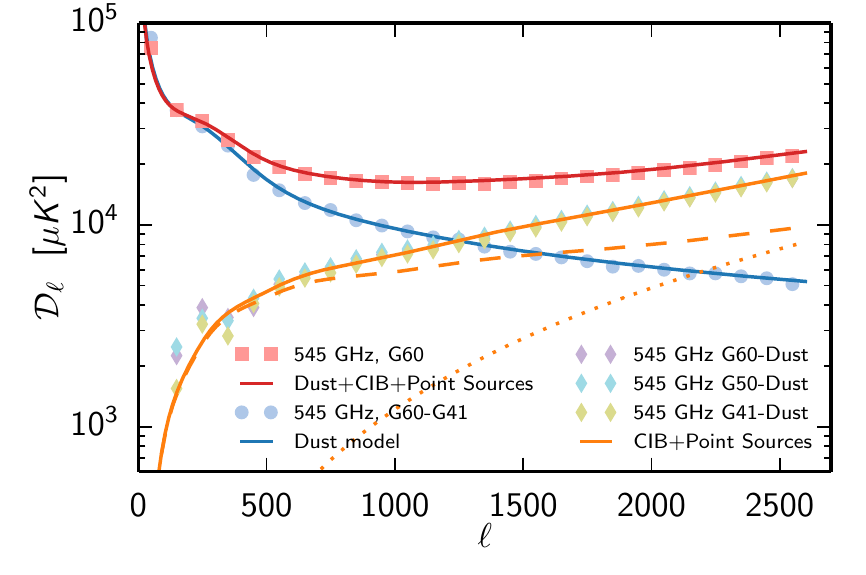}
  \caption{Dust model at 545\, GHz. The dust template is based on the G60-G41 mask difference of the 545\GHz\ half-mission cross-spectrum (blue line and circles, rescaled to the dust level in mask G60). Coloured diamonds display the difference between this model (rescaled in each case) and the cross half-mission spectra in the G41, G50, and G60 masks. The residuals are all in good agreement (less so at low $\ell$, because of sample variance) and are well described by the CIB+point source prediction (orange line). Individual CIB and point sources contributions are shown as dashed and dotted orange lines. The red line is the sum of the dust model, CIB, and point sources for the G60 mask, and is in excellent agreement with the 545\,GHz cross half-mission spectrum in G60 (red squares). In all cases, the spectra were computed by using different Galactic masks supplemented by the single combination of the 100\,GHz, 143\,GHz, and 217\,GHz point sources, extended objects and CO masks.}
  \label{fig:hil:dust545model}
\end{figure}
 
Galactic dust is the main foreground contribution at large scales and thus deserves close attention. This section describes how we model its power spectra. We express the dust contribution to the power spectrum calculated from map $X$ at frequency $\nu$ and map $Y$ at frequency $\nu'$ as 
\begin{equation}
\left( C^{XY,\rm dust}_{\nu\times\nu'}\right)_{\ell} = A^{XY,\rm dust}_{\nu\times\nu'}\times C_{\ell}^{XY,\rm dust}\;,
\label{eq:galactic-dust-intensity}
\end{equation}
where $XY$ is one of $\TT$, $\EE$, or $\TE$, and  $C_{\ell}^{XY,\rm dust}$ is the template dust power spectrum, with corresponding amplitude $A^{XY,\rm dust}_{\nu\times\nu'}$. We assume that the dust power spectra have the same spatial dependence across frequencies and masks, so the dependence on sky fraction and frequency is entirely encoded in the amplitude parameter $A$. We do not try to enforce any a priori scaling with frequency, since using different masks at different frequencies makes determination of this scaling difficult. When  both frequency maps $\nu$ and $\nu'$ are used in the likelihood with the same mask, we simply assume that the amplitude parameter can be written as
\begin{equation}
A^{XY,\rm dust}_{\nu\times\nu'} = a^{XY,\rm dust}_{\nu}\times a^{XY,\rm dust}_{\nu'}.
\end{equation}
This is clearly not exact when $XY=TE$ and $\nu\neq\nu'$. Similarly the multipole-dependent weight used to  combine $\TE$ and $\ET$ for different frequencies breaks the assumption of an invariant dust template. These approximations do not appear to be the limiting factor of the current analysis.

In contrast to the choice we made in 2013, when all Galactic contributions were fixed and a dust template had been explicitly subtracted from the data, we now fit for the amplitude of the dust contribution in each cross-spectrum, in both temperature and polarization. This enables exploration of the possible degeneracy between the dust amplitude and cosmological parameters. A comparison of the two approaches is given in Sect.~\ref{sec:mspec} and Fig.~\ref{fig:217cleaning}.

In the following, we describe how we build our template dust power spectrum from high-frequency data and evaluate the amplitude of the dust contamination at each frequency and for each mask. 

As we shall see later in Sect.~\ref{sec:robust-fsky}, the cosmological values recovered from $\TT$ likelihood explorations do not depend on the dust amplitude priors, as shown by the case ``No gal.\ priors'' in Fig.~\ref{fig:wiskerTT} and discussed in Sect.~\ref{sec:robust-fsky}. The polarization case is discussed in Sect.~\ref{sec:pol-robust}. Section~\ref{sec:hal-params} and Figs.~\ref{fig:params_correl} and \ref{fig:params_correl_T_ext} show the correlation between the dust and the cosmological or other foreground parameters. The dust amplitudes are found to be nearly uncorrelated with the cosmological parameters except for $\TE$. However, the priors do help to break the degeneracies between foreground parameters, which are found to be much more correlated with the dust. In Appendix\ \ref{app:hal} we further show that our results are insensitive to broader changes in the dust model.

\paragraph{\textbf{Galactic \textit{TT} dust emission.}}\label{sub:galactic-dust:TT}

We use the 545\ghz\ power spectra as templates for Galactic dust spatial fluctuations. The 353\ghz\ detectors also have some sensitivity to dust, along with a significant contribution from the CMB, and hence any error in removing the CMB contribution at 353\ghz\ data translates
into biases on our dust template. This is much less of an issue at 545\ghz, to the point where entirely ignoring the CMB contribution does not change our estimate of the template. Furthermore, estimates using 545\ghz\ maps tend to be more stable over a wider range of multipoles than those obtained from 353\ghz\ or 857\ghz\ maps.  

 \begin{figure}
  \includegraphics[angle=0,width=0.5\textwidth]{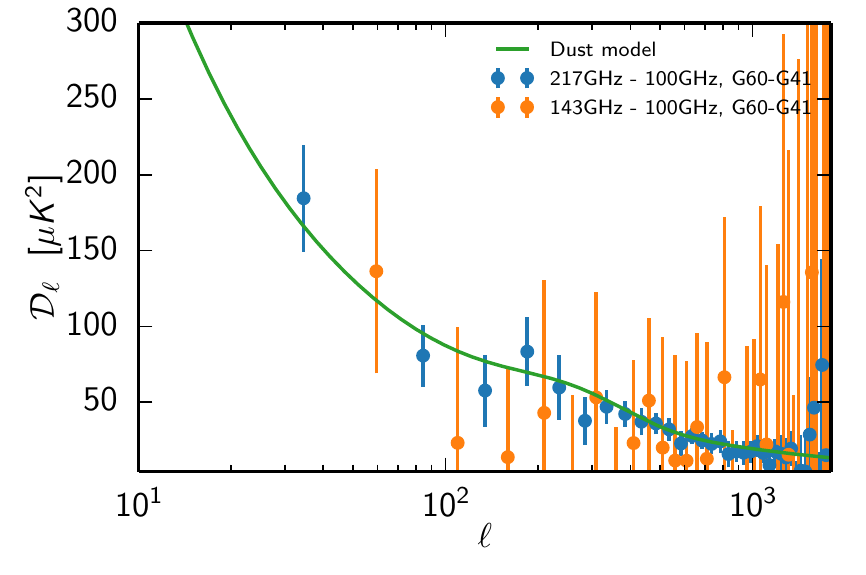}
  \caption{Dust model versus data. In blue, the power spectrum of the double mask difference between 217\ghz\ and 100\ghz\ half-mission cross-spectra in masks G60 and G41 (complemented by the joint masks for CO, extended objects, and point sources). In orange, the equivalent spectrum for 143 and 100\ghz. The mask difference enables us to remove the contribution from all the isotropic components (CMB, CIB, and point sources) in the mean. But simple mask differences are still affected by the difference of the CMB in the two masks due to cosmic variance. Removing the 100\ghz\ mask difference, which is dominated by the CMB, reduces the scatter significantly. The error bars are computed as the scatter in bins of size $\Delta\ell=50$. The dust model (green) based on the 545\ghz\ data has been rescaled to the expected dust contamination in the 217\ghz\ mask difference using values from Table~\ref{table:hil:dustTT}. The 143\ghz\ double mask difference is also rescaled to the level of the 217\ghz\ difference; \ie it is multiplied by approximately $14$. Different multipole bins are used for the 217\ghz\ and 143\ghz\ data to improve readability.}
  \label{fig:hil:dust143217}
\end{figure}

We aggressively mask the contribution from point sources in order to minimize their residual, the approximately white spectrum of which is substantially correlated with the value of some cosmological parameters (see the discussion of parameter correlations in Sect.~\ref{sec:hal-params}). The downside of this is that the point-source masks remove some of the brightest Galactic regions that lie in regions not covered by our Galactic masks. This means that we cannot use the well-established power-law modelling advocated in \citet{planck2013-p06b} and must instead compute an effective dust (residual) template.

All of the masks that we use in this section are combinations of the joint point-source, extended-object, and CO masks used for 100\ghz, 143\ghz, and 217\ghz\ with Galactic masks of various sizes.  In the following discussion we refer only to the Galactic masks, but in all cases the masks contain the other components as well. The half-mission cross-spectra at 545\ghz\  provide us with a good estimate of the large-scale behaviour of the dust. Small angular scales, however, are sensitive to the CIB, with the intermediate range of scales dominated by the clustered part and the smallest scales by the Poisson distribution of infrared point sources. These last two terms are statistically isotropic, while the dust amplitude depends on the sky fraction. Assuming that the shapes of the dust power spectra outside the masks do not vary substantially as the sky fraction changes, we rely on mask differences to build a CIB-cleaned template of the dust. 

Figure~\ref{fig:hil:dust545model} shows that this assumption is valid when changing the Galactic mask from G60 to G41. It shows that the 545\ghz\ cross-half-mission power spectrum can be well represented by the sum of a Galactic template, a CIB contribution, and a point source contribution. The Galactic template is obtained by computing the difference between the spectra obtained in the G60 and the G41 masks. This difference is fit to a simple analytic model 
\begin{equation}
  C_{\ell}^{TT,\rm dust} \propto (1 + h \, \ell^k \, e^{-\ell/t}) \times (\ell/\ell_p)^n,
  \label{eq:dust:TTtemplate}
\end{equation}
with $h = 2.3\times10^{-11}$, $k = 5.05$, $t = 56$, $n = -2.63$, and  fixing $\ell_p=200$. The model behaves like a $C_{\ell,\rm dust}^{\TT} \propto \ell^{-2.63}$ power law at small scales, and has a bump around $\ell=200$. The CIB model we use is described in Sect.~\ref{sec:hil:cib}.  

We can compare this template model with the dust content in each of the power spectra we use for the likelihood. Of course those power spectra are strongly dominated by the CMB, so, to reveal the dust content, one has to rely on the same trick that was used for 545\ghz. This however is not enough, since the CMB cosmic variance itself is significant compared to the dust contamination. We can build an estimate of the CMB cosmic variance by assuming that at 100\ghz\ the dust contamination is small enough that a mask difference gives us a good variance estimate.

Figure~\ref{fig:hil:dust143217} shows the mask difference (corrected for cosmic variance) between G60 and G41 for the 217\ghz\ and 143\ghz\ half-mission cross-spectra, as well as the dust model from Eq.~(\ref{eq:dust:TTtemplate}). The dust model has been rescaled to the expected mask difference dust residual for the 217\ghz. The 143\ghz\ mask-difference has also been rescaled in a similar way. The ratio between the two is about $14$. Rescaling factors are obtained from Table~\ref{table:hil:dustTT}. Error bars are estimated based on the scatter in each bin. The agreement with the model is very good at 217\ghz, but less good  at 143\ghz\ where the greater scatter is probably dominated at large scales  by the chance correlation between CMB and dust (which, as we see in Eq.~\ref{eq:hil:chancedust}, varies as the square root of the dust contribution to the spectra), and at small scale by noise. We also tested these double differences for other masks, namely G50 $-$ G41 and G60 $-$ G50, and verified that the results are similar (\ie general agreement although with substantial scatter).  

Finally, we can estimate the level of the dust contamination in each of our frequency maps used for CMB analysis by computing their cross-spectra with the 545\ghz\ half-mission maps. Assuming that all our maps $\vec{m}^{\nu}$ have in common only the CMB and a variable amount of dust, and assuming that $\vec{m}^{545} = \vec{m}^{\mathrm{cmb}} + a^{545} \vec{m}^{\mathrm{dust}}$, the cross-spectra between each of our CMB frequencies maps
and the 545\ghz\ map is
\begin{eqnarray}
 \left ( C^{TT}_{545\times \nu}\right)_{\ell} = C^{TT,\rm cmb}_{\ell} &+& a^{TT,\rm dust}_{545}a^{TT,\rm dust}_{\nu}\, C^{TT,\rm dust}_{\ell}\nonumber\\ &+& (a^{TT,\rm dust}_{545}+a^{TT,\rm dust}_{\nu})\, C^{\rm chance}_{\ell},\label{eq:hil:chancedust}
\end{eqnarray}
where $C^{\rm chance}_{\ell}$ is the chance correlation between the CMB and dust distribution (which would vanish on average over many sky realizations). By using the 100\ghz\ spectrum as our CMB estimate and assuming that the chance correlation is small enough, one can measure the amount of dust in each frequency map by fitting the rescaling factor between the (CMB cleaned) 545\ghz\ spectrum and the cross frequency spectra. This approach is limited by the presence of CIB which has a slightly different emission law than the dust. We thus limit our fits to the multipoles $\ell<1000$ where the CIB is small compared to the dust and we ignore the emission-law differences.

Table~\ref{table:hil:dustTT} reports the results of those fits at each frequency, for each Galactic mask. The error range quoted corresponds to the error of the fits, taking into account the variations when changing the multipole range of the fit from $30\leq\ell\le1000$ to $30\leq\ell\le500$. The values reported correspond to the sum of the CIB and the dust contamination at $\ell=200$. The last column gives the estimate of the CIB contamination at the same multipole from the joint cosmology and foreground fit. From this table, the ratio of the dust contamination at map level between the 217\ghz\  and 100\ghz\  is around 7, while the ratio between the 217\ghz\ and 143\ghz\ is close to 3.7.

We derive our priors on the foreground amplitudes from this table, combining the 545\ghz\ fit with the estimated residual CIB contamination, to obtain the following values: $(7\pm2)\,\muK^2$ for the $100\times100$ spectrum (G70); $(9\pm2)\,\muK^2$ for $143\times143$ (G60);  and $(80\pm20)\,\muK^2$ for $217\times217$ (G50). Finally the $143\times217$ value is obtained by computing the geometrical average between the two auto spectra under the worst mask (G60), yielding $(21\pm8.5)\,\muK^2$.

\begin{table*}[ht!] 
\begingroup 
\newdimen\tblskip \tblskip=5pt
\caption{Contamination level in each frequency, $D_{\ell=200}$.$^{\rm a}$}
\label{table:hil:dustTT}
\vskip -6mm
\footnotesize 
\setbox\tablebox=\vbox{
\newdimen\digitwidth
\setbox0=\hbox{\rm 0}
\digitwidth=\wd0
\catcode`*=\active
\def*{\kern\digitwidth}
\newdimen\signwidth
\setbox0=\hbox{+}
\signwidth=\wd0
\catcode`!=\active
\def!{\kern\signwidth}
\newdimen\decimalwidth
\setbox0=\hbox{.}
\decimalwidth=\wd0
\catcode`@=\active
\def@{\kern\decimalwidth}
\halign{ 
\hbox to 1in{#\leaderfil}\tabskip=2em& 
    \hfil#\hfil\tabskip=1em& 
    \hfil#\hfil&  
    \hfil#\hfil&  
    \hfil#\hfil&  
    \hfil#\hfil\tabskip=0pt\cr 
\noalign{\doubleline}
\omit&\multispan5\hfil\sc Contamination Level [\muK$^2$]\hfil\cr
\noalign{\vskip -3pt}
\omit&\multispan5\hrulefill\cr 
\noalign{\vskip 3pt}
\omit&\multispan4\hfil Mask\hfil\cr
\noalign{\vskip -3pt}
\omit\hfil\sc Frequency\hfil&\multispan4\hrulefill\cr
\omit\hfil [GHz]\hfil& G41& G50& G60& G70& CIB\cr 
\noalign{\vskip 3pt\hrule\vskip 5pt}
100&$*1.6\pm *0.8 $&$*1.6 \pm *0.8 $&$**3.2 \pm *1.2$&$**7.1 \pm *1.6$&$*0.24 \pm 0.04$\cr
143&$*6.0\pm *1.2 $&$*6.6 \pm *1.4 $&$*10@* \pm *1.8$&$*23.5 \pm *4@* $&$*1.0* \pm 0.2*$\cr
217&$84@*\pm 16@* $&$91@* \pm 18@* $&$150@* \pm 20@* $&$312@* \pm 35@* $&$10@** \pm 2@**$\cr
\noalign{\vskip 5pt\hrule\vskip 3pt}
}}
\endPlancktablewide 
\tablenote {{\rm a}} The levels reported in this table correspond to the amplitude of the contamination, $\mathcal{D}_\ell$, at $\ell=200$ in $\muK^2$. They are obtained at each frequency by fitting the 545\ghz\ cross half-mission spectra against the CMB-corrected $545\times 100$, $545\times 143$ and $545\times 217$ spectra over a range of multipoles. The CMB correction is obtained using the 100\ghz\ cross half-mission spectra. This contamination is dominated by dust, with a small CIB contribution. The columns labelled with a Galactic mask name (G41, G50, G60, and G70) correspond to the results when combining those masks with the same CO, extended object, and frequency-combined point-source masks. The CIB contribution is shown in the last column. The errors quoted here include the variation when changing the range of multipoles used from $30\leq\ell\le1000$ to $30\leq\ell\le500$.\par
\endgroup
\end{table*} 
 
\paragraph{\textbf{Galactic $\TE$ and $\EE$ dust emission.}}
\label{sub:galactic-dustPol}
We evaluate the dust contribution in the $\TE$ and $\EE$ power spectra using the same method as for the temperature. However, instead of the 545\ghz\ data we use the maps at 353\ghz, our highest frequency with polarization information. At sufficently high sky fractions, the 353\ghz\ $\TE$ and $\EE$ power spectra are dominated by dust. As estimated in \citet{planck2014-XXX}, there is no other significant contribution from the Galaxy, even at 100\ghz. Following \citet{planck2014-XXX}, and since we do not mask any  ``point-source-like'' region of strong emission, we can use a power-law model as a template for the polarized Galactic dust contribution. Enforcing a single power law for $\TE$ and $\EE$ and our different masks, we obtain an index of $n=-2.4$. We use the same cross-spectra-based method to estimate the dust contamination. The dust contribution being smaller in polarization, removing the CMB from the $353 \times 353$ and the $353 \times \nu$ (with $\nu$ being one of 100, 143 or 217) is particularly important. Our two best CMB estimates in $\EE$ and $\TE$ being 100 and the 143\ghz, we checked that using any of $100 \times 100$, $143 \times 143$, or $100 \times 143$ does not change the estimates significantly. Table~\ref{table:hil:dustPol}  gives the resulting values. As for the $\TT$ case, the cross-frequency, cross-masks estimates are obtained by computing the geometric average of the auto-frequency contaminations under the smallest mask.

\begin{table*}[ht!] 
\begingroup 
\newdimen\tblskip \tblskip=5pt
\caption{$TE$ and $EE$ dust contamination levels,  $D_{\ell=500}$.}
\label{table:hil:dustPol}
\vskip -6mm
\footnotesize 
\setbox\tablebox=\vbox{
\newdimen\digitwidth
\setbox0=\hbox{\rm 0}
\digitwidth=\wd0
\catcode`*=\active
\def*{\kern\digitwidth}
\newdimen\signwidth
\setbox0=\hbox{+}
\signwidth=\wd0
\catcode`!=\active
\def!{\kern\signwidth}
\newdimen\decimalwidth
\setbox0=\hbox{.}
\decimalwidth=\wd0
\catcode`@=\active
\def@{\kern\decimalwidth}
\halign{ 
\hbox to 1.5in{#\leaderfil}\tabskip=0.5em& 
    \hfil#\hfil&
    \hfil#\hfil&
    \hfil#\hfil\tabskip=0pt\cr
\noalign{\doubleline}
\omit&\multispan3\hfil Contamination level [\muK$^2$]\hfil\cr
\noalign{\vskip -3pt}
\omit&\multispan3\hrulefill\cr
\noalign{\vskip 3pt} 
\omit\hfil Spectrum\hfil& 100 GHz (G70)& 143 GHz (G50)& 217 GHz (G41)\cr 
\noalign{\vskip 3pt\hrule\vskip 5pt}
\omit{\boldmath{$D^{TE}_{\ell=500}$}}\cr
\noalign{\vskip 6pt}
\hglue 1.5em 100 GHz (G70)& $0.14\pm 0.042 $&  $0.12 \pm 0.036$&  $0.3 \pm 0.09*$\cr
\hglue 1.5em 143 GHz (G50)&               &    $0.24 \pm 0.072$&  $0.6 \pm 0.018$\cr
\hglue 1.5em 217 GHz (G41)&               &                  &    $1.8 \pm 0.54*$\cr
\noalign{\vskip 8pt}
\omit{\boldmath{$D^{EE}_{\ell=500}$}}\cr 
\noalign{\vskip 6pt}
\hglue 1.5em 100 GHz (G70)& $0.06\pm 0.012 $&  $0.05 \pm 0.015$&  $0.11 \pm 0.033$\cr
\hglue 1.5em 143 GHz (G50)&               &    $0.1* \pm 0.02*$&  $0.24 \pm 0.048$\cr
\hglue 1.5em 217 GHz (G41)&               &                  &    $0.72 \pm 0.14*$\cr
\noalign{\vskip 5pt\hrule\vskip 3pt}
}}
\endPlancktablewide
\tablenote {{\rm a}} Values reported in the table correspond to the evaluation of the contamination level in each frequency by fitting the 353\ghz\ cross half-mission spectra against the CMB-corrected $353\times 100$, $353\times 143$ and $353\times 217$ spectra over a range of multipoles. The CMB correction is obtained using the 100\ghz\ cross half-mission spectra (we have similar results at 143\ghz). Level reported here correspond to the amplitude of the contamination $\mathcal{D}_\ell$ at $\ell=500$ in $\muK^2$.\par
\endgroup
\end{table*} 

\subsubsection{Extragalactic foregrounds}
\label{sec:extragalactic_foregrounds}

The extragalactic foreground model is similar to that of 2013 and in the following we describe the differences. Since we are neglecting any possible contribution in polarization from extragalactic foregrounds, we omit the $\TT$ index in the following descriptions of the foreground models. The amplitudes are expressed as ${\cal D}_\ell$ at $\ell=3000$ so that, for any component, the template, $C^{\rm FG}_{3000}$, satisfies $C^{\rm FG}_{3000}\, {\cal A}_{3000}=1$ with ${\cal A}_\ell = \ell(\ell+1)/(2\pi)$.

\paragraph{\textbf{The Cosmic Infrared Background.}}
\label{sec:hil:cib}

The CIB model has a number of differences from that used in \citetalias{planck2013-p08}. First of all, it is now entirely parameterized by a single amplitude ${\cal D}_{217}^{\rm CIB}$ and a template $C_\ell^{\rm CIB}$:
\begin{equation}
\left( C^{\rm CIB}_{\nu\times\nu'}\right)_{\ell} 
= a^{\rm CIB}_{\nu} a^{\rm CIB}_{\nu'} C_{\ell}^{\rm CIB} 
\times {\cal D}_{217}^{\rm CIB}, 
\label{eq:CIB}
\end{equation}
where the spectral coefficients $a^{\rm CIB}_{\nu}$ represent the CIB emission law normalized at $\nu=217\,\GHz$. 

In 2013, the template was an effective power-law model with a variable index with expected value $n=-1.37$ (when including the  ``highL'' data from ACT and SPT). We did not assume any emission law and fitted the 143\,GHz and 217\,GHz amplitude, along with their correlation coefficient. The Planck Collaboration has studied the CIB in detail in \citet{planck2013-pip56} and now proposes a one-plus-two-halo model, which provides an accurate description of the \Planck\ and IRAS CIB spectra from 3000\ghz\ down to 217\ghz. We extrapolate this model here, assuming it remains appropriate in describing the 143\ghz\ and 100\ghz\ data. The CIB emission law and template are computed following \citet{planck2013-pip56}. The template power spectrum provided by this work has a very small frequency dependence that we ignore. 

At small scales, $\ell>2500$, the slope of the template is similar to the power law used in \citetalias{planck2013-p08}. At larger scales, however, the slope is much shallower. This is in line with the variation we observed in 2013 on the power-law index of our simple CIB model when changing the maximum multipole. The current template is shown as the green line in the $\TT$ foreground component plots in Fig.~\ref{fig:hil:fg}.

In 2013, the correlation between the 143\ghz\ and 217\ghz\ CIB spectra was fitted, favouring a high correlation, greater than 90\,\% (when including the  ``highL'' data). The present model yields a fully correlated CIB between 143\,GHz and 217\,GHz.

We now include the  the CIB contribution at 100\,GHz, which  was ignored in 2013. Another difference with the 2013 model is that the parameter controlling the amplitude at 217\ghz\ now directly gives the amplitude in the actual 217\ghz\ \Planck\ band at $\ell=3000$, \ie it includes the colour correction. The ratio between the two is $1.33$. The 2013 amplitude of the CIB contribution at $\ell=3000$ (including the highL data) was $66 \pm 6.7\  \mu {\rm K}^2$, while our best estimate for the present analysis is $63.9 \pm 6.6\  \mu {\rm K}^2$ ($\planckTT$).

\paragraph{\textbf{Point sources.}} \label{sec:point-sources}

At the likelihood level, we cannot differentiate between the radio- and IR-point sources. We thus describe their combined contribution by their total emissivity per frequency pair, 
\begin{equation}
\left( C^{\rm PS}_{\nu\times\nu'}\right)_{\ell} 
= {\cal D}_{\nu\times\nu'}^{\rm PS}/{\cal A}_{3000}\, ,
\label{eq:point-sources}
\end{equation} 
where  ${\cal D}_{\nu\times\nu'}$ is the amplitude of the point-source contribution in $\mathcal{D}_\ell$ at $\ell=3000$. Contrary to 2013, we do not use a correlation parameter to represent the $143 \times 217$ point-source contribution; instead we use a free amplitude parameter. This has the disadvantage of not preventing a possible unphysical solution. However, it simplifies the parameter optimization, and it is easier to understand in terms of contamination amplitude.

\paragraph{\textbf{Kinetic SZ (kSZ).}} \label{sec:ksz}

We use the same model as in 2013. The kSZ emission is parameterized with a single amplitude and a fixed template from \citet{TBO11},
\begin{equation}
\left( C^{\rm kSZ}_{\nu\times\nu'}\right)_{\ell} = C_{\ell}^{\rm kSZ} \times {\cal D}^\mathrm{kSZ} \,,
\label{eq:kSZ}
\end{equation}
where ${\cal D}^\mathrm{kSZ}$ is the kSZ contribution at $\ell=3000$. 

\paragraph{\textbf{Thermal SZ (tSZ).}}\label{sec:tsz}

Here again, we use the same model as in 2013. The tSZ emission is also parameterized by a single amplitude and a fixed template using the $\epsilon=0.5$ model from \citet{EM012},
\begin{equation}
\left( C^{\rm tSZ}_{\nu\times\nu'}\right)_{\ell} = a^{\rm tSZ}_{\nu} a^{\rm tSZ}_{\nu'}  C_{\ell}^{\rm tSZ} \times {\cal D}_{143}^\mathrm{tSZ} \,,
\label{eq:tSZ}
\end{equation}
where $ a^{\rm tSZ}_{\nu} $ is the thermal Sunyaev-Zeldovich spectrum, normalized to $\nu_0=143\,\ghz$ and corrected for the \Planck\ bandpass colour corrections. Ignoring the bandpass correction, we recall that the tSZ spectrum is given by 
\begin{equation}
a^{\rm tSZ}_{\nu} = \frac{f(\nu)}{f(\nu_0)},\ f(\nu) = \left(x\coth\left(\frac{x}{2}\right)-4 \right),\ x=\frac{h\nu}{k_\mathrm{B} T_\mathrm{cmb}}.
\end{equation}

\paragraph{\textbf{Thermal SZ $\times$ CIB correlation.} }\label{sec:sz-cib}

Following \citetalias{planck2013-p08} the cross-correlation between the thermal SZ and the CIB, $\mathrm{tSZ} \times$ CIB, is parameterized by a single correlation parameter, $\xi$, and a fixed template from \citet{Aetal12b}, 
\begin{equation}
\begin{split}
\left( C^{{\rm tSZ} \times {\rm CIB}}_{\nu\times\nu'}\right)_{\ell}  
&= \xi\, \sqrt{{\cal D}^{\rm tSZ}_{143}\, {\cal D}^{\rm CIB}_{217}} \\
&\times \left( a^{\rm tSZ}_{\nu}\, a^{\rm CIB}_{\nu'} + a^{\rm tSZ}_{\nu'}\, a^{\rm CIB}_{\nu} \right) \\
&\times C_{\ell}^{{\rm tSZ} \times {\rm CIB}}\, , 
\end{split}
\label{eq:tSZCIB}
\end{equation}
where $a^{\rm tSZ}_{\nu}$ is the thermal Sunyaev-Zeldovich spectrum, corrected for the \Planck\ bandpass colour corrections and $a^{\rm CIB}_{\nu}$ is the CIB spectrum, rescaled at $\nu=217\,\mathrm{GHz}$ as in the previous paragraphs.

\paragraph{\textbf{SZ prior.}}\label{sec:szprior}

The kinetic SZ, the thermal SZ, and its correlation with the CIB are not constrained accurately by the \Planck\ data alone. Besides, the tSZ$\times$CIB level is highly correlated with the amplitude of the tSZ. In 2013, we reduced the degeneracy between those parameters and improved their determination by adding the ACT and SPT data. In 2015, we instead impose a Gaussian prior on the tSZ and kSZ amplitudes, inspired by the constraints set by these experiments. From a joint analysis of the \Planck\ 2013 data with those from ACT and SPT, we obtain
\begin{equation}
{\cal D}^{\rm kSZ} + 1.6 {\cal D}^{\rm tSZ} = (9.5 \pm 3)\,\mu\textrm{K}^2,
\end{equation}
in excellent agreement with the estimates from \citet{reichardt12}, once they are rescaled to the \Planck\ frequencies  \citep[see][for a detailed discussion]{planck2014-a15}.

As can be seen in Fig.~\ref{fig:hil:fg}, the kSZ, tSZ, and tSZ$\times$CIB correlations are always dominated by the dust, CIB, and point-source contributions. 


\subsection{Instrumental modelling}
\label{sec:instrument}

The following sections describe the instrument modelling elements of the model vector, addressing the issues of calibration and beam uncertainties in Sects.~\ref{sec:calib-uncert}, \ref{sec:polcalib-uncert}, and \ref{sec:beam-uncert}, and describing the noise properties in Sect.~\ref{sec:noise_model}. For convenience, Table~\ref{tab:fg-params} defines the symbol used for the calibration parameters and the priors later used for exploring them.

\subsubsection{Power spectra calibration uncertainties}\label{sec:calib-uncert}

As in 2013, we allow for a small recalibration of the different frequency power spectra, in order to account for  residual uncertainties in the map calibration process. The mixing matrix in the model vector from Eq.~(\ref{eq:spectrum-model}) can be rewritten as
\begin{eqnarray}
\left ( {M}^{XY}_{ZW,\nu \times \nu'}\right )_\ell(\theta_{\rm inst}) &=&  G^{XY}_{\nu\times\nu'}(\theta_{\rm calib}) \left ( {M}^{XY,\mathrm{other}}_{ZW,\nu \times \nu'}\right )_\ell(\theta_{\rm other}) \,, \nonumber \\
G^{XY}_{\nu\times\nu'}(\theta_{\rm calib})  &=& \frac{1}{\calibM_{\rm P}^2} \left( \frac{1}{2\sqrt{\calibC^{XX}_{\nu}\calibC^{YY}_{\nu'}}} + \frac{1}{2\sqrt{\calibC^{XX}_{\nu'}\calibC^{YY}_{\nu}}} \right) \,, \label{eq:caldef}
\end{eqnarray}
where $\calibC^{XX}_{\nu}$ is the calibration parameter for the $XX$ power spectrum at frequency $\nu$, $X$ being either $T$ or $E$, and $\calibM_{\rm P}$ is the overall \planck\ calibration. We ignore the $\ell$-dependency of the weighting function between the $\TE$ and $\ET$ spectra at different frequencies that are added to form an effective cross-frequency $\TE$ cross-spectrum. As in 2013, we use the $\TT$ at 143\ghz\ as our inter-calibration reference, so that $\calibC^{TT}_{143}=1$. 

We further allow for an overall \planck\ calibration uncertainty, whose variation is constrained by a tight Gaussian prior, 
\begin{equation}
\calibM_{\rm P} = 1 \pm 0.0025.
\end{equation}
This prior corresponds to the estimated overall uncertainty, which is discussed in depth in \citet{planck2014-a01}.

The calibration parameters can be degenerate with the foreground parameters, in particular the point sources at high $\ell$ (for $\TT$) and the Galaxy for 217\ghz\ at low $\ell$. We thus proceed as in 2013, and measure the calibration refinement parameters on the large scales and on small sky fractions near the Galactic poles. We perform the same estimates on a range of Galactic masks (G20, G30, and G41) restricted to different maximum multipoles (up to $\ell=1500$). The fits are performed either by minimizing the scatter between the different frequency spectra, or by using the \smica algorithm \citep[see][section~7.3]{planck2013-p03} with a freely varying CMB and generic foreground contribution. For the $\TT$ spectra, we obtained in both cases very similar recalibration estimates, from which we extracted the conservative Gaussian priors on recalibration factors, 
\begin{eqnarray}
\label{eq:relcal}
\calibC^{TT}_{100} &=& 0.999 \pm 0.001 \,, \\
\calibC^{TT}_{217} &=& 0.995 \pm 0.002 \,.
\end{eqnarray}
These are compatible with estimates made at the map level, but on the whole sky; see \citet{planck2014-a09}.

\subsubsection{Polarization efficiency and angular uncertainty}\label{sec:polcalib-uncert}

We now turn to the polarization recalibration case. 
The signal measured by an imperfect PSB is given by
\begin{equation}
  d = G(1+\gamma)\left[ I + \rho(1+\eta)\left(Q \cos 2(\phi+\omega) + U
\sin 2(\phi+\omega) \right) \right] + n \,,
\end{equation}
where $I$, $Q$, and $U$ are the Stokes parameters; $n$ is the instrumental noise; $G$, $\rho$, and $\phi$ are the nominal photometric calibration factor, polar efficiency, and direction of polarization of the PSB; and  $\gamma$, $\eta$, and $\omega$ are the (small) errors made on each of them \citep[see, e.g.,][]{Jones:2007ji}. Due to these errors, the measured cross-power spectra of maps $a$ and $b$ are then contaminated by a spurious signal given by
\begin{subequations}
\label{eq:spurious_Cl_systematics}
\begin{eqnarray}
  \Delta C_\ell^{TT} &=&   \left( \gamma_a + \gamma_b \right)  C_\ell^{TT},
\label{eq:spurious_Cl_systematics_a}
 \\
  \Delta C_\ell^{TE} &=&  \left(\gamma_a + \gamma_b +   \eta_b - 2 \omega_b^2\right) C_\ell^{TE},
\label{eq:spurious_Cl_systematics_b}
 \\
  \Delta C_\ell^{EE} &=&  \left(\gamma_a + \gamma_b + \eta_a + \eta_b - 2
\omega_a^2 - 2 \omega_b^2 \right) C_\ell^{EE} \nonumber \\
 & & {}+ 2 \left(\omega_a^2 + \omega_b^2\right) C_\ell^{BB},
\label{eq:spurious_Cl_systematics_c}
\end{eqnarray}
\end{subequations}
where $\gamma_x$, $\eta_x$, and $\omega_x$, for $x=a, b$, are the effective instrumental errors for each of the two frequency-averaged maps.
Pre-flight measurements of the \HFI\ polarization efficiencies, $\rho$, had uncertainties $|\eta_x|\approx 0.3\,\%$, while the polarization angle of each PSB is known to $|\omega_x| \approx 1\deg$ \citep{rosset2010}. Analysis of the 2015 maps shows the relative photometric calibration of each detector at 100 to 217\GHz\ to be known to about $|\gamma_x| = 0.16\,\%$ at worst, with an absolute orbital dipole calibration of about 0.2\,\%, while analysis of the Crab Nebula observations showed the polarization uncertainties to be consistent with the pre-flight measurements \citep{planck2014-a09}.

Assuming $C_\ell^{BB}$ to be negligible, and ignoring $\omega^2 \ll |\eta|$ in Eq.~(\ref{eq:spurious_Cl_systematics}), the Gaussian priors on $\gamma$ and $\eta$ for each frequency-averaged polarized map would have rms of $\sigma_{\gamma}=2\times 10^{-3}$ and $\sigma_{\eta}=3\times 10^{-3}$. Adding those uncertainties in quadrature, the auto-power spectrum recalibration $\calibC^{EE}_{\nu}$ introduced in Eq.~(\ref{eq:caldef}) would be given, for an equal-weight combination of $n_\mathrm{d} = 8$ polarized detectors, by
\begin{eqnarray}
\calibC^{EE}_{\nu} = 1 \pm 2 \sqrt{ \frac{\sigma_{\gamma}^2 + \sigma_{\eta}^2}{n_\mathrm{d}} } =  1 \pm 0.0025 .
\label{eq:calib_EE_prior}
\end{eqnarray}
The most accurate recalibration factors for $\TE$ and $\EE$ could therefore be somewhat different from $\TT$. We found, though, that setting the $\EE$ recalibration parameter to unity or implementing those priors makes no difference with respect to cosmology; \ie we recover the same cosmological parameters, with the same uncertainties. Thus, for the baseline explorations, we fixed  the $\EE$ recalibration parameter to unity, 
\begin{equation}
\calibC^{EE}_\nu = 1 \,,
\end{equation}
and the uncertainty on $\TE$ comes only from the $\TT$ calibration parameter through Eq.~\ref{eq:caldef}. 

We also explored the case of much looser priors, and found that best-fit calibration parameters deviate very significantly, and reach values of several percent (between 3\,\% and 12\,\% depending on the frequencies and on whether we fit the $\EE$ or $\TE$ case). This cannot be due to the instrumental uncertainties embodied in the prior. In the absence of an informative prior, this degree of freedom is used to minimize the differences between frequencies that stem from other effects, not included in the baseline modelling. 

The next section introduces one such effect, the temperature-to-polarization leakage, which is due to combining detectors with different beams without accounting for it at the map-making stage (see Sect.~\ref{sec:beam-uncert}). But anticipating the results of the analysis described in Appendix~\ref{sec:pol-robust}, we note that when the calibration and leakage parameters are explored simultaneously without priors, they remain in clear tension with the priors (even if the level of recalibration decreases slightly, by typically 2\,\%, showing the partial degeneracy between the two). In other words, when calibration and leakage parameters are both explored with their respective priors, there is evidence of residual unmodelled systematic effects in polarization ---  to which we will return. 

\subsubsection{Beam and transfer function uncertainties}\label{sec:beam-uncert}

\begin{figure*}[htb]
\centering
  \includegraphics[width=0.89\textwidth]{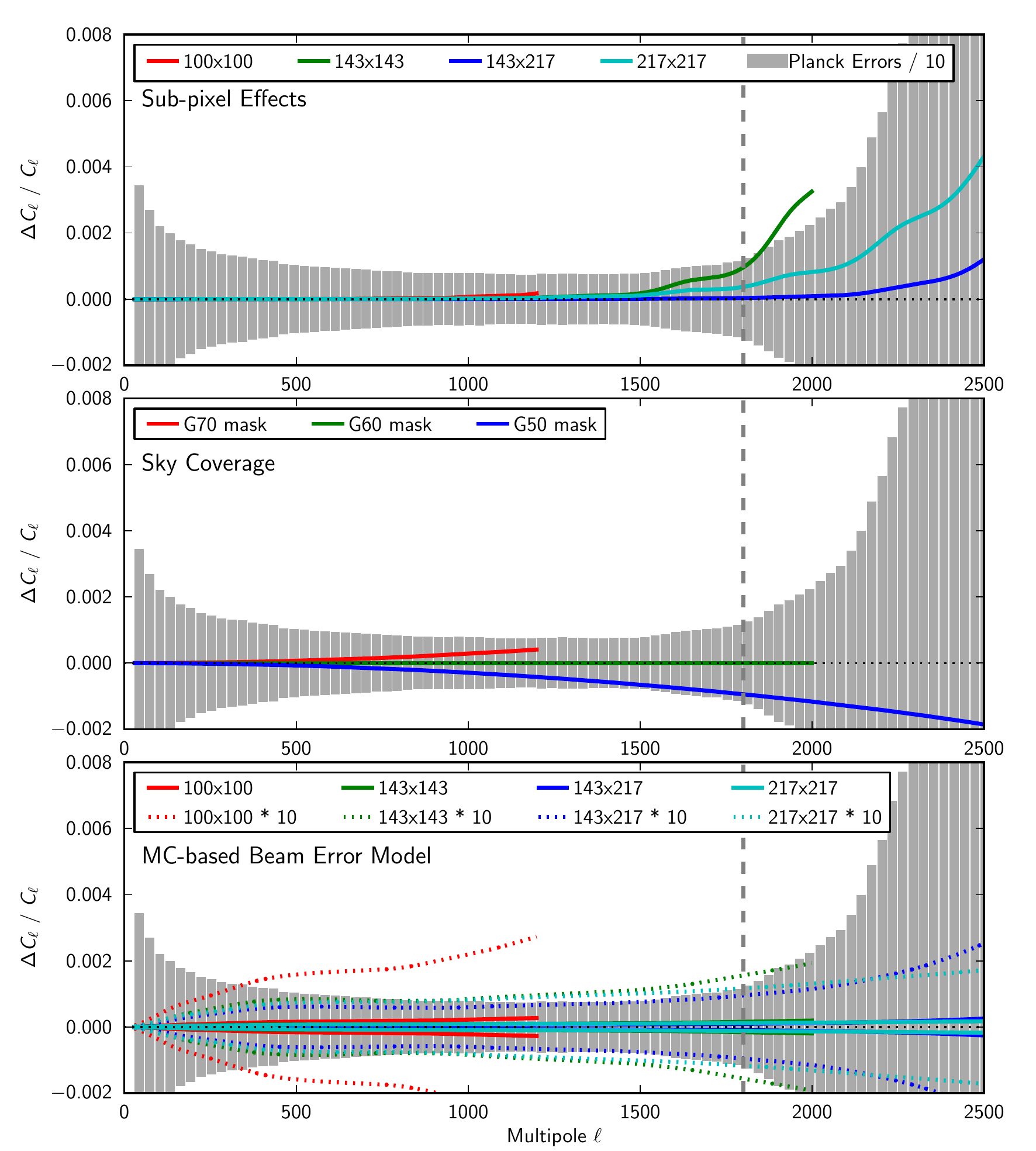}
  \caption{Contribution of various beam-window-function-related errors and
uncertainties to the $C_\ell$ relative error. In each panel, the grey histogram shows the relative statistical error on the \Planck\ CMB $\TT$ binned power spectrum (for a bin width $\Delta \ell=30$) {\em divided by 10}, while the vertical grey dashes delineate the range $\ell < 1800$ that is most informative for base \LCDM. {\it Top}: Estimation of the error made by ignoring the sub-pixel effects for a fiducial $C_\ell$ including the CMB and CIB contributions. {\it Middle}: Error due to the sky mask, for the Galactic masks used in the $\TT$ analysis. {\it Bottom}: Current beam window function error model, shown at $1\,\sigma$ (solid lines) and $10\, \sigma$ (dotted lines).}
  \label{fig:Wl_pse_fsky_MC}
\end{figure*}

The power spectra from map pairs are corrected by the corresponding effective beam window functions before being confronted with the data model. However, these window functions are not perfectly known, and we now discuss various related sources of errors and uncertainties, the impact of which on the reconstructed $C_\ell$s is shown in Fig.~\ref{fig:Wl_pse_fsky_MC}.

\paragraph{\textbf{Sub-pixel effects.}} The first source of error, the so-called ``sub-pixel'' effect, discussed in detail in \citetalias{planck2013-p08}, is a result of the \planck\  scanning strategy and map-making procedure. Scanning along rings with very low nutation levels can result in the centroid of the samples being slightly shifted from the pixel centres; however, the map-making algorithm assigns the mean value of samples in the pixel to the centre of the pixel. This effect, similar to the gravitational lensing of the CMB, has a non-diagonal influence on the power spectra, but the correction can be computed given the estimated power spectra for a given data selection, and recast into an additive, fixed component. We showed in \citetalias{planck2013-p08} that including this effect had little impact on the cosmological parameters measured by \planck. 

\begin{figure}[tb]
\centering
 \includegraphics[angle=90,width=0.47\textwidth]{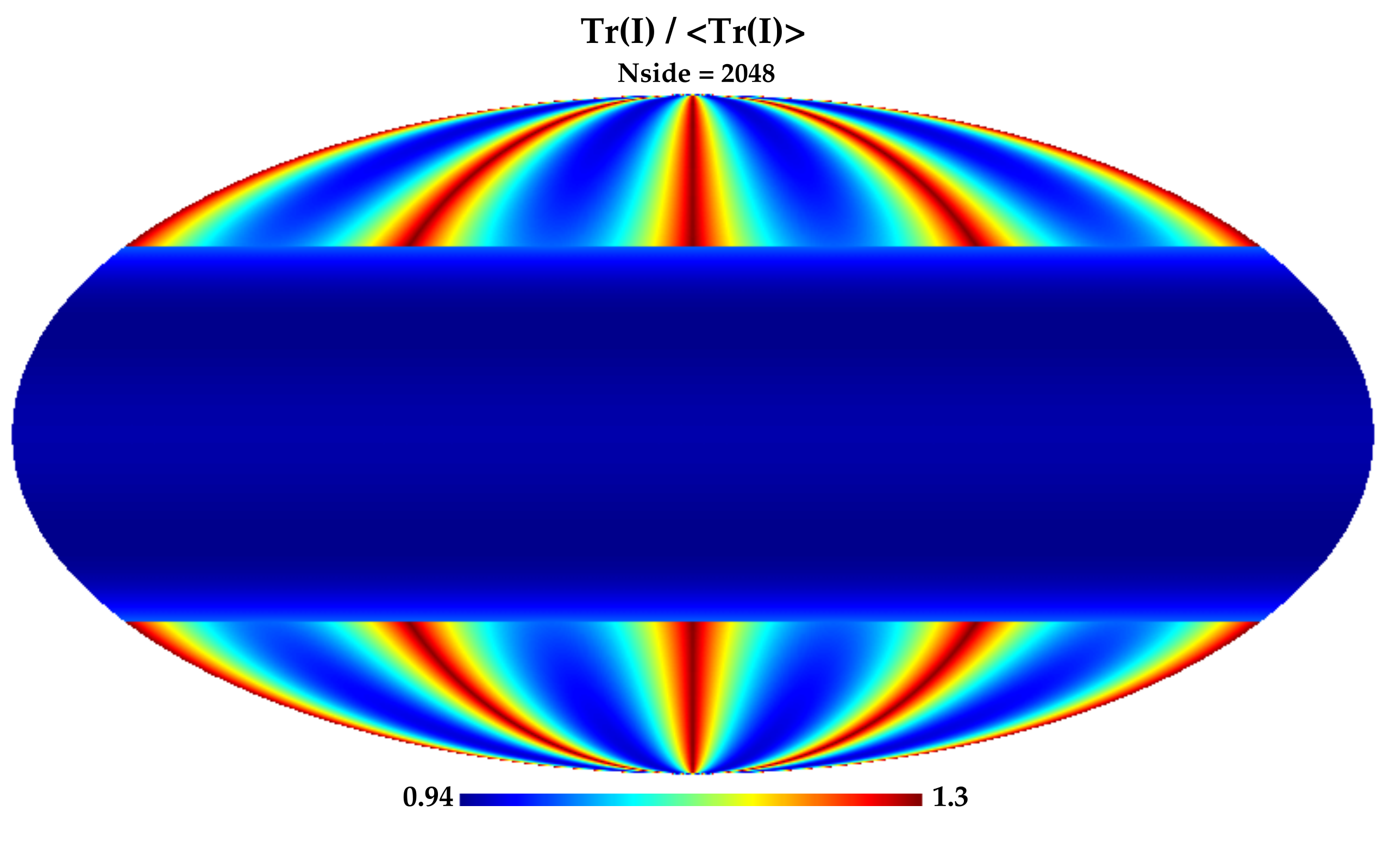}
 \caption{Map of the relative variations of the trace of the \healpix\ pixel moment of inertia tensor
 at $N_{\rm side}=2048$ in Galactic coordinates.}
 \label{fig:trace_inertia_healpix}
\end{figure}

\paragraph{\textbf{Masking effect.}} A second source of error is the variation, from one sky pixel to another,
 of the effective beam width, which is averaged over all samples falling in that pixel. While all the \healpix\ pixels have the same surface area, their shape --- and therefore their moment of inertia (which drives the pixel window function) --- depends on location, as shown in Fig.~\ref{fig:trace_inertia_healpix}, and therefore makes the effective beam window function depend on the pixel mask considered. Of course the actual sampling of the pixels by \planck\ leads to individual moments of inertia slightly different from the intrinsic values shown here, but spot-check comparisons of this semi-analytical approach used by \quickbeam\ with numerical simulations of the actual scanning by \febecop\ showed agreement at the $10^{-3}$ level for $\ell < 2500$ on the resulting pixel window functions for sky coverage varying from 40 to 100\,\%.

In the various Galactic masks used here (Figs.~\ref{fig:hil:mask_stack}--\ref{fig:hil:mask_T}) the contribution of the unmasked pixels to the total effective window function departs from the full-sky average (which is not included in the effective beam window functions), and we therefore expect a different effective transfer function for each mask. We ignored this dependence and mitigated its effect by using transfer functions computed with the Galactic mask G60 which retains an effective sky fraction (including the mask apodization) of $\fsky = 60\,\%$, not too different from the sky fractions $\fsky$ between 41 and 70\,\% (see Sect.~\ref{sec:masks}) used for computing the power spectra.

Figure~\ref{fig:Wl_pse_fsky_MC} compares the impact of these two sources of
uncertainty on the stated \planck\  statistical error bars for $\Delta \ell = 30$. It shows that, for $\ell<1800$ where most of the information on \LCDM\ lies, the error on the $\TT$ power spectra introduced by the sub-pixel effect and by the sky-coverage dependence are less than about $0.1\,\%$, and well below the statistical error bars of the binned $C_\ell$. In the range $1800\le\ell\le2500$, which helps constrain one-parameter extensions to base \LCDM\ (such as $N_{\rm eff}$), the relative error can reach
0.4\,\% (note as a comparison that the high-$\ell$ ACT experiment states a
statistical error of about $3\,\%$ on the  bin $2340\le\ell\le 2540$, \citealt{das/etal:prep}). The bottom panel shows the Monte Carlo error model of the beam window functions, which provides negligible ($\ell$-coupled) uncertainties. Even if this model is somewhat optimistic, since it does not include the effect of the ADC non-linearities and the colour-correction effect of beam measurements on planets \citep{planck2014-a08}, we note that even expanding them by a factor of 10 keeps them within the statistical uncertainty of the power spectra.

\paragraph{\textbf{Modelling the uncertainties.}} As in the 2013 analysis, the beam uncertainty eigenmodes were determined from 100 (improved) Monte Carlo (MC) simulations of each planet observation used to measure the scanning beams, then processed through the same \quickbeam pipeline as the nominal beam to determine their effective angular transfer function $B(\ell)$. Thanks to the use of Saturn and Jupiter transits instead of the dimmer Mars used in 2013, the resulting uncertainties are now  significantly smaller \citep{planck2014-a08}.

For each pair of frequency maps (and frequency-averaged beams) used in the present analysis, a singular-value
decomposition (SVD) of the correlation matrix of 100 Monte Carlo based $B(\ell)$ realizations was performed over the ranges $[0,\lmax]$ with $\lmax=(2000, 3000, 3000)$ at (100, 143, 217\GHz), and the five leading modes were kept, as well as their covariance matrix (since the error modes do exhibit Gaussian statistics). We therefore have, for each pair of beams, five $\ell$-dependent templates, each associated with a Gaussian amplitude centred on 0, and a covariance matrix coupling all of them. 

Including the beam uncertainties in the mixing matrix of Eq.~\ref{eq:spectrum-model} gives 
\begin{eqnarray}
\left ( M^{XY}_{ZW,\nu \times \nu'}\right )_\ell(\theta_{\rm inst}) 
&=&  \left ( M^{XY,\mathrm{other}}_{ZW,\nu \times \nu'}\right )_\ell(\theta_{\rm other})\ \left( \Delta W^{ZW}_{\nu \times \nu'} \right )_\ell(\theta_{\rm beam}) \,, \nonumber \\ 
\left( \Delta W^{ZW}_{\nu \times \nu'} \right )_\ell(\theta_{\rm beam}) 
&=& \exp{\sum_{i=1}^5 2\, \theta^{ZW,i}_{\nu\times\nu'} \left (E^{ZW,i}_{\nu\times\nu'}\right )_\ell} \,,
\end{eqnarray}
where $\left( \Delta W^{ZW}_{\nu \times \nu'} \right )_\ell(\theta_{\rm beam})$ stands for the beam error built from the eigenmodes $\left (E^{ZW,i}_{\nu\times\nu'}\right )_\ell$. The quadratic sum of the beam eigenmodes is
shown in Fig.~\ref{fig:Wl_pse_fsky_MC}. This is much smaller (less than a percent) than the combined $\TT$ spectrum error bars. This contrasts with the 2013 case where the beam uncertainties were greater; for instance, for the 100, 143, and 217\GHz\ channel maps, the rms of the $W(\ell)=B(\ell)^2$
uncertainties at $\ell=1000$ dropped from $(61, 23, 20)\times10^{-4}$ to
$(2.2, 0.84, 0.81)\times10^{-4}$, respectively. The fact that beam uncertainties are sub-dominant in the total error budget is even more pronounced in polarization, where noise is higher. We use the beam modes computed from temperature data, combined with appropriate weights when used as parameters affecting the $\TE$ and $\EE$ spectra.

As in 2013, instead of including the beam error in the vector model, we include its contribution to the covariance matrix, linearizing the vector model so that
\begin{equation}
\left ( C^{XY}_{\nu \times \nu'} \right )_\ell(\theta) 
= \left ( C^{XY}_{\nu \times \nu'} \right )_\ell(\theta,\theta_{\rm beam}=0) + \left( \Delta W^{ZW}_{\nu \times \nu'} \right )_\ell(\theta_{\rm beam}) \left (C^{XY}_{\nu \times \nu'} \right )^*_\ell \,,
\end{equation}
where $\left (C^{XY}_{\nu \times \nu'} \right )^*_\ell$ is the fiducial spectrum $XY$ for the pair of frequencies $\nu \times \nu'$ obtained using the best cosmological and foreground model.
We can then marginalize over the beam uncertainty, enlarging the covariance matrix to obtain
\begin{equation}
\tens{C}_{\mathrm{beam\ marg.}} = \tens{C} + \vec{C}^{*} \left < \Delta \vec{W}\Delta \vec{W}^\tens{T}\right > \vec{C}^{*\tens{T}} \,,
\end{equation}
where $\left < \Delta\vec{W}\Delta\vec{W}^\tens{T}\right >$ is the Monte Carlo based covariance matrix, restricted to its first five eigenmodes. 

In 2013, beam errors were marginalized for all the modes except the two greatest of the $100 \times 100$ spectrum. In the present release we instead marginalize over all modes in $\TT$, $\TE$, and $\EE$. We also performed a test in which we estimated the amplitudes for all of the first five beam eigenmodes in $\TT$, $\TE$, and $\EE$, and found no indication of any beam error contribution (see Sect.~\ref{sec:robust-beam} and Fig.~\ref{fig:wiskerTT}).

\paragraph{\textbf{Temperature-to-polarization leakage.\label{sec:beam_leakage}}} Polarization measurements are differential by nature. Therefore any unaccounted discrepancy in combining polarized detectors can create some leakage from temperature to polarization \citep{HuHedmanZaldarriaga2003}. Sources of such discrepancies in the current \HFI processing include, but are not limited to: differences in the scanning beams that are ignored during the map-making; differences in the noise level, because of the individual inverse noise weighting used in \HFI; and differences in the number of valid samples.

For this release, we did not attempt to model and remove {\it a priori} the form and amplitude of this coupling between the measured $\TT$, $\TE$, and $\EE$ spectra; we rather estimate the residual effect by fitting {\it a posteriori} in the likelihood some flexible template of this coupling, parameterized by some new nuisance parameters that we now describe. 

\begin{figure} [htb]
\begin{center}
\includegraphics[width=\columnwidth,angle=0]{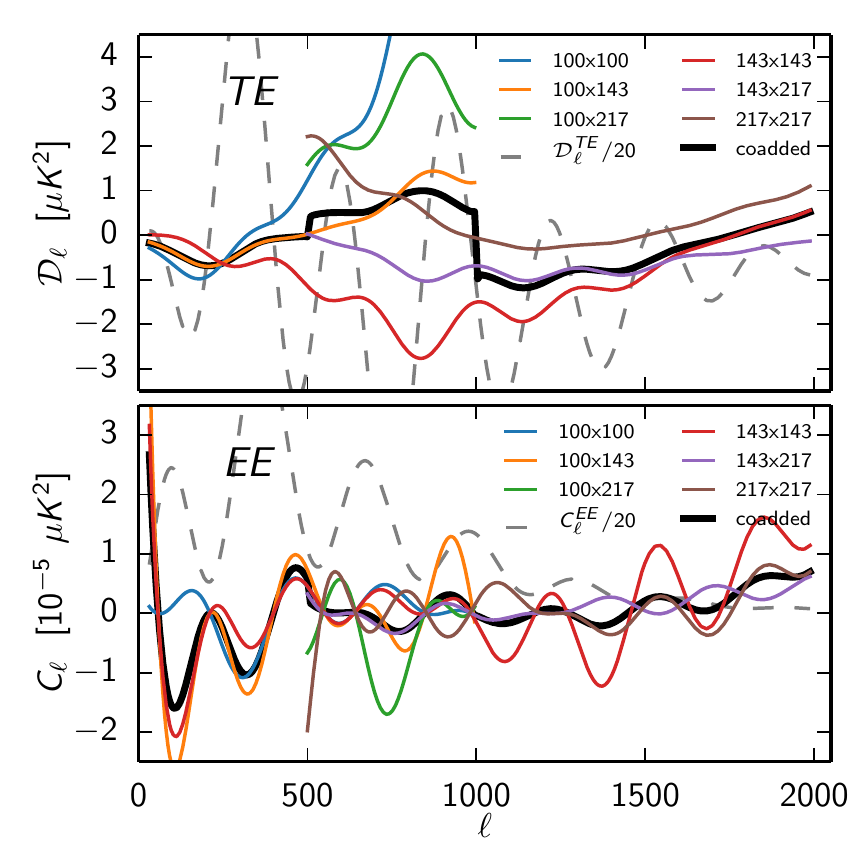}\\
\end{center}
\vspace{-2mm}
\caption {Best fit of the power spectrum leakage due to the beam mismatch for
$\TE$ (Eq.~\ref{eq:bleak_cl_a}, upper panel) and $\EE$ (Eq.~\ref{eq:bleak_cl_b}, lower panel). In each case, we show the correction for individual cross-spectra (coloured thin lines) and the co-added correction (black line). The individual cross-spectra corrections are only shown in the range of multipoles where the data from each particular pair is used. The individual correction can be much higher than the co-added correction. The co-added correction is dominated by the best S/N pair for each multipole. For example, up to $\ell=500$, the $\TE$ co-added correction is dominated by the $100 \times 143$ contribution. The grey dashed lines show the $\TE$ and $\EE$ best-fit spectra rescaled by a factor of 20, to give an idea of the location of the model peaks.}
\label{fig:bleak_fit}
\end{figure}

\newcommand{\vecr}{\ensuremath{{\bf r}}}

The temperature-to-polarization leakage due to beam mismatch is {\em assumed}
to affect the spherical harmonic coefficients via
\begin{subequations}
\label{eq:bleak_alm}
\begin{eqnarray}
	a_{\ell m}^T &\longrightarrow& a_{\ell m}^T \,, \\
	\label{eq:bleak_alm_t}
	a_{\ell m}^E &\longrightarrow& a_{\ell m}^E + \varepsilon({\ell}) a_{\ell m}^T \,,
	\label{eq:bleak_alm_e}
\end{eqnarray}
\end{subequations}
and, for each map, the spurious polarization power spectrum 
$C_\ell^{XY} \equiv \sum_m a^X_{\ell m} a^{Y*}_{\ell m} / (2\ell+1)$
is modelled as
\begin{subequations}
\label{eq:bleak_cl}
\begin{eqnarray}
  \Delta C_{\ell}^{TE} &=& \varepsilon({\ell})\, C_{\ell}^{TT}, 
  \label{eq:bleak_cl_a}
  \\
  \Delta C_{\ell}^{EE} &=& \varepsilon^2({\ell})\, C_{\ell}^{TT} + 2
\varepsilon({\ell})\, C_{\ell}^{TE}.
  \label{eq:bleak_cl_b}
\end{eqnarray}
\end{subequations}
Here $\varepsilon_\ell$ is a polynomial in multipole $\ell$ determined by the
effective beam of the detector-assembly measuring the polarized signal. 
Considering an effective beam map $b(\hat{\vec n})$ (rotated so that it is centred on the north pole), its spherical harmonic coefficients are defined as $b_{\ell m} \equiv \int d{\hat{\vec n}}\, b(\hat{\vec n})\, Y^*_{\ell m}(\hat{\vec n})$.
As a consequence of the \Planck\ scanning strategy, pixels are visited approximately every six months, with a rotation of the focal plane by $180\deg$, and we expect $b_{\ell m}$ to be dominated by even values of $m$, and especially the modes $m=2$ and $4$, which describe the beam ellipticity. As noted by, e.g., \citet{SouradeepRatra2001} for elliptical Gaussian beams, the \planck-\HFI\ beams for a detector $d$ obey
\begin{equation}
  b_{\ell m}^{(d)} \simeq \beta_m^{(d)} \ell^m b_{\ell 0}^{(d)} \,.
  \label{eq:bleak_lpower}
\end{equation}
We therefore fit the spectra using a fourth-order polynomial
\begin{equation}
\varepsilon(\ell) =  \varepsilon_0 +  \varepsilon_2 \ell^2 +  \varepsilon_4 \ell^4\,,   \label{eq:bleak_pol}
\end{equation}
treating the coefficients $\varepsilon_0$, $\varepsilon_2$, and $\varepsilon_4$ as nuisance parameters in the MCMC analysis.  Tests performed on detailed simulations of \planck\ observations with known mismatched beams have shown that Eqs.~(\ref{eq:bleak_cl}) and (\ref{eq:bleak_pol}) describe the power leakage due to beam mismatch with an accuracy of about 20\,\% in the $\ell$ range 100--2000. 

The equations above suggest that the same polynomial $\varepsilon$ can describe the contamination of the $\TE$ and $\EE$ spectra for a given pair of detector sets. But in the current \plik analysis, the $\TE$ cross-spectrum of two different maps $a$ and $b$ is the inverse-variance-weighted average of the cross-spectra $T_a E_b$ and $T_b E_a$, while $\EE$ is simply $E_aE_b$. In addition, the temperature maps include the signal from SWBs, which is obviously not the case for the $E$ maps. We therefore describe the $\TE$ and $\EE$ corrections by different $\varepsilon$ parameters. Similarly, we treated the parameters for the $EE$ cross-frequency spectra as being uncorrelated with the parameters for the auto-frequency ones.

The leakage is driven by the discrepancy between the individual effective beams $b_{\ell m}^{(d)}$ making up a detector assembly, coupled with the details of the scanning strategy and relative weight of each detector. If we assumed a perfect knowledge of the beams, precise --- but not necessarily accurate --- numerical predictions of the leakage would be possible. However, we preferred to adopt a more conservative approach in which the leakage was free to vary over a range wide enough to enclose the true value. On the other hand, in order to limit the unphysical range of variations permitted by so many nuisance parameters, we need priors on the $\varepsilon_m$ terms used in the Monte Carlo explorations. We assume Gaussian distributions of zero mean with a standard deviation $\sigma_m$ representative of the dispersion found in simulations of the effect with realistic instrumental parameters. We found $\sigma_0 = 1 \times 10^{-5}$, $\sigma_2 = 1.25 \times 10^{-8}$, and $\sigma_4 = 2.7 \times 10^{-15}$. This procedure ignores correlations between terms of different $m$, and is therefore likely substantially too permissive.     

Another way of deriving the beam leakage would be to use a cosmological prior, \ie by finding the best fit when holding the cosmological parameters fixed at their best-fit values for base \LCDM. Figure~\ref{fig:bleak_fit} shows the result of this procedure for the cross-frequency pairs. The figure also shows the implied correction for the co-added spectra. This correction is dominated by the pair with the highest S/N at each multipole. The fact that different sets are used in different $\ell$-ranges leads to discontinuities in the correction template of the co-added spectrum. As can be seen in the figure, the co-added beam-leakage correction, of order $\muKsq$, is much smaller than the individual corrections,
which partially compensate each other on average (but improve the agreement between the individual polarized cross-frequency spectra). 

It is shown in Appendix~\ref{sec:pol-robust} that neither procedure is fully satisfactory. The cosmological prior leads to nuisance parameters that vastly exceed the values allowed by the physical priors, and the physical priors are clearly overly permissive (leaving the cosmological parameters unchanged but with doubled error bars for some parameters). In any case, the agreement between the different cross-spectra remains much poorer in polarization than in temperature (see Sect.~\ref{sec:pol-rob}, Fig.~\ref{fig:respol}, and Appendix~\ref{sec:pol-robust}); they present oscillatory features similar to the ones produced by our beam leakage model, but the model is clearly not sufficient. For lack of a completely satisfactory global instrumental model, this correction is only illustrative and it is not used in the baseline likelihood.

\subsubsection{Noise modelling} \label{sec:noise_model}

To predict the variance of the empirical power spectra, we need to
model the noise properties of all maps used in the construction of the likelihood. As described in detail in \citet{planck2014-a08} and \citet{planck2014-a09}, the \planck\ \HFI\ maps have complicated noise properties, with noise levels varying spatially and with correlations between neighbouring pixels along the scanning direction.

\begin{figure}[htb] 
  \includegraphics[width=0.49\textwidth]{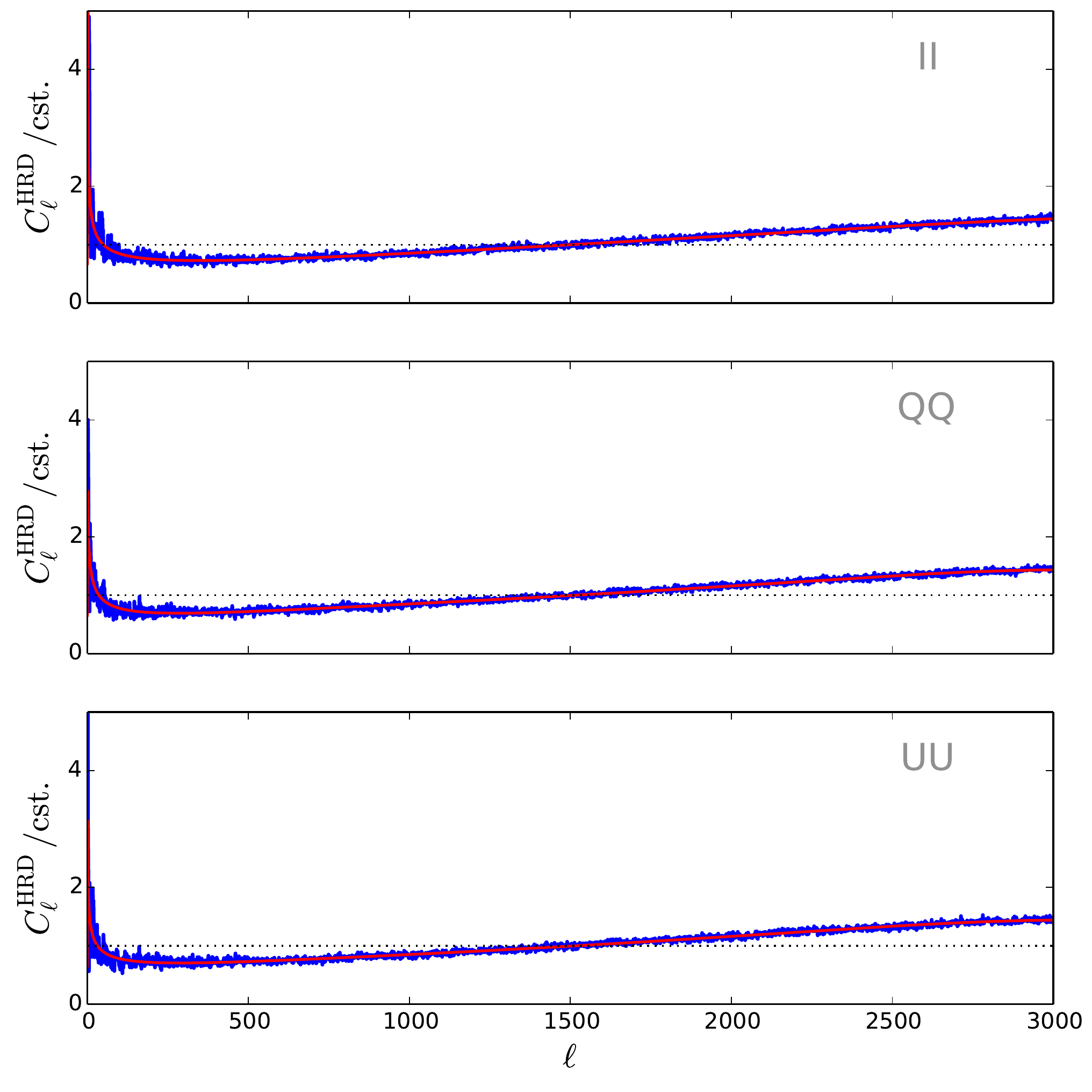}
  \caption{Deviations from a white noise power spectrum induced by noise correlations. We show half-ring difference power spectra for 100\,GHz half-mission 1 maps ({blue lines}) of Stokes parameters $I$ ({top panel}), $Q$ ({middle panel}), and $U$
    ({bottom panel}). The best-fitting analytical model of the form Eq.~(\ref{eq:plik_hrd_noise_fit}) is over-plotted in red.}
  \label{fig:plik_noise_hrd_100_hm1}
\end{figure}

\begin{figure}[htb] 
  \includegraphics[width=0.49\textwidth]{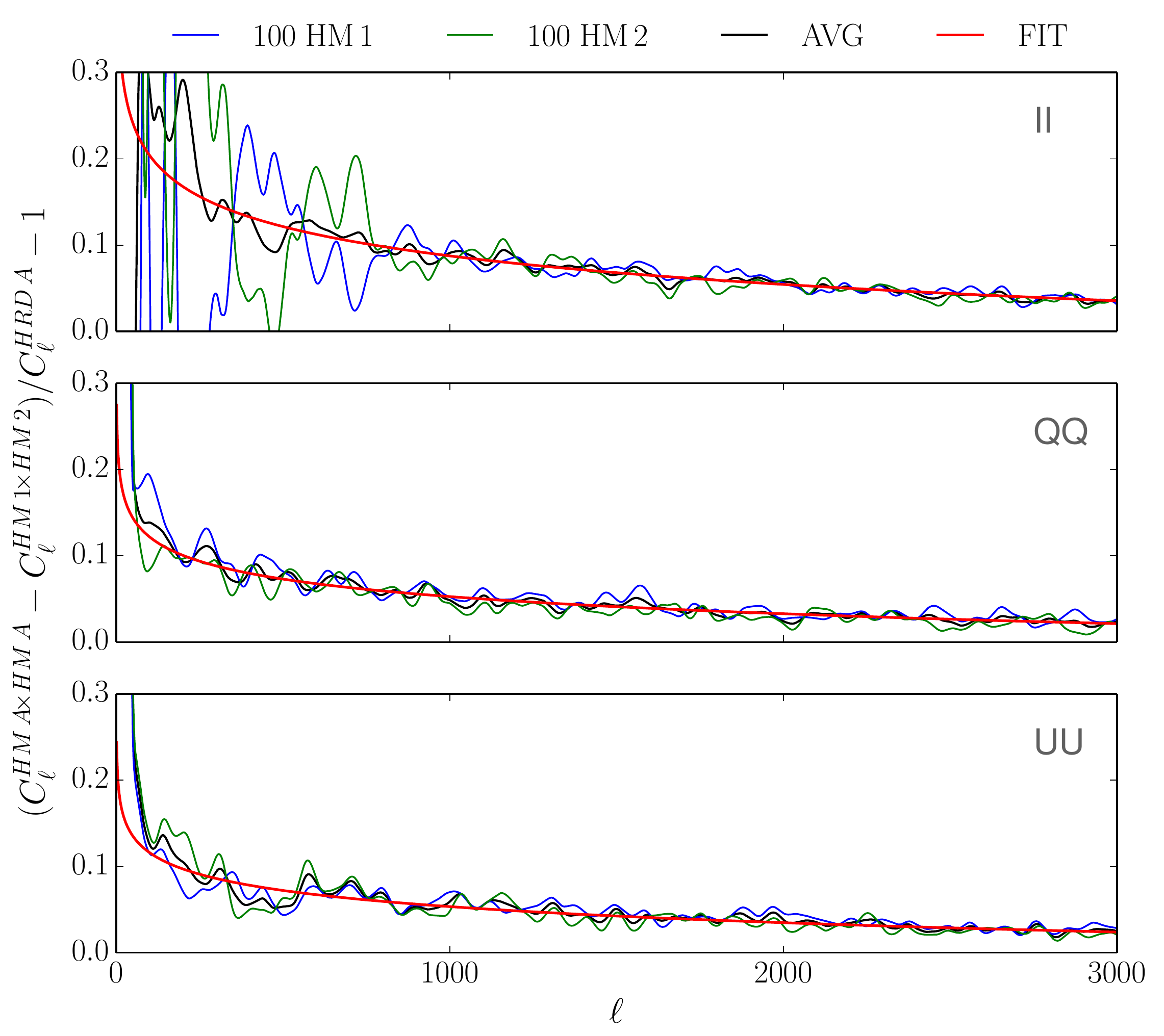}
  \caption{Difference between auto and cross-spectra for the 100\,GHz half-mission maps, divided by the noise estimate from half-ring difference maps ({blue and green lines}). Noise estimates derived from half-ring difference maps are biased low.  We fit the average of both half-mission curves ({black line}) with a power law model ({red line}). The analysis procedure is applied to the Stokes parameter maps $I$, $Q$, and $U$ ({top to bottom}). All data power spectra are smoothed.}
  \label{fig:plik_noise_hrd_bias_100}
\end{figure}
For each channel, full-resolution noise variance maps are constructed during the map-making process \citep{planck2014-a09}. They provide an approximation to the diagonal elements of the true $n_\mathrm{pix} 
\times n_\mathrm{pix}$ noise covariance matrix for Stokes parameters
$I$ (temperature only), or $I$, $Q$, and $U$ (temperature and
polarization). While it is possible to capture the anisotropic
nature of the noise variance with these objects, noise correlations
between pixels remain unmodelled. To include deviations from a white-noise power spectrum, we therefore make use of half-ring difference
maps. Choosing the 100\,GHz map of the first half-mission as an
example, we show the scalar (spin-0) power spectra of the three
temperature and polarization maps in
Fig.~\ref{fig:plik_noise_hrd_100_hm1}, rescaled by arbitrary
constants. We find that the logarithm of the \HFI\ noise power spectra
as given by the half-ring difference maps can be accurately
parameterized using a fourth-order polynomial with an additional logarithmic term,
\begin{equation}
  \log(C^{\mathrm{HRD}}_{\ell}) = \sum_{i = 0}^{4} \alpha_i \,
  \ell^{i} + \alpha_5 \log(\ell + \alpha_6) \, .
  \label{eq:plik_hrd_noise_fit}
\end{equation}

Since low-frequency noise and processing steps like deglitching leave residual correlations between both half-ring maps, noise estimates derived from their difference are biased low, at the percent level at high-$\ell$ (where it was first detected and understood, see \citealt{planck2013-p03}). We correct for this effect by comparing the difference of auto-power-spectra and  cross-spectra (assumed to be free of noise bias) at a given frequency
with the noise estimates obtained from half-ring difference maps. As
shown in Fig.~\ref{fig:plik_noise_hrd_bias_100}, we use a a power-law model with free spectral index to fit the average of the ratios of the first and second half-mission results to the half-ring difference spectrum, using the average to nullify chance correlations between signal and noise:
\begin{equation}
  C^{\mathrm{bias}}_{\ell} = \alpha_0 \, \ell^{\alpha_1} + \alpha_2 \, .
  \label{eq:plik_hrd_noise_bias}
\end{equation}
At a multipole moment of $\ell = 1000$, we obtain correction factors
for the temperature noise estimate obtained from half-ring difference
maps of 9\,\%, 10\,\%, and 9\,\% at 100, 143, and 217\,GHz, respectively.

In summary, our \HFI\ noise model is obtained as follows. For each
map, we capture the anisotropic nature of the noise amplitude by using the diagonal elements of the pixel-space noise covariance matrix. The corresponding white-noise power spectrum is then modulated in harmonic space using the product of the two smooth fitting functions given in Eqs.~(\ref{eq:plik_hrd_noise_fit}) and (\ref{eq:plik_hrd_noise_bias}). 

\begin{figure}[htb]
  \includegraphics[width=0.49\textwidth]{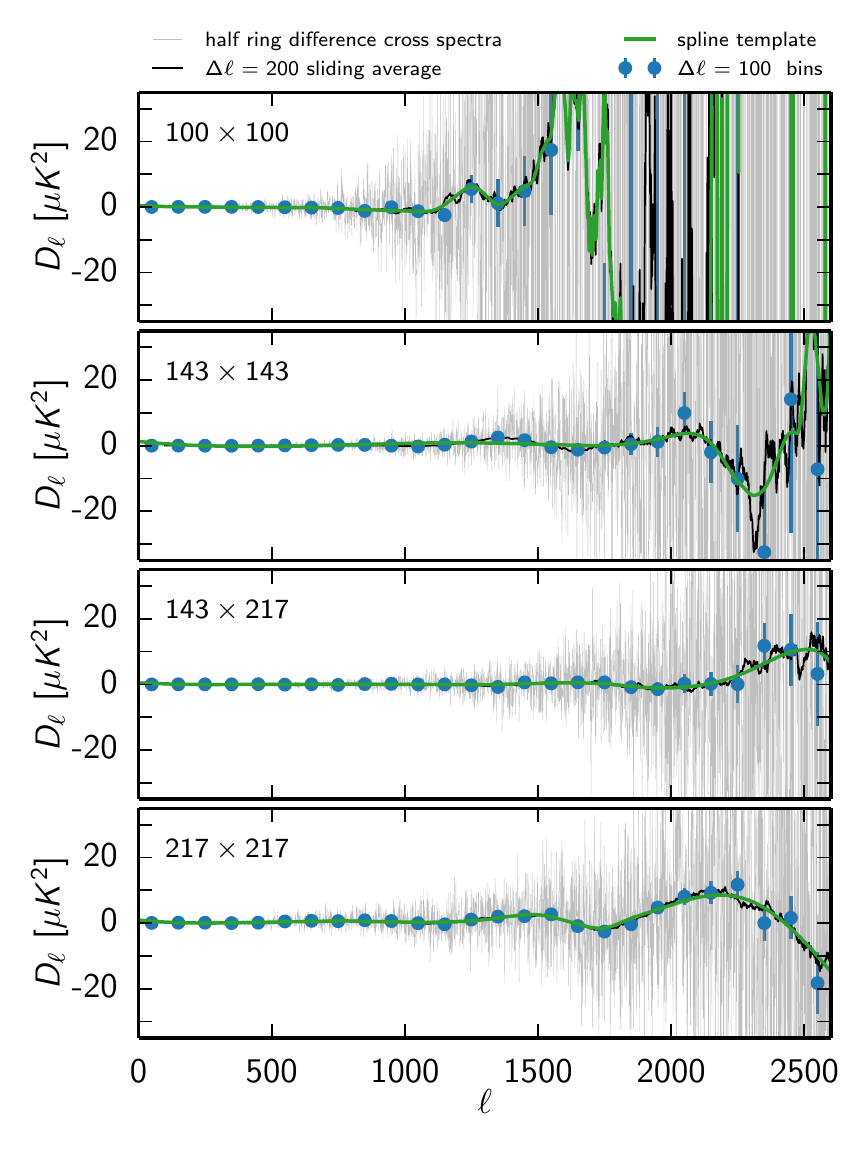}
  \caption{Correlated noise model. In grey are shown the cross-detector $\TT$ spectra of the half-ring difference maps. 
  The black line show the same, smoothed by a $\Delta\ell=200$ sliding average, while the blue data points are a $\Delta\ell=100$ binned version of the grey line. Error bars simply reflect the scatter in each bin. The green line is the spline-smoothed version of the data that we use as our correlated noise template.}
  \label{fig:hil:cornoise}
\end{figure}

\paragraph{\textbf{Correlated noise between detectors.}} If there is some correlation between the noise in the different cuts in our data, the trick of only forming effective frequency-pair power spectra from cross-spectra to avoid the noise biases fails. In 2013, we evaluated the amplitude of such correlated noise between different detsets. The correlation, if any, was found to be small, and we estimated its effect on the cosmological parameter fits to be negligible. 
As stated in Sect.~\ref{sec:hil:detcomb}, the situation is different for the 2015 data. Indeed, we now detect a small but significant correlated noise contribution between the detsets. This is the reason we change our choice of data to estimate the cross-spectra, from detsets to half-mission maps. The correlated noise appears to be much less significant in the latter.

To estimate the amount of correlated noise in the data, we measured the cross-spectra between the half-ring difference maps of all the individual detsets. The cross-spectra are then summed using the same inverse-variance weighting that we used in 2013 to form the effective frequency-pair spectra. Figure~\ref{fig:hil:cornoise} shows the spectra for each frequency pair. All of these deviate significantly from zero. We build an effective correlated noise template by fitting a smoothing spline on a $\Delta \ell=200$ sliding average of the data. Given  the noise level in polarization, we did not investigate the possible contribution of correlated noise in $\EE$ and $\TE$.

Section~\ref{sec:robust-detsets} shows that when these correlated noise templates are used, the results of the detsets likelihood are in excellent agreement with those based on the baseline, half-mission one.


\subsection{Covariance matrix structure}
\label{sec:covariance}

The construction of a Gaussian approximation to the likelihood function requires building covariance matrices for the pseudo-power spectra. Mathematically exact expressions exist, but they are prohibitively expensive to calculate numerically at \planck resolution
\citep{2001PhRvD..64h3003W}; we thus follow the approach taken in \citetalias{planck2013-p08} and make use of analytical approximations (\citealp{2002MNRAS.336.1304H, 2003ApJS..148..135H, Efstathiou2004,   2005MNRAS.360..509C}). 

For our baseline likelihood, we calculate covariance matrices for all 45 unique detector combinations that can be formed out of the six frequency-averaged half-mission maps at 100, 143, and 217\,GHz. To do so, we assume a fiducial power spectrum that includes the data variance induced by the CMB and all foreground components described in Sect.~\ref{sec:foreground_modelling}; this variance is computed assuming these components are Gaussian-distributed. The effect of this approximation regarding Galactic foregrounds is tested by means of simulations in Sect.~\ref{sec:valid-sims}. The fiducial model is taken from the best-fit cosmological and foreground parameters; since they only become available after a full exploration of the likelihood, we iteratively refine our initial guess. As discussed in Sect.~\ref{sec:highl:intro}, the data vector used in the likelihood function of Eq.~(\ref{eq:basic-likelihood}) is constructed from frequency-averaged power spectra. Following \citetalias{planck2013-p08}, for each polarization combination, we therefore build averaged covariance matrices for the four frequencies $\nu_1, \nu_2, \nu_3, \nu_4$,
\begin{multline}
\mathrm{Var}(\hat{C}_{\ell}^{XY \ \nu_1, \nu_2}, \hat{C}_{\ellp}^{ZW \ \nu_3, \nu_4})
= \sum_{\substack{(i, j) \in (\nu_1, \nu_2) \\ (p, q) \in (\nu_3, \nu_4)}}
w^{XY \ i,j}_{\ell} w^{ZW \ p, q}_{\ellp} \\
\times \mathrm{Var}(\hat{C}_{\ell}^{XY \ i, j}, \hat{C}_{\ellp}^{ZW \ p, q}) \, ,
\label{eq:hil_cl_cov_avg}
\end{multline}
where $X,Y,Z,W \in \{ T, E \}$, and $w^{XY \ i,j}$ is the inverse-variance weight for the combination $(i, j)$, computed from
\begin{equation}
w^{XY \ i,j}_{\ell} \propto 1/\mathrm{Var}(\hat{C}_{\ell}^{XY \ i, j},
\hat{C}_{\ell}^{XY \ i, j}) \, ,
\end{equation}
and normalized to unity. For the averaged $XY = TE$ covariance (and
likewise for $ZW = TE$), the sum in Eq.~(\ref{eq:hil_cl_cov_avg}) must be taken over the additional permutation $XY = ET$. That is, the two cases where the temperature map of channel $i$ is correlated with the polarization map of channel $j$ and vice versa are combined into a single frequency-averaged covariance matrix. These matrices are then combined to form the full covariance used in the likelihood,
\begin{equation}
\tens{C} = \begin{pmatrix} C^{TTTT} & C^{TTEE} & C^{TTTE} \\ 
  C^{EETT} & C^{EEEE} & C^{EETE} \\
  C^{TETT} & C^{TEEE} & C^{TETE} \end{pmatrix} \, ,
\end{equation}
where the individual polarization blocks are constructed from the frequency-averaged covariance matrices of Eq.~(\ref{eq:hil_cl_cov_avg}) \citepalias{planck2013-p08}. 

Appendix~\ref{app:hil_eqs} provides a summary of the equations used to compute temperature and polarization covariance matrices and presents a validation of the implementation through direct simulations. Let us note that, for the approximations used in the analytical computation of the covariance matrix to be precise, the mask power spectra have to decrease quickly with multipole moment $\ell$; this requirement gives rise to the apodization scheme discussed in Sect.~\ref{sec:masks}. In the presence of a point-source mask, however, the condition may no longer be fulfilled, reducing the accuracy of the approximations assumed in the calculation of the covariance matrices. We discuss in Appendix~\ref{app:hil_pts_mask_correction} the heuristic correction we developed to restore the accuracy, which is based on direct simulations of the effect.



\subsection{FFP8 Simulations}\label{sec:valid-sims}

In order to validate the overall implementation and our approximations, we generated 300 simulated \HFI\ half-mission map sets in the frequency range $100$ to $217$ GHz, which we analysed like the real data. For the CMB, we created realizations of the \LCDM\ model with the best-fit parameters obtained in this paper. After convolving the CMB maps with beam and pixel window functions, we superimposed CIB, dust, and noise realizations from the FFP8 simulations \citep{planck2014-a14} that capture both the correlation structure and anisotropy of foregrounds and noise. We then computed power spectra using the set of frequency-dependent masks described in Sect.~\ref{sec:masks} and created the corresponding {\plik} $\TT$ likelihood. We modified the shape of the foreground spectra to fit the FFP8 simulations, but kept the parameterization used on the data. In the case of dust, we used priors similar to those used on data. Furthermore, in the following the dust amplitude parameter is named $\mathrm{gal}^{\nu\times\nu'}_{545}$.  We then ran an MCMC sampler to derive the cosmological and foreground parameters posterior distributions for all dataset realizations.
\begin{figure*}[h!] 
\begin{center}
\includegraphics[width=0.49\columnwidth]{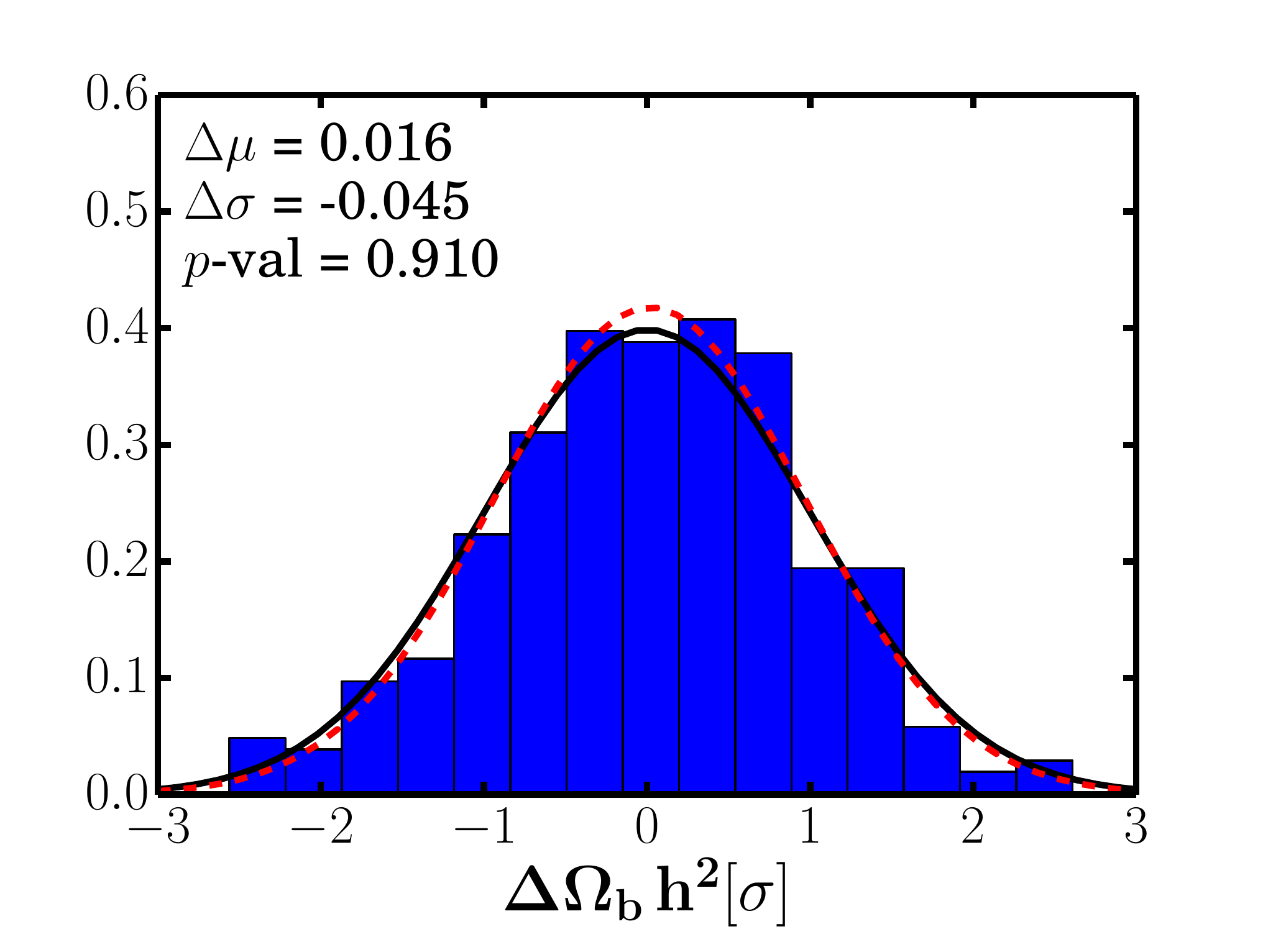}\includegraphics[width=0.49\columnwidth]{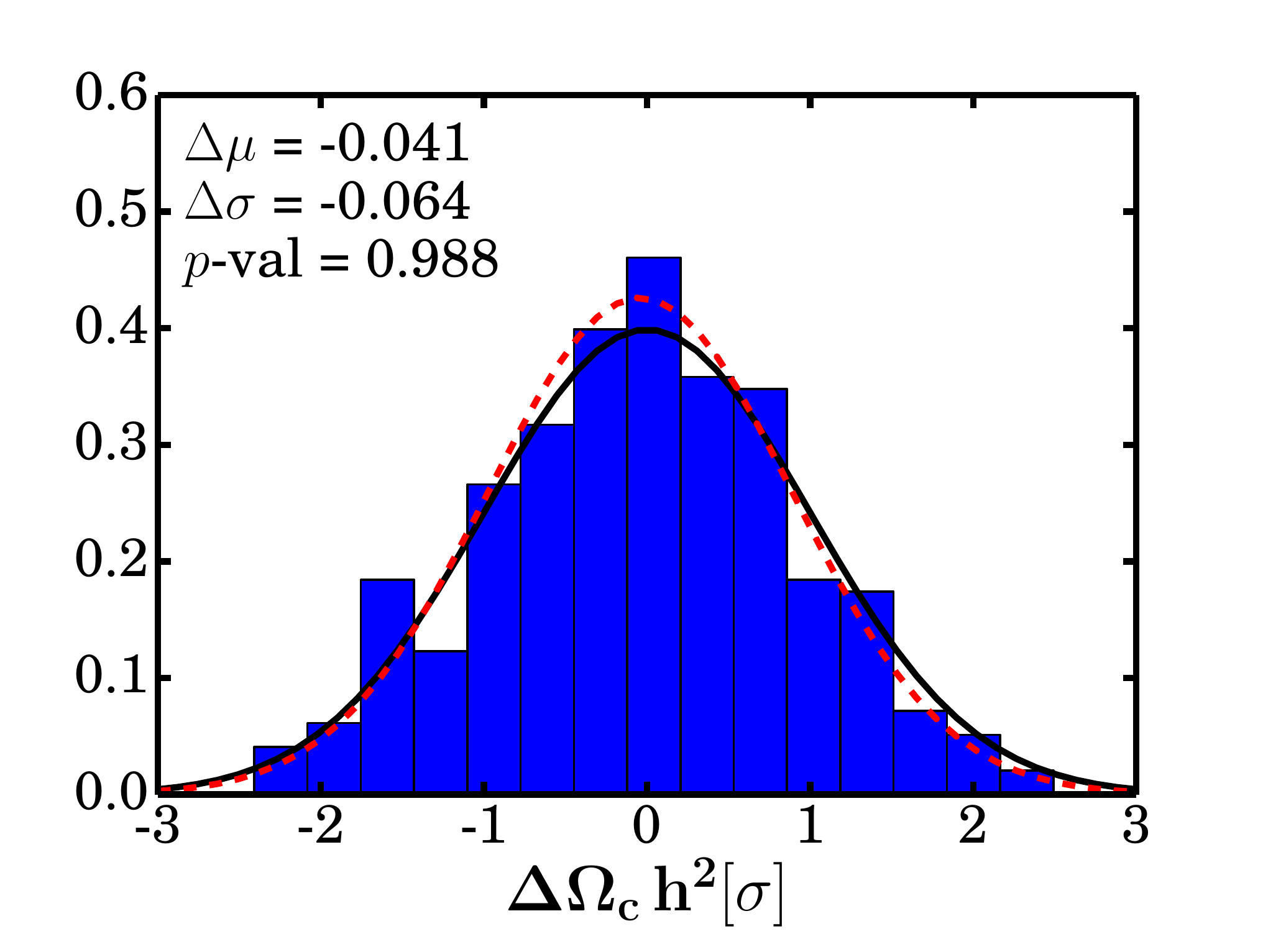}\includegraphics[width=0.49\columnwidth]{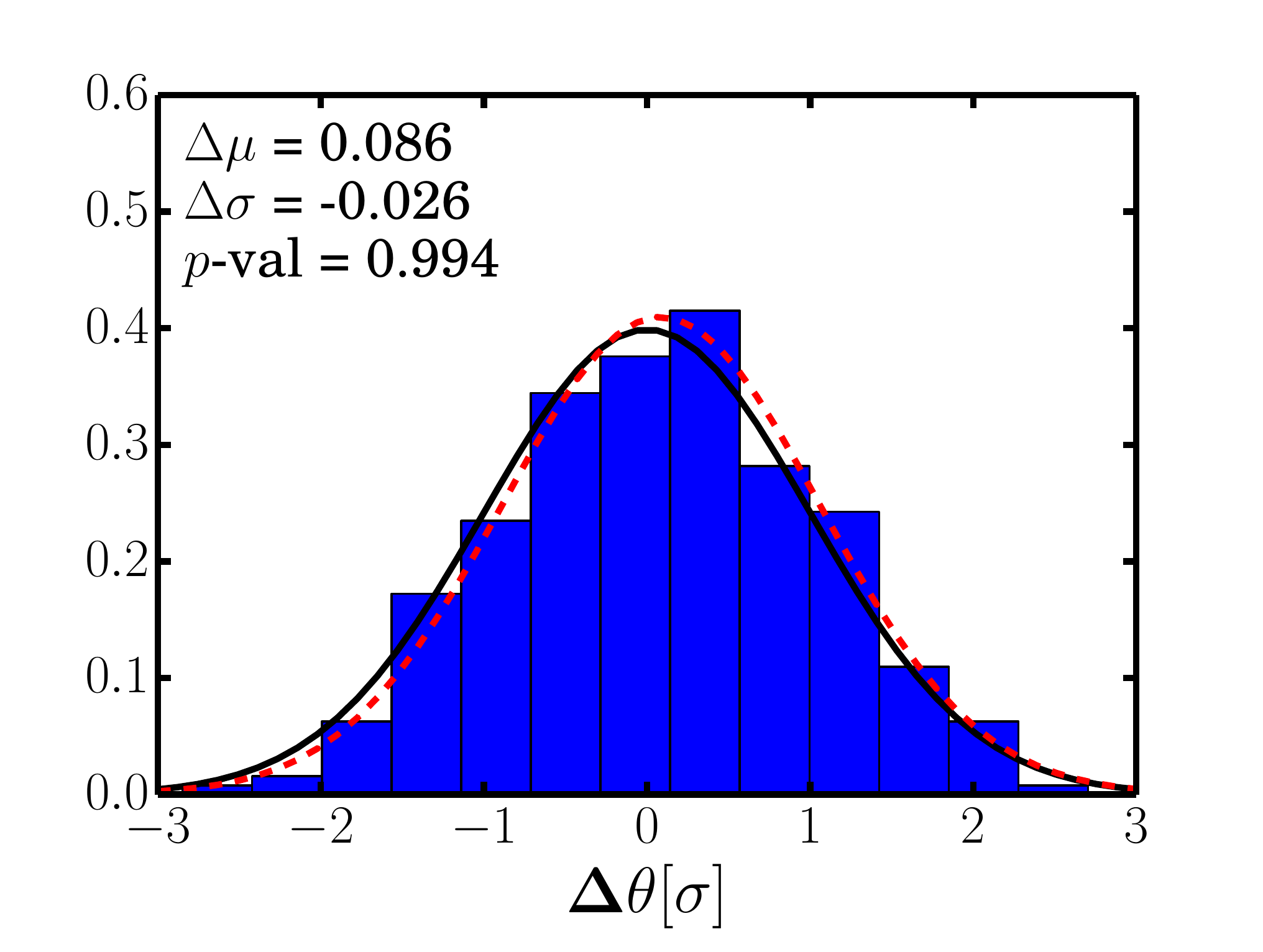}\includegraphics[width=0.49\columnwidth]{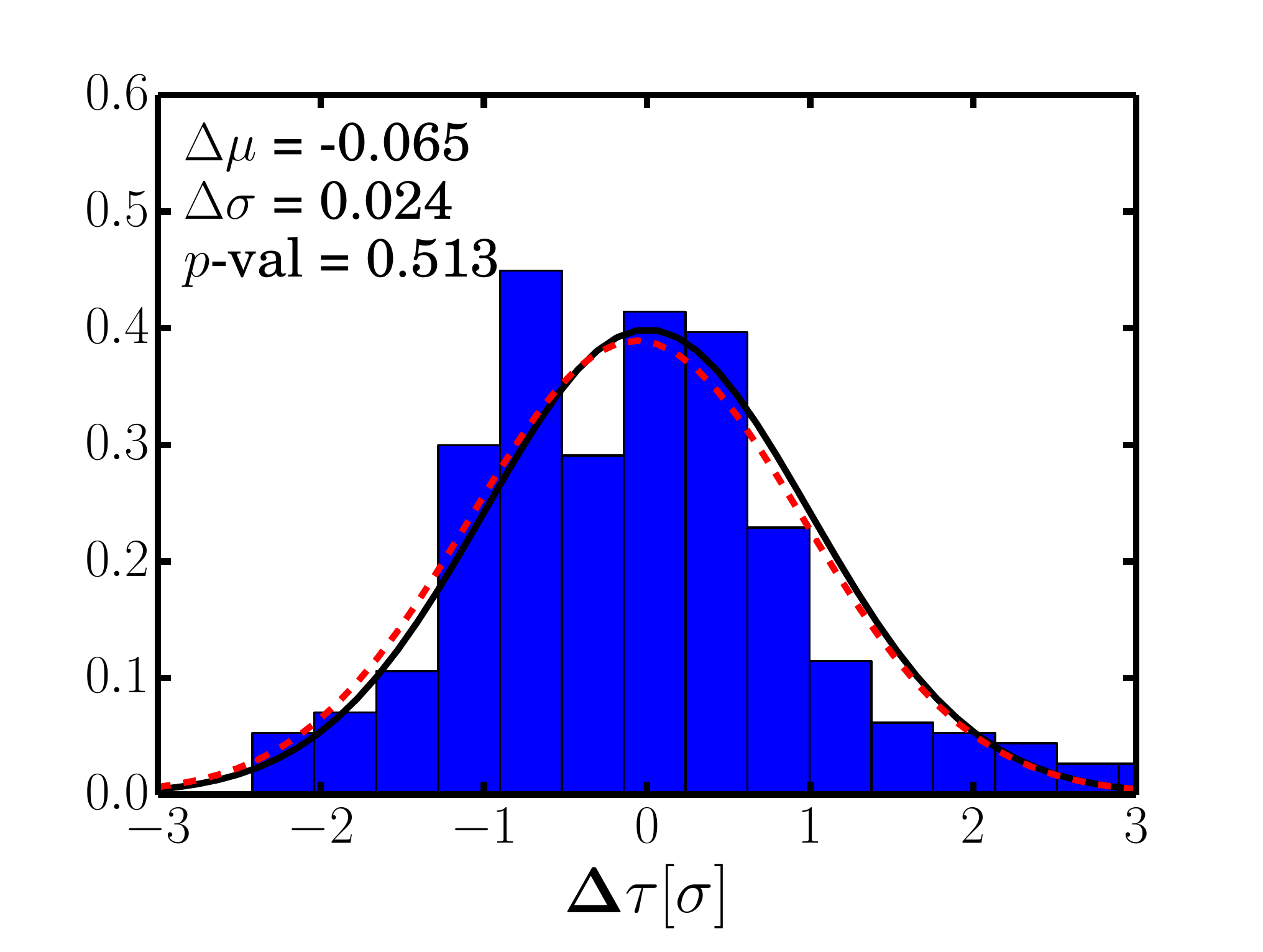}\\
\includegraphics[width=0.49\columnwidth]{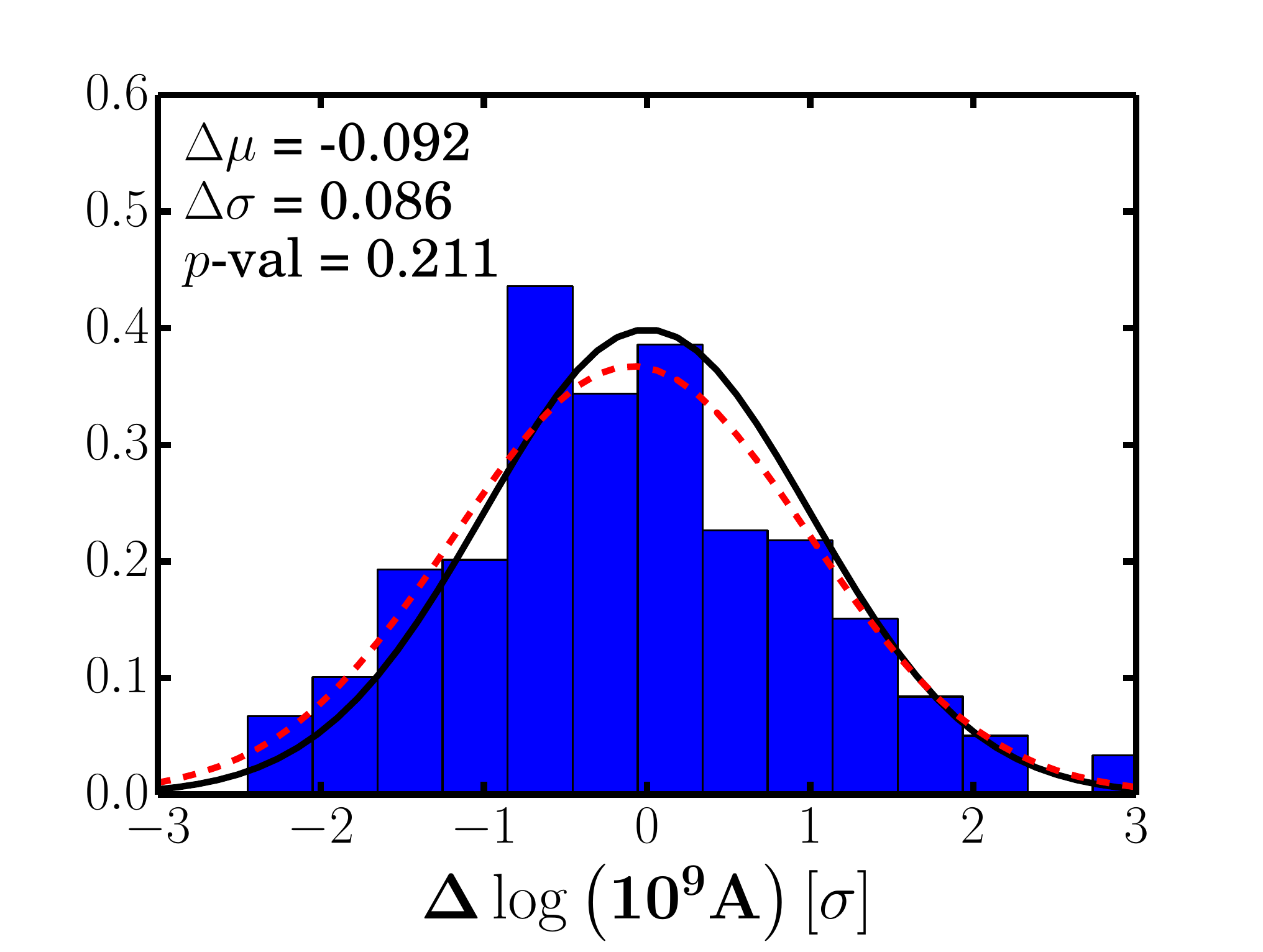}\includegraphics[width=0.49\columnwidth]{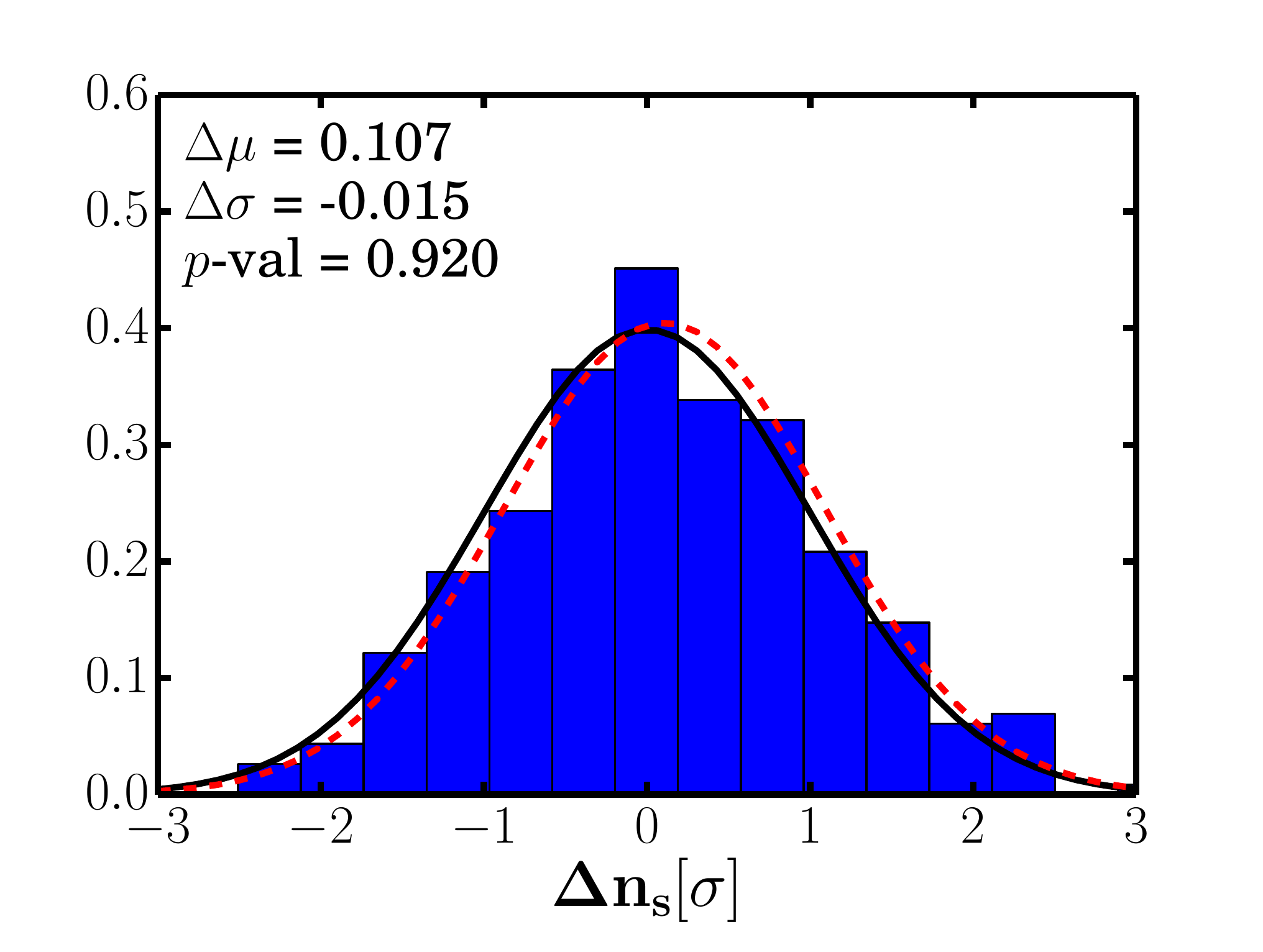}\includegraphics[width=0.49\columnwidth]{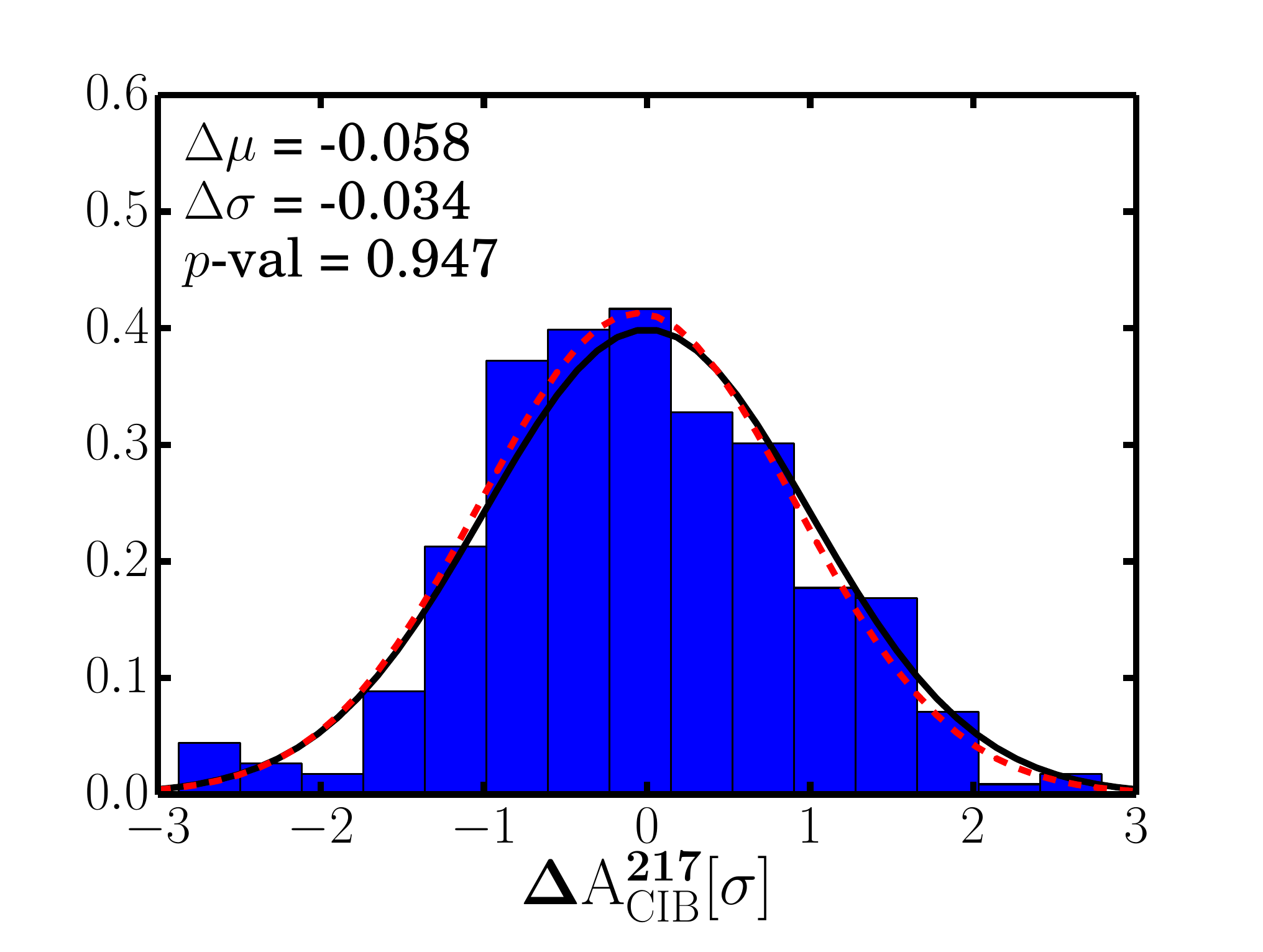}\includegraphics[width=0.49\columnwidth]{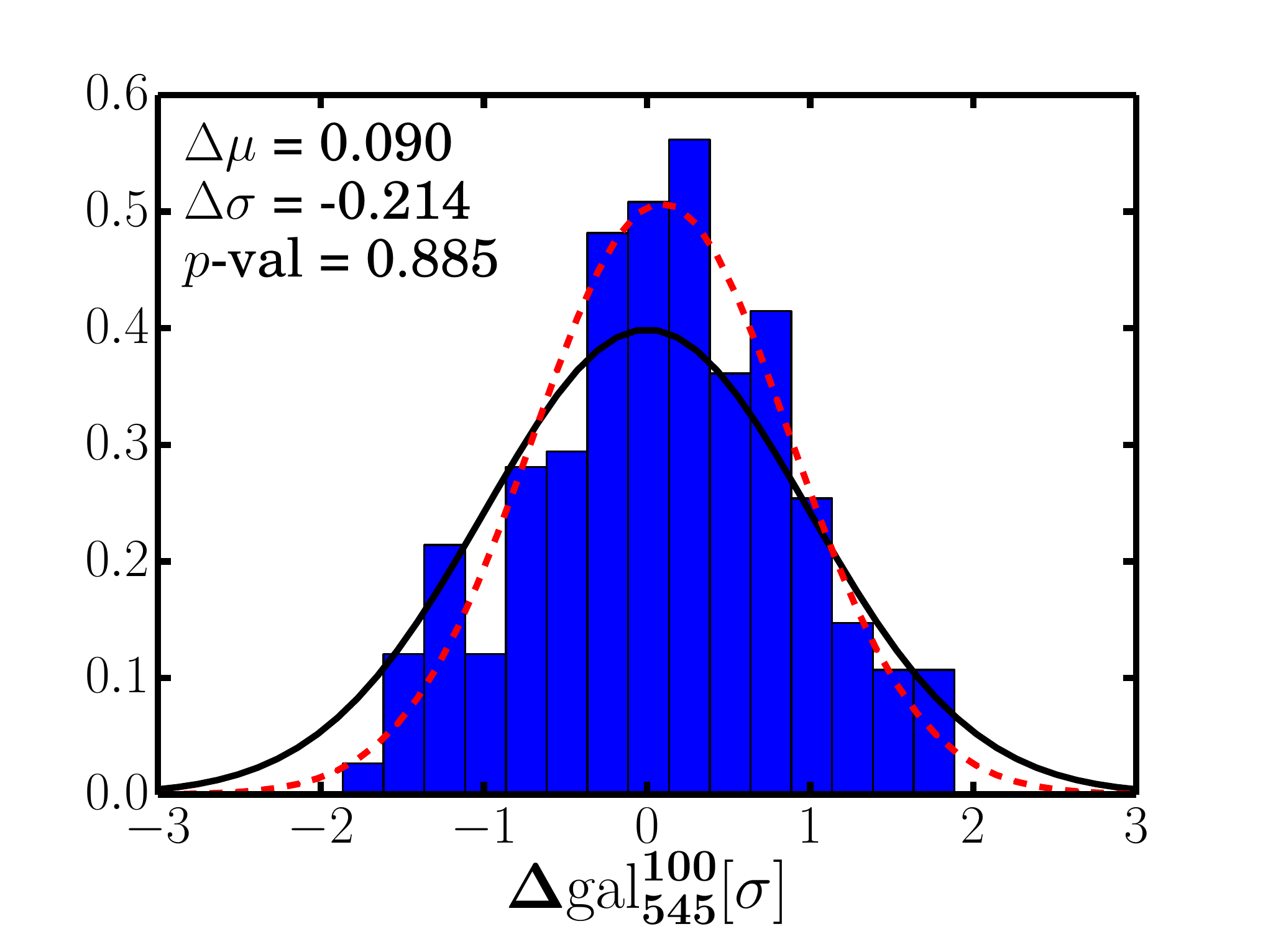}\\
\includegraphics[width=0.49\columnwidth]{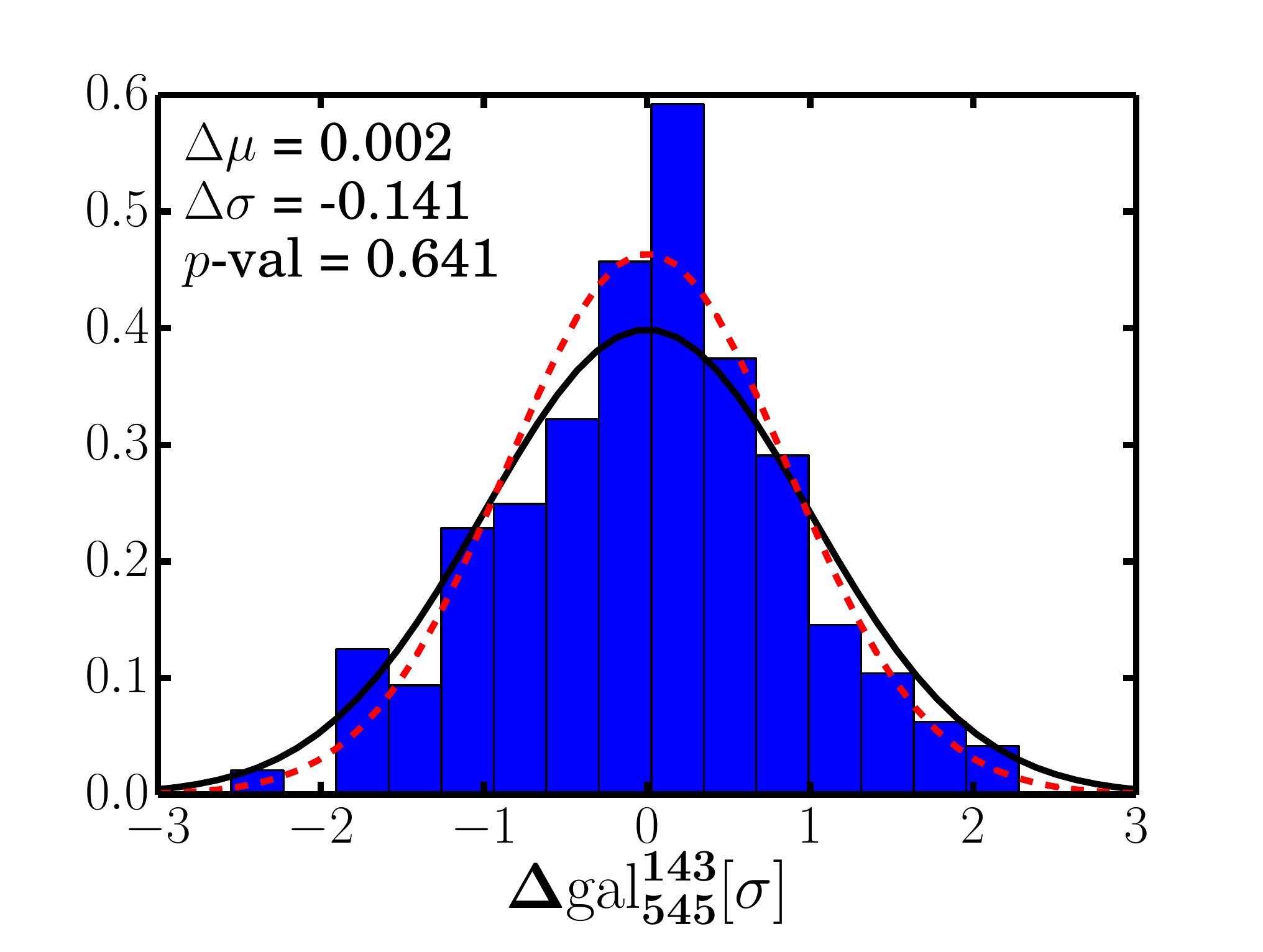}\includegraphics[width=0.49\columnwidth]{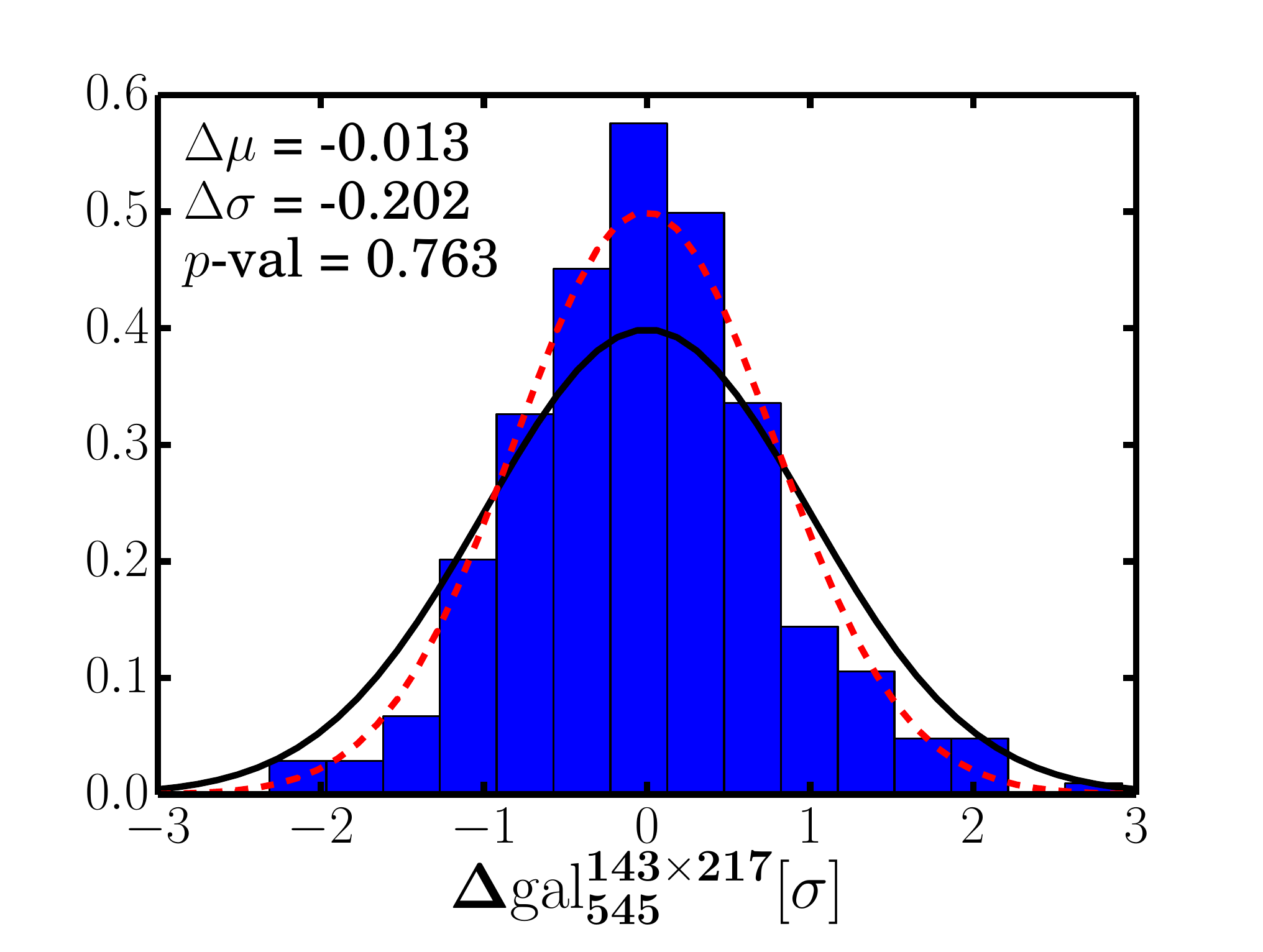}\includegraphics[width=0.49\columnwidth]{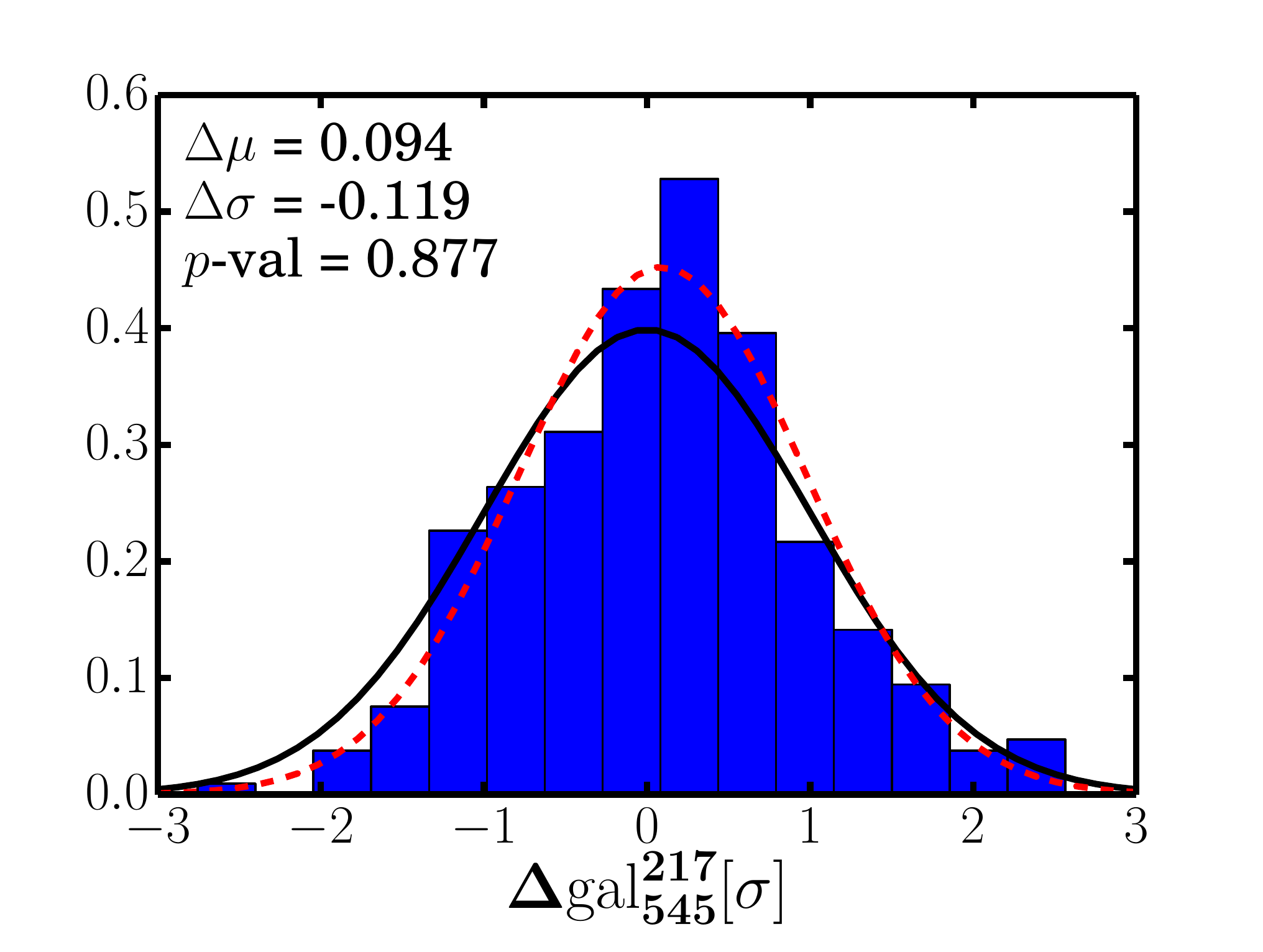}
\caption{\plik parameter results on 300 simulations for the six baseline cosmological parameters, as well as the FFP8 CIB and Galactic dust amplitudes. The simulations   include quite realistic CMB, noise, and foregrounds (see text). The distributions of inferred posterior mean parameters are centred around their input values with the expected scatter. Indeed the dotted red lines show the best-fit Gaussian for each distribution, with a mean shift, $\Delta\mu$, and a departure $\Delta\sigma$ from unit standard deviation given in the legend;  both are close to zero. These best fits are thus very close to Gaussian distributions with zero shift and unit variance, which are displayed for reference as black lines. The legend gives the numerical value of $\Delta\mu$ and $\Delta\sigma$, as well as the $p$-values of a Kolmogorov--Smirnov test of the histograms against a Gaussian
distribution shifted from zero by $\Delta\mu$ and with standard deviation shifted from unity by $\Delta \sigma$. This confirms that the distributions are consistent with Gaussian distributions with zero mean and unit standard deviation, with a small offset of the mean.}
 \label{fig:plik_sims_par}
\end{center}
\end{figure*}

For each simulation, we computed the shift of the derived posterior mean parameters with respect to the input cosmology, normalized by their posterior widths $\sigma_{\mathrm{post}}$.  When a Gaussian prior with standard deviation $\sigma_{\mathrm{prior}}$ is used, we rescale $\sigma_{\mathrm{post}}$ by
$[1-\sigma_{\mathrm{post}}^2 / \sigma_{\mathrm{prior}}^2]^{1/2}$;
this is the case for $\tau$ and for the Galactic dust amplitudes
$\mathrm{gal}^\nu_{545}$ in the four cross-frequency channels used. In
Fig.~\ref{fig:plik_sims_par}, we show  histograms of the shifts we
found for all $300$ simulations for the six baseline cosmological
parameters, as well as the FFP8 CIB and galactic dust amplitudes. As
shown in the figure, we recover the input parameters with little bias
and a scatter of the normalized parameter shifts around unity. The
$p$-values of the Kolmogorov--Smirnov test  that we ran are given in the
legend and we do not detect significant departures from normality.  The
average reduced $\chi^2$ for the histograms of
Fig.~\ref{fig:plik_sims_par} is equal to 1.02.

Table~\ref{tab:simsresults}  (second column) compiles the average shifts of Fig.~\ref{fig:plik_sims_par}, but in order to gauge whether they are as small as expected for this number of simulations (assuming no bias), the shifts are expressed in units of the posterior width rescaled by $1/\sqrt{300}$. We note that the shift of the average is above one
(scaled) $\sigma$ in three cases out of a total of 11 parameters ($68\,\%$ of the $\Delta$s would be expected to lie within $1\,\sigma$ if the parameters were uncorrelated), with $\theta$, $\ns$, and $\mathrm{gal}_{545}^{217}$ at the 1.7, 2.0, and 1.5 (scaled) $\sigma$
level, respectively. 

\begin{table}[ht!]
\begingroup 
\newdimen\tblskip \tblskip=5pt
\caption{Shifts of parameters over 300 $\TT$ simulations.$^{\rm a}$}
\label{tab:simsresults}
\vskip -6mm
\footnotesize
\setbox\tablebox=\vbox{
\newdimen\digitwidth
\setbox0=\hbox{\rm 0}
\digitwidth=\wd0
\catcode`*=\active
\def*{\kern\digitwidth}
\newdimen\signwidth
\setbox0=\hbox{+}
\signwidth=\wd0
\catcode`!=\active
\def!{\kern\signwidth}
\newdimen\decimalwidth
\setbox0=\hbox{.}
\decimalwidth=\wd0
\catcode`@=\active
\def@{\kern\decimalwidth}
\openup 3pt
\halign{ 
\hbox to 0.85in{$#$\leaderfil}\tabskip=2em&
    \hfil$#$\hfil\tabskip=9pt&
    \hfil$#$\hfil\tabskip=9pt&
    \hfil$#$\hfil\tabskip=9pt&
    \hfil$#$\hfil\tabskip=0pt\cr
\noalign{\doubleline}
\omit\hfil Parameter\hfil& \mathrm{300\ sims}& r_{\rm A}^{30}& r_{\rm A}^{65}& r_{\rm A}^{100}\cr
\noalign{\vskip 3pt\hrule\vskip 5pt}
\Omb h^2&                              			\rev{!0.27}& 	\rev{!0.62}& 	\rev{!0.50}& 	\rev{!0.52} \cr
\Omc h^2&                              			\rev{-0.71}&  	\rev{-0.65}& 	\rev{-0.44}& 	\rev{!0.00} \cr 
\theta&                                				\rev{!1.48}&  	\rev{!1.67}& 	\rev{!1.67}& 	\rev{!1.29} \cr
\tau&                                  				\rev{-0.57}&  	\rev{-0.38}& 	\rev{-0.56}& 	\rev{-0.38} \cr
\ln\left( 10^{10}A_{\mathrm{s}}\right)&	\rev{-0.70}&  	\rev{-0.52}& 	\rev{-0.65}& 	\rev{-0.35} \cr
n_{\mathrm{s}}&                        			\rev{!1.86}&  	\rev{!1.87}& 	\rev{!1.46}& 	\rev{!0.78} \cr
A_{\mathrm{CIB}}^{217}&                		\rev{-0.99}&  	\rev{-1.09}& 	\rev{-1.44}& 	\rev{-1.34} \cr
\mathrm{gal}_{545}^{100}&              		\rev{!0.31}&  	\rev{!0.13}& 	\rev{-0.09}& 	\rev{!0.04} \cr
\mathrm{gal}_{545}^{143}&              		\rev{!0.40}&  	\rev{-0.21}& 	\rev{-0.23}& 	\rev{-0.19} \cr
\mathrm{gal}_{545}^{143-217}&          	\rev{-0.22}&  	\rev{-0.35}& 	\rev{!0.36}& 	\rev{!0.22} \cr
\mathrm{gal}_{545}^{217}&              		\rev{!1.61}&  	\rev{!1.48}& 	\rev{!2.19}& 	\rev{!2.04} \cr
\noalign{\vskip 5pt\hrule\vskip 3pt}
}}
\endPlancktable 
\tablenote {{\rm a}} Shifts are given in units of the posterior width rescaled by $1/\sqrt{300}$. If the parameters were uncorrelated, $68\,\%$ of the shifts would be expected to lie within $\pm1\,\sigma$.  The effect of varying the value of $\lmin$ is measured on the likelihood of the average spectra over 300 realizations, labelled $r_{\rm A}^{\lmin}$. A significant decrease of the bias on $n_{\mathrm{s}}$ is obtained by not including low-$\ell$ multipoles, at the cost, however, of a degradation in the determination of the foreground amplitudes $A_{\mathrm{CIB}}^{217}$, $\mathrm{gal}_{545}^{143-217}$, and $\mathrm{gal}_{545}^{217}$. 
\par
\endgroup
\end{table}

Before proceeding, let us note that an estimate (third column) of these shifts is obtained by simply computing the shift from a single likelihood using as input the average spectra of the 300 simulations. This effectively reduces cosmic variance and noise
amplitude by a factor $\sqrt{300}$ and, more importantly, it decreases the cost and length of the overall computation, enabling additional tests. These shift estimates are noted $r_{\rm A}$. The table shows that significant improvement in the determination of
$n_{\mathrm{s}}$ is obtained by removing low-$\ell$ multipoles. Indeed, columns 4 and 5 of Table~\ref{tab:simsresults} show the variation of the shift when the $\lmin$ of the high-$\ell$ likelihood is increased from 30 to 65 and 100. The shift in $\ns$ is decreased by a factor two, while the decrease in the number of bins per cross-frequency spectrum is only reduced from 199 to 185 (having little impact on the size of the covariance matrix of cosmological parameters). 

These changes with $\lmin$ therefore trace the small biases back to the lowest-$\ell$ bins. It suggests that the Gaussian approximation used in the high-$\ell$ likelihood starts to become mildly inaccurate at $\ell=30$. Indeed, even if noticeable, this effect would contribute at most a $0.11\,\sigma$ bias on $\ns$. This is
further confirmed by the lack of a detectable effect found in Sect.~\ref{sec:hal-hybrid} when varying the hybridization scale in $\TT$ between \texttt{Commander} and \plik.  However,  the exclusion of low-$\ell$ information degrades our ability to accurately
reconstruct the foreground amplitudes $A_{\mathrm{CIB}}^{217}$, $\mathrm{gal}_{545}^{143-217}$, and   $\mathrm{gal}_{545}^{217}$.  Indeed, the dust spectral amplitudes in the $143\times217$ and $217\times217$ channels are highest at low multipoles, and the CIB spectrum in the range $30\leq \ell \leq 100$ also adds substantial information.

In spite of this low-$\ell$ trade-off between an accurate determination of $\ns$ on the one hand and $A_{\mathrm{CIB}}^{217}$, $\mathrm{gal}_{545}^{143-217}$, and  $\mathrm{gal}_{545}^{217}$ on the other, we can conclude that the \plik implementation is behaving as expected and can be used for actual data analysis. 

Appendix~\ref{app:plik-sims} extends this conclusion to the joint \plikTTTEEE\ likelihood case. 

\subsection{\rev{End-to-End simulations}}
\label{sec:e2e}

\begin{table*}[ht!]
\begingroup 
\newdimen\tblskip \tblskip=5pt
\caption{\rev{End-to-end parameter shifts for a single realization of CMB and foregrounds, along with five different noise realizations. Shifts are computed with respect to those obtained without noise and with instrumental effects turned off.$^{\rm a}$}}
\label{tab:e2e1}
\vskip -6mm
\footnotesize
\setbox\tablebox=\vbox{
\newdimen\digitwidth
\setbox0=\hbox{\rm 0}
\digitwidth=\wd0
\catcode`*=\active
\def*{\kern\digitwidth}
\newdimen\signwidth
\setbox0=\hbox{+}
\signwidth=\wd0
\catcode`!=\active
\def!{\kern\signwidth}
\newdimen\decimalwidth
\setbox0=\hbox{.}
\decimalwidth=\wd0
\catcode`@=\active
\def@{\kern\decimalwidth}
%
\halign{\hbox to 0.85in{$#$\leaderfil}\tabskip=2em&
    \hfil$#$\hfil\tabskip=9pt&
    \hfil$#$\hfil\tabskip=9pt&
    \hfil$#$\hfil\tabskip=9pt&
    \hfil$#$\hfil\tabskip=9pt&
    \hfil$#$\hfil\tabskip=9pt&
    \hfil$#$\hfil\tabskip=9pt&
    \hfil$#$\hfil\tabskip=9pt&
    \hfil$#$\hfil\tabskip=0pt\cr
\noalign{\doubleline}
\omit\hfil Parameter \hfil & 1 & 2 & 3 & 4 & 5 & \mathrm{Mean} & \mathrm{Median} & \sigma_{\mathrm{FFP8}}\cr
\noalign{\vskip 3pt\hrule\vskip 5pt}
\Omb h^2 & !0.32  & !0.09 & -0.3  & -0.02  & -0.17  & -0.01& -0.02 & !0.42\cr                   
\Omc h^2 & -0.07  & -0.20 & -0.22  & -0.22  & !0.30 & -0.08 & -0.20 & !0.35\cr                                  
\theta & !0.18  & !0.08 & !0.24 & -0.12 & !0.15  & !0.10 & !0.14 & !0.45\cr       
\tau &  !0.08 & !0.19 & -0.02 & !0.006  & -0.21 & !0.01 & !0.005 & !0.25\cr     
\ln\left( 10^{10}A_{\mathrm{s}}\right) & !0.04 & !0.10 & -0.11 & -0.11 & -0.21 & -0.05& -0.11 & !0.25\cr
n_{\mathrm{s}} & !0.11 & !0.06 & !0.22 & -0.13 & -0.64 & -0.07 & !0.05 & !0.40\cr
\noalign{\vskip 5pt\hrule\vskip 3pt}
}}
\endPlancktablewide 
\tablenote {{\rm a}} \rev{In columns labelled 1 to 5: shifts of the cosmological parameters of the five noise realizations of the end-to-end simulations with respect to those obtained for the simulations without noise and with instrumental effects turned off, normalized by the end-to-end simulations' posterior widths.  In columns labelled ``Mean'' and ``Median'': the corresponding mean and median. In column labelled ``$\sigma_{\mathrm{FFP8}}$'': the standard deviation of the distribution obtained from the cosmological parameter shifts of 100 FFP8 simulations, varying the noise only with respect to the cosmological parameters of the CMB only.} 
\par
\endgroup
\end{table*}

\begin{table*}[ht!]
\begingroup 
\newdimen\tblskip \tblskip=5pt
\caption{\rev{End-to-end parameter shifts for four different CMB realizations but comprising four pairs of realizations with the same noise realization with respect to those obtained without noise and with instrumental effects turned off.$^{\rm a}$}}
\label{tab:e2e2}
\vskip -6mm
\footnotesize
\setbox\tablebox=\vbox{
\newdimen\digitwidth
\setbox0=\hbox{\rm 0}
\digitwidth=\wd0
\catcode`*=\active
\def*{\kern\digitwidth}
\newdimen\signwidth
\setbox0=\hbox{+}
\signwidth=\wd0
\catcode`!=\active
\def!{\kern\signwidth}
\newdimen\decimalwidth
\setbox0=\hbox{.}
\decimalwidth=\wd0
\catcode`@=\active
\def@{\kern\decimalwidth}
%
\halign{\hbox to 0.85in{$#$\leaderfil}\tabskip=2em&
    \hfil$#$\hfil\tabskip=9pt&
    \hfil$#$\hfil\tabskip=9pt&
    \hfil$#$\hfil\tabskip=9pt&
    \hfil$#$\hfil\tabskip=9pt&
    \hfil$#$\hfil\tabskip=9pt&
    \hfil$#$\hfil\tabskip=9pt&
    \hfil$#$\hfil\tabskip=9pt&
    \hfil$#$\hfil\tabskip=9pt&
    \hfil$#$\hfil\tabskip=9pt&
    \hfil$#$\hfil\tabskip=0pt\cr
\noalign{\doubleline}
\omit\hfil Parameter \hfil & 4 & 5 & 6 & 7 & 8 & \Delta_{5-6} & \Delta_{4-7} & \Delta_{4-8} & \Delta_{7-8} & \sigma_{\mathrm{FFP8}}\cr
\noalign{\vskip 3pt\hrule\vskip 5pt}
\Omb h^2 & -0.02  & -0.17 & -0.25 & -0.53 & !0.29 & !0.08 & !0.51 & !0.31 & !0.83 & !0.34 \cr                   
\Omc h^2 & -0.22  & !0.30 & -0.20  & !0.52 & -0.59 & !0.50 & !0.74 & !0.37 & !1.11 & !0.30 \cr 
\theta        & -0.12  & !0.15 & -0.34 & -1.06  & !0.41 & !0.48 & !0.94 & !0.52 &!1.47 & !0.35 \cr 
\tau           & 0.006  & -0.21 & -0.16 & -0.17 & -0.16 & !0.04 & !0.18 & !0.16 & !0.01 & !0.175 \cr 
\ln\left( 10^{10}A_{\mathrm{s}}\right) & -0.11  & -0.21 & -0.29 & -0.15  & -0.34 & !0.07 & !0.04 & !0.24 & !0.20 & !0.19 \cr 
n_{\mathrm{s}} & -0.13  & -0.64  & -0.39  & -0.86   & !0.24  & !0.24 & !0.72 & !0.37 & !1.03 & !0.35 \cr 
\noalign{\vskip 5pt\hrule\vskip 3pt}
}}
\endPlancktablewide 
\tablenote {{\rm a}} \rev{In columns labelled 4 to 8: shifts of the cosmological parameters of the end-to-end simulations with respect to those obtained for the simulations without noise and with instrumental effects turned off, normalized by the end-to-end simulations' posterior widths.  We point out that realizations numbered 4 and 5 are common to the sets of Tables~\ref{tab:e2e1} and \ref{tab:e2e2}.  In columns labelled ``$\Delta_{5-6}$'' to ``$\Delta_{7-8}$'', the absolute differences in the shifts within pairs of realizations having different CMB but the same foregrounds and noise realizations. In column labelled ``$\sigma_{\mathrm{FFP8}}$'': the standard deviation of the distribution obtained from the cosmological parameter shifts of 100 FFP8 simulations, varying the CMB only but keeping the same FFP8 noise realization, with respect to the cosmological parameters of the CMB only.} 
\par
\endgroup
\end{table*}

\rev{While the previous section validated our methodology, our approximations, and the overall implementation, this does not yet give the sensitivity to residual systematic uncertainties undetected by data consistency checks. These are by their very nature very much more difficult to address realistically, since, when an effect is detected and sufficiently well understood, it can be modelled and is corrected for, in general at the TOI-processing stage; only the uncertainty of the correction needs to be addressed. Still, \HFI has developed a  complete model of the instrument which contains all identified systematic effects and enables realistic simulation of the instrumental response. We have therefore generated a number of full-mission time streams which we have then processed with the DPC TOI processing pipeline in order to create map datasets as close to instrumental reality as we can in order to assess the possible impact of low-level residual instrumental systematics, the effects of which might have remained undetected otherwise.}

\rev{In this section, we report on the shifts in the values of the cosmological and foreground parameters  induced by these specific residual systematic effects, comparing the results of a $\TT$ likelihood analysis for two overlapping sets of five simulations:}\\

\begin{enumerate}
 \item \rev{five simulations of maps at 100\,GHz, 143\,GHz and 217\,GHz, for a single realization of the CMB and of the foregrounds but for five different realizations of the noise,}
\item \rev{five simulations of maps at 100\,GHz, 143\,GHz and 217\,GHz, composed of four CMB realizations, two noise realizations, a single realization of the foregrounds, but forming four pairs of realizations having the same noise but different CMB.}
\end{enumerate}

\rev{These simulations sum to a total of eight distinct simulations and are numbered from 1 to 8 in Tables~\ref{tab:e2e1} and \ref{tab:e2e2}.  To be more explicit, in the second set, among simulations numbered 4 to 8, simulations 4, 7 and 8 have different CMB, but the same noise as each other. Simulations 5 and 6 have different CMB, and the same noise as each other, but different from simulations 4, 7 and 8. Realizations 4 and 5, having the same CMB but different noise, are common to the two sets of five realizations.}

\rev{Each of these have been performed twice, with the end-to-end (instrument plus TOI processing) pipeline and noise contribution switched either on or off.  End-to-end simulations are computationally very costly, (typically a week for each simulated mission dataset) and hence only a few realizations were generated).}

\rev{As explained in Section~5.4 of \citet{planck2014-a08}, the end-to-end simulations are created by feeding the TOI processing pipeline with simulated data to evaluate and characterize the overall transfer function and the respective contribution of each individual effect on the determination of the cosmological parameters. Simulated TOIs are produced by applying the real mission scanning strategy to a realistic input sky specified by the Planck Sky Model \citep[PSM;][]{delabrouille2012} containing a lensed CMB realization, galactic diffuse foregrounds, and the dipole components. To this sky-scanned TOI, we add a white-noise component, representing the phonon and photon noises. The very low-temporal-frequency thermal drift seen in the real data is also added to the TOI. The noisy sky TOI is then convolved with the appropriate bolometer transfer functions. Another white-noise component, representing Johnson noise and read-out noise, is also added. Simulated cosmic rays using the measured glitch rates, amplitudes, and shapes are added to the TOI. This TOI is interpolated to the electronic HFI fast-sampling frequency. It is then converted from analogue to digital using a simulated non-linear analogue-to-digital converter (ADC). Identified 4\,K cooler spectral lines are added to the TOI. Both effects (ADC and 4\,K lines) are derived from the measured in-flight behaviour. The TOI finally goes through the data compression/decompression algorithm used for communication between the \Planck\ satellite and Earth. The simulated TOI is then processed in the same way as the real mission data for cleaning and systematic error removal, calibration, destriping, and map-making. 
} 

\rev{Some limitations of the current end-to-end approach follow. No pointing error is included, although previous (dedicated) simulations suggest that this has negligible effect. In addition, this effect was included in the dedicated simulations performed to assess the precision of the beam recovery procedure. The first step of the TOI processing is to correct the ADC non-linearity (ADC NL). For the flight data, the ADC NL was determined by using HFI's measured signal at the end of the HFI mission, with the instrument's cooling system switched off and an instrument temperature equal to 4\,K. This determination relied on supposing the signal to be perfect white noise and therefore to correspond to the distortions brought in by the ADC. In the current implementation, we assume perfect knowledge of ADC NL and 4\,K lines.  This is of course not true for the real data and future end-to-simulations, accompanying \Planck's next data release, will  improve our model of this effect. 
After ADC NL correction, the signal is converted to volts.  Deglitching is then performed by flagging glitch heads and by using glitch tail time-lines.  This enables the creation of the thermal baseline which is used for signal demodulation.  The thermal baseline and glitch tails are subtracted, the signal is converted to watts, and the 4\,K lines are removed.  The resulting signal is then deconvolved by the bolometer transfer functions. We do not include uncertainties in the glitch tail shape used in the deglitching procedure, i.e., the templates are the same for the simulations and the processing; but here again, previous studies suggest any difference is a small effect.} 

\rev{
The analysis of these sets of end-to-end simulations, and of their counterparts for which all instrumental effects are turned off, is performed similarly to that of the simulations described in Sect.~\ref{sec:valid-sims}. Angular power spectra for all cross-half-missions and for all frequency combinations are computed using the \Planck\ masks described in Sect.~\ref{sec:masks} and with the appropriate beam functions.  Noise levels are evaluated as described in Sect.~\ref{sec:noise_model}.  Templates for galactic foregrounds (CO, free-free, synchrotron, thermal and spinning dust), the kinetic and thermal SZ effects, the cosmic infrared background, and radio and IR point sources are constructed based on the PSM input foreground maps.  The covariance matrix is computed with the method outlined in Appendix~\ref{app:hil_covmat} with the aforementioned input CMB power spectrum, input foreground spectra, noise levels, beam functions and masks.
} 

\rev{
All sets of power spectra and the inverse covariance matrix are then binned and used in the likelihood analysis performed using an MCMC sampler together with \plik and PICO in order to determine the best fit cosmological parameters.  The shifts in cosmological parameter values induced by the imperfect correction of instrumental effets by the TOI processing pipeline are then computed for the end-to-end simulations with respect to those obtained for the simulations without noise and with instrumental effects turned off, normalized by the end-to-end simulations' posterior widths.  Comparing shifts computed in this way cancels out cosmic variance and chance correlations between the CMB and the foregrounds and are thus fully attributable to the instrument and to the noise, which cannot be disentangled, as well as to CMB-noise chance correlations. That is, those shifts probe directly the scatter and possible biases induced by residual systematics effects.}

\rev{The mean and median shifts for the five simulations with a single CMB realization and a single foreground realization, but different noise realizations, are given in Table \ref{tab:e2e1}. In order to verify that these shifts are within expectations, we computed the shifts in cosmological parameters for 100 FFP8 simulations, each with identical CMB signal but different FFP8 noise, with respect to the cosmological parameters obtained for the CMB only, normalized by their posterior widths. The standard deviations of the resulting distributions are given in the column labelled ``$\sigma_{\mathrm{FFP8}}$'' of Table~\ref{tab:e2e1} and can be compared with the shifts obtained for the five end-to-end simulations. All shifts are within 1$\sigma$ of the shifts expected from FFP8. In addition, there is no indication of any detectable bias. All shifts are thus compatible with scatter introduced by noise.\\
} 

\rev{The shifts for the five realizations with four different CMB realizations, the same foregrounds, but comprising four pairs with the same noise realization, are given in columns 4 to 8 of Table~\ref{tab:e2e2}. As mentioned at the beginning of this Section, realizations numbered 4 and 5 are common to the sets of Tables~\ref{tab:e2e1} and \ref{tab:e2e2}.  In the columns labelled ``$\Delta_{5-6}$'' to ``$\Delta_{7-8}$'', we computed the absolute differences in the shifts within pairs of realizations having different CMB but the same foreground and noise realizations. We compare these differences to the standard deviations of the distributions of cosmological parameter shifts of 100 FFP8 simulations, varying the CMB but keeping the same FFP8 noise realization, with respect to the cosmological parameters of the corresponding CMB but without noise (column labelled ``$\sigma_{\mathrm{FFP8}}$'').  These distributions quantify the impact of CMB-noise correlations on the determination of the cosmological parameters. 
The Table shows that among all $\Delta$'s, 11 are within $1\sigma_{\mathrm{FFP8}}$, 7 are within 1 to $2\sigma_{\mathrm{FFP8}}$, 3 are within 2 to $3\sigma_{\mathrm{FFP8}}$, 2 are within 3 to $4\sigma_{\mathrm{FFP8}}$. 50\% of the differences are within $1\sigma$ and 78\% within $2\sigma$.  At the very worst, taking the example of $\Delta_{7-8}$ a cosmological parameter ($\theta$) moves a total of 4$\sigma$, from $-3\sigma$ to $1\sigma$ in units of $\sigma_{\mathrm{FFP8}}$ when the CMB is changed but the noise is left the same.  This is rare but can be expected in a few percent of simulations. As in the case of the shifts listed in Table~\ref{tab:e2e1}, there is thus no detectable bias, with all shifts compatible with those expected from FFP8.
}

\rev{In summary, we have detected no sign as yet of systematic biases of the cosmological parameters due to known low-level instrumental effects as corrected by the current HFI TOI processing pipeline. An increase in the significance of these tests is left for further work once the simulation chain is further optimized for more massive numerical work.} 



\begin{figure*}[htb] 
\centering
\includegraphics[width=0.97\textwidth, clip=true, trim=5mm 7mm 16mm 7mm]{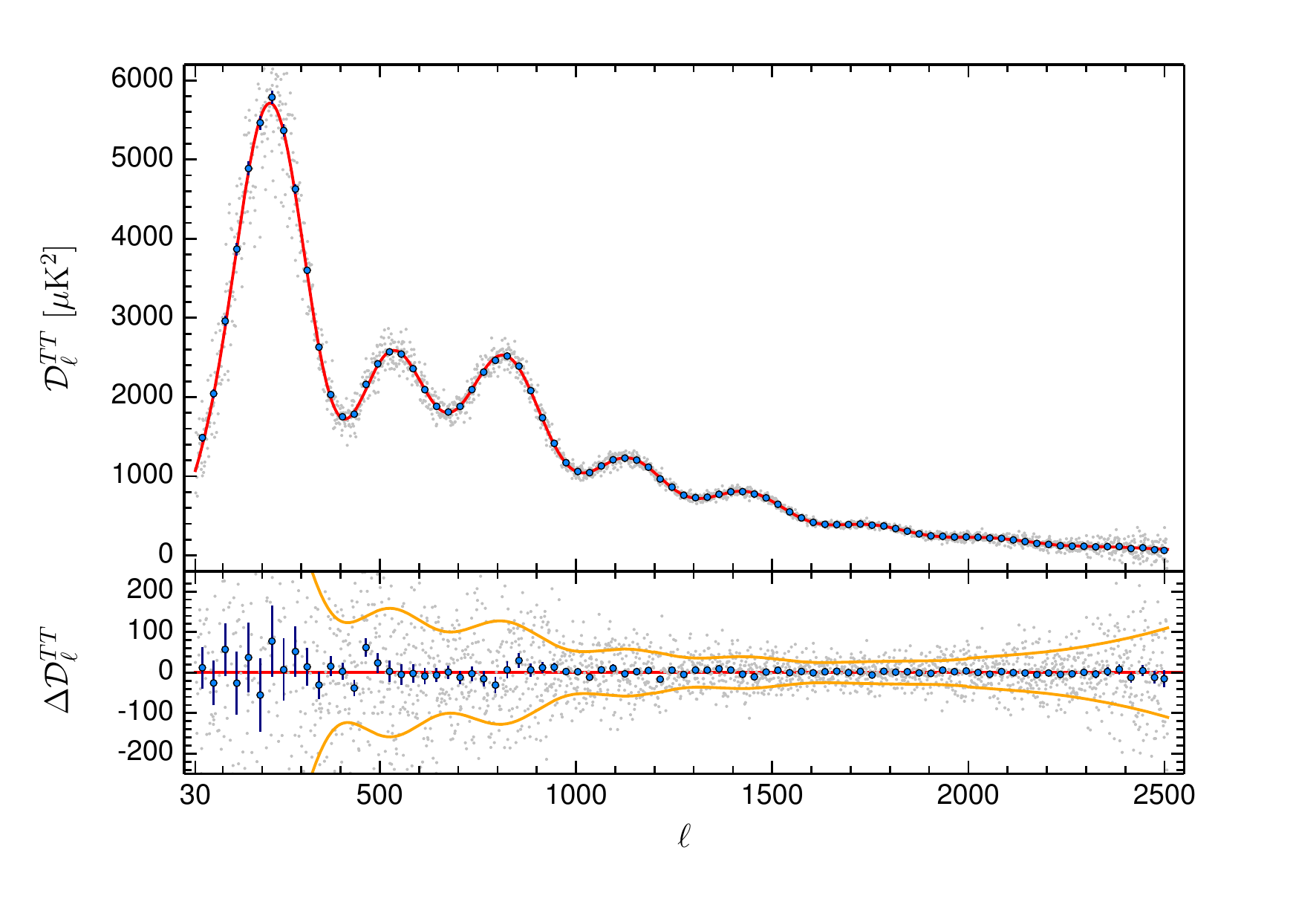}
\includegraphics[width=0.485\textwidth, clip=true, trim=1mm 1mm 7mm 4mm]{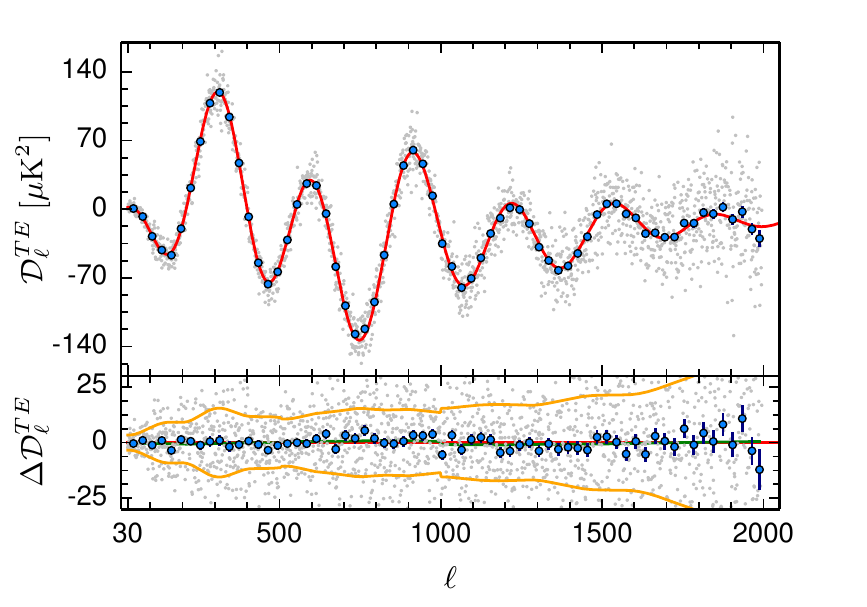}
\includegraphics[width=0.485\textwidth, clip=true, trim=1mm 1mm 7mm 4mm]{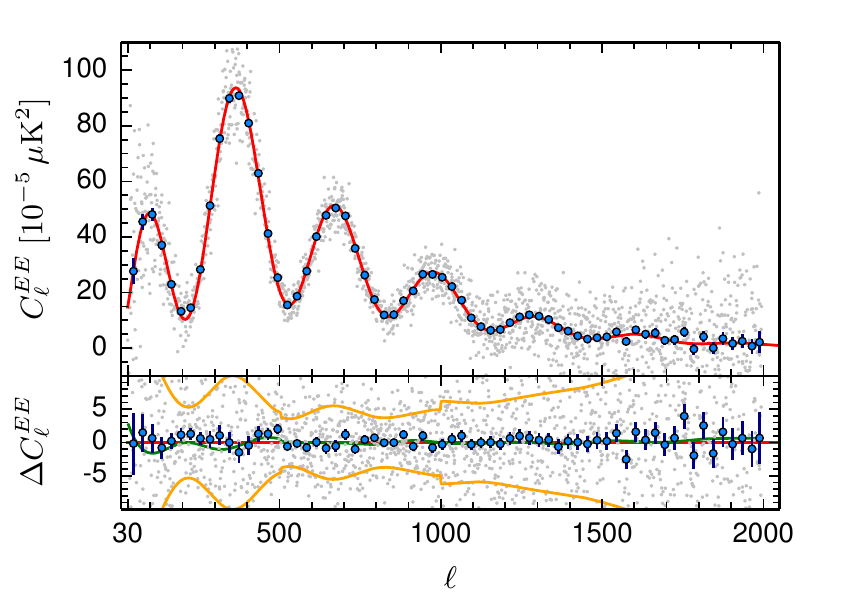}
\caption{\plik\ 2015 co-added $\TT$, $\TE$, and $\EE$ spectra. The blue points are for bins of $\Delta\ell=30$, while the grey points are unbinned. The lower panels show the residuals with respect to the best fit \plikTTtau\ $\Lambda$CDM model. The yellow lines show the 68\,\% unbinned error bars. For $\TE$ and $\EE$, we also show the best-fit beam-leakage correction (green line; see text and Fig.~\ref{fig:bleak_fit}).}
\label{fig:Cl_TE_EE}
\end{figure*}

\begin{figure*}[htb]
\centering
\includegraphics{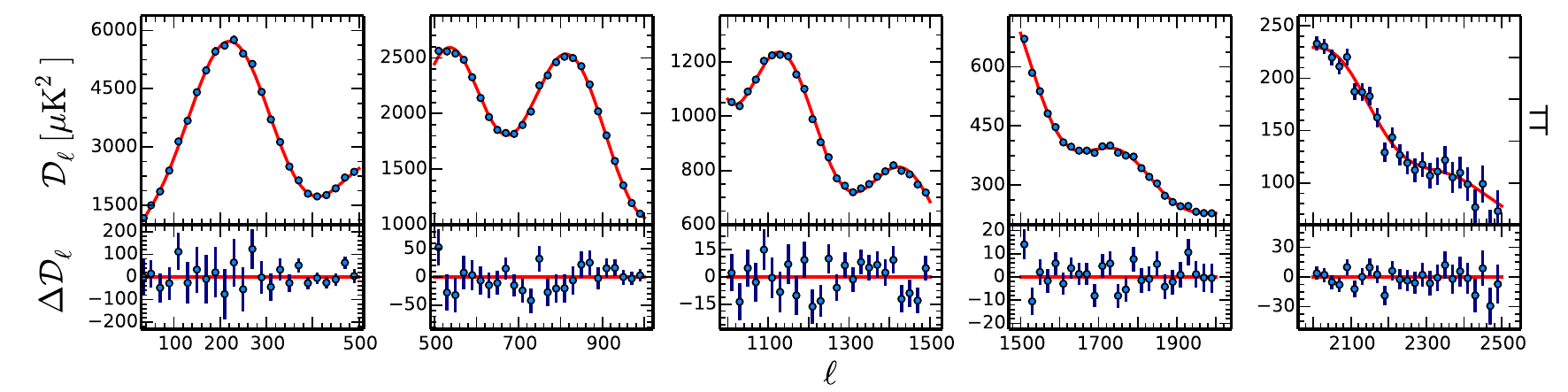}
\includegraphics{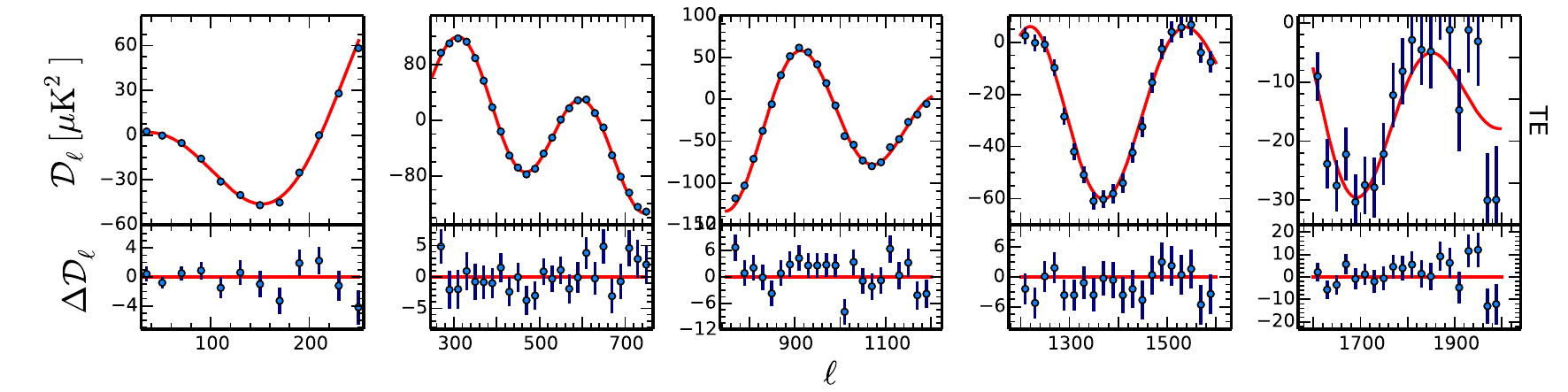}
\includegraphics{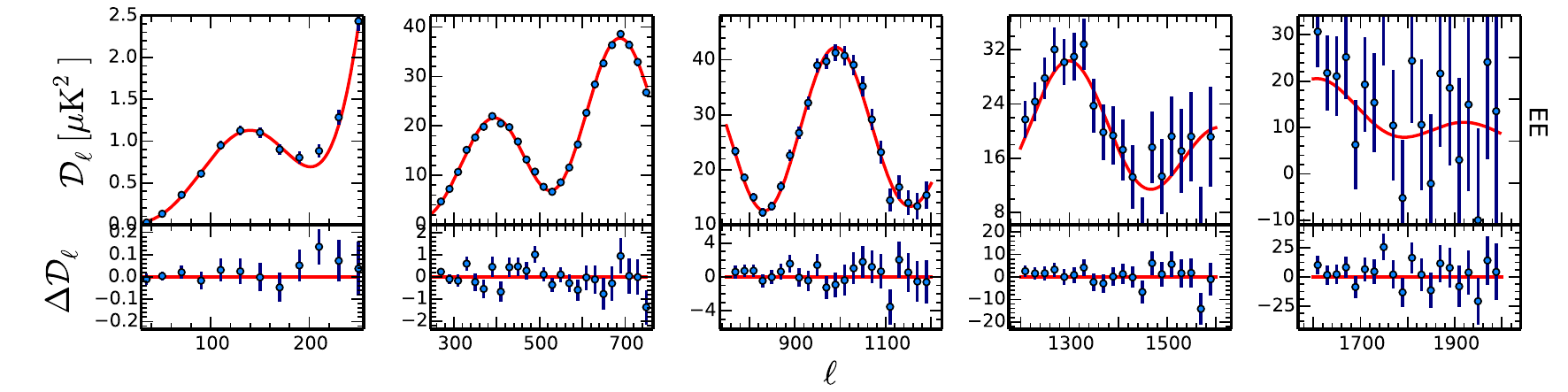}
\caption{Zoom in to various $\ell$ ranges of the HM co-added power spectra, together with the \plikTTtau\ $\Lambda$CDM best-fit model (red line). We show the $\TT$ (top), $\TE$ (centre) and $\EE$ (bottom) power spectra.  The lower panels in each plot show the residuals with respect to the best-fit model. 
\label{fig:resTTcmbzoom} }
\end{figure*}

\subsection{High-multipole reference results}
\label{sec:highlbase}

This section describes the results obtained using the baseline \plik likelihood, in combination with a prior on the optical depth to reionization, $\tau=0.07\pm0.02$ (referred to, in $\TT$, as \plikTTtau). The robustness and validation of these results (presented in Sect.~\ref{sec:hil-ass}) can therefore be assessed independently of any potential low-$\ell$ anomaly, or hybridization issues. The full low-$\ell$ + high-$\ell$ likelihood will be discussed in Sect.~\ref{sec:hal}.

\begin{figure}[htb]
\centering
\includegraphics[clip=true, trim=1mm 0mm 6mm 5mm]{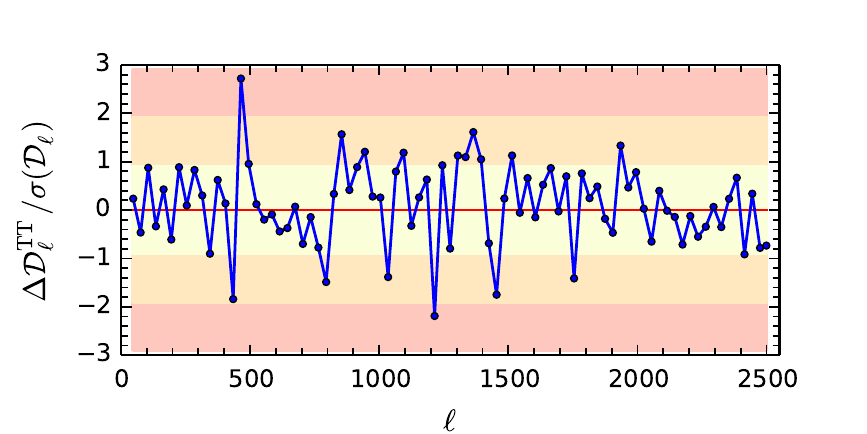}
\includegraphics[clip=true, trim=1mm 0mm 6mm 5mm]{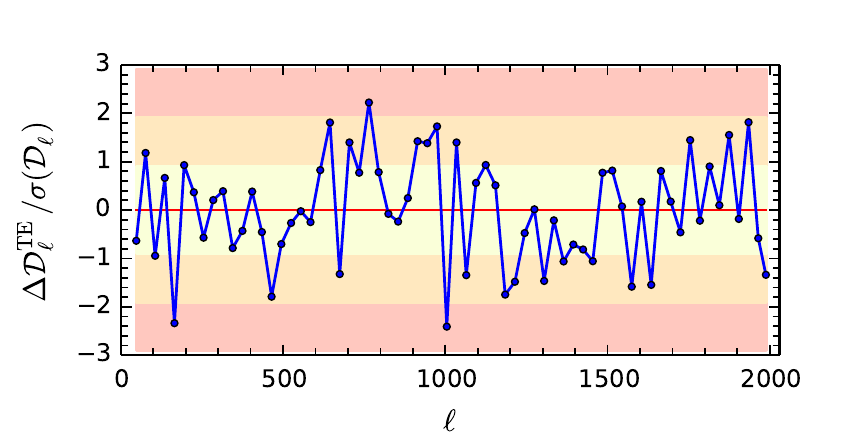}
\includegraphics[clip=true, trim=1mm 0mm 6mm 5mm]{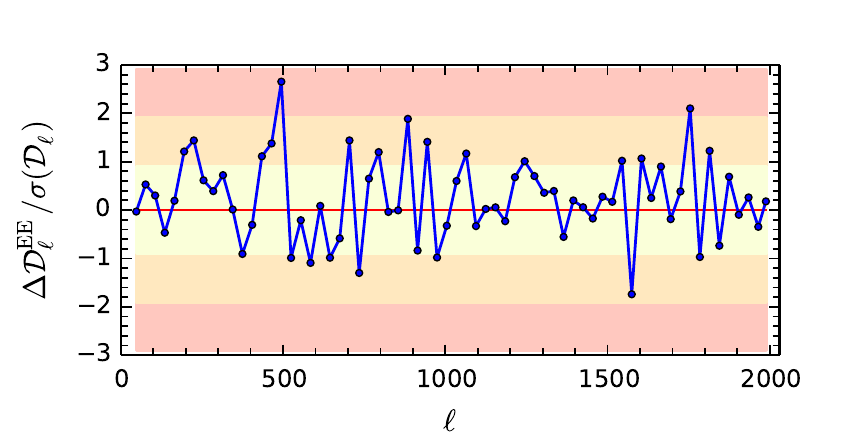}
\caption{Residuals of the co-added CMB $\TT$ power spectra, with respect to the \plikTTtau\ best-fit model, in units of standard deviation. The three coloured bands (from the centre, yellow, orange, and red)  represent the $\pm 1$, $\pm2$, and $\pm3\,\sigma$ regions. }
\label{fig:resTTcmb}
\end{figure}

\begin{figure}[!htbp] 
\centering
\includegraphics[width=\columnwidth, clip=true, trim=4mm 3mm 6mm 2mm]{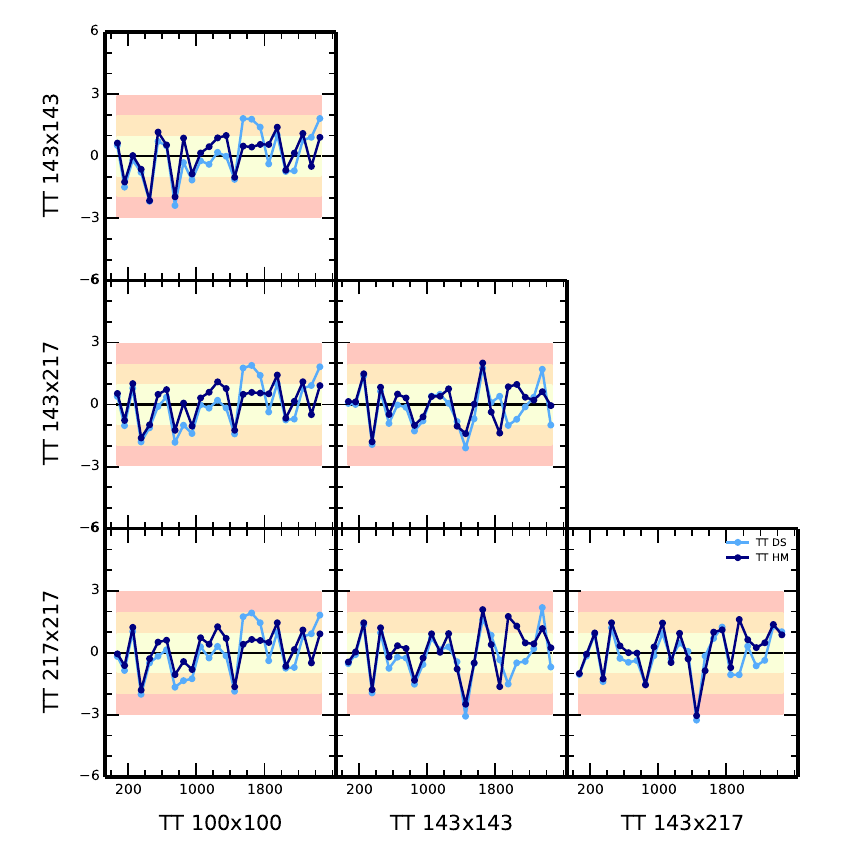}
\caption{Inter-frequency foreground-cleaned $\TT$ power spectra differences, in $\muKsq$. Each of the sub-panels shows the difference, after foreground subtraction, between pairs of frequency power spectra (the spectrum named on the vertical axis minus the one named on the horizontal axis), in units of standard deviation. The coloured bands identify deviations that are smaller than one (yellow), two (orange), or three (red) standard deviations. We show the differences for both the HM power spectra (blue points) and the DS power spectra (light blue points) after correlated noise correction. Figure~\ref{fig:respol-diff} displays the same quantities for the $\TE$ and $\EE$ spectra. \label{fig:nudiffTT} } 
\end{figure}

\begin{figure*}[!htbp] 
\centering
\includegraphics[width=\textwidth]{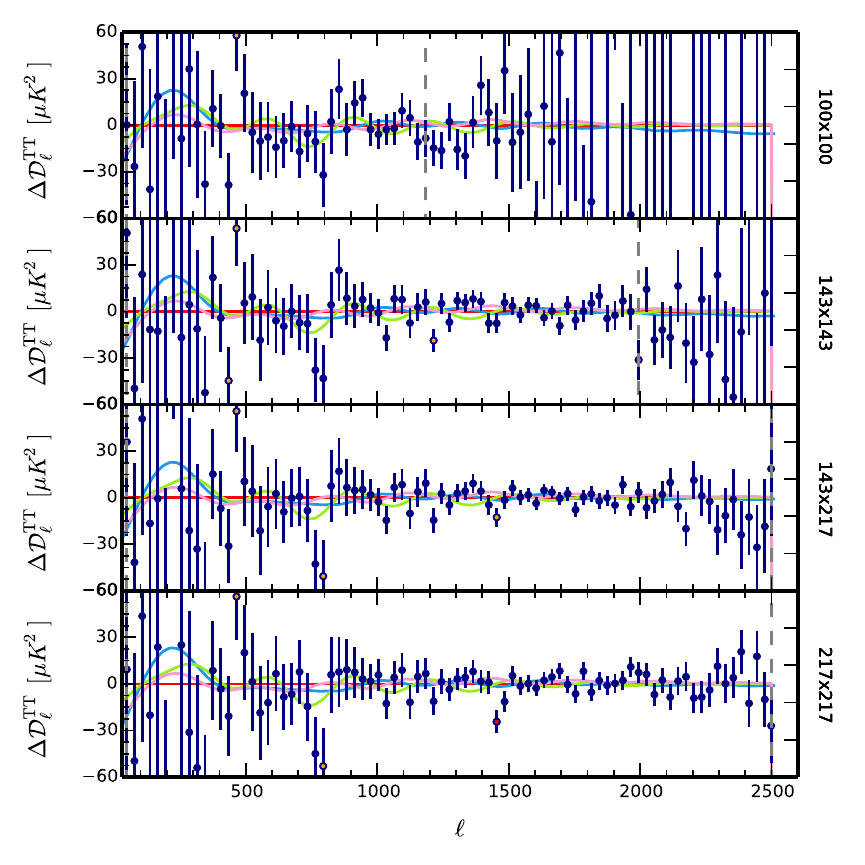} 
\caption{Residuals in the half-mission $\TT$ power spectra after subtracting the \plikTTtau\ \LCDM\ best-fit model (blue points, except for those which differ by at least 2 or $3\,\sigma$, which are coloured in orange or red, respectively). 
The light blue line shows the difference between the best-fit model obtained assuming a $\Lambda$CDM+$\Alens$ model and the $\Lambda$CDM
best-fit baseline; the green line shows the difference of best-fit models using the $\lmax=999$ likelihood (fixing the foregrounds to the baseline solution) minus the baseline best-fit (both in the $\Lambda$CDM framework); while the pink line is the same as the green one but for $\lmax=1404$ instead of $\lmax=999$; see text in Sect.~\ref{sec:hil_par_stability}. For the $\TE$ and $\EE$ spectra, see Fig.~\ref{fig:respol}. \label{fig:resTT} } 
\end{figure*}

\begin{figure*}[htb] 
\centering
\includegraphics[width=\textwidth]{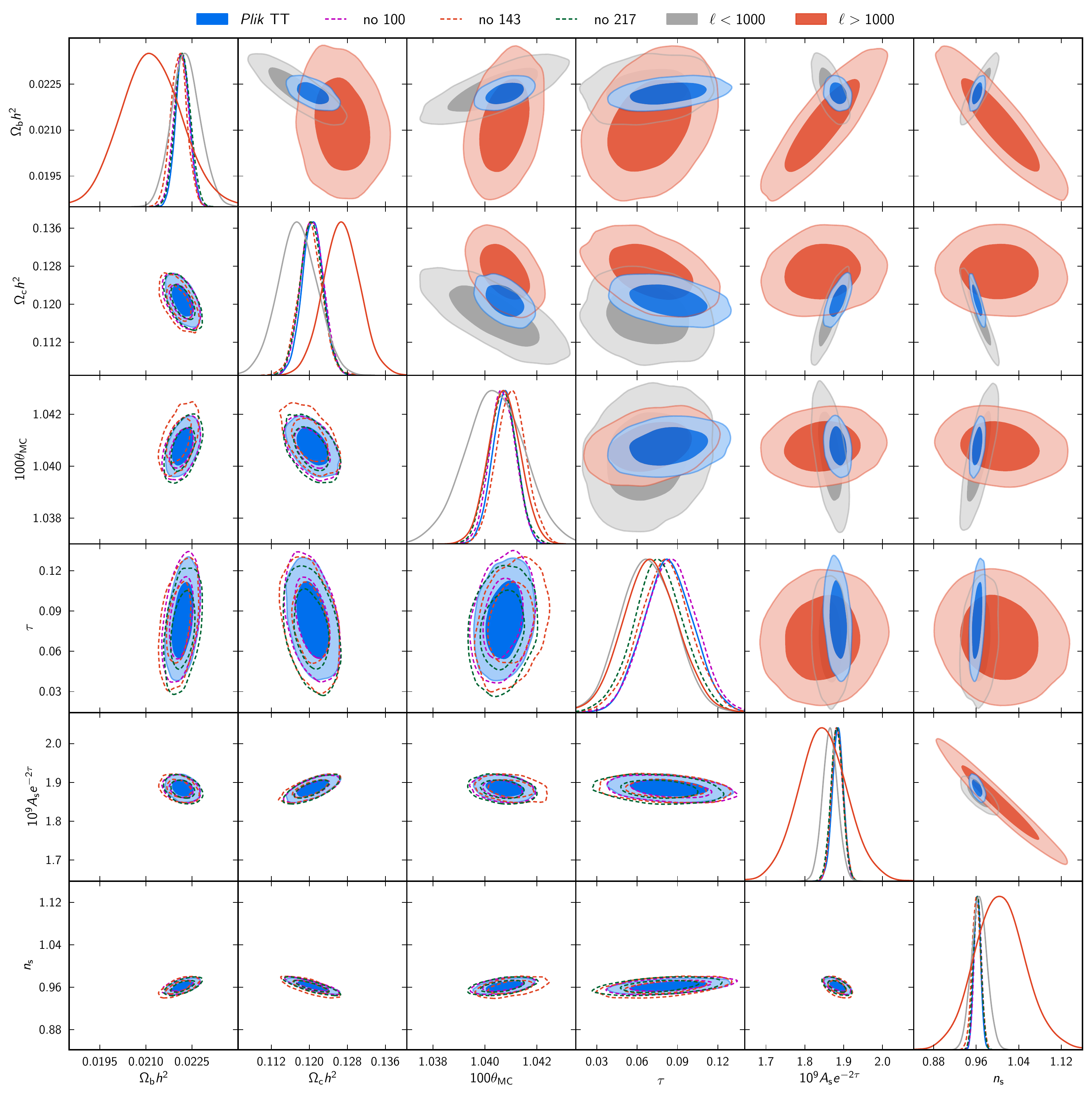}
\caption{\LCDM\ parameters posterior distribution for \plikTTtau. The lower left triangle of the matrix displays how the constraints are modified when the information from one of the frequency channels is dropped. The upper right triangle displays how the constraints are modified when the information from multipoles $\ell$ greater or less than 1000 is dropped. All the results shown in this figure were obtained using the \CAMB\ code.
\label{fig:plik_contours} }
\end{figure*}

Figure~\ref{fig:Cl_TE_EE} shows the high-$\ell$ co-added CMB spectra in $\TT$, $\TE$, and $\EE$, and their residuals with respect to the best-fit \LCDM\ model in $\TT$ (red line), both $\ell$-by-$\ell$ (grey points) and binned (blue circles). The blue error bars
per bin are derived from the diagonal of the covariance matrix computed with the best-fit CMB as fiducial model. The bottom sub-panels with residuals also show (yellow lines) the diagonal of the $\ell$-by-$\ell$ covariance matrix, which may be compared to the dispersion of the individual $\ell$ determinations. Parenthetically, it provides graphical evidence that $\TT$ is dominated by cosmic variance through $\ell \approx 1600$,  while $\TE$ is cosmic-variance dominated at $\ell\lesssim 160$ and $\ell\approx 260$--$460$. The jumps in the  polarization diagonal-covariance error-bars come from the variable $\ell$ ranges retained at different frequencies, which therefore vary the amount of data included discontinuously with $\ell$. Figure~\ref{fig:resTTcmbzoom} zooms in to five  adjacent $\ell$ ranges on the co-added spectra to allow close inspection of the data distribution around the model.     

More quantitatively, Table~\ref{tab:highl:lrange} shows the $\chi^2$  values with respect to the \LCDM\ best fit to the \plikTTtau\ data combination for the unbinned CMB co-added power spectra (obtained as described in Appendix~\ref{app:co-added}). The $\TT$ spectrum has a reduced $\chi^2$ of $1.03$ for 2479 degrees of freedom, corresponding to a probability to exceed (PTE) of 17.2\,\%; the base $\Lambda$CDM model is therefore in agreement with the co-added data. The best-fit \LCDM\ model in $\TT$ also provides an excellent description of the co-added polarized spectra, with a PTE of 12.8\,\% in $\TE$ and 34.6\,\% in $\EE$. This already suggests that extensions with, \eg isocurvature modes can be severely constrained. 

Despite this overall agreement, we note that the PTEs are not uniformly good for all cross-frequency spectra (see in particular the $100\times100$ and $100\times217$ in $\TE$). This shows that the baseline  instrumental model needs to include further effects to describe all of the data in detail, even if the averages over frequencies appear less affected. The green line in Fig.~\ref{fig:Cl_TE_EE} (mostly visible in the $\Delta C_\ell^{EE}$ plot) shows the best-fit leakage correction (shown on its own in Fig.~\ref{fig:bleak_fit}), which is obtained when fixing the cosmology to the $\TT$-based model. Let us recall, though, that this correction is for illustrative purposes only, and it is set to zero for all actual parameter searches. Indeed, we shall see that these leakage effects are not enough to bring all the data into full concordance with the model.  

In more quantitative detail, Fig.~\ref{fig:resTTcmb} shows the binned ($\Delta\ell=100$) residuals for the co-added CMB spectra in units of the standard deviation of each data point, (data\,$-$\,model)/error. For $\TT$, we find the greatest deviations  at $\ell\approx 434$ ($-1.8\,\sigma$), $464$ $(2.7\,\sigma)$, $1214$ ($-2.1\,\sigma$), and $1450$  ($-1.8,\sigma$).  At $\ell=1754$, where we previously reported a deficit due to the imperfect removal of the \HeJT\ cooler line  \citep[see][section~3]{planck2014-a15}, there is now a less significant fluctuation, at the level of $-1.4\,\sigma$. The residuals in polarization show similar levels of discrepancy.

In order to assess whether these deviations are specific to one particular frequency channel or appear as a common signal in all the spectra, Fig.~\ref{fig:nudiffTT} shows foreground-cleaned $\TT$ power spectra differences across all frequencies, in units of standard deviations (details on how this is derived can be found in Appendix~\ref{sec:freq}). The agreement between $\TT$ spectra is clearly quite good. Figure~\ref{fig:resTT} then shows the residuals per frequency for the $\TT$ power spectra with respect to the $\Lambda$CDM \plikTTtau\ best-fit model (see also the zoomed-in residual plots in Fig.~\ref{fig:Cl_TT_perfreq_zoomed}). The $\ell\approx 434$, $464$, and $1214$ deviations from the model appear to be common to all frequency channels, with differences between the frequencies smaller than $2\,\sigma$. However, the deviation at $\ell\approx 1450$ is higher at $217\times217$ than in the other channels. In particular, the inter-frequency differences (Fig.~\ref{fig:nudiffTT}) between the $217\times 217$ power spectrum and the  $100\times 100$, $143\times 143$, and $143\times 217$ ones show deviations  at $\ell\approx 1450$ at the roughly $1.7$, $2.6$, and $3.4\,\sigma$ levels, respectively. 

\rev{This inter-frequency difference is due to a deficit in the residuals of the $217\times 217$ channel of about $-3.4\sigma$ in the bin centred at $1454$ in Figure~\ref{fig:resTT}. To better quantify this deviation, we also fit for a feature of the type $\cos^2((\pi/2)(\ell-\ell_p)/(\Delta\ell))$, with maximum amplitude centred at $\ell_p=1460$, width $\Delta\ell=25$ (we impose the feature to be zero at $|\ell-\ell_p|>\Delta\ell$) and with an independent amplitude in each frequency channel. At   $217\times 217$, we find an amplitude of $(-37.44\pm 9.5) \muKsq$, while in the other channels we find $(-15.0\pm 7.8)\muKsq$ at $143\times 143$ and $(-19.7\pm 7.9)\muKsq$ at $143\times 217$.} \rev{This outlier seems to be at least in part due to chance correlation between the CMB and dust. Indeed, the amplitude of the feature in the different spectra is in rough agreement with the dust emission law. Moreover, the feature can also be found when varying the retained sky fraction in the galactic mask, again with an amplitude scaling compatible with a dust origin. We discuss below the impact on cosmological parameters, see the case ``CUT $\ell$=1404-1504'' in Fig.~\ref{fig:wiskerTT}.} 

Finally we note that there is a deficit in the $\ell=500$--$800$ region (in particular between $\ell=700$ and $800$) in the residuals of all the frequency spectra, roughly in correspondence with the position of the second and third peaks. Section~\ref{sec:hil_par_stability} is dedicated to the study of these deviations and their impact on cosmological parameters.
In spite of these marginally significant deviations from the model, the $\chi^2$ values shown in Table~\ref{tab:highl:lrange} indicate that the \LCDM\ model is an acceptable fit to each of the unbinned individual frequency power spectra, with PTEs always $\mathcal{P}\gtrsim 10\,\%$ in $\TT$. We therefore proceed to examine the parameters of the best-fit model.

\begin{figure*} 
\centering{}
\includegraphics[width=0.495\textwidth]{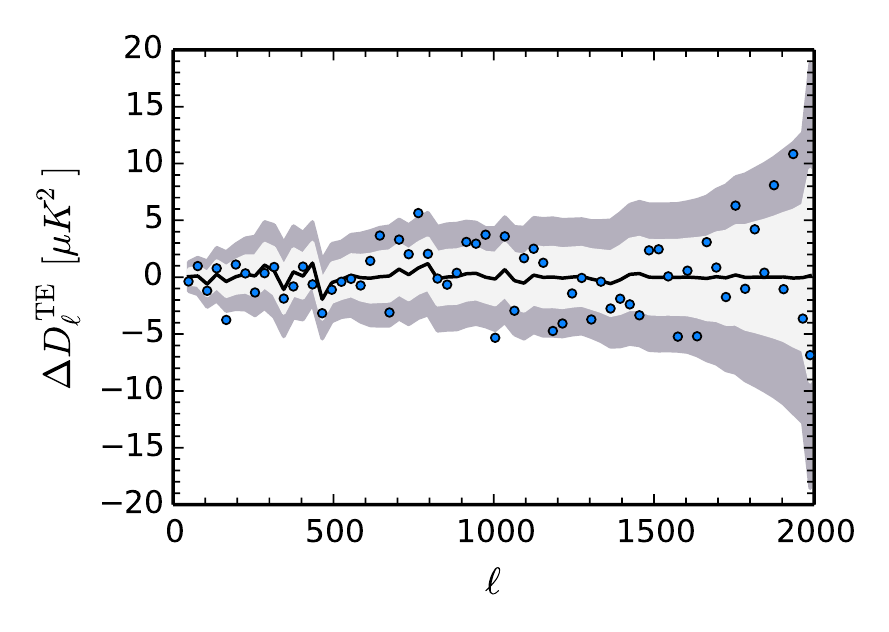} 
\includegraphics[width=0.495\textwidth]{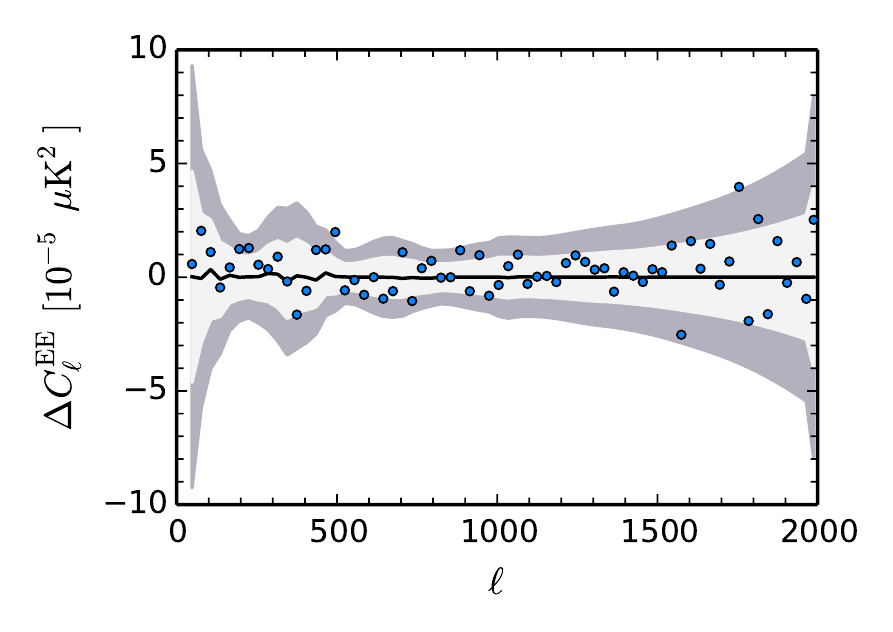}
\caption{$\TE$ ({left}) and $\EE$ ({right}) residuals conditioned on the $\TT$ spectrum (black line) with 1 and $2\,\sigma$ error bands. The blue points are the actual $\TE$ and $\EE$ residuals. We do not include any beam-leakage correction here.
\label{fig:hil:condP} }
\end{figure*}

The cosmological parameters of interest are summarized in Table~\ref{tab:cosmo-params}. Let us note that the cosmological parameters inferred here are obtained using the same codes, priors, and assumptions as in \citet{planck2014-a15}, except for the fact that we use the much faster \PICO\ \citep{pico} code instead of \CAMB\ when estimating cosmological parameters\footnote{The definition of $\Alens$ differs in \PICO\ and \CAMB; see Appendix~\ref{app:pico}.} from $\TT,\TE$ or $\TT$,$\TE$,$\EE$ using high-$\ell$ \Planck\ data. Appendix~\ref{app:pico} establishes that the results obtained with the two codes only differ by small fractions of a standard deviation (less than 15\,\% for most parameters, with a few more extreme deviations). However, we still use the  \CAMB\ code for results from $\EE$ alone, since in this case the parameter space explored is so wide that it includes regions outside the \PICO\ interpolation region (see Appendix~\ref{app:pico} for further details).

Figure~\ref{fig:plik_contours} shows the posterior distributions of each pair of parameters of the base \LCDM\ model from \plikTTtau. The upper-right triangle compares the 1\,$\sigma$ and 2\,$\sigma$ contours for the full likelihood with those derived from only the $\ell <1000$ or the $\ell\ge 1000$ data. Section~\ref{sec:changes_lmax} addresses the question of whether the results from these different cases are consistent with what can be expected statistically. The lower-left triangle further shows that the results are not driven by the data from a specific channel, \ie dropping any of the 100, 143, or 217\GHz\ map data from the analysis does not lead to much change. The next section provides a quantitative analysis of this and other jack-knife tests.

\begin{table*}[ht!] 
\begingroup 
\newdimen\tblskip \tblskip=5pt
 \caption{Goodness-of-fit tests for the \plik\ temperature and polarization spectra at high~$\ell$.}
  \label{tab:highl:lrange}
\vskip -6mm
\footnotesize
\setbox\tablebox=\vbox{
\newdimen\digitwidth
\setbox0=\hbox{\rm 0}
\digitwidth=\wd0
\catcode`*=\active
\def*{\kern\digitwidth}
\newdimen\signwidth
\setbox0=\hbox{+}
\signwidth=\wd0
\catcode`!=\active
\def!{\kern\signwidth}
\newdimen\decimalwidth
\setbox0=\hbox{.}
\decimalwidth=\wd0
\catcode`@=\active
\def@{\kern\decimalwidth}
\halign{\hbox to 1.3in{#\leaderfil}\tabskip=2em&
    \hfil#\hfil\tabskip=1em&
    \hfil#\hfil\tabskip=1.7em&
    \hfil$#$\hfil\tabskip=1.4em&
    \hfil$#$\hfil\tabskip=1.4em&
    \hfil$#$\hfil\tabskip=0.4em&
    \hfil$#$\hfil\tabskip=0.4em&
    \hfil$#$\hfil\tabskip=1em&
    \hfil$#$\hfil\tabskip=1em&
    \hfil$#$\hfil\tabskip=0pt\cr
\noalign{\doubleline}
\omit&&Multipole\cr
\omit\hfil Frequency [GHz]\hfil&\fsky [\%]$^{\rm a}$&range&\chi^2&\chi^2/N_\ell&N_\ell&\Delta \chi^2\sqrt{2N_\ell}^{\rm b}&{\rm PTE} [\%]^{\rm c}&{\chi_\mathrm{norm}}^{\rm d}&{\rm PTE}_\chi [\%]^{\rm e}\cr
\noalign{\vskip 3pt\hrule\vskip 5pt}
\omit{\boldmath{$TT$}}\hfil\cr
\noalign{\vskip 3pt}
$\quad 100\times100$& 66&*30--1197& 1234.91&  1.06&  1168&  !1.38& *8.50& -0.30& 76.44\cr
$\quad 143\times143$& 57&*30--1996& 2034.59&  1.03&  1967&  !1.08& 14.09& -0.39& 69.91\cr
$\quad 143\times217$& 49&*30--2508& 2567.11&  1.04&  2479&  !1.25& 10.63& -1.07& 28.25\cr
$\quad 217\times217$& 47&*30--2508& 2549.40&  1.03&  2479&  !1.00& 15.87& -0.17& 86.72\cr
\noalign{\vskip 2pt}
$\quad$Co-added&    &   *30--2508& 2545.50&  1.03&  2479&  !0.94& 17.22& -0.16& 87.17\cr
\noalign{\vskip 3pt}
\omit{\boldmath{$TE$}}\hfil\cr
\noalign{\vskip 3pt}
$\quad 100\times100$& 67&*30--999*& 1089.75&  1.12&  *970&  !2.72& *0.43& !3.70& *0.02\cr
$\quad 100\times143$& 50&*30--999*& 1033.38&  1.07&  *970&  !1.44& *7.72& !0.92& 35.66\cr
$\quad 100\times217$& 41&505--999*& *527.85&  1.07&  *495&  !1.04& 14.85& !5.05& *0.00\cr
$\quad 143\times143$& 50&*30--1996& 2028.18&  1.03&  1967&  !0.98& 16.45& -2.21& *2.69\cr
$\quad 143\times217$& 41&505--1996& 1606.06&  1.08&  1492&  !2.09& *2.02& -0.75& 45.19\cr
$\quad 217\times217$& 41&505--1996& 1431.65&  0.96&  1492&  -1.10& 86.60& !1.33& 18.20\cr
\noalign{\vskip 2pt}
$\quad$Co-added&    &   *30--1996& 2038.54&  1.04&  1967&  !1.14& 12.76& !0.09& 93.09\cr
\noalign{\vskip 3pt}
\omit{\boldmath{$EE$}}\hfil\cr
\noalign{\vskip 3pt}
$\quad 100\times100$& 70&*30--999*& 1027.14&  1.06&  *970&  !1.30& *9.89& !1.13& 25.88\cr
$\quad 100\times143$& 52&*30--999*& 1048.77&  1.08&  *970&  !1.79& *3.94& !1.77& *7.72\cr
$\quad 100\times217$& 43&505--999*& *479.49&  0.97&  *495&  -0.49& 68.33& -3.01& *0.26\cr
$\quad 143\times143$& 50&*30--1996& 2001.48&  1.02&  1967&  !0.55& 28.87& !3.74& *0.02\cr
$\quad 143\times217$& 43&505--1996& 1430.95&  0.96&  1492&  -1.12& 86.89& -0.71& 47.70\cr
$\quad 217\times217$& 41&505--1996& 1409.48&  0.94&  1492&  -1.51& 93.66& -1.39& 16.45\cr
\noalign{\vskip 2pt}
$\quad$Co-added&    &   *30--1996& 1991.37&  1.01&  1967&  !0.39& 34.55& !1.88& *6.00\cr
\noalign{\vskip 5pt\hrule\vskip 3pt}
}}
\endPlancktablewide
\tablenote {{\rm a}} Effective fraction of the sky retained in the analysis. For the $\TE$ cross-spectra between two different frequencies, we show the smaller \fsky\ of the $\TE$ or $ET$ combinations.\par
\tablenote {{\rm b}} $\Delta \chi^2=\chi^2-N_\ell$ is the difference from the mean, assuming the best-fit $\TT$  base-\LCDM\ model is correct, here expressed in units of  the expected dispersion, $\sqrt{2N_\ell}$.\par
\tablenote {{\rm c}} Probability to exceed the tabulated value of $\chi^2$.\par
\tablenote {{\rm d}} Weighted linear sum of deviations, scaled by the standard deviation, as defined in Eq.\ (\ref{chinorm}).\par
\tablenote {{\rm e}} Probability to exceed the absolute value $|\chi_\mathrm{norm}|$.\par
\endgroup
\end{table*}

\begin{table*}[ht!] 
\caption{Cosmological parameters used in this analysis.$^{\rm a}$}
\label{tab:cosmo-params}
\begingroup
\newdimen\tblskip \tblskip=5pt
\nointerlineskip
\vskip -3mm
\footnotesize
\setbox\tablebox=\vbox{
\newdimen\digitwidth
\setbox0=\hbox{\rm 0}
\digitwidth=\wd0
\catcode`*=\active
\def*{\kern\digitwidth}
\newdimen\signwidth
\setbox0=\hbox{+}
\signwidth=\wd0
\catcode`!=\active
\def!{\kern\signwidth}
\halign{\hbox to 2.7cm{$#$\leaderfil}\tabskip=0.4cm&
  $#$\hfil\tabskip=0.6cm&
  \hfil#\hfil\tabskip=0.6cm&
  #\hfil\tabskip=0pt\cr
\noalign{\doubleline}
\omit\hfil Parameter\hfil&\omit\hfil Prior range\hfil&\omit\hfil Baseline\hfil&\omit\hfil Definition\hfil\cr
\noalign{\vskip 3pt\hrule\vskip 5pt}
\omb \equiv \Omb h^2&    [0.005, 0.1]&  \dots& Baryon density today\cr
\omc \equiv \Omc h^2&    [0.001, 0.99]& \dots& Cold dark matter density today\cr
\theta\;\; \equiv 100\theta_{\mathrm{MC}}& [0.5, 10.0]&   \dots& $100\,{\times}$ approximation to $\rstar/D_{\rm A}$ (used in CosmoMC)\cr
\tau&                    [0.01, 0.8]&   \dots& Thomson scattering optical depth due to reionization\cr
\tau&                    (0.07\pm 0.02)\cr
\neff&                   [0.05, 10.0]&  3.046& Effective number of neutrino-like relativistic degrees of freedom (see text)\cr
\yhe&                    [0.1, 0.5]&      BBN& Fraction of baryonic mass in helium\cr
\Alens&                  [0.0, 10]&         1& Amplitude of the lensing power relative to the physical value\cr
\ns&                     [0.8, 1.2]&    \dots& Scalar spectrum power-law index ($k_0 = 0.05~\Mpc^{-1}$)\cr
\ln(10^{10}\As)&         [2, 4.0]&      \dots& Log power of the primordial curvature perturbations ($k_0 = 0.05~\Mpc^{-1}$)\cr
\noalign{\vskip 5pt\hrule\vskip 5pt}
\Oml&                    \dots&         \dots& Dark energy density divided by the critical density today\cr
{\rm Age}&               \dots&         \dots& Age of the Universe today (in Gyr)\cr
\Omm&                    \dots&         \dots& Matter density (inc.\ massive neutrinos) today divided by the critical density\cr
\zre&                    \dots&         \dots& Redshift at which Universe is half reionized\cr
H_0&                     [20,100]&\dots& Current expansion rate in $\rm{km}\, \rm{s}^{-1}\Mpc^{-1}$\cr
100\theta_{\rm D}&  \dots& \dots& $100\,\times$ angular extent of photon diffusion at last scattering\cr
100\theta_{\rm eq}&  \dots& \dots& $100\,\times$ angular size of the comoving horizon at matter-radiation equality\cr
\noalign{\vskip 5pt\hrule\vskip 3pt}}}
\endPlancktablewide
\tablenote {{\rm a}} The columns indicate the cosmological parameter symbol, their uniform prior ranges in square brackets, or between parenthesis for a Gaussian prior, the baseline values if fixed for the standard \LCDM\ model, and their definition. These parameters are the same as for the previous release. The top block lists the estimated parameters, while  the lower block lists derived parameters.\par
\endgroup
\end{table*}

We now turn to polarization results. Inter-frequency comparisons and residuals for $\TE$ and $\EE$ spectra are analysed in detail in Sect.~\ref{sec:pol-rob}. Suffice it to say here that the results are less satisfactory than in $\TT$, both in the consistency between frequency spectra and in the detailed $\chi^2$ results. This shows that the  instrumental data model for polarization is less complete than for temperature, with residual effects at the $\muKsq$ level. The model thus needs to be further developed to take full advantage of the \HFI data in polarization, given the level of noise achieved. We thus consider the high-$\ell$ polarized likelihood as a ``beta'' version. Despite these limitations, we include it in the product delivery, to allow external reproduction of the results, even though the tests that we show indicate that it should not be used when searching for weak deviations (at the $\muKsq$ level) from the baseline model.  

Nevertheless, we generally find agreement between the $\TT$, $\TE$, and $\EE$ spectra. Figure~\ref{fig:hil:condP} shows the $\TE,$ and $\EE$ residual spectra conditioned on $\TT$, which are close to zero. This is particularly the case for $\TE$ below $\ell=1000$, which gives some confidence in the polarization model. Most of the data points for $\TE$ and $\EE$ lie in the $\pm 2\,\sigma$ range. As for all $\chi^2$-based evaluations, the interpretation of this result depends crucially on the quality of the error estimates, \ie on the quality of our noise model (see Sect.~\ref{sec:noise_model}). We further note that the agreement is consistent with the finding that unmodelled instrumental effects in polarization are at the $\muKsq$ level.

\section{Assessment of the high-multipole likelihood}\label{sec:hil-ass}

This section describes tests that we performed to assess the accuracy and robustness of  the reference results of the high-$\ell$ likelihood that were presented above. First we establish the robustness of the $\TT$ results using \plik alone in Sect.~\ref{sec:hil_par_stability} and  with other likelihoods in Sect.~\ref{sec:comparison}. We verify in Sect.~\ref{sec:astro} that the amplitudes of the compact-source contributions derived at various frequencies are consistent with our current knowledge of source counts. We then summarize in Sect.~\ref{sec:pol-rob} the results of the detailed tests of the robustness of the polarization results, which are expanded upon in Appendix~\ref{sec:pol-robust}. \rev{The paper \citet{planck2014-a18} examines the dependence of the power spectrum on angular direction.}

\subsection{$\TT$ robustness tests} \label{sec:hil_par_stability}

\begin{figure*}[htb] 
\centering
\includegraphics[width=\textwidth, trim=1mm 4mm 0mm 3mm, clip=true]{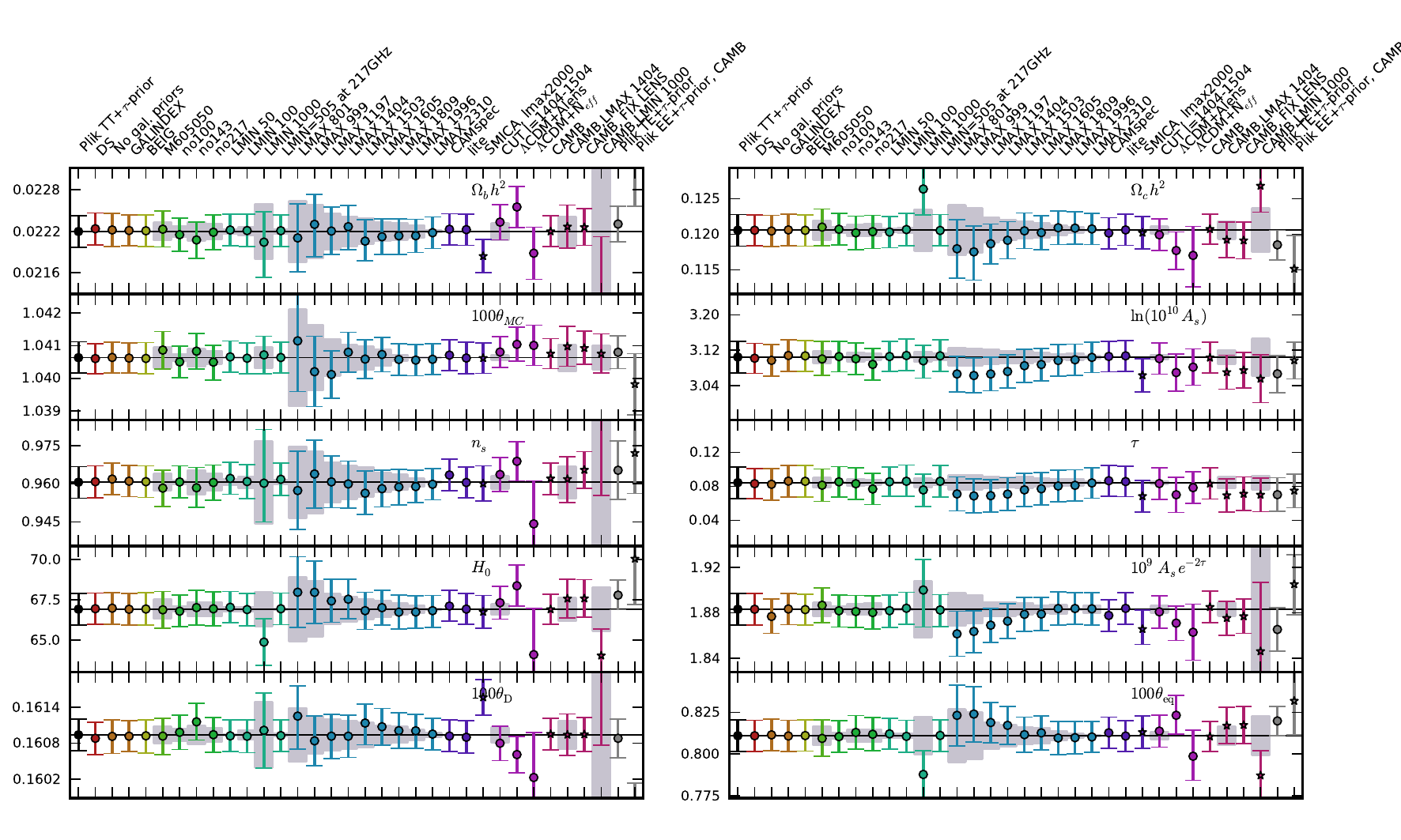}
\caption{Marginal mean and 68\,\% CL error bars on cosmological parameters estimated with different data choices for the \plik likelihood, in comparison with results from alternate approaches or model. We assume a  $\Lambda$CDM model and use variations of the {\plik}TT likelihood in most of the cases, in combination with a prior $\tau=0.07\pm0.02$ (using neither low-$\ell$ temperature nor polarization data). The ``\plikTTtau'' case (black dot and thin horizontal black line) indicates the baseline (HM, $\ell_{\mathrm{min}}=30$, $\lmax=2508$), while the other cases are described in Sect.~\ref{sec:hil_par_stability} (and \ref{sec:comparison}, \ref{sec:planck_compressed}, \ref{sec:mapCheck}).  The grey bands show the standard deviation of the expected parameter shift, for those cases where the data used is a sub-sample of the baseline likelihood (see Eq.~\ref{eq:covresult}). \rev{All the results were run with \PICO\ except for few ones that were run with \CAMB, as indicated in the labels.}   }
\label{fig:wiskerTT}
\end{figure*}

\begin{figure*}[htb]
\centering
\includegraphics[width=0.495\textwidth]{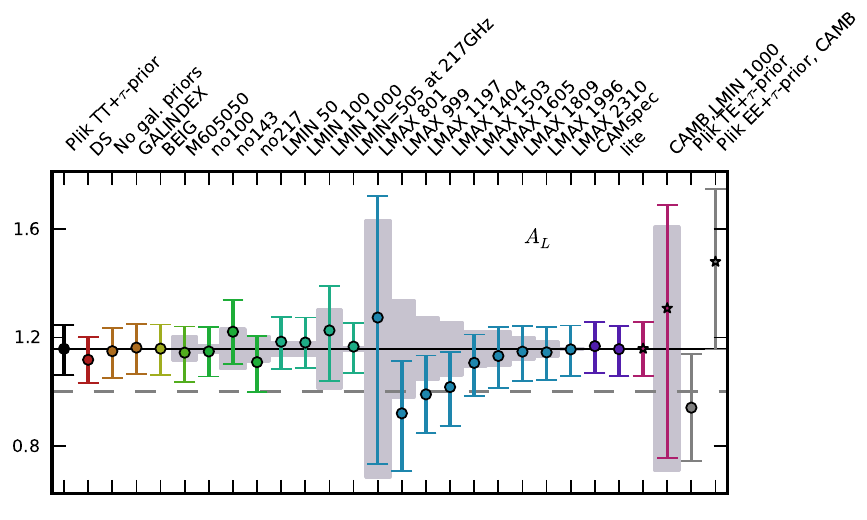}
\includegraphics[width=0.495\textwidth]{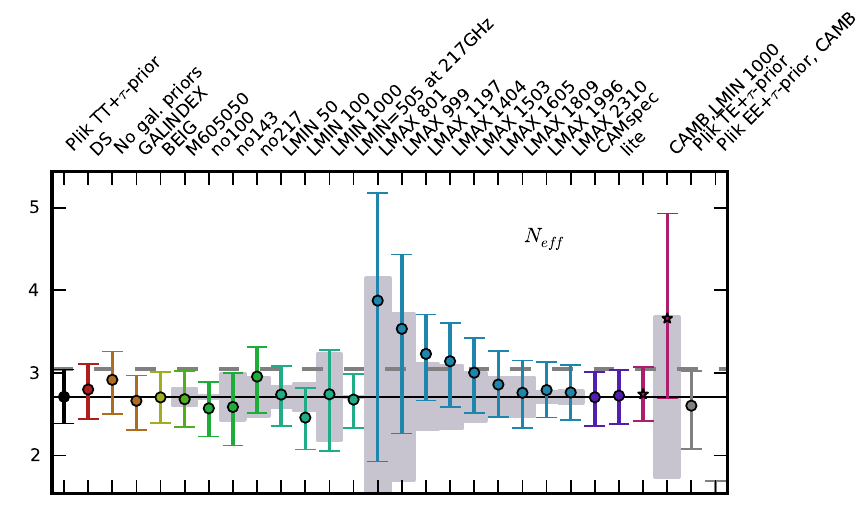}
\caption{Marginal mean and 68\,\% CL error bars on the parameters $\Alens$ ({left}) and $\nnu$ ({right}) in \LCDM\ extensions,  estimated with different data choices for the {\plik}TT likelihood in comparison with results from alternate approaches or model, combined with a Gaussian prior on $\tau=0.07\pm0.02$ (\ie neither low-$\ell$ temperature nor polarization data). The ``\plikTTtau'' case  indicates the baseline (HM, $\ell_{\mathrm{min}}=30$, $\lmax=2508$), while the other cases are described in subsections of Sect.~\ref{sec:hil_par_stability}. The thin horizontal black line shows the baseline result and the thick dashed grey line displays the  \LCDM\ value ($\Alens=1$ and $\nnu=3.04$). The grey bands show the standard deviation of the expected parameter shift, for those cases where the data used is a sub-sample of the baseline likelihood (see Eq.~\ref{eq:covresult}).}
\label{fig:wiskerTText}
\end{figure*}

Figure~\ref{fig:wiskerTT} shows the marginal mean and the 68\,\% CL error bars for cosmological parameters calculated assuming different data choices, likelihoods, parameter combinations, and data combinations.
The 31 cases shown assume a base-$\Lambda$CDM framework, except when otherwise specified. The reference case uses the \plikTTtau\ data combination. Figure~\ref{fig:wiskerTText} adds the specific results for the lensing parameter $\Alens$ (left) in a $\Lambda$CDM$+\Alens$ framework and for the effective number of relativistic species $\nnu$ (right) in a $\Lambda$CDM$+\nnu$ extended framework. 

In both figures, the grey bands show the standard deviation of the parameter shifts  relative to the baseline likelihood expected when using a sub-sample of the data (\eg excising $\ell$-ranges or frequencies). Because the data sets used to make inferences about a model are changed, one would naturally expect the inferences themselves to change, simply because of the effects of noise and cosmic variance. The inferences could also be influenced by inadequacies in the model, deficiencies in the likelihood estimate, and systematic effects in the data. Indeed, one may compare posterior distributions from different data subsets with each other and with those from the full data set, in order to assess the overall plausibility of the analysis.    

To this end it is useful to have some idea about the typical variation in posteriors that one would expect to see even in the ideal case of an appropriate model being used to fit data sets with correct likelihoods and no systematic errors.  It can be shown (Gratton and Challinor, in preparation) that if $Y$ is a subset of a data set $X$, and $\vec{P}_X$ and $\vec{P}_Y$ are vectors of the maximum-likelihood parameter values for the two data sets, then the sampling distribution of the differences of the parameter values is given by
\begin{equation}
\overline{\left(\vec{P}_{Y}-\vec{P}_{X}\right)\left(\vec{P}_Y-\vec{P}_{X}\right)^\tens{T}}=
\mathrm{cov}(\vec{P}_Y)- \mathrm{cov}(\vec{P}_X), \label{eq:covresult}
\end{equation}
i.e., the covariance of the differences is simply the difference of their covariances.  Here the covariances are approximated by the inverses of the appropriate Fisher information matrices evaluated for the true model. One might thus expect the scatter in the modes of the posteriors to follow similarly, and to be able, if the parameters are well-constrained by the data, to use covariances of the appropriate posteriors on the right-hand side. 

\subsubsection{Detset likelihood}\label{sec:robust-detsets}

We have verified (case ``DS'') that the results obtained using the half-mission cross-spectra likelihood are in agreement with those obtained using the detset (DS) cross-spectra likelihood. As explained in Sect.~\ref{sec:noise_model}, the main difficulty in using the DS likelihood is that the results might depend on the accuracy of the correlated noise correction. Reassuringly, we find that the results from the HM and DS likelihoods agree within $0.2\,\sigma$. This is an important cross-check, since we expect the two likelihoods to be sensitive to different kinds of temporal systematics. Direct differences of half-mission versus detset-based $\TT$ cross-frequency spectra are compared in Fig.~\ref{fig:nudiffTT} (Fig.~\ref{fig:respol-diff} shows similar plots for the $\TE$ and $\EE$ spectra.).

When using the detsets, we fit the calibration coefficients of the various detector sets with respect to a reference. The resulting best-fit values are very close to one,\footnote{The fitted values are
 1.0000, 0.9999, 1.0000, 1.0000, 0.9987, 0.9986, 0.9992, 0.9989, 0.9989, 0.9981, 0.9989, 1.0000, and 0.9999 for  detsets 100-ds1, 100-ds2, 143-ds1, 143-ds2, 143-5, 143-6, 143-7, 217-1, 217-2, 217-3, 217-4, 217-ds1, and 217-ds2, respectively.}
with the greatest calibration refinement being less than 0.2\,\%, in line with the accuracy expected from the description of the data processing in \citet{planck2014-a09}. This verifies that the maps produced by the \HFI\ DPC and used for the half-mission-based likelihood come from the aggregation of well-calibrated and consistent data.  

\subsubsection{Impact of Galactic mask and dust modelling}\label{sec:robust-fsky}
We have tested the robustness of our results with respect to our model of the Galactic dust contribution in various ways.

\paragraph{Galactic masks} We have  examined the impact of retaining a smaller fraction of the sky, less contaminated by Galactic emission. The baseline $\TT$ likelihood uses the G70, G60, and G50 masks (see Appendix~\ref{app:masks}) at $100$, $143$, and $217$\,GHz, respectively. We have tested the effects of using  G50, G41, and G41 (corresponding to $\fsky^\textrm{noap}=0.60$, 0.50, and 0.50 before apodization, case ``M605050'' in Fig.~\ref{fig:wiskerTT}), and of the priors on the Galactic dust amplitudes relative to these masks described in Table~\ref{table:hil:dustTT}. We find stable results as we vary these sky cuts, with the greatest shift in $\theta_\textrm{MC}$ of $0.5\,\sigma$, \rev{compatible with the expected shift of $0.57\,\sigma$ calculated using Eq.~\ref{eq:covresult}}. Going to higher sky fraction is more difficult. Indeed, the improvement in  the parameter determination from increasing the sky fraction at 143\ghz\ and 217\ghz\ would be modest, as we would only gain information in the small-scale regime, which is not probed by 100\ghz. Increasing the sky fraction at 100\,\GHz\ is also more difficult because our estimates have shown that adding as little as $5\,\%$ of the sky closer to the Galactic plane requires a change in the dust template and more than doubles the dust contamination at 100\,\GHz. 

\paragraph{Amplitude priors} We have tested the impact of not using any prior (\ie using arbitrarily wide, uniform priors) on the Galactic dust amplitudes (case ``No gal. priors'' in Fig.~\ref{fig:wiskerTT}). Again, cosmological results are stable, with the greatest shifts in $\lnAs$ of $0.23\,\sigma$ and in $\ns$ of $0.20\,\sigma$. The values of the dust amplitude parameters, however, do change, and their best-fit values increase by about $15\,\muKsq$ for all pairs of frequencies, while at the same time the error bars of the dust amplitude parameters increase very significantly. All of the amplitude levels obtained from the 545\ghz\ cross-correlation are within $1\,\sigma$ of this result. The dust levels from this experiment are clearly unphysically high, requiring $22\,\muKsq\ (\mathcal{D}_\ell, \ell=200)$ for the $100\times100$ pair. This level of dust contamination is clearly not allowed by the $545\times100$ cross-correlation, demonstrating that the prior deduced from it is informative. Nevertheless, the fact that cosmological parameters are barely modified in this test indicates that the values of the dust amplitudes are only weakly correlated with those of the cosmological parameters, consistent with the results of Figs.~\ref{fig:params_correl} and \ref{fig:params_correl_T_ext} below, which show the parameter correlations quantitatively.

\paragraph{Galactic dust template slope} 
\label{sec:galslope}
We have allowed for a variation of the Galactic dust index $n$, defined in Eq.\ (\ref{eq:dust:TTtemplate}), from its default value $n=-2.63$, imposing a Gaussian prior of $-2.63\pm 0.05$ (``GALINDEX'' case in Fig.~\ref{fig:wiskerTT}). We find no shift in cosmological parameters (smaller than $\sim0.1\, \sigma$) and recover a value for the index of $n=-2.572\pm0.038$, consistent with our default choice.

\paragraph{Impact of $\ell\lesssim {}$500 at 217\,GHz}
We have analysed the impact of excising the first 500 multipoles (``LMIN=505 at 217\,GHz'' in Fig.~\ref{fig:wiskerTT}) in the $143\times217$ and $217\times217$ spectra, where the Galactic dust contamination is the strongest. We find very good stability in the cosmological parameters, with the greatest change being a $0.16\,\sigma$ increase in $\ns$. This is compatible with the expectations estimated from Eq.~\ref{eq:covresult} of $0.14\,\sigma$. 
The inclusion of the first 500 multipoles at $217$ GHz in the baseline \plik\ likelihood is one of the sources of the roughly $0.45\,\sigma$ difference in $\ns$ observed when using the \camspec\ code, since the latter excises that range of multipoles; for further discussion  see \citet[][table 1 and section 3.1]{planck2014-a15}, as well as Sect.~\ref{sec:comparison}.

\subsubsection{Impact of beam uncertainties}\label{sec:robust-beam}

The case labelled ``BEIG''  in Fig.~\ref{fig:wiskerTT} corresponds to the exploration of beam eigenvalues with priors 10 times higher than indicated by the analysis of our MC simulation of beam uncertainties (which indicated by dotted lines in Fig.~\ref{fig:Wl_pse_fsky_MC}). This demonstrates that these beam uncertainties are so small in this data release that they do not contribute to the parameter posterior widths. They are therefore not enabled by default.

\subsubsection{Inter-frequency consistency and redundancy}\label{sec:robust-interfreq}

We have tested the effect of estimating parameters while excluding one frequency channel at a time. In Figs.~\ref{fig:plik_contours} and \ref{fig:wiskerTT}, the ``no100'' case shows the effect of excluding the $100\times100$ frequency spectrum, the ``no143'' of excluding the $143\times 143$ and $143\times217$ spectra, and the ``no217'' of excluding the $143\times217$ and $217\times217$ spectra.

\begin{figure}[h] 
\centering
\includegraphics[width=0.495\textwidth]{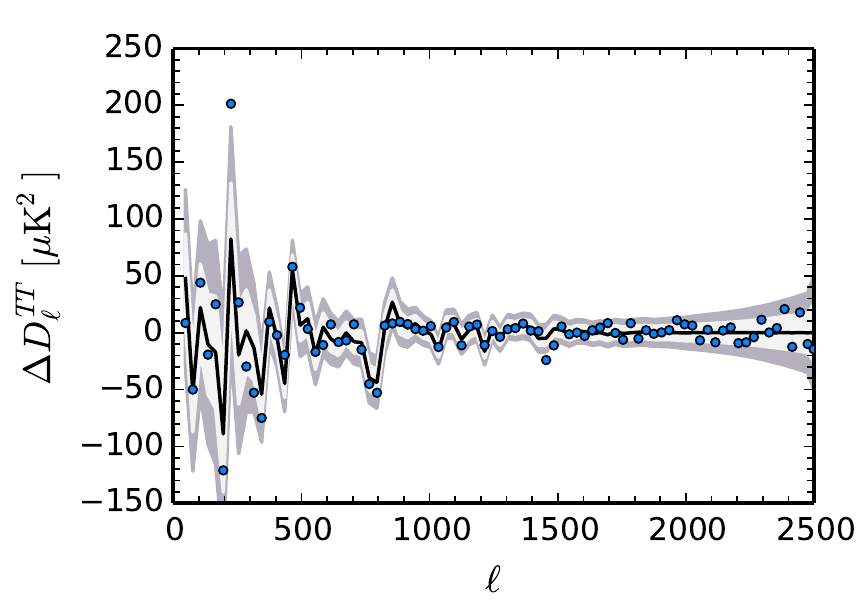}
\caption{$217\times217$ spectrum conditioned on the joint result from the $100\times100$ and $143\times143$ spectra. The most extreme outliers are at $\ell=200$ and $\ell=1450$.}
\label{fig:hil:TT217cond}
\end{figure}

We obtain the greatest deviations in the ``no217'' case for $\lnAs$ and $\tau$, which shift to lower values by $0.53\,\sigma$ and $0.47\,\sigma$, about twice the expected shift calculated using Eq.~\ref{eq:covresult}, $0.25\,\sigma$ and $0.23\,\sigma$ respectively (in units of standard deviations of the ``no217'' case). The value of $\Omega_{\rm c}h^2$ decreases by only $-0.1\,\sigma$. Figure~\ref{fig:hil:TT217cond} further shows the $217\times217$ spectrum conditioned on the $100\times100$ and $143\times143$ ones. This conditional deviates significantly in two places, at $\ell=200$ and $\ell=1450$. The $\ell=1450$ case was already discussed in Sect.~\ref{sec:highlbase} and is further analysed in Sect.~\ref{sec:changes_lmax}. Around $\ell=200$, we see some excess scatter (both positive and negative) in the data around a jump between two consecutive bins of the conditional. This corresponds to the two bins around the first peak (one right before and the other almost at the location of the first peak), as can be seen in Fig.~\ref{fig:Cl_TE_EE}. All of the frequencies exhibit a similar behaviour (see Fig.~\ref{fig:resTT}); however, it is most pronounced in the 217\ghz\ case. This multipole region is also near the location of the bump in the effective dust model. The magnitude of this excess power in the model is not big enough or sharp enough to explain this excess scatter (see Fig.~\ref{fig:hil:fg}). Finally, note that the best-fit CMB solution at large scales is dominated by the $100\times100$ data, which are measured on a greater sky fraction (see Fig.~\ref{fig:SN}).

This test shows that the parameters of the \LCDM\ model do not rely on any specific frequency map, except for a weak pull of the higher resolution 217\,GHz data towards higher values of both $\As$ and $\tau$ (but keeping $\As\exp(-2\tau)$ almost constant).

\subsubsection{Changes  of parameters with $\lmin$}\label{sec:robust-lmin}

We have checked the stability of the results when changing $\lmin$ from the baseline value of $\lmin=30$ to $\lmin=50$ and $100$ (and $\lmin=1000$, which is discussed in Sect.~\ref{sec:changes_lmax}). 
These correspond to the cases labelled ``LMIN 50'' and ``LMIN 100'' in Fig.~\ref{fig:wiskerTT} (to be compared to the reference case ``\plikTTtau''). This check is important, since the Gaussian approximation assumed in the likelihood is bound to fail at very low $\ell$ (for further discussion, see  Sect.~\ref{sec:valid-sims}). 

The results are in good agreement, with shifts in parameters smaller than $0.2\,\sigma$, \rev{well within expectations calculated from Eq.~\ref{eq:covresult}}. This is also confirmed in Fig.~\ref{fig:wiskerTTlowl}, where the $\TT$ hybridization scale of the full likelihood is varied (i.e., the multipole where the low-$\ell$ and high-$\ell$ likelihoods are joined).

\subsubsection{Changes of parameters with $\lmax$} \label{sec:changes_lmax}

We have tested the stability of our results against changes in the maximum multipole $\lmax$ considered in the analysis. We test the restriction to $\lmax$ in the range $\lmax=999$--$2310$, with the baseline likelihood having $\lmax=2508$.
For each frequency power spectrum we choose $\lmax ^{\mathrm{freq}}=\mathrm{min}(\lmax ,\lmax ^{\mathrm{ freq,\,base} })$,  where $\lmax ^{\mathrm{freq,\,  base}} $  is the baseline $\lmax$ at each frequency as reported in Table~\ref{tab:highl:lrange}.
 The results shown in Fig.~\ref{fig:wiskerTT} use the same settings as the baseline likelihood (in particular, we leave the same nuisance parameters free to vary) and always use a prior on $\tau$.

The results in Fig.~\ref{fig:wiskerTT} suggest there is a shift in the mean values of the parameters when using low $\lmax$; e.g., for $\lmax=999$, $\lnAs$, $\tau$, and $\Omc h^2$ are lower by $1.0$, $0.8$,  and $0.8\,\sigma$ with respect to the baseline parameters. These parameters then converge to the baseline values for $\lmax\gtrsim 1500$. Following the arguments given earlier (Eq.~\ref{eq:covresult}), when using these nested sub-samples of the baseline data we expect shifts of the order of $0.5$, $0.4$, and $0.8\,\sigma$ respectively, in units of the standard deviation of the $\lmax=999$ results. 
We further note that the value of  $\theta$ for $\lmax\lesssim 1197$ is lower compared to the baseline value. In particular, at $\lmax=1197$, its value is  $0.8\,\sigma$ low, while the expected shift is of the order of  $0.7\,\sigma$, in units of the standard deviation of the $\lmax=1197$ results. The value of $\theta$ then rapidly converges to the baseline for $\lmax\gtrsim 1300$. 
Figure~\ref{fig:wiskerTTFG} in Appendix~\ref{sec:temp-robust} also shows that these shifts are related to a change in the  amplitude of the foreground parameters. In particular, the overall level of foregrounds at each frequency decreases with increasing $\lmax$, partially compensating for the increase in $\lnAs$ and $\Omch$.
Although all these shifts are compatible with expectations within a factor of  $2$, we performed some further investigations in order to understand the origin of these changes.
In the following, we provide a tentative explanation.

\begin{table}[ht!] 
\begingroup 
\newdimen\tblskip \tblskip=5pt
\caption{Difference of $\chi^2$ values between pairs of best-fit models in different $\ell-$ranges for the co-added $\TT$ power spectrum.$^{\rm a}$}
\label{tab:deltachi2}
\nointerlineskip
\vskip -3mm
\footnotesize
\setbox\tablebox=\vbox{
\newdimen\digitwidth 
\setbox0=\hbox{\rm 0}
\digitwidth=\wd0
\catcode`*=\active
\def*{\kern\digitwidth}
\newdimen\signwidth
\setbox0=\hbox{+}
\signwidth=\wd0
\catcode`!=\active
\def!{\kern\signwidth}
\newdimen\decimalwidth
\setbox0=\hbox{.}
\decimalwidth=\wd0
\catcode`@=\active
\def@{\kern\decimalwidth}
\halign{ 
\hbox to 1in{#\leaderfil}\tabskip=2em& 
    \hfil$#$\hfil\tabskip=2em&
    \hfil$#$\hfil&
    \hfil$#$\hfil\tabskip=0pt\cr
\noalign{\doubleline}
\omit\hfil Multipole range\hfil&\Delta_{\lmax=999}&\Delta_{\lmax=1404}&\Delta_{\Alens}\cr
\noalign{\vskip 3pt\hrule\vskip 5pt}
**30--*129& !0.1*& !0.31&	!0.4*\cr
*130--*229& !0.07& !0.05&	!0.3*\cr
*230--*329& -0.4*& -0.22&	-0.45\cr
*330--*429& !0.34& -0.09&	!0.22\cr
*430--*529& -0.01& !0.17&	!0.26\cr
*530--*629&	 !0.61& -0.26&	-0.2*\cr
*630--*729& -1.66& -0.8*&	-0.8*\cr
*730--*829& -1.15& -0.13&	-0.79\cr
*830--*929& -0.45& !0.01&	!0.91\cr
*930--1029& -0.87& !0.41&	!0.58\cr
1030--1129& !2.17& -0.94&	-0.24\cr
1130--1229& !1.65& !1.47&	-0.17\cr
1230--1329& !0.87& !0.17&	-0.08\cr
1330--1429& !6.21& -1.46&	-0.64\cr
1430--1529& -0.2*& !3.35&	-0.62\cr
1530--1629& !0.78& !0.27&	-0.44\cr
1630--1729& !0.73& !0.9*&	!0.06\cr
1730--1829& !0@**& !1.18&	-0.01\cr
1830--1929& !0.59& -0.08&	-0.31\cr
1930--2029& !0.21& !0.04&	-0.04\cr
2030--2129& !0@**& !0.57&	-0.12\cr
2130--2229& !0.11& !0.19&	-0.18\cr
2230--2329& -0.17& !0.25&	-0.2*\cr
2330--2429& !0.06& -0.16&	!0.09\cr
2430--2508& !2.63& !2.66&	-0.19\cr
\noalign{\vskip 5pt\hrule\vskip 3pt}
}}
\endPlancktable 
\tablenote {{\rm a}} The first column shows the $\ell$-range, the second shows the difference $\Delta_{\lmax=999}$  between the $\chi^2$ values for a $\Lambda$CDM best-fit model obtained using either a likelihood with $\lmax=999$ or the baseline, \ie $\Delta_{999} \equiv \left(\chi^2_{\lmax=999}-\chi^2_{\mathrm{BASE}}\right)_{\Lambda\mathrm{CDM}}$. The $\lmax=999$ case was run fixing the foreground parameters to the best fit of the baseline case. The third column is the same as the second, but for $\lmax=1404$. The fourth column shows the difference $\Delta_{\Alens}$ between the $\chi^2$ values obtained in the $\Lambda$CDM+$\Alens$ and the $\Lambda$CDM frameworks. In this case, all the foreground and nuisance parameters were free to vary in the same way as in the baseline case.\par
\endgroup
\end{table}

Table~\ref{tab:deltachi2} shows the difference in $\chi^2$ between the best-fit model obtained using $\lmax=999$ (or $\lmax=1404$) and the baseline \plikTTtau\ best-fit solution in different multipole intervals. For this test, we ran the $\lmax$ cases fixing the nuisance parameters to the  baseline  best-fit solution. This is required in order to be able to ``predict'' the power spectra at multipoles higher than $\lmax$, since otherwise the foreground parameters, which are only weakly constrained by the low-$\ell$ likelihood, can converge to unreasonable values. We note that fixing the foregrounds has an impact on cosmological parameters, which can differ from the ones shown in Fig.~\ref{fig:wiskerTT} (see Appendix~\ref{sec:lmaxlowl} for a direct comparison). Nevertheless, since the overall behaviour with $\lmax$ is similar, we use this simplified scenario to study the origin of the shifts. 

The $\chi^2$ differences in Table~\ref{tab:deltachi2} indicate that the cosmology obtained using $\lmax=999$ is a better fit in the region between $\ell=630$ and $829$. In particular, the low value of $\theta$ preferred by the $\lmax=999$ data set shifts the position of the third peak to smaller scales. This enables a better fit to the low points at $\ell\approx 700$--$850$ (before the third peak), followed by the high points at $\ell\approx 850$--$950$ (after the third peak). This is also clear from the residuals and the green solid line in Fig.~\ref{fig:resTT}, which shows the difference in best-fit models between the $\lmax=999$ case and the reference case. However, the values in Table~\ref{tab:deltachi2} also show that the $\lmax=999$ cosmology is disfavoured by the multipole region between $\ell\approx 1330$--$1430$, before the fifth peak. The $\lmax=999$ model predicts too little  power in this multipole range, which can be better fit if the position of the fifth peak moves to lower multipoles. As a consequence, $\theta$ shifts to higher values when including $\lmax\gtrsim1400$.

Concerning the shifts in $\Omega_ch^2$, $\As$ and $\tau$, Fig.~\ref{fig:wiskerTT} shows that these parameters  converge to the full baseline solution between $\lmax=1404$ and $\lmax=1505$.
The $\Delta \chi^2$ values in Table~\ref{tab:deltachi2} between the best-fit $\lmax=1404$ case and the baseline suggest that the $\lmax=1404$ cosmology is disfavoured by the multipole region $\ell=1430$-$1530$ (fifth peak), and --- at somewhat lower significance --- by the regions close to the fourth peak ($\ell\approx 1130$--$1230$)  and the sixth peak ($\ell\approx 1730$--$1829$). The pink line in Fig.~\ref{fig:resTT} shows the differences between  the  $\lmax=1404$ best-fit model and the baseline, and it suggests that the $\lmax=1404$ cosmology predicts an amplitude of the high-$\ell$ peaks that is too large.

This effect can be compensated by more lensing, which can be obtained with greater values of $\Omega_{\rm c} h^2$ and $\lnAs$, as well as a greater value of $\tau$ to compensate for the increase in $\As$ in the normalization of the spectra, as observed when considering $\lmax\gtrsim 1500$. This also explains why the baseline ($\lmax=2508$)  best-fit solution prefers a value of the optical depth which is $0.8\,\sigma$ higher than the mean value of the Gaussian prior ($\tau=0.07\pm 0.02$), $\tau=0.085\pm0.018$.
In order to verify this interpretation, we performed the following test (using the \CAMB\ code instead of \PICO). We fixed the theoretical lensing power spectrum to the best-fit parameters preferred by the $\lmax=1404$ cosmology, and estimated cosmological parameters using the baseline likelihood. This is the ``CAMB, FIX LENS'' case in Fig.~\ref{fig:wiskerTT}, which shows that cosmological parameters shift back to the values preferred at $\lmax=1404$ (``CAMB, $\ell$max=1404'') if they cannot alter the amount of lensing in the model.

 Since the $\ell\approx 1400$--$1500$ region is also affected by the deficit at $\ell=1450$ (described in Sect.~\ref{sec:highlbase}), we tested whether excising this multipole region from the baseline likelihood (with $\lmax=2508$) has an impact on the determination of cosmological parameters. The results in Fig.~\ref{fig:wiskerTT} (case ``CUT $\ell$=1404-1504'') show that the parameter shifts are at the level of $0.47$, $-0.29$, $0.38$, and $0.45\,\sigma$ on $\Ombh$, $\Omch$, $\theta$, and $\ns$, respectively ($0.39$, $0.09$, $0.24$, and $0.29\,\sigma$ expected from Eq.~\ref{eq:covresult}), confirming that this multipole region has some impact on the parameters, although it cannot completely account for the shift  between the  $\lmax\approx 1400$ case and the baseline.

We also estimated cosmological parameters including only multipoles  $\ell>1000$ (``LMIN 1000'' case), and compared them to the ``LMAX 999'' case\footnote{
\rev{During the revision of this paper, we noticed that the $\ell>1000$ case explores regions of parameter space that are outside the optimal \PICO\ interpolation region, as also remarked by \citet{Addison:2015wyg}. This inaccuracy mainly affected this particular test for constraints on $\ns$ and $\Ombh$: the error bars for these parameters were underestimated by a factor of about $2$ while the  mean values were misestimated by about $0.8\sigma$ with respect to runs performed with \CAMB. Nevertheless, we found that for all other parameters, and in all other likelihood tests presented in this Section, this problem did not arise, since the explored parameter space was entirely contained in the \PICO\ interpolation region so as to guarantee accurate results, as also detailed in Section \ref{app:pico}. Furthermore, this inaccuracy does not change any of the conclusions of this paper. We therefore decided to keep in Fig.~\ref{fig:wiskerTT} the results obtained with  \PICO\, but we have added results for the $\ell>1000$ case obtained with \CAMB\ (case ``CAMB, lmin=1000''). }} (see also Appendix~\ref{app:pico}).
The two-dimensional posterior distributions in Fig.~\ref{fig:plik_contours} show the complementarity of the information from $\ell\leq 999$ and $\ell\geq 1000$, with degeneracy directions between pairs of parameters changing in these two multipole regimes. The $\lmin=1000$ likelihood sets constraints on the  amplitude of the spectra $\As e^{-2\tau}$ and  on $\ns$ that are almost a factor of $2$ weaker than the ones obtained with the baseline likelihood, and somewhat higher than the ones obtained with $\lmax=999$. The value of $\tau$ is thus more effectively determined by its prior and shifts downward  by $0.59\,\sigma$ with respect to the baseline. The value of $\Omch$ shifts upward  by $1.7\,\sigma$ (cf.\ $0.8\,\sigma$ expected from Eq.~\ref{eq:covresult}). Whether this change is just due to a statistical fluctuation is still a matter of investigation.

\rev{However, since parameter shifts are correlated, we evaluated whether the ensemble of the shifts in all cosmological parameters between the $\lmax=999$ and $\lmin=1000$ cases are compatible with statistical expectations. 
In order to do so, we computed the $\chi_\Delta^2$ statistic of the shift as}
\begin{equation}
	\chi_\Delta^2 = \sum_{ij} \Delta_i \Sigma^{-1}_{ij} \Delta_j\, ,
\end{equation}
\rev{where $\Delta_i$ is the difference in best-fit value of the $i$th parameter between the $\lmax=999$ and $\lmin=1000$ cases  and $\Sigma$ is the covariance matrix of the expected shifts, calculated as the sum of the parameter covariance matrices obtained in each of the two cases, ignoring correlations between the two datasets. We include in this calculation  the $\Lambda$CDM parameters ($\Ombh$, $\Omch$, $\theta$, $\ns$, $\clamp$), excluding $\tau$,  since the constraints on this parameter are dominated by the same prior in both cases, and using $\clamp$ instead of $\lnAs$, since the latter is very correlated with $\tau$ and the $\TT$ power spectrum is mostly sensitive to the combination $\clamp$. Finally, we estimate the $\chi_\Delta^2$ both in the case where we leave the foregrounds free to vary or in the case where we fix them to the best fit of the baseline $\plikTTtau$ solution.
Assuming that $\chi_\Delta^2$ has a $\chi^2$ distribution for $5$ degrees of freedom, we find that the shifts observed in the data are consistent with simulations at the $1.2\,\sigma$ ($1.1\,\sigma$ with fixed foregrounds) level for the case where we do not include the low-$\ell$ $\TT$ likelihood at $\ell<30$ to the $\lmax=999$ case, and at the $1.5\,\sigma$ ($1.4\,\sigma$ with fixed foregrounds) level for the case where we include the low-$\ell$ $\TT$ likelihood. We also find that the use of $\clamp$ instead of $\lnAs$ changes these significances only in the case where we include the low-$\ell$ $\TT$ likelihood to the $\lmax=999$ case and leave the foregrounds free to vary, in which case we find consistency at the level of $1.8\,\sigma$, in agreement with the findings of \citet{Addison:2015wyg} (although in this case the use of $\lnAs$ and the exclusion of $\tau$ makes this test less indicative of the true significance of the shifts).  In all cases, we do not find evidence for a discrepancy between the two datasets. A more  precise and extended evaluation and discussions of these shifts, based on numerical simulations, will be presented in a future publication.}

\subsubsection{Impact of varying $\Alens$}

Figure~\ref{fig:wiskerTText} (left) displays the impact of various choices on the value of the lensing parameter $\Alens$ in the $\Lambda$CDM+$\Alens$ framework.
The baseline likelihood prefers a value of $\Alens$ that is about $2\,\sigma$ greater than the physical value, $\Alens=1$. It is clear that this  preference only arises when data with $\lmax \gtrsim 1400$ are included, and it is caused by the same effects as we proposed in Sect.~\ref{sec:changes_lmax} to explain the shifts in parameters at $\lmax\gtrsim 1400$ in the $\Lambda$CDM case.  More lensing helps  to fit the data in the $\ell\approx 1300$--$1500$ region, as indicated by the $\chi^2$ differences between the $\Lambda$CDM+$\Alens$ best-fit and the $\Lambda$CDM one in Table~\ref{tab:deltachi2}. This  drives the value  of $\Alens$ to $1.159\pm 0.090$ with {\plikTTtau}, $1.8\sigma$ higher than expected. 
The case ``\LCDM+$\Alens$'' of Fig.~\ref{fig:wiskerTT} also shows that opening up this unphysical degree of freedom shifts  the other cosmological parameters  at the  $1\,\sigma$ level;
e.g., $\Omega_c h^2$ and $\As$ shift closer to the values preferred in the $\Lambda$CDM case when using $\lmax\lesssim 1400$. While in the $\Lambda$CDM case high values of these parameters allow increasing lensing, in the $\Lambda$CDM+$\Alens$ case this is already ensured by a high value of $\Alens$, so $\Omega_c h^2$ and $\As$ can adopt values that better fit the $\ell\lesssim 1400$ range.
When using {\plik}TT in combination with the \lowTEB\ likelihood, the deviation increases to $2.4\,\sigma$, $\Alens=1.204\pm0.086$,\footnote{These results were obtained with the \PICO\ code, and are thus close to but not identical to those obtained with \CAMB\ and reported in \citet{planck2014-a15}.} due to the fact that more lensing  allows smaller values of $\Omega_{\rm c}h^2$ and $\As$ and a greater value of $\ns$, better fitting  the deficit at $\ell\approx20$ in the temperature power spectrum \citep[see][section 5.1.2 and figure 13]{planck2014-a15}.

\subsubsection{Impact of varying $\neff$}

We have investigated the effect of  opening up the $\neff$ degree of freedom in order to assess the robustness of the constraints on the \LCDM\ extensions, which rely heavily on the high-$\ell$ tail of the data. Figure~\ref{fig:wiskerTText} (right) shows that $\neff$ departs from the standard 3.04 value by about $1\,\sigma$ when using \plikTTtau, $\nnu=2.7\pm 0.33$. The $\chi^2$ improvement for this model over $\Lambda$CDM is only $\Delta \chi^2=1.5$. \rev{We note that when the \lowTEB\ likelihood (or alternatively, the low-$\ell$ $\TT$ likelihood plus the prior on $\tau$) is used in combination with {\plik}\TT,  the value of $\neff$ shifts higher by about $1\,\sigma$, $\neff=3.09 \pm  0.29$. This shift is about a factor $2$ more than the one expected from Eq.~\ref{eq:covresult}, $0.5\,\sigma$, between the  \plikTTtau\ and \plikTTtau+low-$\ell$ $\TT$ cases. This shift is due to the fact that the deficit at $\ell\approx 20$ is better fit by higher $\ns$ and, as a consequence, an increase in $\neff$ helps decreasing the enhanced power at high~$\ell$.}

Figure~\ref{fig:wiskerTText} also shows that, not surprisingly, the most extreme variations as compared to the reference case (less than $1\,\sigma$) arise when the high-resolution data are dropped (by reducing $\lmax$ or by removing the 217\,GHz channel), owing to the strong dependence of the $\neff$ constraints on the damping tail. 

Having opened up this degree of freedom, the standard parameters are now about $1\,\sigma$ away (see case ``\LCDM+$\neff$'' of Fig.~\ref{fig:wiskerTT}), and such a model would prefer quite a low value of $H_0$, which would then be at odds with priors derived from direct measurements \citep[see][for an in-depth analysis]{planck2014-a15}.

\subsection{Intercomparison of likelihoods} \label{sec:comparison}


In addition to the baseline high-$\ell$ \plik likelihood, we have developed four other high-$\ell$ codes, \camspec, \hil, \mspec, and \xfaster. \camspec and \xfaster have been described in separate papers \citep{planck2013-p08,rocha2009}, and  brief descriptions of \mspec and \hil are given in Appendix~\ref{sec:alternatives}. These codes have been used to perform data consistency tests, to examine various analysis choices, and to cross-check each other by comparing their results and ensuring that they are the same. In general, we find good agreement between the codes, with only minor differences in cosmological parameters. 

The \camspec, \hil, and \mspec codes are, like \plik, based on pseudo-$C_\ell$ estimators and an analytic calculation of the covariance \citep{Efstathiou2004,Efstathiou2006}, with some differences in the approximations used to calculate this covariance. The \xfaster code \citep{rocha2009} is an an approximation to the iterative, maximum likelihood, quadratic band-power estimator based on a diagonal approximation to the quadratic Fisher matrix estimator \citep{rocha2009,rocha2010b}, with noise bias estimated using difference maps, as described in \citet{planck2014-a11}. For temperature, all of the codes use the same Galactic masks, but they differ in point-source masking: \hil uses a mask based on a combination of ${\rm S/N} > 7$ and cuts based on flux, while the others use the baseline ${\rm S/N} > 5$ mask described in Appendix~\ref{app:masks}. The codes also differ in foreground modelling, in the choice of data combinations, and in the $\ell$-range. For the comparison presented here, all make use of half-mission maps.
 
\begin{figure*}[htb] 
\centering
\includegraphics[width=7in]{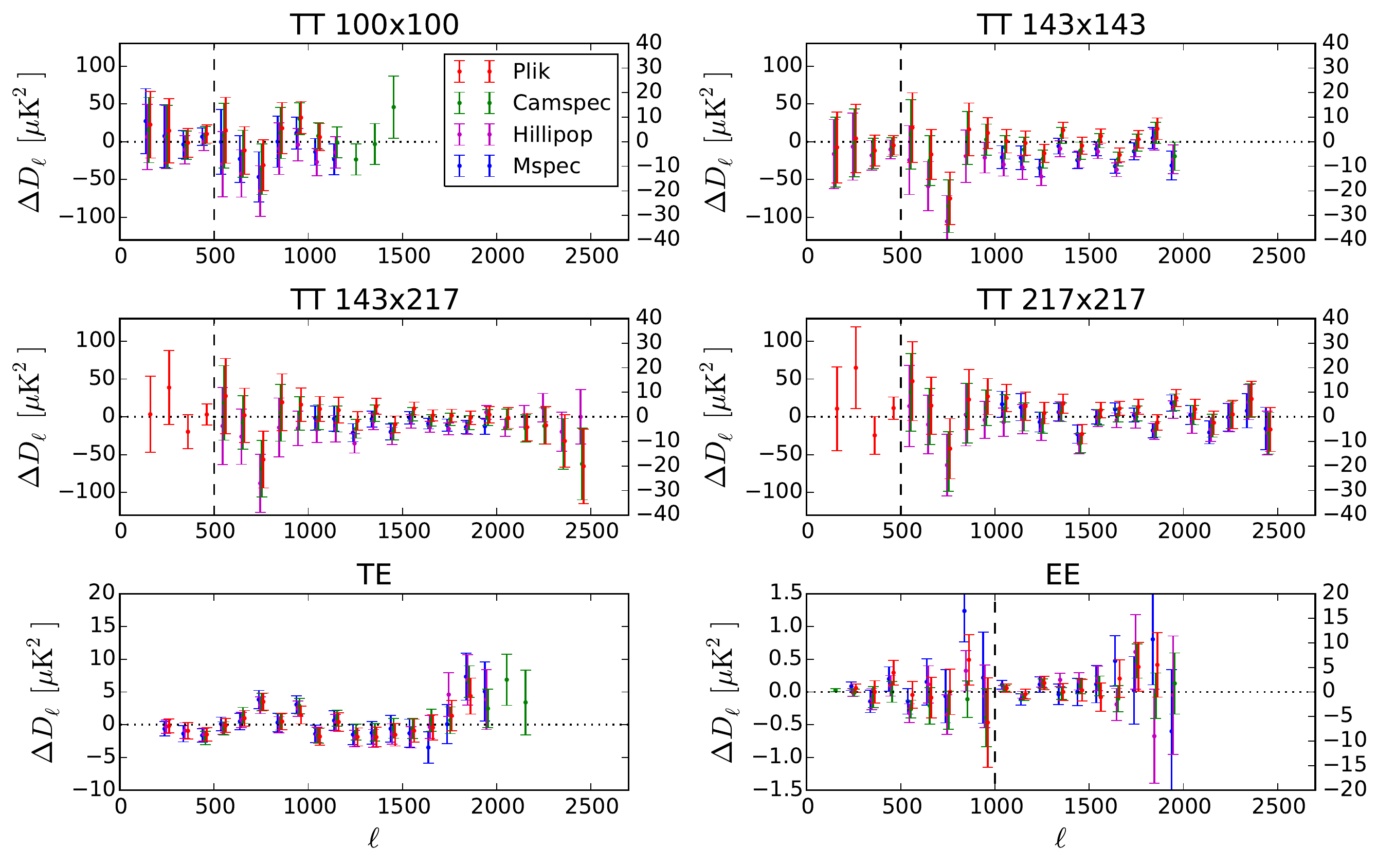}
\caption{Comparison of power spectra residuals from different high-$\ell$ likelihood codes. The figure shows ``data/calib $-$ FG $-$ {\plik}$_\mathrm{CMB}$'', where ``data'' stands for the empirical cross-frequency spectra, ``FG'' and ``calib'' are the best-fit foreground model and recalibration parameter for each individual code at that frequency, and the best-fit model {\plik}$_\mathrm{CMB}$ is subtracted for visual presentation. These plots thus show the difference in the amount of  power each code attributes to the CMB. The power spectra are binned in bins of width $\Delta\ell=100$. The $y$-axis scale changes at $\ell=500$ for $\TT$ and $\ell=1000$ for $\EE$ (vertical dashes). \label{fig:spec_comparisons}}
\end{figure*}

Figure~\ref{fig:spec_comparisons} shows a comparison of the power spectra residuals and error bars from each code, while Fig.~\ref{fig:cl_CMBmaps_dx11_spectra_Xgalcleaned} in Appendix~\ref{sec:mapCheck} compares the combined spectra with the best-fit model. In temperature, the main feature visible in these plots is an overall nearly constant shift, up to $10\,\muKsq$ in some cases. This represents a real difference in the best-fit power each code attributes to foregrounds. For context, it is useful to note the statistical uncertainty on the foregrounds; for example, the $1\,\sigma$ error on the total foreground power at 217\,GHz at $\ell=1500$ is $2.5\,\muKsq$ (calculated here with \mspec, but similar for the other codes). Shifts of this level do not lead to very large differences in cosmological parameters except in a few cases that we discuss. 

\begin{figure*} 
\centering
\includegraphics[width=7in]{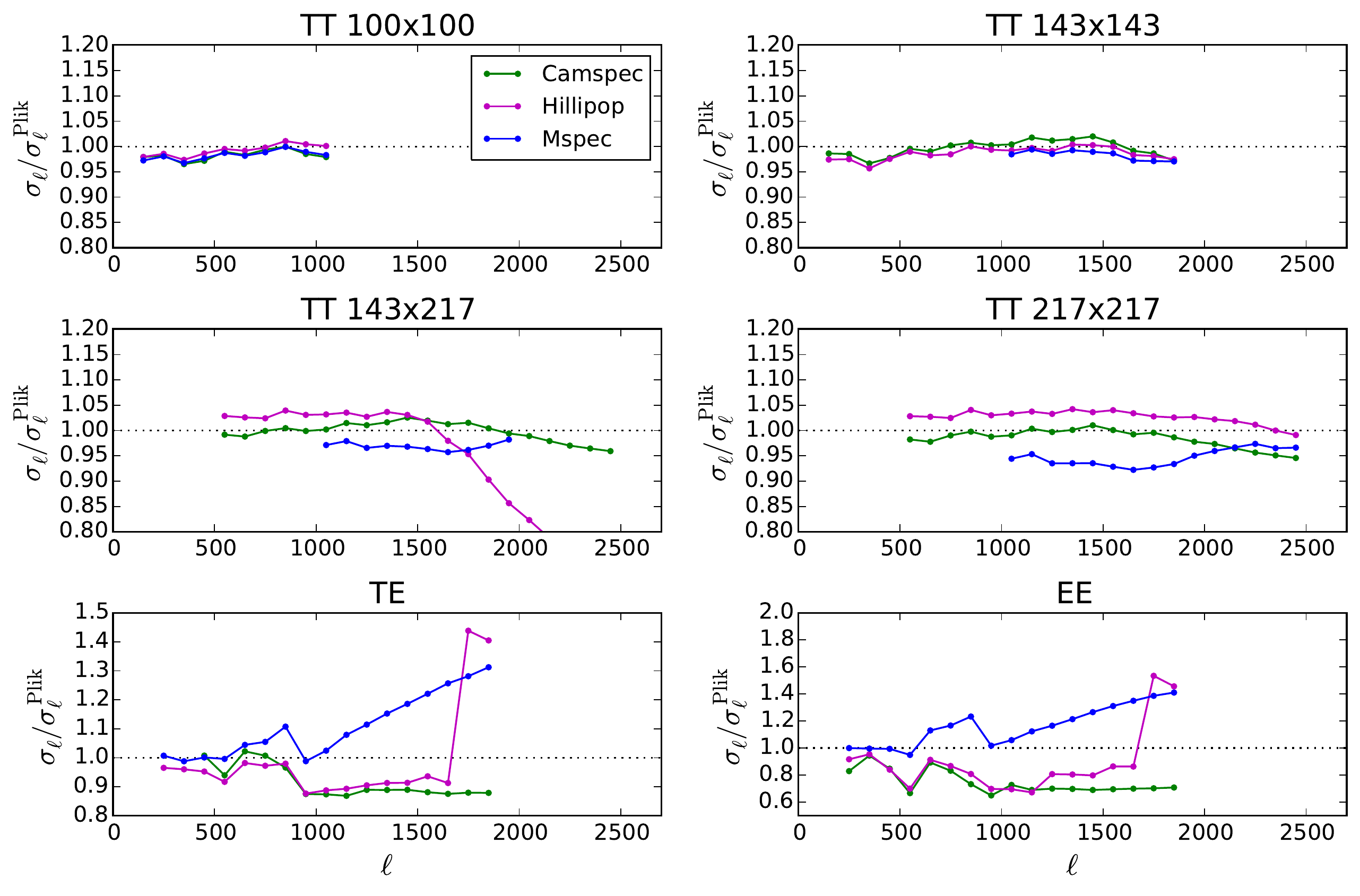}
\caption{Comparison of error bars from the different high-$\ell$ likelihood codes. The quantities plotted are the ratios of each code's error bars to those from \plik, and are for bins of width $\Delta\ell=100$. Results are shown only in the $\ell$ range common to  \plik\ and the code being compared. \label{fig:errorbar_comparisons}}
\end{figure*}

For easier visual comparison of error bars, we show in Fig.~\ref{fig:errorbar_comparisons} the ratios of each code's error bars to those from \plik. These have been binned in bins of width $\Delta\ell=100$, and are thus sensitive to the correlation structure of each code's covariance matrix, up to 100 multipoles into the off-diagonal. For all the codes and for both temperature and polarization, the correlation between multipoles separated by more than $\Delta\ell=100$ is less than $3\,\%$, so Fig.~\ref{fig:errorbar_comparisons} contains the majority of the relevant information about each code's covariance.

A few differences are visible, mostly at high frequency, when the 217\GHz\ data are used. First, the \hil error bars in $\TT$ for 143$\times$217 become increasingly tighter than the other codes at $\ell > 1700$. This is because \hil, unlike the other codes, gives non-zero weight to $143\times 217$ spectra when both the 143 and the 217\,GHz maps come from the same half-mission. This leads to a slight increase in power at high $\ell$ compared to \plik, as can be seen in Fig.~\ref{fig:spec_comparisons}. \rev{Conversely, the \hil error bars are slightly larger by a few percent at $\ell < 1700$; however the source of this difference is not understood.} Second, the \mspec error bars in temperature are increasingly tighter towards higher frequency, as compared to other codes; for $217\times 217$, \mspec uncertainties are smaller by 5--10\,\% for $\ell$ between 1000 and 2000. This arises from the \mspec map-based Galactic cleaning procedure, which removes excess variance due to CMB--foreground correlations by subtracting a scaled 545\,GHz map. However, for polarization, where one must necessarily clean with the noisier 353\,GHz maps, the \mspec error bars for $\TE$ and $\EE$ become larger. \camspec, which also performs a map cleaning for low-$\ell$ polarization, switches to a power-spectrum cleaning at higher $\ell$ to mitigate this effect.

The differences in $\Lambda$CDM parameters from $\TT$ are shown in Table~\ref{tab:param_code_compare}. Generally, parameters agree to within a fraction of $\sigma$, but with some differences we discuss. One thing to keep in mind in interpreting this comparison is that these differences are not necessarily indicative of systematic errors. Some of the differences are expected due to statistical fluctuations because different codes weight the data  differently.

One of the biggest differences with respect to the baseline code is in $\ns$, which is higher by about $0.45\,\sigma$ for \camspec, with a related downward shift of $\As e^{-2\tau}$. To put these shifts into perspective, we refer to the whisker plots of Figs.~\ref{fig:wiskerTT} and \ref{fig:wiskerTText} which compare \camspec $\TT$ results with \plik in the \lcdm\ case (base and extended). 
A difference in $\ns$ of about \rev{$0.16\,\sigma$} between \plik \  and \camspec \  can be attributed to the inclusion in \plik\ of the first $500$ multipoles for $143\times 217$ and $217\times 217$; these multipoles are excluded in \camspec\  (see also Section~\ref{sec:robust-fsky}). \rev{Indeed, cutting out those multipoles in \plik brings $\ns$ closer by $0.16\,\sigma$ to the \camspec value and slightly degrades  the constraint on $\ns$ compared to the full \plik result. Using Eq.~\ref{eq:covresult}, we see that the shift and degradation in constraining power are consistent with expectations.}
\rev{A similar $0.16\,\sigma$ shift can be attributed to different dust templates. \camspec\ uses a steeper power law index ($-2.7$). Using the \camspec template in \plik brings $\ns$ closer to the \camspec value. Allowing the power law index of the galactic template to vary when exploring cosmological parameters yields a slightly shallower slope (see Sect.~\ref{sec:galslope}). The slope of the dust template is mainly} determined at relatively high~$\ell$, \ie in the regime where it is hardest to determine the template accurately since the dust contribution is only a small fraction of the CIB and point-source contributions (see the $\ell \gtrsim 1000$ parts of Figs.~\ref{fig:hil:dust545model} and \ref{fig:hil:dust143217}). \rev{The remaining difference of $0.13\,\sigma$ arises from differences in data preparation (maps, calibration, binning) and covariance estimates}. We \rev{therefore believe that a $0.2\,\sigma$ is a conservative upper bound} of the {systematic} error in $\ns$ associated with the uncertainties in the modelling of foregrounds, {which is the biggest systematic uncertainty in $\TT$}.

A shift that is less well understood is the $\approx1\,\sigma$ shift in $\As e^{-2\tau}$ between \plik\ and \hil. The preference for a lower amplitude from \hil\ is sourced by the lower power attributed to the CMB, seen in Fig.~\ref{fig:spec_comparisons}. With $\tau$ partially fixed by the prior, this implies lower $\As$ and hence a smaller lensing potential envelope, explaining the somewhat lower value of $\Alens$ found by \hil. Tests performed with the same code suggest that $1\,\sigma$ is too great a shift to be explained simply by the different foreground models, so some part of it must be due to the different data weighting; as can be seen in Fig.~\ref{fig:errorbar_comparisons}, \hil\ gives less weight to $500\lesssim \ell \lesssim 1500$, and slightly more outside of this region. 


This comparison also shows the stability of the results with respect to the  Galactic cleaning procedure. \mspec and \plik use different procedures, yet their parameter estimates agree to better than $0.5\,\sigma$ (see Appendix~\ref{sec:mspec}). But we note that the \plik--\camspec differences are higher in the polarization case, and can reach  $1\,\sigma$, as can be judged from the whisker plot in polarization of Fig.~\ref{fig:wiskerpol}.

\begin{table*}[ht!] 
\begingroup
\newdimen\tblskip \tblskip=5pt
\caption{Comparison between the parameter estimates from different high-$\ell$ codes.$^{\rm a}$}
\label{tab:param_code_compare}
\nointerlineskip
\vskip -3mm
\footnotesize
\setbox\tablebox=\vbox{
   \newdimen\digitwidth
   \setbox0=\hbox{\rm 0}
   \digitwidth=\wd0
   \catcode`*=\active
   \def*{\kern\digitwidth}
   \newdimen\signwidth
   \setbox0=\hbox{+}
   \signwidth=\wd0
   \catcode`!=\active
   \def!{\kern\signwidth}
\halign{\hbox to 1.0in{$#$\leaderfil}\tabskip 1.2em&
\hfil$#$\hfil&
\hfil$#$\hfil&
\hfil$#$\hfil&
\hfil$#$\hfil&
\hfil$#$\hfil\tabskip 0pt\cr
\noalign{\doubleline}
\omit\hfil Parameter\hfil& \plik&\camspec&\hil&\mspec&\xfaster\ (\smica)\cr
\noalign{\vskip 3pt\hrule\vskip 5pt}
\Omega_{\rm b}h^2&        0.02221 \pm 0.00023& 0.02224 \pm 0.00023& 0.02218 \pm 0.00023& 0.02218 \pm 0.00024& 0.02184 \pm 0.00024\cr 
\Omega_{\rm c}h^2&        0.1203 \pm 0.0023&   0.1201 \pm 0.0023&   0.1201 \pm 0.0022&   0.1204 \pm 0.0024&   0.1202 \pm 0.0023\cr 
100\theta_{\rm MC}& 1.0406 \pm 0.00047&  1.0407 \pm 0.00048&  1.0407 \pm 0.00046&  1.0409 \pm 0.00050&  1.041 \pm 0.0005\cr 
\tau&               0.085 \pm 0.018&     0.087 \pm 0.018&     0.075 \pm 0.019&     0.075 \pm 0.018&     0.069 \pm 0.019\cr
10^9\As e^{-2\tau}& 1.888 \pm 0.014&     1.877 \pm 0.014&     1.870 \pm 0.011&     1.878 \pm 0.012&     1.866 \pm 0.015\cr 
n_{\rm s}&          0.962 \pm 0.0063&    0.965 \pm 0.0066&    0.961 \pm 0.0072&    0.959 \pm 0.0072&    0.960 \pm 0.0071\cr
\noalign{\vskip 10pt}
\Omega_{\rm m}&           0.3190 \pm 0.014&    0.3178 \pm 0.014&   0.3164 \pm 0.014       &    0.3174 \pm 0.015&    0.3206 \pm 0.015\cr 
H_0&                67.0 \pm 1.0&        67.1 \pm 1.0&     67.1 \pm 1.0 &    67.1 \pm 1.1&        66.8 \pm 1.0\cr 
\noalign{\vskip 5pt\hrule\vskip 3pt}}}
\endPlancktablewide 
\tablenote {{\rm a}} Each column gives the results for various high-$\ell$ $\TT$ likelihoods at $\ell>50$ when  combined with a prior of $\tau=0.07\pm0.02$. The \smica parameters were obtained for $\lmax=2000$.\par
\endgroup
\end{table*}

\subsection{Consistency of Poisson amplitudes with source counts}
\label{sec:astro}


The Poisson component of the foreground model is sourced by shot-noise from astrophysical sources. In this section we discuss the consistency between the measured Poisson amplitudes and other probes and models of the source populations from which they arise. The Poisson amplitude priors that we calculate are not used in the main analysis, because they improve uncertainties on the cosmological parameters by at most 10\,\%, and only for a few extensions; instead they serve as a self-consistency check.

This type of check was also performed in \citetalias{planck2013-p08}, which we update here by:
\begin{enumerate}
\item developing a new method for calculating these priors that is accurate enough to give realistic uncertainties on Poisson predictions (for the first time);
\item including a comparison of more theoretical models;  
\item taking into account the 2015 point-source masks.
\end{enumerate} 
In \citetalias{planck2013-p08} the Poisson power predictions were calculated via
\begin{align}
\label{eq:clpoissonold}
C_\ell = \int_0^{S_{\rm cut}} dS \, S^2 \frac{dN}{dS} ,
\end{align}
where $dN/dS$ is the differential number count, $S_{\rm cut}$ is an effective flux-density cut above which sources are masked, and the integral was evaluated independently at each frequency. Although it is adequate for rough consistency checks, Eq.~(\ref{eq:clpoissonold}) ignores the facts that the 2013 point-source mask was built from a union of sources detected at different frequencies, and that the \planck flux-density cut varies across the sky, and it also ignores the effect of Eddington bias. In order to accurately account for all of these effects, we now calculate the Poisson power as
\begin{align}
\label{eq:clpoisson}
C_\ell^{ij} = \int_0^\infty dS_1 \ldots dS_n \; S_i S_j \frac{dN(S_1,\ldots,S_n)}{dS_1\ldots dS_n} \, I(S_1,\ldots,S_n) ,
\end{align}
where the frequencies are labelled $1\ldots n$, the differential source count model, $dN/dS$, is now a  function of the flux densities at {\it each} frequency, and $I(S_1,\ldots,S_n)$ is the joint ``incompleteness'' of our catalogue for the particular cut that was used to build the point-source mask. 

\begin{table*}[ht!]
\begingroup
\newdimen\tblskip \tblskip=5pt
\caption{Priors on the Poisson amplitudes given a number of different point-source masks and models.$^{\rm a}$}
\label{tab:poisson_priors}
\vskip -5mm
\footnotesize
\setbox\tablebox=\vbox{
 \newdimen\digitwidth
 \setbox0=\hbox{\rm 0}
 \digitwidth=\wd0
 \catcode`*=\active
 \def*{\kern\digitwidth}
 \newdimen\signwidth
 \setbox0=\hbox{+}
 \signwidth=\wd0
 \catcode`!=\active
 \def!{\kern\signwidth}
 \newdimen\pointwidth
 \setbox0=\hbox{\rm .}
 \pointwidth=\wd0
 \catcode`?=\active
 \def?{\kern\pointwidth}
 \halign{\hbox to 1.3in{#\leaderfil}\tabskip=1.5em&
 \hfil#\hfil\tabskip=1.7em&
 #\hfil\tabskip=3.0em&
 \hfil$#$\hfil\tabskip=1.8em&
 \hfil$#$\hfil&
 \hfil$#$\hfil&
 \hfil$#$\hfil\tabskip=0pt\cr
\noalign{\doubleline}
\omit&&&\multispan4\hfil Power spectrum\hfil\cr
\noalign{\vskip -3pt}
\omit&&&\multispan4\hrulefill\cr
\noalign{\vskip 3pt}
\omit\hfil Mask\hfil& Type&\omit\hfil Model\hfil& 100\times100& 143\times143& 143\times217& 217\times217\cr
\noalign{\vskip 3pt\hrule\vskip 5pt}
Baseline 2013& Radio& Power-law&            *84\pm3**&             29\pm1*&             16\pm1*&           **9\pm1**\cr
\noalign{\vskip 4pt}
\omit&         Dusty& Bethermin&            **4\pm1**&             13\pm3*&             41\pm8*&           129\pm25*\cr
\noalign{\vskip 10pt}
Baseline 2015&    Radio& Power-law&         148\pm7**&             40\pm1*&             16\pm1*&           *10\pm1**\cr
\omit&          &    Tucci& 139\phantom{*\pm***}&*40\phantom{*\pm***}&*16\phantom{*\pm***}&*11\phantom{*\pm***}\cr
\noalign{\vskip 4pt}
\omit&            Dusty& Bethermin&         **4\pm1**&             13\pm3*&             41\pm8*&           129\pm25*\cr
\noalign{\vskip 4pt}
\omit&  \multispan2\hfil\plik\hfil&         260\pm28*&             44\pm8*&             39\pm10&           *97\pm11*\cr
\noalign{\vskip 4pt}
\omit& \multispan2\hfil\mspec\hfil&         317\pm46*&             22\pm13&             12\pm7*&           *21\pm9**\cr
\noalign{\vskip 10pt}
\hil 2015& Radio& Power-law&         150\pm7**&             47\pm2*&             18\pm1*&           *11\pm1**\cr
\omit&          &    Tucci& 141\phantom{*\pm***}&*47\phantom{*\pm***}&*18\phantom{*\pm***}&*12\phantom{*\pm***}\cr
\noalign{\vskip 4pt}
\omit&            Dusty& Bethermin&         **4\pm1**&             13\pm3*&             41\pm8*&           129\pm25*\cr
\noalign{\vskip 4pt}
\omit&\multispan2\hfil\hil\hfil&            372\pm38*&             58\pm21&             53\pm24&           105\pm18*\cr
\noalign{\vskip 5pt\hrule\vskip 3pt}
}}
\endPlancktablewide
\tablenote {{a}} Entries are $\mathcal{D}_\ell$ at $\ell=3000$ in $\rm \mu K^2$ and are given at the effective band centre for each component. Uncertainties on the ``power-law'' model are statistical errors propagated from uncertainties in the \citet{2013ApJ...779...61M} source-count data. Priors on the dust component have formally been calculated only for the \hil mask, but they are repeated for the other masks, for which they are accurate to better than $1\,\%$. The results from different codes, to which these predictions should be compared, use $\TT$ $\ell>50$ data with a prior of $\tau=0.07\pm0.02$. For \mspec about 90\,\% of the dusty contribution is cleaned out at the map level, \rev{hence the measured values above are in some cases far less than prior value}.\par
\endgroup
\end{table*}

The joint incompleteness was determined by injecting simulated point sources into the \Planck\ sky maps, using the procedure described in \citet{planck2014-a35}.  The same point-source detection pipelines that were used to produce the Second \Planck\ Catalogue of Compact Sources (PCCS2) were run on the injected maps, producing an ensemble of simulated \planck\ sky catalogues with realistic detection characteristics.  The joint incompleteness is defined as the probability that a source would not be included in the mask as a function of the source flux density, given the specific masking thresholds being considered. The raw incompleteness is a function of sky location, because the \Planck\ noise varies across the sky. The incompleteness that appears in Eq.~(\ref{eq:clpoisson}) is integrated over the region of the sky used in the analysis; the injection pipeline estimates this quantity by injecting sources only into these regions.

Equation~(\ref{eq:clpoisson}) can be applied to any theoretical model which makes a prediction for the multi-frequency $dN/dS$.We have adopted the following models.

1. For radio galaxies we have two models. The first is the \citet{tucci11} model, updated to include new source-count measurements from \citet{2013ApJ...779...61M}. We also consider a phenomenological model that is a power law in flux density and frequency, and assumes that the sources' spectral indices are Gaussian-distributed with mean $\bar\alpha$ and   standard deviation $\sigma_\alpha$; we use  different values for $\bar\alpha$ and $\sigma_\alpha$ above and below 143\,GHz. We shall refer to this second model as the ``power-law'' model, and the differential source counts are given by 
\begin{align}
\label{eq:dNdS}
\frac{dN(S_1,S_2,S_3)}{dS_1 dS_2 dS_3} &= \frac{A(S_1 S_2 S_3)^{\gamma-1}}{2\pi \sigma_{12} \sigma_{23}} \\ &\times \exp \left[ -\frac{(\alpha(S_1,S_2) - \bar\alpha_{12})^2}{2\sigma_{12}^2} -\frac{(\alpha(S_2,S_3) - \bar\alpha_{32})^2}{2\sigma_{32}^2} \right] ,\nonumber
\end{align}
where labels 1--3 refer to \planck 100, 143, and 217\,GHz and $\alpha(S_i,S_j) = \ln(S_j/S_i)/\ln(\nu_j/\nu_i)$. Both radio models are excellent fits to the available source-count data, and we take the difference between them as an estimate of model uncertainty. With the power-law model we are additionally able to propagate uncertainties in the source count data to the final Poisson estimate via MCMC. 

2. For dusty galaxies we use the \citet{Bethermin2012} model, as in \citet{planck2013-pip56}. The model is in good agreement with the number counts measured with te \textit{Spitzer} Space Telescope and the \textit{Herschel} Space Observatory. It also gives a reasonable CIB redshift distribution, which is important for cross-spectra, and is a very good fit to CIB power spectra \citep[see][]{Bethermin2013}. In contrast to the radio-source case, the major contribution to the dusty galaxy Poisson power arises from sources with flux densities well below the cuts; for example, we note that decreasing the flux-density cuts by a factor of 2 decreases the Poisson power by less than $1\,\%$ at the relevant frequencies. In this case, Eq.~(\ref{eq:clpoissonold}) is a sufficient and more convenient approximation, and we make use of it when calculating Poisson levels for dusty galaxies. 

We give predictions for Poisson levels for three different masks: (1) the 2013 point-source mask, which was defined for sources detected at $\mathrm{S/N}>5$ at any frequency between 100 and 353\,GHz; (2) the 2015 point-source mask, which is frequency-dependent and includes $\mathrm{S/N}>5$ sources detected only at each individual frequency (used by \plik, \camspec, and \mspec in this work); and (3) the \hil mask, which is also frequency-dependent and involves both a S/N cut and a flux-density cut.\footnote{\rev{We note that the \hil mask was constructed partly so that Eq.~(\ref{eq:clpoissonold}) would be an accurate approximation. We find that for the radio contribution is is accurate to 2\,\%, or $1\,\sigma$, and for the dust contribution it is essentially exact.}}

Table~\ref{tab:poisson_priors} summarizes the main results of this section. Generally, we find good agreement between the priors from source counts and the posteriors from chains, with the priors being much more constraining. The exception to the good agreement is at 100\,GHz where the prediction is lower than the measured value by around $4\,\sigma$ for the baseline 2015 mask and $6\,\sigma$ for the \hil mask. This is a sign either of a foreground modelling error or (perhaps more likely) of a residual unmodelled systematic in the data. We note that this disagreement was not present in \citetalias{planck2013-p08}, where the Poisson amplitude at 100\,GHz was found to be smaller. We also note that removing the relative calibration prior (Eq.~\ref{eq:relcal}) or increasing the $\ell_{\rm max}$ at 100\,GHz by a few hundred reduces the tension in the \mspec results.  In any case, it is unlikely to affect parameter estimates at all, since very little cosmological information comes from the multipole range at 100\,GHz that constrains the Poisson amplitude.

\subsection{$\TE$ and $\EE$ test results} \label{sec:pol-rob}

\subsubsection{Residuals per frequency and inter-frequency differences}\label{sec:polfreq}

\begin{figure}
\centering
\includegraphics[width=0.498\textwidth]{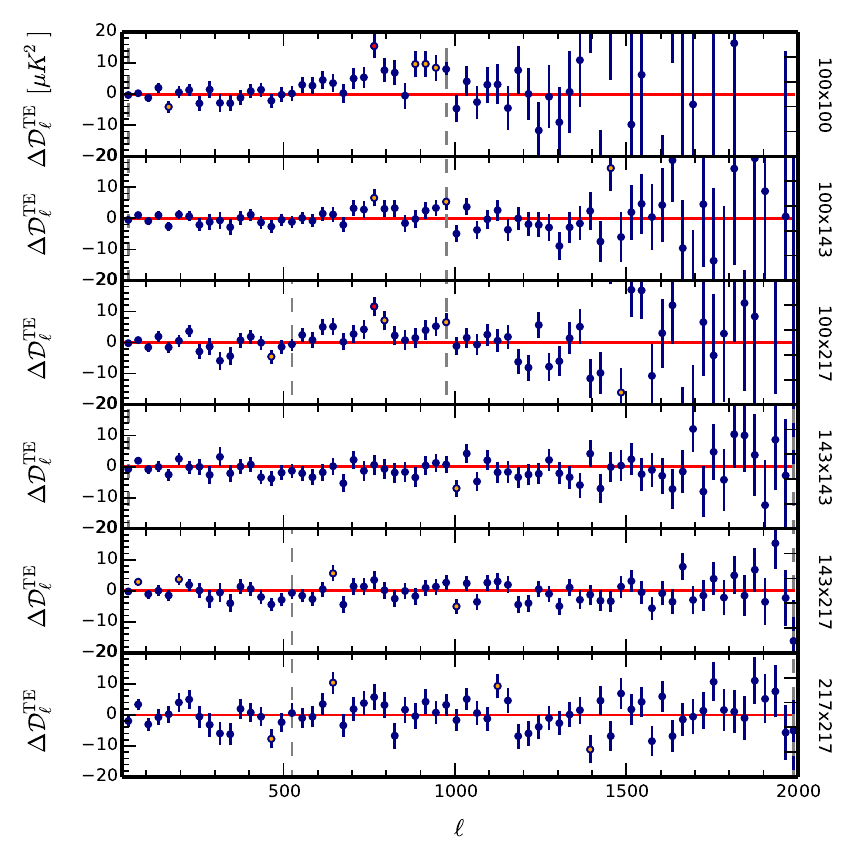}
\vspace{-12pt}
\includegraphics[width=0.498\textwidth]{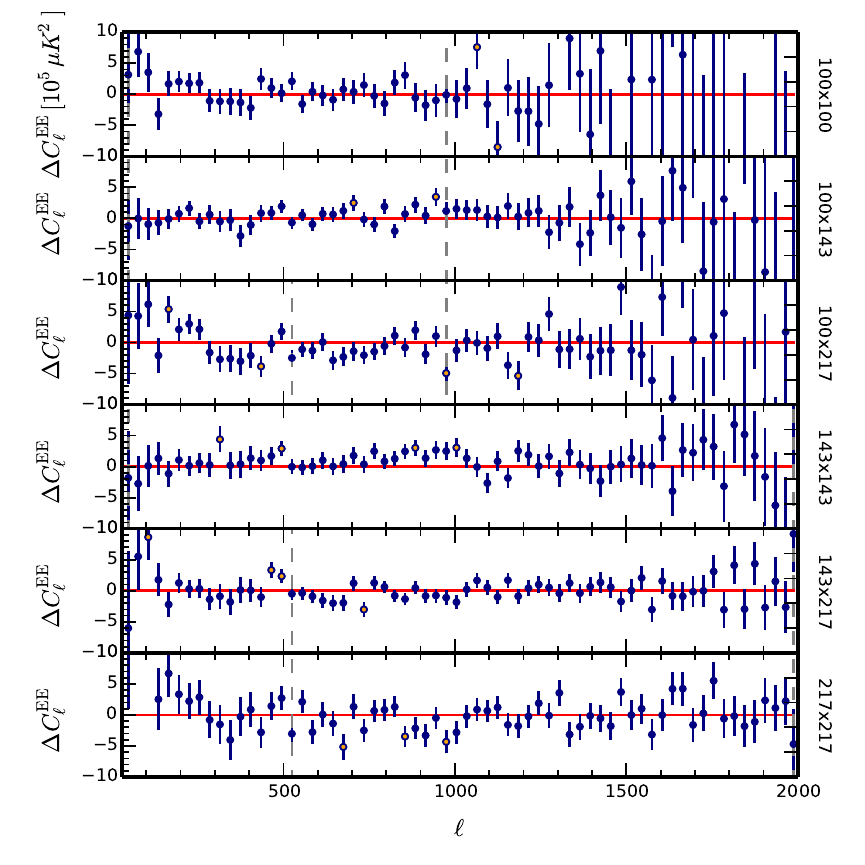}
\caption{Residual frequency power spectra after subtraction of the \plikTTtau\ best-fit model. We clean Galactic dust from the spectra from using the best-fit solution of \plikTTEETEtau. The residuals are relative to the baseline HM power spectra (blue points, except for those that deviate by at least 2 or 3\,$\sigma$,  which are shown in orange or red, respectively). The vertical dashed lines delimit the $\ell$ ranges retained in the likelihood.
\textit{Upper}: $\TE$ power spectra. 
\textit{Lower}: $\EE$ power spectra. }
\label{fig:respol}
\end{figure}

\begin{figure}
\centering
\includegraphics[width=0.498\textwidth]{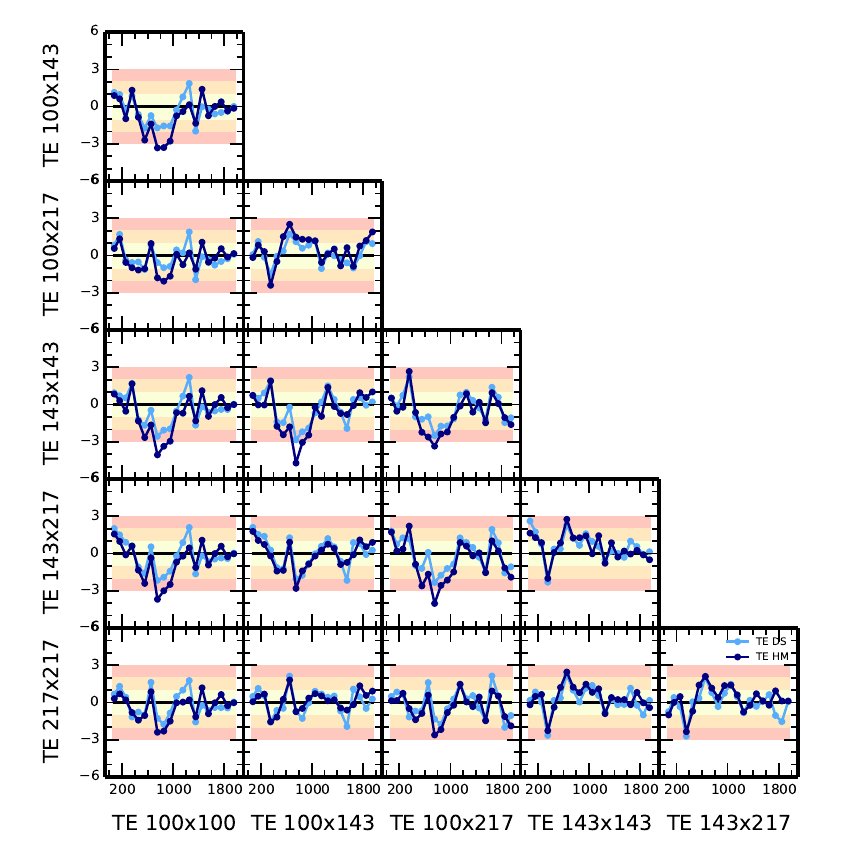}
\includegraphics[width=0.498\textwidth]{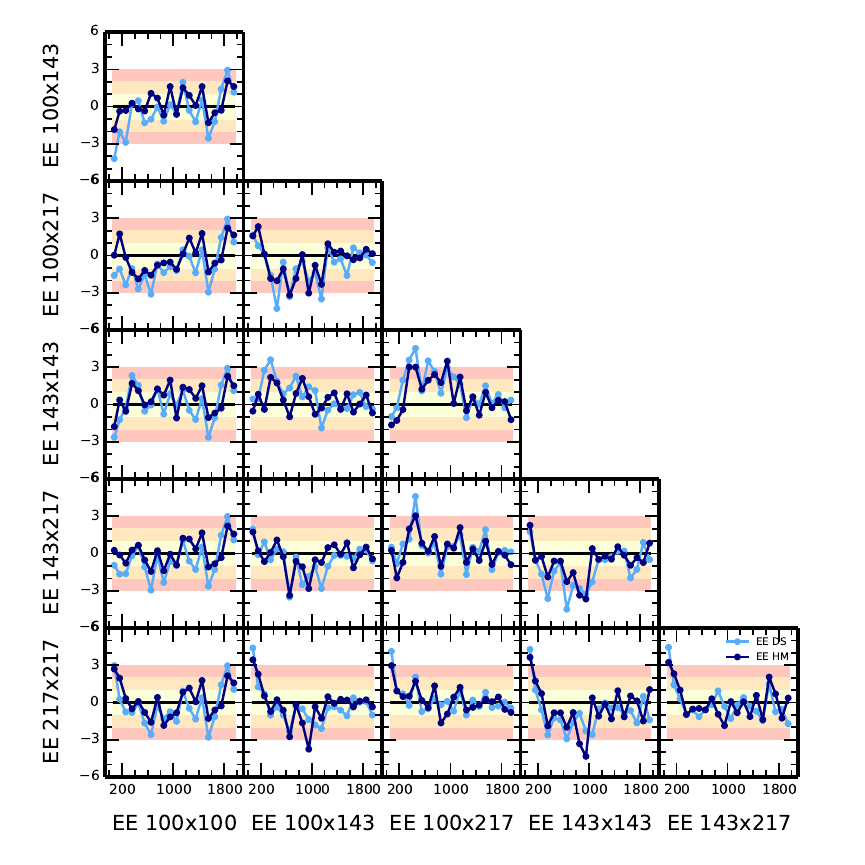}
\caption{Inter-frequency foreground-cleaned power-spectra differences.  Each  panel shows the difference of two frequency power spectra, that indicated on the left axis minus that on the bottom axis, after subtracting foregrounds using the best-fit \planckTT\  foreground solutions. Differences are shown for both the HM power spectra (dark blue) and the DS power spectra (light blue).}
\label{fig:respol-diff}
\end{figure}

Figure~\ref{fig:respol} shows the residuals for each frequency and Fig.~\ref{fig:respol-diff} shows the differences between frequencies of the $\TE$ and $\EE$ power spectra (the procedure is explained in Appendix~\ref{sec:freq}). The residuals are calculated with respect to the best-fit cosmology as preferred by \plikTTtau, although we use the best-fit solution of the \plikTTEETEtau\ run to subtract the polarized Galactic dust contribution. 


The binned inter-frequency residuals show  deviations at the level of a few $\muKsq$ from the best-fit model. These deviations do not necessarily correspond to high values of the $\chi^2$ calculated on the unbinned data (see Table~\ref{tab:highl:lrange}). This is because some of the deviations are relatively small for the unbinned data and correctly follow the expected $\chi^2$ distribution. However, if the deviations are biased (\eg have the same sign) in some $\ell$ range, they can result in larger deviations (and large $\chi^2$) after binning. Thus, the $\chi^2$ calculated on unbinned data is not always sufficient to identify these type of biases. We therefore also use a second quantity, $\chi$, defined as the weighted linear sum of residuals, to diagnose biased multipole regions or frequency spectra:
\begin{eqnarray}
\chi = \vec{w}^\tens{T}(\vec{\hat{C}}-\vec{C}) \quad \text{with}\quad \vec{w} = (\mathrm{diag}\,\tens{C})^{-1/2} ,
\end{eqnarray}
where $\vec{\hat{C}}$ is the unbinned vector of data in the multipole region or frequency spectrum of interest, $\vec{C}$)  is the corresponding model, and $\vec{w}$ is a vector of weights,   equal to the inverse standard deviation evaluated from the diagonal of the corresponding covariance matrix $\tens{C}$.
The $\chi$ statistic is distributed as a Gaussian with zero mean and standard deviation equal to
\begin{eqnarray}
\sigma_\chi=\sqrt{\vec{w}^\tens{T}\tens{C}\vec{w}} .
\end{eqnarray}
We then define the normalized $\chi_{\mathrm{norm}}$ as the $\chi$ in units of standard deviation, 
\begin{equation}
\chi_{\mathrm{norm}}=\chi/\sigma_\chi .
\label{chinorm}
\end{equation}
The $\chi_\mathrm{norm}$ values that we obtain for different frequency power spectra are given in Table~\ref{tab:highl:lrange}.

For $\EE$, the worst-behaved spectra from the $\chi_\mathrm{norm}$ point of view are $143\times143$ ($3.7\,\sigma$ deviation) and $100\times217$ ($-3.0\,\sigma$), while from the $\chi^2$ point of view, the worst is $100\times143$ (${\rm PTE}=3.9\,\%$).
For $\TE$, the worst  from the $\chi_\mathrm{norm}$ point of view are  $100\times217$  ($5\,\sigma$), $100\times100$ ($3.7\,\sigma$), and \ $143\times143$ ($-2.2\,\sigma$), while from the $\chi^2$ point of view the worst is $100\times100$ (${\rm PTE}=0.43\,\%$). The extreme deviations from the expected distributions show that the frequency spectra are not described very accurately by our data model. This is also clear from Fig.~\ref{fig:respol-diff}, which shows that there are differences of up to $5\,\sigma$ between pairs of foreground-cleaned spectra.

However, as the co-added residuals in Fig.~\ref{fig:resTTcmbzoom} show, systematic effects in the different frequency spectra appear to average out, leaving relatively small residuals with respect to the \plikTTtau\ best-fit cosmology. In other words, these effects appear not to be dominated by common modes between detector sets or across frequencies. This is also borne out by the good agreement between the data and the expected polarization power spectra conditioned on the temperature ones, as shown in the conditional plots of Fig.~\ref{fig:hil:condP}.

\subsubsection{$\TE$ and $\EE$ robustness tests}

For $\TE$ and $\EE$, we ran tests of robustness similar  to those applied earlier to $\TT$. These are presented in Appendix~\ref{sec:pol-robust}, and the main conclusions are the following. We find that the \plik cosmological results are affected by less than 1\,$\sigma$ when using detset cross-spectra instead of half-mission ones. This is also the case when we relax the dust amplitude priors, when we marginalize over beam uncertainties, or when we change $\lmin$ or $\lmax$. The alternative \camspec likelihood has larger shifts, but still smaller than $1\,\sigma$ in $\TE$ and $0.5\,\sigma$ in $\EE$. However, we also see larger shifts (more than $2\,\sigma$ in $\TE$) with \plik when some frequency channels are dropped; and, when they are varied, the beam leakage parameters adopt  much higher values than expected from the prior, while still leaving some small discrepancies between individual cross-spectra that have yet to be explained.

These results shows that our data model  leaves residual instrumental systematic errors and is not yet sufficient to take advantage of the full potential of the \HFI\ polarization information. Indeed, the current data model and likelihood code do not account satisfactorily for deviations at the $\muKsq$ level, even if they can be captured in part by our beam leakage modelling.  Nevertheless, the results for the \LCDM\ model obtained from the \plikTEtau\ and \plikEEtau\ runs are in good agreement with the results from \plikTTtau\ (see Appendix~\ref{agreementpol}). This agreement between temperature and polarization results within \LCDM\ is not a proof of the accuracy of the co-added polarization spectra and their data model, but rather a check of consistency at the $\muKsq$ level. This consistency is, of course, a very interesting result in itself. But this comparison of probes cannot yet be pushed further to check for the potential presence of a physical inconsistency within the base model that the data could in principle detect or constrain. 

\section{The full \Planck\ spectra and likelihoods} \label{sec:hal}

This section discusses the results that are obtained by using the full \Planck\ likelihood. Section~\ref{sec:hal-hybrid} first addresses the question of robustness with respect to the choice of the hybridization scale (the multipole at which we transition from the low-$\ell$ likelihood to the high-$\ell$ likelihood). Sects.~\ref{sec:hal-spectra} and \ref{sec:hal-params} then present the full results for the power spectra and the baseline cosmological parameters. \rev{Section~\ref{sub:systematics} summarizes the full systematic error budget.} Sect.~\ref{sec:hal-anomaly} concentrates on the significance of the possibly anomalous structure around $\ell \approx 20$ in this new release. We then introduce in Sect.~\ref{sec:planck_compressed} a useful compressed \Planck\ high-$\ell$ temperature and polarization CMB-only likelihood, \pliklite, which, when applicable, enables faster parameter exploration. Finally, in Sect.~\ref{sec:hal-other}, we compare the \Planck\ 2015 results with the previous results  from \WMAP, ACT, and SPT.

\subsection{Insensitivity to hybridization scale} \label{sec:hal-hybrid}

\begin{figure}[!htb] 
\centering
\includegraphics[width=0.9\columnwidth, clip=true, trim=1mm 4mm 93mm 9mm]{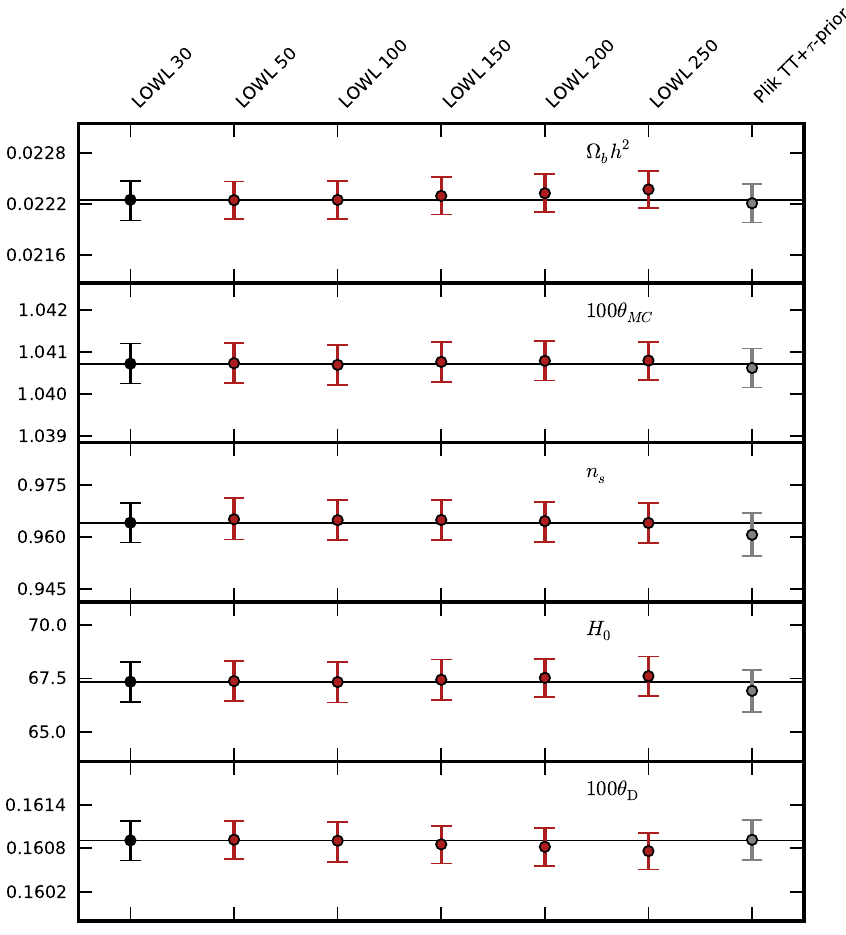}
\includegraphics[width=0.9\columnwidth, clip=true, trim=87mm 4mm 8mm 9mm]{figures_high_ell/wisker_base_lowlTT_w180mm.pdf}
\caption{Marginal mean and 68\,\% CL error bars on cosmological parameters estimated with different multipoles for the transition between the low-$\ell$  and the high-$\ell$ likelihood. Here we use only the $\TT$ power spectra and a Gaussian prior on the optical depth $\tau=0.07\pm 0.02$, within the base-$\Lambda$CDM model. ``{\plikTTtau}'' refers to the case where we use the \plik\ high-$\ell$ likelihood only.}
\label{fig:wiskerTTlowl}
\end{figure}
Before we use the low-$\ell$ and high-$\ell$ likelihoods together, we address the question of the hybridization scale, 
$\lhyb$, at which we switch from one to the other (neglecting correlations between the two regimes, as we did and checked in \citetalias{planck2013-p08}). 
To that end, we focus on the $\TT$ case and use a likelihood based on the Blackwell-Rao estimator and the \commander\ algorithm \citep{chu2005,rudjord2009} as described in Sect.~\ref{sec:commander_lowl}, since this likelihood can be used to much higher $\lmax$ than  the full pixel-based $T,E,B$ one. 
For this test without polarization data, we assume the same $\tau=0.07\pm0.02$ prior as before.

The whisker plot of Fig.~\ref{fig:wiskerTTlowl} shows the marginal mean and the 68\,\% CL\ error bars for base-\LCDM\ cosmological parameters when $\lhyb$ is varied from the baseline value of 30 (case ``LOWL 30'') to $\lhyb =$50, 100, 150, 200, and 250, and compared to the \plikTTtau\ case. The difference between the ``LOWL 30''  and ``\plikTTtau''  values shows the effect of the low-$\ell$ dip at $\ell\approx 20$, which reaches $0.5\,\sigma$ on $\ns$. The plot shows that the effect of varying $\lhyb$ from 30 to 150 is a shift in $\ns$ by less than 0.1\,$\sigma$. This is the result of the Gaussian approximation pushed to $\lmin =30$, already discussed in the simulation section (Sect.~\ref{sec:valid-sims}). It would have been much too slow to run the full low-$\ell$ $TEB$ likelihood with $\lmax$ substantially greater than 30, and we decided against the only other option, to leave a gap in polarization between $\ell=30$ and the hybridization scale chosen in $\TT$.

\subsection{The \Planck\ 2015 CMB spectra} \label{sec:hal-spectra}

The visual appearance of \Planck\ 2015 CMB co-added spectra in $\TT$, $\TE$, and $\EE$ can be seen in Fig.~\ref{fig:Planck_Cl_TT}. Goodness-of-fit values can be found in Table~\ref{tab:goodness} of Appendix~\ref{app:hal}. These differ somewhat from those given previously in Table~\ref{tab:highl:lrange} for \plik alone, because the inclusion of low $\ell$ in temperature brings in the $\ell\approx 20$ feature (see Sect.~\ref{sec:hal-anomaly}). Still, they remain acceptable, with PTEs all above 10\,\% (16.8\,\% for $\TT$). 

With this release, \Planck\ now detects 36 extrema in total, consisting of 19 peaks and 17 troughs. Numerical values for the positions and amplitudes of these extrema may be found in Table~\ref{table_peaks_and_troughs} of Appendix~\ref{ssub:peaks}, which also provides details of the steps taken to derive them. We provide in Appendix~\ref{ssub:TE_corr} an alternate display of the correlation between temperature and ($E$-mode) polarization by showing their Pearson correlation coefficient and their decorrelation angle versus scale (Figs.~\ref{fig:Pearson} and \ref{fig:theta}).

\subsection{\Planck\ 2015 model parameters} \label{sec:hal-params}

\begin{table}[ht!] 
\caption{Constraints on the basic six-parameter $\Lambda$CDM model using \Planck\ angular power spectra.$^{\rm a}$}
\label{tab:params-ref} 
\begingroup
\nointerlineskip
\vskip -6mm
\setbox\tablebox=\vbox{
    \newdimen\digitwidth
    \setbox0=\hbox{\rm 0}
    \digitwidth=\wd0
    \catcode`*=\active
    \def*{\kern\digitwidth}
    \newdimen\signwidth
    \setbox0=\hbox{+}
    \signwidth=\wd0
    \catcode`!=\active
    \def!{\kern\signwidth}
\halign{\hbox to 1.0in{$#$\leaderfil}\tabskip=1.2em&
   \hfil$#$\hfil\tabskip=1.6em&
   \hfil$#$\hfil\tabskip=0pt\cr
\noalign{\doubleline}
\multispan1\hfil \hfil&\multispan1\hfil \planckTT\hfil&\multispan1\hfil \planckall\hfil\cr
\omit\hfil Parameter\hfil&\omit\hfil 68\,\% limits\hfil&\omit\hfil 68\,\% limits\hfil\cr
\noalign{\vskip 3pt\hrule\vskip 5pt}
\Omega_{\mathrm{b}} h^2&           0.02222\pm 0.00023&0.02225\pm 0.00016\cr
\Omega_{\mathrm{c}} h^2&           0.1197\pm 0.0022&0.1198\pm 0.0015\cr
100\theta_{\mathrm{MC}}&           1.04085\pm 0.00047&1.04077\pm 0.00032\cr
\tau&                              0.078\pm 0.019&0.079\pm 0.017\cr
\ln(10^{10} A_\mathrm{s})&         3.089\pm 0.036&3.094\pm 0.034\cr
n_\mathrm{s}&                      0.9655\pm 0.0062&0.9645\pm 0.0049\cr
\noalign{\vskip 8pt}
H_0&                               67.31\pm 0.96*&67.27\pm 0.66*\cr
\Omega_\Lambda&                    0.685\pm 0.013&0.6844\pm 0.0091\cr
\Omega_{\mathrm{m}}&               0.315\pm 0.013&0.3156\pm 0.0091\cr
\Omega_{\mathrm{m}} h^2&           0.1426\pm 0.0020&0.1427\pm 0.0014\cr
\Omega_{\mathrm{m}} h^3&           0.09597\pm 0.00045&0.09601\pm 0.00029\cr
\sigma_8&                          0.829\pm 0.014&0.831\pm 0.013\cr
\sigma_8\Omega_{\mathrm{m}}^{0.5}& 0.466\pm 0.013&0.4668\pm 0.0098\cr
\sigma_8\Omega_{\mathrm{m}}^{0.25}&0.621\pm 0.013&0.623\pm 0.011\cr
z_{\mathrm{re}}&                   9.9^{+1.8}_{-1.6}&10.0^{+1.7}_{-1.5}\cr
10^9 A_{\mathrm{s}}&               2.198^{+0.076}_{-0.085}&2.207\pm 0.074\cr
10^9 A_{\mathrm{s}} e^{-2\tau}&    1.880\pm 0.014&1.882\pm 0.012\cr
\mathrm{Age}/\mathrm{Gyr}&         13.813\pm 0.038*&13.813\pm 0.026*\cr
z_\ast&                            1090.09\pm 0.42***&1090.06\pm 0.30***\cr
r_\ast&                            144.61\pm 0.49**&144.57\pm 0.32**\cr
100\theta_\ast&                    1.04105\pm 0.00046&1.04096\pm 0.00032\cr
z_{\mathrm{drag}}&                 1059.57\pm 0.46***&1059.65\pm 0.31***\cr
r_{\mathrm{drag}}&                 147.33\pm 0.49**&147.27\pm 0.31**\cr
k_{\mathrm{D}}&                    0.14050\pm 0.00052&0.14059\pm 0.00032\cr
z_{\mathrm{eq}}&                   3393\pm 49**&3395\pm 33**\cr
k_{\rm{eq}}&                       0.01035\pm 0.00015&0.01036\pm 0.00010\cr
100\theta_{\rm{s,eq}}&             0.4502\pm 0.0047&0.4499\pm 0.0032\cr
\noalign{\vskip 8pt}
f_{2000}^{143}&                    29.9\pm 2.9*&29.5\pm 2.7*\cr
\noalign{\vskip 4pt}
f_{2000}^{143\times217}&           32.4\pm 2.1*&32.2\pm 1.9*\cr
\noalign{\vskip 4pt}
f_{2000}^{217}&                    106.0\pm 2.0**&105.8\pm 1.9**\cr
\noalign{\vskip 5pt\hrule\vskip 3pt}}}
\endPlancktable
\tablenote {{\rm a}} The top group contains constraints on the six primary parameters included directly in the estimation process.  The middle group contains constraints on derived parameters. The last group gives a measure of the total foreground amplitude (in $\muK^2$) at $\ell=2000$ for the three high-$\ell$ temperature spectra used by the likelihood. These results were obtained using the \CAMB\ code, and are identical to the ones reported in Table~3 in \citet{planck2014-a15}.\par
\endgroup
\end{table}

\begin{figure*}[htbp] 
\centering
\includegraphics[width=0.95\textwidth]{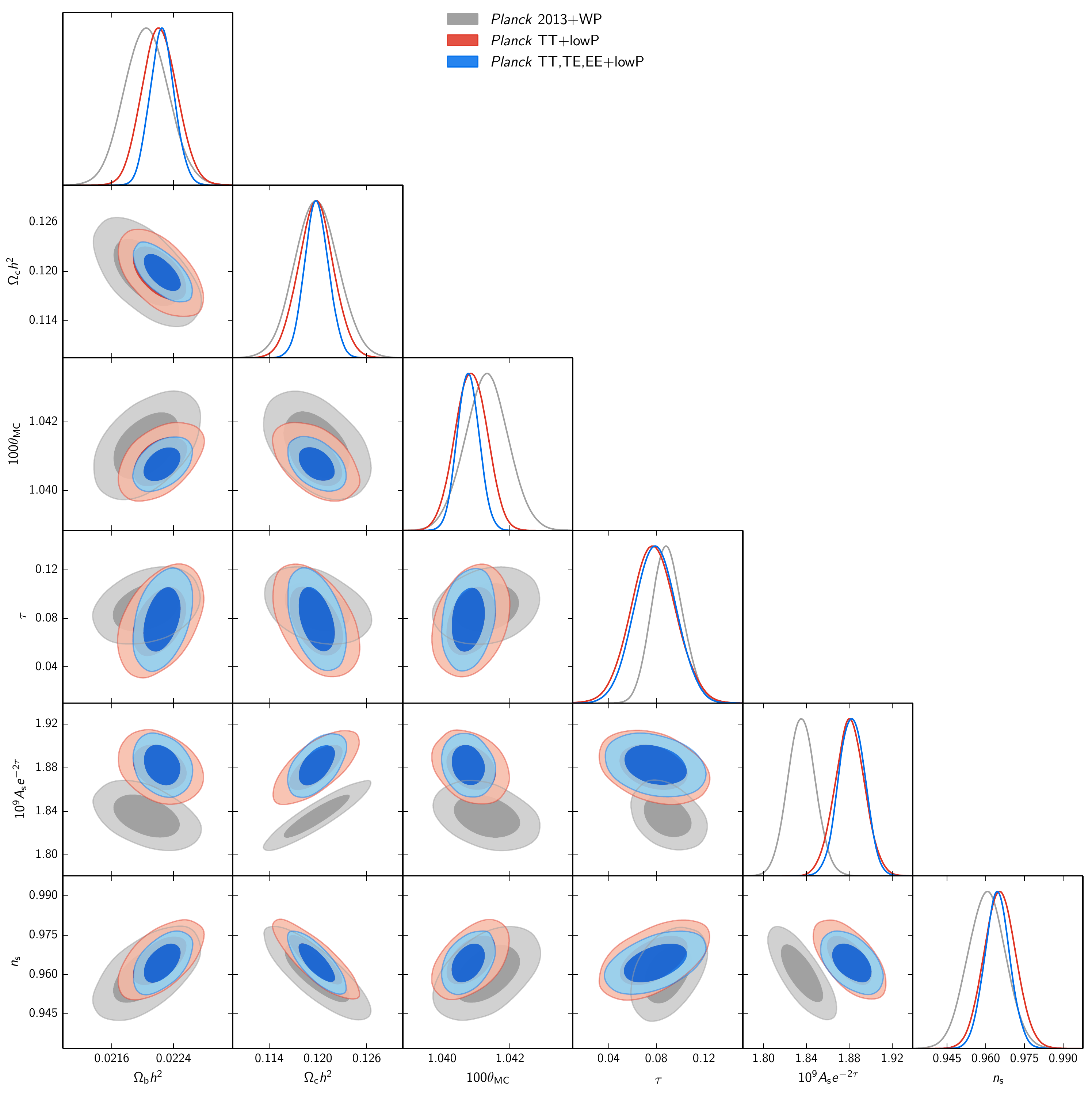}
\caption{\LCDM\ parameter constraints. The grey contours show the 2013 constraints, which can be compared with the current ones, using either $TT$ only at high $\ell$ (red) or the full likelihood (blue). Apart from further tightening, the main difference is in  the amplitude, $A_\textrm{s}$, due to the overall calibration shift.}
\label{fig:params_triangle}
\end{figure*}

Figure~\ref{fig:params_triangle} compares constraints on pairs of parameters as well as their individual marginals for the base-\LCDM\ model. The grey contours and lines correspond to the results of the 2013 release \citepalias{planck2013-p08}, which was based on $\TT$ and \WMAP polarization at low $\ell$ (denoted by WP), using only the data from the nominal mission. The blue contours and lines are derived from the 2015 baseline likelihood, \shortTT\ (``\planckTT'' in the plot), while the red contours and line are obtained from the full \shortall\ likelihood (``\planckall'' in the plot, see Appendix~\ref{sec:ALL-robust} for the relevant robustness tests). In most cases the 2015 constraints are in quite good agreement with the earlier constraints, with the exception of the normalization $\As$, which is higher by about 2\%, reflecting the 2015 correction of the \planck\ calibration which was indeed revised upward by about 2\% in power. The figure also illustrates the consistency and further tightening of the parameter constraints brought by adding the $E$-mode polarization at high $\ell$.  The numerical values of the \Planck\ 2015 cosmological parameters for base \LCDM\ are given in Table~\ref{tab:params-ref}.

\begin{figure*}[htbp] 
\centering
\includegraphics[height=0.55\textheight, clip=true, trim=5mm 18mm 4mm 11mm]{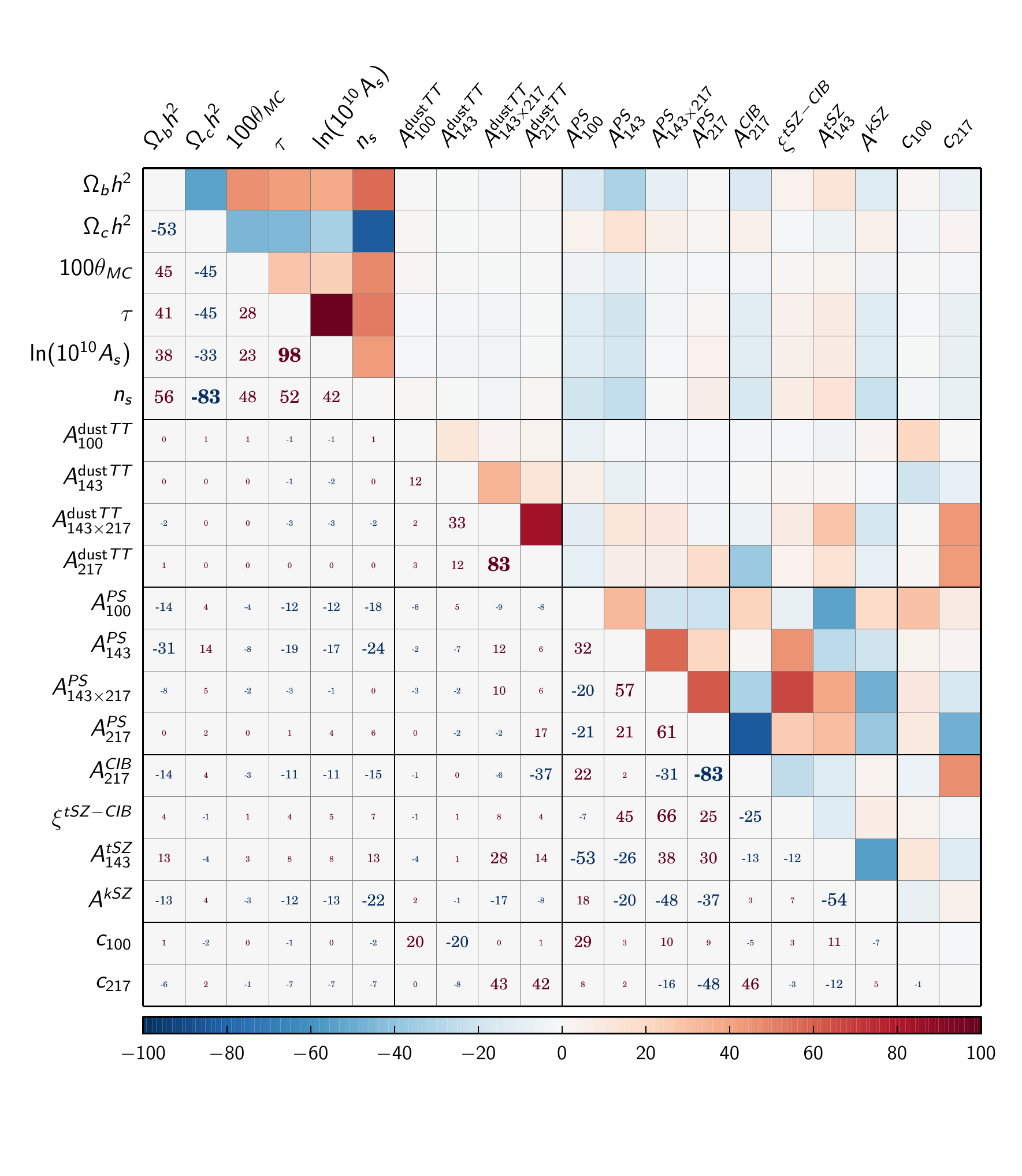}
\includegraphics[width=0.495\textwidth, clip=true, trim=2mm 9mm 1mm 7mm]{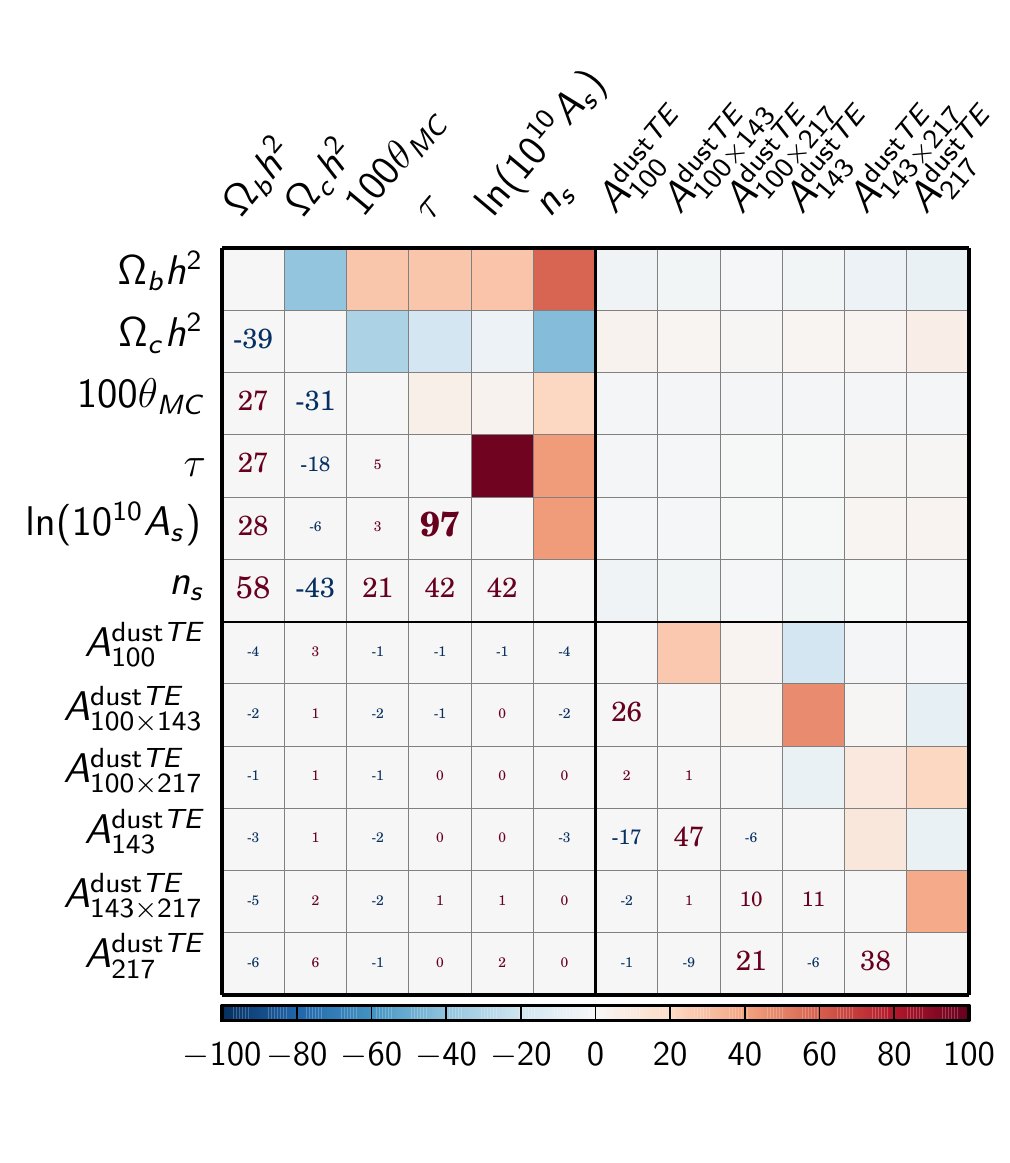} 
\includegraphics[width=0.495\textwidth, clip=true, trim=2mm 9mm 1mm 7mm]{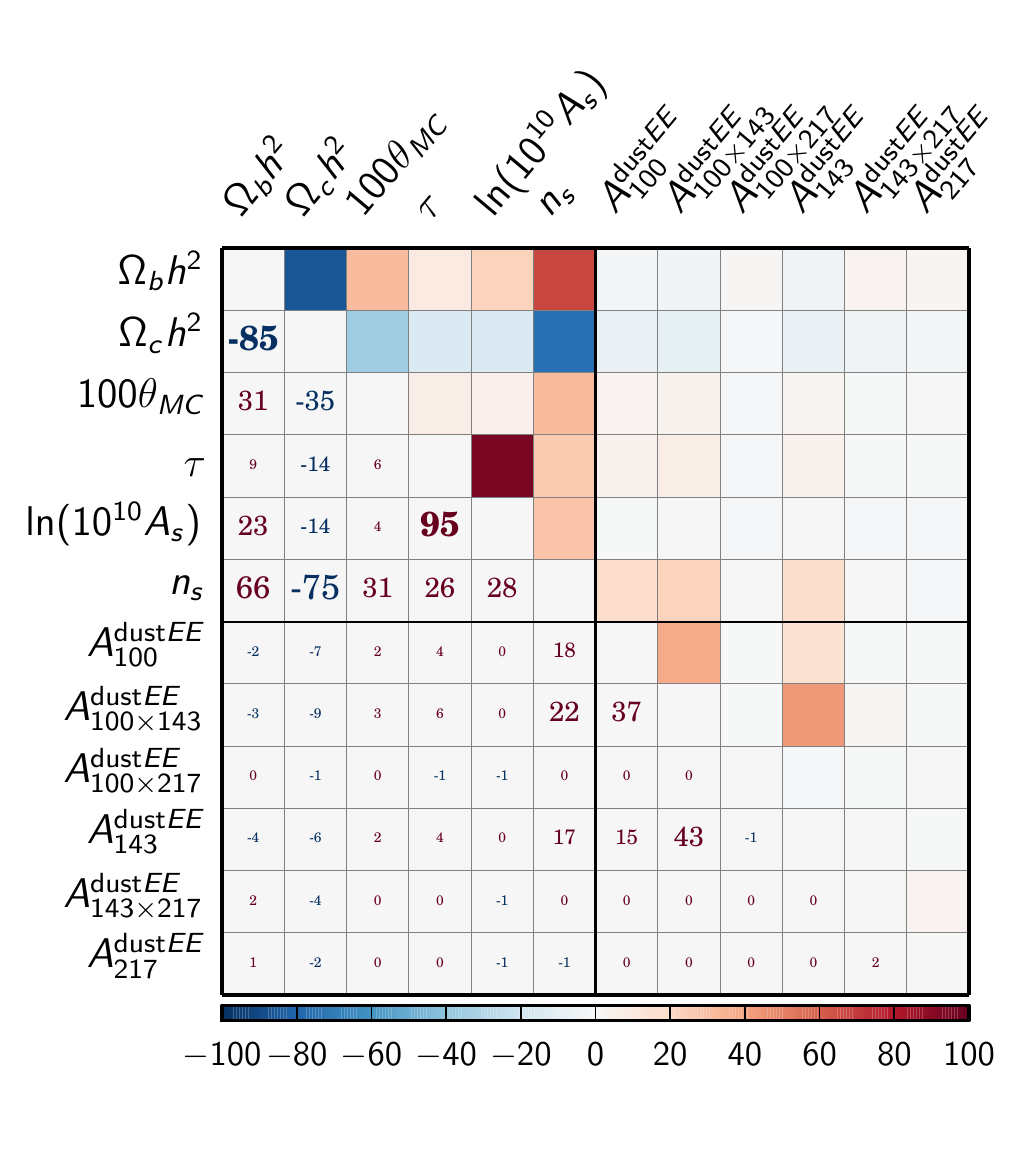}

\caption{Parameter correlations for \shortTT\ (top), \shortTE\  ({bottom left}), and \shortEE\  ({bottom right}). The degeneracies between foreground and calibration parameters do not strongly affect the determination of the cosmological parameters. In these figures the lower triangle gives the numerical values of the correlations in percent (with values  below $10\,\%$ printed at the smallest size), while the upper triangle represents the same values using a colour scale.}
\label{fig:params_correl}
\end{figure*}

As shown in Fig.~\ref{fig:params_correl}, the degeneracies between foreground and calibration parameters generally do not affect the determination of the cosmological parameters. In the \shortTT\ case (top panel), the dust amplitudes appear to be nearly uncorrelated with the basic \lcdm\  parameters. Similarly, the 100 and 217\ghz\ channel calibration is only relevant for the level of foreground emission. Cosmological parameters are, however, mildly correlated with the point-source and kinetic SZ amplitudes. Correlations are strongest (up to $30\,\%$)  for the baryon density ($\Omega_{\rm b} h^2$) and spectral index ($\ns$). We do not show correlations with the \Planck\ calibration parameter ($\calibM_{\rm P}$), which is uncorrelated with all the other parameters except the amplitude of scalar fluctuations ($\As$). The bottom panel shows the correlation for the \shortTE\ and \shortEE\ cases, which do not affect the cosmological parameters, except for $20\,\%$ correlations in $\EE$ between the spectral index ($\ns$) and the dust contamination amplitude in the 100 and 143\ghz\ maps. 

\begin{figure*}[htbp] 
\centering
\includegraphics[width=1.32\columnwidth, clip=true, trim=0mm 9mm 3mm 2mm]{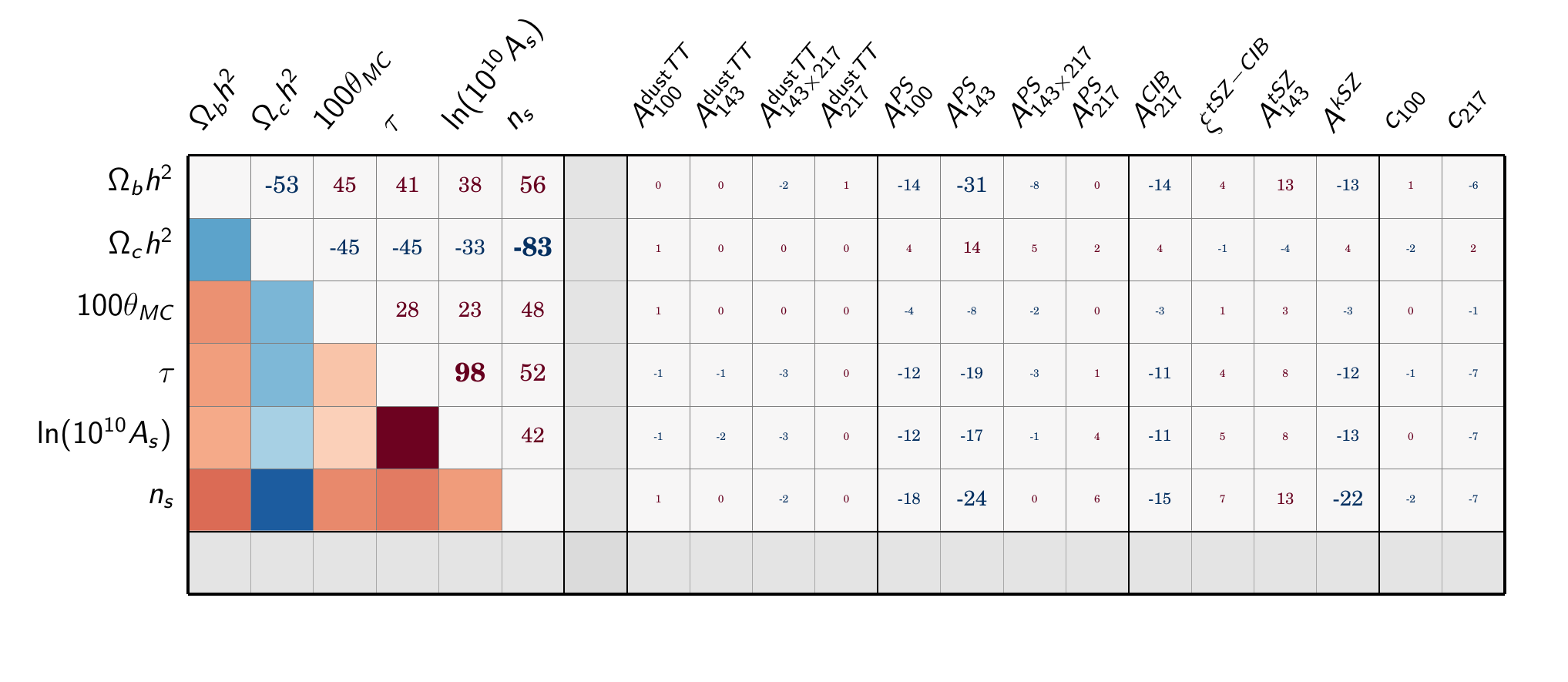}
\includegraphics[width=1.32\columnwidth, clip=true, trim=0mm 9mm 3mm 2mm]{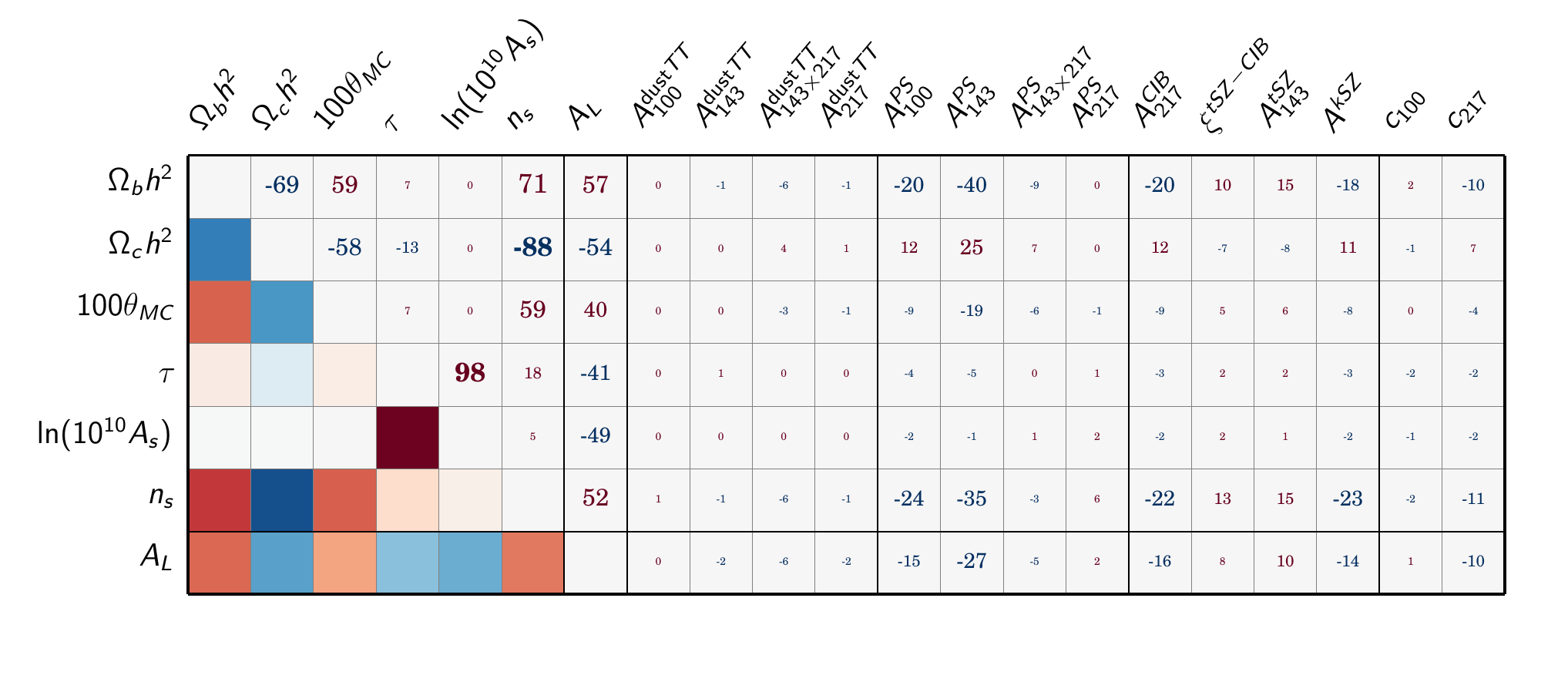}
\includegraphics[width=1.32\columnwidth, clip=true, trim=0mm 9mm 3mm 2mm]{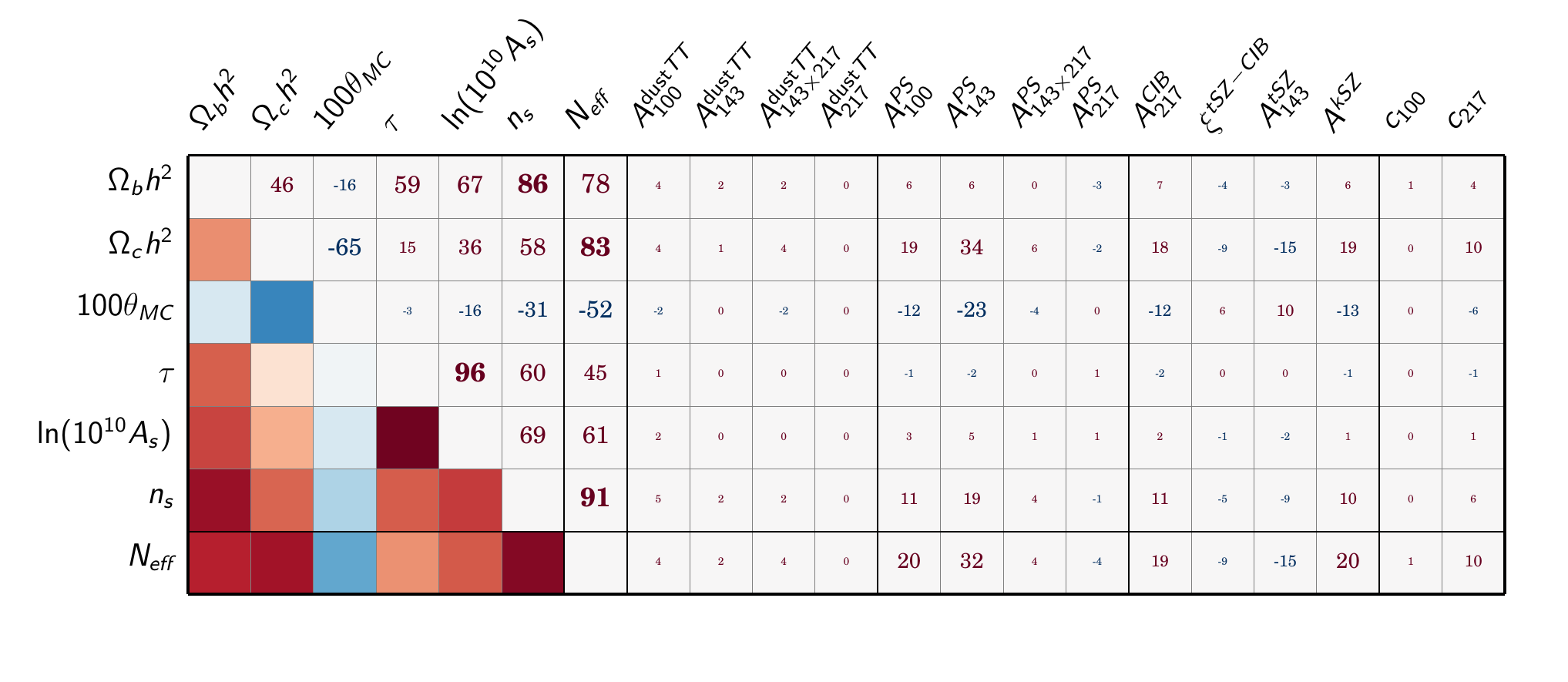}
\includegraphics[width=1.32\columnwidth, clip=true, trim=0mm 9mm 3mm 2mm]{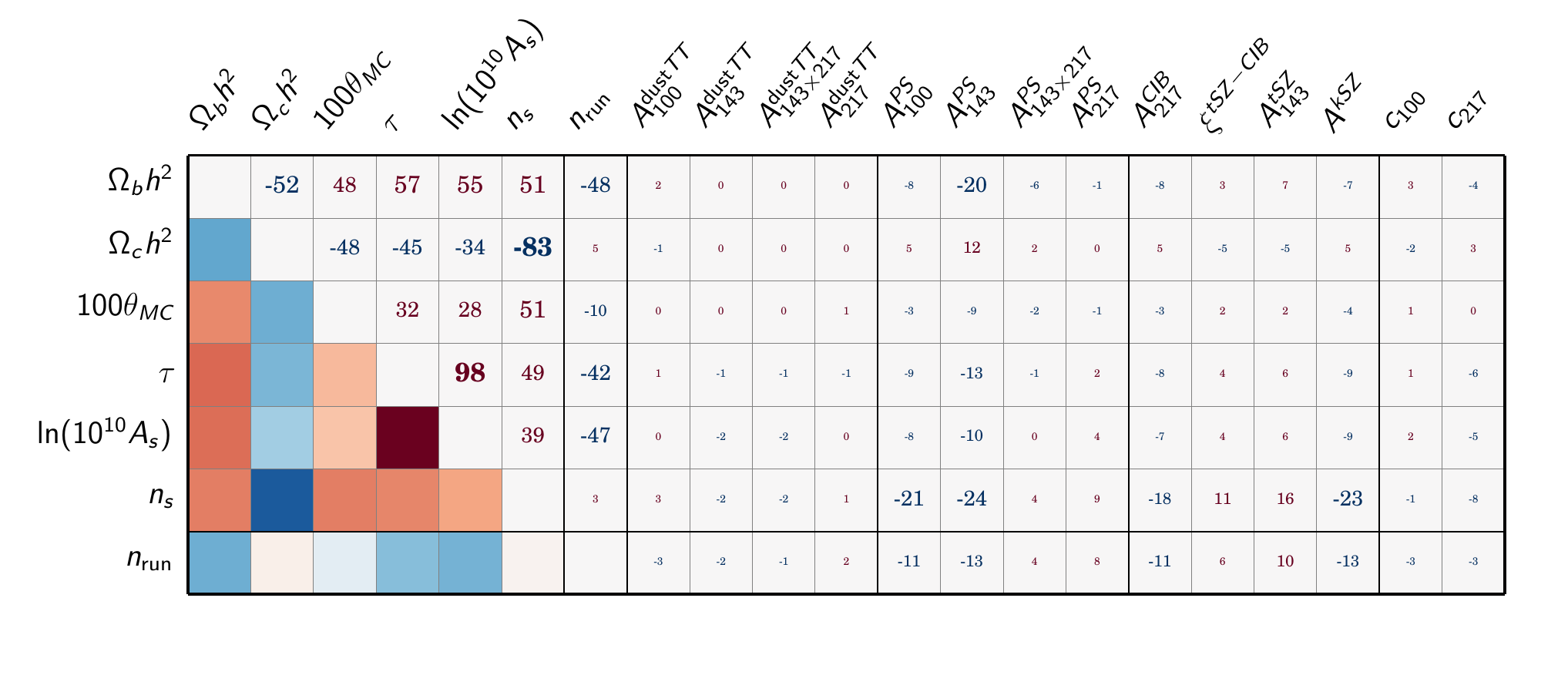}
\includegraphics[width=1.32\columnwidth, clip=true, trim=0mm 9mm 3mm 2mm]{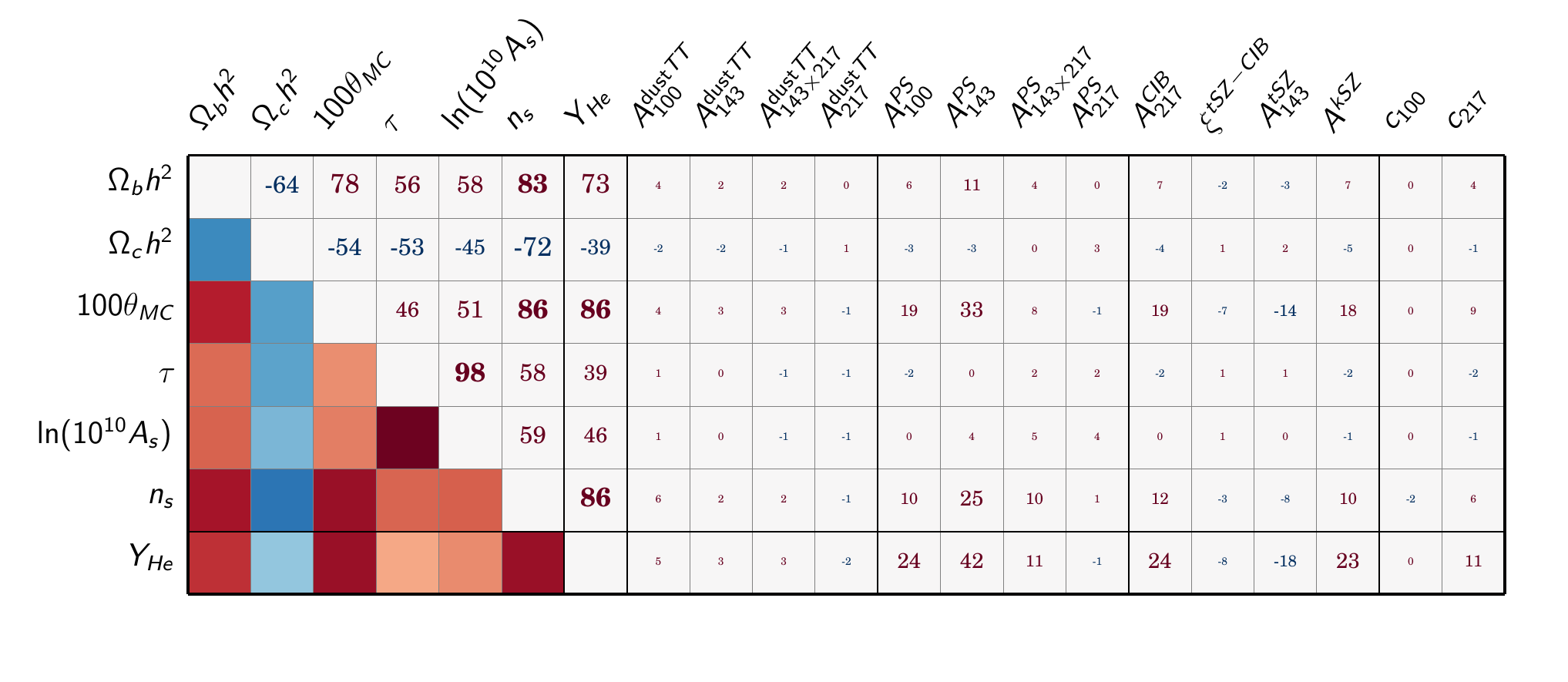}
\caption{Parameter correlations for \shortTT, including some \lcdm\ extensions. 
The leftmost column is identical to Fig.~\ref{fig:params_correl} and is repeated here to ease comparison. Including extensions to the \lcdm\ model changes the correlations between the cosmological parameters, sometimes dramatically, as can be seen in the case of $A_{\rm L}$. 
There is no correlation between the cosmological parameters (including the extensions) and the dust amplitude parameters. In most cases, the extensions are correlated with the remaining foreground parameters (and in particular with the point-source amplitudes at 100 and 143\ghz, and with the level of CIB fluctuations) with a strength similar to those of the other cosmological parameters (\ie less than $30\,\%$). $Y_{\rm He}$ exhibits a stronger sensitivity to the point-source levels.}
\label{fig:params_correl_T_ext}
\end{figure*}

We also display in Fig.~\ref{fig:params_correl_T_ext} the correlations between the foreground parameters and the cosmological parameters in the  \shortTT\ case when exploring classical extensions to the \lcdm\ model. While $n_{\rm run}$ seems reasonably insensitive to the foreground parameters, some extensions do exhibit a noticeable correlation, up to $40\%$ in the case of $Y_{\rm He}$ and the point-source level at 143\ghz.

Finally, we note that power spectra and parameters derived from CMB maps obtained by the component-separation methods described in \citet{planck2014-a11} are generally consistent with those obtained here, at least when restricted to the $\ell < 2000$ range in $\TT$; this is detailed in Sect.~\ref{sec:mapCheck}.

\subsection{\rev{Overall systematic error budget assessment}} 
\label{sub:systematics}

\rev{The tests presented throughout this paper and its appendices documented our numerous tests of the \planck\ likelihood code and its outputs. Here, we summarize those results and attempt to isolate the dominant sources of systematic uncertainty. This assessment is of course a difficult task. Indeed, all known systematics are normally corrected for, and when relevant, the uncertainty on the correction is included in the error budget and thus in the error bar we report. In that sense, except for a very few cases where we decided to leave a known uncertainty in the data, this section tries to deal with the more difficult task of evaluating the unknown uncertainty!}

\rev{This section summarizes the contribution of the known systematic uncertainties along with these potential unknown unknowns, specifically highlighting both internal consistency tests based on comparing subsets of the data, along with those using end-to-end instrumental simulations.}

\subsubsection{Low-$\ell$ budget}

\rev{The low-$\ell$ likelihood has been validated using both internal consistency tests and simulation-based, tests. Here we summarize only the main result of the analysis, which has been set forth in Sect.~\ref{sec:low-ell} above.} 

\rev{A powerful consistency test of the polarization data, described in Sect.~\ref{sec:lowl_pse_params}, is derived by rotating some of the likelihood components by $\pi/4$.  Specifically, the rotation is applied to the data maps and only to the noise covariance matrix (the likelihood being a scalar function, applying the same rotation to the signal matrix as well would be equivalent to not performing the rotation). The net effect is a conversion of $E \rightarrow -B$ and $B \rightarrow E$ for the signal, but leaving unaffected the Gaussian noise described in the covariance matrix. Under these circumstances, we do not expect to pick up any reionization signal, since it would then be present in $BB$ or $TB$: the operation should result in a null $\tau$ detection. This is precisely what happens (see the blue dashed curve in Fig.~\ref{fig:lowl-rotparams}). It is of course possible --- though unlikely --- that systematics are only showing up in the $E$ channel, leaving $B$ modes unaffected. Indeed, this possibility is further challenged by the fact that we do not detect anomalies in any of the six polarized power spectra; as detailed in Fig.~\ref{fig:lowl_harmchi2}, they are consistent with a $\Lambda$CDM signal and noise as described by the final 70\,GHz covariance matrix.} 

\rev{These tests are specific to \Planck\ and aimed at validating the internal consistency of the datasets employed to build the likelihood. As a further measure of consistency, we have carried out a null test employing the \WMAP\ data, detailed in Sect.~\ref{Planck_wmap_lowP}. In brief, we have taken \WMAP's K$_\mathrm{a}$, Q and V channels and cleaned them from any polarized foreground contributions using a technique analogous to the one used to clean the LFI 70\,GHz maps, employing the \Planck\ 353\,GHz map to minimize any dust contribution, but relying on \WMAP's K channel to remove any synchrotron contribution. The resulting LFI 70\,GHz and \WMAP\ maps separately lead to compatible $\tau$ detections; their half-difference noise estimates are compatible with their combined noise and do not exhibit a reionization signal, as shown in Fig.~\ref{fig:tau_WMAP_LFI}.}  

\rev{We learn from these tests that if the $EE$ and $TE$ signal we measure at 70\,GHz is due to systematics, then these systematics should affect only the above spectra in such a way to mimic a genuine reionization signal, and one that is fully compatible in the maps with that present in (cleaned) \WMAP data. This is extremely unlikely and conclude that \Planck\ 70\,GHz is dominated by a genuine contribution from the sky, compatible with a signal from cosmic reionization.}  

\rev{The tests described so far do not let us accurately quantify  the magnitude of a possible systematic contribution, nor to exclude artefacts arising from the data pipeline itself and, specifically, from the foreground cleaning procedure. These can be only controlled through detailed end-to-end tests, using the FFP8 simulations \citep{planck2014-a14}. As detailed in Sect.~\ref{sec:lowl_consistency}, we have performed end-to-end validation with 1000 simulated frequency maps containing signal, noise, and foreground contributions as well as specific systematics effects, mimicking all the steps in the actual data pipeline. Propagating to cosmological parameters ($\tau$ and $A_\textrm{s}$, which are most relevant at low $\ell$ in the $\Lambda$CDM model), we detect no bias within the simulation error budget. The total impact of any unknown systematics on the final $\tau$ estimate is at most $0.1\,\sigma$. This effectively rules out any detectable systematic contribution from the data pipeline or or from the  instrumental effects considered in the FFP8 simulations. A complementary analysis has been performed in \citet{planck2014-a04}, including further systematic contributions not incorporated into FFP8. This study, which should be taken as a worst-case scenario, limits the possible contribution to final $\tau$ of all known systematics at $0.005$, \ie about 0.25\,$\sigma$. We conclude that we were unable to detect any systematic contribution to the 2015 \Planck\ $\tau$ measurement as driven by low $\ell$, and have limited it to well within our final statistical error budget.}
 
 \rev{Finally, since the submission of this paper, dedicated work on \HFI\ data at low $\ell$ leads to a higher-precision determination of $\tau$ \citep{planck-pip121} which is consistent with the one described in this paper. This latest work paves the way towards a future release of improved \Planck\ likelihoods.   
}
\subsubsection{High-$\ell$ budget}
\rev{We now turn to the high-$\ell$ likelihood. The approximate statistical model from which we build the likelihood function may turn out to be an unfaithful representation of the data for three main reasons. First the equations describing the likelihood or the parameters of those equations can be inaccurate. They are, of course, since we are relying on approximations, but we expect that in the regime where they are used our approximations are good enough not to bias the best fit or strongly alter the estimation of error bars. We call such errors due to a breakdown of the approximations a {``methodological systematic''}. We may also lump into this any coding errors. Second, our data model must include a faithful description of the relation between the sky and the data analysed, \ie one needs to describe the transfer function and/or additive biases due to the non-ideal instrument and data processing. Again, an error in this model or in its parameters translates into possible errors that we call  {``instrumental systematics''}. Finally, to recover the properties of the CMB, the contribution of astrophysical foregrounds must be correctly modelled and accounted for. Errors in this model or its parameters is denoted {``astrophysical systematics''}. When propagating each of these systematics to cosmological parameters, this is always within the framework of the \LCDM\ model, as systematic effects can project differently into parameters depending on the details of the model.}

\rev{We investigated the possibility of {methodological systematics} with massive Monte-Carlo simulations. One of the main technical difficulties of the high-$\ell$ likelihood is the computation of the covariance of the band powers. Appendix~\ref{app:hil_cov_mat_validation} describes how we validated the covariance matrix, through the use of Monte-Carlo simulations, to better than a percent accuracy. This includes a first-order correction for the excess scatter due to point source masks, which can induce a systematic error in the covariance reaching a maximum of around $10\,\%$ near the first peak and  the largest scales ($\ell<50$), somewhat lower (about $5\,\%$ or less) at other scales. In Section~\ref{sec:valid-sims} we propagated the effect of those possible methodological systematics to the cosmological parameters and found a $0.1\,\sigma$ systematic shift on $n_{\mathrm{s}}$, when using the temperature data, which decreases when cutting the largest and most non-Gaussian modes. This is further demonstrated on the data in Sect.~\ref{sec:hal-hybrid} where we vary the hybridization scale. At this stage it is unclear whether this is a sign of a breakdown of the Gaussian approximation at those scales, or if it is the result of the limitations of our point source correction to the covariance matrix. We did not try to correct for this bias in the likelihood and we assess this $0.1\sigma$ effect on $\ns$ to be the main contribution to the methodological systematics error budget.}

\rev{{Instrumental systematics} are mainly assessed in three ways. First, given a foreground model, we estimate the consistency between frequencies and between the $\TT$, $\EE$ and $\TE$ combinations at the spectrum and at the parameter level (removing some cross-spectra). For $\TT$, the agreement is excellent, with shifts between parameters that are always compatible with the extra cosmic variance due to the removal of data when compared to the baseline solution (see Figs.~\ref{fig:nudiffTT} and \ref{fig:wiskerTTlowl}). $\TE$ and $\EE$ inter-frequency tests reveal discrepancies between the different cross spectra that we assigned to leakage from temperature to polarization (see Fig.~\ref{fig:respol}). In co-added spectra, these discrepancies tend to average out, leaving  a few-$\mu\mathrm{K}^2$-level residual in the difference between the co-added $TE$ and $EE$ spectra and their theoretical predictions based on the $TT$ parameters. Section~\ref{sec:beam_leakage} describes an effective model that succeeds in capturing some of that mismatch, in particular in $\TE$. But as argued in Sect.~\ref{sec:beam_leakage} and Appendix~\ref{sec:pol-robust} one cannot, at this stage, use this model as-is to correct for the leakage, or to infer the level of systematic it may induce on cosmological parameters, due to a lack of a good prior on the leakage model parameters. However, cosmological parameters deduced from the current polarization likelihoods are in perfect agreement with those calculated from the temperature, within the uncertainty allowed by our covariance. The second way we assess possible instrumental systematics is by comparing the detset (DS) and the half-mission (HM) results. As argued in Sect.~\ref{sec:noise_model}, the DS cross spectra are known to be affected by a systematic noise correlation that we correct for. Ignoring any uncertainty in this correction (which is difficult to assess), the overall shift between the HM- and DS-based parameters is of the order of $0.2\,\sigma$ (on $\omb$) at most on $\TT$ (Sect.~\ref{sec:robust-detsets} and Fig.~\ref{fig:wiskerTT}), similar in $\TE$ and slightly worse in $\EE$, particularily for $\ns$. Since the uncertainty on the correlated noise correction is not propagated, those shifts are only {upper bounds} on possible instrumental systematics (at least those which would manifest differently in these two data cuts which are completely different as regards temporal systematics). Finally, in Sect.~\ref{sec:e2e}, we evaluate the propagation of all known instrumental effects to parameters. Due to the cost of the required massive end-to-end simulations, this test can only reveal large deviations; no such instrumental systematic bias is detected in this test. To summarize, our instrumental systematics budget is at most $0.2\sigma$ in temperature, slightly higher in $\EE$, and there is no sign of bias due to temperature-to-polarization leakage that would not be compatible with our covariance (within the \LCDM\ framework).}

\rev{Finally, we assess the contribution of {astrophysical systematics}. Given the prior findings on polarisation, we only discuss the case of temperature here. The uncertainty on the faithfulness of the astrophysical model is relatively high, and we know from the DS/HM comparison that our astrophysical components certainly absorb part of the correlated noise that is not entirely captured by our model. In that sense, the recovered astrophysical parameters may be a biased estimate of the real astrophysical foreground contribution (due to the flexibility of the model which may absorb residual instrumental systematics provided they are sufficiently small). At small scales, the dominant astrophysical component is the point source Poisson term. We checked in Sect.~\ref{sec:astro} that the recovered point source contributions are in general agreement with models of their expected level. This is much less the case at 100\,GHz and we argued in Sect.~\ref{sec:astro} that, nonetheless, an error in the description of the Poisson term is unlikely to translate into a bias in the cosmological parameters, as the point source contribution is negligible at all scales where the 100\,GHz spectrum dominates the CMB solution. At large scales, the dust is our strongest foreground. We checked in Fig.~\ref{fig:wiskerTT} the effect of either marginalizing out the slope of the dust spectrum or removing the amplitude priors (i.e., making them arbitrarly wide). When marginalizing over the slope, one recovers a value compatible with the one in our model ($-2.57\pm0.038$ whereas our model uses $-2.63$) and the cosmological parameters do not change (Sect.~\ref{sec:robust-fsky}). When comparing the baseline likelihood result to \camspec which uses a slightly different template we find a $0.16\,\sigma$ systematic shift in $\ns$ that can be attributed to the steeper dust template slope ($-2.7$) (Sect.~\ref{sec:comparison}). When ignoring the amplitude priors, a $0.2\,\sigma$ shift appears on $\ns$ (and $\As$, due to its correlation with $\ns$). However, in this case the level of dust contribution increases by about $20\mu\mathrm{K}^2$ in all spectra, which corresponds to more than doubling the $100\times100$ dust contribution. This level is completely ruled out by the $100\times545$ cross spectrum, which enables estimation of the dust contribution in the 100\,GHz channel. The parameter shift can hence be attributed to a degeneracy between the dust model and the cosmological model broken by the prior on the amplitude parameters. We also use the fact that the dust distribution is anisotropic on the sky and evaluate the cosmological parameters on a smaller sky fraction. On $\TT$ there is no shift in the parameters that cannot be attributed to the greater cosmic variance on the smaller sky fraction. We are also making a simplifying assumption by describing the dust as a Gaussian field with a specific power spectrum. The numerical simulations (FFP9 and End-to-end) that include a realistic, anisotropic template for the dust contribution do not uncover any systematic effect due to that approximation.  In the end, we believe that $0.2\,\sigma$ on $\ns$ is a conservative upper bound of our astrophysical systematic bias on the cosmological parameters. There is, however, a possibility of a residual instrumental bias affecting foreground parameters (but not cosmology), but we cannot, at this stage, provide quantitative estimates.}  

\rev{To summarize, our systematic error budget consists of a $0.1\,\sigma$ methodology bias on $\ns$ for $TT$, at most a $0.2\,\sigma$ instrumental bias on $TT$ (on $\omb$), $TE$ and possibly a slightly greater one on $EE$. The few-$\mu\mathrm{K}^2$-level leakage residual in polarization does not appear to project onto biases on the \LCDM\ parameters. We conservatively evalute our astrophysical bias to be $0.2\,\sigma$ on $\ns$. The astrophysical parameters might suffer from instrumental biases.}

\subsection{The low-$\ell$ ``anomaly''} \label{sec:hal-anomaly}

In \citetalias{planck2013-p08} we noted that the \Planck\ 2013 low-$\ell$ temperature power spectrum exhibited a tension with the \Planck\ best-fit model, which is mostly determined by high-$\ell$ information. In order to quantify such a tension, we performed a series of tests, concluding that the low-$\ell$ power anomaly was mainly driven by multipoles between $\ell=20$ and $30$, which happen to be systematically low with respect to the model. \rev{The effect was shown to be also present (although less pronounced) using WMAP data (again, see \citetalias{planck2013-p08}  and
\citet{WMAP-3yrsPol})}. The statistical significance of this anomaly was found to be around 99\,\%, with slight variations depending on the \Planck\ CMB solution or the estimator considered. This anomaly has drawn significant attention as a potential tracer of new physics (e.g., \citealt{2015arXiv150304483K,2014JCAP...04..017K,2012JCAP...05..012D}; see also \citealt{2008PhRvD..78b3013D}), so it is worth checking its status in the 2015 analysis.

\begin{figure}[htbp] 
 \includegraphics[width=\columnwidth]{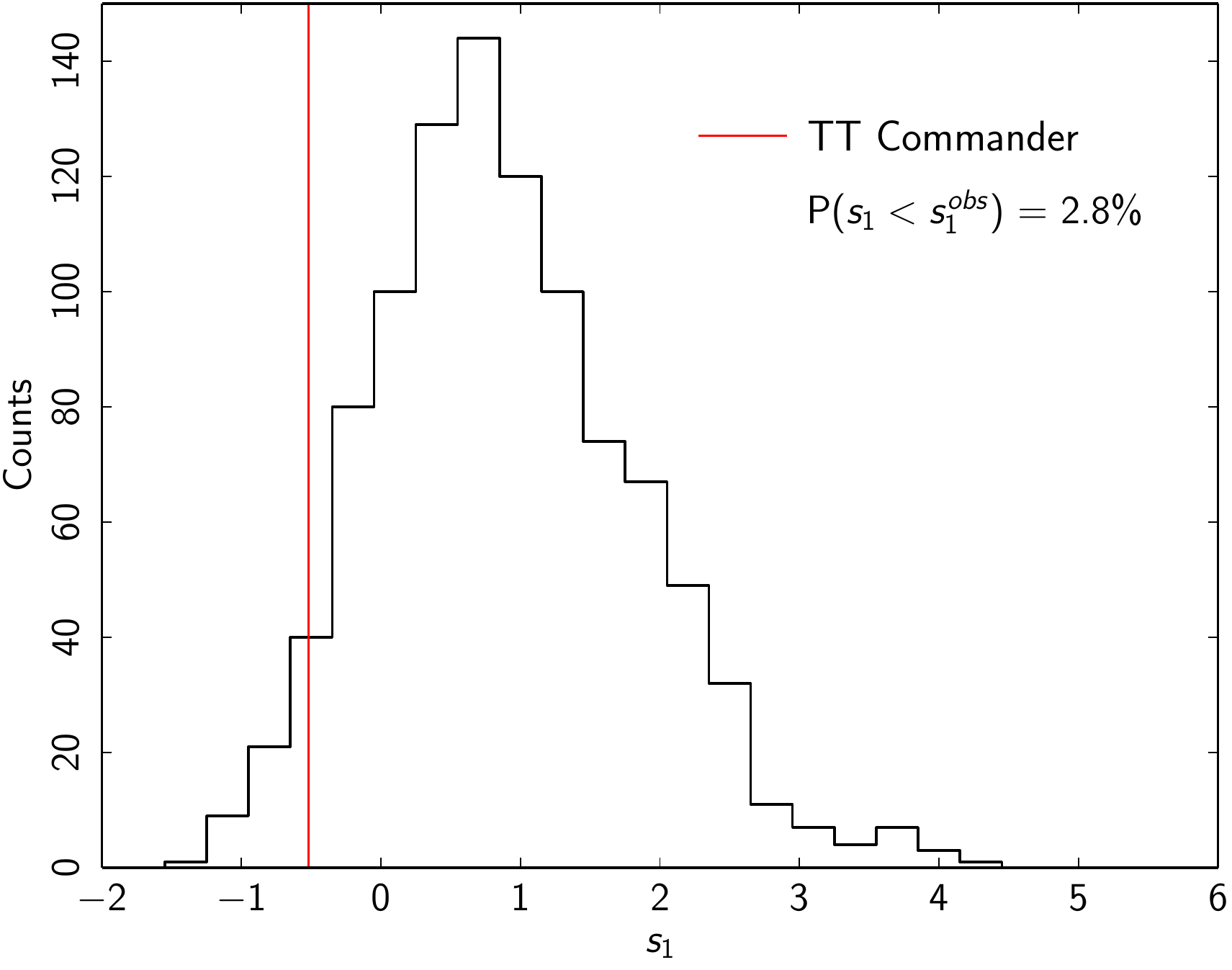}
  \includegraphics[width=\columnwidth]{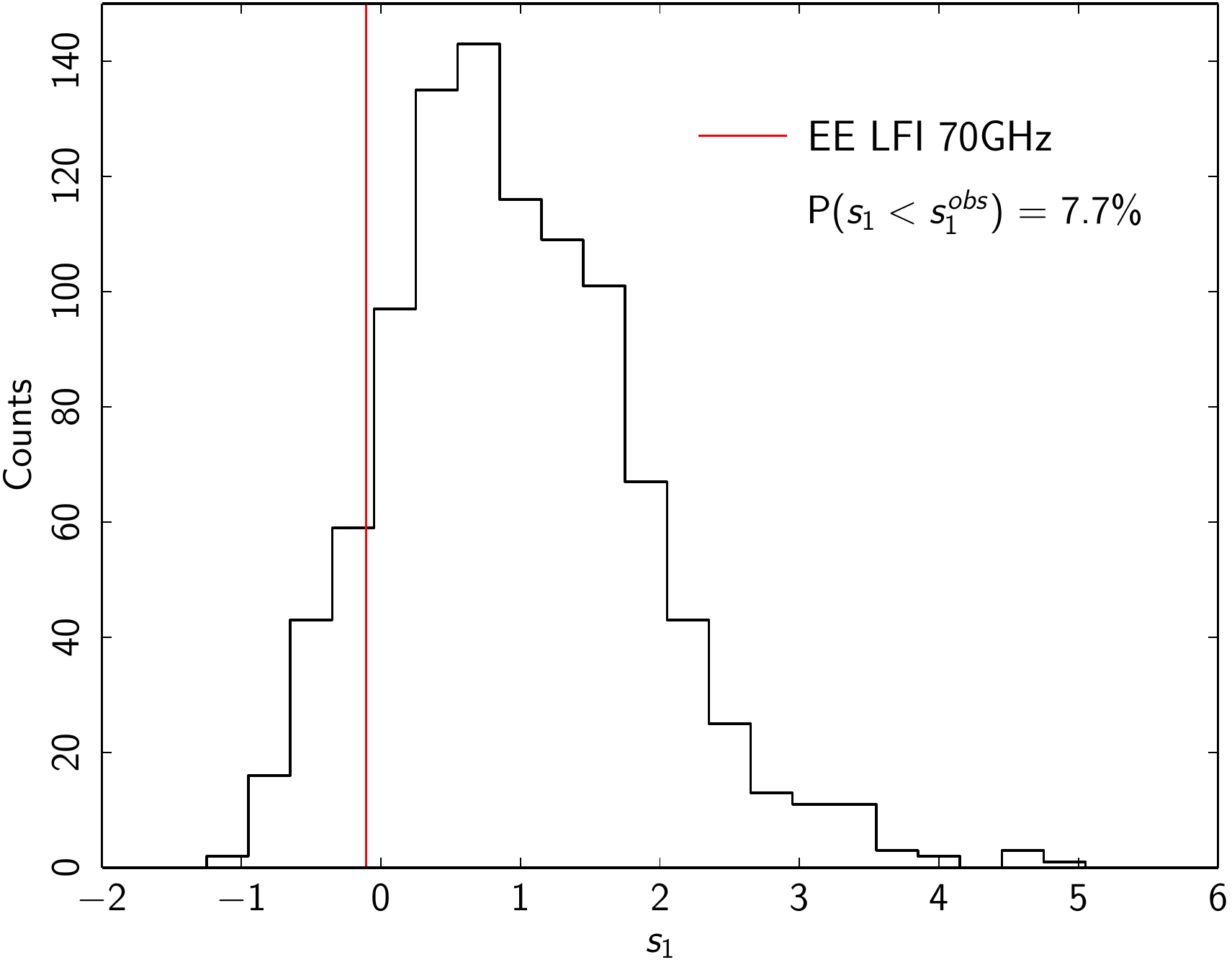}
 \caption{\emph{Top}: Empirical distribution for the Hausman $s_1$ statistic for $\TT$ derived from simulations; the vertical bar is the observed value for the \Planck\ \texttt{Commander} map. \emph{Bottom}: The empirical distribution of $s_1$ for $\EE$ and the \Planck\ 70 GHz polarization maps described in Sect.~\ref{sec:low-ell}.}
  \label{fig:haus_lowlvshighl}
\end{figure}

We present here updated results from a selection of the tests performed in 2013. While in \citetalias{planck2013-p08} we only concentrated on temperature, we now also consider low-$\ell$ polarization, which was not available as a \Planck\ product in 2013. We first perform an analysis through the Hausman test \citep{Polenta_CrossSpectra}, modified as in  \citetalias{planck2013-p08} for the statistic $s_{1} = \mathrm{sup}_{r}B(\ell_\mathrm{max},r)$, with  $\ell_{\textrm{max}}=29$ and
\begin{align}
B(\ell_\mathrm{max},r)&=\frac{1}{\sqrt{\ell_\mathrm{max}}}\sum_{\ell=2}^{\textrm{int}(\ell_\mathrm{max}r)}H_{\ell},\ \ \ r\in\left[0,1\right] \label{eq:hausman1} \,,  \\
H_{\ell}&=\frac{\hat{C_{\ell}}-C_{\ell}}{\sqrt{\textrm{Var}\,\hat{C_{\ell}}}} \,,
\end{align}
where $\hat{C}_{\ell}$ and $C_{\ell}$ denote the observed and model
power spectra, respectively. Intuitively, this statistic measures the relative bias between the observed spectrum and model, expressed in units of standard deviations, while taking the so-called ``look-elsewhere effect'' into account by maximizing $s_1$ over multipole ranges. We use the same simulations as described in Sect.~\ref{sec:70ghz_pol}, which are based on FFP8, for the likelihood validation. We plot in Fig.~\ref{fig:haus_lowlvshighl} the empirical distribution for $s_{1}$ in temperature and compare it to the value inferred from the \Planck\ \commander 2015 map described in Sect.~\ref{sec:low-ell} above. The significance for the \commander map has weakened from $0.7\,\%$ in 2013  to $2.8\,\%$ in 2015. This appears consistent with the changes between the 2013 and 2015 \commander power spectra shown in Fig.~\ref{fig:lowlTpowspec}, where we can see that the estimates in the range $20 < \ell < 30$ were generally shifted upwards (and closer to the \Planck\ best-fit model) due to revised calibration and improved analysis on a larger portion of the sky. We also report in the lower panel of Fig.~\ref{fig:haus_lowlvshighl} the same test for the $\EE$ power spectrum, finding that the observed \Planck\ low-$\ell$ polarization maps are anomalous only at the $7.7\,\%$ level.  

\begin{figure}[htbp] 
 \includegraphics[width=\columnwidth]{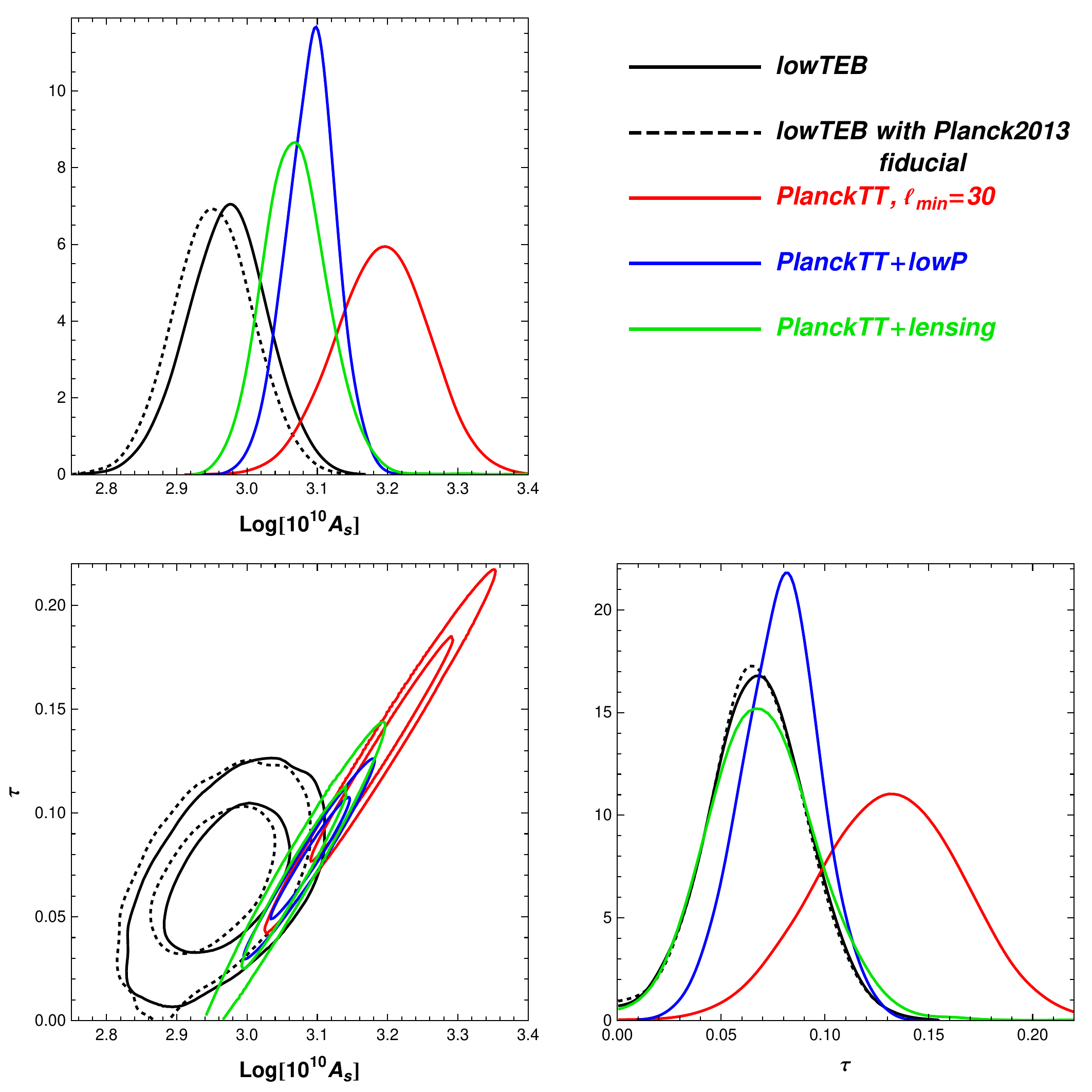}
 \caption{Joint estimates of primordial amplitude $\As$ and $\tau$ for the data sets indicated in the legend. For low-$\ell$ estimates, all other parameters are fixed to the 2015 fiducial values, except for the dashed line, which uses the \Planck\ 2013 fiducial. \rev{The \planckTT\ estimates  fall roughly half way between the low- and high-$\ell$ only ones.}}
  \label{fig:lowellanomaly}
\end{figure}

As a further test of the low-$\ell$ and high-$\ell$ \Planck\ constraints, we compare the estimate of the primordial amplitude $\As$ and the optical depth $\tau$, first separately for low and high multipoles, and then jointly. Results are displayed in Fig.~\ref{fig:lowellanomaly}, showing that the $\ell < 30$ and the $\ell \ge 30$ data posteriors in the primordial amplitude are separated by $2.6\,\sigma$, where the standard deviation is computed as the square root of the sum of the variances of each posterior. We note that a similar separation exists for $\tau$, but it is only significant at the $1.5 \,\sigma$ level. Fixing the value of the high-$\ell$ parameters to the \Planck\ 2013 best-fit model slightly increases the significance of the power anomaly, but has virtually no effect on $\tau$. A joint analysis using all multipoles retrieves best-fit values in $\As$ and $\tau$ which are between the low and high-$\ell$ posteriors. This behaviour is confirmed when the \Planck\ 2015 lensing likelihood \citep{planck2014-a17} is used in place of low-$\ell$ polarization.

Finally, we note a similar effect on $\neff$, which, in the high-$\ell$ analysis with a $\tau$ prior is about $1\,\sigma$ off the canonical value of 3.04, but is right on top of the canonical value once the lowP and its $\ell=20$ dip is included.




\subsection{Compressed CMB-only high-$\ell$ likelihood}\label{sec:planck_compressed}

We extend the Gibbs sampling scheme described in \citet{Dunkley:2013} and \citet{Calabrese:2013} to construct a compressed temperature and polarization \Planck\ high-$\ell$ CMB-only likelihood, {\tt Plik\_lite}, estimating CMB band-powers and the associated covariances after marginalizing over foreground contributions.
Instead of using the full multi-frequency likelihood to directly estimate cosmological parameters and nuisance parameters describing other foregrounds, we take the intermediate step of using the full likelihood to extract CMB temperature and polarization power spectra, marginalizing over possible Galactic and extragalactic contamination. In the process, a new covariance matrix is generated for the marginalized spectra, which therefore includes foreground uncertainty. We refer to Appendix~\ref{app:plik_lite} for a description of the methodology and to Fig.~\ref{fig:spec_cmbonly} for a comparison between the multi-frequency data and the extracted CMB-only band-powers for $\TT$, $\TE$, and $\EE$.

By marginalizing over nuisance parameters in the spectrum-estimation step, we decouple the primary CMB from non-CMB information. We use the extracted marginalized spectra and covariance matrix in a compressed, high-$\ell$, CMB-only likelihood. No additional nuisance parameters, except the overall \Planck\ calibration $y_{\rm P}$, are then needed when estimating cosmology, so the  convergence of the MCMC chains is significantly faster. To test the performance of this compressed likelihood, we compare results using both the full multi-frequency likelihood and the CMB-only version, for the $\Lambda$CDM six-parameter model and for a set of six \lcdm\ extensions. 

We show in Appendix~\ref{app:plik_lite} that the agreement between the results of the full likelihood and its compressed version is excellent, with consistency to better than 0.1$\,\sigma$ for all parameters. We have therefore included this compressed likelihood, \pliklite, in the \Planck\ likelihood package that is available in the \Planck\ Legacy Archive.\footnote{\url{http://pla.esac.esa.int/pla/}}

\subsection{\Planck\ and other CMB experiments} \label{sec:hal-other}

\subsubsection{\WMAP-9} \label{sec:wmap9}

In Sect.~\ref{Planck_wmap_lowP} we presented the \WMAP-9-based low-$\ell$ polarization likelihood, which uses the \Planck\ 353\,GHz map  as a dust tracer, as well as the \Planck\ and \WMAP-9 combination. Results for these likelihoods are presented in Table~\ref{table:params_WMAP_LFI_highell}, in conjunction with the \Planck\ high-$\ell$ likelihood. 
Parameter results for the joint \Planck\ and \WMAP\ data set in the union mask are further discussed in \citet{planck2014-a15} and \citet{planck2014-a24}. 

\begin{table}[ht!]
\begingroup
\newdimen\tblskip \tblskip=5pt
\caption{Selected parameters estimated from \Planck, \WMAP, and their noise-weighted combination in low-$\ell$ polarization, assuming \Planck\ in temperature at all multipoles.$^{\rm a}$}
\label{table:params_WMAP_LFI_highell}
\nointerlineskip
\vskip -6mm
\footnotesize
\setbox\tablebox=\vbox{
   \newdimen\digitwidth 
   \setbox0=\hbox{\rm 0} 
   \digitwidth=\wd0 
   \catcode`*=\active 
   \def*{\kern\digitwidth}
   \newdimen\signwidth 
   \setbox0=\hbox{+} 
   \signwidth=\wd0 
   \catcode`!=\active 
   \def!{\kern\signwidth}
 \openup 4pt
\halign{\hbox to 0.9in{#\leaderfil}\tabskip=2em&
  \hfil#\hfil& 
  \hfil#\hfil\tabskip=1.2em&
  \hfil#\hfil \tabskip=0pt\cr
\noalign{\doubleline}
\omit\hfil Parameter\hfil& \Planck& \WMAP& \Planck+\WMAP\cr
\noalign{\vskip 3pt\hrule\vskip 5pt}
$\tau$& $0.077^{+0.019}_{-0.018}$& $0.071^{+0.012}_{-0.012}$& $0.074^{+0.012}_{-0.012}$\cr
$z_\textrm{re}$& $9.8^{+1.8}_{-1.6}$& $9.3^{+1.1}_{-1.1}$&  $9.63^{+1.1}_{-1.0}$\cr
 $\log[10^{10}A_\textrm{s}]$& $3.087^{+0.036}_{-0.035}$& $3.076^{+0.022}_{-0.022}$& $3.082^{+0.021}_{-0.023}$\cr
 $r$& $[0,\,0.11]$& $[0,\,0.096]$& $[0,\,0.10]$\cr
 $A_\textrm{s} e^{-2\tau}$& $1.878^{+0.010}_{-0.010}$& $1.879^{+0.011}_{-0.010}$& $1.879^{+0.010}_{-0.010}$\cr 
\noalign{\vskip 5pt\hrule\vskip 3pt}}}
\endPlancktable
\tablenote {{\rm a}} The \Planck\ \texttt{Commander} temperature map  is always used at low $\ell$, while the \texttt{Plik} $\TT$ likelihood is used at high $\ell$. All the base-$\Lambda$CDM parameters and $r$ are sampled. \par
\endgroup
\end{table}

\begin{figure*}[tbp] 
\centering
  \includegraphics[width=0.88\textwidth]{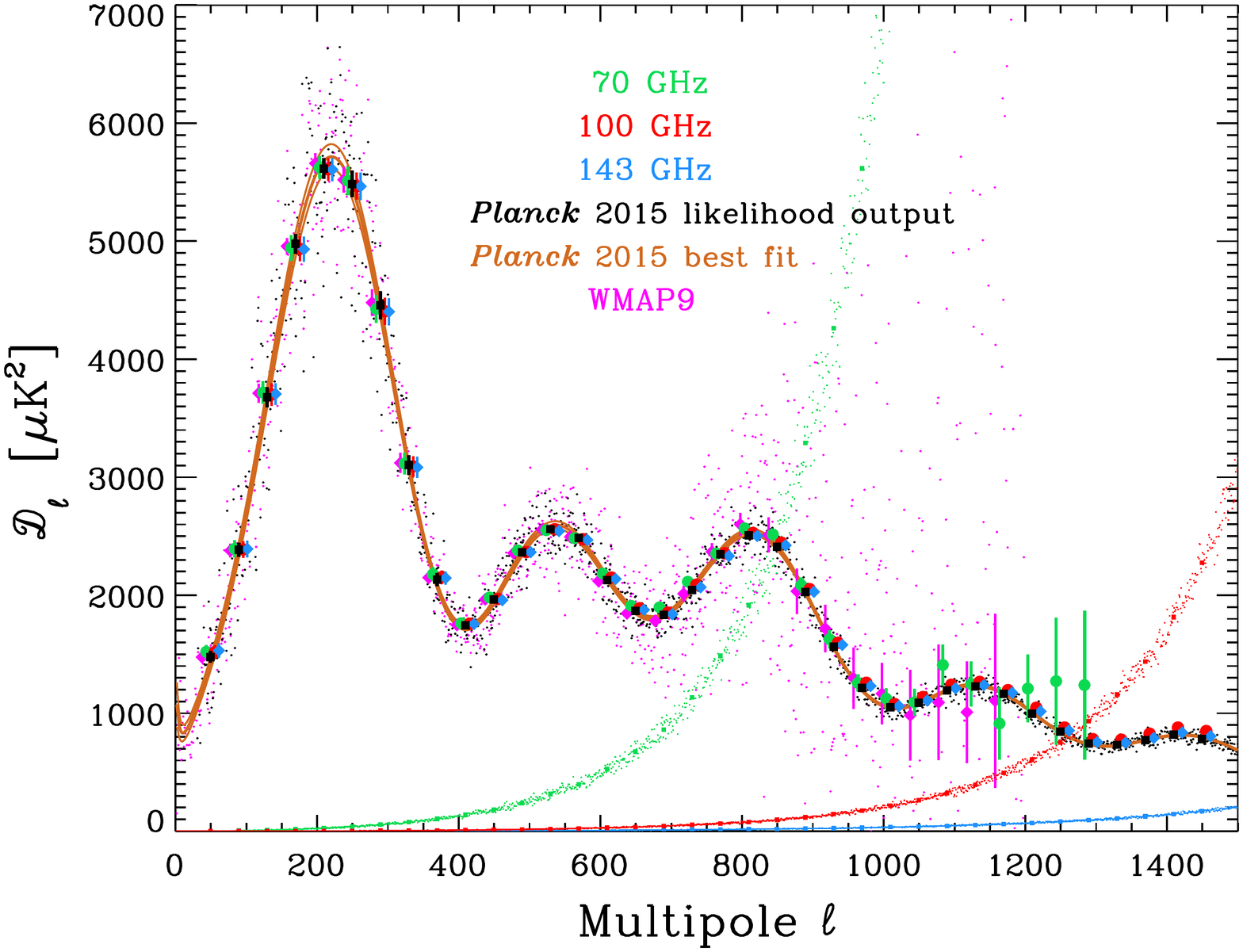}\vspace{-24pt}
  \includegraphics[width=0.88\textwidth]{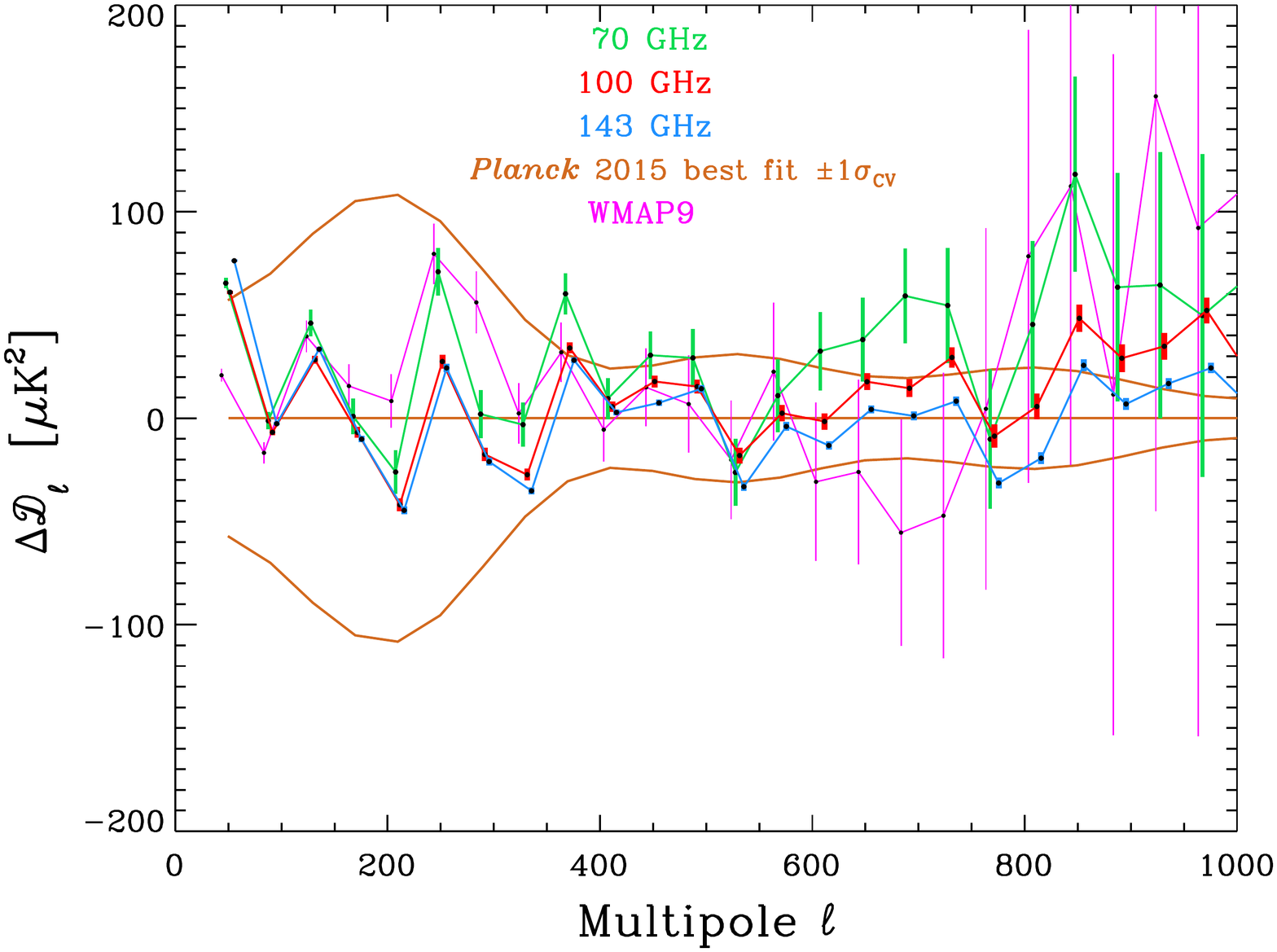}
  \caption{Comparison of \planck\ and \WMAP-9 CMB power spectra. \emph{Top}: Direct comparison. Noise spectra are derived from the half-ring difference maps. \emph{Bottom}: Residuals with respect to the \planck\ \LCDM\ best-fit model. \rev{The error bars do not include the cosmic variance contribution (but the (brown) 1-$\sigma$ contour lines for the Likelihood best fit model do).} 
}
  \label{fig:planck-vs-wmap}
\end{figure*}

We now illustrate the state of agreement reached between the \Planck\ 2015 data, in both the raw and likelihood processed form, and the final cosmological power spectra results from \WMAP-9. In 2013 we noted that the difference between \WMAP-9 and \Planck\ data was mostly related to calibration, which is now resolved with the upward calibration shift in the \Planck\ 2015 maps and spectra, as discussed in \citet{planck2014-a01}. This leads to the rather impressive agreement that has been reached between the two \Planck\ instruments and \WMAP-9.

Figure~\ref{fig:planck-vs-wmap} (top panel) shows all the spectra after correction for the effects of sky masking, with different masks used in the three cases of the \Planck\ frequency-map spectra, the spectrum computed from the \Planck\ likelihood, and the \WMAP-9 final spectrum. The \Planck\ 70, 100, and 143\,GHz spectra (which are shown as green, red, and blue points, respectively) were derived from the raw frequency maps (cross-spectra of the half-ring data splits for the signal, and spectra of the difference thereof for the noise estimates) 
on approximately 60\,\% of the sky (with no apodization), where the sky cuts include the Galaxy mask, and a concatenation of the 70, 100, and 143\,GHz point-source masks.      

The  spectrum computed from the \Planck\ likelihood (shown in black as both individual and binned $C_\ell$ values in Fig.~\ref{fig:planck-vs-wmap}) was described earlier in the paper. We recall that it was derived with no use of the 70\,GHz data, but including the 217\,GHz data. Importantly, since it illustrates the likelihood output, this spectrum has been corrected (in the spectral domain) for the residual effects of diffuse foreground emission, mostly in the low-$\ell$ range, and for the collective effects of several components of discrete foreground emission (including tSZ, point sources, CIB, etc.). This spectrum effectively carries the information that drives the likelihood solution of the \Planck\ 2015 best-fit CMB anisotropy, shown in brown. Our aim here is to show the conformity between this \Planck\ 2015 solution and the raw \Planck\ data (especially at 70\,GHz) and the \WMAP-9 legacy spectrum.

The \WMAP-9 spectrum (shown in magenta as both individual and binned $C_\ell$ values) is the legacy product from the \WMAP-9 mission, and it represents the final results of the \WMAP team's efforts to clean the  residual effects of foreground emission from the cosmological anisotropy spectrum.

All these spectra are binned the same way, starting at $\ell=30$ with $\Delta \ell = 40$ bins, and the error-bars represent the  error on the mean within each bin. 
In the low-$\ell$ range, especially near the first peak, the error calculation includes the cosmic variance contribution from the multipoles within each bin, which vastly exceeds any measurement errors (all the measurements shown here have high S/N over the first spectral peak), so we would expect good agreement between the errors derived for all the spectra in the completely signal-dominated range of the data.

The figure shows how \WMAP-9 loses accuracy above $\ell \approx 800$ due to its inherent beam resolution and instrumental noise, and shows how the \LFI\ 70\,GHz data achieve improved fidelity over this range. \HFI\ was designed to improve over both \WMAP-9 and \LFI\ in both noise performance and angular resolution, and the gains achieved are clearly visible, even over the relatively modest range of $\ell$ shown here, in the tiny spread of the individual $C_\ell$ values of the \Planck\ 2015 power spectrum. While the overall agreement of the various spectra, especially in the low-$\ell$ range, is noticeable in this coarse plot, it is also clear that the \Planck\ raw frequency-map spectra do show excess power over the \Planck\ best-fit spectrum at the higher end of the $\ell$-range shown --- the highest level at 70\,GHz and the lowest at 143\,GHz. This illustrates the effect of uncorrected discrete foreground residuals in the raw spectra.

A better view of these effects is seen in the bottom panel of Fig.~\ref{fig:planck-vs-wmap}. Here we plot the binned values from the top panel as deviations from the best-fit model. Naturally, the black bins of the likelihood output fit well, since they were derived jointly with the best-fit spectrum, while correcting for foreground residuals. The \WMAP-9 points show good agreement, given their errors, with the \Planck\ 2015 best fit, and illustrate very tight control of the large-scale residual foregrounds (at the low-$\ell$ range of the figure); beyond $\ell\sim600$ the \WMAP-9 spectrum shows an increasing loss of fidelity.
\Planck\ raw 70, 100, and 143\,GHz spectra show excess power in the lowest $\ell$ bin due to diffuse foreground residuals. The higher-$\ell$ range now shows more clearly the upward drift of power in the raw spectra, growing from 143\,GHz to 70\,GHz. This is consistent with the  well-determined integrated discrete foreground contributions to those spectra. 
As previously shown in \citet[figure~8]{planck2013-p01a}, the unresolved discrete foreground power (computed with the same sky masks as used here) can be represented in the bin near $\ell=800$ as levels of approximately $40\,\muKsq$ at 70\,GHz, $15\,\muKsq$ at 100\,GHz, and $5\,\muKsq$ at 143\,GHz, in good agreement with the present figure.

\subsubsection{\act and \spt} \label{sec:planck_actspt}

\Planck\ temperature observations are complemented at finer scales by measurements from the ground-based Atacama Cosmology Telescope (ACT) and South Pole Telescope (SPT). The ACT and SPT high-resolution data help \Planck\ in separating the primordial cosmological signal from other Galactic and extragalactic emission, so as not to bias cosmological reconstructions in the damping-tail region of the spectrum. In 2013 we combined \Planck\ with ACT \citep{das/etal:prep} and SPT \citep{reichardt12} data in the multipole range $1000<\ell<10\,000$, defining a common foreground model and extracting cosmological parameters from all the data sets. Our updated ``highL'' temperature data include ACT power spectra at 148 and 218\,GHz \citep{das/etal:prep} with a revised binning \citep{Calabrese:2013} and final beam estimates \citep{Hasselfield:2013}, and SPT measurements in the range $2000<\ell<13\,000$ from the $2540\,{\rm deg}^2$ SPT-SZ survey at 95, 150, and 220\,GHz \citep{George:2014}.
However, in this new analysis, given the increased constraining power of the \Planck\ full-mission data, we do not use ACT and SPT as primary data sets. Using the same $\ell$ cuts as the 2013 analysis (\ie ACT data at $1000<\ell<10\,000$ and SPT at $\ell>2000$) we only check for consistency and retain information on the nuisance foreground parameters that are not well constrained by \Planck\ alone. 

To assess the consistency between these data sets, we extend the \Planck\ foreground model up to $\ell=13\,000$ with additional nuisance parameters for ACT and SPT, as described in \citet[section 4]{planck2014-a15}. 
Fixing the cosmological parameters to the best-fit \planckTT\, base-\lcdm\ model and varying the ACT and SPT foreground and calibration parameters, we find a reduced $\chi^2=1.004$ (PTE\,{=}\,0.46), showing very good agreement between \Planck\ and the highL data.

\begin{figure}[tbp] 
\centering
 \includegraphics[width=\columnwidth]{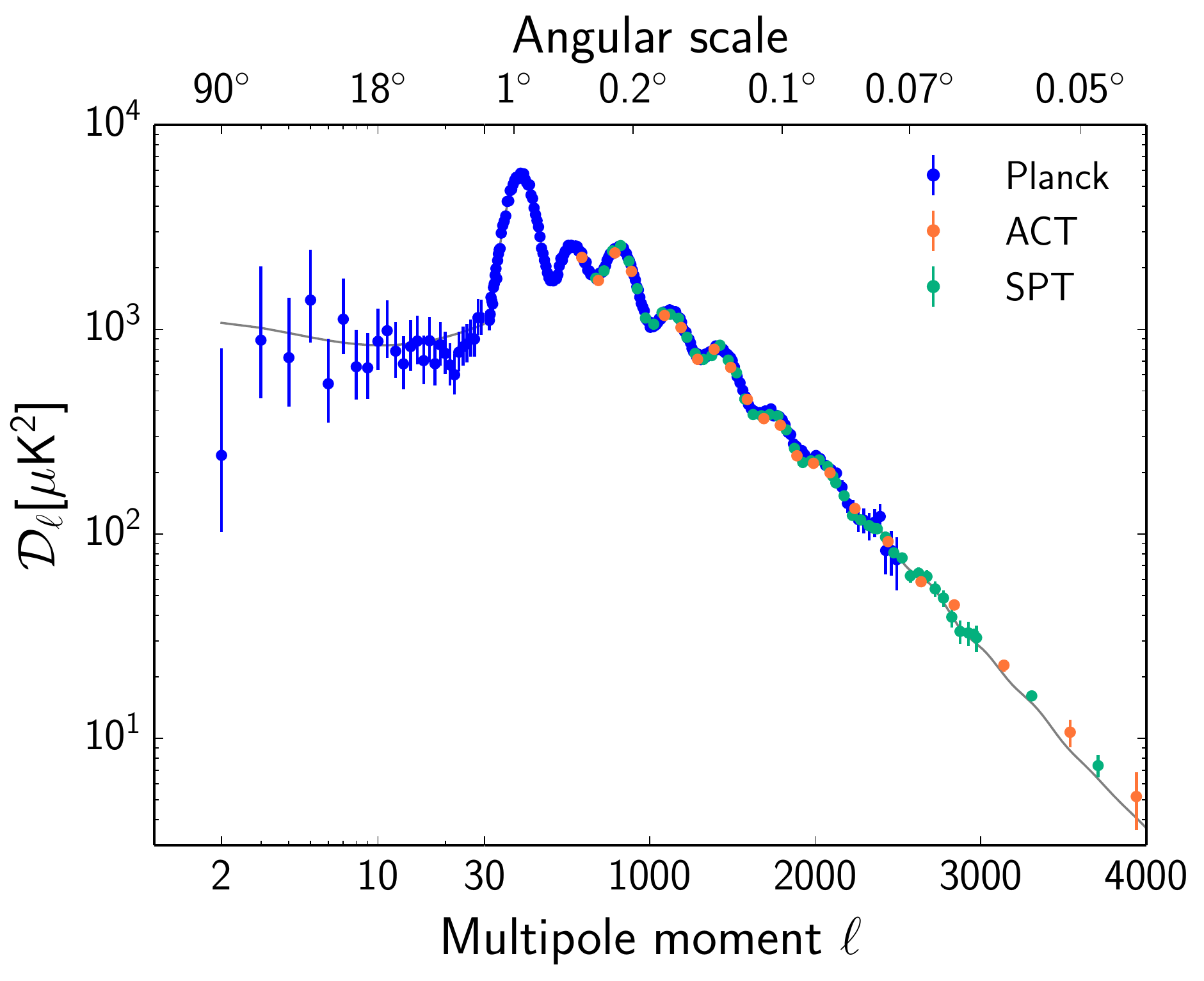}
\caption{CMB-only power spectra measured by \Planck\ (blue), ACT (orange), and SPT (green). The best-fit \planckTT\, $\Lambda$CDM model is shown by the grey solid line. ACT data at $\ell>1000$ and SPT data at $\ell>2000$ are marginalized CMB band-powers from multi-frequency spectra presented in \citet{das/etal:prep} and \citet{George:2014} as extracted in this work. Lower multipole ACT ($500<\ell<1000$) and SPT ($650<\ell<3000$) CMB power extracted by \citet{Calabrese:2013} from multi-frequency spectra presented in \citet{das/etal:prep} and \citet{story/etal:prep} are also shown. The binned values in the range $3000< \ell <4000 $ appear higher than the unbinned best-fit line because of the binning (this is numerically confirmed by the residual plot in \citealt[figure 9]{planck2014-a15}). } 
\label{fig:cmbcls}
\end{figure}

As described in \citet{planck2014-a15}, we then take a further step and extend the Gibbs technique presented in \citet{Dunkley:2013} and \citet{Calabrese:2013} (and applied to \Planck\ alone in Sect.~\ref{sec:planck_compressed}) to extract independent CMB-only band-powers from \planck, ACT, and SPT. The extracted CMB spectra are reported in Fig.~\ref{fig:cmbcls}. We also show ACT and SPT band-powers at lower multipoles as extracted by \citet{Calabrese:2013}. This figure shows the state of the art of current CMB observations, with \Planck\ covering the low-to-high-multipole range and ACT and SPT extending into the damping region. We consider the CMB to be negligible at $\ell>4000$ and note that these ACT and SPT band-powers have an overall calibration uncertainty (2\,\% for ACT and 1.2\,\% for SPT). 

The inclusion of ACT and SPT improves the full-mission \Planck\ spectrum extraction presented in Sect.~\ref{sec:planck_compressed} only marginally. The main contribution of ACT and SPT is to constrain small components (\eg the tSZ, kSZ, and tSZ$\times$CIB) that are not well determined by \Planck\ alone.
However, those components are sub-dominant for \Planck\ and are well described by the prior based on the 2013 \Planck+highL solutions imposed in the \Planck-alone analysis. The CIB amplitude estimate improves by $40\,\%$ when including ACT and SPT, but the CIB power is also reasonably well constrained by \Planck\ alone. The main \Planck\ contaminants are the Poisson sources, which are treated as independent and do not benefit from ACT and SPT. As a result, the errors on the extracted \Planck\ spectrum are only slightly reduced, with little additional cosmological information added by including ACT and SPT for the baseline \LCDM\ model (see also \citealt[section 4]{planck2014-a15}).

\section{Conclusions}\label{sec:conc}

\begin{figure*}[!htbp] 
\centering
\includegraphics[height=0.34\textheight]{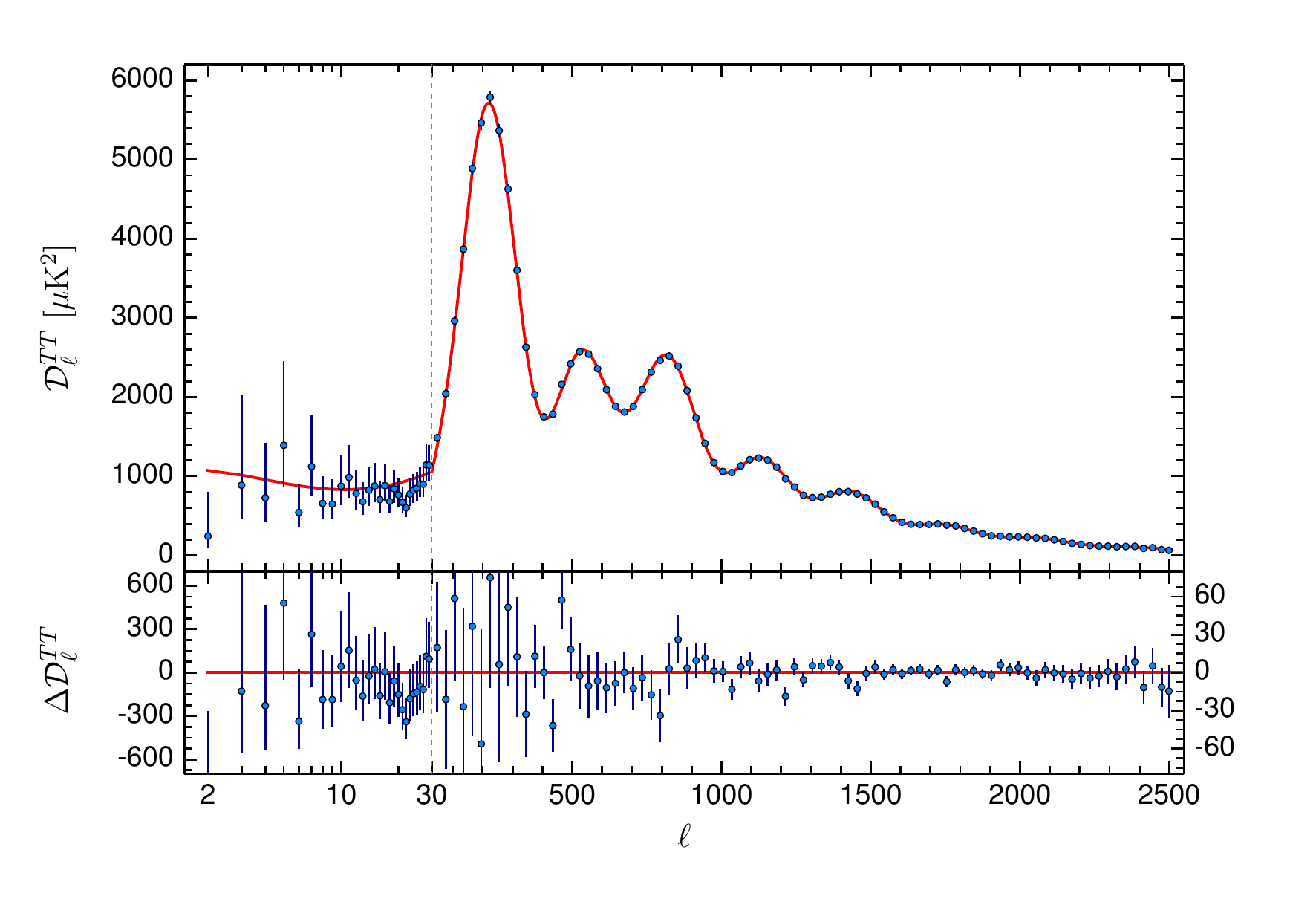}\vspace{-20pt}
\includegraphics[height=0.34\textheight]{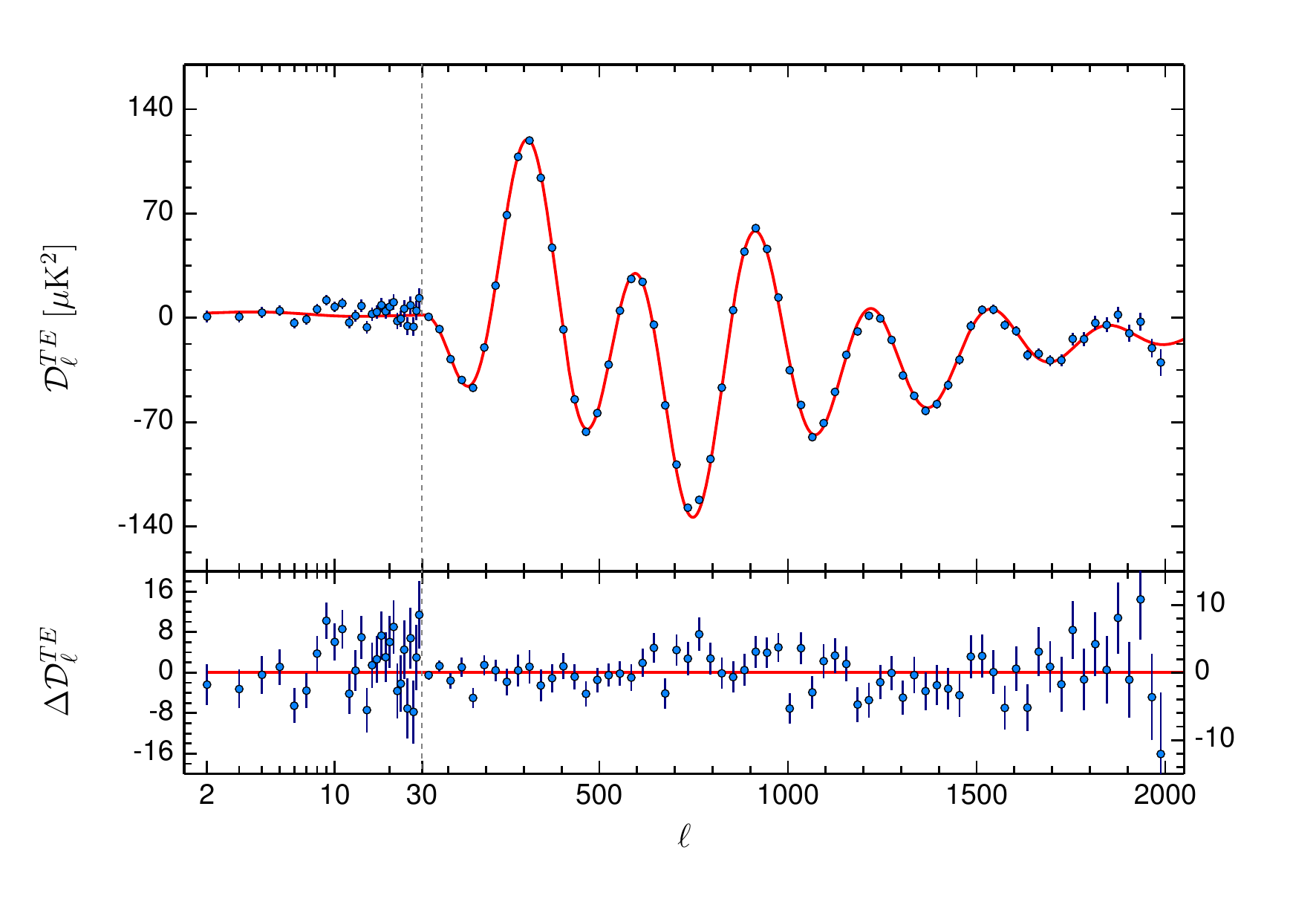}\vspace{-20pt}
\includegraphics[height=0.34\textheight]{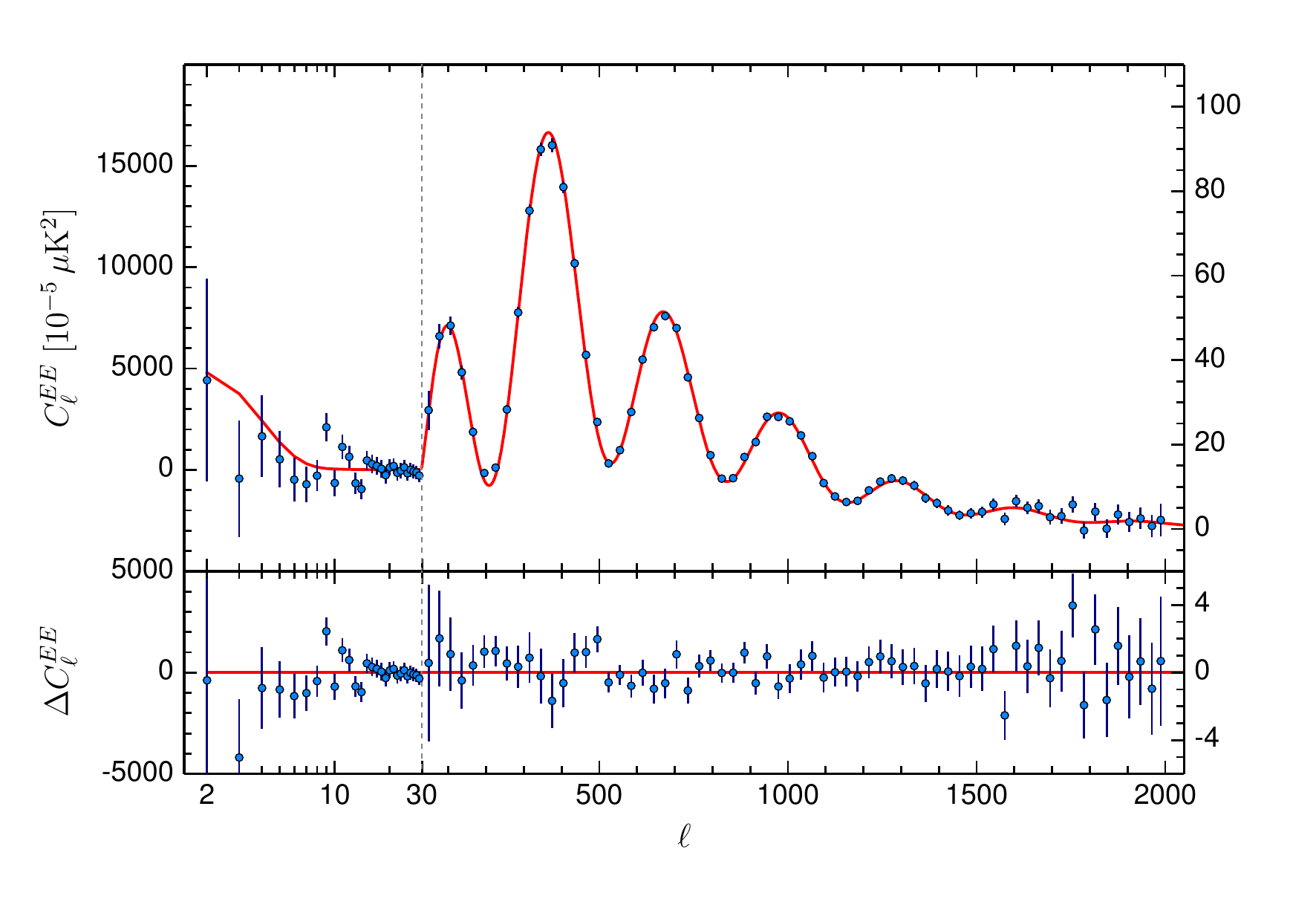}\vspace{-20pt}
\caption{\Planck\ 2015 CMB spectra, compared with the base \LCDM\ fit to \planckTT\ data (red line). The upper panels show the spectra and the lower panels the residuals. In all the panels, the horizontal scale changes from logarithmic to linear at the ``hybridization'' scale, $\ell=29$ (the division between the low-$\ell$ and high-$\ell$ likelihoods). For the residuals, the vertical axis scale changes as well, as shown by different left and right axes. We show ${\cal D}_\ell=\ell(\ell+1)C_\ell/(2\pi)$ for $\TT$ and $\TE$, but $C_\ell$ for $\EE$, which also has different vertical scales at low- and high-$\ell$.}
\label{fig:Planck_Cl_TT}
\end{figure*}

The \Planck\ 2015 angular power spectra of the cosmic microwave background derived in this paper are displayed in Fig.~\ref{fig:Planck_Cl_TT}. These spectra in $\TT$ (top), $\TE$ (middle), and $\EE$ (bottom) are all quite consistent with the best-fit base-\LCDM\ model obtained from $\TT$ data alone (red lines). The horizontal axis is logarithmic at $\ell < 30$, where the spectra are shown for individual multipoles, and linear at $\ell\ge30$, where the data are binned. The error bars correspond to the diagonal elements of the covariance matrix. The lower panels display the residuals, the data being presented with different vertical axes, a larger one at left for the low-$\ell$ part and a zoomed-in axis at right for the high-$\ell$ part. 

The 2015 \Planck\ likelihood presented in this work is based on more temperature data than in the 2013 release, and on new polarization data. It benefits from several improvements in the processing of the raw data, and in the modelling of astrophysical foregrounds and instrumental noise. Apart from a revision of the overall calibration of the maps, discussed in \citet{planck2014-a01}, the most significant improvements are in the  likelihood procedures:
\begin{enumerate}
\item[(i)] a joint temperature-polarization pixel-based likelihood at $\ell \leq 29$, with more high-frequency information used for foreground removal, and smaller sky masks (Sects.~\ref{sec:lowl_algorithm} and \ref{sec:commander_lowl});
\item[(ii)] an improved Gaussian likelihood at $\ell \geq 30$ that includes
a different strategy for estimating power spectra from data-subset cross-correlations, using half-mission data instead of detector sets (which enables us to reduce the effect of correlated noise between detectors, see Sects.~\ref{sec:hil:detcomb} and \ref{sec:beam-uncert}), and  better foreground templates, especially for Galactic dust (Sect.~\ref{sec:dust}) that lets us mask a smaller fraction of the sky (Sect.~\ref{sec:masks}) and to retain large-angle temperature information from the 217\,GHz map that was neglected in the 2013 release (Sect.~\ref{sec:cuts}).
\end{enumerate}

We performed several consistency checks of the robustness of our likelihood-making process, by introducing more or less freedom and nuisance parameters in the modelling of foregrounds and instrumental noise, and by including different assumptions about the relative calibration uncertainties across frequency channels and about the beam window functions. 

For temperature, the reconstructed CMB spectrum and error bars are remarkably insensitive to all these different assumptions. Our final high-$\ell$ temperature likelihood, referred to as ``\planckTTonly'' marginalizes over 15 nuisance parameters (12 modelling the foregrounds, and 3 for calibration uncertainties). Additional nuisance parameters (in particular, those associated with beam uncertainties) were found to have a negligible impact, and can be kept fixed in the baseline likelihood. \rev{Detailed end-to-end simulations of the instrumental response to the sky analysed like the real data did not uncover hidden low-level residual systematics.}

For polarization, the situation is different. Variation of the assumptions leads to scattered results,
with greater deviations than would be expected due to changes in the
data subsets used, and at a level that is significant compared to
the statistical error bars. This suggests that further systematic
effects need to be either modelled or removed. In particular, our attempt to model calibration errors and temperature-to-polarization leakage suggests that the $\TE$ and $\EE$ power spectra are affected by systematics at a level of roughly $1\,\mu\mathrm{K}^2$. Removal of polarization systematics at this level of precision requires further work, beyond the scope of this release. The 2015 high-$\ell$ polarized likelihoods, referred to as ``\plikTE'' and ``\plikEE'', or ``\plikTTTEEE'' for the combined version, ignore these \rev{uncertain} corrections. They only include 12 additional nuisance parameters accounting for polarized foregrounds.
Although these likelihoods are distributed in the \Planck\ Legacy Archive,\footnote{\url{http://pla.esac.esa.int/pla/}} we stick to the \planckTT\ choice in the baseline analysis of this paper and the companion papers such as \citet{planck2014-a15}, \citet{planck2014-a16}, and \citet{planck2014-a24}. 

We developed internally several likelihood codes, exploring not only different assumptions about foregrounds and instrumental noise, but also different algorithms for building an approximate Gaussian high-$\ell$ likelihood (Sect.~\ref{sec:comparison}). We compared these codes to check the robustness of the results, and decided to release: 
\begin{enumerate}
\item[(i)] A baseline likelihood called \plik (available for $\TT$, $\TE$, $\EE$, or combined observables), in which the data are binned in multipole space, with a bin-width increasing from $\Delta \ell=5$ at $\ell \approx 30$ to $\Delta \ell=33$ at $\ell \approx 2500$. 
\item[(ii)] An unbinned version which, although slower, is preferable when investigating models with sharp features in the power spectra.
\item[(iii)] A simplified likelihood called \pliklite\ in
which the foreground templates and calibration errors are
marginalized over, producing a marginalized spectrum and covariance
matrix. This likelihood does not allow investigation of correlations between cosmological and foreground/instrumental parameters, but speeds up parameter extraction, having no nuisance parameters to marginalize over. 
\end{enumerate}

In this paper we have also presented an investigation of the measurement of cosmological parameters in the minimal six-parameter $\Lambda$CDM model and  a few simple seven-parameter extensions, using both the new baseline \Planck\ likelihood and  several alternative likelihoods relying on different assumptions. The cosmological analysis of this paper does not replace the investigation of many extended cosmological models presented, e.g., in \citet{planck2014-a15}, \citet{planck2014-a16}, and \citet{planck2014-a24}. However, the careful inspection of residuals presented here addresses two questions:
\begin{enumerate}
\item[(i)] a priori, is there any indication that an alternative model to $\Lambda$CDM could provide a significantly better fit?
\item[(ii)] if there is such an indication, could it come from caveats in the likelihood-building (imperfect data reduction, foreground templates or noise modelling) instead of new cosmological ingredients?
\end{enumerate}
Since this work is entirely focused on the power-spectrum likelihood, it can only address these questions at the level of 2-point statistics; for a discussion of higher-order statistics, see \citet{planck2014-a18} and \citet{planck2014-a19}.

The most striking result of this work is the impressive consistency of different cosmological parameter extractions, performed with different versions of the \plikTT+tauprior or \planckTT\ likelihoods, with several assumptions concerning: data processing (half-mission versus detector set correlations); sky masks and foreground templates; beam window functions; the use of two frequency channels instead of three; different cuts at low $\ell$ or high $\ell$; a different choice for the multipole value at which we switch from the pixel-based to the Gaussian likelihood;  different codes and algorithms; the inclusion of external data sets like \WMAP-9, ACT, or SPT; and the use of foreground-cleaned maps (instead of fitting the CMB+foreground map with a sum of different contributions). In all these cases, the best-fit parameter values drift by only a small amount, compatible with what one would expect on a statistical basis when some of the data are removed (with a few exceptions summarized below). 

The cosmological results are stable when one uses the simplified \pliklite\ likelihood. We checked this by comparing \planckTT\ results from \plik\ and \pliklite\ for $\Lambda$CDM, and for six examples of seven-parameter extended models. 

Another striking result is that, despite evidence for small unsolved systematic effects in the high-$\ell$ polarization data, the cosmological parameters returned by the \plikTT, \plikTE, or \plikEE\ likelihoods (in combination with a $\tau$ prior or \Planck\ lowP) are consistent with each other, and the residuals of the (frequency combined) $\TE$ and $\EE$ spectra after subtracting the temperature $\Lambda$CDM best-fit are consistent with zero. As has been emphasized in other \Planck\ 2015 papers, this is a tremendous success for cosmology, and an additional proof of the predictive power of the standard cosmological model. It also suggests that the level of temperature-to-polarization leakage (and possibly other systematic effects) revealed by our consistency checks is low enough \rev{ (on average over all frequencies)} not to \rev{significantly} bias  parameter extraction, at least for the minimal cosmological model. We do not know yet whether this conclusion applies also to extended models, especially those in which the combination of temperature and polarization data has stronger constraining power than  temperature data alone, e.g., dark matter annihilation \citep{planck2014-a15} or isocurvature modes \citep{planck2014-a24}. One should thus wait for a future \Planck\ release before applying the \Planck\ temperature-plus-polarization likelihood to such models. However, the fact that we observe a significant reduction in the error bars when including polarization data is very promising, since this reduction is expected to remain after the removal of systematic effects.

Careful inspection of residuals with respect to the best-fit $\Lambda$CDM model has revealed a list of anomalies in the \Planck\ CMB power spectra, of which the most significant is still the low-$\ell$ temperature anomaly in the range $20\leq \ell \leq 30$, already discussed at length in the 2013 release. In this 2015 release, with more data and with better calibration, foreground modelling, and sky masks, its significance has decreased from the 0.7\,\% to the 2.8\,\% level for the $TT$ spectrum (Sect.~\ref{sec:hal-anomaly}). This probability is still small (although not very small), and the feature remains unexplained. We have also investigated the $EE$ spectrum, where the anomaly, if any, is significant only at the 7.7\,\% level.

Other ``anomalies'' revealed by inspection of residuals (and of their dependence on the assumptions underlying the likelihood) are much less significant. There are a few bins in which the power in the $\TT$, $\TE$, or $\EE$ spectrum lies 2--3\,$\sigma$ away from the best-fit $\Lambda$CDM prediction, but this is not statistically unlikely and we find acceptable probability-to-exceed (PTE) levels. Nevertheless, in Sects.~\ref{sec:highlbase} and \ref{sec:hil_par_stability}, we presented a careful investigation of these features, to see whether they could be caused by some imperfect modelling of the data. We noted that a deviation in the $\TT$ spectrum at $\ell \approx 1450$ is somewhat suspicious, since it is driven \rev{mostly} by a single channel (217\,GHz), and since it depends on the foreground-removal method. But this deviation is too small to be worrisome ($1.8\,\sigma$ with the baseline \plik\ likelihood). As in the 2013 release, the data at intermediate $\ell$ would be fitted slightly better by a model with more lensing than in the best-fit $\Lambda$CDM model (to reduce the peak-to-trough contrast), but more lensing generically requires higher values of $A_\textrm{s}$ and $\Omega_{\rm c} h^2$ that are disfavoured by the rest of the data, in particular when high-$\ell$ information is included. This mild tension is illustrated by the preference for a value greater than unity for  the unphysical parameter $A_{\rm L}$, a conclusion that is stable against variations in the assumptions underlying the likelihoods. However, $A_{\rm L}$ is compatible with unity at the $1.8\,\sigma$ level when using the baseline \planckTTonly\ likelihood with a conservative $\tau$ prior (to avoid the effect of the low-$\ell$ dip), so what we see here could be the result of statistical fluctuations.

This absence of large residuals in the \Planck\ 2015 temperature and polarization spectra further establishes the robustness of the $\Lambda$CDM model, even with about twice as much data as in the \Planck\ 2013 release. This conclusion is supported by several companion papers, in which many non-minimal cosmological models are investigated but no significant evidence for extra physical ingredients is found. The ability of the  temperature results to pass several demanding consistency tests, and the evidence of excellent agreement down to the $\muKsq$ level between the temperature and polarization data, represent an important milestone set by the \Planck\ satellite. The \Planck\ 2015 likelihoods are the best illustration to date of the predictive power of the minimal cosmological model, and, at the same time, the best tool for constraining interesting, physically-motivated deviations from that model.  

\begin{acknowledgements}
The Planck Collaboration acknowledges the support of: ESA; CNES, and CNRS/INSU-IN2P3-INP (France); ASI, CNR, and INAF (Italy); NASA and DoE (USA); STFC and UKSA (UK); CSIC, MINECO, JA and RES (Spain); Tekes, AoF, and CSC (Finland); DLR and MPG (Germany); CSA (Canada); DTU Space (Denmark); SER/SSO (Switzerland); RCN (Norway); SFI (Ireland); FCT/MCTES (Portugal); ERC and PRACE (EU). A description of the Planck Collaboration and a list of its members, indicating which technical or scientific activities they have been involved in, can be found at 
\href{url}{http://www.cosmos.esa.int/web/planck/planck-collaboration}.

We further acknowledge the use of the \texttt{CLASS} Boltzmann code \citep{2011arXiv1104.2932L} and the \texttt{Monte Python} package \citep{2013JCAP...02..001A} in earlier stages of this work. The likelihood code and some of the validation work was built on the library \texttt{pmclib} from the \texttt{CosmoPMC} package \citep{2011arXiv1101.0950K}.

This research used resources of the IN2P3 Computer Center (http://cc.in2p3.fr) as well as of the Planck-HFI DPC infrastructure hosted at the Institut d'Astrophysique de Paris (France) and financially supported by CNES.
\end{acknowledgements}

\alltwentyfifteenresultspapers

\bibliographystyle{aat}
\bibliography{Planck_bib,0_Likelihood_bib,cmbonly,hillipop,actspt,peak_ells_table} 

\appendix

\clearpage


\section{Sky masks}  \label{app:masks}

\begin{figure}[tbp] 
 \centering
  \includegraphics[angle=0,width=0.45\textwidth]{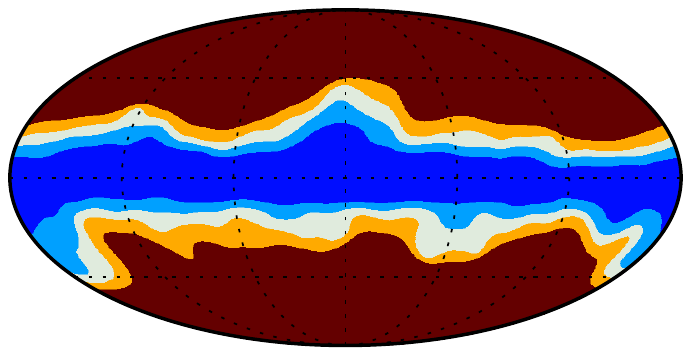}
  \caption{Unapodized Galactic masks \GM50, \GM60, \GM70, and \GM80, from orange to dark blue.}
  \label{fig:gal_masks}
\end{figure}

This appendix provides details of the way we build sky masks for the high-$\ell$ likelihood. Since it is based on data at frequencies between 100 and 217\,GHz, Galactic dust emission is the main diffuse foreground to minimize. We subtract the \smica CMB temperature map \citep{planck2014-a11} from the 353\,GHz map and we adopt the resulting CMB-subtracted 353\,GHz map as a tracer of dust. After smoothing the map with a $10\deg$ Gaussian kernel, we threshold it to generate a sequence of masks with different sky coverage. Galactic masks obtained in this way are named \GM80 to \GM50, where the number gives the retained sky fraction $f_{\rm sky}$ in percent (Fig.~\ref{fig:gal_masks}). 

For the likelihood analysis, we aim to find a trade-off between maximizing the sky coverage and having a simple, but reliable, foreground model of the data. The combination of masks and frequency channels retained is given in Table~{\ref{tab:mask_comb}}. In order to get $C_{\ell}$-covariance matrices for the cosmological analysis that are accurate at the few percent level (cf.~Sect.~\ref{sec:covariance}), we actually use apodized versions of the Galactic masks. The apodization corresponds to a Gaussian taper of width $\sigma=2\deg$.\footnote{We use the routine {\texttt{process\_mask}\xspace} of the {\texttt{HEALPix}} package to obtain a map of the distance of each pixel of the mask from the closest null pixel. We then use a smoothed version of the distance map to build the Gaussian apodization. The smoothing of the distance map is needed to avoid  sharp edges in the final mask.} Apodized Galactic masks are also used for the polarization analysis. The effective sky fraction of an apodized mask is $f_{\rm sky}=\sum_i w^2_i \Omega_i/(4\pi)$, where $w_i$ is the value of the mask in pixel $i$ and $\Omega_i$ is the solid angle of the pixel.

\begin{figure}[tb] 
 \centering
 \includegraphics[angle=0,width=0.45\textwidth]{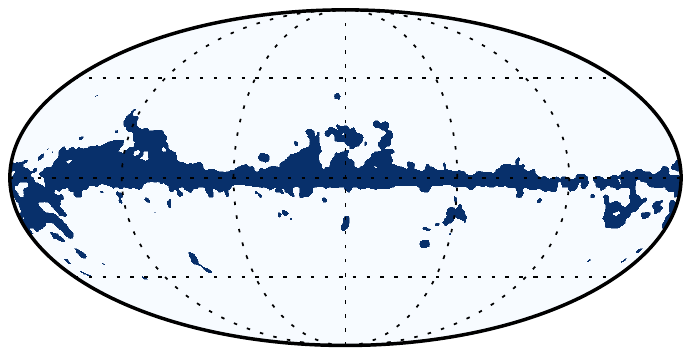}
  \caption{Unapodized CO mask ($f_{\rm sky}=87\,\%$).}
  \label{fig:co_mask}
\end{figure}

All the \HFI frequency channels, except 143\,GHz, are also contaminated by CO emission from rotational transition lines. Here we are concerned with emission around 100 and 217\,GHz, associated with the CO $J=1\rightarrow0$ and $J=2\rightarrow1$ lines, respectively. Most of the emission is concentrated near the Galactic plane and is therefore masked out by the Galactic dust masks. However, there are some emission regions at intermediate and low latitudes that are outside the quite small \GM80 mask we use at 100\,GHz. We therefore create a mask specifically targeted at eliminating CO emission. The Type 3 CO map, part of the \Planck\ 2013 product delivery \citep{planck2013-p03a}, is sensitive to low-intensity diffuse CO emission over the whole sky. It is a multi-line map, derived using prior information on line ratios and a multi-frequency component separation method. Of the three types of \Planck\ CO maps, this has the highest S/N. We smooth this map with a $\sigma=120\arcm$ Gaussian and mask the sky wherever the CO line brightness exceeds $1\;{\rm K_{RJ}\;km\;s}^{-1}$. The mask is shown in Fig.~\ref{fig:co_mask}, before apodization with a Gaussian taper of ${\rm FWHM}=30\arcm$. 

Finally, we include extragalactic objects in our temperature masks, both point sources and nearby extended galaxies. The nearby galaxies that are masked are listed in Table~\ref{tab:ngal_list}, together with the corresponding cut radii. For point sources, we build conservative masks for 100, 143, and 217\,GHz separately. At each frequency, we mask sources that are detected above ${\rm S/N}=5$ in the 2015 point-source catalogue~\citep{planck2014-a35} with holes of radius three times the $\sigma=\mathrm{FWHM}/\sqrt{\ln8}$ of the effective Gaussian beam at that frequency. We take the FWHM values from the elliptic Gaussian fits to the effective beams\citep{planck2014-a35}, i.e., FWHM values of $9\parcm66$, $7\parcm22$, and $4\parcm90$ at 100, 143, and 217\,GHz, respectively. We apodize these masks with a Gaussian taper of ${\rm FWHM}=30\arcm$. As already noted, these masks are designed to reduce the contribution of diffuse and discrete Galactic and extragalactic foreground emission in the ``raw'' (half-mission and detset) frequency maps used for the baseline high-$\ell$ likelihood.


\begin{table}[t!] 
\begingroup 
\newdimen\tblskip \tblskip=5pt
\caption{Galactic masks used for the high-$\ell$ analysis.} 
\vskip -6mm
\footnotesize 
\setbox\tablebox=\vbox{
\newdimen\digitwidth
\setbox0=\hbox{\rm 0}
\digitwidth=\wd0
\catcode`*=\active
\def*{\kern\digitwidth}
\newdimen\signwidth
\setbox0=\hbox{+}
\signwidth=\wd0
\catcode`!=\active
\def!{\kern\signwidth}
\newdimen\decimalwidth
\setbox0=\hbox{.}
\decimalwidth=\wd0
\catcode`@=\active
\def@{\kern\signwidth}
\halign{ 
\hbox to 1in{#\leaderfil}\tabskip=2em& 
    \hfil#\hfil\tabskip=2em& 
    \hfil#\hfil\tabskip=3em&
    \hfil#\hfil\tabskip=2em&
    \hfil#\hfil\tabskip=0pt\cr
\noalign{\doubleline}
\omit\hfil Frequency [GHz]\hfil&\multispan2\hfil Temperature\hfil&\multispan2 \hfil Polarization\hfil\cr
\noalign{\vskip 2pt\hrule\vskip 2pt}
$100$&\GM80&G70&\GM80&G70\cr
$143$&\GM70&G60&\GM60&G50\cr
$217$&\GM60&G50&\GM50&G41\cr
\noalign{\vskip 2pt\hrule\vskip 2pt}
}}
\endPlancktable 
{For each frequency channel, the Galactic and apodized Galactic masks are labelled by their ``\GM'' and ``G'' prefixes, followed by the retained sky fraction (in percent).}
\label{tab:mask_comb}
\endgroup
\end{table} 

\begin{table}[t!] 
\begingroup
\newdimen\tblskip \tblskip=5pt
\caption{Masked nearby galaxies and corresponding cut radii. \label{tab:ngal_list}}
\vskip -6mm
\footnotesize 
\setbox\tablebox=\vbox{
\newdimen\digitwidth 
\setbox0=\hbox{\rm 0}
\digitwidth=\wd0
\catcode`*=\active
\def*{\kern\digitwidth}
\newdimen\signwidth
\setbox0=\hbox{+}
\signwidth=\wd0
\catcode`!=\active
\def!{\kern\signwidth}
\newdimen\decimalwidth
\setbox0=\hbox{.}
\decimalwidth=\wd0
\catcode`@=\active
\def@{\kern\signwidth}
\halign{\hbox to 1.5in{#\leaderfil}\tabskip=2em&
   \hfil#\hfil\tabskip 0pt\cr
\noalign{\doubleline}
\omit&Radius\cr
\omit\hfil Galaxy\hfil&[arcmin]\cr
\noalign{\vskip 3pt\hrule\vskip 3pt}
LMC&250\cr
SMC&110\cr
SMC ext$^{\rm a}$&*50\cr
M31 F1$^{\rm b}$&*80\cr
M31 F2&*80\cr
M33& *30\cr
M81& *30\cr
M101&*18\cr
M82& *15\cr
M51& *15\cr
CenA&*15\cr
\noalign{\vskip 3pt\hrule\vskip 5pt}
}}
\endPlancktable 
\tablenote {{\rm a}} Inspection of the SMC at 857\,GHz reveals an extra signal, localized in a small area near the border of the excised disk, which we mask with a disk centred at $(l, b) = (299\pdeg85, -43\pdeg6)$.\par
\tablenote {{\rm b}} M31 is elongated. Therefore, instead of cutting an unnecessarily large disk, we use two smaller disks centred at the focal points of an elliptical fit to the galaxy image (F1, F2).\par
\endgroup
\end{table} 

The masks described in this appendix are used in the papers on cosmological parameters \citep{planck2014-a15}, inflation \citep{planck2014-a24}, dark energy \citep{planck2014-a16}, and primordial magnetic fields \citep{planck2014-a22}, which are notable examples of the application of the high-$\ell$ likelihood. However, the masks differ from those adopted in some of the other \Planck\ papers.  For example, reconstructions of gravitational lensing \citep{planck2014-a17} and
integrated Sachs-Wolfe effect \citep{planck2014-a26}, constraints on isotropy and statistics \citep{planck2014-a18}, and searches for primordial non-Gaussianity \citep{planck2014-a19} mainly rely on the high-resolution foreground-reduced CMB maps presented in \citet{planck2014-a11}.  Those maps have been derived by four component-separation methods that combine data from different frequency channels to extract ``cleaned'' CMB maps.  For each method, the corresponding confidence masks, for both temperature and polarization, remove regions of the sky where the CMB solution is not trusted.  This is described in detail in appendices A--D of \citet{planck2014-a11}. The masks recommended for the analysis of foreground-reduced CMB maps are constructed as the unions of the confidence masks of all the four component separation methods.  
Their sky coverages are $f_{\rm sky} =0.776$ in temperature and $f_{\rm sky} =0.774$ in polarization.  Since component separation mitigates the foreground contamination even at relatively low Galactic latitudes, those masks feature a thinner cut along the Galactic plane than the ones described in this appendix.  Nevertheless, propagation of noise, beam, and  extragalactic foreground uncertainties in foreground-cleaned CMB maps is more difficult, and this is the main reason why we do not employ them in the baseline high-$\ell$ likelihood.  We also note that the recommended mask for temperature foreground-reduced maps has a greater number of compact object holes than the masks used here.  This is due to the fact that some component separation techniques can introduce contamination of sources from a wider range of frequencies than the approach considered here for the high-$\ell$ power spectra.  According to the tests provided in
Sect.~\ref{app:hil_pts_mask_correction}, such masks would result in sub-optimal performance of the analytic $C_{\ell}$-covariance matrices.


\section{Low-$\ell$ likelihood supplement \label{app:low-ell} }

\subsection{Sherman--Morrison--Woodbury formula}\label{app:lol-speed}

In the \Planck\ 2015 release we follow a pixel-based approach to the joint low-$\ell$ likelihood (up to $\ell = 29$) of $T$, $Q$, and $U$. This approach treats temperature and polarization maps  consistently at \texttt{HEALPix} resolution $N_\mathrm{side} = 16$, as opposed to the \WMAP\ low-$\ell$ likelihood, which incorporates polarization information from lower-resolution maps to save computational time \citep{page2007}. The disadvantage of a consistent-resolution, brute-force approach lies in its computational cost \citepalias{planck2013-p08}, which may require massively parallel coding (and adequate hardware) in order to be competitive in execution time with the high-$\ell$ part of the CMB likelihood  (see, e.g., \citealt{2013MNRAS.431.2961F} for one such implementation).  Such a choice, however, would hamper the ease of code distribution across a community not necessarily specialized in massively parallel computing. Luckily,  the Sherman--Morrison--Woodbury formula and the related matrix determinant lemma provide a means to achieve good timing without resorting to supercomputers.  To see how this works, rewrite the covariance matrix from  Eq.~(\ref{eq:pbLike}) in a form  that explicitly separates the $C_\ell$ to be varied from those that stay fixed at the reference model:
\begin{align}
\tens{M} &= \sum_{XY}\sum_{\ell =2}^{\ell_{\rm cut}} C_\ell^{XY} \tens{P}_\ell^{XY} +\sum_{XY} \sum_{\ell=\ell_{\rm cut}+1}^{\ell_{\rm max}} C_\ell^{XY,\rm ref} \tens{P}_\ell^{XY}   +\tens{N}\\ 
&\equiv \sum_{XY} \sum_{\ell =2}^{\ell_{\rm cut}} C_\ell^{XY} \tens{P}_\ell^{XY} +\tens{M}_0,
\end{align}
where we have effectively redefined the fixed multipoles as
``high-$\ell$ correlated noise,'' as far as the varying low-$\ell$
multipoles are concerned. Next, note that for fixed $\ell$,
$\tens{P}_{\ell}^{TT}$ has rank\footnote{Masking can in principle
  reduce the effective rank, but for the high sky fractions used in
  the \Planck\ analysis, this is not an issue.} $\lambda = 2 \ell +1$,
and this matrix may therefore be decomposed as $\tens{P}_{\ell}^{TT} =
(\tens{V}_\ell^{TT})^\tens{T}\, \tens{A}_\ell^{TT} \,
\tens{V}_\ell^{TT}$, where $\tens{A}_\ell^{TT}$ and
$\tens{V}_\ell^{TT}$ are $(\lambda \times \lambda)$ and $(\lambda
\times N_{\textrm{pix}})$ matrices, respectively, which depend only upon the unmasked pixel locations.  A similar
decomposition holds for the $\tens{P}_{\ell}^{EE,BB}$ matrices, while
$\tens{P}_{\ell}^{TE}$ can be expanded in the
$[\tens{V}_\ell^{TT},\tens{V}_\ell^{EE}]$ basis for the corresponding
$\ell$. We can then write
\begin{equation}
\tens{M} = \tens{V}^\tens{T} \tens{A}(C_\ell) \tens{V} +\tens{M}_0,
\end{equation}
where $\tens{V} =
[\tens{V}_2^{TT},\tens{V}_2^{EE},\tens{V}_2^{BB},\dots\tens{V}_{\ell_{\rm
      cut}}^{BB}]$ is an $(n_\lambda \times N_{\textrm{pix}})$ matrix with $n_\lambda
= 3[(\ell_{ \rm cut} +1)^2 -4]$, and $\tens{A}(C_\ell)$ is an
$(n_\lambda \times n_\lambda)$ block-diagonal matrix (accounting for four modes removed in monopole and dipole subtraction). Each
$\ell$-block in the latter matrix reads
\begin{equation}
\begin{bmatrix}
C_\ell^{TT} \tens{A}_\ell^{TT} & C_\ell^{TE} \tens{A}_\ell^{TE}  & \tens{0} \\
C_\ell^{TE} \tens{A}_\ell^{TE} & C_\ell^{EE} \tens{A}_\ell^{EE}  & \tens{0} \\
\tens{0} &   \tens{0} & C_\ell^{BB} \tens{A}_\ell^{BB}
\end{bmatrix}.
\end{equation}
Finally, using the Sherman--Morrison--Woodbury identity and the matrix
determinant lemma, we can rewrite the inverse and determinant of
$\tens{M}$ as
\begin{align}
& \tens{M}^{-1}  &=& \, \tens{M}_0^{-1} - \tens{M}_0^{-1} \tens{V}^\tens{T}(\tens{A}^{-1} + \tens{V} \tens{M}_0^{-1} \tens{V}^\tens{T})^{-1}\tens{V} {\tens{M}_0^{-1}} \\
& |\tens{M}|       &=& \, |\tens{M}_0| \, |\tens{A}| \, |\tens{A}^{-1} + \tens{V} \tens{M}_0^{-1} \tens{V}^\tens{T}|~.
\end{align}
Because neither $\tens{V}$ nor $\tens{M}_0$ depends on $C_\ell$, all terms
involving only their inverses, determinants, and products may be
precomputed and stored. Evaluating the likelihood for a new set of
$C_\ell$ then requires only the inverse and determinant of an
$(n_\lambda \times n_\lambda)$ matrix, not an $(N_{\textrm{pix}} \times N_{\textrm{pix}})$
matrix. For the current data selection, described in
Sects.~\ref{sec:commander_lowl} and \ref{sec:70ghz_pol}, we find
$n_\lambda = 2688$, which is to be compared to $N_{\textrm{pix}} =
6307$, resulting in an order-of-magnitude speed-up compared to the
brute-force computation.

\subsection{\texttt{Lollipop}}


\label{sec:lollipop} 

We performed a complementary analysis of low-$\ell$ polarization using the \HFI data, in order to check the consistency with the \LFI-based baseline result. The level of systematic residuals in the \HFI maps at low $\ell$ is quite small, but comparable to the \HFI noise \citep[see][]{planck2014-a09}, so these residuals should be either corrected, which is the goal of a future release, or accounted for by a complete analysis including parameters for all relevant systematic effects, which we cannot yet perform. Instead, we use \texttt{Lollipop}, a low-$\ell$ polarized likelihood function based on cross-power spectra. The idea behind this approach is that the systematics are considerably reduced in cross-correlation compared to auto-correlation. 

At low multipoles and for incomplete sky coverage, the $C_\ell$ statistic is not simply distributed and is correlated between modes. \texttt{Lollipop} uses the approximation presented in \citet{HL08}, modified as described in \citet{2015arXiv150301347M} to apply to cross-power spectra. We restrict ourselves to the one-field approximation to derive a likelihood function based only on the $\EE$ power spectrum at very low multipoles. The likelihood function of the $C_\ell$ given the data $\tilde{C}_\ell$ is then
\begin{equation}
	-2\ln P(C_\ell|\tilde{C}_\ell)=\sum_{\ell \ell'} [X_g]^\tens{T}_\ell [M_f^{-1}]_{\ell \ell'} [X_g]_{\ell'},
\end{equation}
with the variable
\begin{equation}
	\left[X_g\right]_\ell = \sqrt{ C_\ell^{f} + O_\ell} \,\, g{\left(\frac{\tilde{C}_\ell + O_\ell}{C_\ell + O_\ell}\right)} \,\, \sqrt{ C_\ell^{\rm fid} + O_\ell} ,
\end{equation}
where $g(x)=\sqrt{2(x-\ln x -1)}$, $C_\ell^{\rm fid}$ is a fiducial model and $O_\ell$ is the offset needed in the case of cross-spectra. This likelihood has been tested on Monte Carlo simulations including both realistic signal and noise. In order to extract cosmological information on $\tau$ from the $\EE$ spectrum alone, we  restrict the analysis to the cross-correlation between the \HFI 100 and 143\,GHz maps, which exhibits the lowest variance.

At large angular scales, the \HFI maps are contaminated by systematic residuals coming from temperature-to-polarization leakage \citep[see][]{planck2014-a09}. We used our best estimate of the $Q$ and $U$ maps at 100 and 143\,GHz, which we correct for residual leakage coming from destriping uncertainties, calibration mismatch, and bandpass mismatch, using templates as described in \citet{planck2014-a09}. Even though the level of systematic effects is thereby significantly reduced, we still have residuals above the noise level in null tests at very low multipoles ($\ell \leqslant 4$). To mitigate the effect of this on the likelihood, we restrict the range of multipoles to $\ell=5$--$20$.

Cross-power spectra are computed on the cleanest  50\,\% of the sky by using a pseudo-$C_\ell$ estimate (\texttt{Xpol}, an extension to polarization of the code described in \citealt{tristram2005}). The mask corresponds to thresholding a map of the diffuse polarized Galactic dust at large scales. In addition, we also removed pixels where the intensity of diffuse Galactic dust and CO lines is strong. This ensures that bandpass leakage from dust and CO lines does not bias the polarization spectra \citep[see][]{planck2014-a09}.

We construct the $C_\ell$ correlation matrix using simulations including CMB signal and realistic inhomogeneous and correlated noise. In order to take into account the residual systematics, we derive the noise level from the estimated $\BB$ auto-spectrum where we neglect any possible cosmological signal. This over-estimates the noise level and ensures conservative errors. However, this estimate assumes by construction a Gaussian noise contribution, which is not a full description of the residuals.

\begin{figure}[h!] 
	\center
	\includegraphics[width=\columnwidth]{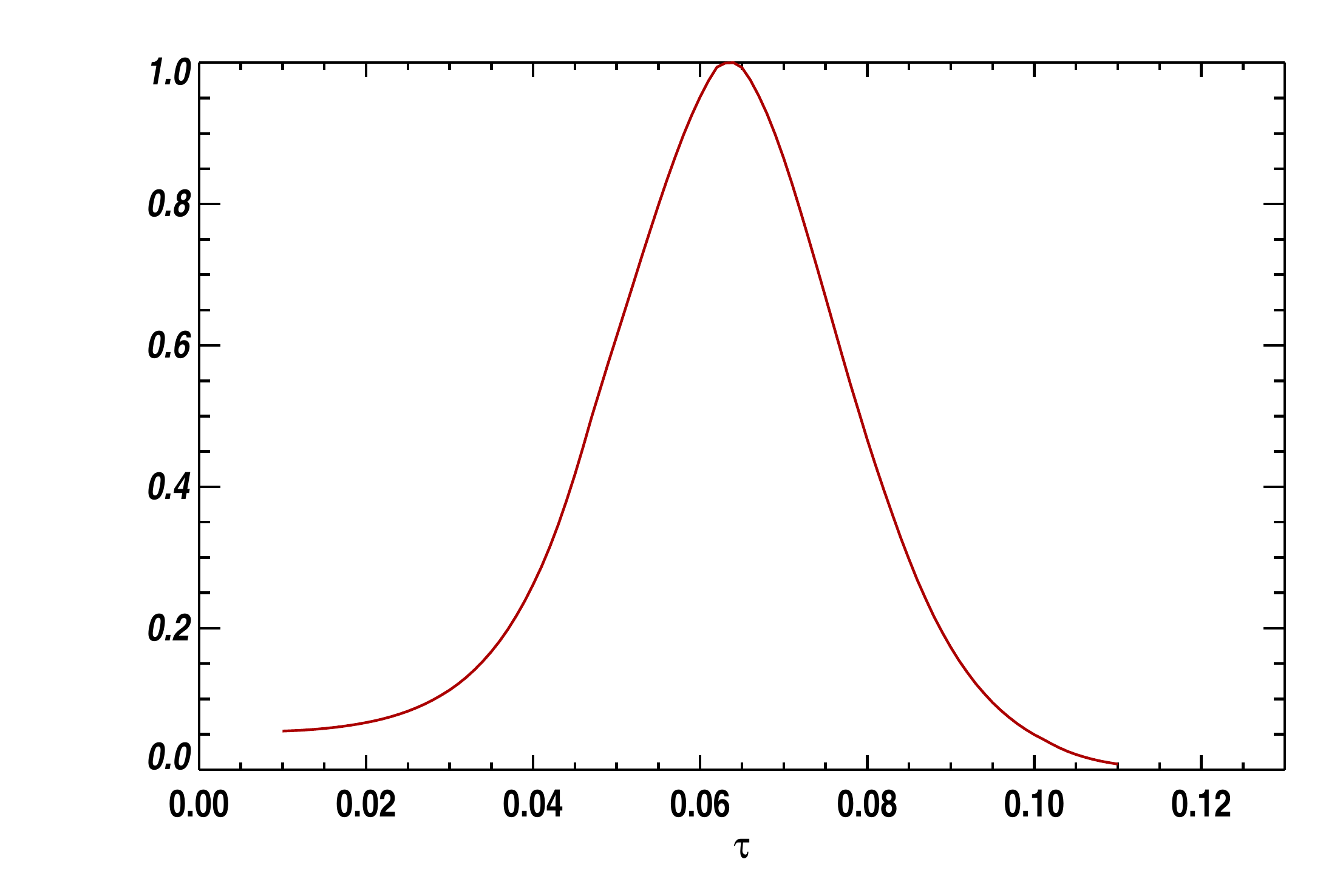}
	\caption{Distribution of the reionization optical depth $\tau$ using the \texttt{Lollipop} likelihood, based on the cross-correlation of the 100 and 143\,GHz channels.}
	\label{fig:lollipop}
\end{figure}

We then sample the reionization optical depth $\tau$ from the likelihood, with all other parameters fixed to the \Planck\ 2015 best-fit values \citep{planck2014-a15}. Without any other data, the degeneracy between $A_\textrm{s}$ and $\tau$ is broken by fixing the amplitude of the first peak of the $\TT$ spectrum (directly related to $A_\textrm{s}e^{-2\tau}$) at $\ell=200$.
The resulting distribution is plotted in Fig.~\ref{fig:lollipop}. The best fit is at 
\begin{equation}
	\tau=0.064_{-0.016}^{+0.015},\qquad z_\mathrm{re} = 8.7_{-1.6}^{+1.4}	\,,
\end{equation}
in agreement with the current \Planck\ low-$\ell$ baseline (see Table~\ref{table:lowl_params}), even though this result only relies on the $\EE$ spectrum between $\ell=5$ and 20.

\section{High-$\ell$ baseline likelihood: \plik\label{app:hilbase} }

In this appendix, we provide detailed information on the \plik baseline likelihood used at high $\ell$. First we describe in Sect.~\ref{app:hil_covmat} the \plik covariance matrix, by providing the equations we have implemented, by giving results from some of the numerical tests we carried out, and by describing our procedure to deal with the excess variance (as compared to the prediction of our approximate analytical model) due to the point source mask. Section~\ref{app:plik-sims} validates the overall \plik implementation with Monte Carlo simulations of the full mission. For reference, Sect.~\ref{app:plik-val} gives the results of a large body of validation and stability tests on the actual data, including polarization in particular. We also discuss the numerical agreement of the temperature- and polarization-based results on base-\LCDM\ parameters. Section~\ref{app:co-added} describes how we calculate co-added CMB spectra from foreground-cleaned frequency power spectra. Section~\ref{app:pico} compares \plik cosmological results obtained using the \PICO\ or \CAMB\ codes. Finally, Sect.~\ref{app:margin-like} details how we marginalize over nuisance parameters to provide a fast but accurate CMB-only likelihood.

\subsection{Covariance matrix \label{app:hil_covmat} }

\subsubsection{Structure of the covariance matrix \label{app:hil_eqs} }

Here we summarize the mathematical formalism implemented
to calculate the pseudo-power spectrum covariance matrices for
temperature and polarization.

In the following, the fiducial power spectra $C_{\ell}$ are assumed to
be the smooth theory spectra multiplied by beam ($b$) and pixel window
function ($p$) for detectors $i$ and $j$,
\begin{equation}
C_{\ell}^{i, j} = b_{\ell}^{i} \, b_{\ell}^{j} \, p_{\ell}^2 
\left( C_{\ell}^{\mathrm{CMB}} + C_{\ell}^{\mathrm{FG}}(f_{i}, f_{j})
\right) \, ,
\end{equation}
where the $f_{k}$ denote the frequency dependence of the foreground
contribution.

We now present  the equations used to compute all the unique covariance matrix
polarization blocks that can be formed from temperature and $E$-mode
polarization maps (\citealp{2002MNRAS.336.1304H, 2003ApJS..148..135H,
  Efstathiou2004, 2005MNRAS.360..509C}; \citetalias{planck2013-p08}). They
approximate the variance of the biased pseudo-power spectrum
coefficients, before correcting for the effects of pixel window
function, beam, and mask.

\paragraph{$TTTT$ block:}
\begin{align}
\Var&(\hat{C}_{\ell}\hiTTij, \hat{C}_{\ellp}\hiTTpq) \nonumber\\
& \approx \sqrt{C_{\ell}\hiTTip C_{\ellp}\hiTTip C_{\ell}\hiTTjq
  C_{\ellp}\hiTTjq} \ \Xi_{\TT}^{\none, \none} \! \left[(i, p)\hiTT, (j,
  q)\hiTT \right]_{\ell \ellp} \nonumber\\
& + \sqrt{C_{\ell}\hiTTiq C_{\ellp}\hiTTiq C_{\ell}\hiTTjp
  C_{\ellp}\hiTTjp} \ \Xi_{\TT}^{\none, \none} \! \left[(i,q)\hiTT, (j,
  p)\hiTT\right]_{\ell \ellp} \nonumber\\
& + \sqrt{C_{\ell}\hiTTip C_{\ellp}\hiTTip} \ \Xi_{\TT}^{\none,
    } \! \left[(i, p)\hiTT, (j, q)\hiTT\right]_{\ell \ellp} \nonumber\\
& + \sqrt{C_{\ell}\hiTTjq C_{\ellp}\hiTTjq} \ \Xi_{\TT}^{\none,
  \TT} \! \left[(j, q)\hiTT, (i, p)\hiTT\right]_{\ell \ellp} \nonumber\\
& + \sqrt{C_{\ell}\hiTTiq C_{\ellp}\hiTTiq} \ \Xi_{\TT}^{\none,
  \TT} \! \left[(i, q)\hiTT, (j, p)\hiTT\right]_{\ell \ellp} \nonumber\\
& + \sqrt{C_{\ell}\hiTTjp C_{\ellp}\hiTTjp} \ \Xi_{\TT}^{\none, \TT} \! \left[(j,
   p)\hiTT, (i, q)\hiTT\right]_{\ell \ellp} \nonumber\\
& + \Xi_{\TT}^{\TT, \TT} \! \left[(i, p)\hiTT, (j, q)\hiTT\right]_{\ell \ellp} +
\Xi_{\TT}^{\TT, \TT} \! \left[(i, q)\hiTT, (j, p)\hiTT\right]_{\ell \ellp} \,.
\label{eq:hil_cov_mat_tttt_block}
\end{align}

\paragraph{$TTTE$ block:}
\begin{align}
\Var&(\hat{C}_{\ell}\hiTTij, \hat{C}_{\ellp}\hiTEpq) \nonumber\\
& \approx \frac{1}{2} \sqrt{C_{\ell}\hiTTip C_{\ellp}\hiTTip}
\left( C_{\ell}\hiTEjq + C_{\ellp}\hiTEjq \right) \ \Xi_{\TT}^{\none,
  \none} \! \left[(i, p)\hiTT, (j, q)\hiTP\right]_{\ell \ellp} \nonumber\\
& + \frac{1}{2} \sqrt{C_{\ell}\hiTTjp C_{\ellp}\hiTTjp}
\left( C_{\ell}\hiTEiq + C_{\ellp}\hiTEiq \right) \ \Xi_{\TT}^{\none,
  \none} \! \left[(i, q)\hiTP, (j, p)\hiTT\right]_{\ell \ellp} \nonumber\\
& + \frac{1}{2} \left( C_{\ell}\hiTEjq + C_{\ellp}\hiTEjq \right)
\ \Xi_{\TT}^{\none, \TT} \! \left[(j, q)\hiTP, (i, p)\hiTT\right]_{\ell \ellp}
\nonumber\\
& + \frac{1}{2} \left( C_{\ell}\hiTEiq + C_{\ellp}\hiTEiq \right)
\ \Xi_{\TT}^{\none, \TT} \! \left[(i, q)\hiTP, (j, p)\hiTT\right]_{\ell \ellp} \,.
\label{eq:hil_cov_mat_ttte_block}
\end{align}

\paragraph{$TETE$ block}
\begin{align}
\Var&(\hat{C}_{\ell}\hiTEij, \hat{C}_{\ellp}\hiTEpq) \nonumber\\
& \approx \sqrt{C_{\ell}\hiTTip C_{\ellp}\hiTTip C_{\ell}\hiEEjq
  C_{\ellp}\hiEEjq} \ \Xi_{\TE}^{\none, \none} \! \left[(i, p)\hiTT, (j,
  q)\hiPP\right]_{\ell \ellp} \nonumber\\
& + \frac{1}{2} \left( C_{\ell}\hiTEiq C_{\ellp}\hiTEjp +
C_{\ell}\hiTEjp C_{\ellp}\hiTEiq \right) \ \Xi_{\TT}^{\none,
  \none} \! \left[(i, q)\hiTP, (j, p)\hiPT\right]_{\ell \ellp} \nonumber\\
& + \sqrt{C_{\ell}\hiTTip C_{\ellp}\hiTTip} \ \Xi_{\TE}^{\none,
  \PP} \! \left[(i, p)\hiTT, (j, q)\hiPP\right]_{\ell \ellp} \nonumber\\
& + \sqrt{C_{\ell}\hiEEjq C_{\ellp}\hiEEjq} \ \Xi_{\TE}^{\none,
  \TT} \! \left[(j, q)\hiPP, (i, p)\hiTT\right]_{\ell \ellp} \nonumber\\
& + \Xi_{\TE}^{\TT, \PP} \! \left[(i, p)\hiTT, (j, q)\hiPP\right]_{\ell \ellp} \,.
\label{eq:hil_cov_mat_tete_block}
\end{align}

\paragraph{$TTEE$ block:}
\begin{align}
\Var&(\hat{C}_{\ell}\hiTTij, \hat{C}_{\ellp}\hiEEpq) \nonumber\\
& \approx \frac{1}{2} \left( C_{\ell}\hiTEip C_{\ellp}\hiTEjq +
C_{\ell}\hiTEjq C_{\ellp}\hiTEip \right) \ \Xi_{\TT}^{\none,
  \none} \! \left[(i, p)\hiTP, (j, q)\hiTP\right]_{\ell \ellp} \nonumber\\
& + \frac{1}{2} \left( C_{\ell}\hiTEiq C_{\ellp}\hiTEjp +
C_{\ell}\hiTEjp C_{\ellp}\hiTEiq \right) \ \Xi_{\TT}^{\none,
  \none} \! \left[(i,q)\hiTP, (j, p)\hiTP\right]_{\ell \ellp}
\label{eq:hil_cov_mat_ttee_block}
\end{align}

\paragraph{$TEEE$ block:}
\begin{align}
\Var&(\hat{C}_{\ell}\hiTEij, \hat{C}_{\ellp}\hiEEpq) \nonumber\\
& \approx \frac{1}{2} \sqrt{C_{\ell}\hiEEjq C_{\ellp}\hiEEjq}
\left(C_{\ell}\hiTEip + C_{\ellp}\hiTEip \right) \ \Xi_{\EE}^{\none,
  \none} \! \left[(i, p)\hiTP, (j, q)\hiPP\right]_{\ell \ellp} \nonumber\\
& + \frac{1}{2} \sqrt{C_{\ell}\hiEEjp C_{\ellp}\hiEEjp}
\left(C_{\ell}\hiTEiq + C_{\ellp}\hiTEiq \right) \ \Xi_{\EE}^{\none,
  \none} \! \left[(i,q)\hiTP, (j, p)\hiPP\right]_{\ell \ellp} \nonumber\\
& + \frac{1}{2} \left( C_{\ell}\hiTEip + C_{\ellp}\hiTEip \right)
\ \Xi_{\EE}^{\none, \PP} \! \left[(i, p)\hiTP, (j, q)\hiPP\right]_{\ell \ellp} \nonumber\\
& + \frac{1}{2} \left( C_{\ell}\hiTEiq + C_{\ellp}\hiTEiq \right)
\ \Xi_{\EE}^{\none, \PP} \! \left[(i, q)\hiTP, (j, p)\hiPP\right]_{\ell \ellp} \,.
\label{eq:hil_cov_mat_teee_block}
\end{align}

\paragraph{$EEEE$ block:}
\begin{align}
\Var&(\hat{C}_{\ell}\hiEEij, \hat{C}_{\ellp}\hiEEpq) \nonumber\\
& \approx \sqrt{C_{\ell}\hiEEip C_{\ellp}\hiEEip C_{\ell}\hiEEjq
  C_{\ellp}\hiEEjq} \ \Xi_{\EE}^{\none, \none} \! \left[(i, p)\hiPP, (j,
  q)\hiPP\right]_{\ell \ellp} \nonumber\\
& + \sqrt{C_{\ell}\hiEEiq C_{\ellp}\hiEEiq C_{\ell}\hiEEjp
  C_{\ellp}\hiEEjp} \ \Xi_{\EE}^{\none, \none} \! \left[(i, q)\hiPP, (j,
  p)\hiPP\right]_{\ell \ellp} \nonumber\\
& + \sqrt{C_{\ell}\hiEEip C_{\ellp}\hiEEip} \ \Xi_{\EE}^{\none, \PP} \! \left[(i,
  p)\hiPP, (j, q)\hiPP\right]_{\ell \ellp} \nonumber\\
& + \sqrt{C_{\ell}\hiEEjq C_{\ellp}\hiEEjq} \ \Xi_{\EE}^{\none, \PP} \! \left[(j,
  q)\hiPP, (i, p)\hiPP\right]_{\ell \ellp} \nonumber\\
& + \sqrt{C_{\ell}\hiEEiq C_{\ellp}\hiEEiq} \ \Xi_{\EE}^{\none, \PP} \! \left[(i,
  q)\hiPP, (j, p)\hiPP\right]_{\ell \ellp} \nonumber\\
& + \sqrt{C_{\ell}\hiEEjp C_{\ellp}\hiEEjp} \ \Xi_{\EE}^{\none, \PP} \! \left[(j,
  p)\hiPP, (i, q)\hiPP\right]_{\ell \ellp} \nonumber\\
& + \Xi_{\EE}^{\PP, \PP} \! \left[(i, p)\hiPP, (j, q)\hiPP\right]_{\ell \ellp} +
\Xi_{\EE}^{\PP, \PP} \! \left[(i, q)\hiPP, (j, p)\hiPP\right]_{\ell \ellp} \,.
\label{eq:hil_cov_mat_eeee_block}
\end{align}

In Eqs.~\ref{eq:hil_cov_mat_tttt_block}--\ref{eq:hil_cov_mat_eeee_block}, we have introduced the projector functions
$\Xi_{\TT}, \, \Xi_{\EE}$, and $\Xi_{\TE}$ to describe the coupling
between multipoles induced by the mask,
\begin{multline}
\Xi_{\TT}^{X, Y} \! \left[(i, j)^{\alpha}, (p, q)^{\beta}\right]_{\ell_1 \ell_2} =
\sum_{\ell_3} \frac{2 \ell_3 + 1}{4 \pi}
\wigner{\ell_1}{\ell_2}{\ell_3}{0}{0}{0}^2 \\
\times W^{X, Y} \! \left[(i, j)^{\alpha}, (p,
  q)^{\beta}\right]_{\ell_3} \, ,
\label{eq:hil_cov_mat_projector_tt}
\end{multline}
\begin{multline}
\Xi_{\EE}^{X, Y} \! \left[(i, j)^{\alpha}, (p, q)^{\beta}\right]_{\ell_1 \ell_2} =
\sum_{\ell_3} \frac{2 \ell_3 + 1}{16 \pi} \left(1 + (-1)^{\ell_1 +
  \ell_2 + \ell_3} \right)^2 \\
\times \wigner{\ell_1}{\ell_2}{\ell_3}{-2}{2}{0}^2  W^{X, Y} \!
\left[(i, j)^{\alpha}, (p, q)^{\beta}\right]_{\ell_3} \, ,
\label{eq:hil_cov_mat_projector_ee}
\end{multline}
and
\begin{multline}
\Xi_{\TE}^{X, Y} \! \left[(i, j)^{\alpha}, (p, q)^{\beta}\right]_{\ell_1 \ell_2} =
\sum_{\ell_3} \frac{2 \ell_3 + 1}{8 \pi} \left(1 + (-1)^{\ell_1 +
  \ell_2 + \ell_3} \right) \\
\times \wigner{\ell_1}{\ell_2}{\ell_3}{0}{0}{0}
\wigner{\ell_1}{\ell_2}{\ell_3}{-2}{2}{0} W^{X, Y} \! \left[(i,
  j)^{\alpha} (p, q)^{\beta}\right]_{\ell_3} \, ,
\label{eq:hil_cov_mat_projector_te}
\end{multline}
where $X, Y \in \{\none, \TT, \PP \}$, and $\alpha, \beta \in \{\TT,
\TP, \PT, \PP \}$. They make use of window functions $W$,
\begin{equation}
W^{\none, \none} \! \left[(i, j)^{\alpha}, (p,
  q)^{\beta}\right]_{\ell} = \frac{1}{2\ell + 1} \sum_{m}
w^{\none}_{\ell m}(i, j)^{\alpha} w^{\ast \, \none}_{\ell m}(p,
q)^{\beta} \, ,
\end{equation}
\begin{equation}
W^{\none, \TT} \! \left[(i, j)^{\alpha}, (p, q)\hiTT\right]_{\ell} =
\frac{1}{2\ell + 1} \sum_{m} w^{\none}_{\ell m}(i, j)^{\alpha}
w^{\ast \, \II}_{\ell m}(p, q)\hiTT  \, ,
\end{equation}
\begin{multline}
W^{\none, \PP} \! \left[(i, j)^{\alpha}, (p, q)\hiPP\right]_{\ell} =
\frac{1}{2\ell + 1} \sum_{m} \\ \frac{1}{2} \left( w^{\none}_{\ell
  m}(i, j)^{\alpha} w^{\ast \, \QQ}_{\ell m}(p, q)\hiPP \right.
 + \left. w^{\none}_{\ell m}(i, j)^{\alpha} w^{\ast \, \UU}_{\ell
   m}(p, q)\hiPP \right)  \, ,
\end{multline}
\begin{equation}
W^{\TT, \TT} \! \left[(i, j)\hiTT, (p, q)\hiTT\right]_{\ell} =
\frac{1}{2\ell + 1} \sum_{m} w^{\II}_{\ell m}(i, j)\hiTT  w^{\ast \,
  \II}_{\ell m}(p, q)\hiTT  \, ,
\end{equation}
\begin{multline}
W^{\TT, \PP} \! \left[(i, j)\hiTT, (p, q)\hiPP\right]_{\ell} =
\frac{1}{2\ell + 1} \sum_{m} \\ \frac{1}{2} \left( w^{\II}_{\ell m}(i,
j)\hiTT w^{\ast \, \QQ}_{\ell m}(p, q)\hiPP + w^{\II}_{\ell m}(i,
j)\hiTT w^{\ast \, \UU}_{\ell m}(p, q)\hiPP \right)  \, ,
\end{multline}
and
\begin{align}
W^{\PP, \PP} &\! \left[(i, j)\hiPP, (p, q)\hiPP\right]_{\ell} =
\frac{1}{2\ell + 1} \sum_{m} \nonumber\\
& \frac{1}{4} \left( w^{\QQ}_{\ell m}(i, j)\hiPP
w^{\ast \, \QQ}_{\ell m}(p, q)\hiPP + w^{\UU}_{\ell m}(i, j)\hiPP
w^{\ast \, \UU}_{\ell m}(p,q)\hiPP \right. \nonumber\\
& \left. + w^{\QQ}_{\ell m}(i, j)\hiPP w^{\ast \, \UU}_{\ell m}(p,
q)\hiPP + w^{\UU}_{\ell m}(i, j)\hiPP w^{\ast \, \QQ}_{\ell m}(p,
q)\hiPP \right) \, .
\end{align}

In the above expressions, we defined the spherical harmonic
coefficients of the effective weight maps $w^{\none}$,
\begin{equation}
w^{\none}_{\ell m}(i, j)\hiTT = \sum_{p=1}^{N_{\mathrm{pix}}} m_p^{i,
  \singleT} m_p^{j, \singleT} Y^{\ast}_{\ell m}(\hat{\vec{n}}_p) \Omega_p
\, ,
\end{equation}
\begin{equation}
w^{\none}_{\ell m}(i, j)\hiPP = \sum_{p=1}^{N_{\mathrm{pix}}} m_p^{i, \singleP}
m_p^{j, \singleP} Y^{\ast}_{\ell m}(\hat{\vec{n}}_p) \Omega_p \, ,
\end{equation}
\begin{equation}
w^{\none}_{\ell m}(i, j)\hiTP = \sum_{p=1}^{N_{\mathrm{pix}}} m_p^{i, \singleT}
m_p^{j, \singleP} Y^{\ast}_{\ell m}(\hat{\vec{n}}_p) \Omega_p \, ,
\end{equation}
and
\begin{equation}
w^{\none}_{\ell m}(i, j)\hiPT = \sum_{p=1}^{N_{\mathrm{pix}}} m_p^{i, \singleP}
m_p^{j, \singleT} Y^{\ast}_{\ell m}(\hat{\vec{n}}_p) \Omega_p \, ,
\end{equation}
where $m^{\singleT}$ is the temperature mask (Stokes $I$),
$m^{\singleP}$ the polarization mask (Stokes $Q$ and $U$), and
$\Omega_p$ the solid angle of pixel $p$.

 Accordingly, the noise-variance-weighted maps $w^{\II}$, $w^{\QQ}$, and
$w^{\UU}$ are
\begin{equation}
w^{\II}_{\ell m}(i, j)\hiTT = \delta_{i, j} \sum_{p=1}^{N_{\mathrm{pix}}}
\left( \sigma^{II}_p \right)^2 m_p^{i, \singleT} m_p^{j, \singleT}
Y^{\ast}_{\ell m}(\hat{\vec{n}}_p) \Omega_p^2 \, ,
\end{equation}
\begin{equation}
w^{\QQ}_{\ell m}(i, j)\hiPP = \delta_{i, j} \sum_{p=1}^{N_{\mathrm{pix}}}
\left( \sigma^{QQ}_p \right)^2 m_p^{i, \singleP} m_p^{j, \singleP}
Y^{\ast}_{\ell m}(\hat{\vec{n}}_p) \Omega_p^2 \, ,
\end{equation}
and
\begin{equation}
w^{\UU}_{\ell m}(i, j)\hiPP = \delta_{i, j} \sum_{p=1}^{N_{\mathrm{pix}}}
\left( \sigma^{UU}_p \right)^2 m_p^{i, \singleP} m_p^{j, \singleP}
Y^{\ast}_{\ell m}(\hat{\vec{n}}_p) \Omega_p^2 \, ,
\end{equation}
where the $\sigma_p^2$s are the noise variances in pixel $p$ in the given
Stokes map, and the Kronecker symbols $\delta_{i, j}$ ensure that there is
only a noise contribution if the two detectors $i$ and $j$ are
identical.

In the spherical harmonic representation of the noise-variance-weighted
window functions that appear  in Eqs.~(\ref{eq:hil_cov_mat_projector_tt}--\ref{eq:hil_cov_mat_projector_te}) it is possible to
take into account noise correlations approximately. Following
\citetalias{planck2013-p08} and given the characterization of the observed
noise power spectra discussed in Sect.~\ref{sec:noise_model}, we
multiply the projector functions $\Xi^{X,Y}_{\ell_1 \ell_2}$ for each
factor of $X, Y \in \{\TT, \PP \}$ by an additional rescaling
coefficient,
\begin{equation}
r_{\ell_1 \ell_2} = \sqrt{\frac{N^{\mathrm{data}}_{\ell_1}
    N^{\mathrm{data}}_{\ell_2}}{N^{\mathrm{white}}_{\ell_1}
      N^{\mathrm{white}}_{\ell_2}}} \, .
\end{equation}
Here, $N^{\mathrm{data}}_{\ell} / N^{\mathrm{white}}_{\ell}$ is the
ratio of the observed noise power spectrum to the white-noise power
spectrum predicted by the pixel noise variance values $\sigma_p^2$.

\subsubsection{Mask deconvolution}

In a last step, we correct the individual covariance matrix blocks for
the effect of pixel window function, beam, and mask. Using the
coupling matrices \citep{Hietal02, kogut2003},
\begin{equation}
M\hiTT(i, j)_{\ell_1 \ell_2} = (2\ell_2 + 1) \sum_{\ell_3} \frac{2
  \ell_3 + 1}{4 \pi} \wigner{\ell_1}{\ell_2}{\ell_3}{0}{0}{0}^2
V\hiTT(i, j)_{\ell_3} \, ,
\end{equation}
\begin{multline}
M\hiEE(i, j)_{\ell_1 \ell_2} = (2\ell_2 + 1) \sum_{\ell_3} \frac{2
  \ell_3 + 1}{16 \pi} \left(1 + (-1)^{\ell_1 + \ell_2 + \ell_3}
\right)^2 \\ \times \wigner{\ell_1}{\ell_2}{\ell_3}{-2}{2}{0}^2
V\hiPP(i,j)_{\ell_3} \, ,
\end{multline}
\begin{multline}
M\hiTE(i, j)_{\ell_1 \ell_2} = (2\ell_2 + 1) \sum_{\ell_3} \frac{2
  \ell_3 + 1}{8 \pi} \left(1 + (-1)^{\ell_1 + \ell_2 + \ell_3} \right) \\
\times \wigner{\ell_1}{\ell_2}{\ell_3}{0}{0}{0}
\wigner{\ell_1}{\ell_2}{\ell_3}{-2}{2}{0} V\hiTP(i,j)_{\ell_3} \, ,
\end{multline}
\begin{multline}
M\hiET(i, j)_{\ell_1 \ell_2} = (2\ell_2 + 1) \sum_{\ell_3} \frac{2
  \ell_3 + 1}{8 \pi} \left(1 + (-1)^{\ell_1 + \ell_2 + \ell_3} \right) \\
\times \wigner{\ell_1}{\ell_2}{\ell_3}{0}{0}{0}
\wigner{\ell_1}{\ell_2}{\ell_3}{-2}{2}{0} V\hiPT(i,j)_{\ell_3} \, ,
\end{multline}
where
\begin{equation}
V\hiTT(i, j)_{\ell} = \frac{1}{2\ell + 1} \sum_{m} m_{\ell m}^{i,
  \singleT} m_{\ell m}^{\ast \, j, \singleT} \, ,
\end{equation}
\begin{equation}
V\hiPP(i, j)_{\ell} = \frac{1}{2\ell + 1} \sum_{m} m_{\ell m}^{i,
  \singleP} m_{\ell m}^{\ast \, j, \singleP} \, ,
\end{equation}
\begin{equation}
V\hiTP(i, j)_{\ell} = \frac{1}{2\ell + 1} \sum_{m} m_{\ell m}^{i,
  \singleT} m_{\ell m}^{\ast \, j, \singleP} \, ,
\end{equation}
and
\begin{equation}
V\hiPT(i, j)_{\ell} = \frac{1}{2\ell + 1} \sum_{m} m_{\ell m}^{i,
  \singleP} m_{\ell m}^{\ast \, j, \singleT} \, ,
\end{equation}
we obtain the final result for the deconvolved covariance matrix,
\begin{multline}
\Var(\hat{C}_{\ell}^{XY \ i, j}, \hat{C}_{\ellp}^{ZW \ p,
  q})^{\mathrm{dec}} = \left[ M^{XY}(i, j)^{-1} \,
  \Var(\hat{C}^{XY \ i, j}, \hat{C}^{ZW \ p, q}) \right. \\
\times \left. \left(M^{ZW}(p, q)^{-1}\right)^{\dagger} \right]_{\ell \ellp}
\left/ \left( b^{X \ i}_{\ell} \, b^{Y \ j}_{\ell} \, b^{Z \ p}_{\ellp} \,
b^{W \ q}_{\ellp} \, p_{\ell}^{2 \ XY} \, p_{\ellp}^{2 \ ZW} \right)
\right. \, .
\end{multline}

\subsubsection{Validation of the implementation \label{app:hil_cov_mat_validation}}

\begin{figure*}[htbp] 
 \includegraphics[width=0.49\textwidth]{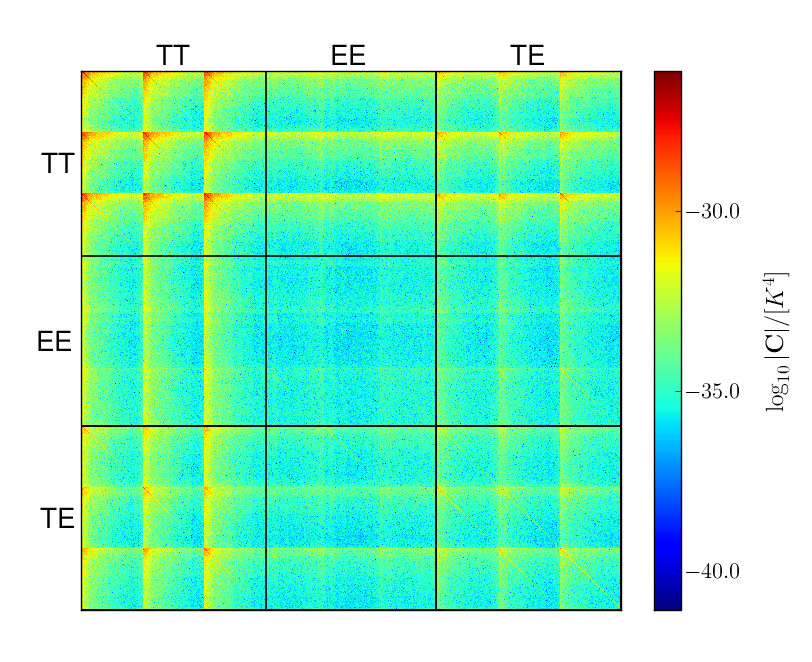}
 \includegraphics[width=0.49\textwidth]{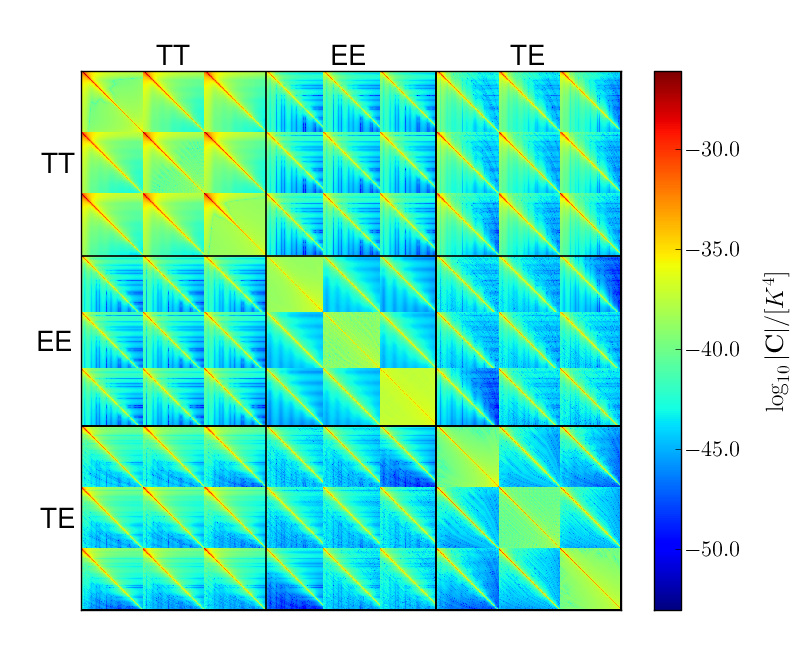}
 \caption{Combined $C_{\ell}$-covariance matrices comprising the
   $TTTT$ ({upper left sub-block}), $EEEE$ ({middle sub-block}), and
   $TETE$ ({lower right sub-block}) covariances and their
   cross-correlations. \emph{Left}: Empirical
   covariance. \emph{Right}: Analytic covariance. We note the different scales; 
   despite visual appearance, the diagonals are in good
   agreement.}
  \label{fig:plik_cov_mat_sim_comparison}
\end{figure*}

 We verified the numerical implementation of the pipeline used to
compute covariance matrices by means of Monte Carlo simulations.
Specifically, we generated a set of $10 \, 000$ simulated maps for the
four \HFI detector sets 143-ds1, 143-ds2, 217-ds1, and 217-ds2. The
simulations included CMB and an isotropic frequency-dependent
foreground component, convolved with effective beam and pixel window
functions. To each map, we added a realization of anisotropic,
correlated noise.

In this test, we used a Galactic mask that leaves 40\,\% of the sky for
analysis at both frequencies and neglected the point source mask
usually applied to temperature data. We then computed a total of $120
\, 000$ cross power spectra and constructed empirical covariance
matrices for the 21 unique detector combinations that can be built
from the four channels. Being based on at least $10 \, 000$
simulations each, the covariance matrix estimates reach an intrinsic
relative precision of 1\,\% or better.

\begin{figure}[htbp] 
  \includegraphics[width=0.49\textwidth]{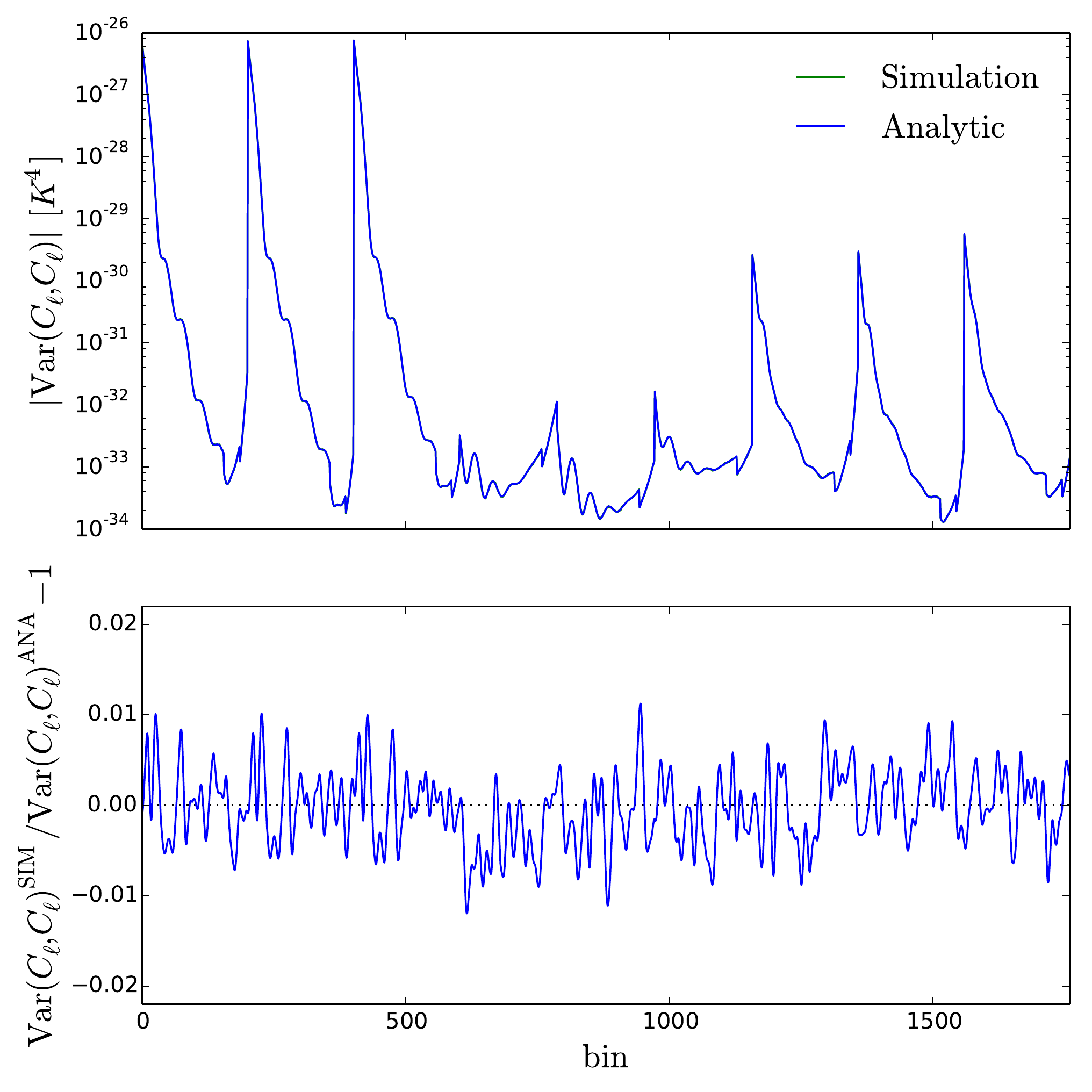}
  \caption{\textit{Top}: Diagonal elements of the empirical (green line)
    and analytic (blue line) covariance matrices; the two lines are
    indistinguishable. \textit{Bottom}: Ratio of the two estimates:
    the ratios differ from unity by $<1$\,\% over the full multipole
    range for all frequency combinations and polarization blocks.}
  \label{fig:plik_cov_mat_diag_comparison}
\end{figure}

We then compared the empirical covariance matrix to its approximate
analytic counterpart computed using identical input parameters. To do
so, we applied the standard post-processing procedure discussed in
Sect.~\ref{sec:covariance} to produce frequency averaged covariance
matrices for all frequency combinations at 143 and 217\,GHz. For the
analysis, we adopted frequency-independent multipole ranges $100 \le
\ell \le 2500$ for $\TT$ and $\TE$, and $100 \le \ell \le 2000$ for $\EE$. In a
final step, we reduced the size of the matrices by binning. The
temperature and polarization blocks were then combined into the single
matrices shown in Fig.~\ref{fig:plik_cov_mat_sim_comparison}. We note that, owing to the Monte Carlo noise floor, the colour scales are different, which may be misleading, since the diagonals appear to be fairly different, which is actually not the case. Indeed, Fig.~\ref{fig:plik_cov_mat_diag_comparison} compares the
diagonal elements of the covariance matrix, and shows that
for all polarization components and over the full multipole range, there is good agreement between the two covariance matrices, verifying the implementation of the equations summarized in the previous section, and their accuracy.

\subsubsection{Excess variance induced by the point-source mask  \label{app:hil_pts_mask_correction}}

The approximations used in the calculation of the covariance matrix
assume that the power spectra of the masks decline rapidly, and therefore require a conservative
apodization scheme at the expense of a reduction in the sky fraction
available for analysis. The point-source masks used in the temperature
analysis excise large numbers of sources with an approximately
isotropic distribution. Owing to their high number, only a severely
reduced apodization of individual holes is feasible in practice
(cf.\ Sect.~\ref{sec:masks}). As a consequence, the power spectrum of the
combined Galactic and point-source mask flattens and the precision of
the approximation deteriorates noticeably, leading to systematic
errors in the calculated analytical covariance matrices.

\begin{figure}[htbp] 
  \includegraphics[width=0.49\textwidth]{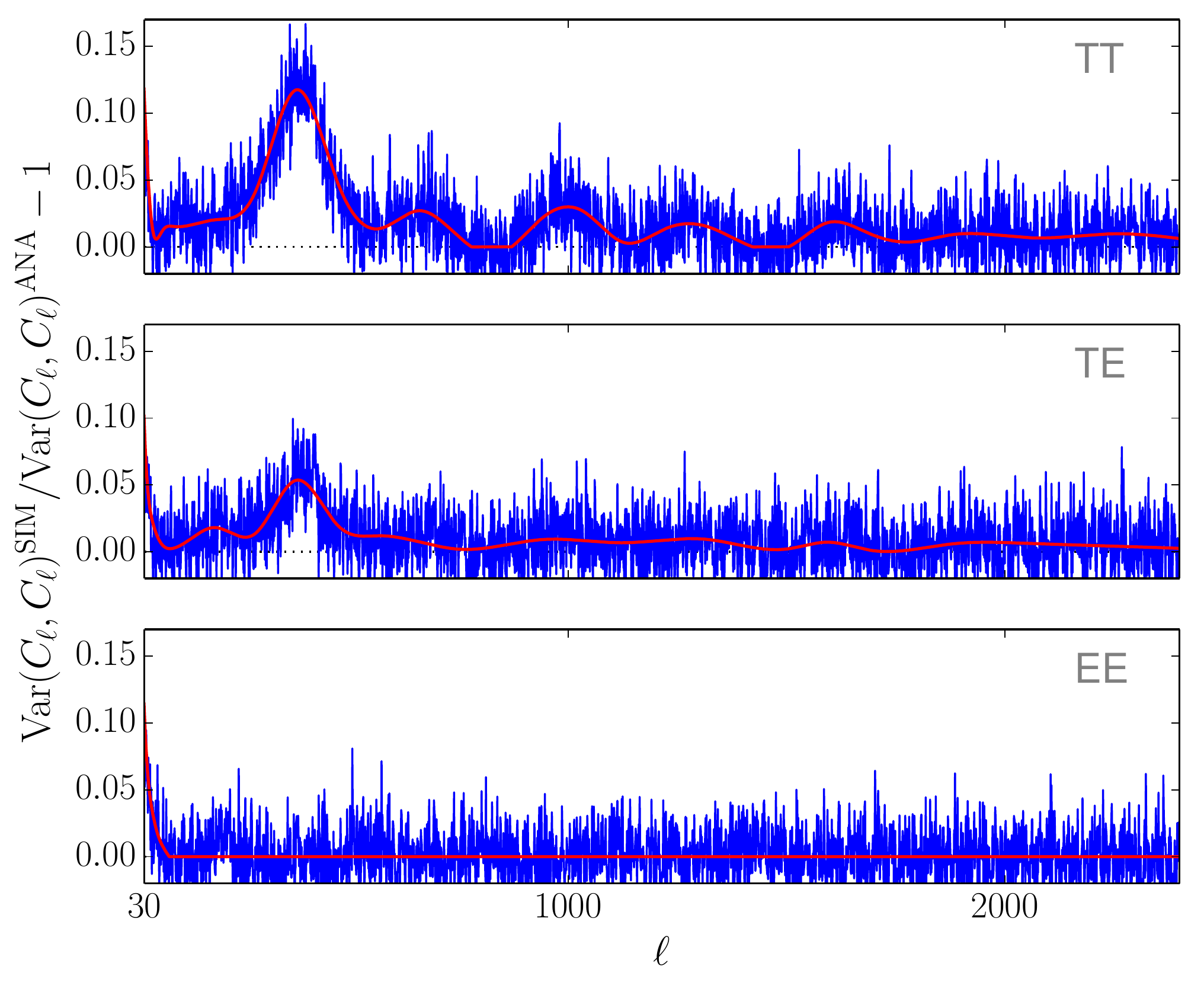}
  \caption{ Excess
    variance induced by the temperature point-source mask. The graphs compare the diagonal elements of the empirical and analytical
    power spectrum covariance matrices ({blue lines}) for $TT$
    ({upper panel}), $TE$ ({middle panel}), and $EE$ ({lower
      panel}), and show deviations at the 10\,\% level. The red lines are
    smooth fits based on cubic splines.}
  \label{fig:plik_cov_pts_correction}
\end{figure}

Here, we propose a heuristic approach to capture the variance modulations
introduced by the point-source masks. In a first step, we use Monte
Carlo simulations to quantify the level of mismatch between analytical
and empirical power spectra variances. Since the point-source mask is
frequency dependent, we simulate 5000 realizations of the six
half-mission CMB and foreground maps, without noise contribution, at
100, 143, and 217\,GHz. Using the reference Galactic and point-source
masks in temperature, and Galactic masks in polarization
(Sect.~\ref{sec:masks}), we compute power spectra and construct
empirical covariance matrices.

A comparison with the analytic covariance matrices reveals that the
point-source mask has introduced excess variance that is not fully
captured by the analytical approximation. In
Fig.~\ref{fig:plik_cov_pts_correction} we plot results for the $217
\times 217$\,GHz power spectrum variance, finding a deviation of up to
about 10\,\% at $\ell \approx 400$, with characteristic oscillating
features in the $\TT$ and, to a lesser extent, in the $\TE$ power spectrum
variance. Furthermore, on large scales ($\ell \la 50$), the
approximations start to break down in both temperature and polarization, a known feature of pseudo-power spectrum estimators
\citep[\eg][]{Efstathiou2004}.

In the signal-dominated regime, the analytical approximations of the
covariance matrices are proportional to the square of the fiducial
power spectrum $C_\ell$ (Eqs.~\ref{eq:hil_cov_mat_tttt_block}--\ref{eq:hil_cov_mat_eeee_block}). Using spline
fits to the variance ratios, we obtain correction factors that
describe the excess scatter introduced by the point-source masks. We then
multiply the fiducial power spectrum by the square-root of this ratio,
cancelling the observed mismatch in the variance to first order.

\subsection{\plik joint likelihood simulations} \label{app:plik-sims}

In Sect.~\ref{sec:valid-sims} we discussed the 300 simulations
performed to validate the overall implementation and our
approximations for \plik \TT. Here we complement that section with
additional results for the full \plikTTTEEE\ joint likelihood \rev{\st{from
which we have excluded the $EE$ $100\times 100$, $143\times 143$, and
$217 \times 217$ spectra and corresponding covariance matrix blocks
due to the presence of low-$\ell$ correlated noise in the
half-mission-1 $\times$ half-mission-2 cross-spectra from simulated
maps. This as-yet unaccounted-for correlated noise signal in the
simulations is likely attributable to a destriping effect. In order to
account for this source of bias consistently,  we would need to
introduce an additional correlated noise component in \plik, and we leave this to future work. We note that another
improvement to the joint likelihood analysis might be made by
incorporating the possibility of $\TE$ detector-noise covariance terms
which are assumed to vanish in the existing covariance matrix
calculation (although they are not necessarily completely negligible in the numerical covariance matrix calculable from the existing set of 300 simulations)}}.

\begin{figure*}[h!] 
\begin{center}
\includegraphics[width=0.49\columnwidth]{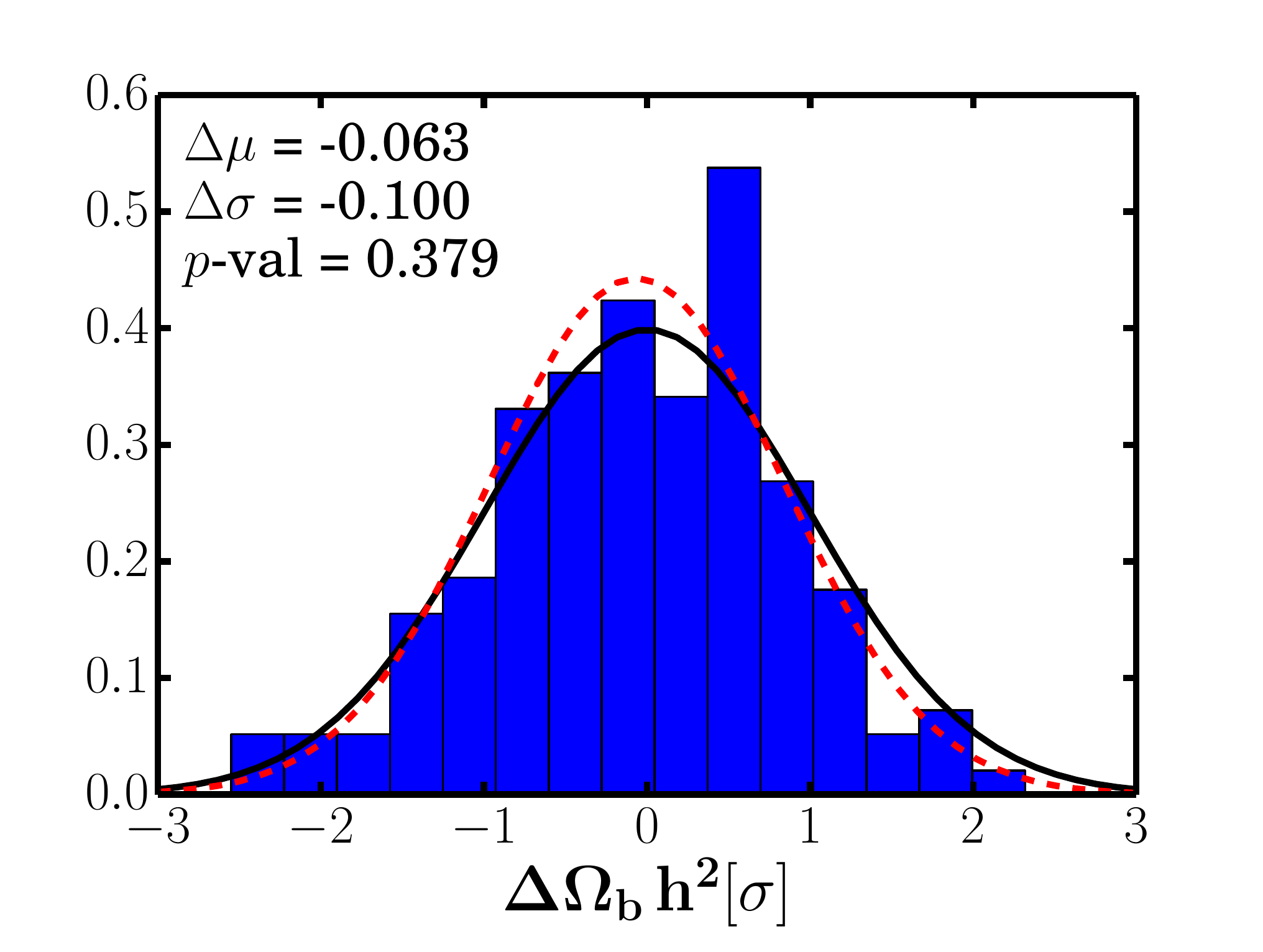}
\includegraphics[width=0.49\columnwidth]{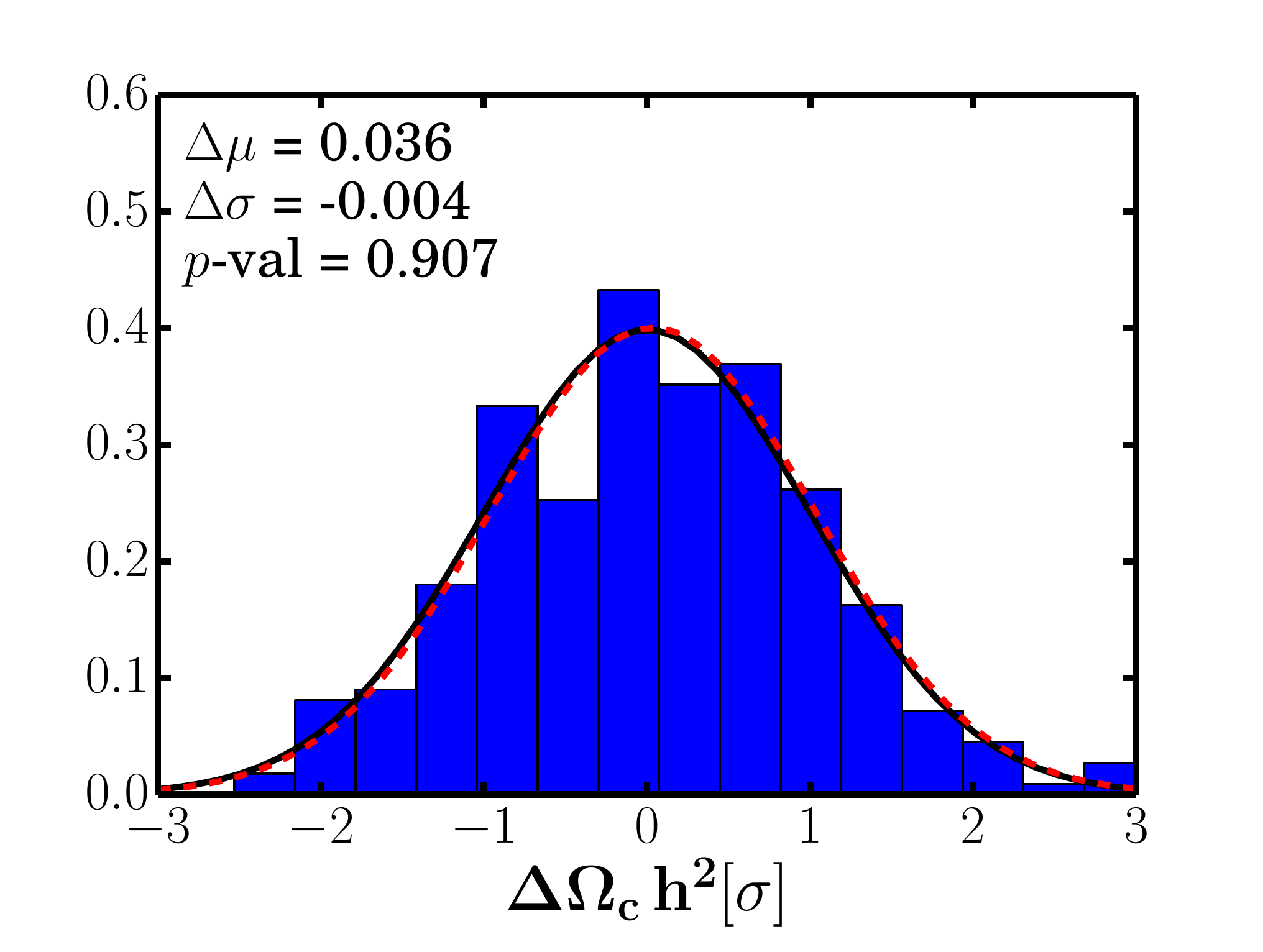}
\includegraphics[width=0.49\columnwidth]{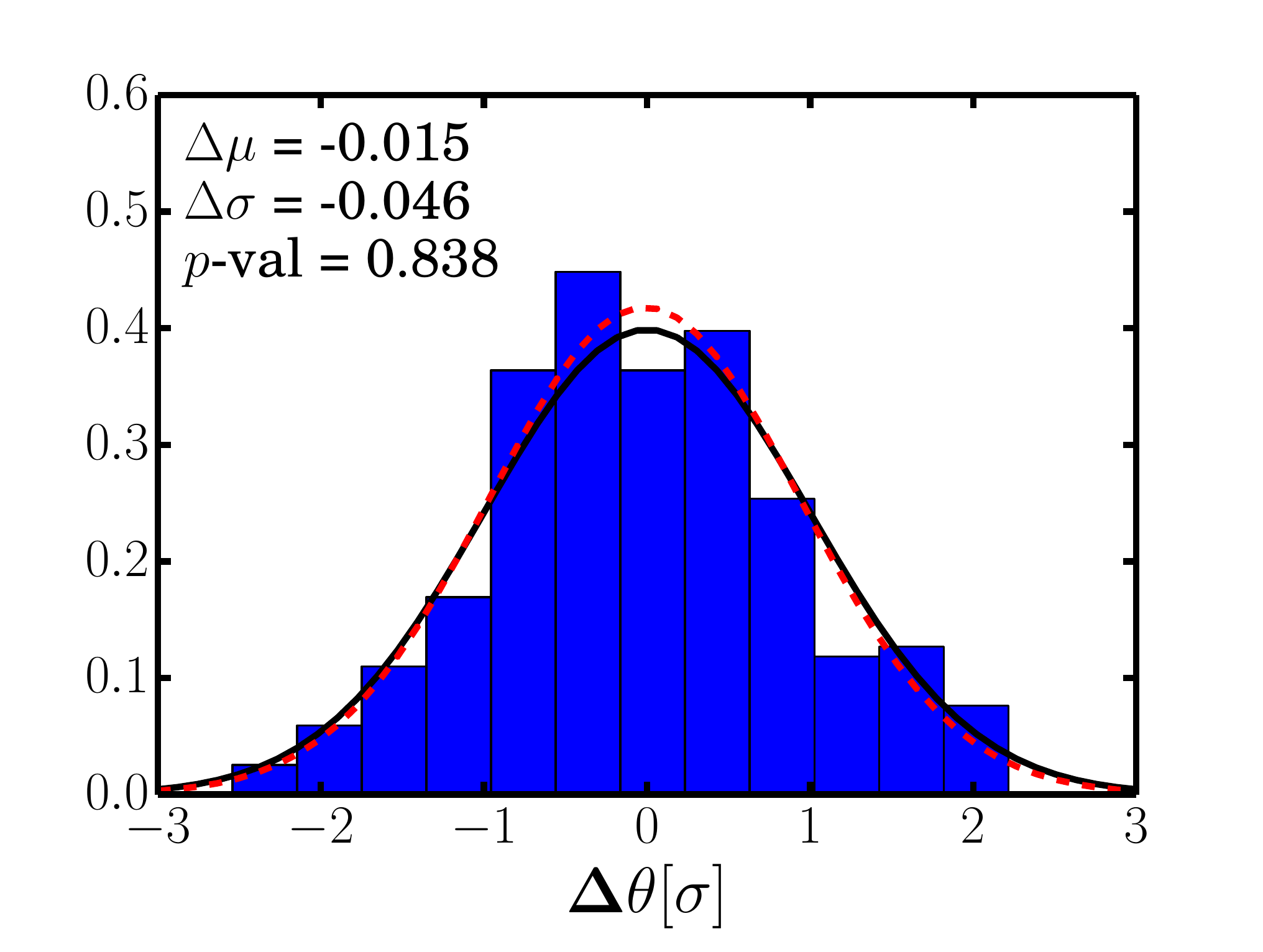}
\includegraphics[width=0.49\columnwidth]{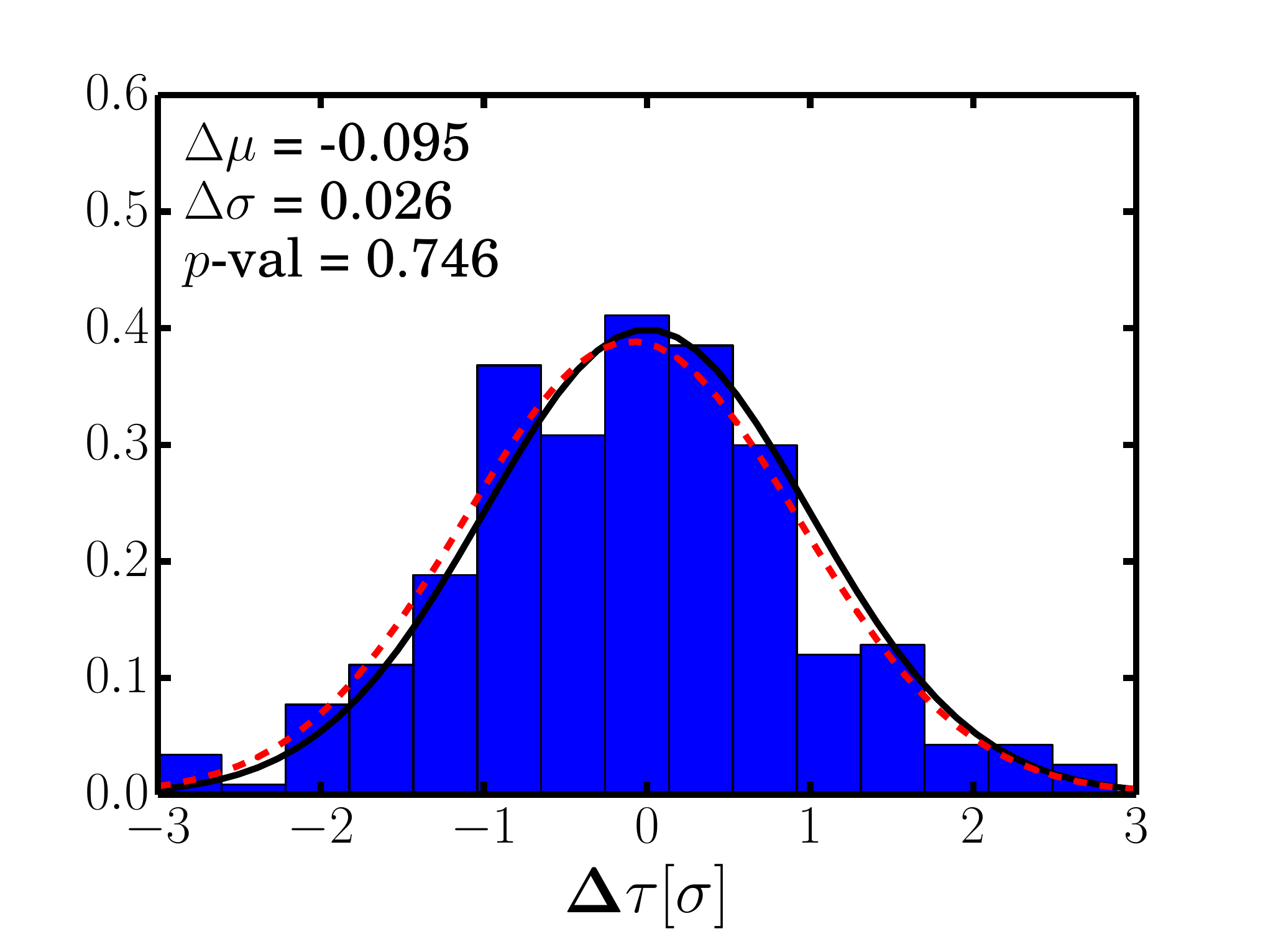}\\
\includegraphics[width=0.49\columnwidth]{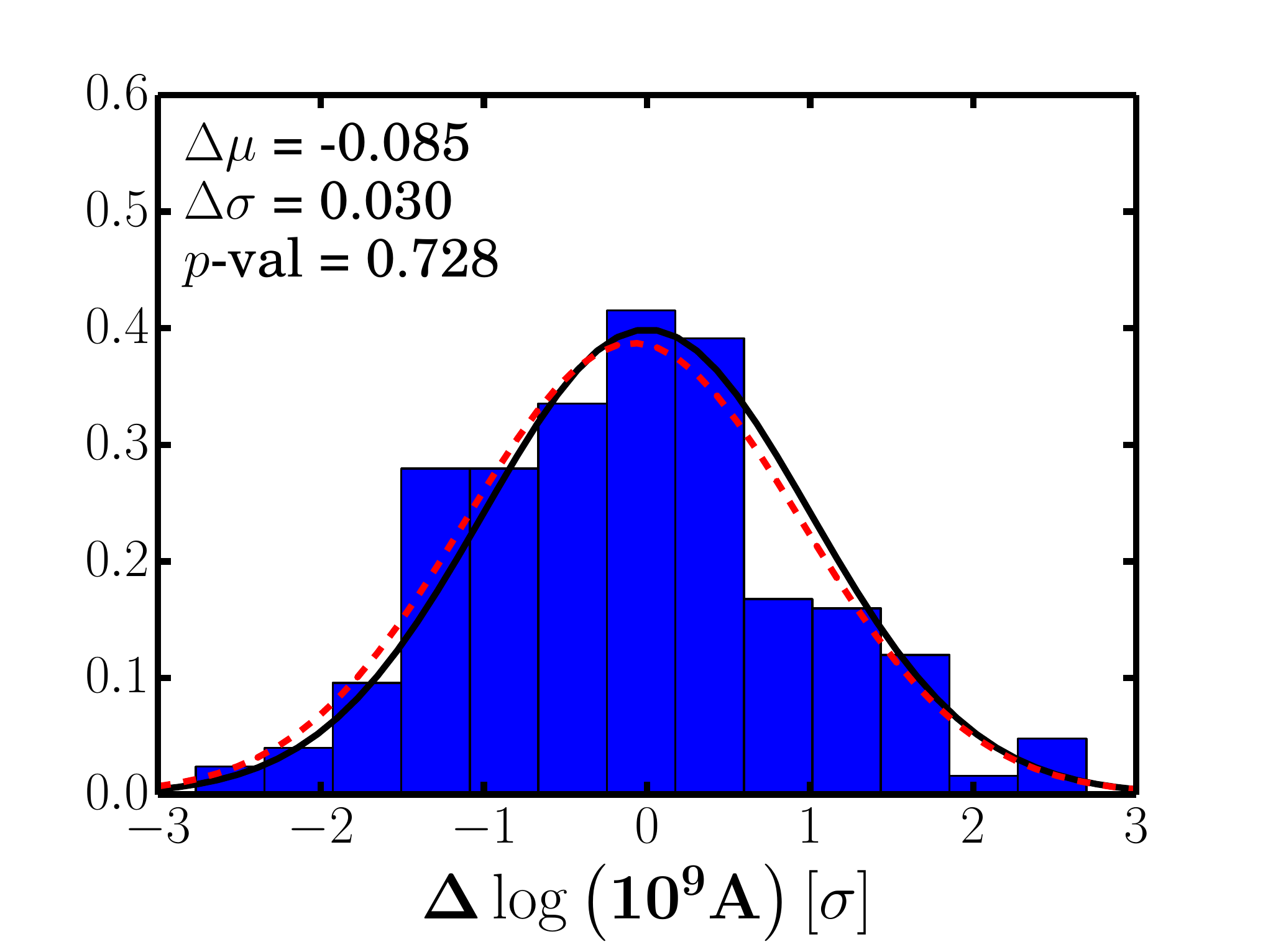}
\includegraphics[width=0.49\columnwidth]{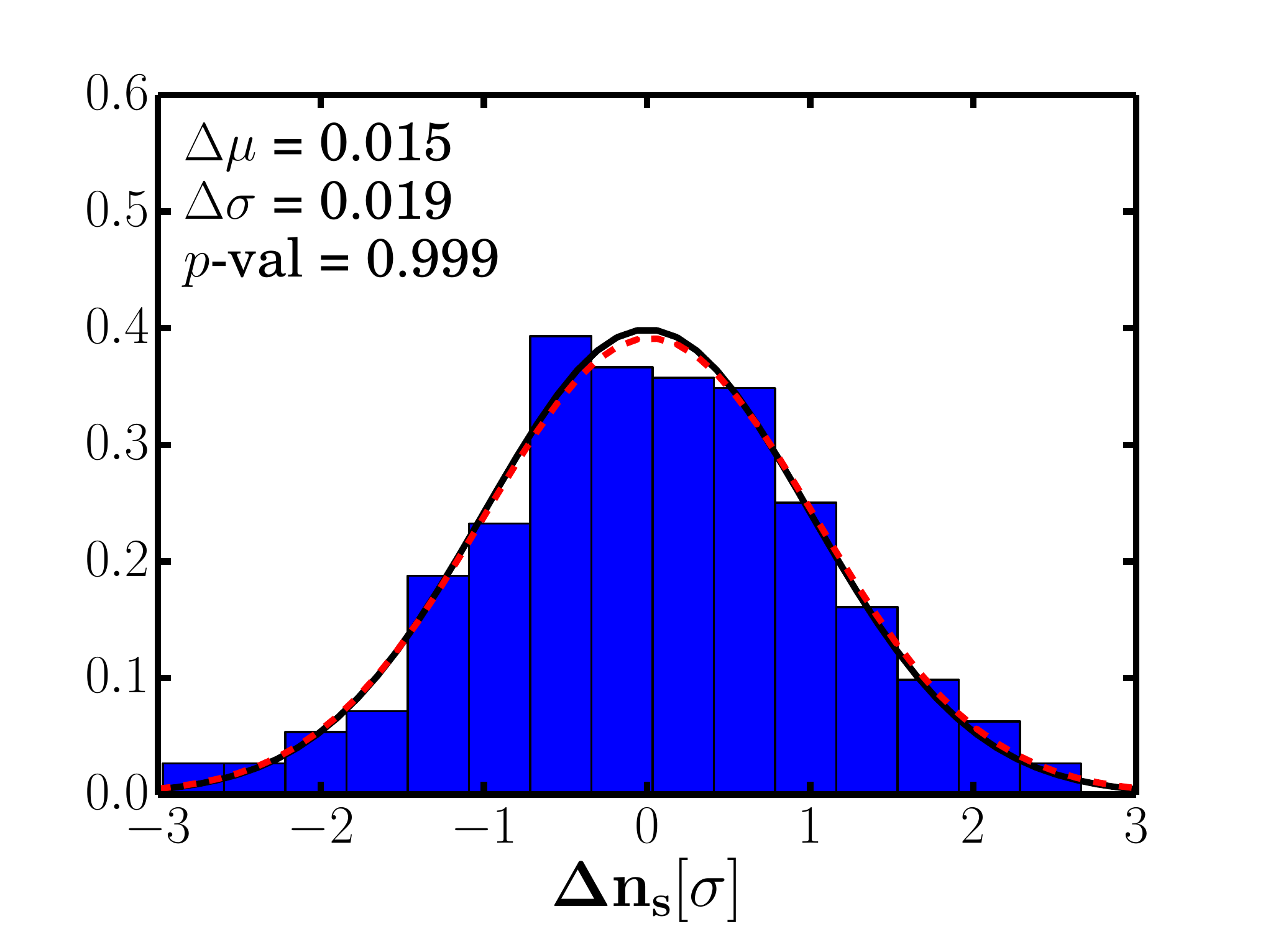}
\includegraphics[width=0.49\columnwidth]{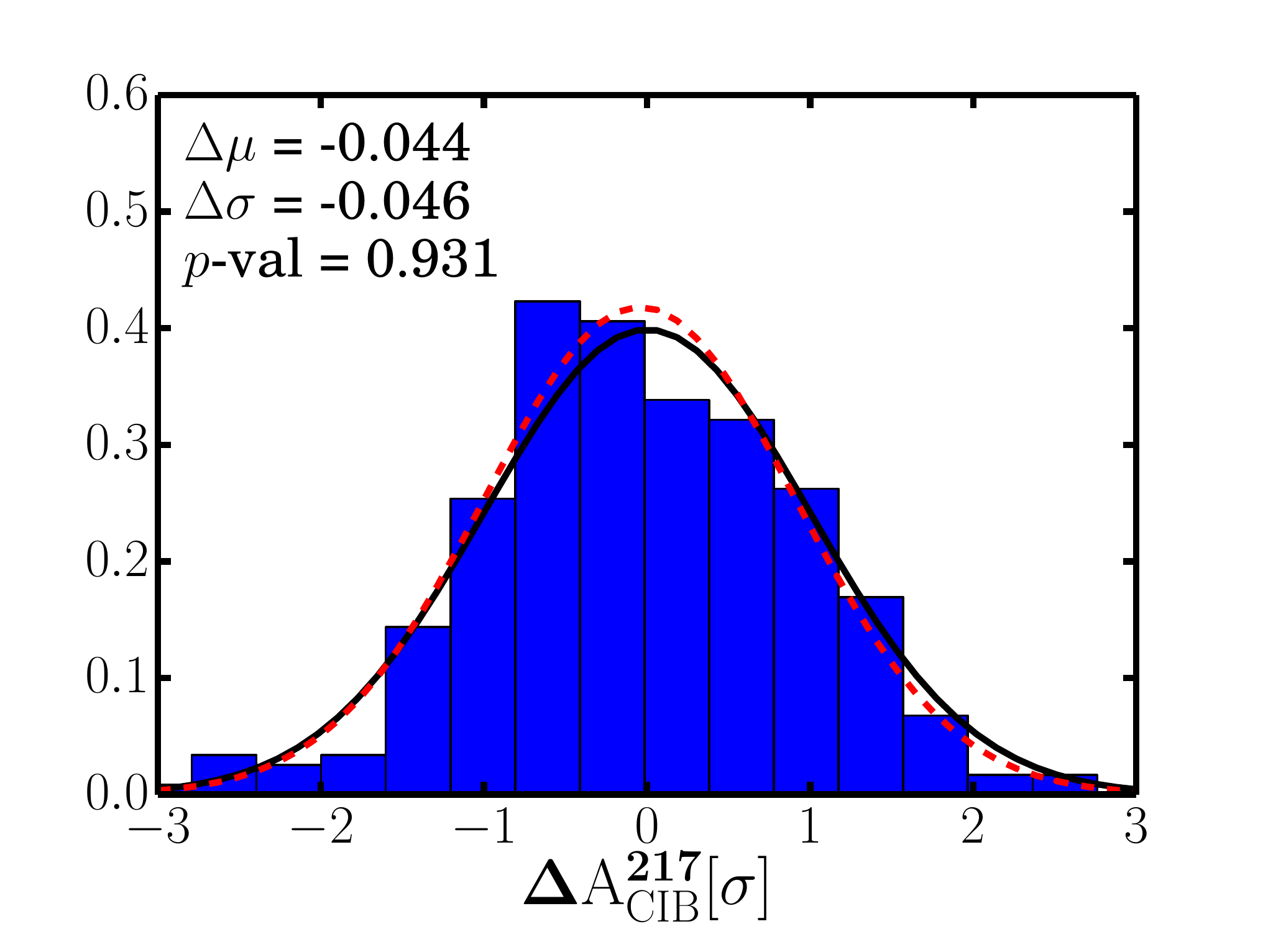}
\includegraphics[width=0.49\columnwidth]{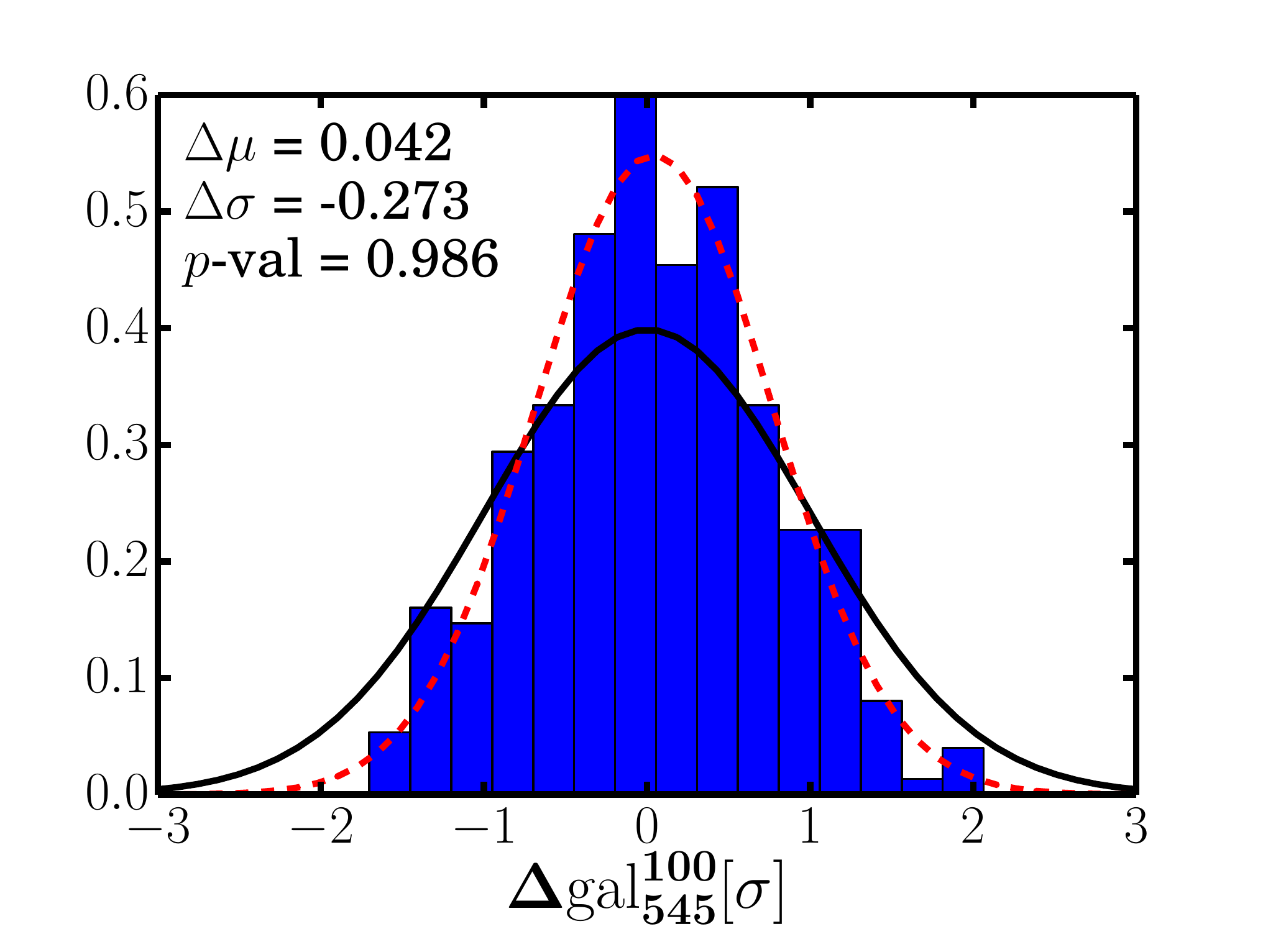}\\
\includegraphics[width=0.49\columnwidth]{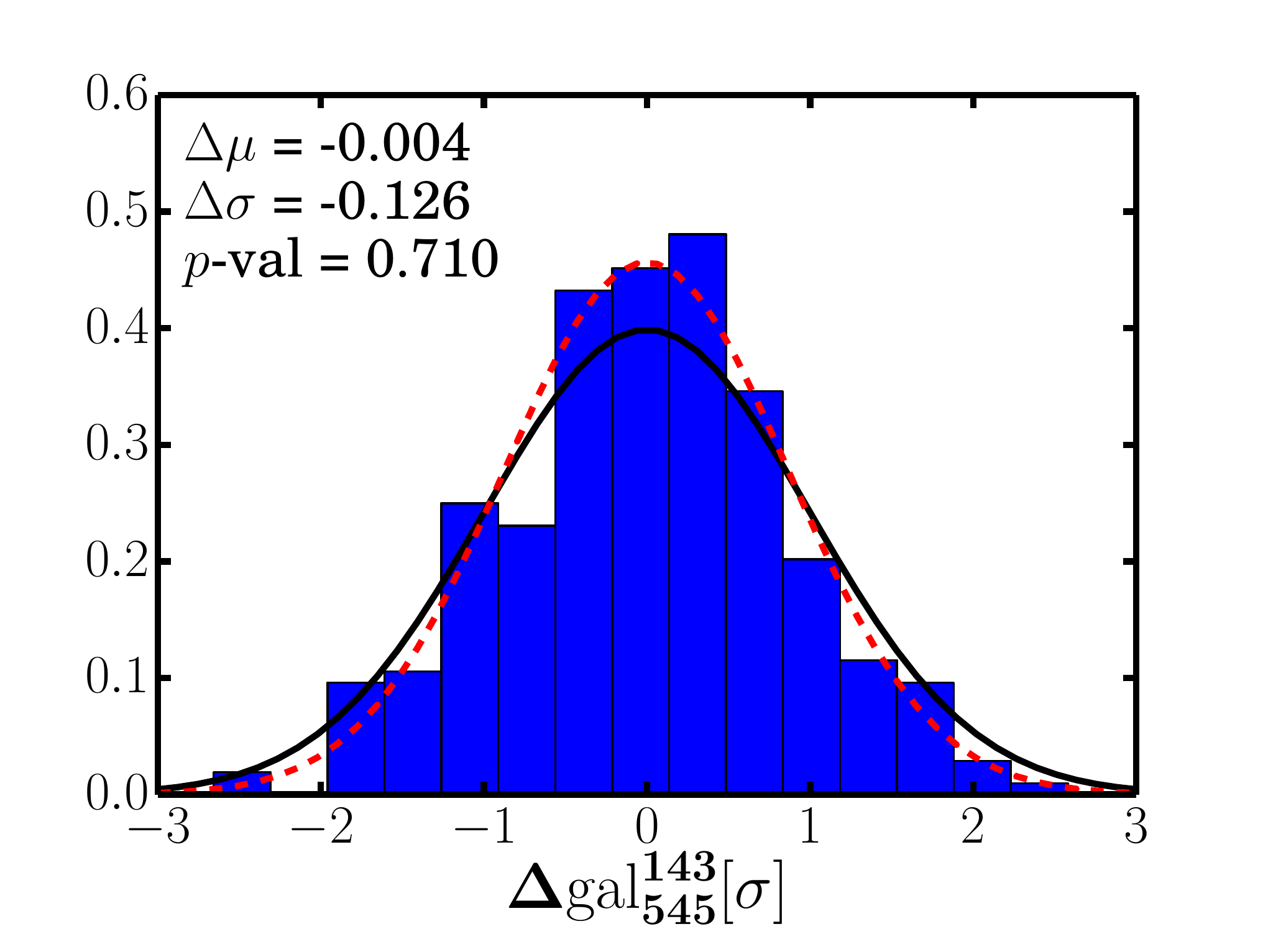}
\includegraphics[width=0.49\columnwidth]{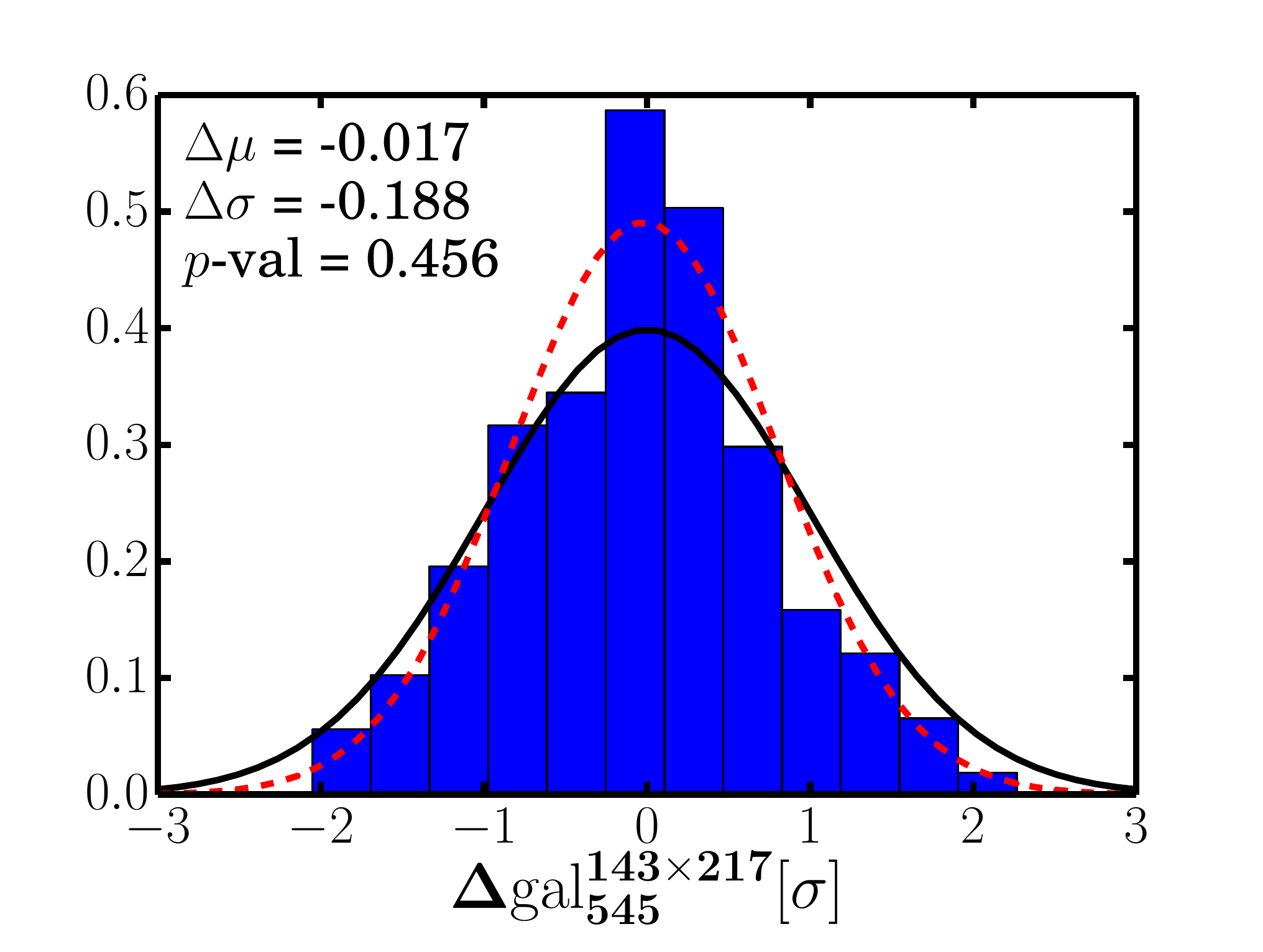}
\includegraphics[width=0.49\columnwidth]{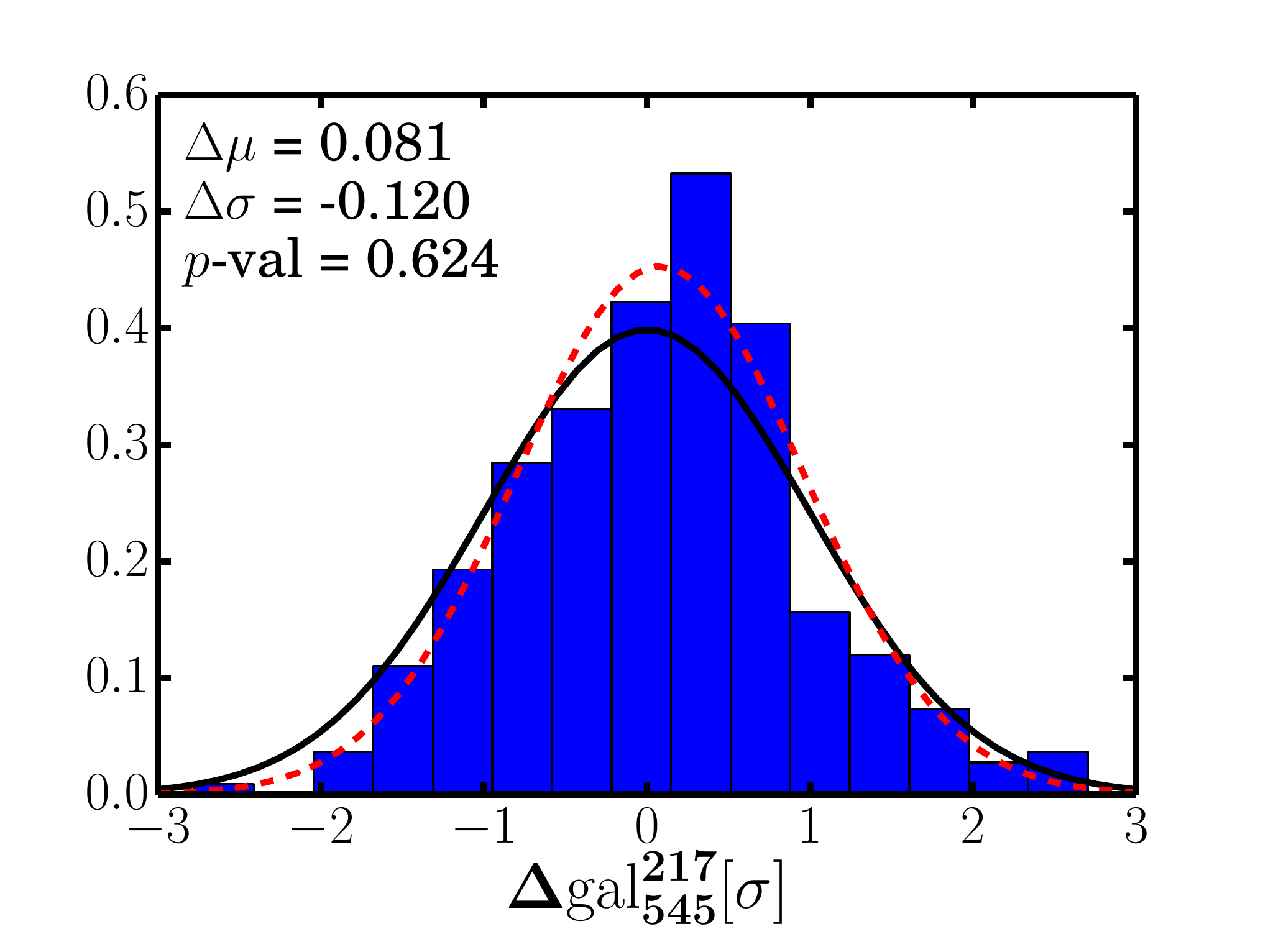}
\includegraphics[width=0.49\columnwidth]{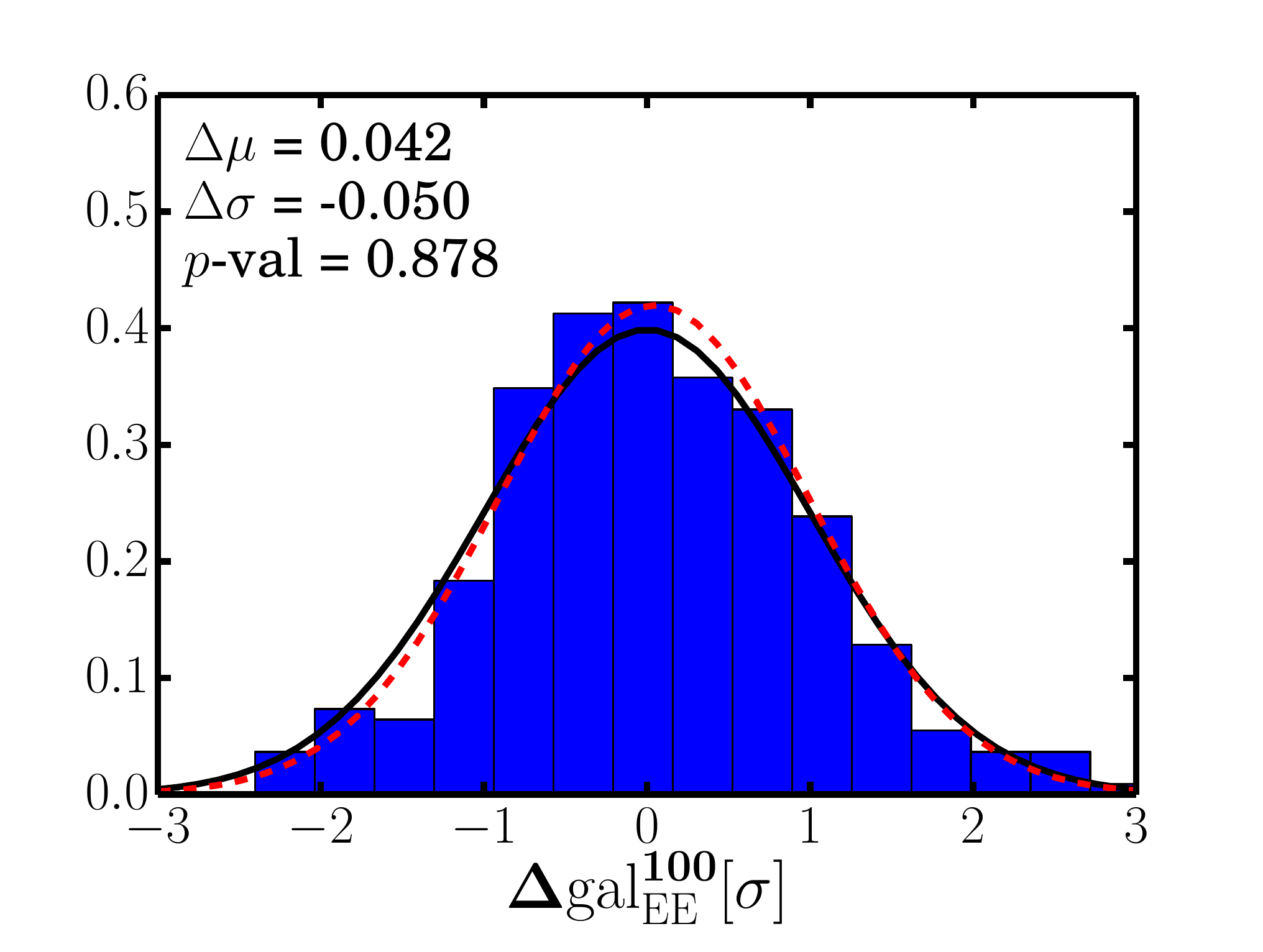}\\
\includegraphics[width=0.49\columnwidth]{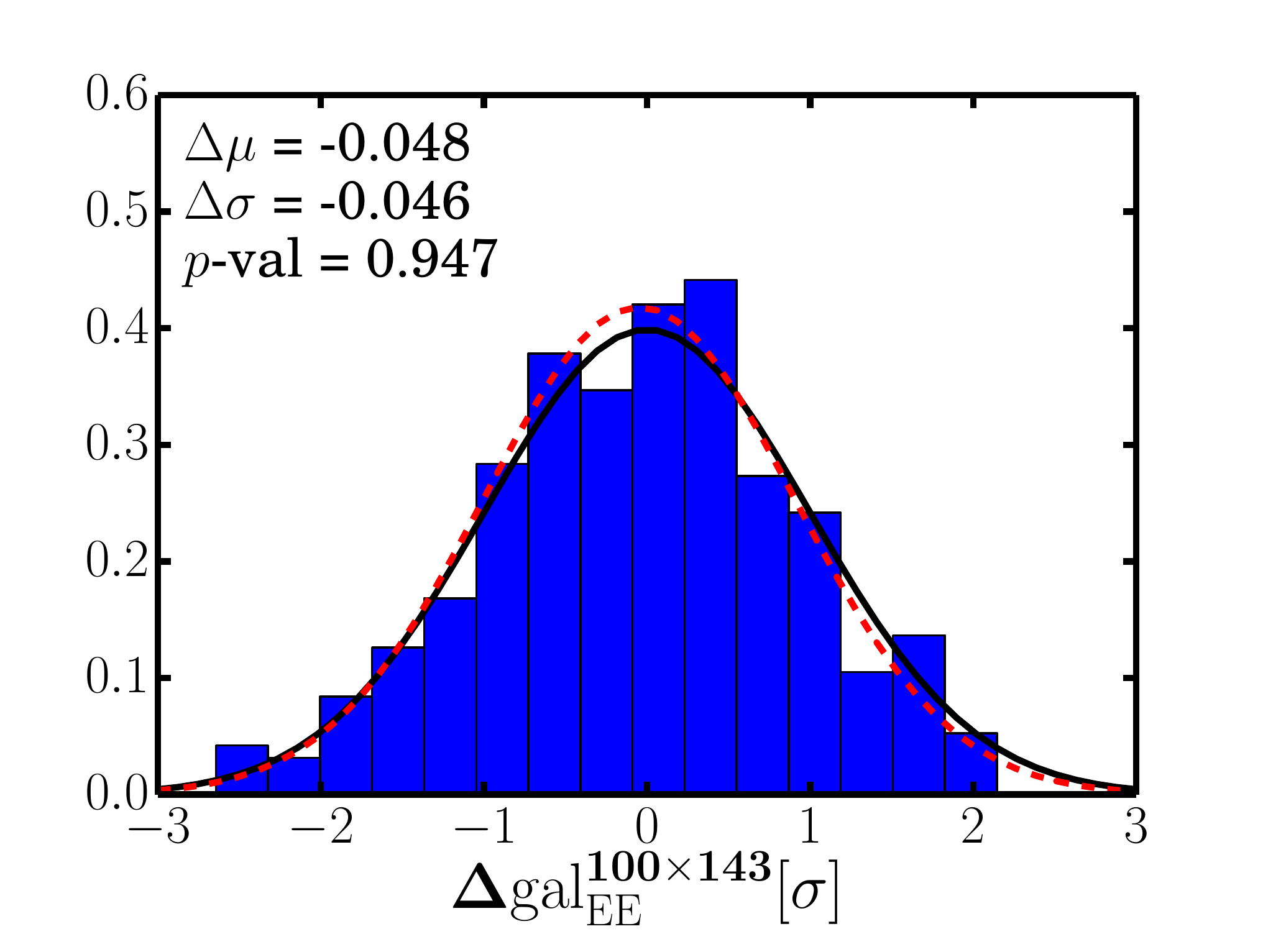}
\includegraphics[width=0.49\columnwidth]{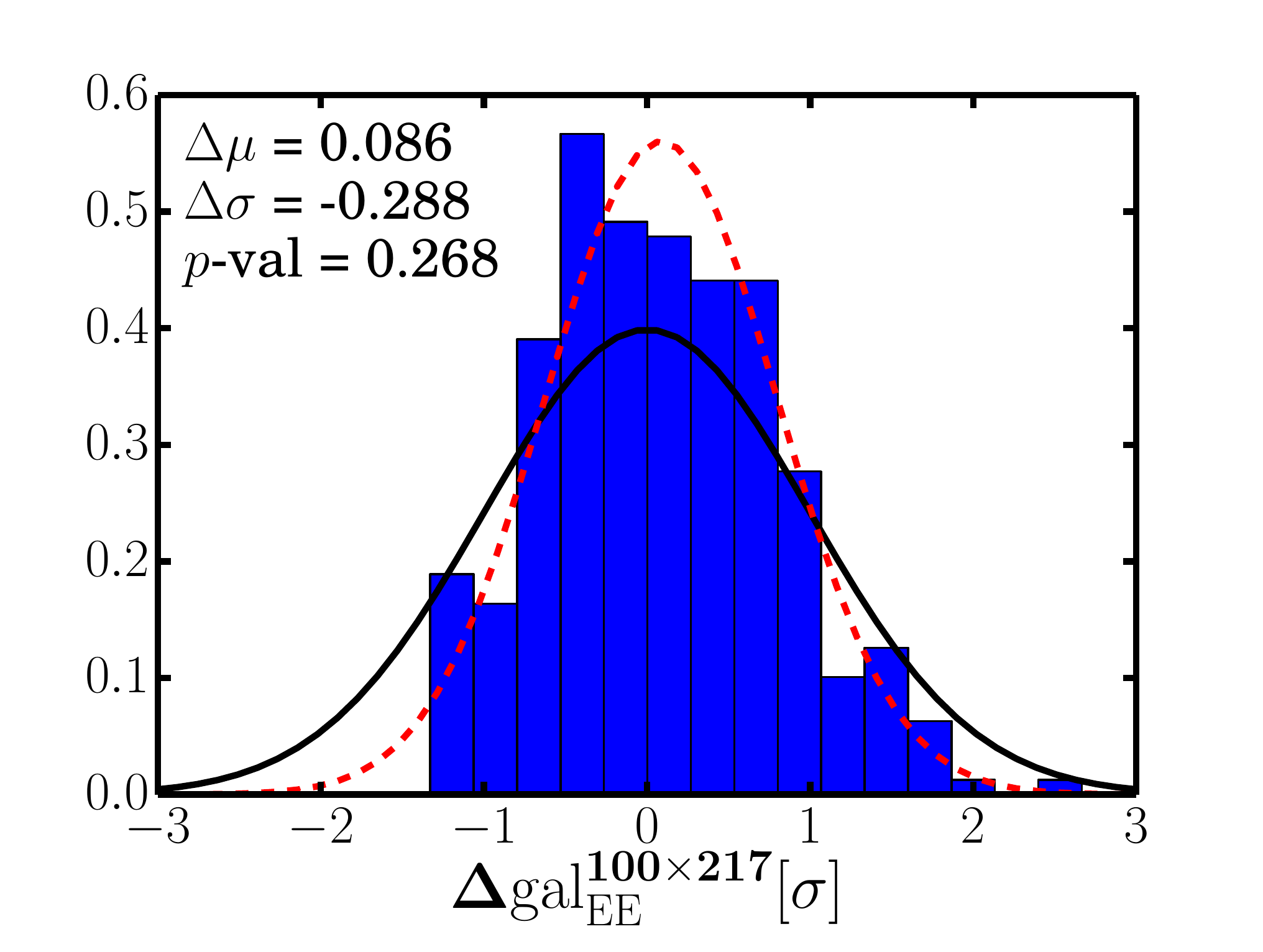}
\includegraphics[width=0.49\columnwidth]{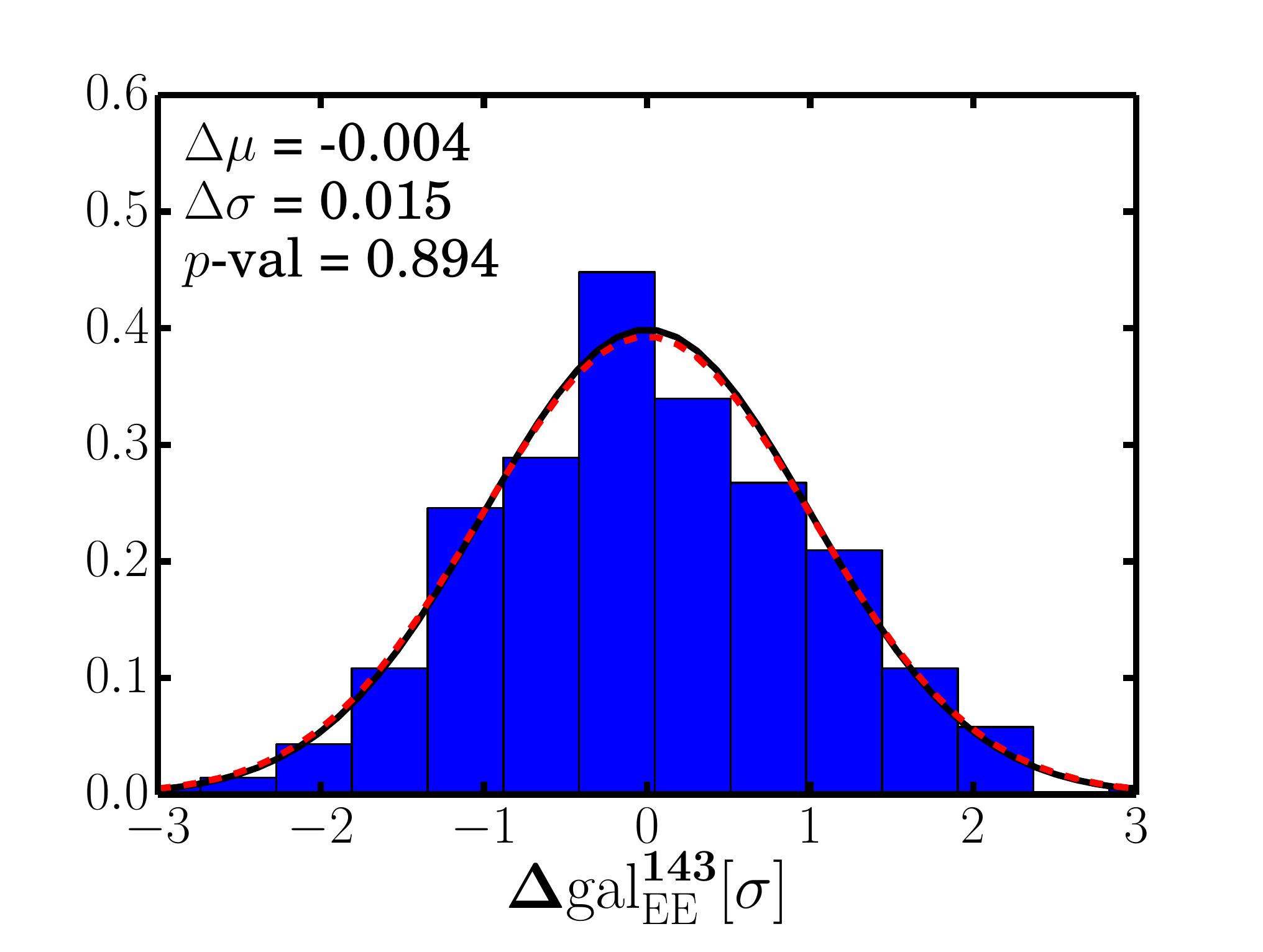}
\includegraphics[width=0.49\columnwidth]{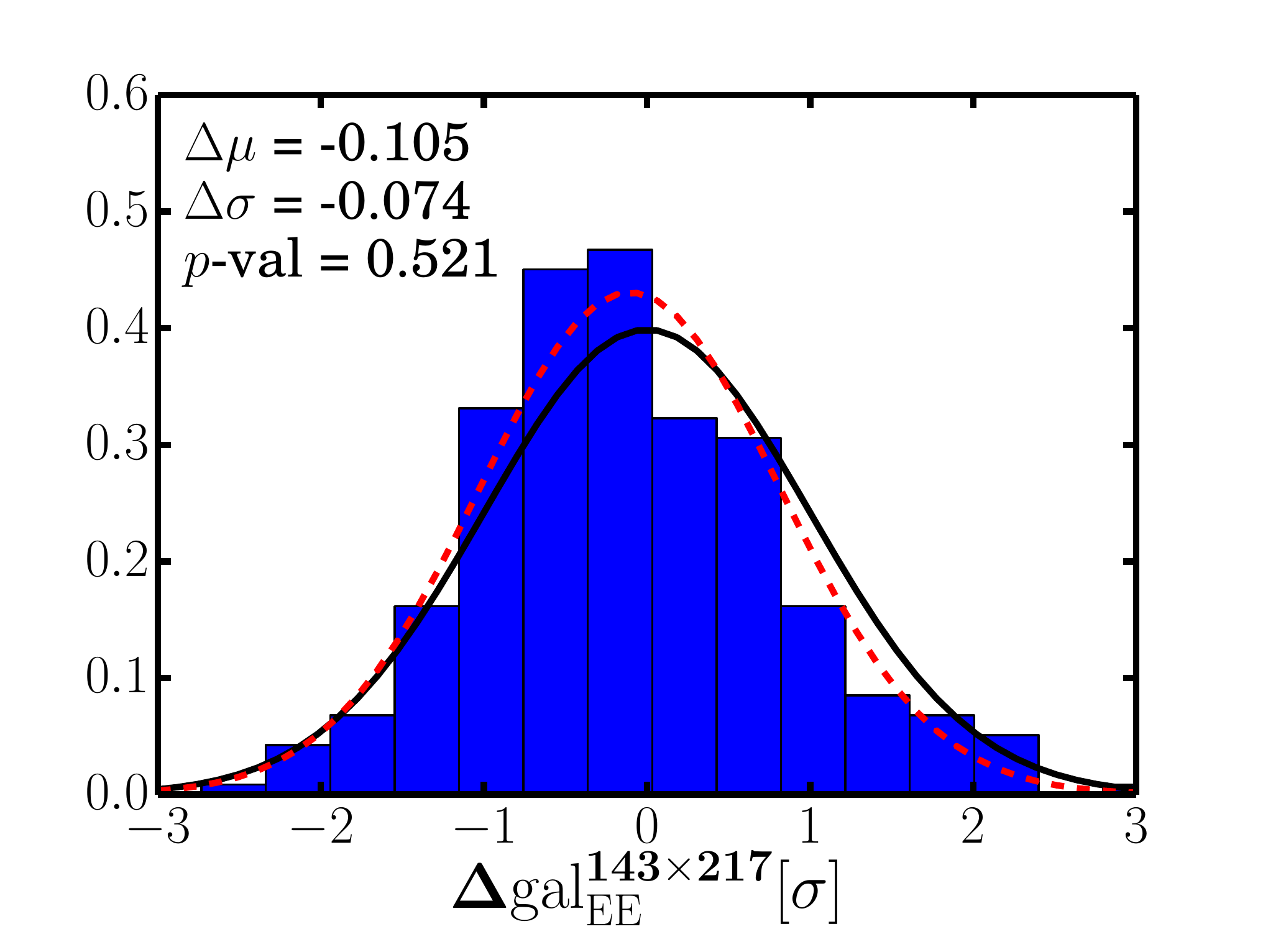}\\
\includegraphics[width=0.49\columnwidth]{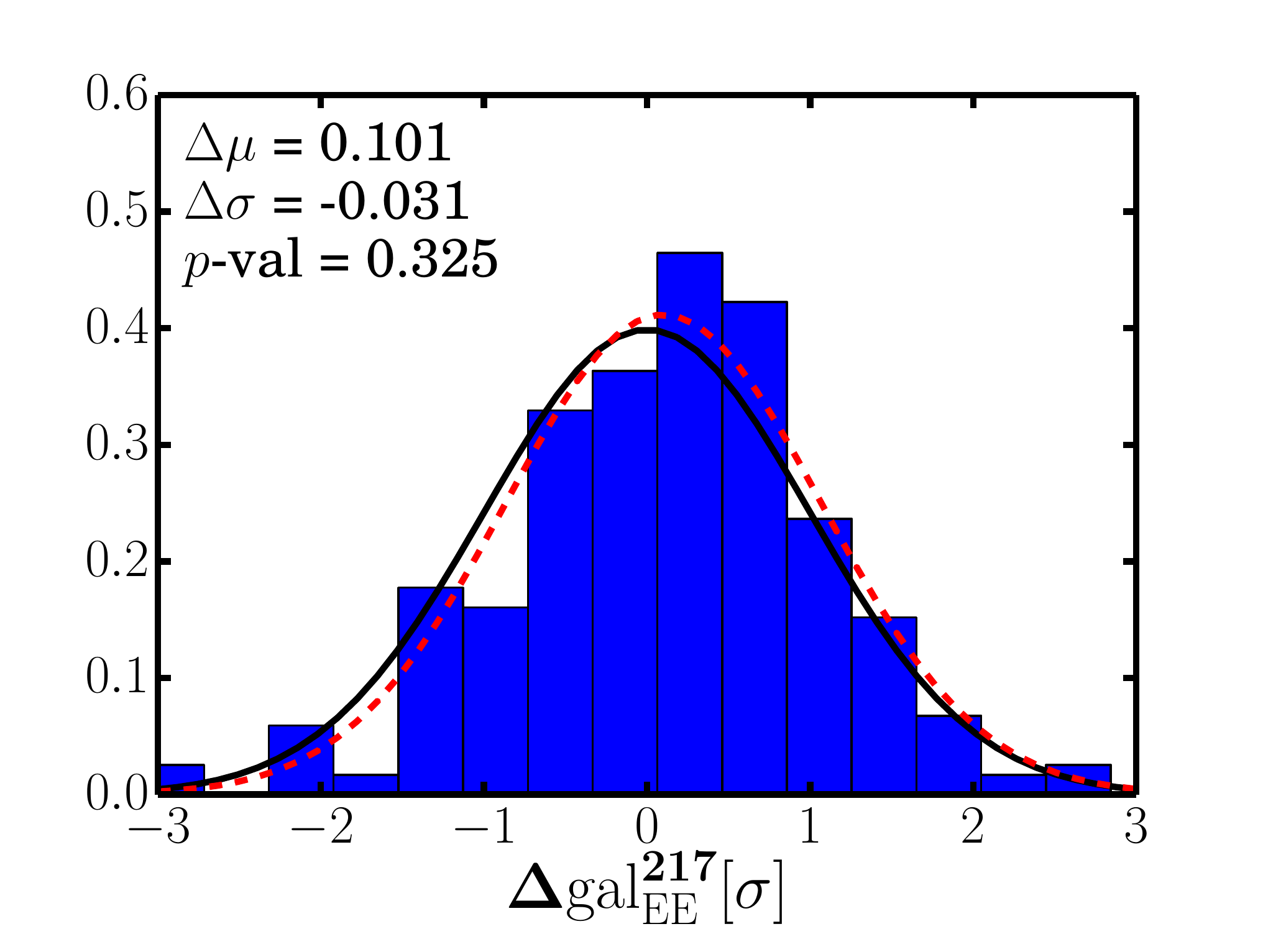}
\includegraphics[width=0.49\columnwidth]{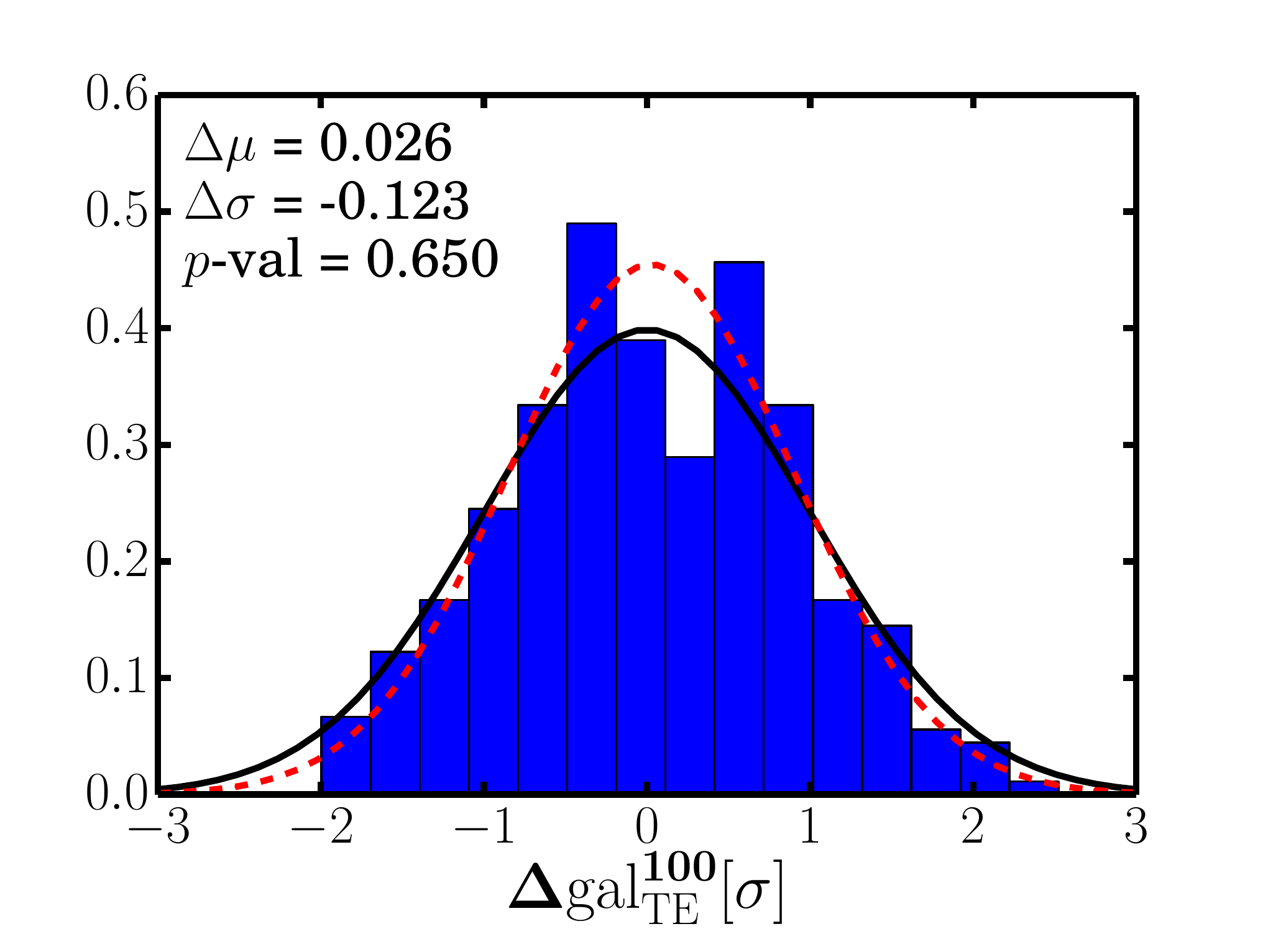}
\includegraphics[width=0.49\columnwidth]{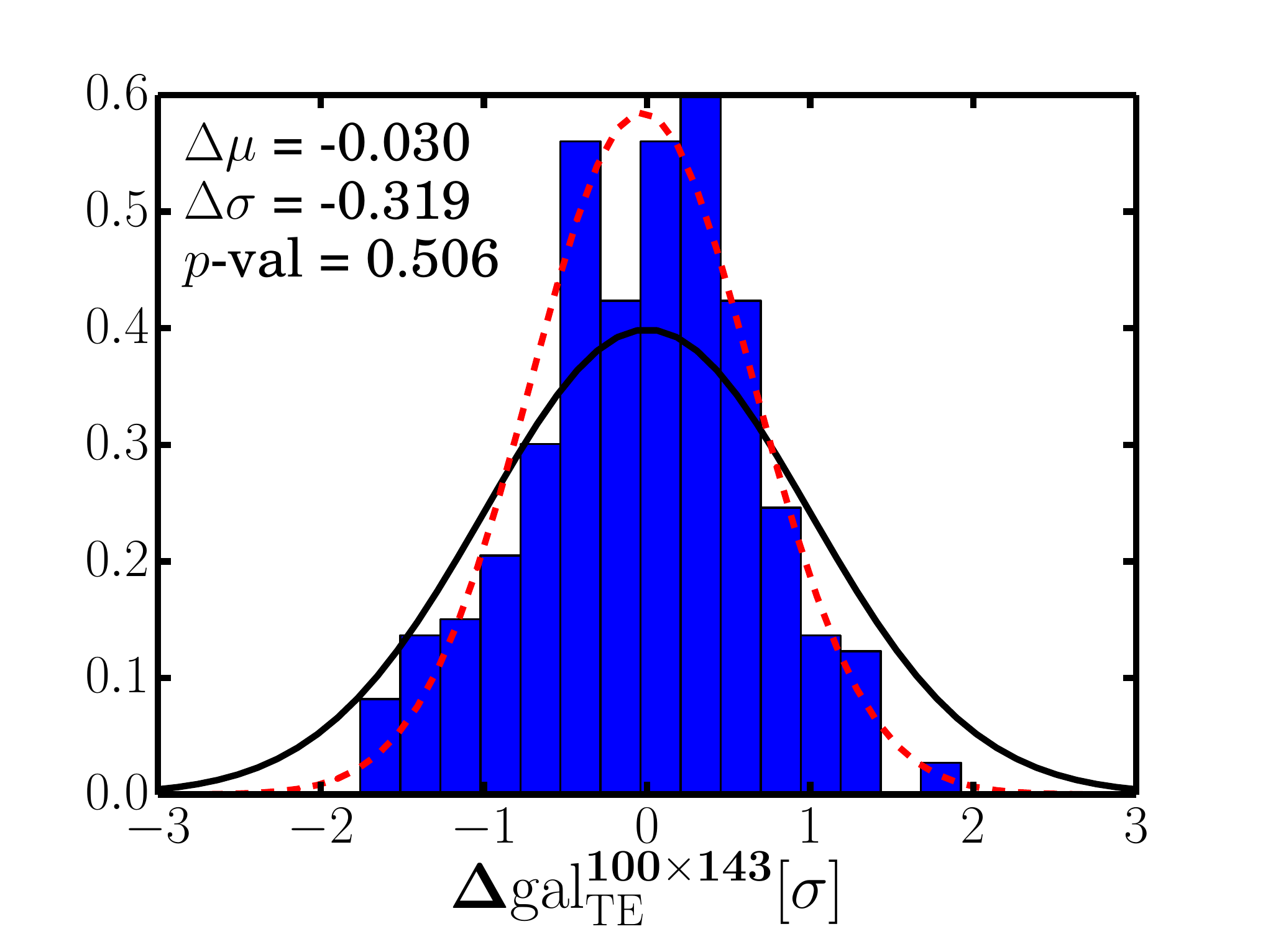}
\includegraphics[width=0.49\columnwidth]{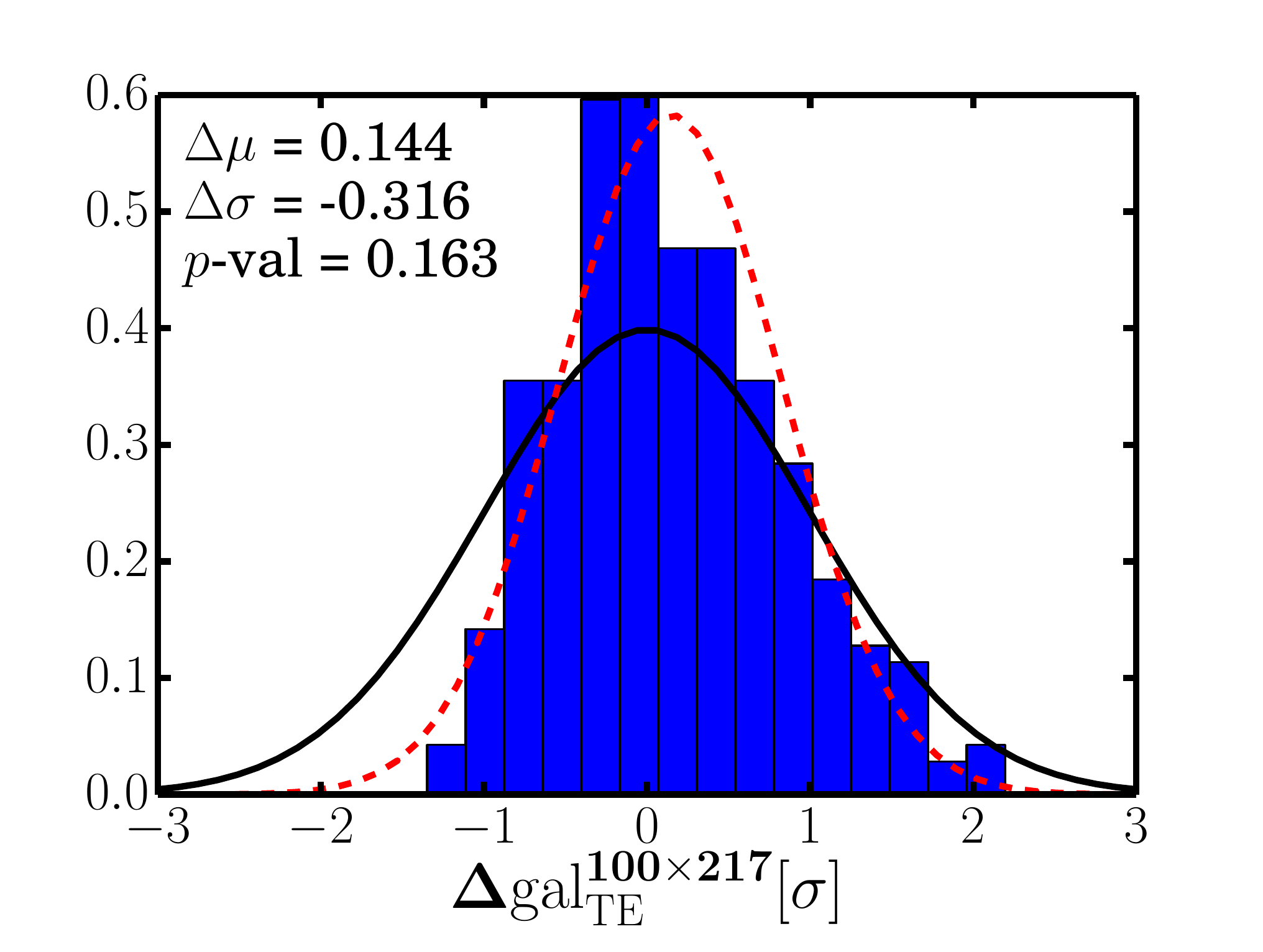}\\
\includegraphics[width=0.49\columnwidth]{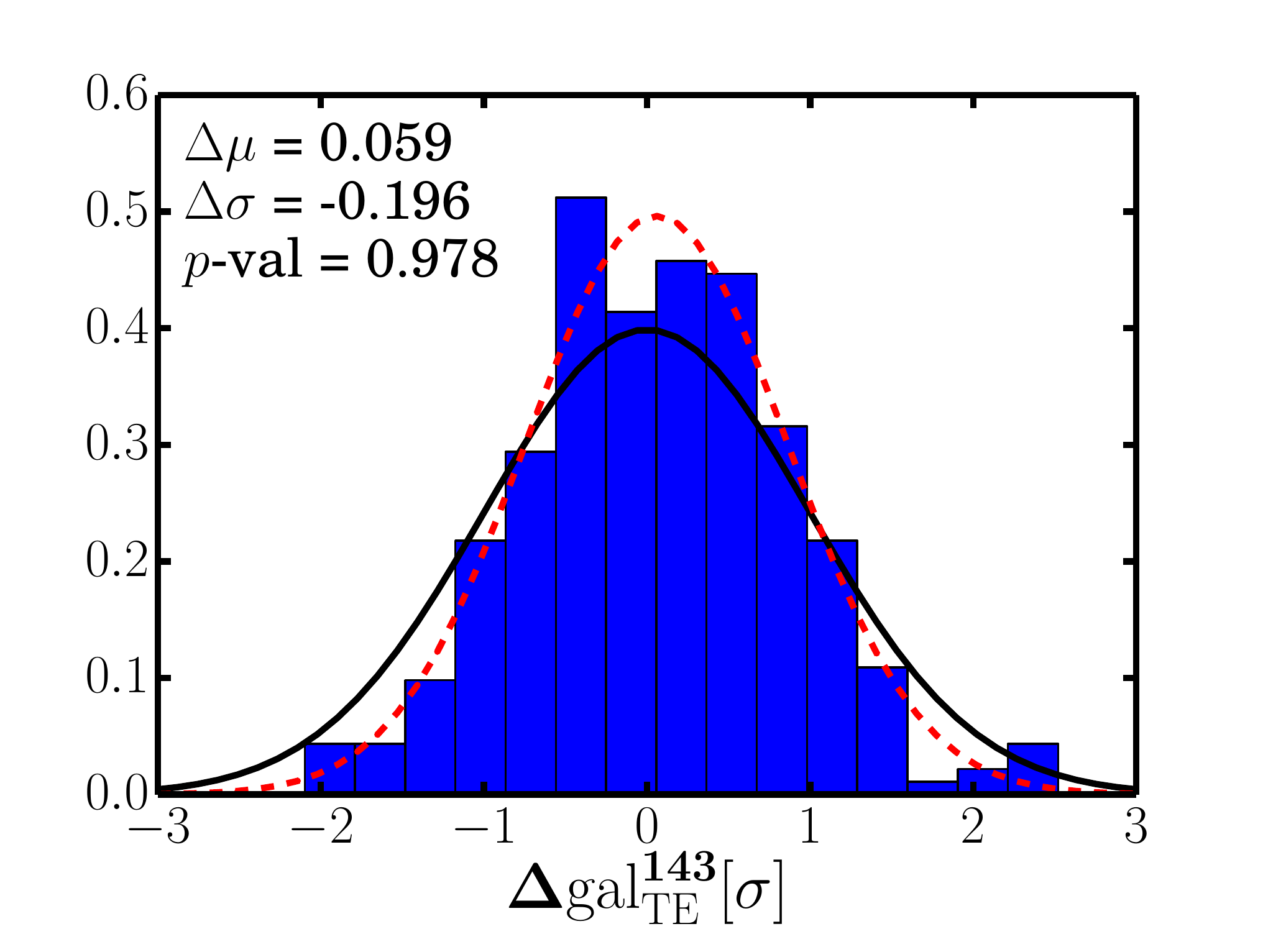}
\includegraphics[width=0.49\columnwidth]{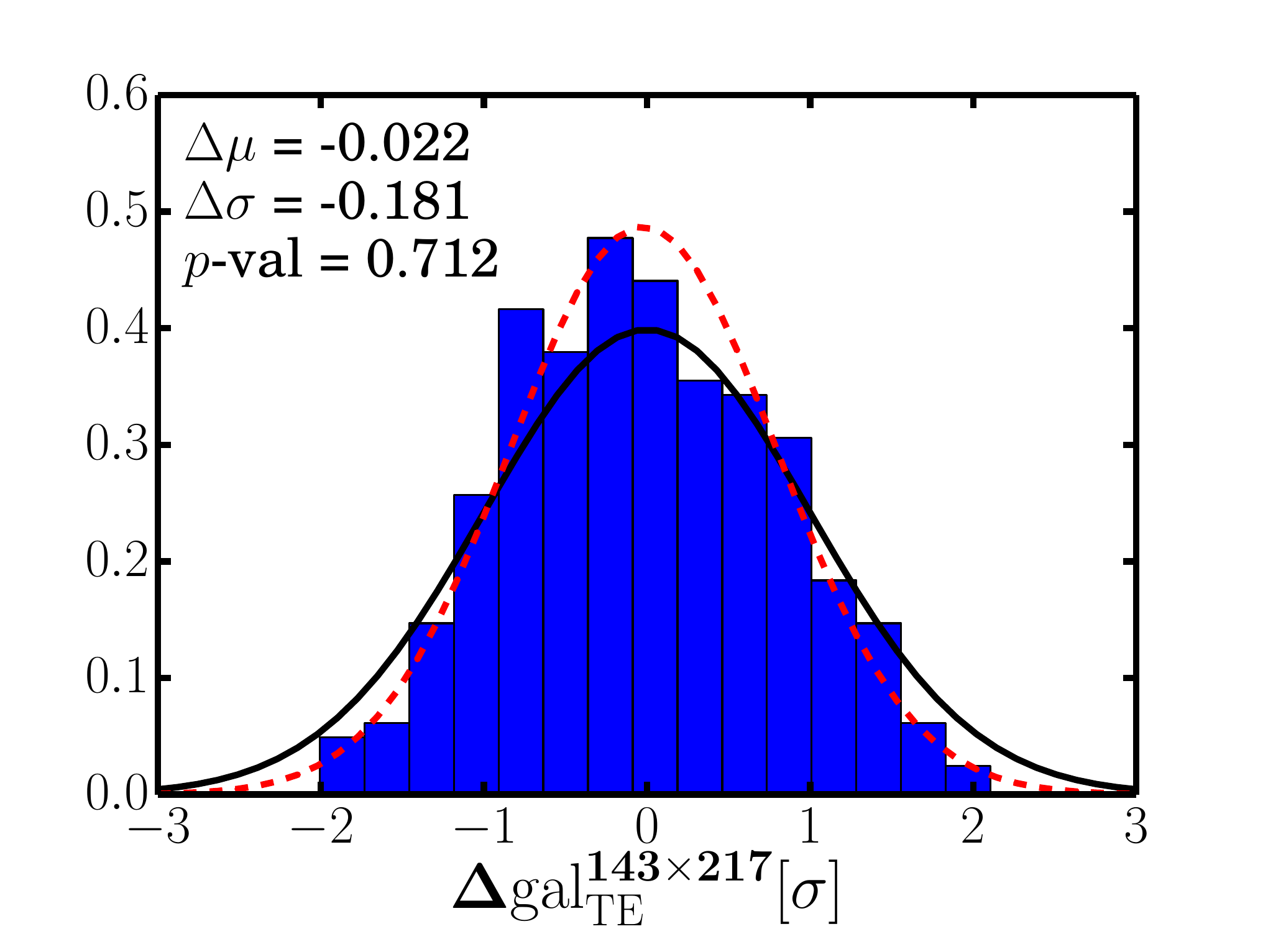}
\includegraphics[width=0.49\columnwidth]{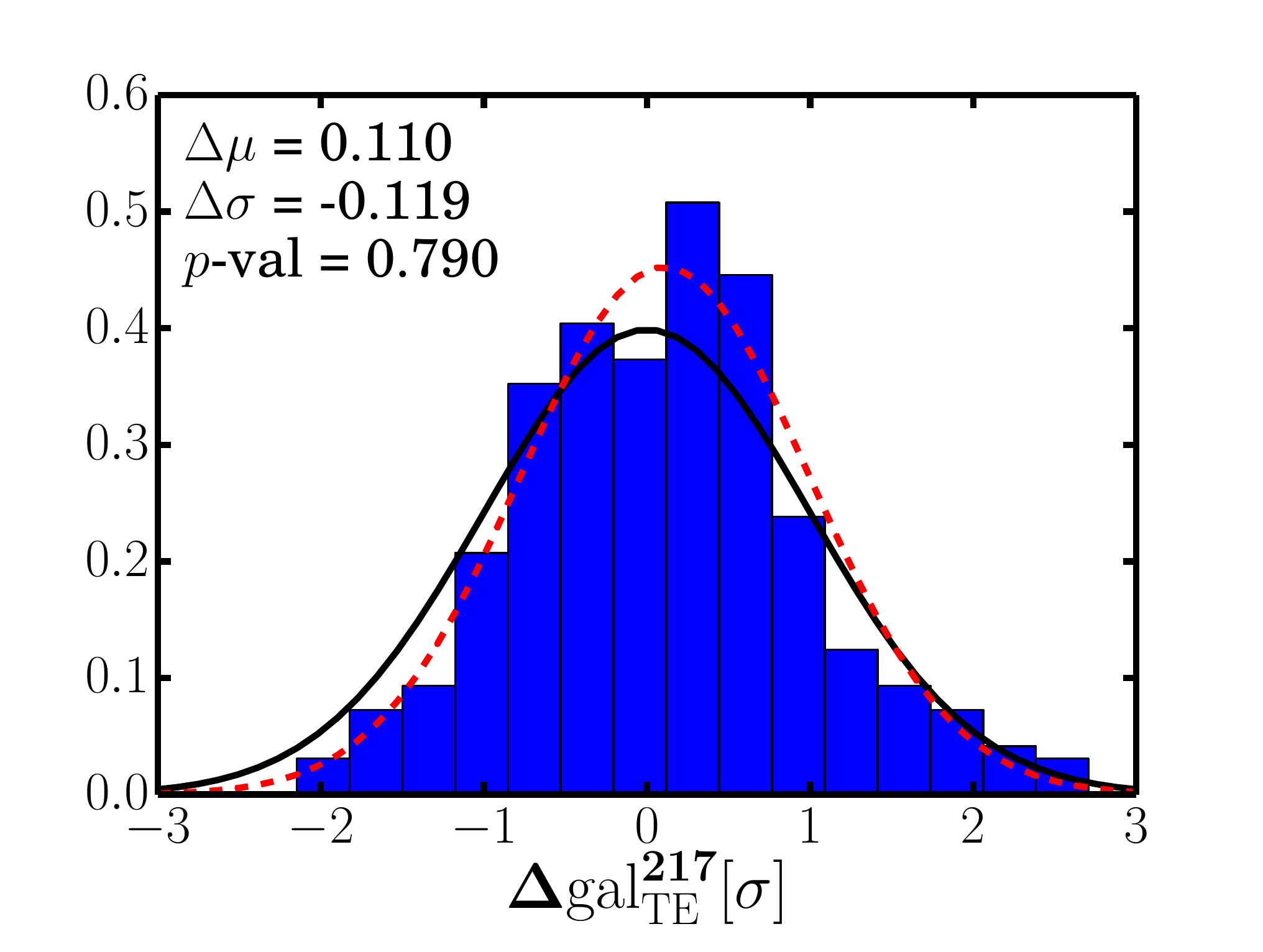}
\caption{\rev{\plik parameter results from 300 simulations for the six baseline cosmological parameters, as well as the FFP8 CIB and Galactic dust amplitudes, as in Fig.~\ref{fig:plik_sims_par}, but for the joint \plikTTTEEE\ likelihood.}}
\label{fig:plik_fullsims_par}
\end{center}
\end{figure*}

Figure~\ref{fig:plik_fullsims_par} and Table~\ref{tab:fullsimsresults}
show the full-likelihood parameter results; these are companions to
Fig.~\ref{fig:plik_sims_par} and Table~\ref{tab:simsresults} of
Sect.~\ref{sec:valid-sims}, which were devoted to the $\TT$
case. The average reduced $\chi^2$ corresponding to the histograms of
Fig.~\ref{fig:plik_fullsims_par} is equal to 1.01.  
Compared to $\TT$, the inclusion of $\EE$ and $\TE$ provides a
significant improvement in the determination of several cosmological
parameters, in particular $\ns$, $\theta$, and $\tau$.  It also reduces
the small bias in $\ns$ already discussed in the main text, since the
entire $\ell$ range is used in the joint analysis. \rev{\st{$A_\mathrm{CIB}^{217}$ is degenerate with $\mathrm{gal}_{545}^{217}$, a likely explanation for their relatively big shifts (of respectively -0.991 and 1.62 in units of rescaled posterior widths).  The high value of the shift in $\mathrm{gal}_{TE}^{217}$ is possibly attributable to the slightly underestimated covariance amplitude in blocks involving $TE$, which, as mentioned above, is caused by the neglected $TE$ detector-noise terms}}. \rev{All but three foreground parameters are at or well within one rescaled $\sigma$, lending support to the \plik\ likelihood methodology}.

\begin{table}[ht!]
\begingroup
\newdimen\tblskip \tblskip=5pt
\caption{Shifts of parameters for the joint \plikTTTEEE\ likelihood.$^{\rm a}$}
\label{tab:fullsimsresults}
\vskip -6mm
\footnotesize 
\setbox\tablebox=\vbox{
\newdimen\digitwidth
\setbox0=\hbox{\rm 0}
\digitwidth=\wd0
\catcode`*=\active
\def*{\kern\digitwidth}
\newdimen\signwidth
\setbox0=\hbox{+}
\signwidth=\wd0
\catcode`!=\active
\def!{\kern\signwidth}
\newdimen\decimalwidth
\setbox0=\hbox{.}
\decimalwidth=\wd0
\catcode`@=\active
\def@{\kern\decimalwidth}
\openup 4pt
\halign{
\hbox to 1.0in{$#$\leaderfil}\tabskip=3em&
    \hfil$#$\hfil\tabskip=0pt\cr
\noalign{\doubleline}
\omit\hfil Parameter\hfil&\omit\hfil 300 sims\hfil\cr
\noalign{\vskip 3pt\hrule\vskip 5pt}
\Omb h^2&                               			\rev{-1.09}\cr
\Omc h^2&                               			\rev{!0.62}\cr
\theta&                                     			\rev{-0.25}\cr
\tau&                                                           	\rev{-0.88}\cr
\ln\left( 10^{10}A_{\mathrm{s}}\right)& 	\rev{-0.76}\cr
n_{\mathrm{s}}&                                  		\rev{!0.25}\cr
A_{\mathrm{CIB}}^{217}&                 		\rev{-0.75}\cr
\mathrm{gal}_{545}^{100}&               		\rev{-0.03}\cr
\mathrm{gal}_{545}^{143}&               		\rev{-0.05}\cr
\mathrm{gal}_{545}^{143-217}&      		\rev{-0.28}\cr
\mathrm{gal}_{545}^{217}&               		\rev{!1.38}\cr
\mathrm{gal}_{EE}^{100}&               		\rev{!0.69}\cr
\mathrm{gal}_{EE}^{100-143}&        		\rev{-0.80}\cr
\mathrm{gal}_{EE}^{100-217}&         		\rev{!0.02}\cr
\mathrm{gal}_{EE}^{143}&        			\rev{-0.07}\cr
\mathrm{gal}_{EE}^{143-217}&         		\rev{-1.21}\cr
\mathrm{gal}_{EE}^{217}&        			\rev{!1.08}\cr
\mathrm{gal}_{TE}^{100}&                  		\rev{-0.11}\cr
\mathrm{gal}_{TE}^{100-143}&           	\rev{-0.39}\cr
\mathrm{gal}_{TE}^{100-217}&           	\rev{!0.32}\cr
\mathrm{gal}_{TE}^{143}&                   	\rev{!0.55}\cr
\mathrm{gal}_{TE}^{143-217}&           	\rev{-0.47}\cr
\mathrm{gal}_{TE}^{217}&                   	\rev{!1.20}\cr
\noalign{\vskip 5pt\hrule\vskip 3pt}
}}
\endPlancktable
\tablenote {{\rm a}} The shifts are given in units of the posterior width rescaled by $300^{-1/2}$. If the parameters were uncorrelated, $68\,\%$ of the shifts would be expected to lie within $1\,\sigma$ of their fiducial values. \rev{Of a total 23 parameters this would mean that 5 or 6 parameters are over 1 $\sigma$ away. As shown in the table, 3 parameters are in between 1 and 2 $\sigma$, 2 parameters are marginally above 1 $\sigma$ and the remaining 18 parameters are well below 1 $\sigma$.}\par
\endgroup
\end{table}

\subsection{\plik validation and stability tests} \label{app:plik-val}

This section complements the main text with detailed information on \plik results and tests on data, and how they are obtained. We start in Sect.~\ref{app:plik-zooms} with zooms in five adjacent $\ell$-ranges of all the individual frequency cross-spectra, and their residuals with respect to the \plikTTtau\ \lcdm\ best-fit model, both in temperature and polarization. In order to facilitate the search for possible common features across frequency spectra, we compute inter-frequency power spectra differences, according to a procedure discussed in Sect.~\ref{sec:freq}. Section~\ref{sec:polfreq} presents the corresponding results in polarization, which show that there are sizeable differences between pairs of foreground-cleaned spectra, much greater than those described in the main text for temperature. We proceed in Sect.~\ref{sec:pol-robust} to assess the robustness of the polarization results. Finally, we present in Sect.~\ref{agreementpol} simulations to quantify whether the level of agreement between temperature- and polarization-based cosmological parameters is as expected.  

\subsubsection{Zoomed-in frequency power spectra and residuals} \label{app:plik-zooms}
\begin{figure*} 
\centering
\includegraphics[width=\textwidth,clip=true, trim=23mm 60mm 21mm 55mm]{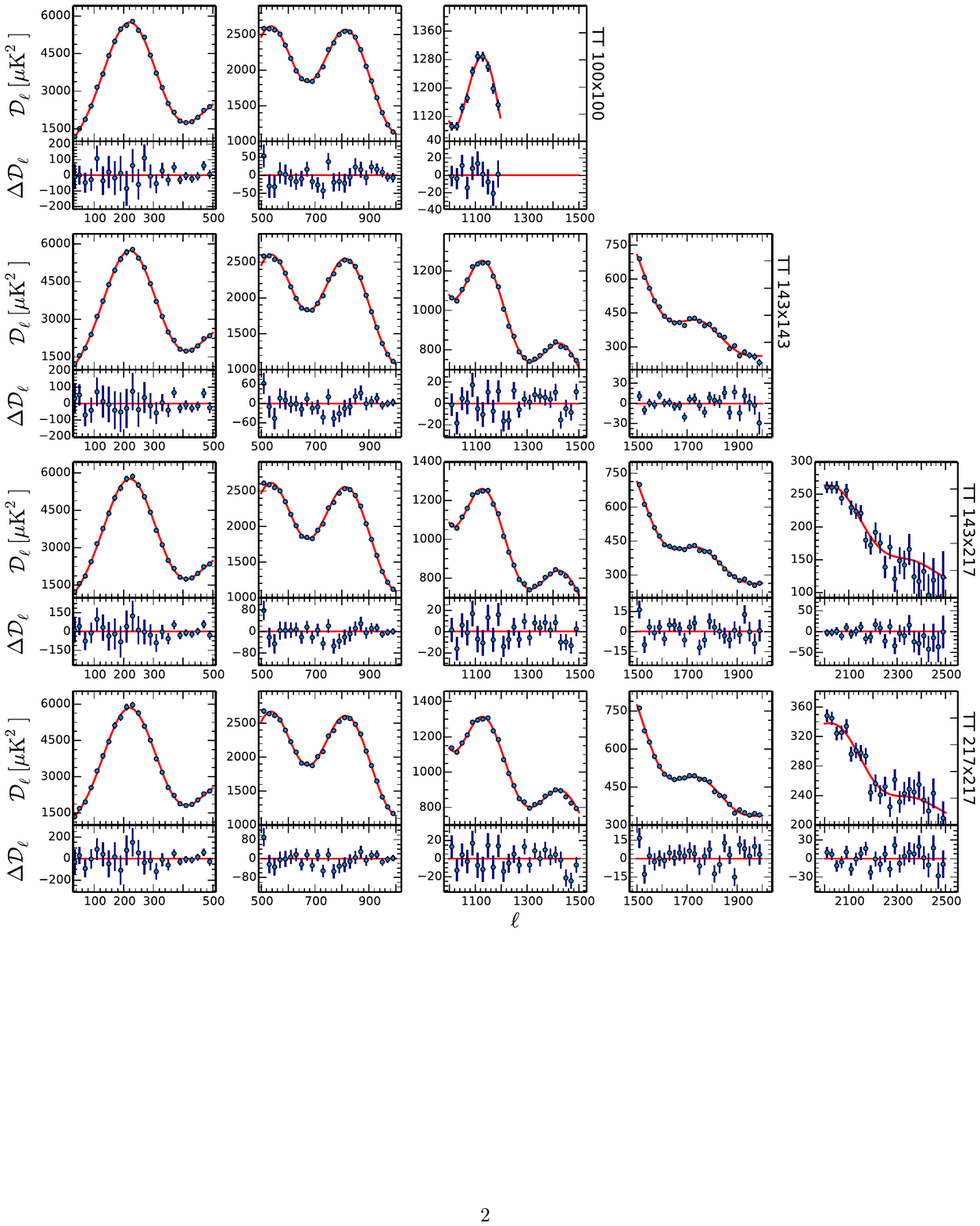}
\caption{Per-frequency zoomed-in $\TT$ power spectra, in $\Delta\ell=20$ bins. The red line shows the \plikTTtau\ $\Lambda$CDM best-fit model. The lower plots show the residuals. We only show the $\ell$ ranges used in the baseline \plik\ likelihood.}
\label{fig:Cl_TT_perfreq_zoomed}
\end{figure*}

\begin{figure*} 
\centering
\includegraphics[height=0.975\textheight,clip=true, trim=23mm 20mm 21mm 14mm]{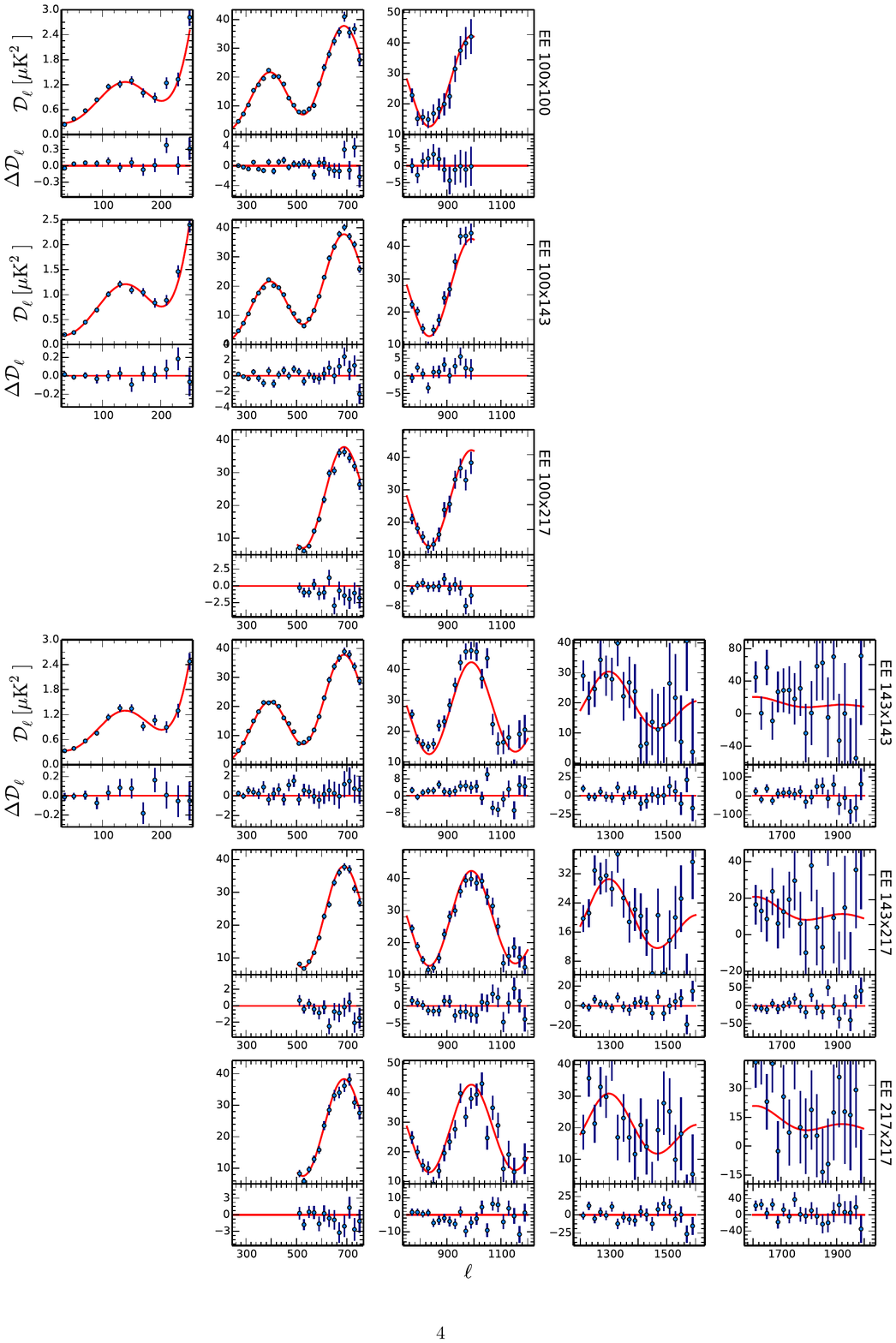}
\caption{Same as Fig.~\ref{fig:Cl_TT_perfreq_zoomed}, but for $\EE$.}
\label{fig:Cl_EE_perfreq_zoomed}
\end{figure*}

\begin{figure*} 
\centering
\includegraphics[height=0.975\textheight,clip=true, trim=23mm 20mm 21mm 14mm]{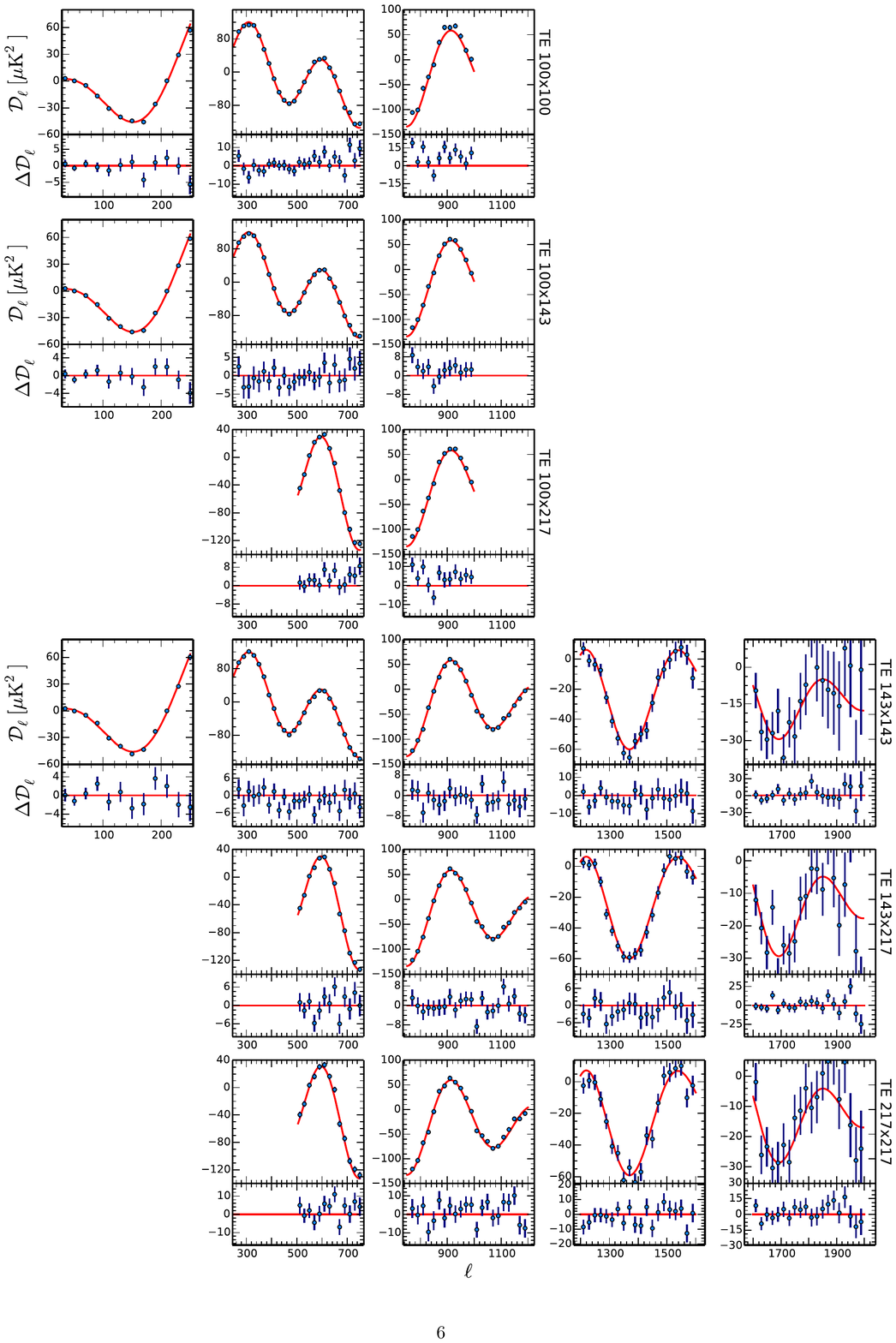}
\caption{Same as Fig.~\ref{fig:Cl_TT_perfreq_zoomed}, but for $\TE$.}
\label{fig:Cl_TE_perfreq_zoomed}
\end{figure*}

Figures~\ref{fig:Cl_TT_perfreq_zoomed}, \ref{fig:Cl_EE_perfreq_zoomed}, and \ref{fig:Cl_TE_perfreq_zoomed} show the frequency zoomed-in $\TT$, $\EE$ and $\TE$ power spectra (respectively), in $\Delta \ell = 20$ bins. The red lines show the \plikTTtau\ \LCDM\ best-fit model. The lower plots show the residuals with respect to this best-fit model. We only show the multipole ranges that are  included in the baseline analysis. These plots are meant to help the visual inspection of the residuals already shown in Fig.~\ref{fig:resTT} and described in Sect.\ \ref{sec:highlbase} for $\TT$; and in Fig.\ \ref{fig:respol} and described in Sect.\ \ref{sec:pol-rob} and Appendix\ \ref{sec:polfreq} for $\TE$ and $\EE$.

\subsubsection{Inter-frequency power spectra differences} \label{sec:freq}

We describe here the procedure followed to obtain the inter-frequency power spectra differences shown in Figs.~\ref{fig:nudiffTT} and \ref{fig:respol-diff}. We first clean the frequency power spectra by subtracting from the data the best-fit foreground solution obtained using the \plikTTtau\ (for $\TT$) or  \plikTTEETEtau\ (for $\TE$ and EE) data combinations, assuming a $\Lambda$CDM framework. 

We then  calculate the difference between a pair of cleaned spectra of length $n$ as $\Delta_\ell^\mathrm{XY-X'Y'}= {C}_\ell^{\mathrm{XY}}-{C}_\ell^{\mathrm{X'Y'}}$ .

 The covariance matrix $\tens{ C}^{\mathrm{\Delta}}$ of the difference $\Delta_\ell^\mathrm{XY-X'Y'}$ is then: 
\begin{eqnarray}
 \tens{ C}^{\mathrm{\Delta}}&=& A\tens{ C}^\mathrm{XY,X'Y'}A^T ,
\end{eqnarray}
where $\tens{C}^\mathrm{XY,X'Y'}$ is the $2n \times 2n$ covariance matrix relative to the XY and X'Y' spectra, and $A$ is a $n\times 2n$ matrix with blocks: 
\begin{eqnarray}
  A&=&\left(\begin{array}{cc}
 \mathbb{1}^{\mathrm{XY}} &  \quad -\mathbb{1}^{\mathrm{X'Y'}}  
  \end{array} \right) ,
\end{eqnarray}
 where $\mathbb{1}^{\mathrm{XY}}$ is the $n\times n$ identity matrix.

\subsubsection{Robustness tests on foreground parameters in $TT$} \label{sec:temp-robust}

\begin{figure*}[htb] 
\centering
\includegraphics[width=\textwidth, trim=7mm 4mm 2mm 7mm]{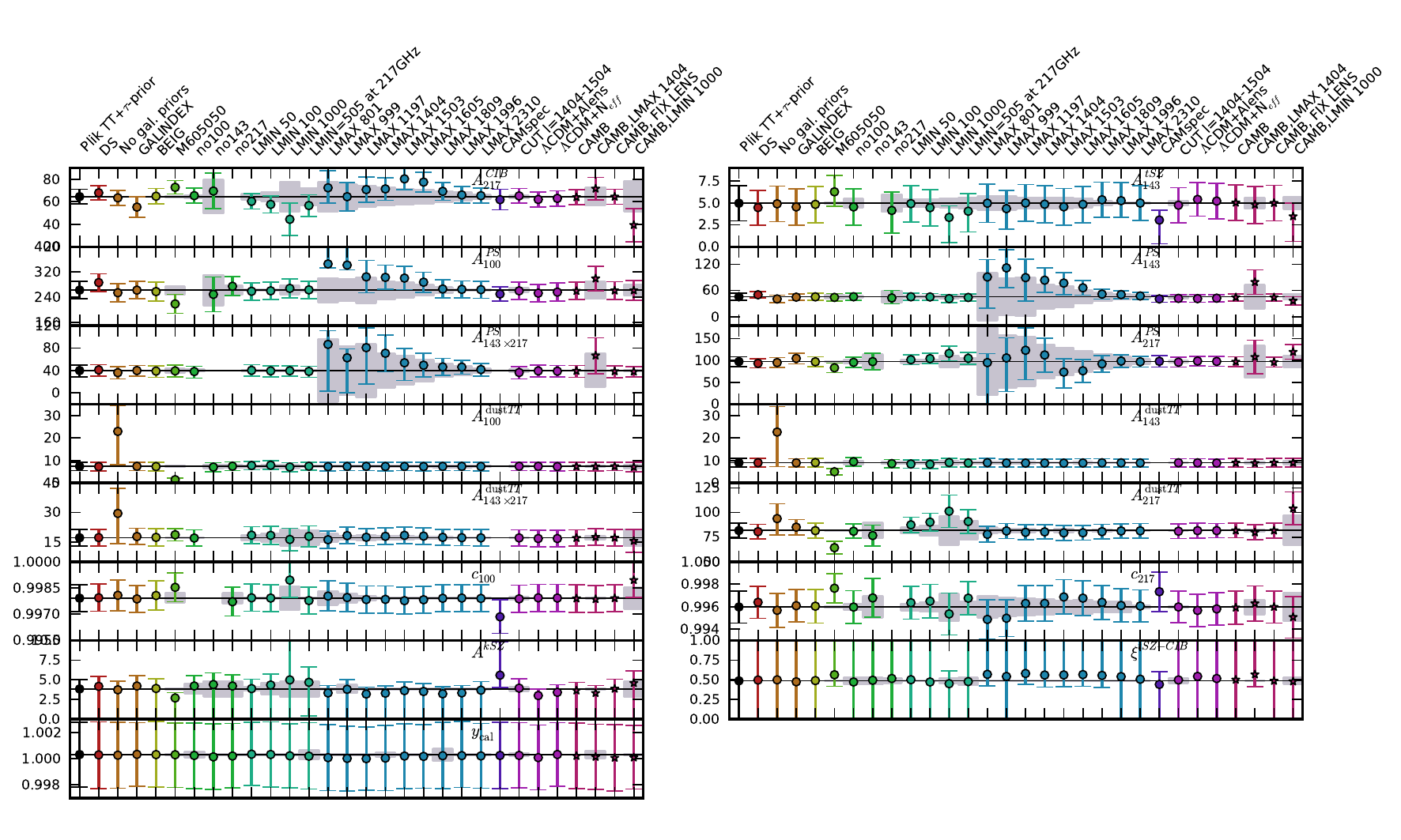}
\caption{Marginal mean and 68\,\% CL error bars on $\TT$ foreground parameters estimated when adopting different data choices for the \plik likelihood, in comparison with results from alternate approaches or models. We assume a  $\Lambda$CDM model and always combine the \plik likelihood with a prior on $\tau=0.07\pm0.02$ (we do not use low-$\ell$ temperature or polarization data here). ``\plikTTtau'' indicates the baseline (HM, $\ell_{\mathrm{min}}=30$, $\lmax=2508$), while the other cases are described in Sect.~\ref{sec:hil_par_stability}. The grey bands show the standard deviation of the expected parameter shift, for those cases where the data used are a sub-sample of the baseline likelihood (see Eq.~\ref{eq:covresult}).}
\label{fig:wiskerTTFG}
\end{figure*}

This section  presents some further checks that we performed to validate the results from $\TT$. 
Figure\ \ref{fig:wiskerTTFG} shows the marginal mean and the 68\,\% confidence level  error bars for the foreground parameters of the \plik $\TT$ high-$\ell$ likelihood under different assumptions about the data selection, foreground model, or treatment of the systematics. The cases considered are the same as those in Sect.~\ref{sec:hil_par_stability}, and the results for cosmological parameters can be found in Fig.~\ref{fig:wiskerTT}. 
We now comment on them in turn. 

\paragraph{{Detset likelihood}} 

In the detsets (``DS'') case, the amplitude of the point sources at $100\times100$\,GHz is higher than in the baseline case. This might indicate a residual correlated noise component in the DS spectra, not corrected by the procedure described in Sect.~\ref{sec:noise_model}.

\paragraph{{Impact of Galactic mask and dust modelling}} 
We recover Galactic dust amplitudes within $1\,\sigma$ of the baseline values when we leave these parameters free to vary without any prior (``No gal priors'') or when we leave the Galactic slope (described in Sect.~\ref{sec:robust-fsky}) free to vary. 
The dust amplitudes for the ``M605050'' case (\ie when we use  more conservative Galactic masks, as detailed in Sect.~\ref{sec:robust-fsky}) cannot be directly compared to the baseline values, since we  expect smaller amplitudes when using reduced sky fractions.

\paragraph{Changes with $\lmin$}
We observe  variations by up to $1\,\sigma$, as well as an increase in the error bars, in the level of dust contamination at $217\times217$ and of the CIB amplitude when we consider $\lmin=50,100$ instead of the baseline $\lmin=30$, or when we excise the first 500 multipoles at $143\times 217$ and $217\times 217$. This is due to the fact that the lowest multipoles help in breaking the degeneracy between these two foreground components, giving tighter constraints when included in the analysis.

\paragraph{Changes with $\lmax$}
We find that the overall amplitude of the foregrounds decreases when increasing the maximum multipole $\lmax$ included in the analysis.\footnote{
We remind the reader that, in this test, at each frequency we always use $\lmax ^{\mathrm{freq}}=\mathrm{min}(\lmax ,\lmax ^{\mathrm{ freq,\,baseline} })$,  with $\lmax ^{\mathrm{freq,\,  baseline}} $  the baseline $\lmax$ at each frequency as reported in Table~\ref{tab:highl:lrange} (\eg in the $\lmax=1404$ case, we still use the $100\times 100$ power spectrum through $\ell=1197$).} This is related to the shift in cosmological parameters observed at different $\lmax$, which is described in Sect.~\ref{sec:changes_lmax}.
In Fig.~\ref{fig:wiskerTTFG} the results for extragalactic foregrounds at $\lmax\lesssim 1200$ are not very meaningful, since these parameters are very weakly constrained in those multipole regions.

\paragraph{\LCDM\ extensions} 
Figure~\ref{fig:wiskerTTFG} also show the level of foregrounds obtained using the baseline likelihood in extensions of the \LCDM\ model. In the \LCDM+$\nnu$ case, the level of foregrounds is very similar to that in the base-\LCDM\ case, while in the \LCDM+$\Alens$ model it is few $\muK^2$ lower at all frequencies.

\paragraph{\camspec} 
The foreground contamination levels determined by the \camspec\ and \plik\ codes differ by a few $\mu$K$^2$. This appears in Fig.~\ref{fig:wiskerTTFG} as differences at the 1\,$\sigma$ level in the sub-dominant (and ill-determined) foreground components ($A^{\rm kSZ},\ A^{\rm tSZ}_{143}$), together with different best-fit recalibration factors ($c_{100}, c_{217}$), a result of the different modelling choices made regarding the $\ell$ ranges retained, and small variations in the dust template (where it is least well determined by the data). As already mentioned earlier in the discussion of cosmological parameters, the strongest effect is in $\ns$, resulting in our estimate of a $0.3\,\sigma$ {systematic} uncertainty on this parameter.

\paragraph{Other cases} 
The remaining cases shown in Fig.~\ref{fig:wiskerTTFG}  are described in Sect.~\ref{sec:hil_par_stability}. We find good agreement in the cases where we excise one frequency at a time, or when we use the \CAMB\ code instead of \PICO.

%
\subsubsection{Further tests of the shift with $\lmax$}
\label{sec:lmaxlowl}

We have investigated whether different data combination choices have an impact on the shift in cosmological parameters we observe when we change the maximum multipole included in the analysis, as described in Sec.~\ref{sec:changes_lmax}.

\begin{figure*}[htb] 
\centering
\includegraphics[width=\textwidth, trim=7mm 4mm 2mm 7mm]{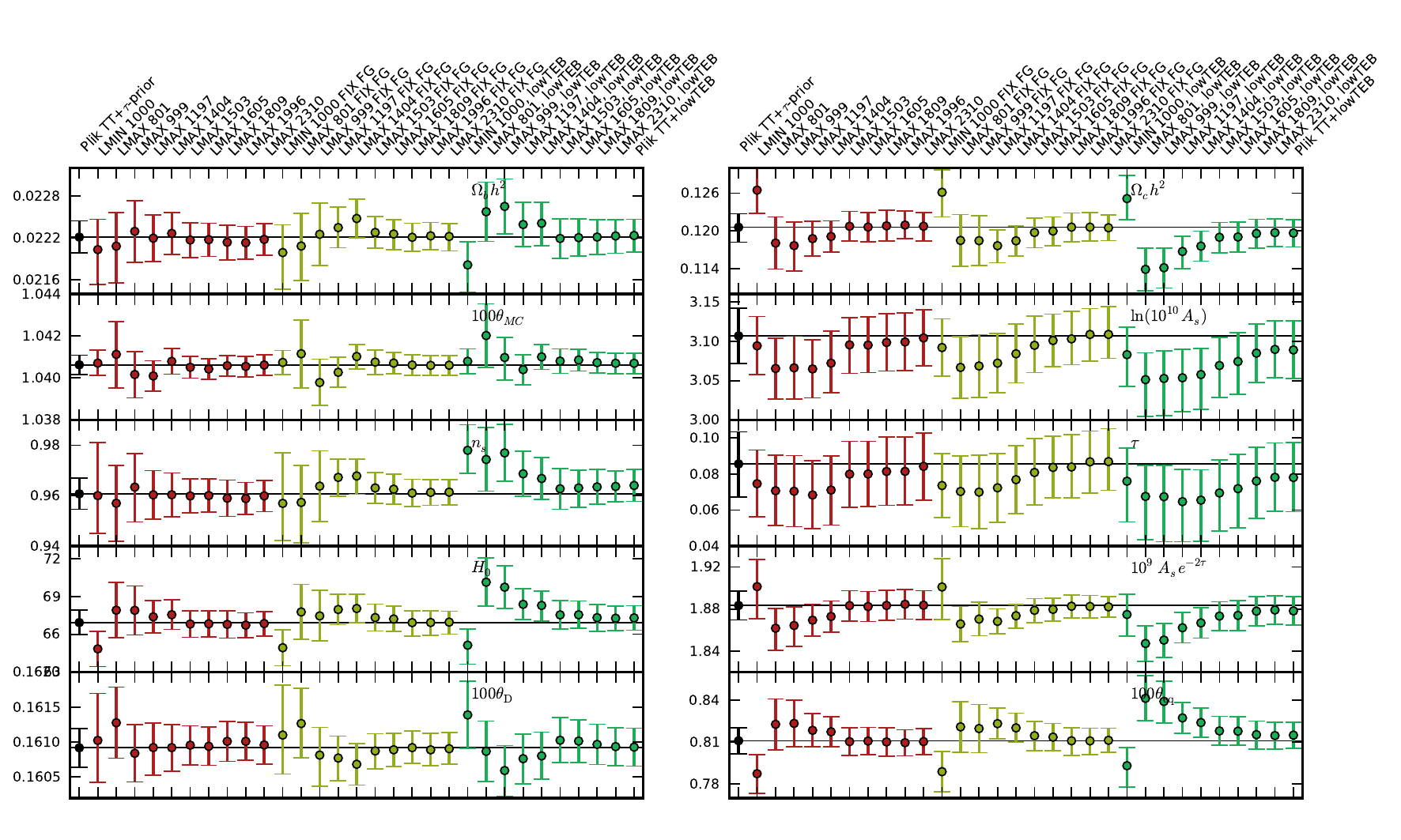}
\caption{Marginal mean and 68\,\% CL error bars on cosmological parameters estimated when adopting different data choices for the \plik\ likelihood. We assume a  $\Lambda$CDM model and calculate parameters using different maximum multipole $\lmax$. The red points show the results for \plikTTtau, with the points specifically labelled ``\plikTTtau'' in black showing the baseline \plikTT\ likelihood at $\lmax=2508$, the yellow points show results for  \plikTTtau\ but fixing the foregrounds to the best-fit of the baseline likelihood (``FIX FG''), and the green points show results for $\plik$TT combined with the low-$\ell$ likelihood in temperature and polarization (``\plikTT+\lowTEB'').}
\label{fig:wiskerTTFGlowl}
\end{figure*}

Figure~\ref{fig:wiskerTTFGlowl} shows the results for different $\lmax$ for three different settings. We show results for \plikTTtau\ (red points), identical to the ones already shown in Fig.~\ref{fig:wiskerTT}; for \plikTTtau, but fixing the foregrounds to the best-fit of the baseline likelihood (yellow points); and for {\plik}TT combined with the low-$\ell$ likelihood in temperature and polarization (green points, {\plik}TT+\lowTEB\ in the plot). This figure shows that in all these three cases we have similar behaviour for $\lnAs$, $\Omch$, and $\tau$, \ie\ they all increase with increasing $\lmax$. However, the evolution of the other parameters differs. While in the \plikTTtau\ case the other parameters do not change significantly (apart from the shift in $\theta$ between $\lmax \approx 1200$--$1300$ already described in Sect.~\ref{sec:changes_lmax}), fixing the foregrounds forces other parameters such as $\ns$ and $\Ombh$ to shift as well. It is interesting to note that all the parameters tend to converge to the baseline solution between $\lmax=1404$ and $1505$, confirming the impact of the fifth peak in determining the final solution, as already described in Sect.~\ref{sec:changes_lmax}.

As far as the {\plik}TT+\lowTEB\ combination is concerned, adding the low-$\ell$ multipoles in temperature pulls $\ns$ to higher values in order to better fit the deficit at $\ell\sim 20$--$30$. This pull is  more effective when excising the high-$\ell$ data (\ie when using low $\lmax$), pushing $\Omch$ to even lower values, following the $\ns$--$\Omch$ degeneracy.

\subsubsection{Polarization robustness tests} \label{sec:pol-robust}

\begin{figure*} 
\centering
\includegraphics[width=\textwidth]{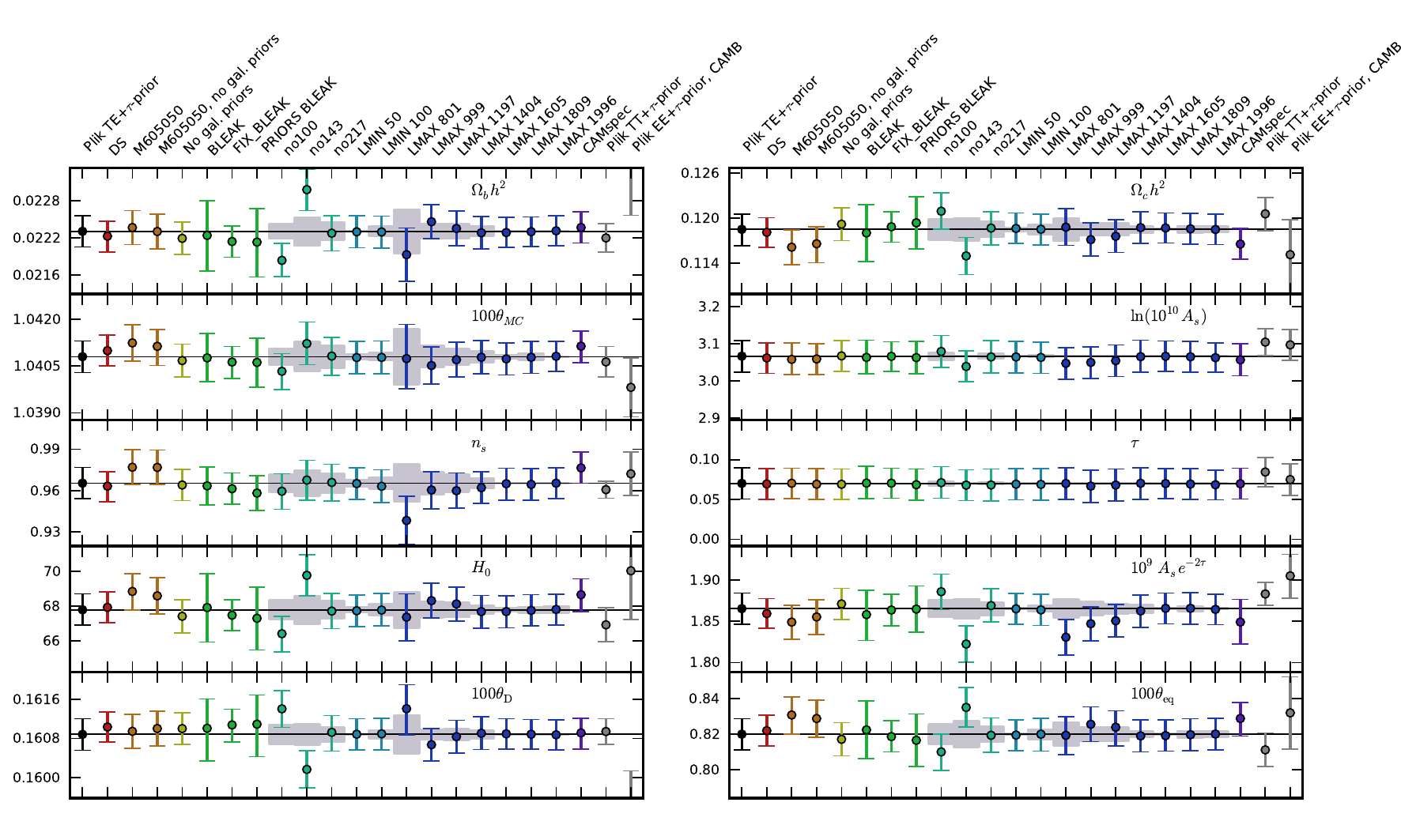}
\includegraphics[width=\textwidth]{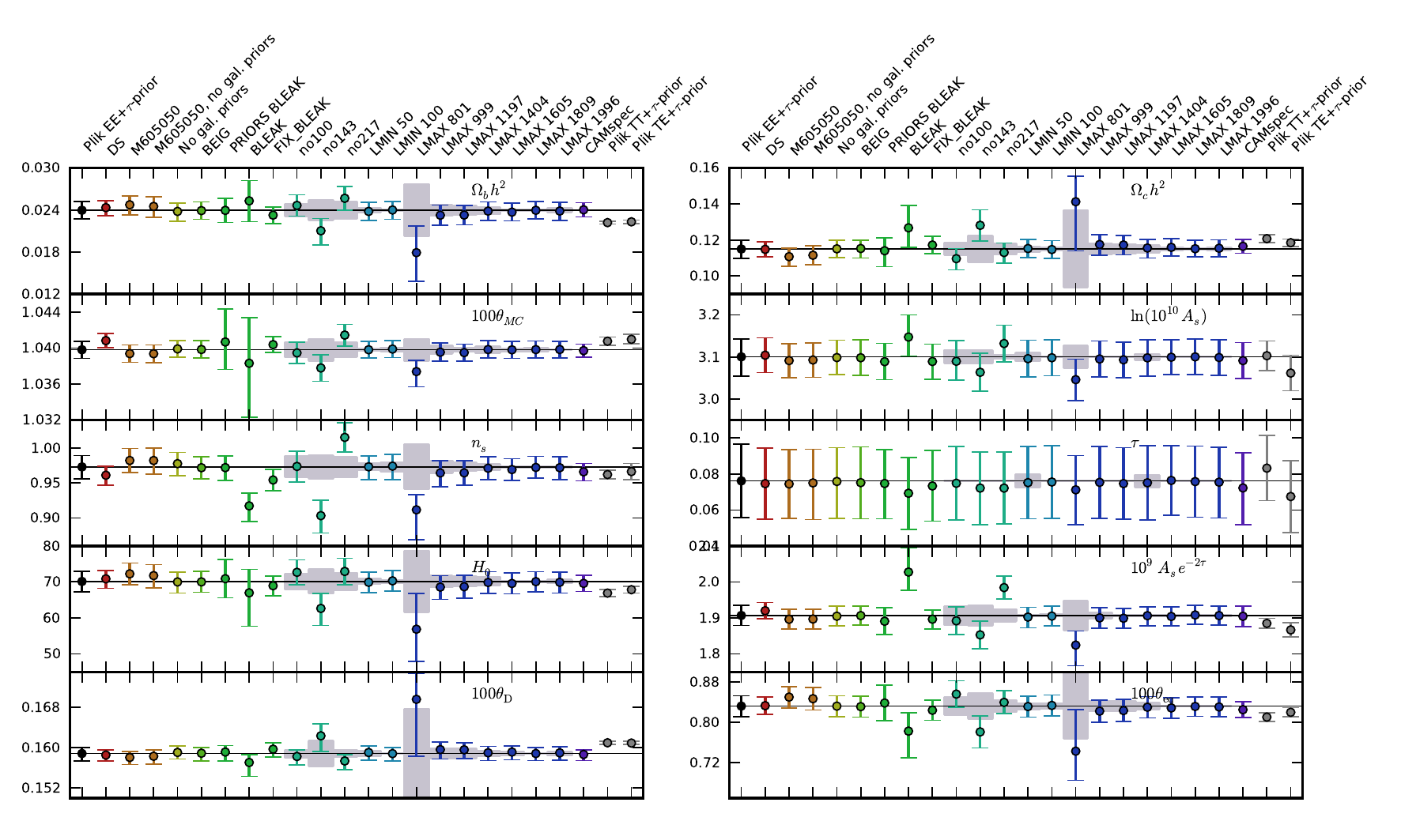}
\caption{Marginal mean and 68\,\% CL error bars on cosmological parameters estimated adopting different data choices for the \plik likelihood, in comparison with results from alternate approaches or model. \emph{Top}: $\TE$ tests;  we assume a  $\Lambda$CDM model and use the \plikTEtau\ likelihood in most of the cases, with a prior on $\tau=0.07\pm0.02$ (we do not use low-$\ell$ temperature or polarization data here.). The ``\plikTEtau'' case (black dot and thin horizontal black line) indicates the baseline (HM, $\ell_{\mathrm{min}}=30$, $\lmax=1996$), while the other cases are described in Appendix~\ref{sec:pol-robust}.  The grey bands show the standard deviation of the expected parameter shift, for those cases where the data used are a sub-sample of the baseline likelihood (see Eq.~\ref{eq:covresult}). All the cases shown in these $\TE$ plots are run with \PICO, except for the ``\plikEEtau, \CAMB'' case, which is run with \CAMB\ (see Appendix~\ref{app:pico} for further details). \emph{Bottom}: $\EE$ tests; the same as the top plots, but for the \plikEEtau\ likelihood. For these $\EE$ plots we used \CAMB\ instead of \PICO\ to run all the cases (including \plikTTtau\ and \plikTEtau). }
\label{fig:wiskerpol}
\end{figure*}

We now present the results of the tests we conducted so far to assess the robustness and accuracy of the polarization results, with the same tools as used for $\TT$ (described in the main text).
In the parameter domain, the results are summarized in Fig.~\ref{fig:wiskerpol}, which shows the  marginal mean and the 68\,\% confidence limit (CL) error bars for cosmological parameters using the {\plik}TE or {\plik}EE high-$\ell$ likelihoods under different assumptions about the data selection, foreground model, or treatment of the systematics. In the following, we comment, in turn, on each of the tests shown in this figure (from left to right). In most of the cases, we use the $\plik$ likelihoods in combination with the usual Gaussian $\tau$ prior, $\tau=0.07\pm 0.02$. The reference \plikTEtau\ and \plikEEtau\ results for the \LCDM\ model are denoted as ``\plikTEtau'' and ``\plikEEtau.'' We note that all the $\TE$ tests are run with the \PICO\ code, while the $\EE$ ones are run with the \CAMB\ code, for the reasons given in  Appendix~\ref{app:pico}.

\paragraph{Detsets}
We find good agreement between the baseline cases based on half-mission spectra and those based on  detset spectra (case ``DS''). We find the greatest deviations in $\EE$, where the DS case shows values of $\Ombh$ and $\theta$ higher than the baseline case by about $1\,\sigma$, while $\ns$ is lower by $1\,\sigma$.

\paragraph{Larger Galactic mask} 
We examined the impact of using a larger Galactic mask (case ``M605050'') with $\fsky=0.50$, 0.41, and 0.41 at $100$, $143$, and $217\,\GHz$, respectively (corresponding to $\fsky^\textrm{noap}=0.60$, $0.50$, and $0.50$ before apodization), instead of the baseline values $\fsky=0.70$, $0.50$, and $0.41$. In $\TE$ we observe substantial shifts in the parameters, at the level of $\la 1\,\sigma$. We did not assess whether this is consistent with cosmic variance, but we note that the results remain compatible with ``\plikTTtau'' at the $1\,\sigma$ level.

\paragraph{Galactic dust priors}
We find that leaving the Galactic dust amplitudes completely free to vary (``No Gal.\ priors''), without applying the priors described in Sect.~\ref{sec:dust}, does not have a significant impact on cosmological parameters. This suggest that our foreground model is satisfactory, despite its simplicity.\footnote{We discovered late in the preparation of this paper  that in some of the tests the prior for the $143\times217$ $\TE$ dust contamination was set inaccurately, with an offset of $-0.3\,\mu{\rm K}^2$ at $\ell=500$. With our cuts, this spectrum contributes only at $\ell>500$ where the dust contamination is already small compared to the signal. We verified that this has no impact on the cosmology and on our conclusions.}

\paragraph{Beam eigenmodes}

We have marginalized over the beam uncertainty eigenmodes (case ``BEIG''), finding, as in $\TT$, no impact on cosmological parameters.

\paragraph{Beam leakage}
Section~\ref{sec:beam_leakage} presented a model for the polarization systematic error induced by assuming identical beams in detsets combined at the map-making stage (when the beams do in fact differ). Here we consider three cases for exploring the impact of the 18 amplitudes of the beam leakage model parameters,  $\varepsilon_m$ (for $m=0$, $2$, and $4$; \ie three parameters per cross-frequency spectrum): when we leave these amplitudes completely free to vary along with all other parameters (case ``BLEAK''); when we apply the priors motivated in Sect.~\ref{sec:beam_leakage} (case ``priors\_BLEAK''); and when we use the best-fit values of these parameters (``FIX\_BLEAK'').  The amplitudes for ``FIX\_BLEAK'' are obtained by a prior exploration while keeping all other parameters ($\TT$ cosmology and foregrounds) fixed.  We find that this case has better goodness of fit without otherwise affecting the model.

When we leave the amplitudes completely free to vary, there is no significant impact on cosmology in $\TE$, with shifts at the level of fractions of $\sigma$, which is reassuring. For $\EE$, though, we find large deviations in the ``BLEAK'' case, suggesting strong degeneracies between the  cosmological and beam leakage parameters in $\EE$. And for both $\TE$ and $\EE$, we find that the beam leakage parameters adopt values in the ``BLEAK'' case that are {much} higher than the values expected from the priors. This shows that other residual systematic effects project substantially onto these template shapes, which is not surprising, given the additional degrees of freedom. 

If we use our so-called cosmological prior (case ``FIX\_BLEAK''), \ie when we fix leakage parameters to their best-fit values, in order to see how they improve the overall goodness of fit, the uncertainties remain close to the reference case (when the $\varepsilon_m$ are set to zero) and of course the results shift slightly towards the ``\plikTTtau'' result.  By using this $\TT$ solution, the fit improves by $\Delta\chi^2=55$ in $\TE$, and only $\Delta\chi^2=26$ in $\EE$, while  opening 18 new parameters (and $\TT$ has 765 bins, while $\TE$ and $\EE$ have 762 bins). For $\TE$ in particular, the corrections are not sufficient to significantly improve the $\chi^2$, which is too large, and dominated by the disagreement between individual spectra. Furthermore, the beam-leakage parameter values that we recover are higher than what we expect from the physical priors.

If instead we apply the physical priors, the best-fit cosmological values are not strongly affected, except for a small shift towards the ``\plikTTtau'' case, and the errors bars are increased substantially compared to the fixed-leakage-parameter cases. But we find that the $\chi^2$ value of the fit does not improve significantly (\ie barely any change in $\EE$, and $\Delta \chi^2 \approx 20$ in $\TE$). The discrepancy between frequencies remains. We also explored the simultaneous variation of the leakage and calibration parameters within their expected physical priors, and found results similar to the case of the variation of the leakage alone. 

In any case, we cannot assign the origin of the frequency-spectra disagreement to beam leakage, alone or in combination with polarization recalibration. The surprisingly high values found for the leakage parameters when they are allowed to vary widely are indicative of the presence of other systematic effects that are absent from our model. We therefore do not include these corrections in the final baseline likelihood; we only use them to estimate the possible amount of residual beam leakage in the co-added spectra, which is around $1\,\mu{\rm K}^2\ (\mathcal{D}_\ell)$ in $\TE$ and $1\times 10^{-5}\mu{\rm K}^2\ ({C}_\ell)$ in $\EE$. 

\paragraph{Cutting out frequency channels}
We have considered the cases where we eliminate all the power-spectra related to one particular frequency at a time, as in the $\TT$ analyses;  e.g., the ``no 100'' case uses only the $143\times143$, $143\times217$, and $217\times217$ spectra. In $\TE$, we see strong shifts (in opposite directions) when either the 100 or the 143\ghz\ data are removed, much more than one would expect due to the change of information (given by the grey bands in Fig.~\ref{fig:wiskerpol}). In $\EE$, we instead see strong shifts in opposite directions when either the 143 or the 217\ghz\ data are dropped. Furthermore, we note in $\EE$ the rather big and similar change in $\EE$ parameters when the 143\ghz\ data are dropped and when the leakage parameters are varied. 

\paragraph{Changing $\lmin$}
We find good stability in the results when changing the minimum multipole $\lmin$ considered in the analysis (``LMIN'' case). The baseline likelihood has $\lmin=30$, and we test the cases of $\lmin=50$ and $100$.

\paragraph{Changing $\lmax$}
We observe small shifts when including maximum multipoles between $\lmax\sim 1000$ and $2000$ (``LMAX'' cases). This is not surprising, since even though the baseline has $\lmax=2000$, most of the constraining power of our polarization spectra comes from $\ell <1000$. When using $\lmax=801$, we find bigger shifts, non-Gaussian parameter posterior distributions (for $EE$), and a significant increase in the error bars. This increase is expected from Fisher-matrix forecasts (see, e.g., figure~8 of \citealt{Galli:2014kla}), which show that the $\EE$ constraint on $\ns$ is expected to be more than a factor of $2$ weaker in the $\lmax=801$ case. This is confirmed by the tests presented here. Also, note that the grey bands in Fig.~\ref{fig:wiskerpol}, which indicate the standard deviation of the expected shifts, are calculated under the assumption of Gaussian parameter posterior distributions, and thus fail to properly describe non-Gaussian cases such as $\EE$ $\lmax=801$ considered here.

\paragraph{Comparison to \camspec}
We find relatively good consistency with the results of the \camspec code, with shifts smaller than about $1\,\sigma$ in $\TE$ and $0.5\,\sigma$ in $\EE$. Let us recall that the \camspec and \plik codes adopt different choices of Galactic mask, Galactic dust treatment, and likelihood codes in polarization. Differences at this level therefore illustrate the good agreement reached for this release, and are useful to gauge the impact of quite different choices in the analysis procedures.

\paragraph{Remaining cases}
As expected, the ``lite'' CMB-only likelihood is in agreement with the \plik\ code (see further discussion in Sect.~\ref{app:margin-like}). 

Finally, we note that in some of the cases discussed above, the calibration parameter for $\calibC^{TT}_{217}$ was wrongly set to unity instead of being to varied within its prior. We checked that this does not change our conclusions on the behaviour of the cosmological parameters and their uncertainties.

\paragraph{Summary} 
While a number of tests have been passed, the behaviour for masks, leakage parameters, and channel-data removal shows that systematic uncertainties are at least comparable to the statistical uncertainties. In the absence of a fully satisfactory data model, it is difficult to assess precisely the extent to which the extensive data averaging in the co-added $\TE$ or $\EE$ spectra effectively suppresses  the residual systematic errors, many of which are detector-specific.

\subsubsection{Agreement between temperature and polarization results} \label{agreementpol}

\begin{figure*} 
\centering
\includegraphics[width=\textwidth]{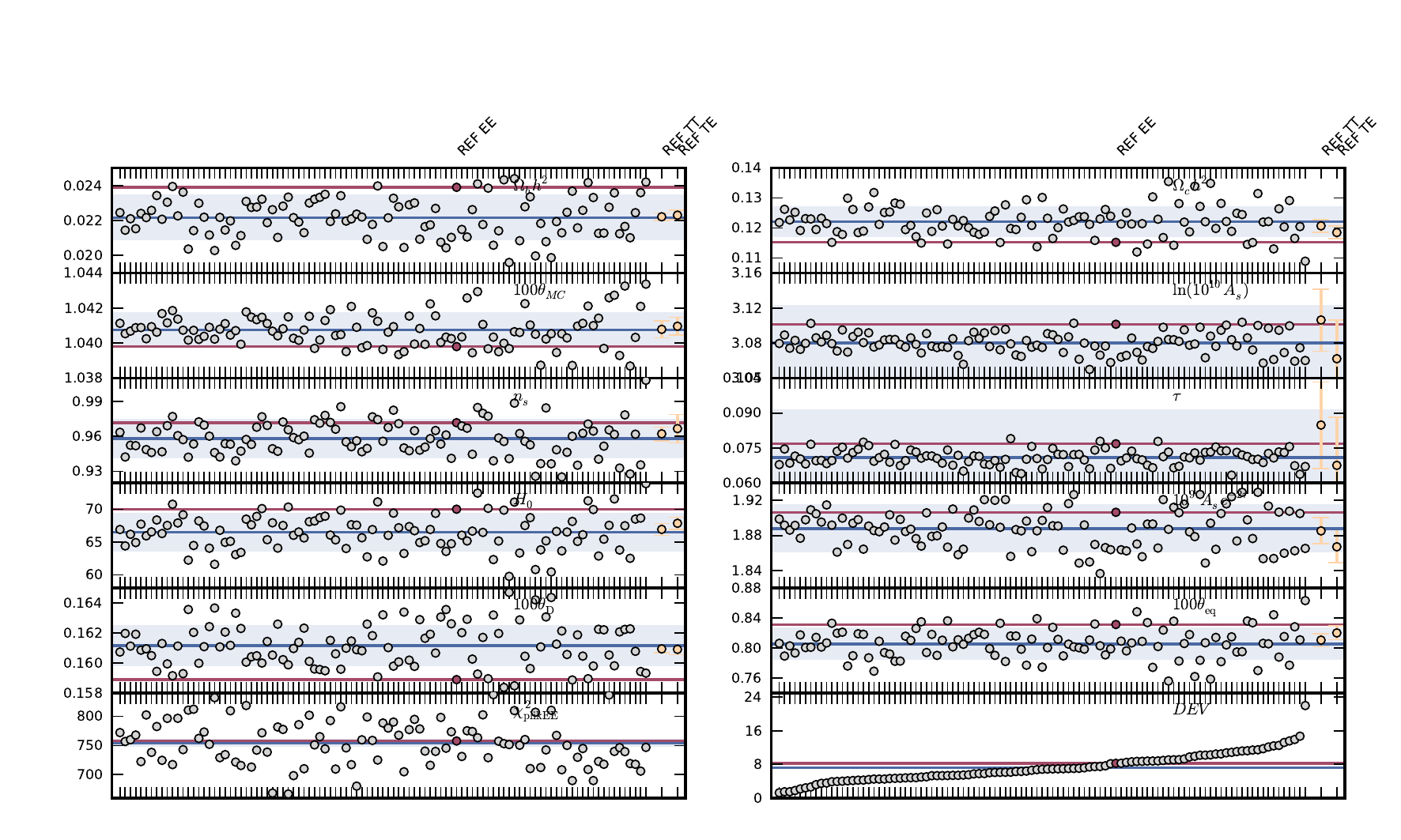}
\includegraphics[width=\textwidth]{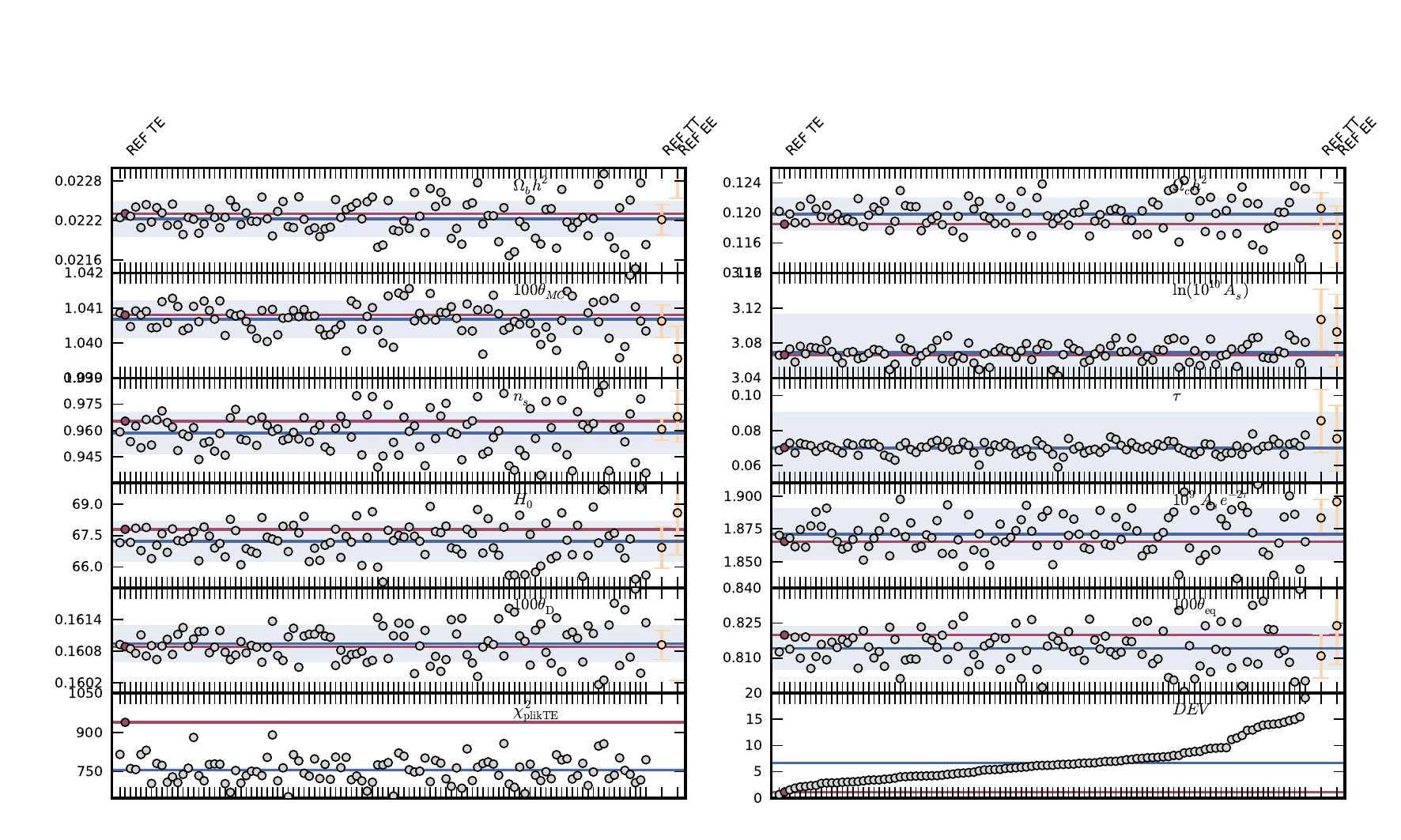}
\caption{Marginal mean and 68\,\% CL error bars on cosmological parameters estimated from 100 $\EE$ (left) or $\TE$ (right) power-spectra simulations conditioned on the $\TT$ power spectrum, assuming as a fiducial cosmology the best-fit of the $\Lambda$CDM {\plikTTtau} results (grey circles). The blue line shows the mean of the simulations, \ie the expected cosmology from the conditioned $\EE$ (or $\TE$) spectra, while the blue band just shows the 68\,\% CL error bar. The different cases are ordered from the least to the most ``deviant'' result according to the $\cal P$ parameter defined in Eq.~\ref{app:DEV} and called `DEV' in the plots. The {\plikEEtau}, {\plikTEtau}, and {\plikTTtau} cases (in red or yellow) show the results from the real data. All the results in the $\EE$ plots were produced using the \CAMB\ code, while those in the $\TE$ plots used the \PICO\ code.}
\label{fig:whiskersims}
\end{figure*}

In order to assess the extent to which the cosmological parameters results that we obtain using the {\plik}EE or {\plik}TE  data alone are compatible with those obtained from {\plik}TT alone, we performed the following test. We simulated 100 sets of $\TE$ or $\EE$ frequency power spectra conditioned on the $\TT$ power spectrum. As a fiducial model, we used the best-fit solution of the $\Lambda$CDM \plikTTtau\ data combination. For all the polarization-related parameters (e.g., Galactic dust amplitudes) we used the best-fit solution of the {\plikTTEETEtau} data combination. We estimated cosmological parameters from each of these simulations, using the same assumptions as were adopted for the real data, and estimated the mean of the parameters obtained from the simulations. We then evaluated the deviation parameter ${\cal P}$ for each of the simulations as
\begin{equation}
{\cal P} = (\vec{P} - \left<\vec{P}\right>)^\tens{T} \tens{P}^{-1}  (\vec{P} - \left<\vec{P}\right>) ,
\label{app:DEV}
\end{equation}
where $\vec{P}$ is the vector of all varied parameters in the run (cosmological and foreground), $\tens{P}$ is the covariance matrix of the parameters, and $\langle\vec{P}\rangle$ is the mean of the parameters over the 100 simulations. The ${\cal P}$ parameter provides us a measure of how much {all} the parameters differ from their means, taking into account the correlations among them.
We  calculate the ${\cal P}$ parameter also for the results obtained from the real data, {\plikEEtau} or {\plikTEtau}, and compare these values to those obtained from the simulations.
For $\EE$, there are 36 simulations with a deviation ${\cal P}$ higher than the {\plikEEtau} case, suggesting that the shifts in parameters we observe between {\plikEEtau} and {\plikTTtau} are in good agreement with expectations.
For $\TE$, there are 99 simulations with a deviation ${\cal P}$ higher than the {\plikTEtau} case, suggesting that for $\TE$ the probability of obtaining parameters so close to the expected ones is only at the level of a few percent (although a more precise statement would require at least an order of magnitude more simulations). We note that this is not statistically very probable, but we could not identify any systematic reason why this should be so in all the tests conducted so far.

Figure~\ref{fig:whiskersims} shows the cosmological parameters obtained from the simulations (grey points), together with their mean (blue line). For clarity, we omit the error bars on the individual points (since they are all the same for each parameter), but show it instead as a light-blue band around the mean of the simulations. The cases shown in the figure are ordered by the ${\cal P}$ parameter from smallest to biggest (most ``deviant'').

It is interesting to note that the mean of the simulations, both for $\EE$ and $\TE$, is very close to the cosmology obtained using the {\plikTTtau} data, as expected. However, for $\As$ and $\tau$, the mean of the simulations is almost 1\,$\sigma$ lower than the value from {\plikTTtau}. As explained in Sect.~\ref{sec:hil_par_stability},  the high value of $\As$ obtained from {\plikTTtau} gives more lensing, better fitting the multipole region $\ell\approx1400$--$1500$. This forces  $\tau$ to adopt values about $1\,\sigma$ higher that those preferred by its Gaussian prior, in order to marginally compensate for the rise in $\As$ in the normalization of the power spectrum, $\As \exp(-2\tau)$.

The high-$\ell$ $\TE$ and $\EE$ likelihoods, however, detect lensing at a much lower significance than in $\TT$, and are thus sensitive only to the combination $\As \exp(-2\tau)$. The individual constraints on $\As$ and $\tau$ are thus completely dominated by the prior on $\tau$, centred on a value lower by about $1\,\sigma$ with respect to the value preferred by the {\plikTTtau} data combination. As a consequence, the constraint on $\As$ from the simulated polarized spectra is lower that that obtained from the temperature data. 

\subsection{Co-added CMB spectra} \label{app:co-added}
This section illustrates the method we use to calculate the co-added CMB spectra. We first produce foreground-cleaned frequency power spectra using a fiducial model for the nuisance (e.g., foreground) parameters. The figures shown in Sects.~\ref{sec:hil}--\ref{sec:hal} use the $\Lambda$CDM {\plikTTtau} ({\planckTT}) best-fit solution as a  fiducial model for the temperature-related nuisance parameters, and {\plikTTEETEtau}  ({\planckall}) for all the other polarization-specific nuisance parameters (e.g., polarized Galactic dust amplitudes).
We then search for the maximum likelihood solution for the CMB power spectrum $C_\ell^\mathrm{CMB}$ that minimizes:
\begin{eqnarray}
-\ln{\cal L}(\vec{\hat C} | \vec{C}^\mathrm{CMB}) &=& \frac{1}{2} \left[\vec{\hat C} - \vec{C}^\mathrm{CMB}\right]^\tens{T} \tens{C}^{-1} \left[\vec{\hat C} - \vec{C}^\mathrm{CMB}\right] + {\rm constant}\, ,\nonumber\\
\label{tominimize}
\end{eqnarray}
where $\vec{\hat C}$ is the foreground-cleaned frequency data vector, $\vec{C}^\mathrm{CMB}$ is the CMB vector we want to determine, and $\tens{C}$ is the covariance matrix. For instance, if we wanted to find the co-added CMB spectrum for $\TT$ alone, the vectors would be: 
\begin{eqnarray}
\vec{\hat{C}} \! &= \left(\vec{\hat{C}}^{TT}_{100 \times 100},
                          \vec{\hat{C}}^{TT}_{143 \times 143}, \vec{\hat{C}}^{TT}_{143\times 217}, \vec{\hat{C}}^{TT}_{217 \times 217}\right) \\
\vec{C}^\mathrm{CMB} \! &= \left(\vec{C}^{TT,\mathrm{CMB}},
\vec{C}^{TT,\mathrm{CMB}}, \vec{C}^{TT,\mathrm{CMB}}, \vec{C}^{TT,\mathrm{CMB}}\right), 
\end{eqnarray}
\rev{which we can rewrite}
\begin{eqnarray}
\vec{C}^\mathrm{CMB} \! &= \tens{J}\ \vec{C}^{TT,\mathrm{CMB}},
\end{eqnarray}
\rev{where $\tens{J}$ is a tall matrix which connects the power spectrum multipoles to the correct locations in the vector $\vec{C}^\mathrm{CMB}$;  each column of the matrix contains only ones and zeros.}

\rev{We minimize Eq.~(\ref{tominimize}) by solving the linear system}
\begin{eqnarray}
\frac{\partial (-\ln{\cal L}(\vec{\hat C}))}{\partial \vec{C}^\mathrm{CMB}}=\frac{1}{2}\left(2\,  \tens{J}^\tens{T} \tens{C}^{-1} \left[\vec{\hat C} - \tens{J}\ \vec{C}^{TT,\mathrm{CMB}}\right]  \right)=0 \,,
\label{app:min}
\end{eqnarray}
where we used the fact that 
\begin{equation}
\tens{J}^\tens{T} \tens{C}^{-1} \left[\vec{\hat C} - \vec{C}^\mathrm{CMB}\right]=\left(\tens{J}^\tens{T} \tens{C}^{-1} \left[\vec{\hat C} - \vec{C}^\mathrm{CMB}\right]\right)^\tens{T}=\left[\vec{\hat C} -\vec{C}^\mathrm{CMB}\right]^\tens{T} \tens{C}^{-1} \tens{J} ,
\end{equation}
since $\tens{C}^{-1}=(\tens{C}^{-1})^T$.
\rev{The solution to Eq.~(\ref{app:min}) is just that of a generalized least-squares problem and is given by}
\begin{eqnarray}
\vec{C}^{TT,\mathrm{CMB}} = \left(\tens{J}^\tens{T} \tens{C}^{-1} \tens{J}\right)^{-1}\ \tens{J}^\tens{T} \tens{C}^{-1}\ \vec{\hat C}.
\end{eqnarray}
We then evaluate the covariance matrix $\tens{C}^\mathrm{CMB}$ of the co-added $\vec{C}^{TT,\mathrm{CMB}}$ spectrum as
\begin{eqnarray}
\tens{C}^\mathrm{CMB}=\left(\tens{J}^\tens{T}\tens{C}^{-1}\tens{J}\right)^{-1}\,.
\end{eqnarray}

\rev{The matrix $\left(\tens{J}^\tens{T} \tens{C}^{-1} \tens{J}\right)^{-1}\ \tens{J}^\tens{T} \tens{C}^{-1}$ mixes the different frequency cross-spectra to compute the co-added solution. This matrix is flat and consists of the concatenation of blocks weighting each a particular cross-spectrum. Taking into account the different $\ell$ ranges for each, one can recast the blocks into diagonal-dominated square matrices. For a given multipole, ignoring the small out-of-band correlations, the relative weights of the cross-spectra in the co-added solution are given by the diagonals of those blocks. This is what we show Fig.~\ref{fig:mixing}.}

\subsection{\PICO} \label{app:pico}

We have used \PICO\ to perform the extensive tests in this paper because it is much faster than \CAMB, which is used in the \Planck\ paper on parameters \citep{planck2014-a15}.  In this section we compare the results obtained using these two codes when evaluating cosmological parameters. 

Table~\ref{tab:picos} shows the parameter shifts (\CAMB\ minus \PICO), in units of standard deviations, assuming a $\Lambda$CDM model and using either code to evaluate cosmological parameters from the  {\plikTTtau}, {\plikTEtau}, and {\plikEEtau} data combinations. For the {\plikTTtau} and the {\plikTEtau}  combinations, the biggest differences are in $\theta$ at about $0.3\,\sigma$, and in $\ns$ at about $0.2\,\sigma$. These differences occur because (1) \PICO\ was trained on the October\ 2012 version of \CAMB\, whereas our \CAMB\ runs use the January 2015 version (relevant differences include minor code changes and a slightly different default value of $T_{\rm CMB}$); (2) \PICO\ assumes three equal-mass neutrinos rather than one single massive one; and (3) a bug in the CosmoMC \PICO\ wrapper  caused a shift in $N_{\rm eff}$ of about 0.015. Despite these differences, the \PICO\ results are sufficient for the  inter-comparisons within this paper. While for {\plikTTtau} and {\plikTEtau} the \PICO\ fitting error is negligible, for {\plikEEtau} runs this is not the case, since the area of parameter space is much greater. For this reason, we actually use \CAMB\ in these cases. 

\rev{During the revision of this paper, we realized that this problem also affects the $\plikTT$ likelihood test that excises the $\ell<1000$ ($\lmin=1000$ case) shown in Fig.~\ref{fig:wiskerTT}. As mentioned in Section~\ref{sec:changes_lmax}, this is due to the fact that this run explores regions of the parameter space that are wider than the $\PICO$ training region; this was also noticed by \citet{Addison:2015wyg}. As a consequence, the results on $\ns$ and $\Ombh$ from this particular case have error bars underestimated by a factor of about two and  mean values mis-estimated by about $0.8\sigma$ with respect to runs performed with \CAMB. We therefore use $\CAMB$ rather than $\PICO$ to calculate results for this particular test.}

Finally, we note that the definition of the $\Alens$ parameter used in \CAMB\ is different from the one in \PICO. The \PICO\ $\Alens$ parameter is defined such that
\begin{align}
C_\ell = \Alens C_\ell^{\rm lensed} + (1-\Alens)C_\ell^{\rm unlensed}\,,
\end{align}
which is identical to \CAMB's definition only to first order. 

\begin{table}[ht!] 
\begingroup 
\newdimen\tblskip \tblskip=5pt
\caption{Differences between cosmological parameter estimates from \CAMB\ and \PICO.$^{\rm a}$}
\label{tab:picos}
\vskip -6mm
\footnotesize
\setbox\tablebox=\vbox{
\newdimen\digitwidth
\setbox0=\hbox{\rm 0}
\digitwidth=\wd0
\catcode`*=\active
\def*{\kern\digitwidth}
\newdimen\signwidth
\setbox0=\hbox{+}
\signwidth=\wd0
\catcode`!=\active
\def!{\kern\signwidth}
\newdimen\decimalwidth
\setbox0=\hbox{.}
\decimalwidth=\wd0
\catcode`@=\active
\def@{\kern\decimalwidth}
\halign{ 
\hbox to 1in{#\leaderfil}\tabskip=2em& 
    \hfil$#$\hfil&  
    \hfil$#$\hfil&  
    \hfil$#$\hfil\tabskip=0pt\cr  
\noalign{\doubleline}
\omit&\multispan3\hfil (\CAMB$-$\PICO)/$\sigma$(\CAMB))\hfil\cr
\noalign{\vskip -3pt}
\omit&\multispan3\hrulefill\cr
\omit\hfil Parameter\hfil&TT&TE&EE\cr
\noalign{\vskip 3pt\hrule\vskip 5pt}%
$\Ombh$&	   !0.00&	!0.00&	!0.63\cr
$\Omch$&	   !0.05&	-0.01&	-0.40\cr
$\theta$&	   !0.32&	!0.31&	!0.26\cr
$\tau$&		   -0.04&	-0.14&	!0.08\cr
$\lnAs$&	   -0.01&	-0.11&	!0.21\cr
$\ns$&		   !0.23&	!0.10&	!0.24\cr
$H_0$&		   !0.01&	!0.07&	!0.52\cr
$\clamp$&	   !0.12&	!0.09&	!0.41\cr
\noalign{\vskip 5pt\hrule\vskip 3pt}
}}
\endPlancktable 
\tablenote {{\rm a}} Parameter shifts, in standard deviations, obtained using \PICO\ or \CAMB. The results assume a \LCDM\ model and the \plikTTtau, \plikTEtau, or \plikEEtau\ data combinations.\par
\endgroup
\end{table}

\subsection{Marginalized likelihood construction}\label{app:margin-like}

\subsubsection{Estimating temperature and polarization CMB-only spectra}

The $\ell$-range selection of the \Planck\ high-$\ell$ likelihood defines $N_{b}=613$ CMB band-powers, $C_b$. The $C_b$ vector is structured in the following way: the first 215 elements describe the \Planck\ $\TT$ CMB power spectrum, followed by 199 elements for the $\EE$ spectrum and 199 for $\TE$. 

The model for the theoretical power for a single cross-frequency spectrum (between frequencies $i$ and $j$) in temperature or polarization, $C_\ell^{{\rm th},ij}$, is written as
\be
C_\ell^{{\rm th},ij}= C_\ell^{\rm CMB} + C_\ell^{{\rm sec},ij}(\theta) ,
\ee
where $C_\ell^{{\rm sec},ij}(\theta)$ is the secondary signal given by thermal and kinetic SZ effects, clustered and Poisson point source emission, and Galactic emission, and is a function of secondary nuisance parameters $\theta$. We convert $C_\ell^{{\rm th},ij}$ to band-powers by multiplying by the binning matrix ${\cal B}_{b\ell}$, \ie $C_b^{{\rm th},ij}=\sum_\ell {\cal B}_{b\ell} C_\ell^{{\rm th},ij}$.
We then write the model for the $C_b$ parameters in vector form as 
\be
C_b^{\rm th}= {\tens A} C_b^{\rm CMB} + C_b^{\rm sec}(\theta),
\ee
where $C_b^{\rm th}$ and $C_b^{\rm sec}$ are multi-frequency spectra, and the mapping matrix ${\tens A}$, with elements that are either 1 or 0, maps the CMB $C_b$ vector (of length $N_b$), which is the same at all frequencies, onto the multi-frequency data. 
We calibrate the model as in the full multi-frequency likelihood, fixing the 143-GHz calibration factor to 1 and sampling the 100 and 217 calibration factors as nuisance parameters (\ie as part of the $\theta$ vector).

We estimate $C_b^{\rm CMB}$, marginalized over the secondary parameters, $\theta$. The posterior distribution for $C_b^{\rm CMB}$, given the observed multi-frequency spectra $C_b$, can be written as
\be
p(C_b^{\rm CMB}|C_b) = \int p(C_b^{\rm CMB},\theta|C_b) p(\theta) d\theta.
\ee
Rather than using, for example,  Metropolis-Hastings, we use Gibbs sampling, which provides an efficient way to map out the joint distribution $p(C_b^{\rm CMB},\theta|C_b$) and to extract the desired marginalized distribution $p(C_b^{\rm CMB}|C_b)$. We do this by splitting the joint distribution into two conditional distributions: $p(C_b^{\rm CMB}|\theta,C_b$), and  $p(\theta|C_b^{\rm CMB},C_b)$.

We write the multi-frequency \textit{Planck} likelihood as 
\ba
-2 \ln\mathscr{L} & = & ({\tens A}C_b^{\rm CMB}+C_b^{\rm sec} - C_b)^{\tens T} 
\Sigma^{-1} ({\tens A}C_b^{\rm CMB}+C_b^{\rm sec}-C_b) \nonumber\\
&&+ \ln \det \Sigma,
\label{eqn:likemf}
\ea
which is a multivariate Gaussian. If $C_b^{\rm sec}$ is held fixed, the conditional distribution for the CMB $C_b$ parameters, $p(C_b^{\rm CMB}|\theta,C_b$),  assuming a uniform prior for $p(C_b^{\rm CMB})$, is then also a Gaussian. It has a distribution given by
\ba
 -2 \ln p(C_b^{\rm CMB}|\theta,C_b) &=& (C_b^{\rm CMB}-{\hat C}_b)^{\tens T} \tens{Q}^{-1} (C_b^{\rm CMB}-{\hat C}_b)\nonumber\\
&& + \ln \det \tens{Q} .
\ea
The mean, $\hat C_b$, and covariance, $\tens Q$, of this conditional distribution are obtained by taking the derivatives of the likelihood in Eq.~(\ref{eqn:likemf}) with respect to $C_b^{\rm CMB}$. This gives mean
\be
{\hat  C}_b= \left[{\tens A}^T{\Sigma}^{-1}{\tens A}\right]^{-1}  \left[{\tens A}^\tens{T}{ \Sigma^{-1}}(C_b- C_b^{\rm sec})\right], 
\ee
and covariance
\be
{\tens  Q}={\tens A}^\tens{T}{\Sigma}^{-1}{\tens A}.
\ee
We draw a random sample from this Gaussian distribution by taking the Cholesky decomposition of the covariance matrix, ${\tens Q}= {\tens L} {\tens L}^\tens{T}$, and drawing a vector of Gaussian random variates $G$. The sample is then given by 
$C_b^{\rm CMB}= {\hat C}_b + {\tens L}^{-1} G$.  

If instead $C_b^{\rm CMB}$ is held fixed, the conditional distribution for the secondary parameters, $p(\theta|C_b^{\rm CMB},C_b$) can be sampled with the Metropolis algorithm in a simple MCMC code. 

To map out the full joint distribution for $\theta$ and $C_b^{\rm CMB}$ we alternate a Gibbs-sampling step, drawing a new vector $C_b^{\rm CMB}$, with a Metropolis step, drawing a trial vector of the secondary parameters $\theta$. About 700\,000 steps are required for convergence of the joint distribution. The mean and covariance of the resulting marginalized CMB powers, $C_b^{\rm CMB}$, are then estimated following the standard MCMC prescription.

\begin{figure}[htbp]
\centering
\includegraphics[width=\columnwidth]{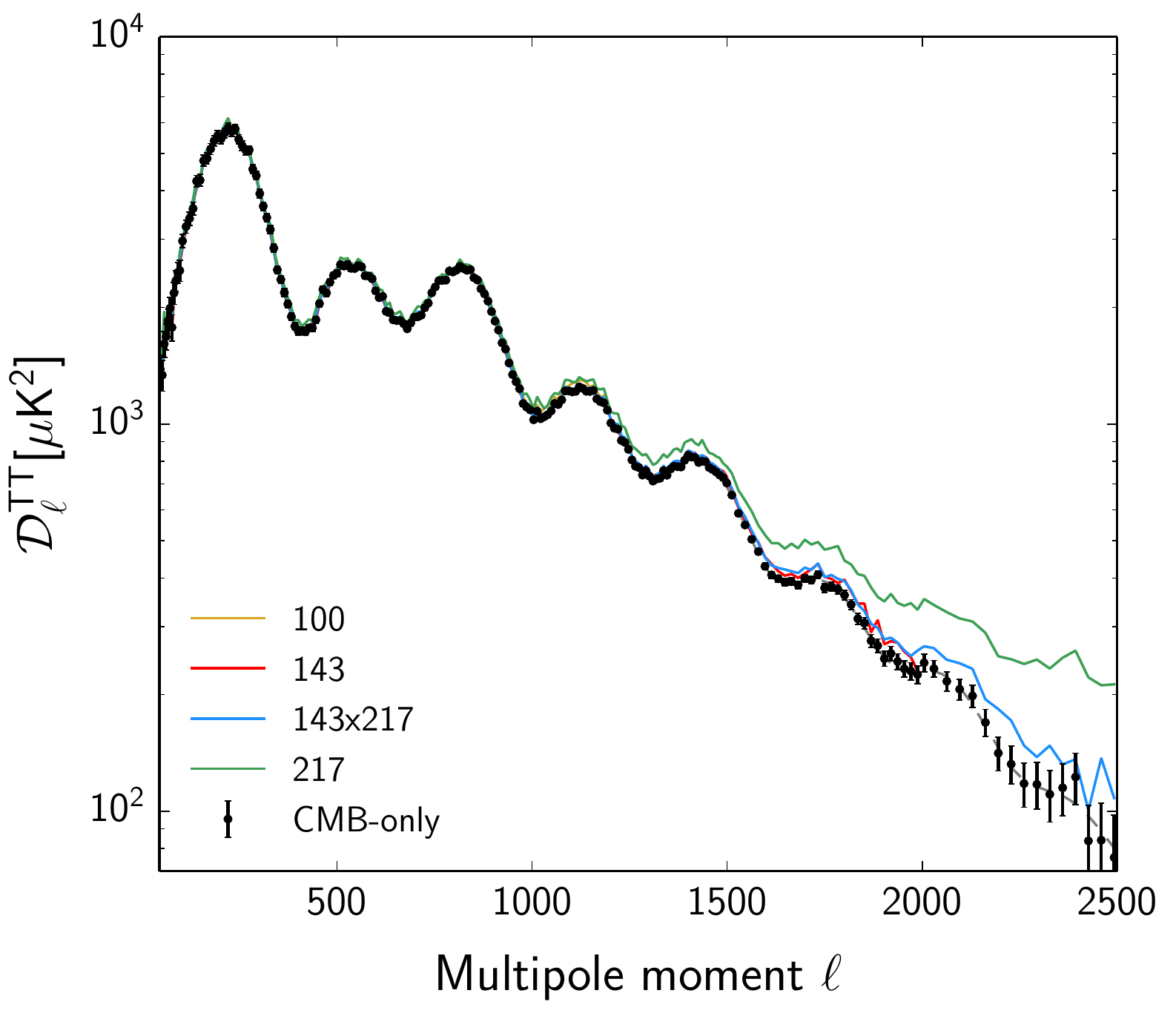}
\includegraphics[width=\columnwidth]{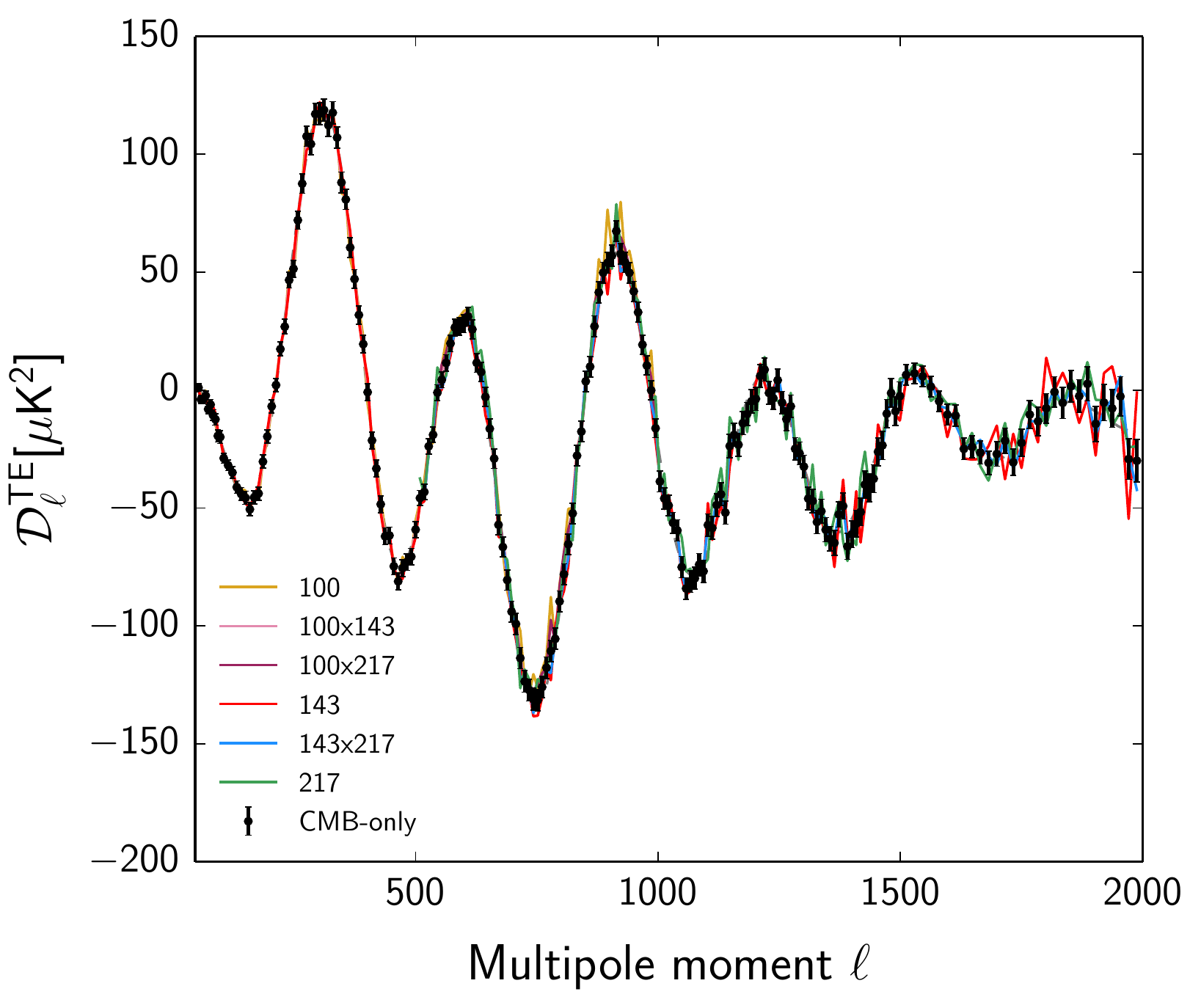}
\includegraphics[width=\columnwidth]{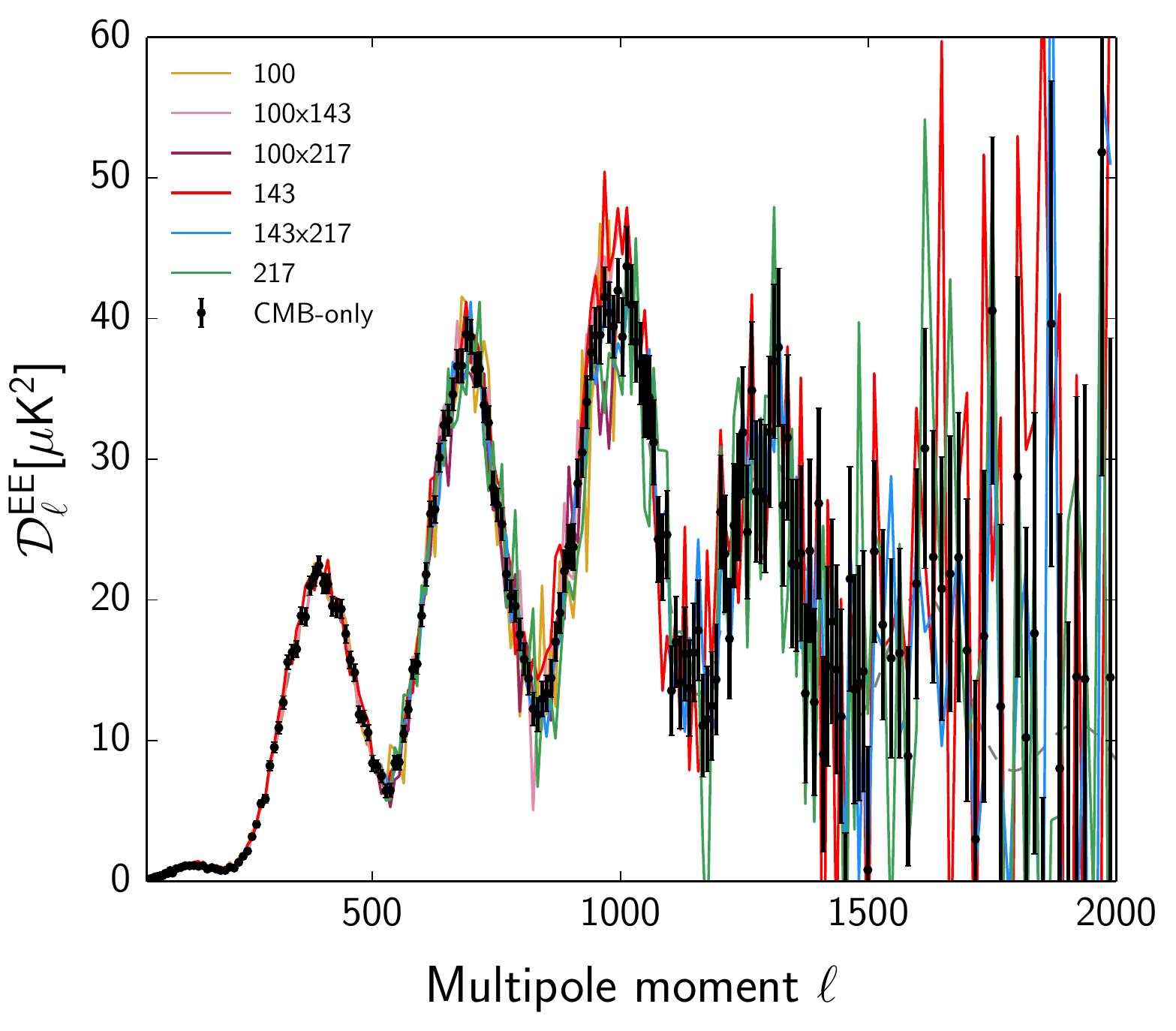}
\caption{Planck multi-frequency power spectra (solid coloured lines) and extracted CMB-only spectra (black points).} 
\label{fig:spec_cmbonly}
\end{figure}

Figure~\ref{fig:spec_cmbonly} shows the multi-frequency data and the extracted CMB-only band-powers for $\TT$, $\EE$, and $\TE$; the CMB is clearly separated out from foregrounds in both temperature and polarization.

\begin{figure*}[htbp]
\centering
 \includegraphics[width=\textwidth]{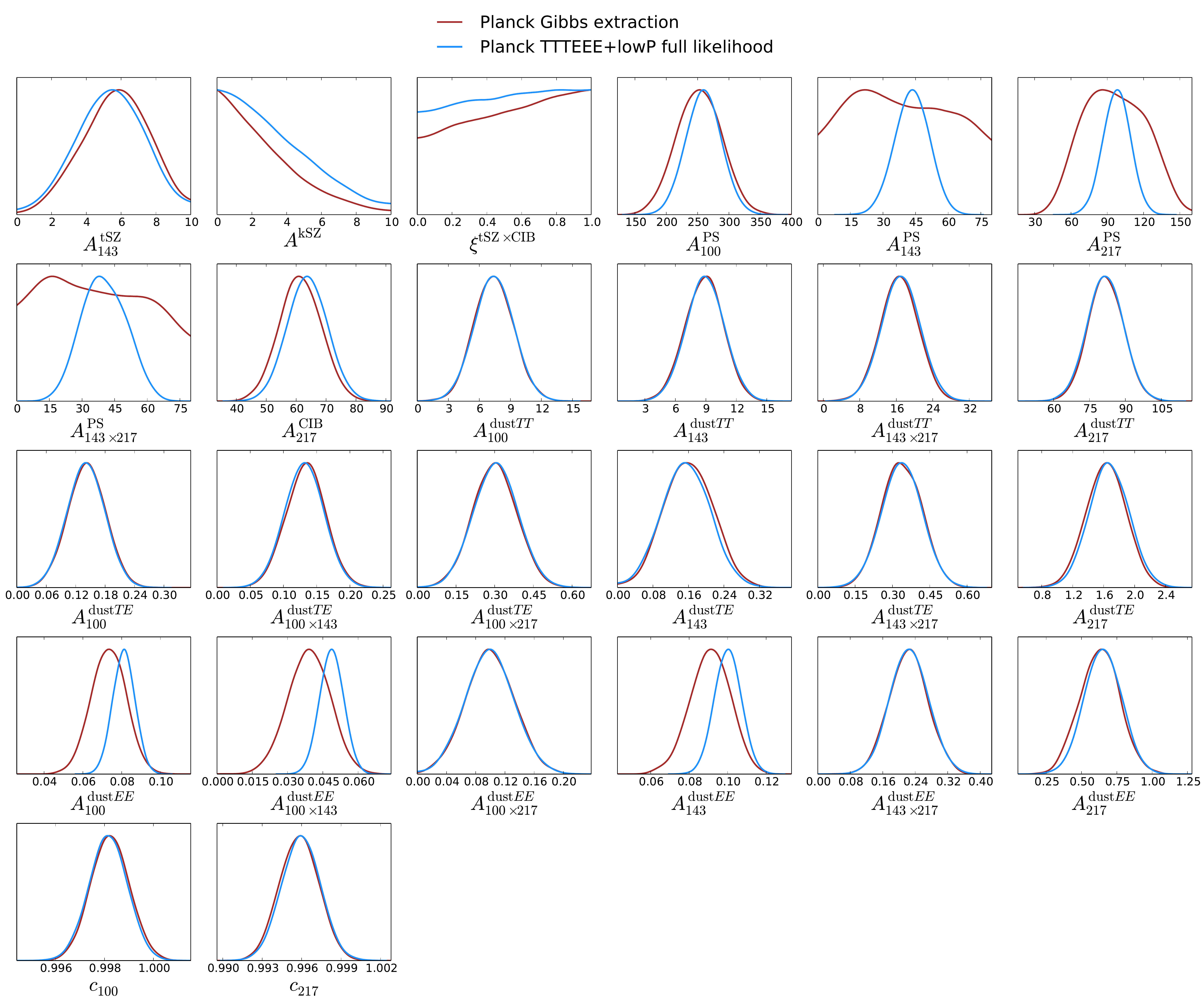}
\caption{Comparison of the nuisance parameters estimated simultaneously with the CMB band-powers (red lines) and the results from the full multi-frequency likelihood (blue lines).} 
\label{fig:fgcomp}
\end{figure*}

Figure~\ref{fig:fgcomp} compares the nuisance parameters $\theta$ recovered in this model-independent sampling and the distributions obtained with the full likelihood. The parameters are consistent, with a broader distribution for the \Planck\ Poisson sources. This degeneracy is observed because the sources can mimic black-body emission and so are degenerate with the freely-varying CMB $C_b$ parameters.

\subsubsection{The {\tt Plik\_lite} CMB-only likelihood}\label{app:plik_lite}

We construct a CMB-only Gaussian likelihood from the extracted CMB C$_b$ bandpowers in the following way:
\be
 -2 \ln\mathscr{L}({\tilde C}_b^{\rm CMB}|C_b^{\rm th}) =  \vec{x}^{\tens T}  {{\tilde \Sigma}}^{-1}\vec{x} \,,
\label{eqn:cmblike}
\ee
where $\vec{x} = {\tilde C}^{\rm CMB}_b/y_p^2 - C_b^{\rm th}$, ${\tilde C}^{\rm CMB}_b$ and ${{\tilde \Sigma}}$ are the marginalized mean and covariance matrix for the $C_b$s, and $C_b^{\rm th}$ is the binned lensed CMB theory spectrum generated from \plik.
The overall \Planck\ calibration $y_p$ is the only nuisance parameter left in this compressed likelihood. The Gaussianity assumption is a good approximation in the selected $\ell$ range, the extracted C$_b$s are well described by Gaussian distributions over the whole multiple range.

\begin{figure}[htbp] 
\centering
 \includegraphics[angle=0,width=\columnwidth]{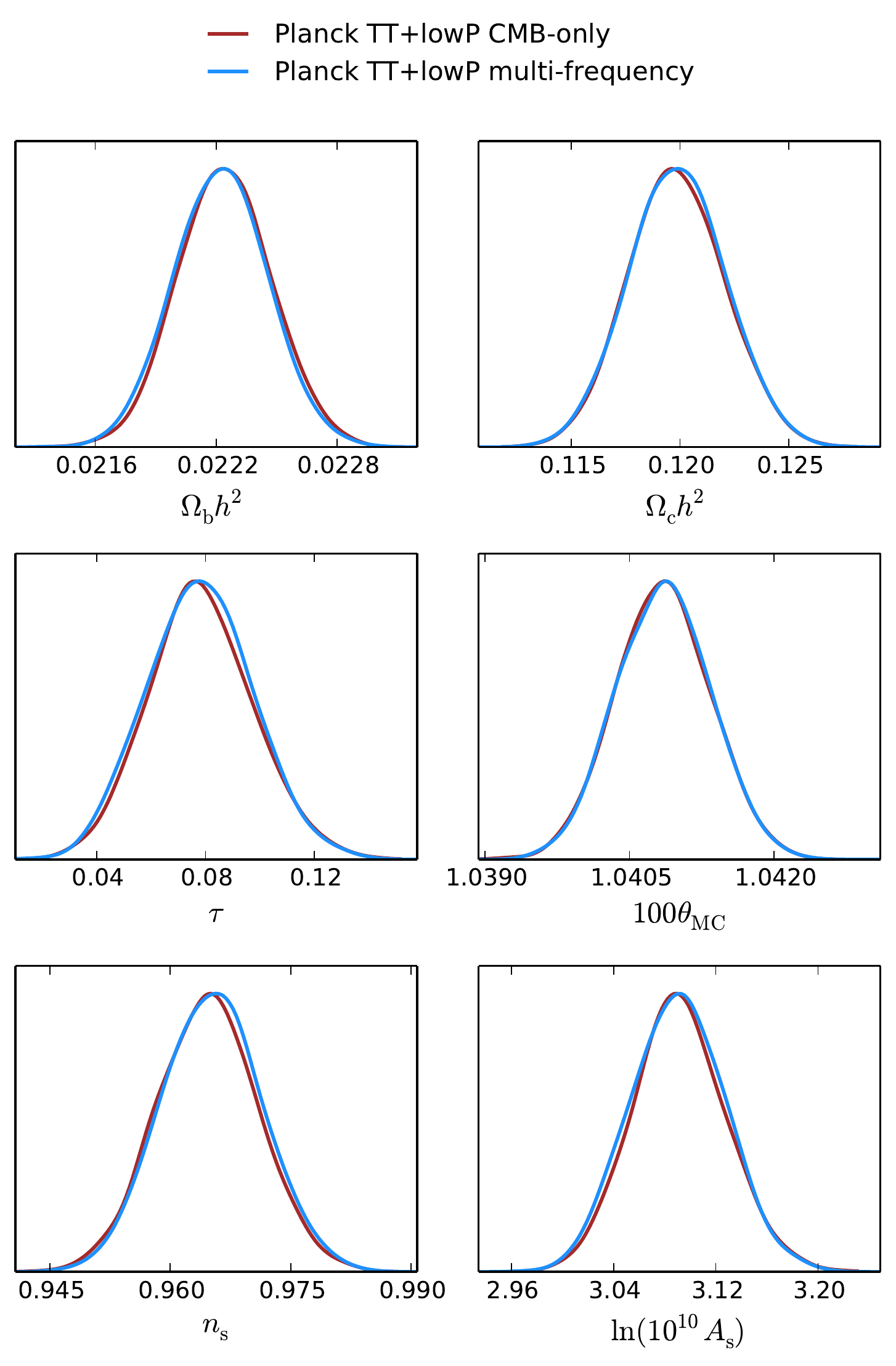}
\caption{Comparison of the six base $\Lambda$CDM parameters estimated with the \Planck\ compressed CMB-only likelihood (red lines) and the full multi-frequency likelihood (blue lines), in combination with \Planck\ lowP data.} 
\label{fig:lcdm}
\end{figure}

To test the performance of this compressed likelihood, we compare results using both the full multi-frequency likelihood and the CMB-only version. We report below examples for the baseline \planckTT\ case. We first estimate cosmological parameters with {\tt Plik\_lite} for the restricted $\Lambda$CDM six-parameter model (see Fig.~\ref{fig:lcdm}) and compare them with  the full-likelihood results. The agreement between the two methods is excellent, showing consistency to better than 0.1$\,\sigma$ for all parameters.

\begin{figure}[htbp] 
\centering
 \includegraphics[width=\columnwidth]{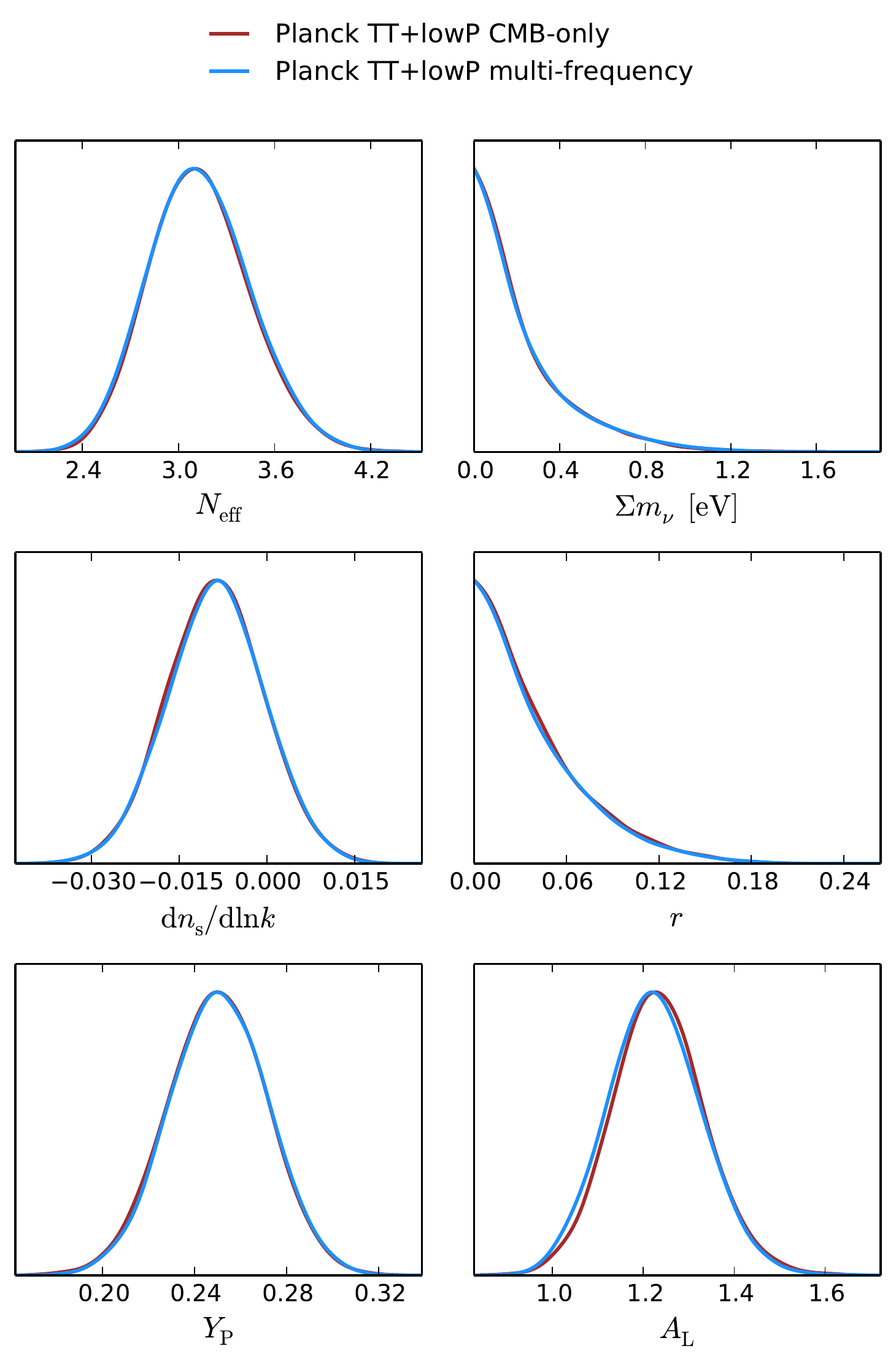}
\caption{Comparison of extensions to the \lcdm\ model from the CMB-only likelihood (red) and the multi-frequency likelihood (blue). There is excellent agreement between the two methods.} 
\label{fig:cmbonly_ext}
\end{figure}

We then extend the comparison to a set of six \lcdm\ extensions, adding one  parameter at a  time to the base-\lcdm\ model: the effective number of neutrino species $\neff$, the neutrino mass $\mnu$, the running of the spectral index $\nrun$, the tensor-to-scalar ratio $r$, the primordial helium fraction $\yhe$, and the lensing amplitude $\Alens$. These parameters affect the damping tail more than the base set, and so are more correlated with the foreground parameters. Distributions for the added parameter in each of the six extensions are shown in Fig.~\ref{fig:cmbonly_ext}. Also in these cases we note that the agreement between the two methods is excellent, with all parameters differing by less than $0.1\,\sigma$. 

We find the same consistency when the polarization data are included in tests using the CMB-only high-$\ell$ $TT$, $TE$, and $EE$ spectra in combination with lowP.


\section{High-$\ell$ likelihood supplement \label{sec:alternatives} }

The \Planck\ team have developed several independent approaches to the high-$\ell$ likelihood problem. These approaches and their implementations differ in several aspects, including  the approximations, the foreground modelling, and the specific aspects that are checked. We have chosen \plik, for which the most supporting tests are available, as the baseline method. The comparison of the approaches  given in the main text gives an indication of how well they agree, and the rather small differences give a feel for the remaining methodological uncertainties. In this appendix, we give a short description of two alternatives to \plik: \mspec\ and \hil. Another alternative,  \camspec, was the baseline for the previous \planck release, and has already been described in detail in \citetalias{planck2013-p08}. Further comparison of \plik and \camspec is provided in the companion paper on cosmological parameters \citep{planck2014-a15}.

\subsection{\mspec}


\label{sec:mspec}

The \mspec likelihood differs from the baseline \plik likelihood mainly in the treatment of Galactic contamination in $\TT$. \mspec results offer a cross-check of the baseline Galactic cleaning method, confirming that Galactic contamination does not have significant impact on the baseline parameters. A second smaller difference is the use of additional covariance approximations that reduce the computation cost while preserving satisfactory accuracy. We now describe these two aspects in more detail.

\paragraph{Galactic cleaning}
Galactic dust cleaning in \mspec is a half-way point between some sophisticated component-separation methods (see Appendix~\ref{sec:mapCheck} and \citealt{planck2013-p06}) and the simple power-spectrum template subtraction or marginalization performed by \plik, \camspec, and \hil. Component-separation methods are flexible and powerful, but propagation of beam and extragalactic-foreground uncertainties into the cleaned maps is difficult, and prohibitive in cost at high $\ell$ even when formally possible (\eg for a Gibbs sampler). On the other hand, the power-spectrum template methods may be  sensitive to errors in template shape and have bigger uncertainties due to signal-dust correlations.   

\mspec cleaning is thus a two-step process. The first step is a simplified component-separation procedure that avoids the above shortcomings: we subtract a single scaled, high-frequency map from each CMB channel. This is very similar to the procedure used in \citet{2013arXiv1312.3313S}, but it is targeted to remove Galactic as opposed to extragalactic contamination. It is also known as a ``two-band ILC,'' and we refer to the procedure as ``map cleaning'' for short. The second step is to {\it then} subtract and marginalize a residual power-spectrum template model akin to the other likelihoods. We now describe each step in more detail.

In the map-cleaning step we subtract a scaled, higher-frequency \Planck\ map from the lower-frequency CMB channels. This is a powerful method of cleaning, because the dust temperature is nearly uniform across the sky, and its intensity increases with frequency. High-frequency maps thus provide essentially noise-free dust maps that are highly correlated with the contamination at lower frequency. We choose to clean temperature maps with 545\,GHz because it is less noisy than 353\,GHz, but more correlated than 857\,GHz. For polarization, only the 353\,GHz detectors are polarization sensitive, and thus we use those. 

\begin{figure*}[htbp] 
\centering
\begin{tabular}{cc}
\includegraphics[width=3in]{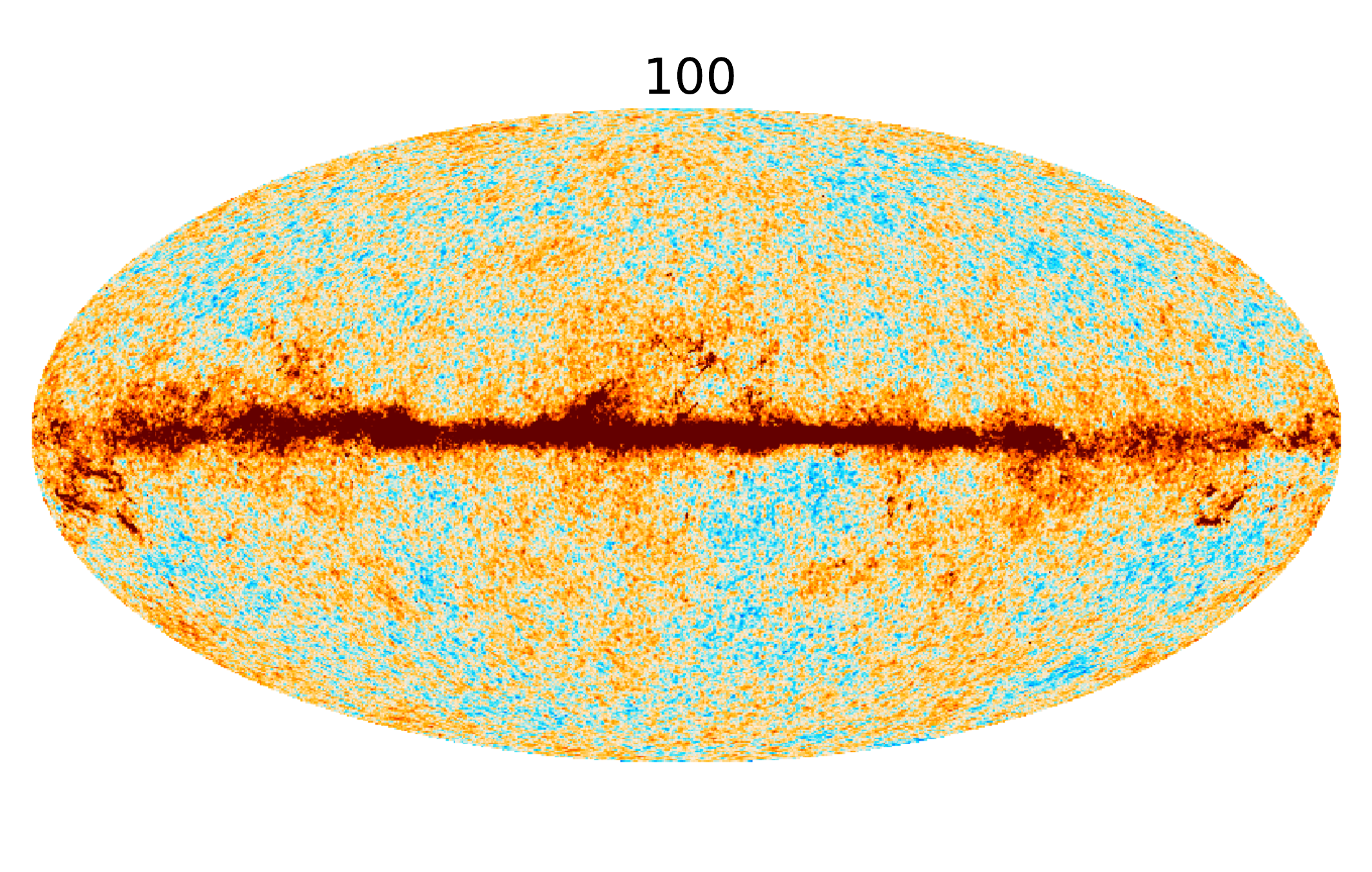} & \includegraphics[width=3in]{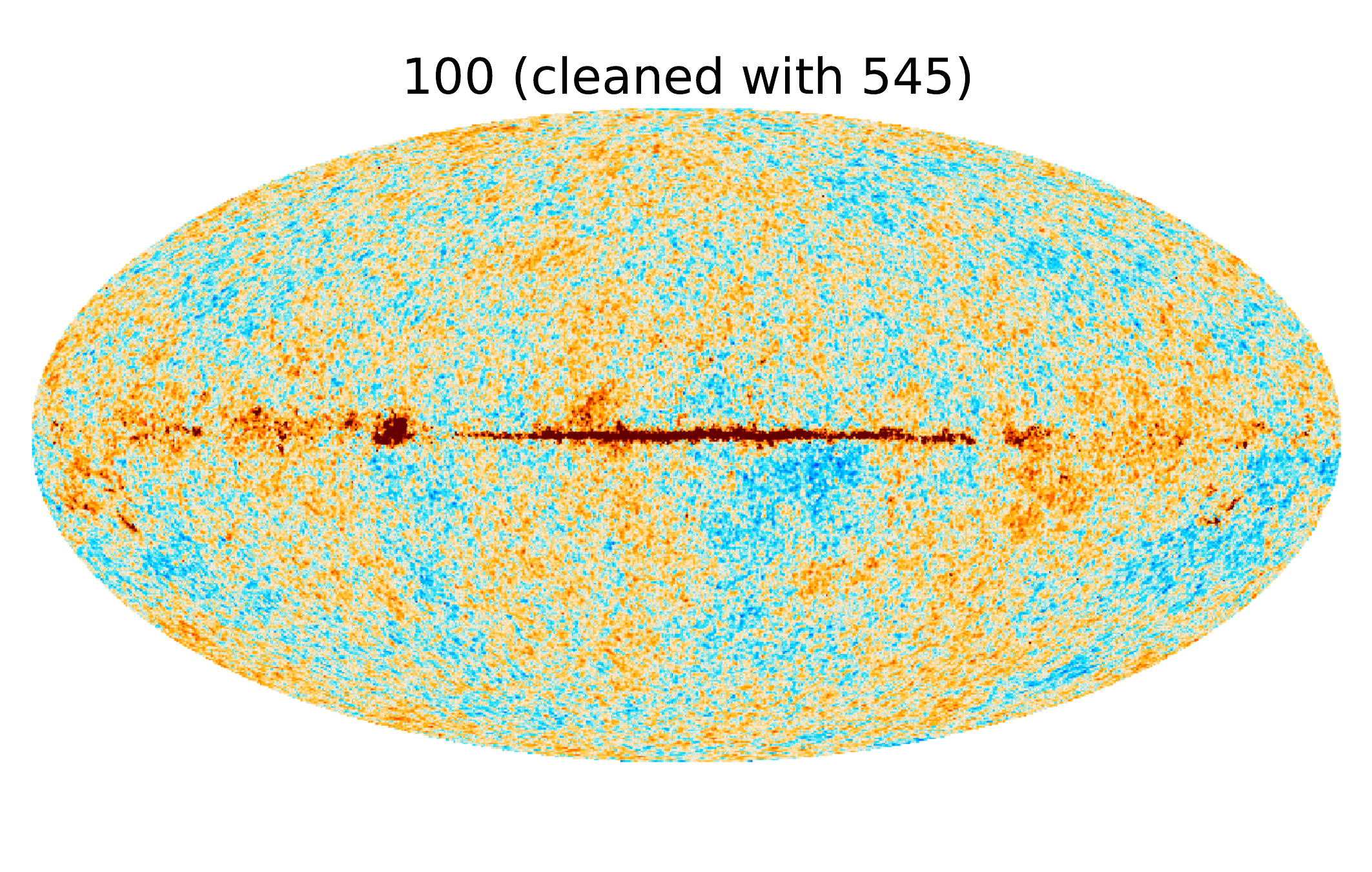} \\
\includegraphics[width=3in]{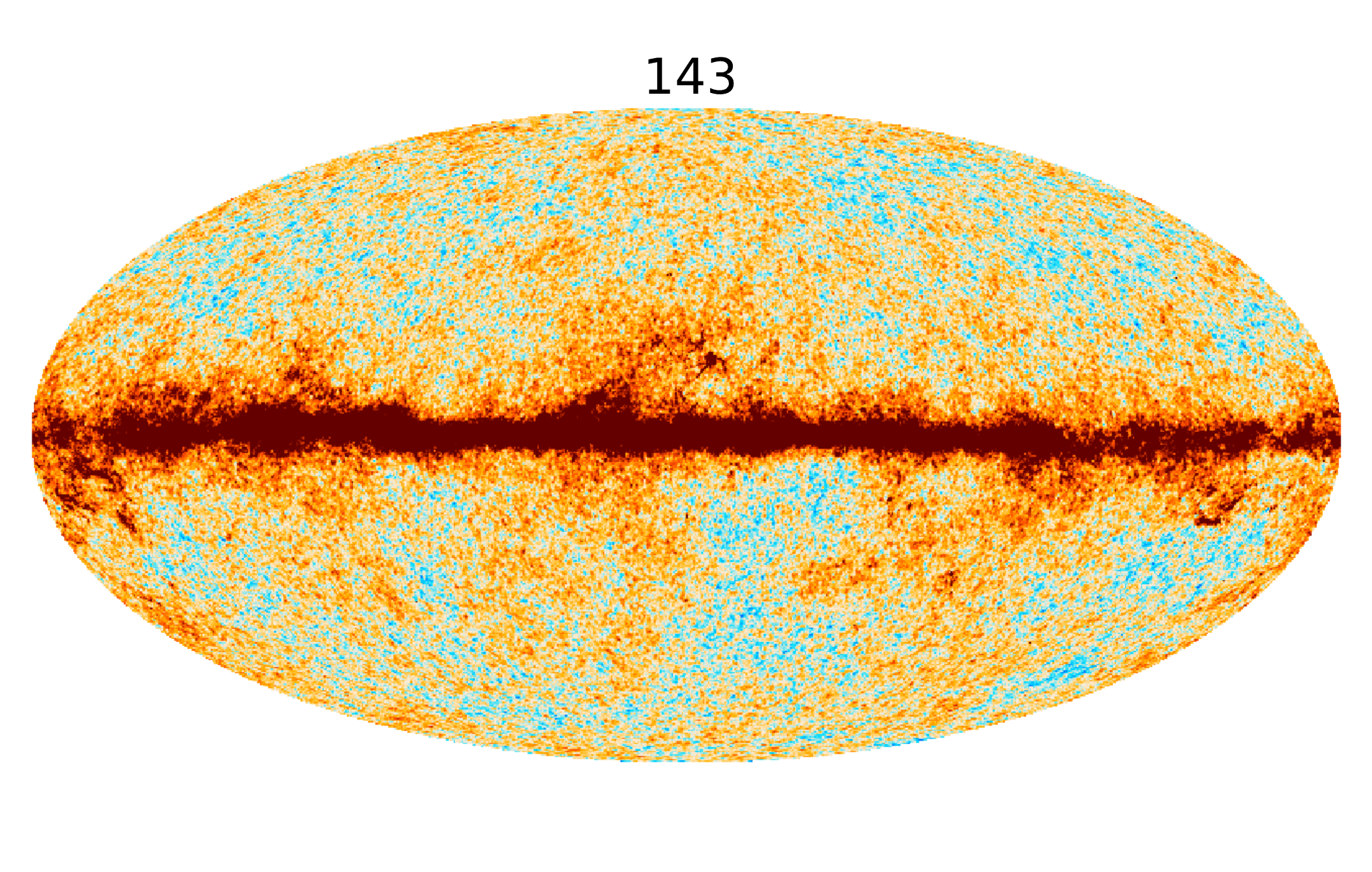} & \includegraphics[width=3in]{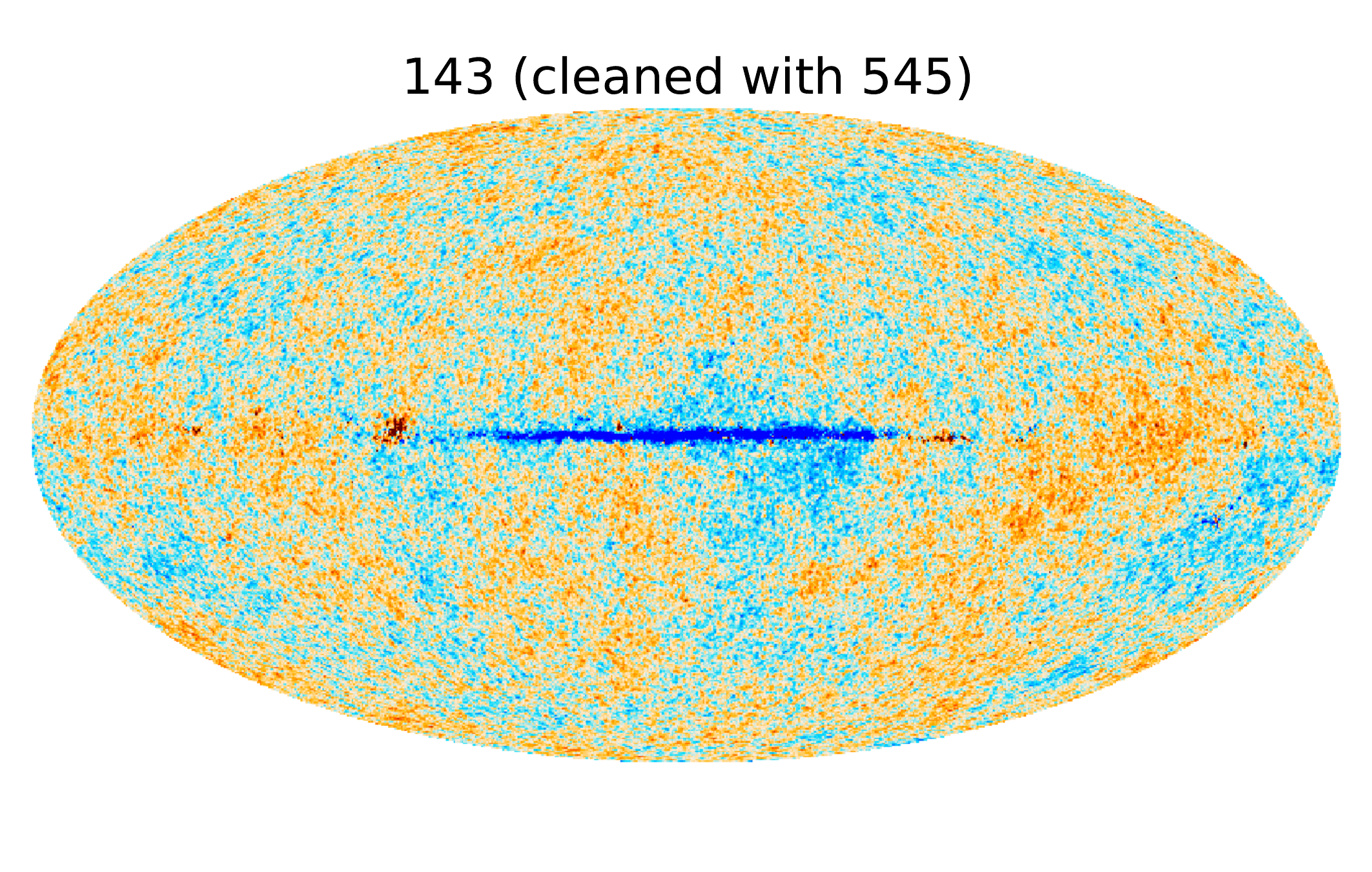} \\
\includegraphics[width=3in]{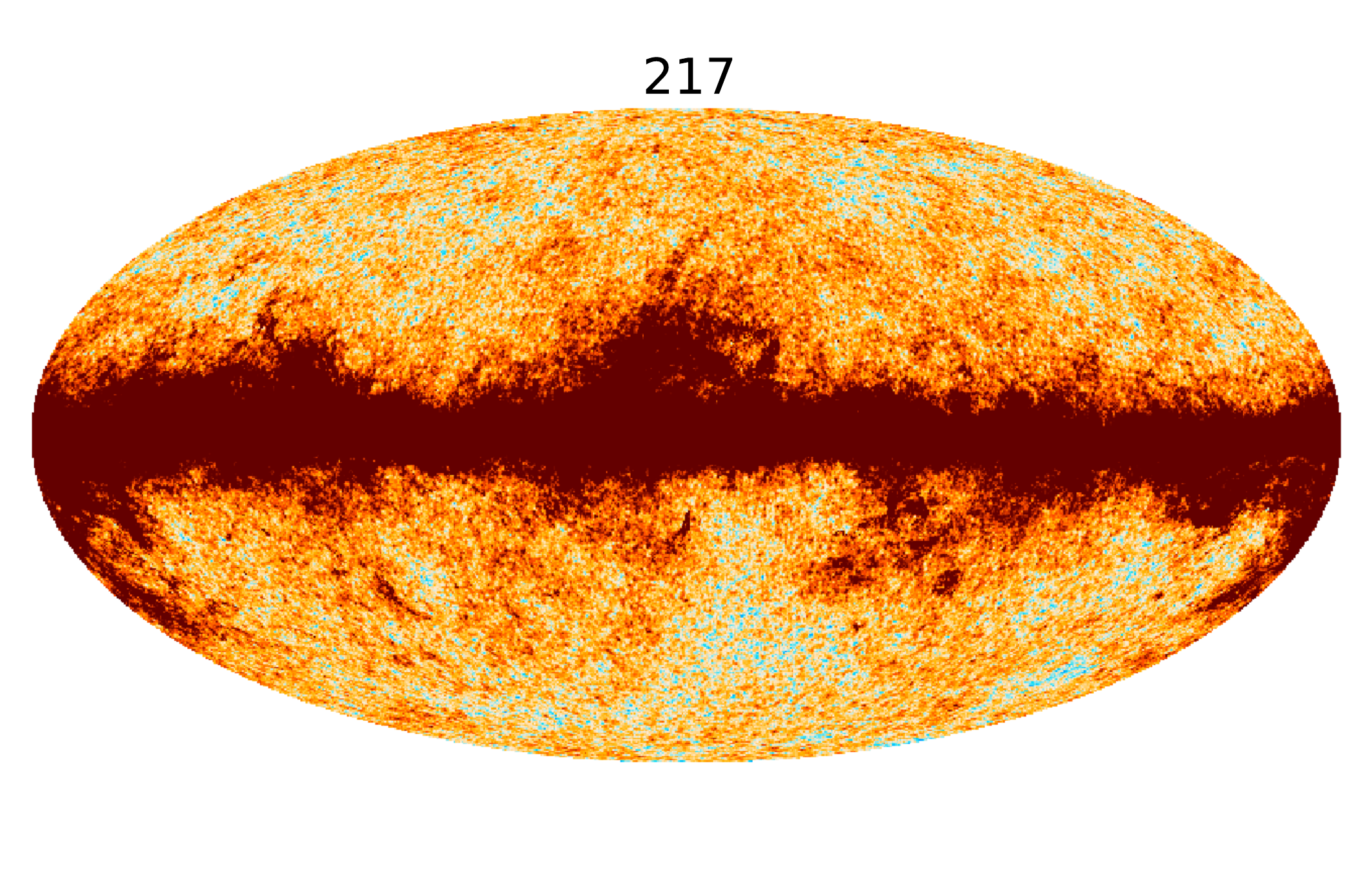} & \includegraphics[width=3in]{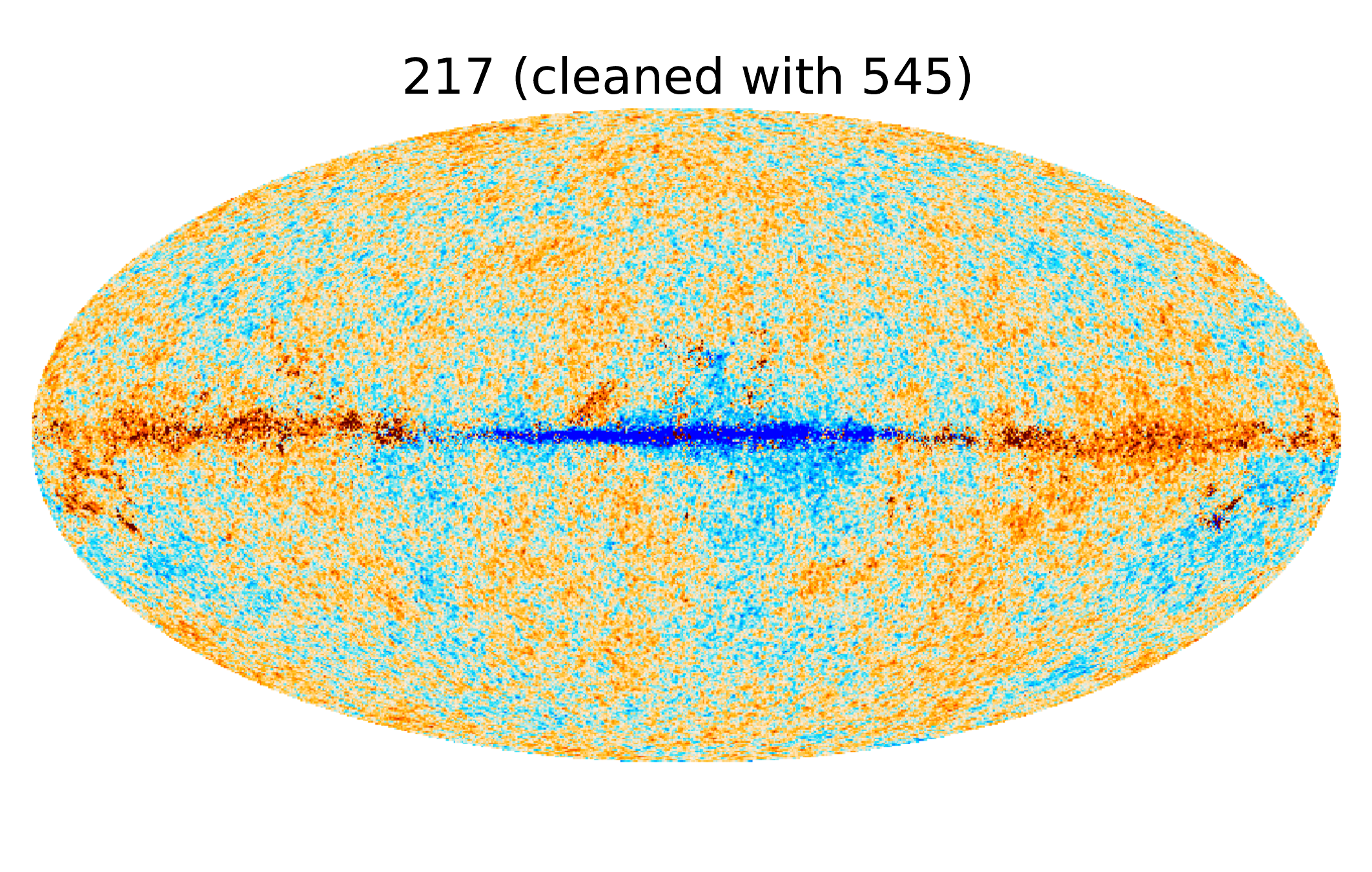}
\end{tabular}
\caption{\Planck\ temperature maps for the three CMB channels computed with \mspec. \emph{Left:} raw frequency maps. \emph{Right:} the same maps, after map cleaning with the 545\,GHz map. The cleaning coefficients are given in Table~\ref{tab:cleancoeff}. Comparison- of power spectra from these maps with different levels of Galactic masking are shown in Fig.~\ref{fig:217cleaning}. 
}
\label{fig:mappics}
\end{figure*}

\begin{figure}[htbp] 
\centering
\includegraphics[width=3in]{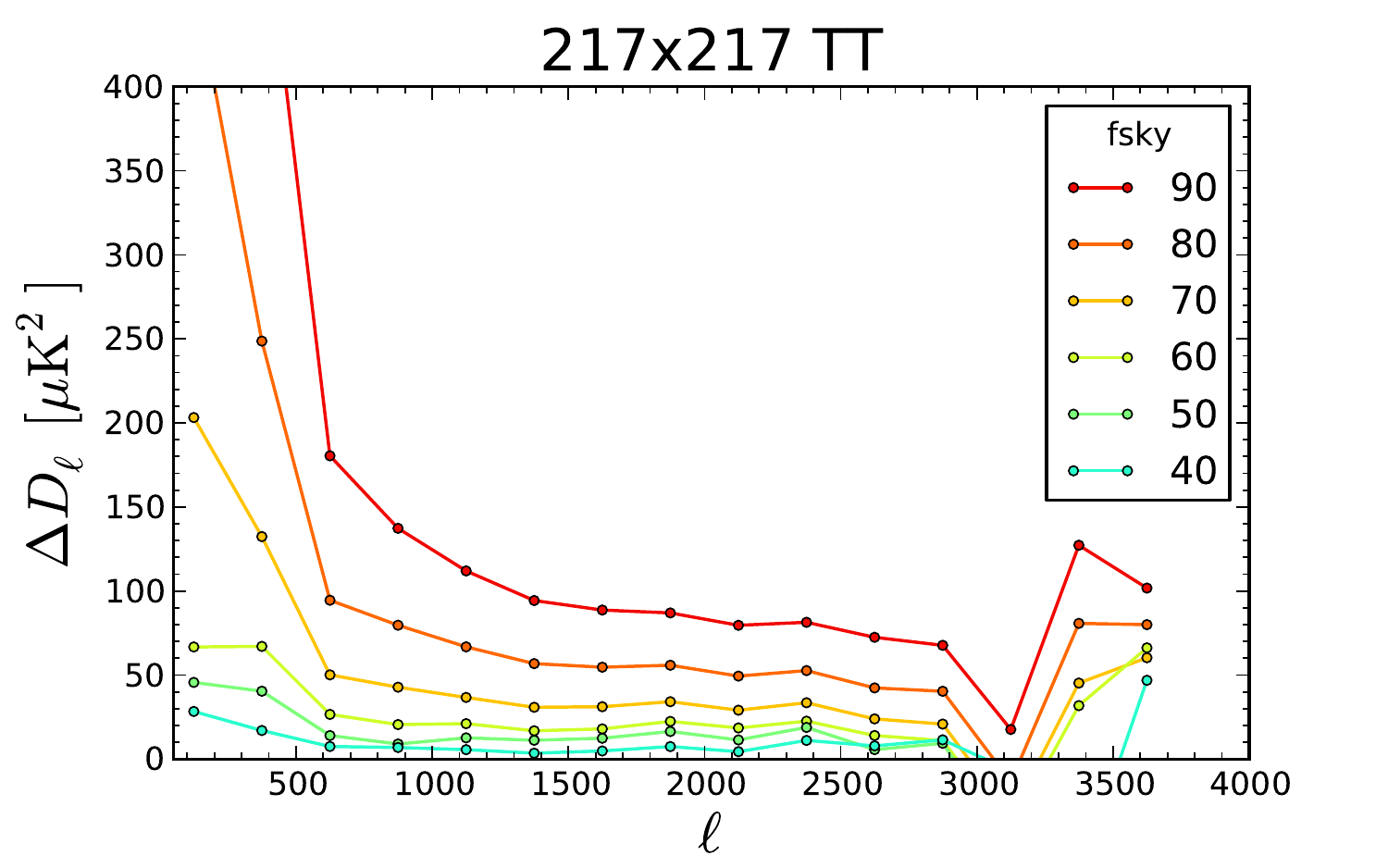}
\includegraphics[width=3in]{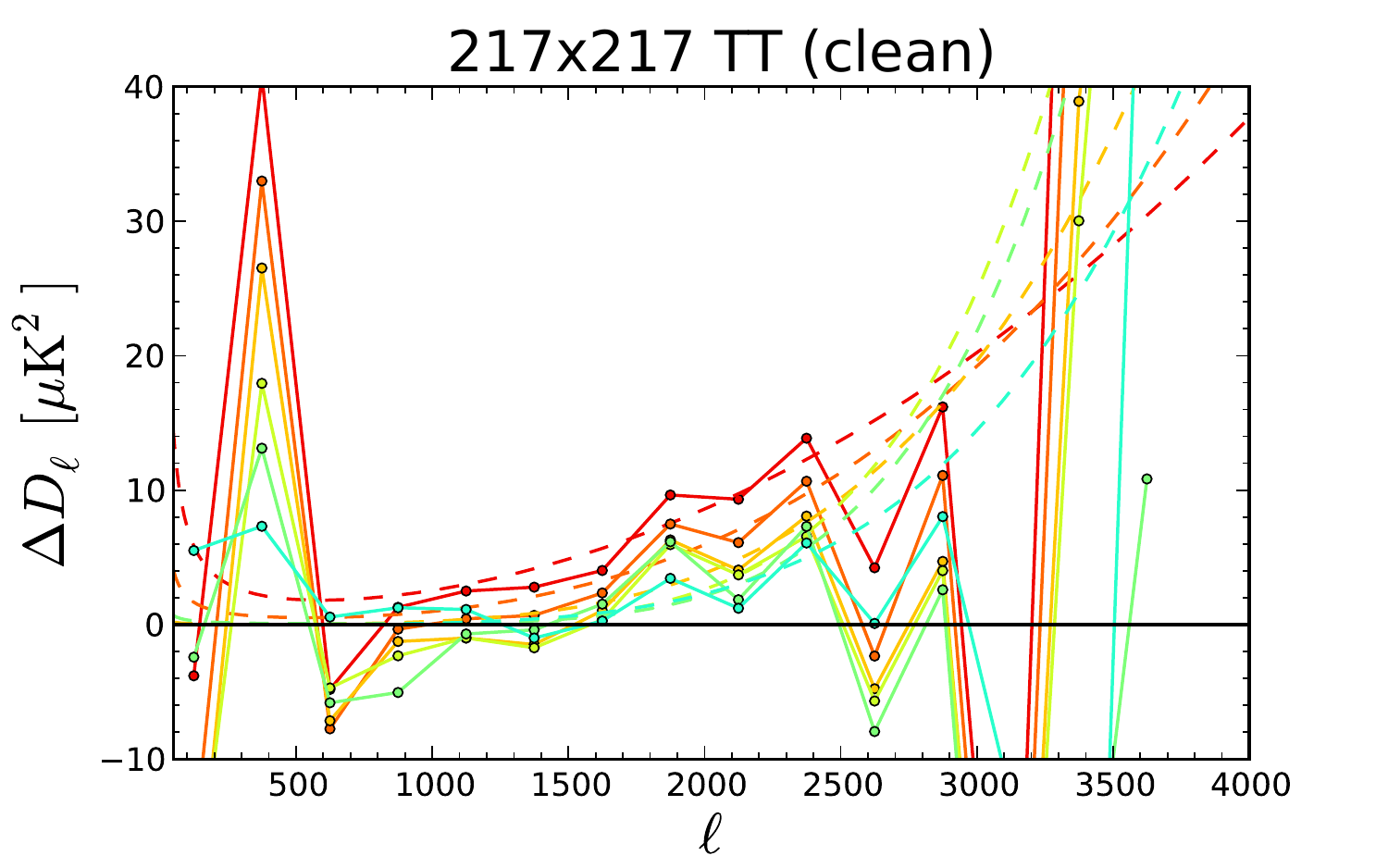}
\includegraphics[width=3in]{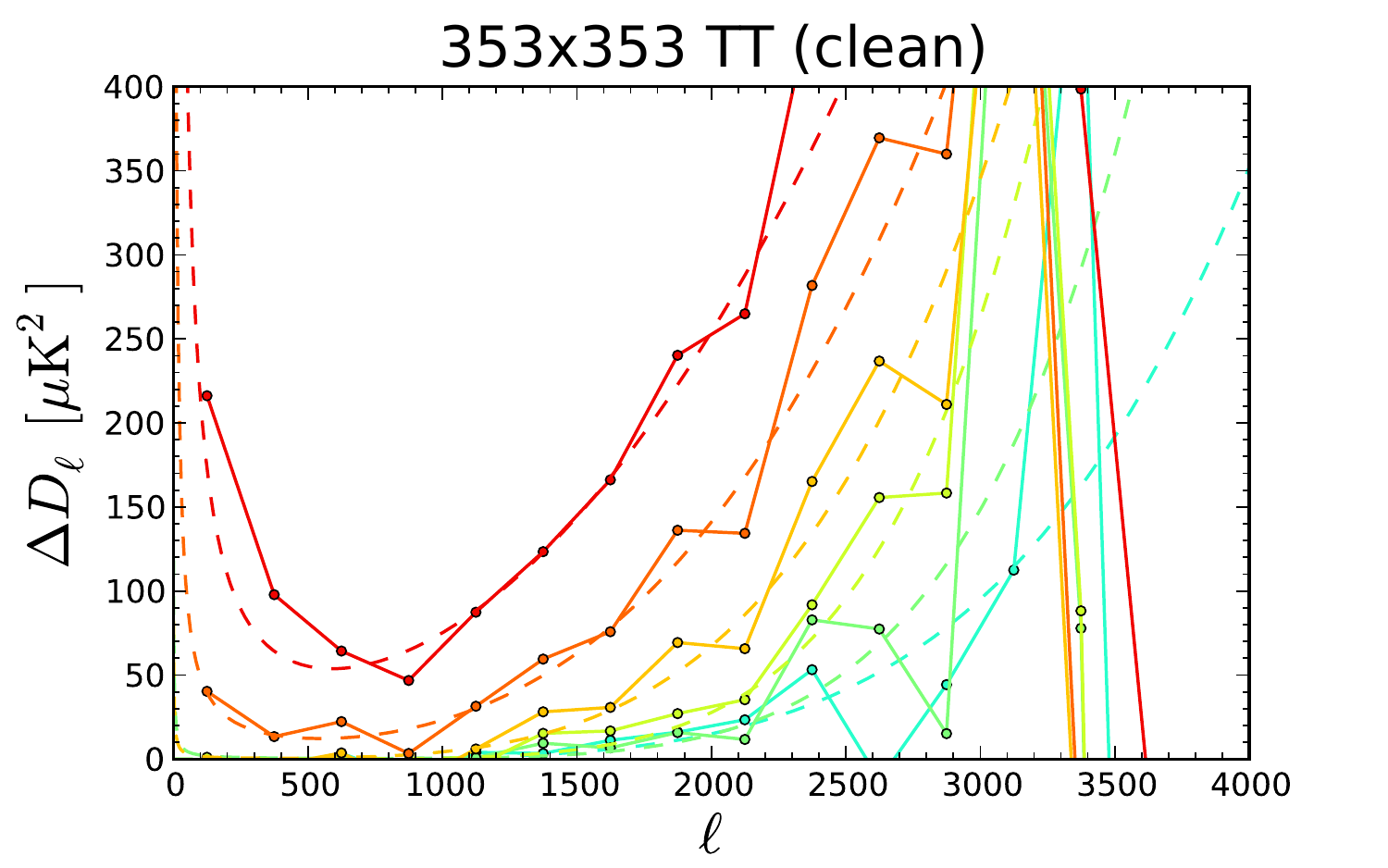}
\caption{\emph{Top}: Power spectra of the 217\,GHz raw temperature maps, with various Galactic masks, differenced against the $f_{\rm sky}=30\,\%$ case. \emph{Middle}: The same, but after map cleaning. The $y$-scale is now 10 times smaller. \emph{Bottom}: The same as the middle panel, but for 353\,GHz. Additionally, we first subtract the 143\,GHz map to remove scatter due to CMB sample variance. A model consisting of the sum of two power laws is then fit through these points. This sets the shape of the residual Galactic template. In the middle panel this template is scaled by an amplitude to fit the residuals seen there, showing that the Galactic residuals after map cleaning do not change shape significantly between 353 and 217\,GHz.}
\label{fig:217cleaning}
\end{figure}

Although we describe it as ``map'' subtraction, in practice it is done at the power-spectrum level by forming the equivalent linear combination of power spectra, 
\begin{align}
\label{eq:undusting}
C_{\ell}^{\nu_1,\rm clean} = (C_{\ell}^{\nu_1} - 2xC_{\ell}^{\nu_1 \times \nu_2} + x^2 C_{\ell}^{\nu_2})/(1-x)^2    \,,
\end{align}
where the $C_\ell$ are mask- and beam-deconvolved power spectra, $x$ is the cleaning coefficient, $\nu_1 \in \{100,143,217\}$ refers to one of the CMB channels, $\nu_2\in\{353,545\}$ is the cleaning frequency, and both have the same mask applied. We obtain the cleaning coefficient by maximizing the reduction in pixel variance due to cleaning, filtered to a given $\ell$-range and on a particular mask. We choose the filter range to be $\ell=(50,500)$ since that is where we expect dust to be dominant over extragalactic CIB. We use the $f_{\rm sky}=90\,\%$ mask to maximize the dust signal, fitting it well while avoiding bias from strong signals from the Galactic plane. In other words, we find $x$ for each frequency $\nu$ by maximizing
\begin{align}
\sum_{\ell=50}^{500} (2\ell+1) ( C_{\ell}^{\nu} - C_{\ell}^{\nu,\rm clean})\,.
\end{align}
The resulting cleaning coefficients are given in Table~\ref{tab:cleancoeff}. Temperature maps cleaned with these coefficients are shown in Fig.~\ref{fig:mappics}, visually demonstrating the remarkable effectiveness of this procedure. This can be seen more quantitatively at the power-spectrum level in Fig.~\ref{fig:217cleaning}, which shows the power spectrum computed on different masks, differenced against the $f_{\rm sky}=30\,\%$ case to cancel out any isotropic components. The middle panel shows that the map-cleaning step leads to about a factor of ten reduction in Galactic power. 

This suppression of  power not only reduces sensitivity to errors in the Galactic power-spectrum modelling, but it also leads directly to a smaller covariance by reducing chance correlations between signal (CMB and extragalactic foregrounds) and Galactic dust. In general, if a map contains a sum of signal and dust, $T(\hat{\vec{n}})=S(\hat{\vec{n}})+D(\hat{\vec{n}})$, its auto-spectrum contains signal, dust, and signal--dust terms, $C_\ell = C^{SS}_\ell + 2 C^{SD}_\ell + C^{DD}_\ell$. The $DD$ term can be modelled and subtracted as in the baseline \plik likelihood. The $SD$ term has zero mean but non-zero variance, so it must be included in the covariance matrix. For high sky fractions this can become important, \eg for $f_{\rm sky}=80$\% at 217\,GHz it is equal to the noise contribution at $\ell\approx1500$. Conversely, the signal--dust term is not present if the dust is removed from the map initially as is done by the map-cleaning procedure. 

\begin{table}[ht!] 
\centering
\begingroup 
\newdimen\tblskip \tblskip=5pt
\caption{\mspec map cleaning coefficients.}
\label{tab:cleancoeff}
\nointerlineskip
\vskip -6mm
\footnotesize
\setbox\tablebox=\vbox{
\newdimen\digitwidth
\setbox0=\hbox{\rm 0}
\digitwidth=\wd0
\catcode`*=\active
\def*{\kern\digitwidth}
\newdimen\signwidth
\setbox0=\hbox{+}
\signwidth=\wd0
\catcode`!=\active
\def!{\kern\signwidth}
\newdimen\decimalwidth
\setbox0=\hbox{.}
\decimalwidth=\wd0
\catcode`@=\active
\def@{\kern\decimalwidth}
\halign{ 
\hbox to 0.9in{#\leaderfil}\tabskip=2em& 
    \hfil#\hfil&
    \hfil#\hfil\tabskip=0pt\cr
\noalign{\doubleline}
\omit\hfil Raw Map\hfil&$x$&Template\cr
\noalign{\vskip 3pt\hrule\vskip 5pt}
$T$\,100&   0.0013& 545\cr
$T$\,143&   0.0024& 545\cr
$T$\,217&   0.0080& 545\cr
$Q,U$\,100& 0.019*& 353\cr
$Q,U$\,143& 0.040*& 353\cr
$Q,U$\,217& 0.128*& 353\cr
\noalign{\vskip 5pt\hrule\vskip 3pt}
}}
\endPlancktable 
\tablenote {{\rm a}} The coefficients for the \mspec map cleaning procedure, with $x$ defined as in Eq.~\ref{eq:undusting}\par
\endgroup
\end{table}

For $\TT$, there is a second step of \mspec cleaning that removes any remaining dust contamination left due to spatial variation of the dust spectral index (or equivalently the decorrelation between the CMB channels and the high-frequency cleaning channel). The level of this residual Galactic contamination can be seen in the single differences in the middle panel of Fig.~\ref{fig:217cleaning}, however there is too much scatter due to sample variance to derive a template at all multipoles. To remedy this, we consider 353\,GHz, where the dust intensity increases relative to the scatter and also subtract 143\,GHz to remove the CMB contribution to the scatter. This gives the lower panel of Fig.~\ref{fig:217cleaning}. The shape of the template is set by fitting a model to these residuals, which we take phenomenologically to be the sum of two power laws. During parameter estimation, this template is added to the foreground model with an amplitude parameter that is marginalized over. We place a tight prior on this amplitude parameter coming from fitting the single-differences directly. For $\TE$ and $\EE$, there is no evidence of residual Galactic contamination for the sky fractions used, and thus we do not perform the second step. 

\begin{figure*} 
\centering
\includegraphics[width=7in]{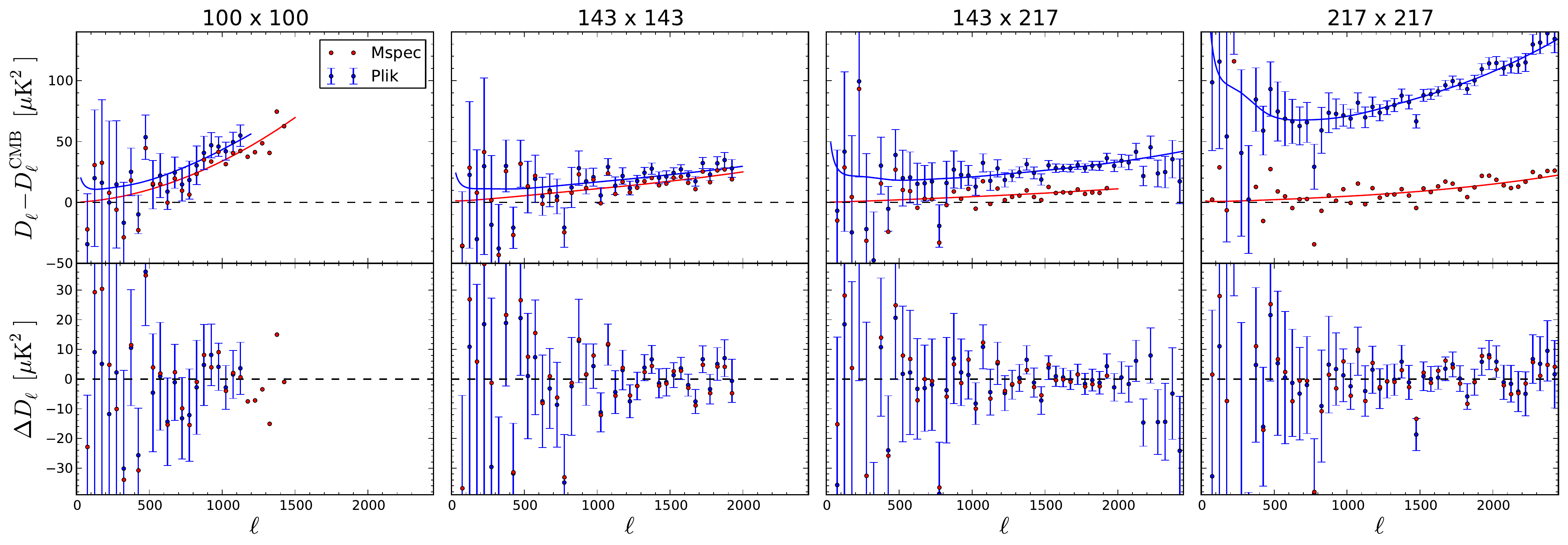}
\caption{Comparison of power spectra on identical masks obtained from the \plik Galactic cleaning procedure and from the \mspec map cleaning. In the top panels we subtract the best-fit CMB spectrum as determined by each code;  in the bottom panel we additionally subtract the Galactic and extragalactic foreground model.}
\label{fig:mspec_plik_cleaning}
\end{figure*}

The end result of the \mspec cleaning procedure for $\TT$ is summarized in Fig.~\ref{fig:mspec_plik_cleaning}, which shows a comparison of \mspec and \plik power spectra. The top row shows that the map-subtraction cleans out a significant amount of foreground power at 143$\times$217 and 217$\times$217, but makes a very small impact at lower frequencies, as expected. In the bottom row we see power spectra from both codes after total-foreground cleaning, and we find excellent point-by-point agreement. The agreement at the parameter level is also very good, as described in Sec.~\ref{sec:comparison}. Figure~\ref{fig:mspec_plik_cleaning} is similar to figure~2 of \citep{planck2014-a15}, which compared \camspec spectra with and without a map-cleaning step, but using a simpler model for the power spectrum of residuals left over after map cleaning. Here, by using the \mspec model instead, we confirm the consistency of our results with respect to Galactic cleaning using a more realistic foreground model.

\paragraph{Covariance approximations}

\mspec computes the covariance of the pseudo-$C_\ell$s in the same way as the baseline likelihood, but with two additional approximations. First, we ignore the fact that the noise is non-uniform across the sky. Using the FFP8 simulations \citep{planck2014-a14}, which include realistic spatial variations of the noise, we find that this is an excellent approximation for $\TT$ and $\TE$, where the noise is only important at high~$\ell$, but leads to about a 5\,\% underestimation of the error bars for low-$\ell$ $\EE$, which we correct for heuristically. Second, we always take the coupling kernel that appears in the pseudo-$C_\ell$ covariance to be the $\TT$ one. This leads to much less than $1\,\%$ changes. The benefit is a huge simplification of the covariance expressions, allowing for all the temperature and polarization entries to be written concisely as
\begin{align}
\langle \Delta \widetilde C^{ab,XY}_{\ell} \Delta \widetilde C^{cd,ZW}_{\ell'} \rangle =
\frac{1}{2} &\left\{ C^{ac,XZ}_{(\ell}C^{bd,YW}_{\ell')} \Xi_{\ell\ell'}^{TT}[W^{ac,bd}] \right. \nonumber
\\        + &\left.  C^{ad,XW}_{(\ell}C^{bc,YZ}_{\ell')} \Xi_{\ell\ell'}^{TT}[W^{ad,bc}] \right\}  \,,
\label{eq:pclcov}
\end{align}
where $a,b,c,d$ label detectors, $X,Y,Z,W$ each label one of $T,E,B$, and the $C_\ell$'s that appear on the right-hand side are the fiducial beam-convolved signal-plus-noise power-spectra.

\subsection{\hil}


\label{annex:hillipop}

\newcommand{\draft}{false}

\texttt{HiLLiPOP} is another high-$\ell$ likelihood procedure, based on a Gaussian approximation, to confront the \Planck\ \HFI data $\ell$-by-$\ell$ with cosmological models. In this approach, the data consist of six maps: two sets of half-mission ($I, Q, U$) maps at 100, 143, and 217\,GHz. Frequency-dependent apodized masks are applied to these maps in order to limit contamination from diffuse Galactic dust, Galactic CO lines, nearby galaxies, and extragalactic point sources. With regard to the latter, unlike the \plik\ masks, which are based on the point source catalogue with a flux-density cut (see Appendix~\ref{app:masks}), the masks used here rely on a more refined procedure that preserves Galactic compact structures and ensures the completeness level at each frequency, but with a higher detection threshold (\ie leaving more extragalactic sources unmasked). \hil retains 72, 62, and 48\,\% of the sky at 100, 143, and 217\,GHz, respectively, and uses the same set of masks for both temperature and polarization. Mask-deconvolved and beam-corrected cross-half-mission power spectra are computed using \texttt{Xpol}, an extension of the \texttt{Xspect} \citep{tristram2005} code to polarization. From the six maps, we can derive 15 sets of power spectra in $\TT$, $\EE$, $\TE$, and $\ET$:  one each for 100$\times$100, 143$\times$143, and 217$\times$217, and four each for 100$\times$143, 100$\times$217, and 143$\times$217.

The covariance matrix, which encompasses the correlations ($\ell$-by-$\ell$) between all these 60 power spectra, is estimated semi-analytically with \texttt{Xpol}. Unlike the \plik\ likelihood, which assumes a model for signal (of cosmological and astrophysical origin) and noise, here the calculation relies on data estimates only by using as input a smooth version of the estimated power spectra. Contributions from noise, sky emission, and the associated cosmic variance are automatically taken into account. Several approximations \citep[as in][]{Efstathiou2006} are needed in this calculation and Monte Carlo simulations have been performed to test their accuracy. A precision better than a few percent is achieved.

In addition to the CMB component, the \texttt{HiLLiPOP} likelihood accounts for foreground residuals and differences in calibration between maps. A differential calibration coefficient, $d_i$, is defined per map, $\tilde{m_i} = (1+d_i)\,m_i$, with the 143 half-mission-1 map calibration taken as reference.\footnote{Therefore, \eg $\tilde{C}_\ell^\mathrm{143h2\times217h1} = (1 + d_\mathrm{143h2} + d_\mathrm{217h1})\,C_\ell^\mathrm{143h2\times217h1}$.}
We use different models for foregrounds in temperature and polarization. The temperature model includes contributions from cosmic infrared background (CIB), Galactic dust, thermal and kinetic Sunyaev-Zeldovich (tSZ and kSZ) effects, Poisson point sources (PS), and the cross-correlation between infrared galaxies and the tSZ effect (tSZ$\times$CIB). The polarization model includes only Galactic dust.  The calibration coefficients are assigned
Gaussian priors reflecting the uncertainties in the half-mission map calibration (from \citealt{planck2014-a09}): $d_\mathrm{100h1} = d_\mathrm{100h2} = d_\mathrm{143h1} = d_\mathrm{143h2} = 0.000 \pm 0.002$ and $d_\mathrm{217h1} = d_\mathrm{217h2} = 0.004 \pm 0.002$.


\begin{figure}[h!tbp] 
\begin{center}
\includegraphics[clip=true,trim=0cm 1.5cm 1cm 3cm,draft=\draft,width=\columnwidth]{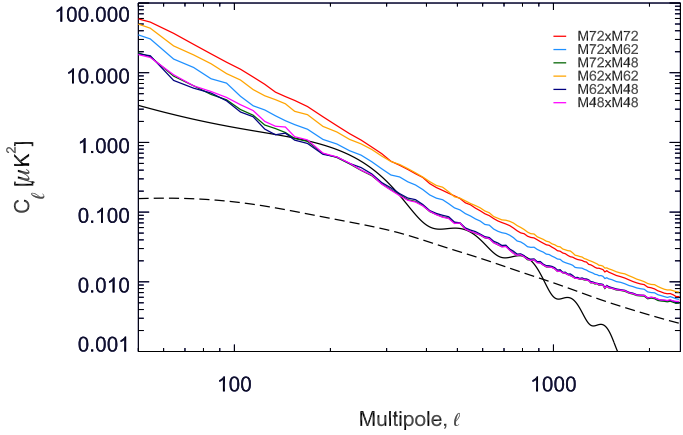}
\caption{Cross-power spectra of half-mission maps at 353\,GHz in each combination of masks after subtraction of CMB ({solid line}) and CIB ({dashed line}) models.}
\label{fig:dust_TT}
\end{center}
\end{figure}

\texttt{HiLLiPOP} uses physically motivated templates of foreground emission power spectra, in both photon frequency and spherical harmonic wave number, based on \Planck\ measurements, as described below.
\begin{itemize}

\item Galactic dust: The $\TT$, $\EE$, and $\TE$ Galactic dust power-spectrum templates are derived following the methodology presented in \citet{planck2014-XXX}. We first estimate the $\TT$, $\EE$, and $\TE$ half-mission cross-power spectra at 353\,GHz in the different combinations of masks (Fig.~\ref{fig:dust_TT}) and then subtract the best-fit CMB power spectrum \citep[see][]{planck2013-p11}. In temperature, the CIB power spectrum \citep[see][]{planck2013-pip56} is also subtracted. A good fit is obtained on the resulting power spectra using the power-law model $A\ell^\alpha + B$, where $B$ describes the Poisson contribution from unresolved point sources in temperature (the contribution of polarized sources is negligible in $\EE$ and $\TE$).
It is worth stressing that the shape of the Galactic dust component is found to be compatible with a power law. This is due to the choice of point-source masks that have minimal effect on Galactic structures or bright cirrus. The masks used by \plik, which are based on lower flux-density cuts to further reduce the compact-source contribution, introduce a knee in the power spectra around $\ell \approx 300$. 

\item tSZ: The power-spectrum template is based on \citet{2008ApJ...688..709T} for the mass function and \citet{2010A&A...517A..92A} for the ``universal pressure profile.'' It contains both the one-halo and the two-halo terms \citep{2011MNRAS.418.2207T}. A full description can be found in \citet{planck2013-p05b}.

\item kSZ: The power spectrum is taken from \citet{Bat13} for the patchy reionization part and \citet{Shaw12} for the Ostriker-Vishniac effect, both normalized to the \Planck\ cosmological parameters. 

\item CIB and tSZ$\times$CIB: The CIB power-spectrum templates are based on a halo model linking directly the galaxies' luminosities to their host dark-matter halo masses. It has been successfully applied in \citet{planck2013-pip56}. For the tSZ$\times$CIB power-spectrum templates, the tSZ power-spectrum template is based on \citet{EM012}.

\item PS: we used a Poisson-like flat power spectrum for the unresolved point source contribution at each cross-frequency.

\end{itemize}
A single free parameter ($A$, scaling all frequencies equally) for each foreground template is used to adjust the amplitude of the power spectra. Each amplitude is assigned a uniform prior.

The model used to describe the \Planck\  $\TT$ power spectra thus reads
\begin{eqnarray} 
	\begin{split}
		\hat{C}_\ell^{T_iT_j} 				& = (1+c_i+c_j) \left(C_\ell^{\mathrm{CMB},\TT} + A_\mathrm{dust}^{TT}C_\ell^{\mathrm{dust},T_iT_j} + A_\mathrm{PS}^{T_iT_j} + \right. \\
		\phantom{\hat{C}_\ell^{T_iT_j}}		& \phantom{ {} = (1+c_i+c_j) \left( \right.} \left. A_\mathrm{CIB}C_\ell^{\mathrm{CIB},T_iT_j} + A_\mathrm{tSZ}C_\ell^{\mathrm{tSZ},T_iT_j} + \right. \\
		\phantom{\hat{C}_\ell^{T_iT_j}}		& \phantom{ {} = (1+c_i+c_j) \left( \right.} \left. A_\mathrm{kSZ}C_\ell^{\mathrm{kSZ}} + A_\mathrm{tSZ\times CIB}C_\ell^{\mathrm{tSZ\times CIB},T_iT_j} \right), 
	\end{split}
	\nonumber
	\label{eq:modelTT}
\end{eqnarray}
while the $\EE$ and $\TE$ power spectrum models simply read
\begin{eqnarray}
	\hat{C}_\ell^{E_iE_j} &=& \left(1+c_i+c_j \right) \left(C_\ell^{\mathrm{CMB},EE}+A_\mathrm{dust}^{EE}C_\ell^{\mathrm{dust},E_iE_j}\right) \, ,
	\nonumber \\
	\hat{C}_\ell^{T_iE_j} &=& \left(1+c_i+c_j \right) \left(C_\ell^{\mathrm{CMB},TE}+A_\mathrm{dust}^{TE}C_\ell^{\mathrm{dust},T_iE_j}\right) \, .
	\nonumber
\end{eqnarray}

Among all the map cross-power spectra, we focus on just the six frequency cross-spectra ($100\times100$, $100\times143$, $100\times217$, $143\times143$, $143\times217$, and $217\times217$) in $\TT$, $\EE$, and $\TE$ and compress the covariance matrix accordingly. 
The \hil likelihood then reads
\begin{equation}
-2 \ln \mathcal{L} = \sum_{\substack{X,Y \\ X^\prime,Y^\prime}}
			       \sum_{\substack{i\leqslant j \\ i^\prime\leqslant j^\prime}}
			       \sum_{\substack{\ell = \ell_{\min}^{X_iY_j} \\ \ell^\prime = \ell_{\min}^{X_{i^\prime}^\prime Y_{j^\prime}^\prime}}}^{\substack{\ell_{\max}^{X_iY_j} \\ \ell_{\max}^{X_{i^\prime}^\prime Y_{j^\prime}^\prime}}}
			       \mathcal{R}_\ell^{X_iY_j}
			       \left[\Sigma_{\ell\ell^\prime}^{X_iY_j,X_{i^\prime}^\prime Y_{j^\prime}^\prime} \right]^{-1}
			       \mathcal{R}_{\ell^\prime}^{X_{i^\prime}^\prime Y_{j^\prime}^\prime}, \label{eq:hillipop}
\end{equation}
where $\mathcal{R} = C_\ell-\hat{C}_\ell$ denotes the residual of the estimated power spectrum ($C_\ell$) with respect to the model ($\hat{C}_\ell$),  and $\Sigma$ is the full covariance matrix, which is symmetric and positive-definite. The frequency band (100, 143, or 217\,GHz) is given by the $i, j$ indices and the CMB modes ($T$, $E$) by $X,Y$. The multipole ranges [$\ell_\mathrm{min}$,$\ell_\mathrm{max}$] are chosen with the goal of limiting contamination  in each power spectrum from diffuse Galactic dust emission at low $\ell$ and noise at high $\ell$.

At the end, we have a total of 5 instrumental, 13 astrophysical, and 6 or more cosmological ($\Lambda$CDM and possible extensions) parameters, \ie a total of 24 (or more) free parameters in the full ($TT$, $\EE$, and $TE$) \texttt{HiLLiPOP} likelihood function. The theoretical CMB power spectra are generated with the \texttt{CLASS} Boltzmann solver \citep{2011arXiv1104.2932L,Blas_2011} or the \texttt{PICO} algorithm \citep{Fendt_2007}. The \texttt{HiLLiPOP} likelihood function is explored using \texttt{Minuit} \citep{James_1975}. Table~\ref{tab:hillipop_chi2} shows the best-fit $\chi^2$ values for $\TT$, $\EE$, $\TE$, and the full data set. The number of degrees of freedom ($n_\ell$) is simply the total number of multipoles considered. 

\begin{table}[htbp]
\begingroup 
\newdimen\tblskip \tblskip=5pt
\caption{\texttt{HiLLiPOP} goodness of fit.}
\label{tab:hillipop_chi2}
\vskip -6mm
\footnotesize
\setbox\tablebox=\vbox{
\newdimen\digitwidth
\setbox0=\hbox{\rm 0}
\digitwidth=\wd0
\catcode`*=\active
\def*{\kern\digitwidth}
\newdimen\signwidth
\setbox0=\hbox{+}
\signwidth=\wd0
\catcode`!=\active
\def!{\kern\signwidth}
\newdimen\decimalwidth
\setbox0=\hbox{.}
\decimalwidth=\wd0
\catcode`@=\active
\def@{\kern\decimalwidth}
\halign{ 
\hbox to 0.7in{#\leaderfil}\tabskip=2em& 
    \hfil#\hfil&
    \hfil#\hfil\tabskip 1em&
    \hfil#\hfil\tabskip=0pt\cr  
\noalign{\doubleline}
\omit\hfil CMB mode\hfil&$\chi^2$&$n_\ell$&$\Delta\chi^2/\sqrt{2n_\ell}$\cr
\noalign{\vskip 3pt\hrule\vskip 5pt}%
		$TT$ & 9949.7 & 9556 & 2.85\cr
		$EE$ & 7309.5 & 7256 & 0.44\cr
		$TE$ & 9322.5 & 8806 & 3.89\cr
\noalign{\vskip 5pt\hrule\vskip 3pt}
}}
\endPlancktable
\tablenote {{\rm a}} The $\Delta\chi^2=\chi^2-n_\ell$ is the
difference from the mean ($n_\ell$), assuming that the model follows a
$\chi^2$ distribution. The quantity $\Delta\chi^2/\sqrt{2n_\ell}$ is
$\Delta\chi^2$ in units of the standard deviation. These numbers
correspond to the global, all-frequency fits, and are therefore not
directly comparable to the \plik numbers for the co-added CMB
spectra.\par
\endgroup
\end{table}

\begin{figure}[ht]
	\begin{center}
	\includegraphics[clip=true,trim=1cm 0cm 0cm 1cm,draft=\draft,width=\columnwidth]{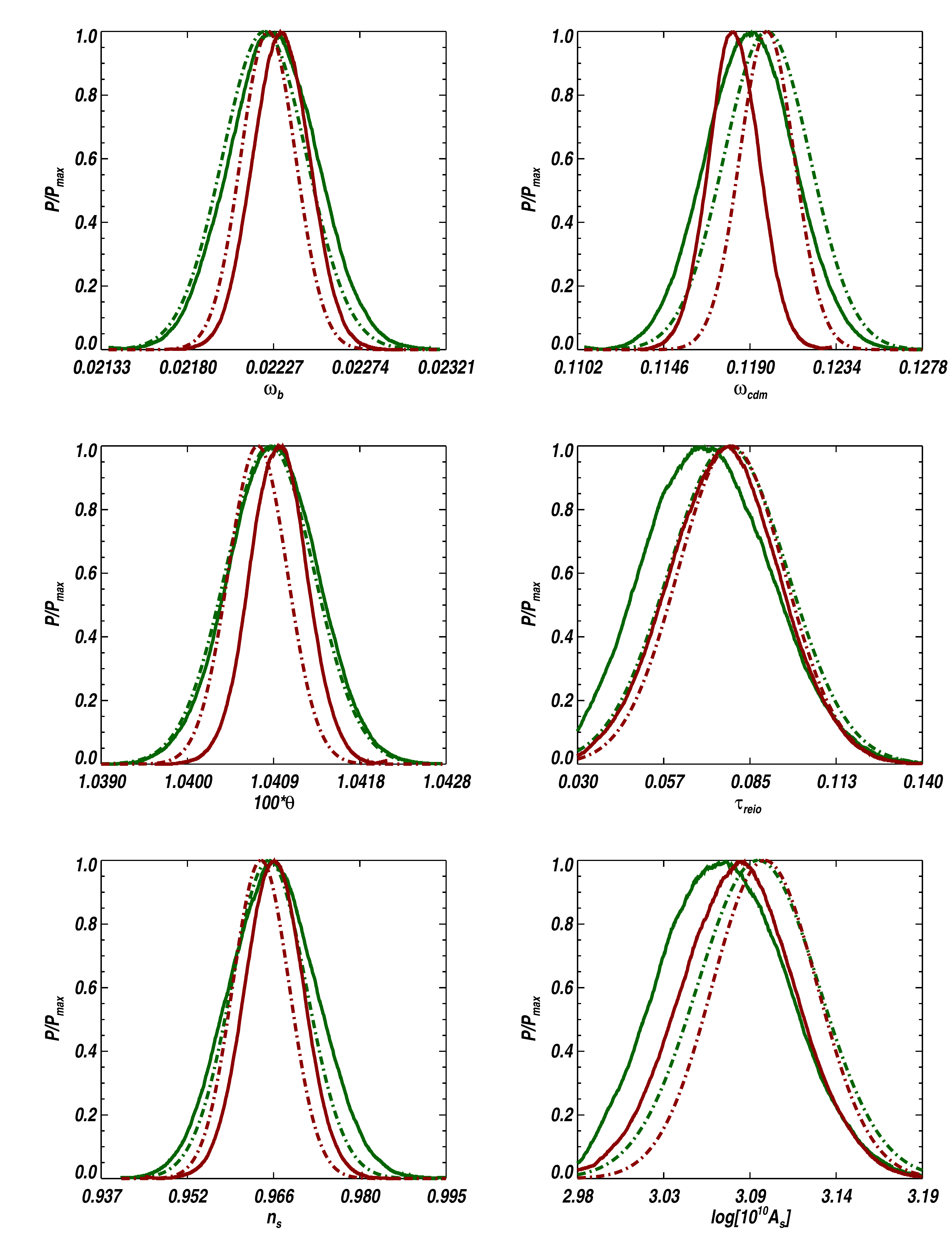}
	\caption{Marginalized constraints for the base-$\Lambda$CDM model obtained with the \texttt{HilliPOP} likelihood using $\TT$ only (green solid line) and the full temperature and polarization data (red solid line). For comparison \plik\ $\TT$ (dashed green) and \plik\ full data (dashed red) results are also shown. The high$-\ell$ information is always complemented with \lowTEB.}
	\label{fig:hillipop-plik}
	\end{center}
\end{figure}

The posterior distributions of the six base $\Lambda$CDM parameters obtained with MCMC sampling are shown in Fig.~\ref{fig:hillipop-plik} in comparison with the baseline \plik results. 
Here the high$-\ell$ information is complemented with lowTEB. When considering the $\TT$ data only, almost all parameters are compatible with the baseline within approximately $0.1\,\sigma$, with the exception of $\Omega_ch^2$ where the shift is slightly higher (about $0.4\,\sigma$). The approximately $0.5\,\sigma$ difference in $\tau$ and $\As$ can be  understood as a mild preference of the \texttt{HiLLiPOP} likelihood for a lower $\Alens$ ($1.20 \pm 0.09$ compared to $1.23 \pm 0.10$ for \plik). The shifted value for $\Alens$ comes in both cases from a tension between high~$\ell$ and lowTEB, the  value from \texttt{HilliPOP} alone for this parameter being compatible with unity at the $1\,\sigma$ level. Error bars from the baseline \plik and \texttt{HiLLiPOP} are nearly identical, with only a slightly bigger error bar for $n_s$ in the latter.

When considering the full data set, the shifts with respect to the baseline \plik\ are more pronounced, but still remain within about $0.5\,\sigma$. We observe the same trend for a lower $\Omega_{\rm c}h^2$,  but the difference in $\tau$ and $A_\textrm{s}$ is instead alleviated by the compatibility of $\Alens$ for the full likelihoods. Again the error bars derived with both likelihood methods are nearly identical.

We checked the robustness of our results with respect to the foreground models. Recall that we adjust each foreground component to data using a single rescaling amplitude. In temperature, the dust amplitude is recovered almost perfectly ($A^{TT}_\mathrm{dust}\approx1$) while the $A_{\mathrm{CIB}}$ estimation lies $1.7\,\sigma$ away. 
Using the full likelihood, the dust $\TT$ remains perfectly compatible with $A^{\TT}_\mathrm{dust}=1$ and the shift in the CIB amplitude is reduced to $0.5\,\sigma$. The polarized dust amplitudes ($A^{\TE}_\mathrm{dust}$ and $A^{\EE}_\mathrm{dust}$) are compatible with unity within about $2\,\sigma$  (this is true also for single $\EE$ and $\TE$ spectra). As described in \citet{planck2014-a15}, using \Planck-only data, we are not very sensitive to SZ components. In any event, the marginalized posteriors on $A_\mathrm{tSZ}$, $A_\mathrm{kSZ}$, and $A_{\mathrm{tSZ}\times\mathrm{CIB}}$ are compatible with the expectations.

Compatibility with $A=1$ for the foreground scaling parameters is a good indication of the consistency of the internal \Planck\ templates. We have also tested the stability of the cosmological results with respect to the choice of priors on these scaling parameters, by considering a set of Gaussian priors, $A=1.0\pm0.2$. The $\chi^2$ values remain unchanged, with almost no shift in cosmological parameters and a slight reduction of the error bar on $\ns$.


\section{The \Planck\ CMB likelihood supplement \label{app:hal} }

\begin{table*}[ht!]
\begingroup
\caption{Goodness-of-fit tests for the 2015 \planck\ temperature and polarization spectra.$^{\rm a}$}
\label{tab:goodness}
\vskip -6mm
\footnotesize
\setbox\tablebox=\vbox{
\newdimen\digitwidth
\setbox0=\hbox{\rm 0}
\digitwidth=\wd0
\catcode`*=\active
\def*{\kern\digitwidth}
\newdimen\signwidth
\setbox0=\hbox{+}
\signwidth=\wd0
\catcode`!=\active
\def!{\kern\signwidth}
\newdimen\decimalwidth
\setbox0=\hbox{.}
\decimalwidth=\wd0
\catcode`@=\active
\def@{\kern\decimalwidth}
\halign{\hbox to 1.3in{#\leaderfil}\tabskip=2em&
    \hfil#\hfil\tabskip=1em&
    \hfil#\hfil\tabskip=1.7em&
    \hfil#\hfil\tabskip=1.4em&
    \hfil#\hfil\tabskip=1.4em&
    \hfil$#$\hfil\tabskip=0.4em&
    \hfil#\hfil\tabskip=0pt\cr
\noalign{\doubleline}
\omit\hfil Frequency [GHz]\hfil&Multipole range$^{\rm a}$&$\chi^2$\,$^{\rm b}$&$\chi^2/N_\ell$&$N_\ell$&{\Delta \chi^2\sqrt{2N_\ell}}^{\,\rm c}&{\rm PTE} [\%]$^{\rm d}$\cr
\noalign{\vskip 3pt\hrule\vskip 5pt}%
\omit{\boldmath{$TT$}}\hfil\cr
\noalign{\vskip 3pt}
$\quad 100\times100$&   *30--1197& 1234.37& 1.06& 1168& !1.37& *8.7\cr
$\quad 143\times143$&   *30--1996& 2034.45& 1.03& 1967& !1.08& 14.1\cr
$\quad 143\times217$&   *30--2508& 2566.74& 1.04& 2479& !1.25& 10.7\cr
$\quad 217\times217$&   *30--2508& 2549.66& 1.03& 2479& !1.00& 15.8\cr
$\quad$Combined $TT$&   *30--2508& 2546.67& 1.03& 2479& !0.96& 16.8\cr
\noalign{\vskip 3pt}
\omit{\boldmath{$TE$}}\hfil\cr
\noalign{\vskip 3pt}
$\quad 100\times100$&   *30--*999& 1088.78& 1.12& *970& !2.70& *0.5\cr
$\quad 100\times143$&   *30--*999& 1032.84& 1.06& *970& !1.43& *7.9\cr
$\quad 100\times217$&   505--*999& *526.56& 1.06& *495& !1.00& 15.8\cr
$\quad 143\times143$&   *30--1996& 2028.43& 1.03& 1967& !0.98& 16.4\cr
$\quad 143\times217$&   505--1996& 1606.25& 1.08& 1492& !2.09& *2.0\cr
$\quad 217\times217$&   505--1996& 1431.52& 0.96& 1492& -1.11& 86.7\cr
$\quad$Combined $TE$&   *30--1996& 2046.11& 1.04& 1967& !1.26& 10.5\cr
\noalign{\vskip 3pt}
\omit{\boldmath{$EE$}}\hfil\cr
\noalign{\vskip 3pt}
$\quad 100\times100$&   *30--*999& 1027.89& 1.06& *970& !1.31& *9.6\cr
$\quad 100\times143$&   *30--*999& 1048.22& 1.08& *970& !1.78& *4.1\cr
$\quad 100\times217$&   505--*999& *479.72& 0.97& *495& -0.49& 68.1\cr
$\quad 143\times143$&   *30--1996& 2000.90& 1.02& 1967& !0.54& 29.2\cr
$\quad 143\times217$&   505--1996& 1431.16& 0.96& 1492& -1.11& 86.8\cr
$\quad 217\times217$&   505--1996& 1409.58& 0.94& 1492& -1.51& 93.6\cr
$\quad$Combined $EE$&   *30--1996& 1986.95& 1.01& 1967& !0.32& 37.2\cr
\noalign{\vskip 5pt\hrule\vskip 3pt}
}}
\endPlancktablewide
\tablenote {{\rm a}} The $\ell$ range used in the high-$\ell$ likelihood.\par
\tablenote {{\rm b}} The $\chi^2$ are with respect to the best-fit from the \plikTT+\lowTEB\ data combination in a $\Lambda$CDM framework.\par 
\tablenote {{\rm c}} $\Delta \chi^2=\chi^2-N_\ell$ is the difference from the mean assuming that the best-fit base $\TT$ \LCDM\ model is correct, expressed in units of  the expected dispersion, $\sqrt{2N_\ell}$.\par
\tablenote {{\rm d}} Probability to exceed the tabulated value of $\chi^2$.\par
\endgroup
\end{table*}

In this appendix, we provide additional material to further characterize the CMB power spectra from \Planck. Table~\ref{tab:goodness} provides goodness-of-fit values for the full likelihood; comparison with the equivalent Table~\ref{tab:highl:lrange} for \plikTTtau\ shows  the effect of replacing a $\tau$ prior by the full low-$\ell$ likelihood, which encompasses the $\ell\approx20$ dip in $\TT$ power. In Sect.~\ref{sec:ALL-robust} we extend our tests to combine low-$\ell$ temperature and polarization information with the baseline \plik\ likelihood at high $\ell$.  In the following Sect.~\ref{ssub:peaks},  we fit the shapes of the spectra as a series of (Gaussian) peaks and troughs, and in Sect.~\ref{ssub:TE_corr} we provide an alternative display of the correlation between temperature and ($E$-mode) polarization. Section~\ref{sec:mapCheck} discusses the spectra and parameters one can derive by analysing the CMB maps cleaned by component-separation methods, and Sect.~\ref{ssub:profile} confirms that the profile likelihood approach \citep{planck2013-XVI} leads to parameter constraints consistent with those discussed in the main text. This offers a breakdown of the uncertainties on the cosmological parameters between those that arise from the finite sensitivity of the experiment and those coming from a lack of knowledge of the nuisance parameters related to the modelled foregrounds and instrumental effects.

\subsection{$TT$, $TE$, $EE$ robustness tests} \label{sec:ALL-robust}

This section presents  the tests conducted so far to assess the robustness and accuracy of the results when using the high-$\ell$ \plikTTTEEE\ likelihood in combination with the low-$\ell$ likelihood in temperature and polarization.

\begin{figure*} 
\centering
\includegraphics[width=\textwidth]{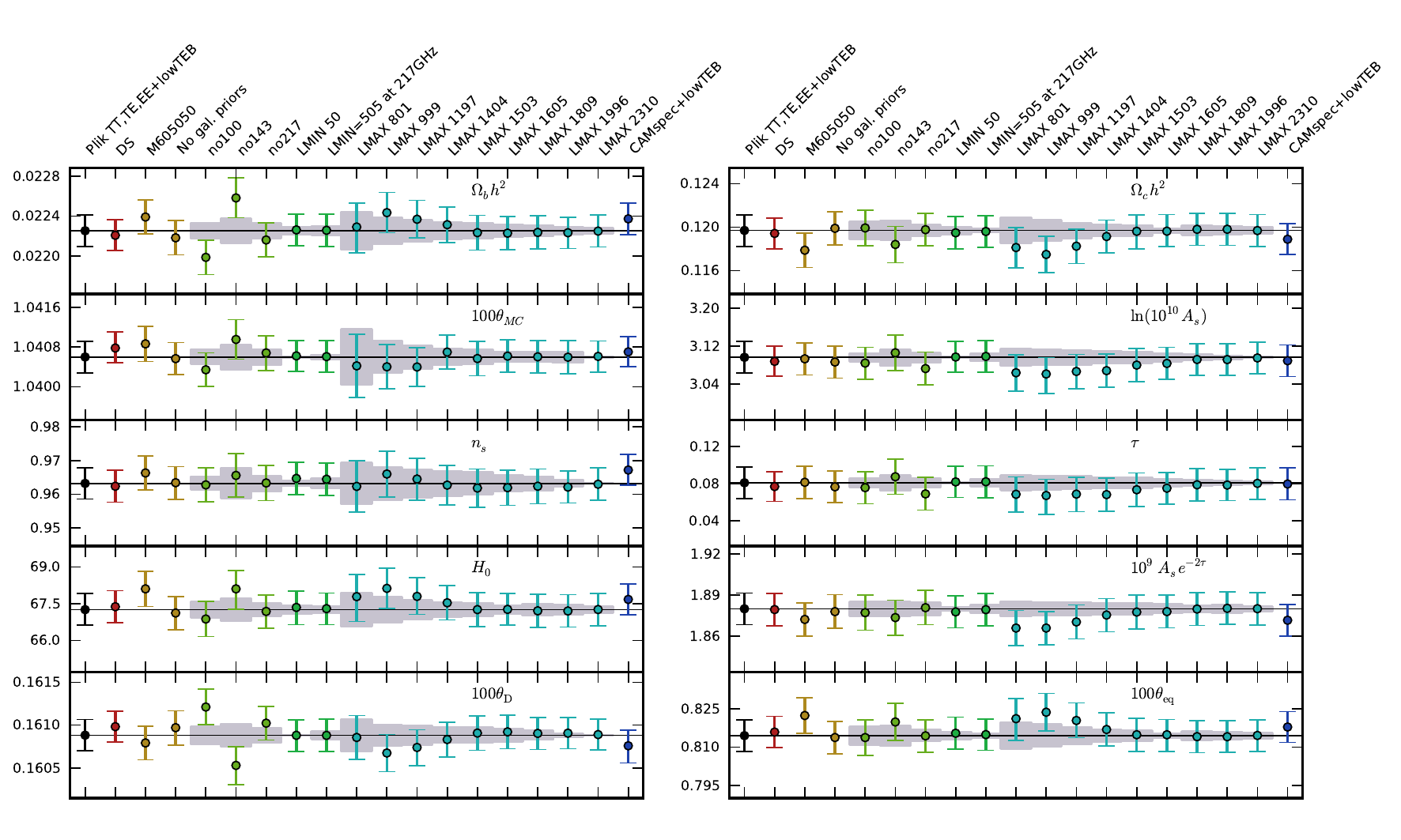}
\caption{Marginal mean and 68\,\% CL error bars on cosmological parameters estimated when adopting different data choices for the \plik likelihood, in comparison with results from alternate approaches or model.  We assume a  $\Lambda$CDM model and use variations of  \plikTTTEEE\ in combination with the \lowTEB\ likelihood for most of the cases. The ``\plikTTTEEE+\lowTEB'' case (black dot and thin horizontal black line) indicates the baseline (HM, $\ell_{\mathrm{min}}=30$, $\lmax=2508$), while the other cases are described in Appendix~\ref{sec:pol-robust}.  The grey bands show the standard deviation of the expected parameter shift, for those cases where the data used are a sub-sample of the baseline likelihood data (see Eq.~\ref{eq:covresult}). All the cases  are run with \PICO.}
\label{fig:wiskerALL}
\end{figure*}

The results are summarized in Fig.~\ref{fig:wiskerALL}, which shows the  marginal mean and the 68\,\% CL error bars for cosmological parameters using the \plikALL\ data combination  and  different assumptions about the data selection, foreground model,  or treatment of the systematics for the \plikTTTEEE\ likelihood. 
In the following, we comment, in turn, on each of the tests shown in this figure (from left to right).  The reference \plikALL\ results for the \LCDM\ model are denoted as ``\plikALL''.\ All the tests are run with the \PICO\ code.
 Most of the shifts we observe in the different cases are consequences of the shifts we observe in the $\TT$, $\TE$, and $\EE$  tests presented in Sect.~\ref{sec:hil-ass} and Appendix~\ref{sec:pol-robust}.

\paragraph{Detsets}
We find good agreement between the baseline cases that use half-mission spectra  and those that use detsets (case ``DS''). The greatest deviation is an upward $0.5\,\sigma$ shift in $\theta$, driven by the upward shifts observed in $\TE$ and $\EE$ in Fig.~\ref{fig:wiskerpol}.

\paragraph{Larger Galactic mask} 
We examined the impact of using larger Galactic masks,  G50, G41, and G41 (case ``M605050''), instead of the baseline ones at $100$, $143$, and $217$\,GHz, respectively (see also Sect.~\ref{sec:masks}).
We observe substantial shifts in the parameters, at the level of $\la 1\,\sigma$. These are mainly driven by the shifts already observed in the $\TE$ test in Fig.~\ref{fig:wiskerpol}.

\paragraph{Galactic dust priors}
We find that leaving the Galactic dust amplitudes completely free to vary (``No gal. priors'') in both temperature and polarization, without applying the priors described in Sect.~\ref{sec:dust}, does not have a significant impact on cosmological parameters.

\paragraph{Cutting out frequency channels}
We have considered the cases where we eliminate all the power-spectra related to one particular frequency at a time. We see shifts at the level of $1$--$2\,\sigma$ when eliminating all the power spectra containing the $100$ GHz (``no100'')  or the $143$ GHz (``no143'') maps. These reflect the analogous shifts observed in the $\EE$ and $\TE$ tests. 

\paragraph{Changing $\lmin$}
We find good stability in the results when changing the minimum multipole $\lmin$ considered in the analysis for the \plikTTTEEE\ likelihood (``LMIN'' case). The baseline likelihood has $\lmin=30$, and we test the cases of $\lmin=50$ and $\lmin=100$. For each frequency power spectrum in temperature or polarization we choose $\lmin ^{\mathrm{freq}}=\mathrm{max}(\lmin ,\lmin ^{\mathrm{ freq,\,base} })$,  with $\lmin ^{\mathrm{freq,\,  base}} $ being the baseline value of $\lmin$ at each frequency, as reported in Table~\ref{tab:highl:lrange}. For example, for the $217\times217$ spectrum in $\EE$ we use the baseline value $\lmin=505$ in all the cases we consider here.

\paragraph{Changing $\lmax$}

We observe shifts when including maximum multipoles between $\lmax\approx 800$ and $2500$ (``LMAX'' cases), consistent with the shifts already observed in $\TT$ in Fig.~\ref{fig:wiskerTT}.
For each frequency power spectrum in temperature or polarization we choose $\lmax ^{\mathrm{freq}}=\mathrm{min}(\lmax ,\lmax ^{\mathrm{ freq,\,base} })$,  with $\lmax ^{\mathrm{freq,\,  base}} $  being the baseline value of $\lmax$ at each frequency, as reported in Table~\ref{tab:highl:lrange}. 
\paragraph{Comparison to \camspec}
We find roughly $1\,\sigma$ shifts when comparing to the results obtained with the \camspec\ code. Furthermore, we note that our ``M605050'' case is in better agreement with \camspec.
This is due to the fact that while \camspec\ uses the same masks as \plik\ in temperature, it uses much larger masks in polarization, retaining about $50\,\%$ of the sky (before apodization). We showed in Sect.~\ref{sec:robust-fsky} and Appendix~\ref{sec:pol-robust} that using a larger Galactic mask does not affect the $\TT$ results, while it does change the $\TE$ and $\EE$ results. We therefore expect the results from the \camspec\ code in the $\TTTEEE$+\lowTEB\ case to be in better agreement with our ``M605050'' test.

\def\cltt{$C_\ell^{TT}$}
\def\clte{$C_\ell^{TE}$}
\def\cltb{$C_\ell^{TB}$}
\def\clee{$C_\ell^{EE}$}
\def\clbb{$C_\ell^{BB}$}
\def\cleb{$C_\ell^{EB}$}
\def\dlee{${\cal D}_\ell^{EE}$}
\def\dlbb{${\cal D}_\ell^{BB}$}
\def\dlte{${\cal D}_\ell^{TE}$}
\def\dltb{${\cal D}_\ell^{TB}$}
\def\dleb{${\cal D}_\ell^{EB}$}
\def\dltt{${\cal D}_\ell^{TT}$}

\subsection{Peaks and troughs in \Planck\ power spectra}
\label{ssub:peaks}

The power spectrum of CMB temperature anisotropies has been measured by \Planck\ to an exquisite level of precision, and the \cltt\ spectrum has now been joined by \clte\ and \clee.  This precision enables us to estimate the underlying cosmological parameters, but the $C_\ell$s are themselves a set of cosmological observables, whose properties can be described independently of any model.  The peaks and troughs in the power spectra reflect their origin in oscillating sound waves.  These features tell us that the Universe once contained a very hot, dense plasma, with the CMB anisotropies mostly originating from acoustic modes in the coupled photon-baryon fluid, driven by dark matter potential
perturbations \citep[e.g.,][]{HuSS1997}. The overall angular structure is determined by $\theta_\ast$, the ratio of the sound horizon to the last-scattering surface distance, the
statistical quantity that is best constrained by \Planck.  However, the positions of the {\it individual\/} peaks and troughs are now well determined in their own right, and this information has become part of the canon of facts now known about our Universe.

\begin{table}[ht!]
\begingroup
\newdimen\tblskip \tblskip=5pt
\caption{Positions and amplitudes of extrema in power spectra using \Planck\ data.}
\label{table_peaks_and_troughs}
\vskip -6mm
\footnotesize
\setbox\tablebox=\vbox{
 \newdimen\digitwidth
 \setbox0=\hbox{\rm 0}
 \digitwidth=\wd0
 \catcode`*=\active
 \def*{\kern\digitwidth}
 \newdimen\signwidth
 \setbox0=\hbox{+}
 \signwidth=\wd0
 \catcode`!=\active
 \def!{\kern\signwidth}
 \newdimen\pointwidth
 \setbox0=\hbox{\rm .}
 \pointwidth=\wd0
 \catcode`?=\active
 \def?{\kern\pointwidth}
 \halign{\hbox to 1.4in{#\leaderfil}\tabskip=2em&
 \hfil$#$\hfil&
 \hfil$#$\hfil\tabskip=0pt\cr
\noalign{\doubleline}
\omit&&\omit\hfil Height\hfil\cr
\omit\hfil Extremum\hfil&\omit\hfil Multipole\hfil& *[\mu{\rm K}^2]\cr
\noalign{\vskip 4pt\hrule\vskip 6pt}
\omit{\boldmath{$TT$}} \bf power spectrum\hfil\cr
\noalign{\vskip 5pt}
Peak 1&               *220.0\pm*0.5& !5717?*\pm35?*\cr
\hglue 1.0em Trough 1&   *415.5\pm*0.8& !1696?*\pm13?*\cr
Peak 2&              *537.5\pm*0.7& !2582?*\pm11?*\cr
\hglue 1.0em Trough 2&  *676.1\pm*0.8& !1787?*\pm12?*\cr
Peak 3&               *810.8\pm*0.7& !2523?*\pm10?*\cr
\hglue 1.0em Trough 3&  *997.7\pm*1.4& !1061?*\pm*5?*\cr
Peak 4&              1120.9\pm*1.0& !1237?*\pm*4?*\cr
\hglue 1.0em Trough 4&  1288.8\pm*1.6& !*737?*\pm*4?*\cr
Peak 5&               1444.2\pm*1.1& !*797.1\pm*3.1\cr
\hglue 1.0em Trough 5&   1621.2\pm*2.3& !*400?*\pm*4?*\cr
Peak 6&               1776?*\pm*5?*& !*377.4\pm*2.9\cr
\hglue 1.0em Trough 6&   1918?*\pm*7?*& !*245?*\pm*4?*\cr
Peak 7&             2081?*\pm25?*& !*214?*\pm*4?*\cr
\hglue 1.0em Trough 7& 2251?*\pm*8?*& !*119.5\pm*3.5\cr
Peak 8&              2395?*\pm24?*& !*105?*\pm*4?*\cr
\noalign{\vskip 10pt}
\omit{\boldmath{$TE$}} \bf power spectrum\hfil\cr
\noalign{\vskip 5pt}
Trough 1&             *150.0\pm*0.8& *-48.0\pm*0.8\cr
\hglue 1.0em Peak 1&     *308.5\pm*0.4& !115.9\pm*1.1\cr
Trough 2&            *471.2\pm*0.4& *-74.4\pm*0.8\cr
\hglue 1.0em Peak 2&    *595.3\pm*0.7& !*28.6\pm*1.1\cr
Trough 3&             *746.7\pm*0.6& -126.9\pm*1.1\cr
\hglue 1.0em Peak 3&     *916.9\pm*0.5& !*58.4\pm*1.0\cr
Trough 4&            1070.4\pm*1.0& *-78.0\pm*1.1\cr
\hglue 1.0em Peak 4&    1224?*\pm*1.0& !**0.7\pm*0.5\cr
Trough 5&             1371.7\pm*1.2& *-60.9\pm*1.1\cr
\hglue 1.0em Peak 5&     1536?*\pm*2.8& !**5.6\pm*1.3\cr
Trough 6&             1693.0\pm*3.3& *-27.6\pm*1.3\cr
\hglue 1.0em Peak 6&     1861?*\pm*4?*& !**1.2\pm*1.0\cr
\noalign{\vskip 10pt}
\omit{\boldmath{$EE$}} \bf power spectrum\hfil\cr
\noalign{\vskip 5pt}
Peak 1&               *137?*\pm*6?*& !*1.15*\pm*0.07*\cr
\hglue 1.0em Trough 1&   *197?*\pm*8?*& !*0.848\pm*0.034\cr
Peak 2&              *397.2\pm*0.5& !22.04*\pm*0.14*\cr
\hglue 1.0em Trough 2&  *525?*\pm*0.7& !*6.86*\pm*0.16*\cr
Peak 3&               *690.8\pm*0.6& !37.35*\pm*0.25*\cr
\hglue 1.0em Trough 3&   *832.8\pm*1.1& !12.5**\pm*0.4**\cr
Peak 4&              *992.1\pm*1.3& !41.8**\pm*0.5**\cr
\hglue 1.0em Trough 4&  1153.9\pm*2.7& !12.3**\pm*0.9**\cr
Peak 5&               1296?*\pm*4?*& !31.6**\pm*1.0**\cr
\noalign{\vskip 4pt\hrule\vskip 6pt}
}}
\endPlancktable
\endgroup
\end{table}

Here we use the \Planck\ data directly to fit for the multipoles of individual features in the measured $\TT$, $\TE$, and $\EE$ power spectra.  We specifically use the CMB-only band-powers described in this paper and available in the Planck Legacy Archive,\footnote{\url{http://pla.esac.esa.int/pla/}} adopting the same weighting scheme within each bin. Fitting for the positions and amplitudes of features in the band-powers is a topic with a long history, with approaches becoming more sophisticated as the fidelity of the data improved \citep[e.g.,][]{Scott94,Hancock97,Knox00,deBernardis02,bond03,page2003b,Benoit03,Durrer03,Readhead04,Jones06,hinshaw2007,Corasaniti08,Pryke09,Naess14}. Following earlier approaches, we fit Gaussian functions to the peaks and troughs in \cltt\ and \clee, but parabolas for the \clte\ features.  We have to remove a featureless damping tail (which we do by using an extreme amount of lensing to wash out the structure) to fit the higher-$\ell$ \cltt\ region and care has to be taken to treat the lowest-$\ell$ ``recombination'' peak in \clee. We do not try to fit the ``reionization'' bumps in the lowest few multipoles of \clte\ and \clee (even though these might technically be the
``first peaks'' in these power spectra). We explicitly focus on features in the conventional quantity ${\cal D}_\ell\equiv\ell(\ell+1)C_\ell/2\pi$; note that other quantities (e.g., $C_\ell$) have maxima and minima at slightly different multipoles, and that the selection of which band-powers to use for fitting each peak is somewhat subjective.

Our numerical values, presented in Table~\ref{table_peaks_and_troughs}, are consistent with previous estimates, but with a dramatically increased number of features measured.
\Planck\ detects 36 extrema in total, consisting of 19 peaks and 17 troughs. The eighth \cltt\ peak is only marginally detected using the released likelihood (although this cannot be seen from the values in the table, since the ``height'' includes the amplitude of the featureless spectrum).  However, by digging further into the \Planck\ data it would be possible to strengthen this detection and perhaps distinguish more features, since using the fact that the foreground power spectra have no structure on the relevant scales, one could be more liberal with foreground contamination for the purposes of feature detection.


\subsection{\textit{T}--\textit{E} correlations in \Planck\ power spectra}
\label{ssub:TE_corr}

The information contained in the primary CMB anisotropies comes from both
temperature and polarization.  It is well known that the $T$ and $E$-mode
fields are correlated, and hence one has to measure \clte, as well as \cltt\
and \clee, in order to extract all the information from $T$ and $E$
\citep[see, e.g.,][]{ZaldarriagaS1997,KamionkowskiKS1997,HuW1997}.

The \clte\ power spectrum has an amplitude that is roughly the geometric
mean of the \cltt\ and \clee\ power spectra and it oscillates in sign
as a function of $\ell$, depending on whether the fields are
correlated or anticorrelated.  Now that \Planck\ has provided high quality
measurements of all three power spectra over a wide range of multipoles, it
is possible to plot the strength of this correlation directly.  In other
words one can form the quantity
\begin{equation}
r_\ell \equiv
 \frac{C_\ell^{TE}}
 {\left({C_\ell^{TT}C_\ell^{EE}}\right)^{1/2}}
= \frac{{\cal D}_\ell^{TE}}
 {\left({{\cal D}_\ell^{TT}{\cal D}_\ell^{EE}}\right)^{1/2}},
\label{eq:Pearson}
\end{equation}
which is the Pearson correlation coefficient for $T$ and $E$ in harmonic space.

\begin{figure}[htpb]
\begin{center}
  \includegraphics[width=\columnwidth]{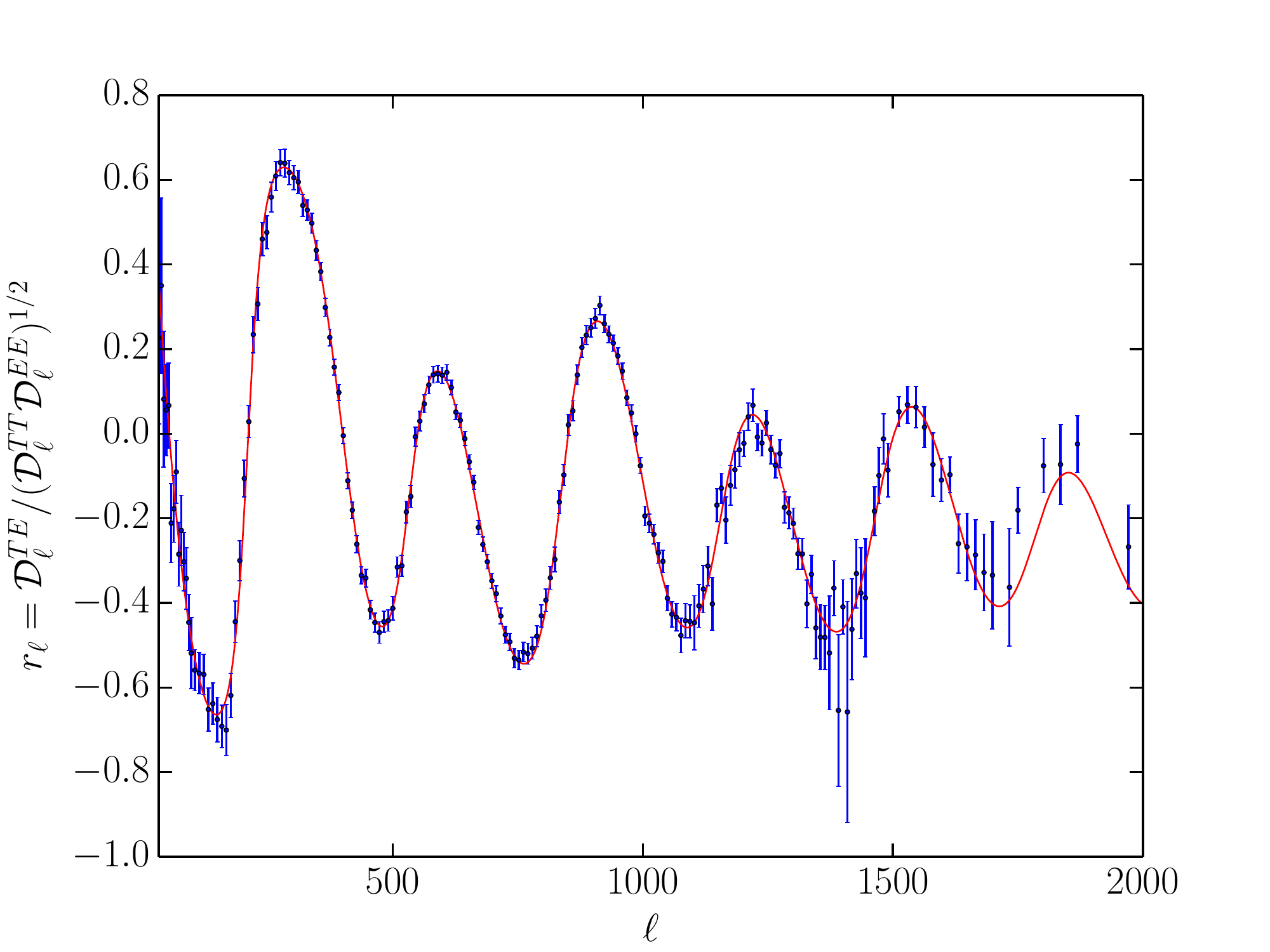}
\caption{\label{fig:Pearson} Pearson correlation coefficient for
$T$ and $E$.  The red curve is the theoretical best-fit to the
\planckall\ data.  The blue points are the binned \Planck\ data
described in this paper.  Data points with negative values for
$\EE$, or low S/N in polarization (specifically ${\rm S/N}<1$)
are omitted, which leads to an apparent bias between the theory curve
and the data at the highest multipoles.
The error bars here are calculated assuming a Gaussian distribution for the
$\TT$, $\TE$, and $\EE$ power values, which should be a reasonable approximation
at high $\ell$.  The Pearson coefficient is also normally
distributed at high $\ell$.}
\end{center}
\end{figure}

\begin{figure}[htpb]
\begin{center}
  \includegraphics[width=\columnwidth]{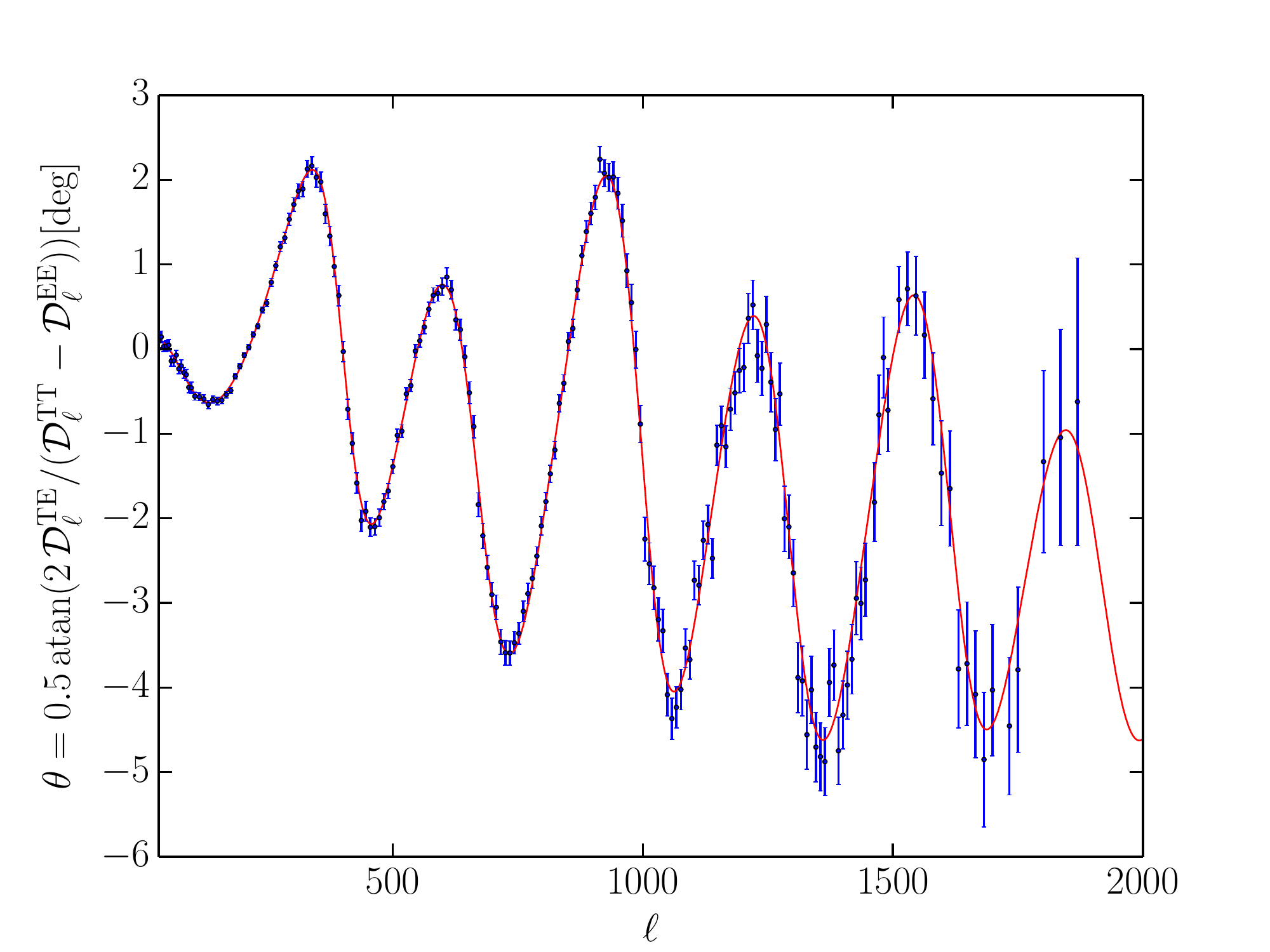}
\caption{\label{fig:theta} Decorrelation angle for $T$ and $E$.  This is the
angle through which we have to rotate in $T$--$E$ space in order to make
two uncorrelated quantities.}
\end{center}
\end{figure}

We plot this in Fig.~\ref{fig:Pearson}, using the binned \dltt, \dlte, and
\dlee\ values from the \Planck\ 2015 data release.  As expected the data fit
the theoretical expectation (except for a slight bias at high multipoles,
since we have to remove $\ell$ bins for which the $\EE$ power is measured to
be quite small).

For the best-fit cosmology one can see that the maximum correlation is at
$\ell\approx300$ (actually $\ell=282$), where $T$ and $E$ are 63\,\% correlated.  The most anti-correlated scale is at $\ell\approx150$ (actually $\ell=148$), where $r=-0.66$. When first observed \citep{kogut2003} the anti-correlation at relatively large scales, i.e., $r<0$ for $50\la\ell\la250$, was seen as a confirmation of the adiabatic nature of the density perturbations on super-Hubble scales \citep{peiris2003}.
The oscillatory behaviour of all three power spectra of course confirms adiabaticity much more dramatically.  But more quantitatively, looking at
the $r_\ell$ plot enables us to directly gauge the strength of the $T$--$E$
correlation as a function of scale.  The fact that the power spectrum is
mostly negative (i.e., the troughs are typically deeper than the heights
of the peaks) is the usual ``baryon drag'' effect \citep[e.g.,][]{HuSS1997}.

Because $-1\leq r\leq 1$, the Pearson correlation coefficient is sometimes
interpreted geometrically as the cosine
of a correlation angle.  Fig.~\ref{fig:Pearson} could hence be plotted with
this angle on the vertical axis, but such a plot would
carry no additional information.  Nevertheless, there is an alternative
geometrical interpretation of the $T$--$E$ correlation that is also worth
examining.  This involves considering
the angle through which one has to rotate in $T$--$E$ space in order to
decorrelate the two quantities.  It is equivalent to a principal component
analysis for two correlated variables, or the Jacobi rotation of a matrix
with the rotation angle given by
\begin{equation}
\tan2\theta = \frac{2{\cal D}_\ell^{TE}}{{\cal D}_\ell^{TT}-{\cal D}_\ell^{EE}}.
\label{eq:theta}
\end{equation}

In Fig.~\ref{fig:theta} we plot the quantity defined in Eq.~(\ref{eq:theta}).
It can be regarded as the angle through which we have to rotate $E$ and $T$
to make the transformed quantity $C_\ell^{T^\prime E^\prime}=0$.  Or
alternatively, it can be thought of in terms of $T$ ``leaking'' into $E$.
Since generally $C_\ell^{TT}\gg C_\ell^{EE}$,
this angle is quite small.  We can see from the figure that the
required angle oscillates, becoming as high as $2^\circ$ and as low as almost
$-5^\circ$; actually the greatest angle required is for $\ell=1992$, where
$\theta=-4\fdg63$.  Again we see that for the best-fit cosmology the plot
is mostly negative.  Dimensionless quantities such as those plotted in
Figs.~\ref{fig:Pearson} and \ref{fig:theta} may have additional value in
being independent of the overall normalization of the power spectra
(\ie $A_{\rm s}$) or the calibration of the data.


\begin{figure*}[htb] 
\vskip 6mm
\begin{center}
  \includegraphics[width=\columnwidth]{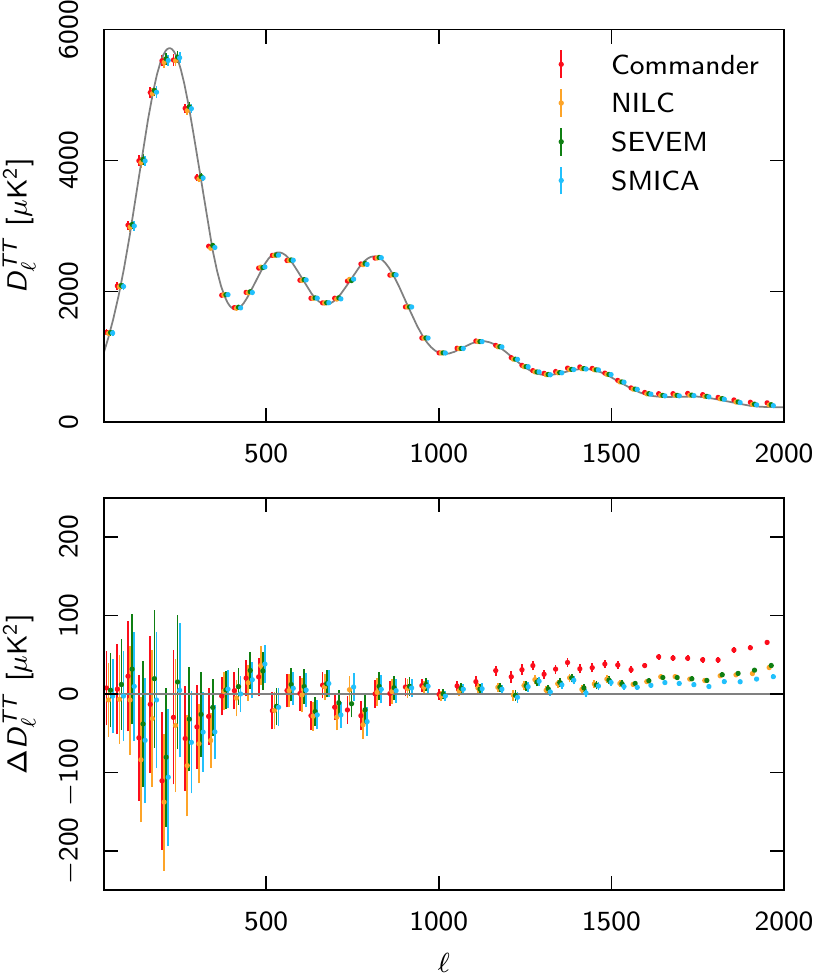}
  \includegraphics[width=\columnwidth]{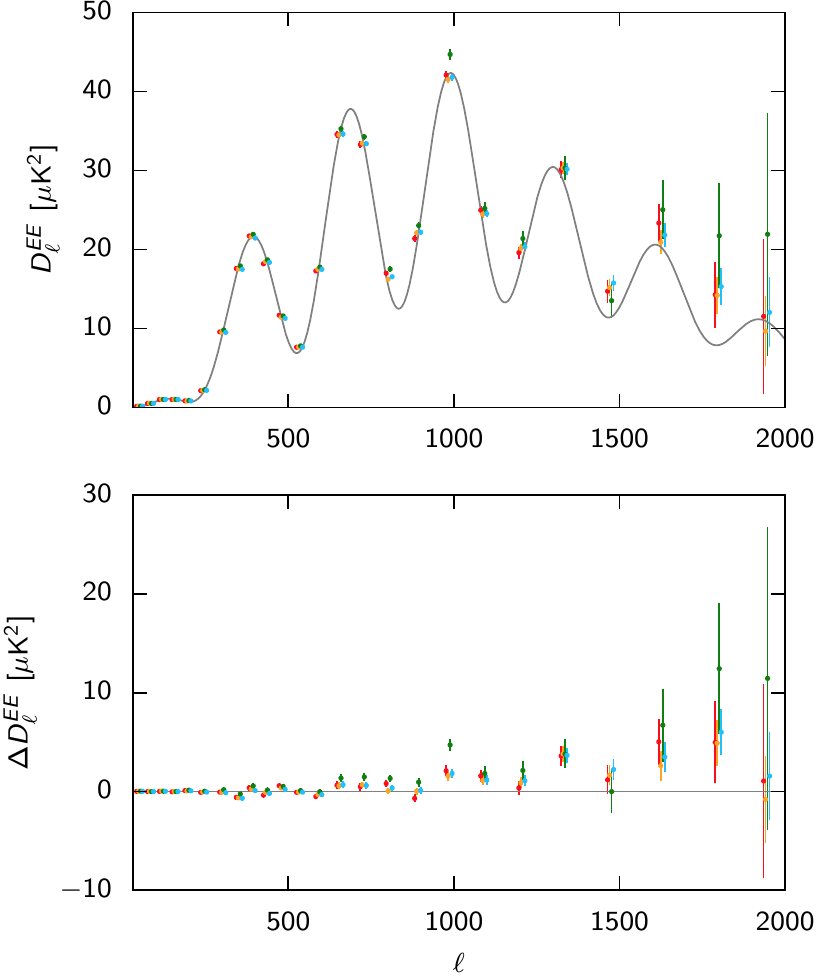}
\end{center}
\caption{CMB $\TT$ ({left}) and $\EE$ ({right}) power spectra for each of the four foreground-cleaned CMB maps. Upper panels show raw band-powers; the grey lines show the best-fit \LCDM\ model from the \Planck\ 2015 likelihood. Lower panels show residual  band-powers after subtracting the best-fit \LCDM\ model, showing the residual extragalactic foreground contribution.}
\label{fig:cl_CMBmaps_dx11_spectra}
\end{figure*}

\begin{figure*}[htb] 
\vskip 6mm
\begin{center}
  \includegraphics[width=\columnwidth]{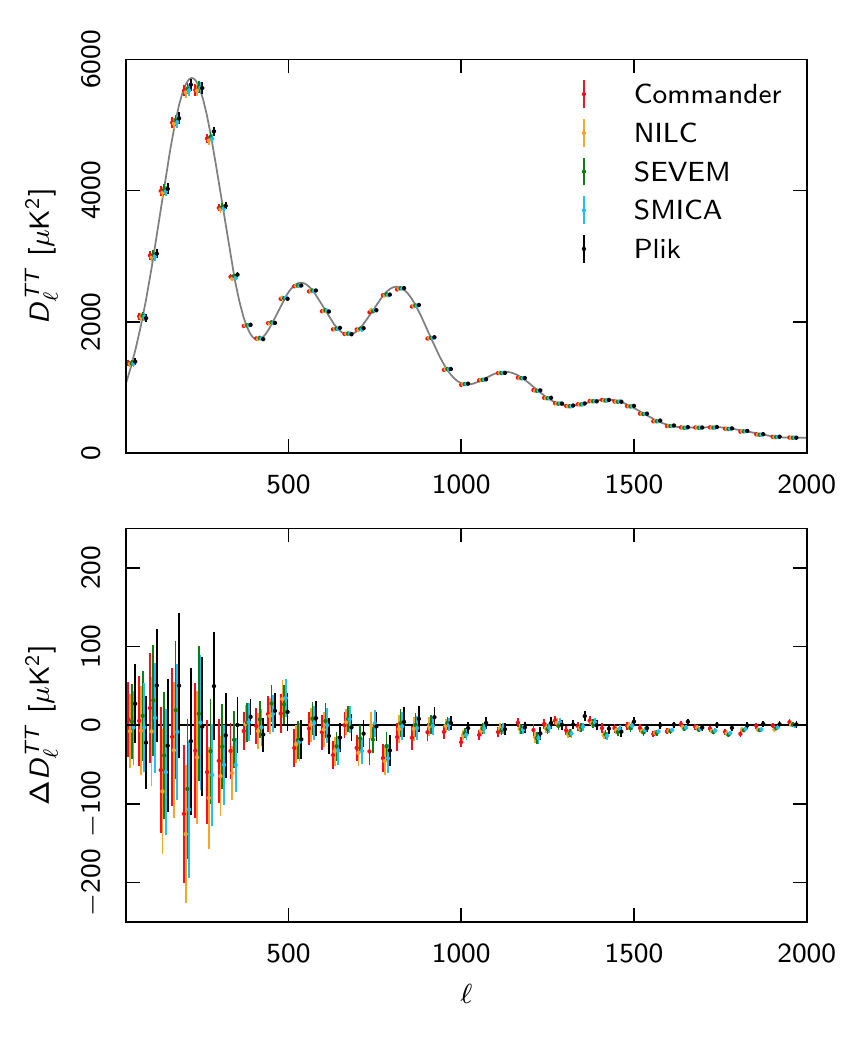}
  \includegraphics[width=\columnwidth]{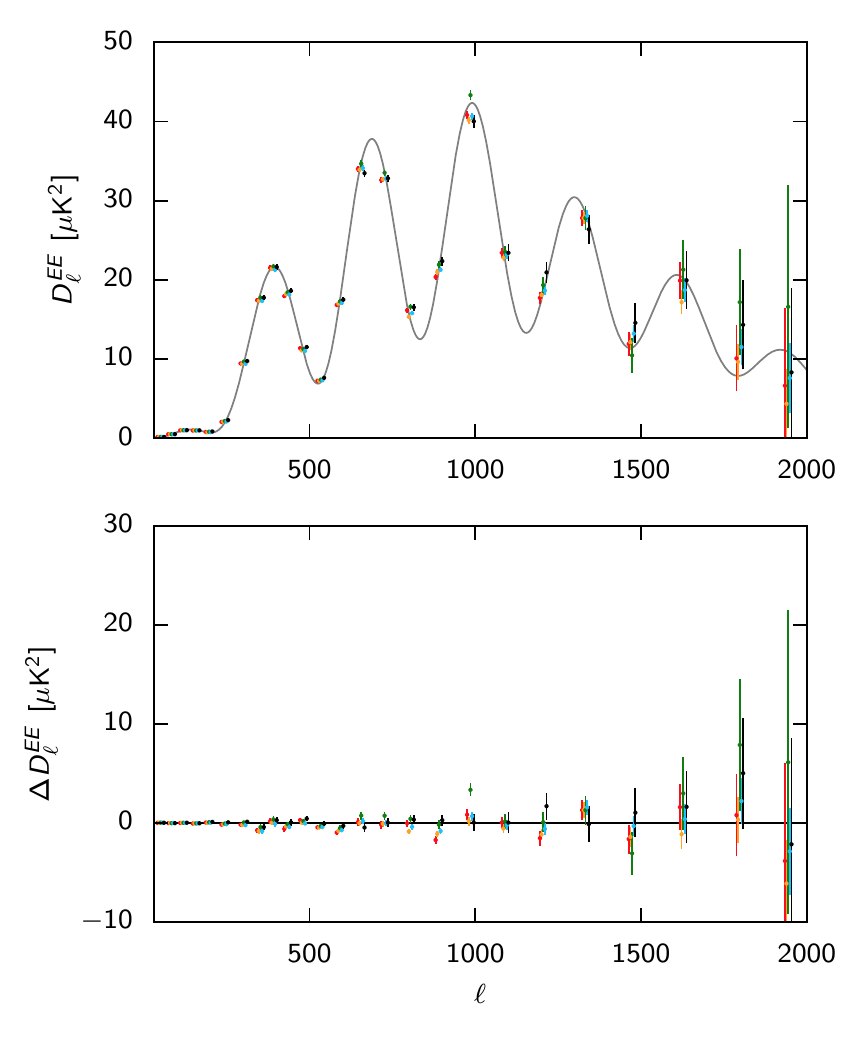}
\end{center}
\caption{CMB $\TT$ ({left}) and $\EE$ ({right}) power spectra for each of the four foreground-cleaned CMB maps, as in Fig.~\ref{fig:cl_CMBmaps_dx11_spectra}, {after} an extra cleaning out the extragalactic residuals at the spectrum level, along with the \plik\ results for comparison. Upper panels show band powers; the grey lines show the best-fit \LCDM\ model from the \Planck\ 2015 likelihood. Lower panels show residual  band powers after subtracting the best-fit \LCDM\ model. }
\label{fig:cl_CMBmaps_dx11_spectra_Xgalcleaned}
\end{figure*}

\subsection{Analysis of CMB maps derived by component-separation methods\label{sec:mapCheck}}

\begin{figure*}[htb] 
\begin{center}
  \includegraphics[width=0.9\textwidth]{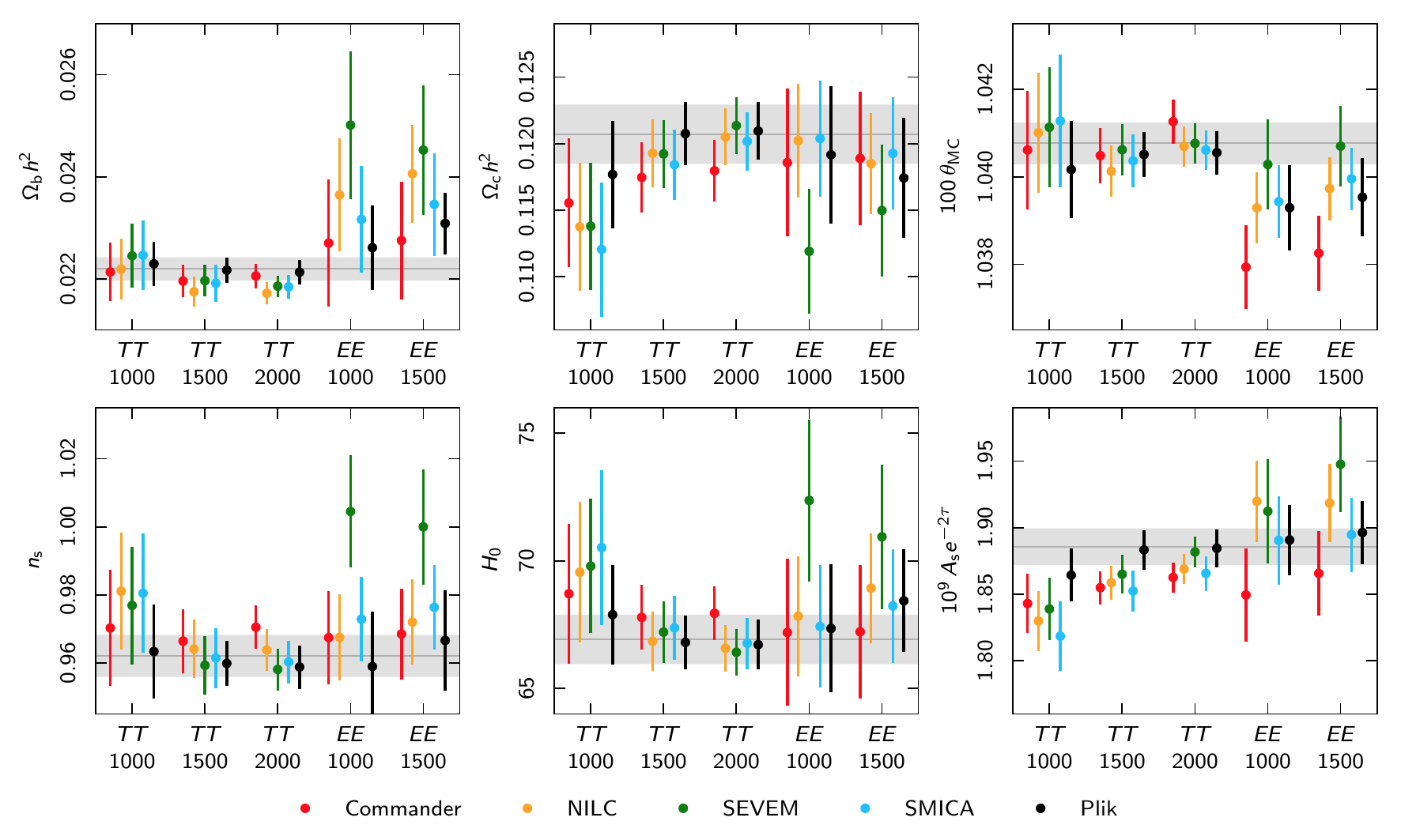}
\end{center}
\caption{Comparison of cosmological parameters estimated from the $\TT$ and $\EE$ spectra computed from the CMB maps and further accounting for foreground residuals through method-tailored templates derived from the FFP8 simulations. The sets of points correspond to various values of $\ell_{\mathrm{max}}$, and lets the four methods and \plik\ to be compared when the same scales are used. 
For comparison, we also show the corresponding parameters obtained with the \Planck\ 2015
likelihood including multipoles up to $\ell_{\mathrm{max}} = 2500$ as the horizontal line surrounded by a grey band giving the uncertainties.}
\label{fig:dx11_params_TT_EE}
\end{figure*}

\begin{table*}[ht!]
\begingroup
\newdimen\tblskip \tblskip=5pt
\caption{Comparison between parameters based on various CMB temperature maps.$^{\rm a}$}
\label{tab:param_cmb_compare2}
\nointerlineskip
\vskip -3mm
\footnotesize
\setbox\tablebox=\vbox{
   \newdimen\digitwidth
   \setbox0=\hbox{\rm 0}
   \digitwidth=\wd0
   \catcode`*=\active
   \def*{\kern\digitwidth}
   \newdimen\signwidth
   \setbox0=\hbox{+}
   \signwidth=\wd0
   \catcode`!=\active
   \def!{\kern\signwidth}
\halign{\hbox to 1.0in{$#$\leaderfil}\tabskip 1.2em&
\hfil$#$\hfil&
\hfil$#$\hfil&
\hfil$#$\hfil&
\hfil$#$\hfil\tabskip 0pt\cr
\noalign{\doubleline}
\omit\hfil Parameter\hfil& \omit\hfil \texttt{Commander}\hfil& \omit\hfil \texttt{NILC}\hfil& \omit\hfil \texttt{SEVEM}\hfil& \omit\hfil \texttt{SMICA}\hfil\cr
\noalign{\vskip 3pt\hrule\vskip 5pt}
\Omega_{\rm b}h^2&        0.02207 \pm 0.00025& 0.02172 \pm 0.00023& 0.02185 \pm 0.00023& 0.02184 \pm 0.00024\cr
\Omega_{\rm c}h^2&        0.1178 \pm 0.0024*&  0.1205  \pm 0.0023*& 0.1214 \pm 0.0023&   0.1202 \pm 0.0023\cr
100\theta_{\rm MC}& 1.041 \pm 0.0005*&   1.041 \pm 0.0005*&   1.040 \pm 0.0005&    1.041 \pm 0.0005\cr
\tau&               0.075 \pm 0.019**&   0.069 \pm 0.018**&   0.069 \pm 0.018&     0.068 \pm 0.019\cr
10^9\As e^{-2\tau}& 1.862 \pm 0.011**&   1.868 \pm 0.011**&   1.882 \pm 0.012&     1.866 \pm 0.015\cr
n_{\rm s}&                0.971 \pm 0.007**&   0.964 \pm 0.007**&   0.958 \pm 0.007&     0.960 \pm 0.007\cr
\noalign{\vskip 8pt}
\Omega_{\rm m}&           0.304 \pm 0.014&     0.322 \pm 0.014&     0.327 \pm 0.014&     0.321 \pm 0.014\cr
H_0&                68.0 \pm 1.1&        66.6 \pm 1.0&        66.4 \pm 1.0&        66.8 \pm 1.0\cr
\noalign{\vskip 5pt\hrule\vskip 3pt}}}
\endPlancktablewide 
\tablenote {{\rm a}} These constraints are obtained from $\TT$ in the range $50<\ell<2000$ combined with a prior of $\tau=0.07\pm0.02$, allowing for residual foreground contributions (see text).\par
\endgroup
\end{table*}

The high-$\ell$ likelihoods considered in this paper perform component separation
at the power-spectrum level, to fully exploit the signal at the smallest scales probed by \Planck\ and to enable full error propagation. In this section, we describe the steps involved in the alternative approach of deriving CMB spectra and cosmological parameters from CMB maps obtained by component-separation techniques. These cleaning techniques, referred to as \texttt{Commander}, \texttt{SMICA}, \texttt{NILC}, and \texttt{SEVEM}, are described in \citet{planck2014-a11}. The maps are weighted with the union of the confidence masks of the different component-separation methods, applying the \texttt{UT78} mask in temperature and the \texttt{UP78} mask in polarization \citep{planck2014-a11}, with a further cosine apodization of $10'$.

In order to look at relatively small differences, we compare the angular power spectra of the four CMB maps and cosmological parameters with those from \plik. To estimate the power spectra we use the \texttt{XFaster} method, an approximation to the iterative, maximum likelihood, quadratic band-power estimator based on a diagonal approximation to the quadratic Fisher matrix estimator \citep{rocha2009,rocha2010b}. The noise
bias is estimated using difference maps, as described in \citet{planck2014-a11}.

We estimate the power spectra of the half-mission half-sum (HMHS) CMB maps. The HMHS spectra contain signal and noise. The noise bias is estimated from the half-mission half-difference (HMHD) data. The HMHD spectra contain only noise and potential systematic
effects. The resulting spectra are shown in Fig.\,\ref{fig:cl_CMBmaps_dx11_spectra}. The top panels compare each of the four power spectra derived from the component-separation maps with the best-fit \LCDM\ power spectrum derived from the \Planck\ likelihood including multipoles up to $\ell = 2500$, with no subtraction of extragalactic foregrounds. The bottom panels show the spectrum differences between the component-separation methods and the best-fit spectrum, showing the residual extragalactic foreground contribution. We note that the \smica map appears to be the least contaminated by foreground residuals at small scales. Still, it is obvious that the foreground residuals are too strong for a direct cosmological parameter analysis. 

We therefore proceed, as in the baseline likelihoods, with component separation at the power-spectrum level (albeit with much smaller non-CMB contributions). One must  nevertheless obtain templates for these residual foregrounds after component separation. To that effect, we propagate the {simulated} full-sky FFP8 foreground maps through the respective pipelines and estimate the resulting power spectra normalized to some pivotal multipole, under the hypothesis that the simulated sky is close enough to the real one for these residual templates to be accurate. \rev{We have indications that this is only marginally the case, and have used a dedicated foreground model for the validation of the \plik\ likelihood on the FFP8 simulations. This can be the source of the variation under $\ell_{\textrm{max}}$ of some of the parameters obtained in this test.}

We then estimate cosmological parameters using a Gaussian correlated likelihood derived from these \texttt{XFaster}  bandpowers, coupled to \texttt{CosmoMC} \citep{cosmomc}.  Specifically, we include multipoles between $\ell_{\textrm{min}} = 50$ and $\ell_{\textrm{max}}$, where $\ell_{\textrm{max}} = 1000$, 1500, or 2000 for temperature, and 1000 or 1500 for polarization. We explore the base six-parameter \LCDM\ model, and, since low-$\ell$ data are not used in the likelihood, impose an informative Gaussian prior of $\tau = 0.070 \pm 0.006$. 

The resulting spectra are shown in Fig.~\ref{fig:cl_CMBmaps_dx11_spectra_Xgalcleaned}, and the corresponding cosmological parameters are summarized in Fig.~\ref{fig:dx11_params_TT_EE} for both $\TT$ (filled symbols) and $\EE$ (unfilled symbols). The comparison of Fig.~\ref{fig:cl_CMBmaps_dx11_spectra} with Fig.~\ref{fig:cl_CMBmaps_dx11_spectra_Xgalcleaned} enables assessment of the efficiency of this extra cleaning step, which brings the component-separation-based spectra into much greater agreement with those derived from the high-$\ell$ likelihood. To be more quantitative, we now turn to the corresponding cosmological parameters in Fig.~\ref{fig:dx11_params_TT_EE}. Starting with the temperature cases, we first observe reasonable overall internal agreement between the four component-separation methods, with almost all differences smaller than $1\,\sigma$ within each multipole band. Second, we also observe acceptable agreement with the best-fit \Planck\ 2015 \LCDM\ model derived from the likelihood, since most of the differences are within 1$\,\sigma$, at least for $\lmax=2000$. One notable exception to this agreement is the power spectrum amplitude, $A_{\rm s} e^{-2\tau}$, which is systematically low by about $2\,\sigma$ for $\ell_{\textrm{max}} = 1000$ for all methods.  We further note that otherwise all approaches find similar shifts of some parameters ($\omc$) between $\lmax =1000$ and $\lmax =1400$. Table~\ref{tab:param_cmb_compare2} provides the numerical $\TT$ constraints at $\ell < 2000$.

\rev{One should keep in mind that the CMB cleaned maps used here have been obtained with a different objective than providing the best/safest estimation of the cosmological parameters. Rather, they are meant to maximize the sky coverage, in particular for non-Gaussianity analyses; this entails a very different set of trade-offs. To reach this goal, the component separation methods use the \LFI channels to clean regions close to the Galactic plane, where foreground sources other than dust need to be accounted for. The efficiency of the cleaning is dominated by the residual in the Galactic plane, where contamination is strong, possibly at the expense of regions far from the plane, where the residual compared to the CMB can then be higher than in a dedicated solution. The separation methods also require the use of a common mask across all frequencies, contrary to our spectrum-based approach where we can tailor the masks per frequency in order to minimize foreground contamination as well as the size and number of the point source holes. In the end, the channel weights in the component-separated CMB map is different from the weighting in \plik, with 143\,GHz dominating the CMB map at the first peak, whereas, thanks to the use of a higher sky fraction at this frequency, the $100\times100$ dominates the \plik\ spectrum (see Figure~\ref{fig:mixing}. Further, the use of a larger point-source mask increases the excess scatter seen in Fig.~\ref{app:hil_pts_mask_correction}, which is not taken into account in  \texttt{XFaster}. The propagation of instrumental uncertainties, like calibration and beam errors, are also handled very differently, if at all. Given all those caveats, it is remarkable to nevertheless find such a general agreement between the different component-separated maps created using the \texttt{XFaster} likelihood, between themselves and with the \plik\ likelihood.}

The $\EE$-based results are generally more discrepant, apart from \smica. For instance \texttt{Commander} gives a quite significantly different value for $\theta_{\rm MC}$, \texttt{NILC} differs in $\Omb$, and \texttt{SEVEM} in $\Omb$, $\ns$, $H_0$, and $\As e^{-2\tau}$.

\subsection{Profile likelihood}\label{ssub:profile}


We have additionally made a comparison of the cosmological parameters determined using a ``profile likelihood'' method as in \citet{planck2013-XVI}, along with an independent Boltzmann code, \texttt{CLASS} \citep{2011arXiv1104.2932L}. The results of this analysis for the \Planck\ \lowTEB\  likelihood on the \lcdm\ model are shown in the first column of Table~\ref{tab:tableprofile}, which summarizes the values of the parameters and their errors at 68\,\% CL for both \plik and \camspec. This table should be compared with the first column of Table~4 of \citet{planck2014-a15}, which shows the \Planck\ 2015 baseline MCMC parameter results; there is excellent agreement with the profile-likelihood fits, the small difference observed for the mean value of $\theta_{\mathrm{MC}}$ being attributed to a slightly different definition of this parameter in \texttt{CLASS} with respect to \texttt{CAMB}.

\begin{table*}[ht!] 
\begingroup
\caption{Profile likelihood estimates of cosmological parameters within the \lcdm\ model for \plik and \camspec.}
\label{tab:tableprofile}
\openup 2pt
\newdimen\tblskip \tblskip=5pt
\nointerlineskip
\vskip -3mm
\footnotesize
\setbox\tablebox=\vbox{
    \newdimen\digitwidth
    \setbox0=\hbox{\rm 0}
    \digitwidth=\wd0
    \catcode`"=\active
    \def"{\kern\digitwidth}
    \newdimen\signwidth
    \setbox0=\hbox{+}
    \signwidth=\wd0
    \catcode`!=\active
    \def!{\kern\signwidth}
\halign{
\hbox to 1.0in{#\leaderfil}\tabskip=2em&
	\hfil#\hfil\tabskip=2.5em&
	\hfil#\hfil\tabskip=2.0em&	
	\hfil#\hfil\tabskip=0.8em&	
    \hfil#\hfil\tabskip=0pt\cr
\noalign{\doubleline}
\omit&&\multispan3\hfil\sc Error [68\,\% CL]\hfil\cr
\noalign{\vskip -3pt}
\omit&&\multispan3\hrulefill\cr
\omit&&&Sensitivity-&Foreground and\cr
\omit\hfil\sc Parameter\hfil&\sc Estimate&Full&related&instrumental\cr
\noalign{\vskip 4pt\hrule\vskip 5pt}
\omit\bf  \plik parameters\hfil\cr
\noalign{\vskip 4pt}
\hglue 1.1em$\Omega_{\mathrm{b}} h^2$&   0.02227& 0.00023& 0.00019& 0.00014\cr
\hglue 1.1em$\Omega_{\mathrm{c}} h^2$&   0.1198&  0.0022&  0.0021&  0.0007\cr
\hglue 1.1em$100\theta_{\mathrm{MC}}$&   1.04184& 0.00044& 0.00044& 0.00003\cr
\hglue 1.1em$\tau$&                      0.082&   0.020&   0.018&   0.009\cr
\hglue 1.1em$\ln(10^{10} A_\mathrm{s})$& 3.098&   0.037&   0.034&   0.014\cr
\hglue 1.1em$n_\mathrm{s}$ &             0.9663&  0.0063&  0.0051&  0.0036\cr
\noalign{\vskip 8pt}
\omit\bf  \camspec parameters\hfil\cr
\noalign{\vskip 4pt}
\hglue 1.1em$\Omega_{\mathrm{b}} h^2$&   0.02229& 0.00023& 0.00018& 0.00014\cr
\hglue 1.1em$\Omega_{\mathrm{c}} h^2$&   0.1194&  0.0022&  0.0021&  0.0006\cr
\hglue 1.1em$100\theta_{\mathrm{MC}}$&  1.04102& 0.00047& 0.00047& 0.00009\cr
\hglue 1.1em$\tau$&                      0.080&   0.020&   0.018&   0.009\cr
\hglue 1.1em$\ln(10^{10} A_\mathrm{s})$& 3.091&   0.037&   0.034&   0.016\cr
\hglue 1.1em$n_\mathrm{s}$&              0.9685&  0.0062&  0.0051&  0.0036\cr
\noalign{\vskip 5pt\hrule\vskip 3pt}
}}
\endPlancktablewide
\endgroup
\end{table*}

\begin{figure}[htbp]
  \includegraphics[angle=0,width=0.4\textwidth]{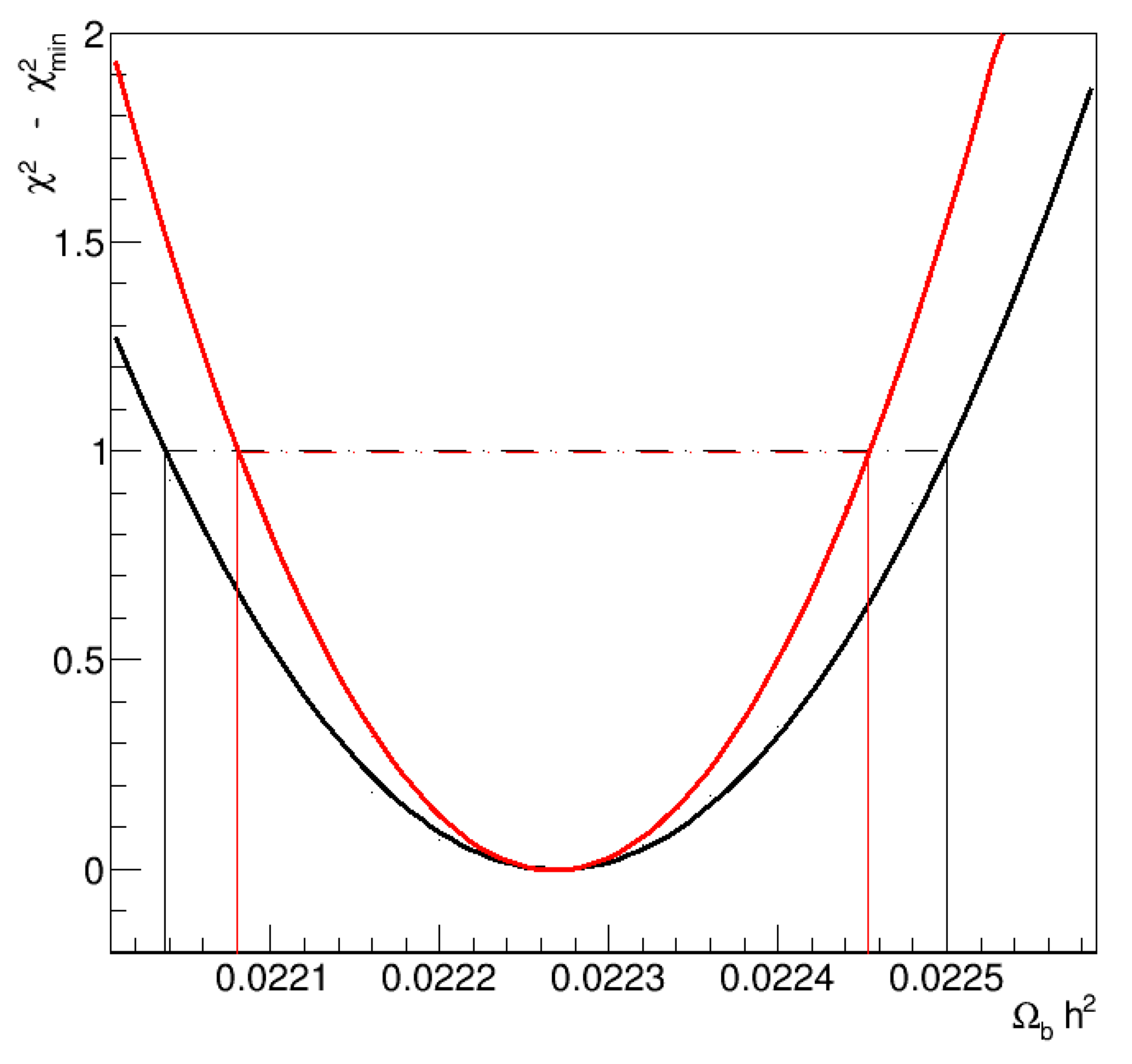}
  \caption{Likelihood (specifically $\chi^2-\chi^2_{\rm min}$) profile of $\Omega_{\mathrm{b}} h^2$ within the \lcdm\ model for \plik. In black is shown the full profile-likelihood fit (to derive the full error of Table~\ref{tab:tableprofile}) and in red the profile-likelihood fit when the nuisance are fixed to the values obtained when maximizing the likelihood function (to derive the ``sensitivity related'' error of Table~\ref{tab:tableprofile}).}
  \label{fig:profileOb}
\end{figure}

We have also estimated the contribution of the fit of the nuisance (foreground and instrumental) parameters to the error on the cosmological parameters; this analysis permits us to assess how much our lack of knowledge of the nuisance parameters impacts the cosmological error budget. First, a profile-likelihood fit is performed to estimate the full error on each parameter, giving, for instance, the black curve of Fig.~\ref{fig:profileOb}; this corresponds to column~3 in  Table~\ref{tab:tableprofile}. In a second step, another profile likelihood is built, fixing the nuisance parameters to their best-fit values; from this we obtain the ``sensitivity-related'' uncertainty (the red curve of Fig.~\ref{fig:profileOb}). 
This corresponds to the ultimate error one would obtain if we knew the nuisance parameters perfectly (and they had these best-fit values). Finally the ``foreground and instrumental'' error is deduced by quadratically subtracting the ``sensitivity-related'' uncertainty from the total error; this procedure has been used, for instance, in \citet{PhysRevD.90.052004}.
The results are shown in columns 4 and 5  of Table~\ref{tab:tableprofile} for \plik and \camspec, and are very similar for both likelihoods. $\Omega_{\rm b}$ and $n_{\rm s}$ are more strongly impacted by the fit of the nuisance parameters, while $100\theta_{\mathrm{MC}}$ is less so, which is consistent with the correlation matrix of the parameters.

\bigskip{}
\clearpage


\raggedright
\end{document}